\documentclass[a4paper,12pt,titlepage]{report}
\usepackage[round,sort]{natbib} 
\usepackage{graphics} 
\usepackage{graphicx} 
\usepackage{latexsym}  
\usepackage{longtable} 
 \usepackage{multirow}      
\usepackage{supertabular,lscape}  
\usepackage{rotating}        
\usepackage{fancyhdr} 
\usepackage[plain]{fancyref} 
\usepackage[sumlimits]{amsmath} 
\usepackage{amssymb} 
\usepackage{amstext} 
\usepackage{rotating} 
\usepackage{url}

\makeatletter
\def\url@leostyle{%
  \@ifundefined{selectfont}{\def\UrlFont{\sf}}{\def\UrlFont{\small\ttfamily}}}
\makeatother
\usepackage{doublespace}
\newcommand{\chapquote}[2]
   {\thispagestyle{empty}
        \vspace*{\fill}
        \begin{center}#1\\{\em #2}%
           \end{center}
        \vspace*{\fill}

}

\begin{document}
\begin{titlepage}
\pagestyle{plain}
\begin{center}
\vspace*{-20mm}
\LARGE{\sc Microlensing and Variability in the Bulge of M31}\\
\vspace*{50mm}
\LARGE{Jonathan Paul Duke} \\
\Large{Astrophysics Research Institute, January 11th 2010} \\
\vspace{10.0cm}
\normalsize{A thesis submitted in partial fulfilment of the requirements of\\
           Liverpool John Moores University\\
           for the degree of\\
           Doctor of Philosophy.\\}
\end{center}
\end{titlepage}

\pagestyle{plain}
\pagenumbering{roman}
\setcounter{page}{2}
\vspace*{\fill}
\begin{flushright}
        \begin{tabular}{p{0.8\hsize}}
        {\footnotesize {The copyright of this thesis rests with the author.
 No quotation from it should be published without his prior written 
consent and information derived from it should be acknowledged \copyright \ J. P. Duke 2010}}
        \end{tabular}
        \end{flushright}

\newpage
\chapter*{Declaration}
\addcontentsline{toc}{chapter}{Declaration}
The work presented in this thesis was carried out in the 
Astrophysics Research Institute, Liverpool John Moores University. 
Unless otherwise stated, it is the original work of the author.

While registered as a candidate for the degree of Doctor of Philosophy,
for which submission is now made, the author has not been registered as a
candidate for any other award. This thesis has not been submitted in
whole, or in part, for any other degree.

\vfill
\begin{center}
J. P. Duke \\
Astrophysics Research Institute \\
Liverpool John Moores University \\
Twelve Quays House \\
Egerton Wharf\\
Birkenhead\\
CH41 1LD \\
U.K. \\
\vspace*{5mm}
January 11th 2010
\end{center}
\newpage
        \vspace{4cm}
        \begin{center}
            {\LARGE\sc Microlensing and Variability in the Bulge of M31}\\
            \vspace{0.5cm}      
            {\Large\bf{\sc J. P. Duke} } \\
             \vspace{0.5cm}
            {\large Submitted for the Degree of Doctor of Philosophy}\\
            \vspace{0.15cm}
            {\large \sc Astrophysics Research Institute \\}
             \vspace{0.15cm}
         \end{center}
               \hspace{5.35cm} January 11th 2010\newline   
        \begin{center}
            \vspace{0.5cm}
            {\LARGE{{\bf Abstract}}\par}
            \vfil
        \end{center}

\addcontentsline{toc}{chapter}{Abstract}
{For the past five seasons, the Angstrom Project, an international microlensing collaboration,
 has been making observations of the central bulge of M31, the Andromeda galaxy, searching for
 microlensing events. This thesis describes the work that has been done to develop an automatic
 candidate selection pipeline which enables lensing candidates to be found even if they are 
blended with a periodic variable baseline, something which has never been attempted before in
 the same way.
As a by-product of this process, many variable stars are found and their properties
are investigated and characterised.
The results of the investigations to date are presented. The final selection of microlensing 
candidates selected from the most recent Angstrom lightcurve data set
is shown, and a separate more detailed investigation into one particularly interesting
 microlensing candidate of very short duration is described.}


\newpage
\chapter*{Acknowledgements}
\addcontentsline{toc}{chapter}{Acknowledgements}

Firstly I must thank my long-suffering wife Lorna for putting up with many things, especially over the last two years, including moving house six times, including twice while pregnant, living with my moodiness which some days bordered on bipolar disorder, not being allowed to spend any money, especially on Derby County memorabilia or penguin artefacts, and never going on holiday anywhere more exciting than Barra airport. Secondly, a big hug and thankyou to my gorgeous daughter Isla whose arrival, while admittedly slowing down the pace of my work, simultaneously gave me all the motivation I might need to keep going and made my life worth living. Without her I might have gone even more insane.
Lorna and I would like to sincerely thank her parents, Kathleen and Stewart, for giving us somewhere to live on our return from Germany, and for putting up with
having squatters for as long as they did. I would also like to thank my own mother for ``not mentioning the PhD'', even though she did anyway.
\paragraph{} I would like to express my gratitude to my supervisor Dr Eamonn Kerins for all his 
assistance over the past (almost) five years, and especially for being patient with me when I am being a bit thick. I am also extremely grateful to my second supervisor Dr Andrew Newsam for being eternally optimistic, positive and upbeat, even when it isn't entirely justified by the facts. Many thanks also go to Dr Matthew Darnley for much assistance and advice throughout my PhD studies, and Dr Daniel Harman for fixing the ARI computing systems when they get broken (again).
I would like to add a big thankyou to my two examiners, Dr Phil James and Prof. Martin Hendry for their hard work in reading my (long) thesis, and their fair and helpful treatment during and after my viva. Their useful comments have undoubtedly resulted in
a much improved document.
I would also like to send warm greetings and thanks to all my other colleagues and friends at ARI Liverpool John Moores University (LJMU), but certainly
NOT Dr Claire Thomas as she cruelly left me out of her Thesis Acknowledgements.
 The author would like to thank the ARI, LJMU for giving him his PhD place and 
providing the location, equipment and working atmosphere necessary to enable his work to progress.
  For the first three years of his research, the author was funded by a
stipend from what was at the time the Particle Physics and Research Council, which has since merged with the Central Laboratories for the Research Council (CLRC) to form the Science and Technology Facilities Council (STFC).
  After this money ran out, the author was supported for seven months by funding from the European 
Community's Sixth Framework Marie Curie Research Training Network Programme, 
Contract No. MRTN-CT-2004-505183 “ANGLES”. During this time he was based at the 
Astronomisches Rechen-Institut (ARI), Heidelberg, which is part of the Zentrum 
f\"{u}r Astronomie der Universit\"{a}t Heidelberg (ZAH). Jonathan would like to thank Professor Dr Joachim Wambsganss for allowing him to join the Institute for a brief period, and to acknowledge the warm welcome he received from everyone at the ARI. In particular, the great deal of assistance readily given by Dr Robert Schmidt to help him to find somewhere to live and to settle into a (very) strange country.
Of those many people from the ARI, Heidelberg, Jonathan would also particularly like to thank Janine Fohlmeister for not saying ``I torld you'' too often, Timo Anguita and Cecile Faure for spending more time outside smoking than they did inside working, and fluffy bunnikins for, well, just being herself. By the way, Timo, I \emph{STILL} don't know much, but I know how that line can really get on one's nerves after a while.
 For the remainder of his studies, the author received no funding from anyone, 
and would like to thank the U.K. government and the banks of Mums and Dads for helping to keep him and his family alive long enough to finish.

\vfill
{\sc Jonathan Duke \hfill\today}

\newpage
\chapquote{"Of course, there is no hope of observing this phenomenon directly"}{Albert Einstein}
\newpage
\addcontentsline{toc}{chapter}{Contents}
\tableofcontents{}
\newpage
\addcontentsline{toc}{section}{List of Tables}
\listoftables{}
\newpage
\addcontentsline{toc}{section}{List of Figures}
\listoffigures{}
\newpage

\pagenumbering{arabic}
\fancypagestyle{lxh}{}%
\fancyhf{} 
\renewcommand{\footrulewidth}{0pt} 
\renewcommand{\headrulewidth}{0.1pt} 
\fancyhead[R]{\thepage} 
\fancyhead[L]{\nouppercase{\slshape \rightmark}}

\pagestyle{lxh}


\chapter{Introduction}
\label{Chapter_1}
\section{Introduction}
In this Chapter, a brief review of the history of microlensing is given,
along with a summary of the basic physics of the area. Next, descriptions
of the major recent microlensing surveys are made, along with highlights of
their major results. Then follows a description of our own Angstrom project,
 and a statement of the aims of this thesis.
 
\subsection{The Geometry of Lensing}

\begin{figure}[!ht]
\vspace*{19.5cm}
   \includegraphics{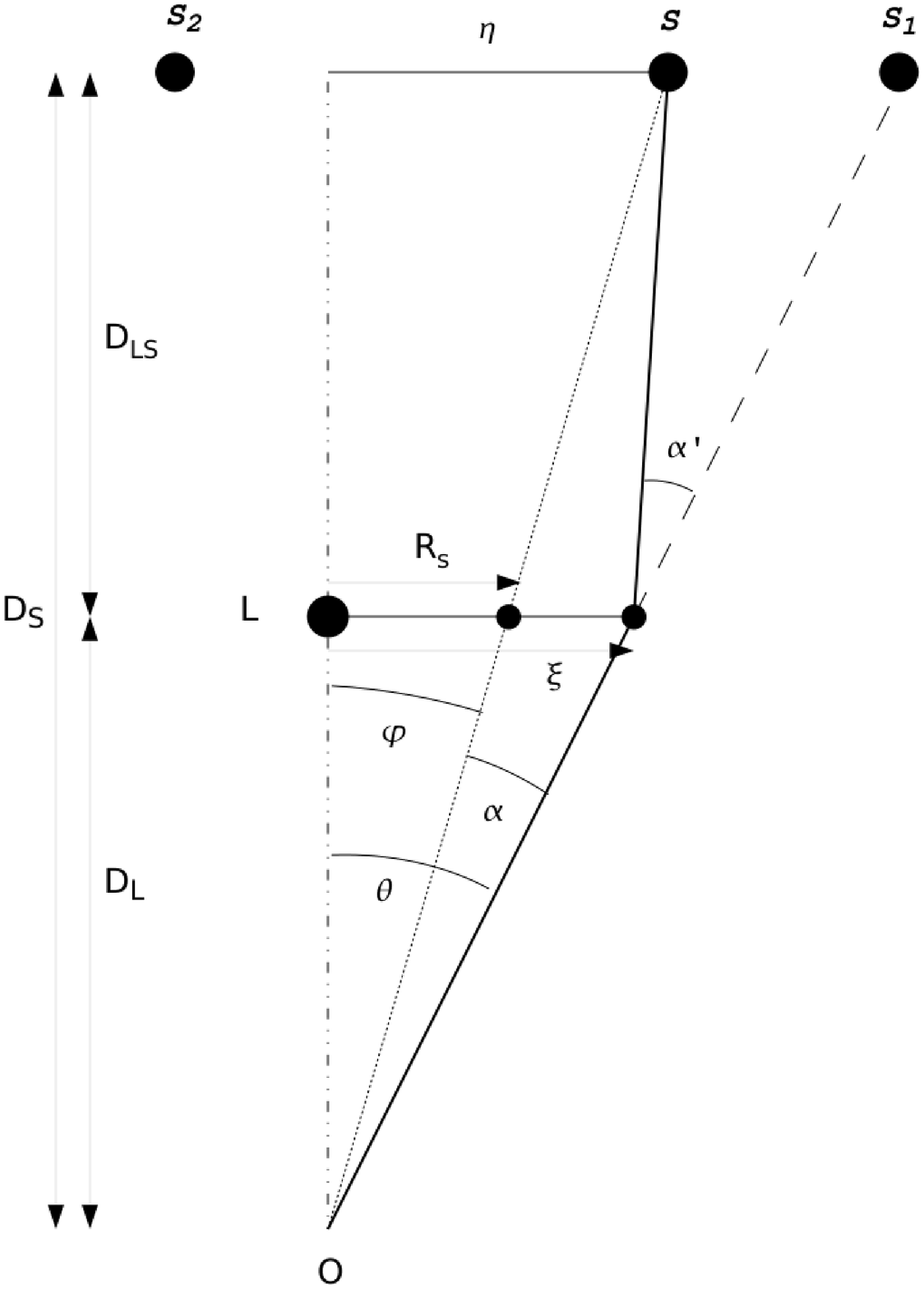}
\caption[The geometry of gravitational lensing.]{Diagram showing the geometry of lensing in the small angle approximation. The source of lensed light is $S$, and has unlensed radial position $\eta$ in the source plane and $R_s$ in the lens plane. The observer is at $O$. $S_1$ and $S_2$ are the apparent positions of the two lensed images in the source plane. $L$ is the massive object which is acting as a gravitational lens. Light from the source $S$ is bent by an angle $\alpha^{\prime}$, giving the image $S_1$ a radial position in the lens plane of $\xi$. The angle $\phi$ corresponds to the angular position on the sky of the source and the angle $\theta$  corresponds to the angular position on the sky of the image $S_1$. The angle $\alpha$ is the angular change in the apparent position of the source caused by lensing. $D_L$ = distance of lens from observer. $D_S$ = distance of source from observer. $D_{LS}$ = distance between lens and source}
\label{lensing_geometry}
\end{figure}

As is illustrated by Figure~\ref{lensing_geometry}, when a massive object (known as the lens, L) comes between a light source (S) such as a galaxy or star and us, the observers (O), the light can be bent around the massive object by its gravity.


 This means that the source is no longer
 seen at S, its true position, but rather at $S_{1}$ (and $S_{2}$, as the light can alternatively be bent around the opposite side of the lensing object). In Figure~\ref{lensing_geometry} the light
 paths are drawn as straight lines, but of course in reality they would be hyperbolae; however since
 the angles involved are actually very small, use of the small angle approximation
is justified and does not cause significant errors in practical situations.
The microlensing regime is that in which the two (or more, for non point-like lenses)
images are unresolved by the observation, as opposed to macrolensing, where all images
are separately observed.

\section{The History of Microlensing}
\label{lensing_hist}
 One of the many good reviews of this subject can be found in \cite{1998LRR.....1...12W}.
 \cite{1804BAJ....161} is generally credited with performing the first calculations of the deflection of light, or rather a body of velocity $c$, by gravity, but because he used classical Newtonian mechanics his result underestimated the relativistic result given in Equation \ref{einsteinangle} by a factor of $2$.
 The possibility of the phenomenon of gravitational lensing was first pointed out by
 \cite{1924AN....221..329C}, but he did not perform any calculations. The first serious
 attempt to calculate the magnitude of the effect was made by
 \cite{1936Sci....84..506E}. In this brief paper, which Einstein states was only 
written because he was persuaded to do so by R.W.Mandle, he clearly showed a 
good understanding of many of the principles thought of today as relatively modern 
discoveries, and already was able to calculate a version of the equation for the magnification of the source star as either the source moves relative to the line joining the observer and lens (or in Einstein's terminology, the observer moves relative to the line joining the source and lens stars). This equation (see Equation \ref{magnification}) is usually 
referred to nowadays as a ``Paczy\'nski curve'' after \cite{1996ARA&A..34..419P}.
Zwicky discussed many aspects of gravitational lensing in \cite{1937PhRv...51..290Z}.
However, Einstein was of the opinion that it would be a very uncommon effect, due mainly to the very small angular separations between source and lens that are required, and the possible ``dazzling'' from the lens star. It seems from the paper that he only considered the motion of the Earth, and not the possibility of significant motion of the sources. One imagines that he had not anticipated advances in technology which allow us to repeatedly monitor thousands of stars
simultaneously. Hence
his famous quote (see the start of this Thesis), which is repeated twice in \cite{1936Sci....84..506E} in slightly different terms, and his apparent opinion that it was not, therefore, of practical interest. However, it is my opinion that the ``this phenomenon'' to which the quote refers is not gravitational lensing of one star by another per se, but more precisely the observation of the Einstein ring or
two resolved images (see Figure~\ref{paczynskifig3}), for Einstein's second reason for his opinion was that the angular radius of the Einstein ring (see Equation \ref{einsteinring} and Figure~\ref{einstein_ring}) would ``defy the resolving power of our instruments", (which is actually still true today) and he had already worked out that the lensing magnification effect could be infinite, for a point source (Equation \ref{magnification_small_impact_param}).
 \cite{Tikhov} calculated the effect of
gravitational lensing on the intensities of the light from the source in the general case, but his
method was apparently not easily followed. The first person to make a simpler explanation of the
 physics and geometry involved and to make an estimate that the passage of one star behind another
would be a common event was \cite{1964MNRAS.128..295R}. He published virtually
at the same time as \cite{1964PhRv..133..835L}, who  discussed similar issues. Refsdal 
utilised a geometrical diagram very similar to Figure~\ref{lensing_geometry} to derive the general
equation for the change in brightness of a lensing event (see Equation \ref{magnification}).
The first observation of a gravitationally lensed object was of a double quasar, 
Q0957 + 561, by \cite{1979Natur.279..381W}. It was later determined that the two main objects are 
 two lensed images of a single background quasar by a foreground galaxy. The difference in the 
light travel time between the two images was measured to be $1.03$ years, by \cite{1986ApJ...300..209S}, 
and this enabled the geometry of the system to be completely established.

\paragraph{}

After Einstein, microlensing was next explicitly discussed by \cite{1979Natur.282..561C}. They referred to the subject as 
``star disturbances''. Later, \cite{1986ApJ...301..503P} was the first to coin the term ``microlensing'', and gave
 a description of the phenomenon:
\paragraph{}
   ``A real galaxy is made of stars and some continuously distributed mass: interstellar matter and
 possibly some exotic particles (dark matter). A detailed structure of a macro-image may depend on 
the masses and surface density of stars which split the macro image into a large number of micro-images,
 separated by some micro-arcsec. Even if we cannot resolve this structure, it may affect brightness 
of the macro image and contribute to its variations in time".
From this quote can be seen the explanation for why the phenomenon is commonly known as ``microlensing'', even though the image separation for stellar lensing is more typically on the scale of milli-arcseconds. The name was originally coined to describe lensing in galaxies at cosmological distances.

Later that year, \cite{1986A&A...166...36K} extended this work in several directions, and thereby 
gave the study of microlensing a firm foundation. The first detection of a microlensing event was 
reported by \cite{1989AJ.....98.1989I} and \cite{1989A&A...215....1V} where in
 the gravitationally lensed quasars QSO 2237+0305 and QSO 0957+561 respectively, the
brightness of one of the lensed images changed relative to the other in a way inconsistent with a simple time delay.

\section{Microlensing Theory}
\label{lensing_theory}

\subsection{Basic lensing theory}

The lensing angle $\alpha'$ is given by Equation \ref{einsteinangle}, which was calculated by Einstein using General Relativity,

\begin{equation}
\label{einsteinangle}
 \alpha' = \frac{4GM}{\xi{c^2}}
\end{equation}

where $\xi$ is the impact parameter of the light ray- the separation on the sky at the point of closest
 approach to the lensing mass.
Note that this deflection angle is only dependent on the lensing mass, and not on any property 
(e.g. frequency)
 of the light involved.
 Assuming this geometry, it can be seen that
\begin{equation}
 \eta = {D_{S}}\phi = {D_{S}}\theta - {D_{LS}}\alpha'
\end{equation}

In the case of $\phi$ = 0 and a circularly symmetric mass distribution, i.e. when the source and
 the lensing mass are perfectly aligned with the
 observer's line of sight,
it is intuitively obvious that positions $S_{1}$ and $S_{2}$ will be symmetrically arranged along
 a straight line on either
 side of S. In this case, some simple arithmetic can show that

\begin{equation}
\label{thetaE}
 \theta_{E} = \sqrt{\frac{4GM}{c^2}}\sqrt{\frac{D_{LS}}{D_{L}D_{S}}}  
\end{equation}

Since this particular situation is rotationally symmetric, the image of the source forms a
 perfect circle around the position of the lensing object. This is called an \textbf{Einstein ring}. 
$\theta_{E}$ is called the \textbf{Einstein angle}. Since 

\begin{equation}
\theta = \frac{\xi}{D_{L}},
\end{equation}

this means that 

\begin{equation}
\label{einsteinring}
 R_{E} = \sqrt{\frac{4GM}{c^2}}\sqrt{\frac{D_{LS}D_{L}}{D_{S}}}  
 \end{equation}

This radius $R_{E}$ is known as the \textbf{Einstein radius} and is the physical radius of the Einstein ring
 in the lens plane. There are many observational examples known of Einstein rings (and, more commonly,
 incomplete ``arcs'', where the alignment between source, lens and observer is not perfect). One example
 of a recently discovered Einstein ring is shown in Figure~\ref{einstein_ring}. The objects causing the lensing in situations such as Figure~\ref{einstein_ring} are galaxies, 
which may be clearly visible or not, and frequently the source objects are galaxies or quasars.

\clearpage
\begin{figure}[!ht]
\vspace*{0cm}
   \leavevmode
\centerline{\includegraphics[height=9.5cm]{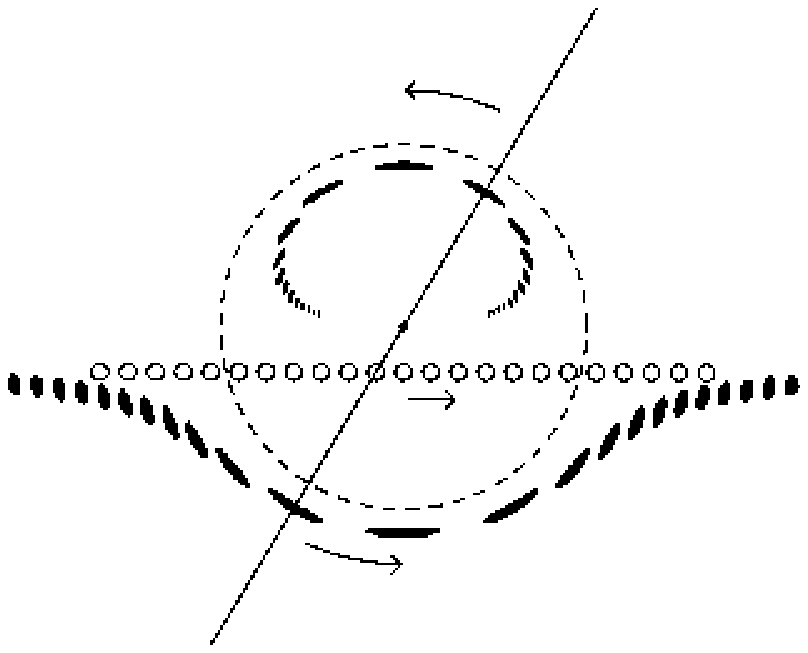}}
\caption[How lensed images change with a moving source] {In this drawing of gravitational lensing,
the lensing mass is indicated with a dot at the centre of the Einstein ring, which is marked with 
a dashed line; the sequential source positions are shown with a series of small open circles; and the locations and shapes
of the two images are shown with a series of dark ellipses. At any instant the two images, source 
and the lens are on a single line, as shown in the figure for one particular instant. Figure is Figure 3
from \cite{1996ARA&A..34..419P}. (Reprinted, with permission, from the Annual Review of Astronomy and Astrophysics, Volume 34 © 1996 by Annual Reviews  www.annualreviews.org) \protect\label{paczynskifig3}}
\end{figure}

\begin{figure}[!ht]
\vspace*{12cm}
  \includegraphics{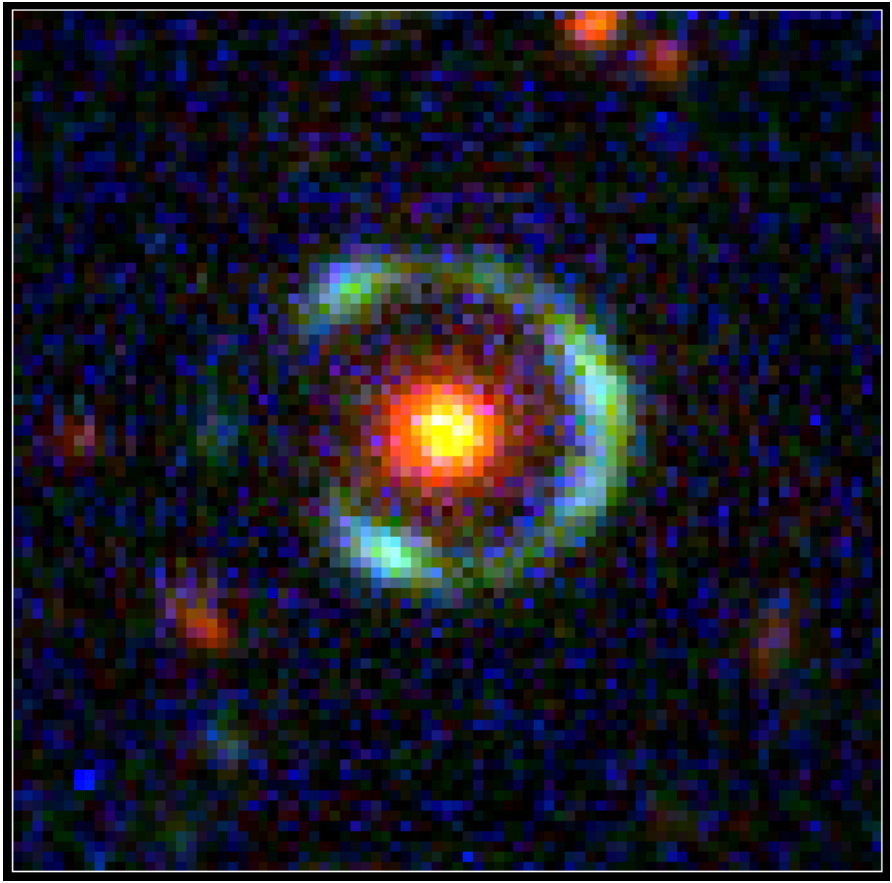}
\caption[An almost complete Einstein ring found around a Giant Luminous Red Galaxy (GLRG). ]{An almost complete Einstein ring found around a Giant Luminous Red Galaxy (GLRG). This is a Sloan Digital Sky Survey (SDSS) view of the sky composed from g, r \& i images around
the so-called ``Cosmic Horseshoe". Most of the objects in the field are faint galaxies. The blueish color of
the ring is striking. Image can be found at \cite{belokurov} (reproduced courtesy of V.Belokurov). The discovery paper can be found at \cite{2007ApJ...671L...9B}.
}
\label{einstein_ring}
\end{figure}

\subsection{Microlensing magnification\label{Microlensing}}

As can be seen in Figure~\ref{paczynskifig3}, for a point-like lens, there will be two lensed images,
one on either side of the lens, one inside and one outside the Einstein ring. Since any lensing
 conserves surface brightness, \citep{schneider},
the ratio of image to source intensity is given by the ratio of their areas.
With the geometry as shown in Figure~\ref{lensing_geometry}, this ratio can be calculated as

\begin{equation}
\label{arearatio}
 A_{+,-} = \left|\frac{R_{+,-}}{R_{s}}\cdot\frac{dR_{+,-}}{dR_{s}}\right| = \frac{u^2+2}{2u\sqrt{(u^2+4)}} \pm 0.5
 \end{equation}

 where $u \equiv \frac{R_s}{R_E}$ ($R_s$ is the position at which the source would appear in the absence of the lens mass and lensing effect) and it is assumed that the source is very small compared to $R_{E}$, i.e. a ``point source''.
 The quantity $A$ is called the magnification. The total magnification of the two images summed together can therefore be calculated as

 \begin{equation}
\label{magnification}
A = \frac{u^2+2}{u\sqrt{(u^2+4)}}
\end{equation}

 This equation has been simplified by assuming a point source and point lens. If either of the above 
are not the case, then the equation becomes much more complicated.

The increase in flux from the baseline value $F_{0}$ in a lensing event
is given by the formula first written down by \cite{1936Sci....84..506E} which is:

 \begin{equation}
\label{flux_difference}
\Delta\frac{F(t)}{F_{0}} = f[u^{2}(t)] ~\rm{where}~ f(x) = \frac{2+x}{\sqrt{x(4+x)}} - 1 
\end{equation}

The two Equations (\ref{magnification}) and (\ref{flux_difference}) are simply related due to Equation \ref{flux_link_equation}

 \begin{equation}
\label{flux_link_equation}
 \Delta\frac{F(t)}{F_{0}} = \frac{(F - F_{0})}{F_{0}} = \frac{F}{F_{0}} - 1 = A - 1 
\end{equation}

Since the original flux of the lensed object is not able to be measured in the pixel lensing regime (see Section \ref{pixel_lensing} for a definition of this)
 only the quantity $f$ can be discerned by measuring the flux.

In the case that $u \rightarrow 0$ then Equation \ref{magnification} can be approximated by

 \begin{equation}
\label{magnification_small_impact_param}
A = \frac{1}{u}
\end{equation}

Taking the limit, $u = 0$ leads to the surprising conclusion that the magnification is infinite, when the source is precisely behind the lens from the point of view of the observer.
 Of course this is only true if the source is 
a point of zero spatial extent, which is unphysical. The magnification is given by the ratio of the summed area of the two images divided by the surface area of the source.
 The reason the infinity occurs is because one must formally divide the
area of an infinitely thin annulus (Einstein ring) by the area of an infinitely small circle, which approaches zero faster than the area of the annulus as $u \rightarrow 0$.

 The dimensionless impact parameter $u$ (in units of $R_E$) of lens and source on the sky varies 
as a function of time as the source passes the lens with a minimum impact parameter $\beta$ as in Equation \ref{uoft}:

 \begin{equation}
\label{uoft}
 u^2(t) = {\beta}^2 + {\left({\frac{(t-t_{0})}{t_{E}} }\right)}^2
\end{equation}

where $t_E={R_{E}}/{v_t}$ is the Einstein radius crossing time (the time taken by the source
 to cross the Einstein radius of the lens with a relative transverse velocity $v_t$)
 and the time of peak flux is given by $t_{0}$. When $u$ from Equation \ref{uoft} is substituted into Equation \ref{magnification} the resulting equation is known as the ``Paczy\'nski curve'', after 
\cite{1996ARA&A..34..419P}, but an equivalent form was known already by \cite{1936Sci....84..506E}.
 For large $u$, the magnification approaches $1$,
and for $u \rightarrow 0$, $A \rightarrow \infty$. For high magnification events,
$A$ approximates to $u^{-1}$.  This is only true for the assumed point source. If the source is 
finite in transverse extent, the divergence at $u = 0$ becomes
``rounded off'' and hence $A$ always remains finite.

It has been shown by \cite{1999ApJ...510L..29G} that the measured full width, half maximum timescale of a lensing event $t_{\rm{FWHM}}$ is given by Equation \ref{tfwhm}

 \begin{equation}
\label{tfwhm}
t_{\rm{FWHM}} = t_{E}\omega(\beta)
\end{equation}

where $\omega(\beta)$ is given by Equation \ref{omega_beta}

 \begin{equation}
\label{omega_beta}
\omega(\beta) = \sqrt{2f[f(\beta^2)] - \beta^2}
\end{equation}

and the function $f$ is that in Equation \ref{flux_difference}.

It was shown by \cite{2000ApJ...530..578B} that there exist useful limiting values of the function $\omega(\beta)$, which are: for $(\beta \ll 1), \omega \simeq \beta\sqrt{3}$ and for $(\beta \gg 1)$, $\omega \simeq \beta({\sqrt{2} -1})^{\frac{1}{2}}$. However, \cite{2000ApJ...530..578B} use a different definition of $t_E$
which is the same as the one used by the MACHO collaboration, namely $t_E={2R_{E}}/{v_t}$.

 Taking the first of these two limits, for the ``high magnification regime'' where $(\beta \ll 1)$, and substituting into Equation (\ref{tfwhm}) results in Equation (\ref{high_mag_approx_2}) in terms of our definition of $t_E$.

 \begin{equation}
\label{high_mag_approx_2}
 t_{\rm{FWHM}} = {2\sqrt{3}t_{E}\beta}\rm{     }  (\beta<<1)
\end{equation}

Since when $u$ is small compared to the Einstein radius the magnification may be approximated by the inverse of Equation \ref{uoft},
substituting for $t_{E}$ using Equation \ref{high_mag_approx_2} leads to Equation \ref{red_pac_curve}:

 \begin{equation}
\label{red_pac_curve}
\delta F(t) = \frac{F_{0}}{u(t)} = \frac{F_{0}}{\beta} \frac{1}{\left({1 + 12\left[{\frac{(t-t_{0})}{t_{\rm{FWHM}}}}\right]^2 }\right)^{\frac{1}{2}}}
\end{equation}

Here the non-degenerate variables have been ``reduced'' from $4$: $t_0$, $F_{0}$, $\beta$ and $t_{E}$ to the $3$ combinations: $t_0$, $\frac{F_{0}}{\beta}$ and 
$t_{\rm{FWHM}}$ (or $2\sqrt{3}\beta t_{E}$), and so this equation is known as the ``reduced Paczy\'nski Curve''.

\subsubsection{Finite Source Effects}

If the angular diameter of the source is comparable to that of the Einstein ring of the lens,
 then the simple Paczy\'nski curve is no longer valid. The central magnification is no longer infinite,
 and the central region becomes broadened to reflect the diameter of the now finite source. The 
lightcurve for a finite source (but still a point lens) becomes the convolution of the Paczy\'nski 
curve appropriate to each point on the star's surface with the luminosity distribution on the face of the star, along the path of the lens as it passes the star.
 Therefore, to derive the full expression, it is necessary to take into account such things as 
the limb darkening of the source star, which necessitates integrating the equations over the 
whole face of the star, with all its luminosity changes. Hence the process may become very
 mathematically complex. The full expression and its derivation is described in 
\cite{1994ApJ...430..505W} and in an alternative but complementary way in \cite{1994ApJ...421L..71G}.

\subsubsection{Microlensing optical depth}

 An important parameter for microlensing is the optical depth, $\tau$. It is defined as
the number of lenses inside a tube along the line of sight with transverse size equal to the
 Einstein radius of the lens. Mathematically, this can be written as
in Equation \ref{opticaldepth}, where $\rho(x)$ is the density distribution along the line of 
sight and $m$ is the mass of the lenses being considered. $D_{S}$ is again the distance to the source and
 hence the length of the line of sight being considered.

 \begin{equation}
\label{opticaldepth}
 \tau = \displaystyle\int^{D_{S}}_{0} \frac{\rho(x)}{m}\pi R_E^2(x)\,dx
\end{equation}

 When the optical depth is much less than one it is equivalent to the probability that a
 source along the line of sight will be lensed. Of course, in real situations the magnitude of the
 integral of $\rho$ along the line of sight will vary depending on the particular line of sight,
 meaning that in order to estimate event rates for a survey such as Angstrom where the function $\rho$ varies significantly across the field the integral
 must also be performed over the two transverse directions as well as over $x$.
In addition, the sources will vary in luminosity, which may mean that not all sources are detectable in real situations with the limitations of physical observations.

\subsection{Pixel Lensing}
\label{pixel_lensing}

When microlensing is observed in the Milky Way, in the great majority of cases it can be
expected that the light from the source will be
the dominant source of light detected in the lightcurve. However, when observations are carried out in other galaxies, even in M31 (the nearest galaxy of comparable size to the Milky Way) it is certain that in every pixel of a CCD image there will exist many stars, whose light is combined. This makes the interpretation of the changes in the flux in any one pixel much more complicated, and not only will the light from many different stars be present, but several of them may even be varying at the same time.

When a lensed source is well resolved, it is possible to observe all four of the physical 
parameters of the lensing event; the baseline flux $\phi_{0}$, the normalised impact parameter
 $\beta$, the time of maximum flux $t_{0}$ and the Einstein timescale $t_{E}$.
 However, as in the case of Andromeda, where the density of stars in the field of view of the
 telescope is very large and the sources are unresolved, there will be many sources in each pixel
 of the image. Therefore, the baseline flux observed before the lensing event takes place is the
 sum of the flux of all the unresolved sources. This means that we do not know the original unlensed
flux of the particular source which is lensed.

 Equation \ref{uoft} can be re-written as Equation \ref{uoft2} if one writes $t_{eff}= \beta t_{E}$.

\begin{equation}
\label{uoft2}
 {(u(t)t_{E})}^2 = {t_{eff}}^2 + {(t-t_{0})}^2
\end{equation}

If $A(t)$ is the magnification given in Equation \ref{magnification}, which is a function of
 time due to the variation in $u$ in Equation \ref{uoft}, and $\phi$ is the total observed flux, $\phi_{bg}$
 is the summed flux of all the background sources including the source star (in the absence of lensing) and $\phi_{\star}$ is the
 original (unknown) flux of the source, then Equation \ref{pixellensing} describes the flux 
variation in the pixels covering a microlensing event.

 \begin{equation}
\label{pixellensing}
   \phi - \phi_{bg} = \phi_{\star}( A(t) - 1 )
\end{equation}

 Since $\phi_{bg} = \phi_{rest} + \phi_{\star}$, where $\phi_{rest}$ is the total flux of all other 
sources in the pixel other than the lensed source, this can be written

 \begin{equation}
\label{pixellensing2}
   \phi = \phi_{\star}A(t) + \phi_{rest}
\end{equation}

 The only information about the lensing event that can be gleaned comes from the change
 $\phi_{\star}(A(t)-1)$ which is observed in the flux, and, experimentally, can only be detected 
if the change in the flux is greater than the statistical error in photon counts $N$ in the pixel,
 which is $\sqrt{N}$ for purely Poisson noise. In other words, if $N_{\star}$ is the number of photon
 counts due to the source, and $N_{gal}$ is the number of photon counts due to galaxy surface brightness, then the criterion
 for detection is that
 
 \begin{equation}
\label{detection}
    (A-1)N_{\star}\simeq u^{-1}N_{\star} > kN^{\frac{1}{2}}_{gal}
\end{equation}
where k is a constant, often taken as $3$. The value of $u$ when the criterion above is just met
 will be referred to as $u_{t}$. Therefore, only the so-called ``visibility timescale'', $t_{v}$, when the lensed flux is visible 
above the noise, rather than the intrinsic Einstein timescale $t_{E}$ of the lensing event, can 
be directly measured. Thus the visibility timescale can be written

 \begin{equation}
\label{tv}
  t_{v} = 2{(u^2_{t} - \beta^2)}^{\frac{1}{2}} \frac{\theta_{E}}{\mu} = 2{(u^2_{t} - u^2_{0})}^{\frac{1}{2}}t_{E}
\end{equation}

where $\theta_{E}$ is as defined in equation \ref{thetaE} and $\mu$ is the relative proper motion of the lens across the line of sight. Therefore the visibility timescale
depends on the square root of the lensing mass.

 The fact that only $t_{v}$ can be observed leads to an ambiguity/degeneracy between the true 
lensing timescale and the true impact parameter $\beta$ (or alternatively the source star flux $\phi_{\star}$), as, for e.g., a long timescale, low magnification event can appear very similar 
to a short timescale, high magnification event, if different values for $\phi_{bg}$ are assumed.

 \cite{1996ApJ...470..201G} took account of this degeneracy when he approximated the Paczy\'nski 
curve with one fewer parameter for the special case of high magnification. His equation is given in Equation \ref{pixel_lensing_equation}:

\begin{equation}
\label{pixel_lensing_equation}
   \Delta\phi = \frac{\phi_{eff}}{\sqrt{(\frac{(t-t_{0})^2}{{t_{eff}}^2} + 1)}}
\end{equation}

where the three free parameters are: $\phi_{eff}=\frac{\phi_{0}}{u_{0}}$, $t_{eff}$ and $t_{0}$.
 $\phi_{0}$ is the peak flux at the peak time $t_{0}$. It can be seen that this is an exactly equivalent formulation to Equation \ref{red_pac_curve}, if $t_{eff}$ is replaced by the correct function of $t_{\rm{FWHM}}$ found by substituting for $t_{eff}$ in Equation \ref{high_mag_approx_2}.

In the same paper, Gould concluded that pixel lensing was potentially a very sensitive and widely applicable technique. He showed that there are two distinct regimes of pixel lensing, which he labelled ``semi-classical'' and ``spike''. The boundary between these two regimes is a certain value $F_{max}$ (which is defined in the paper) of the original unlensed flux of the source star. Stars which have unlensed flux $F_{0} > F_{max}$ are in the ``semi-classical'' regime, which means that they can be recognised even when they have low impact parameters, and it is usually possible to extract information about the timescale $t_{E}$ and $F_{0}$ separately from these events. On the other hand, for source stars which have unlensed flux $F_{0} < F_{max}$, generally no direct information about the timescale $t_{E}$ is available.
In the above paper, various different possible sources of noise and unwanted backgrounds which could occur in the process of performing a survey using difference imaging in the pixel lensing regime were also discussed. Some possible future applications of pixel lensing were given.

\section{Microlensing Surveys}

\subsection{Milky Way surveys}

 \label{milky_way_surveys}

 In 1986, \cite{1986ApJ...304....1P} suggested that a variability search among
 the millions of stars in the Large Magellanic Cloud could be used to detect dark matter in the 
Galactic halo, which the line of sight would pass through. Fortunately, the technology needed to perform such a search became available soon
 afterwards, and the 1986 paper is credited with inspiring the microlensing searches which followed:
 EROS \citep{1993Natur.365..623A}, MACHO, which began in 1992, \citep{1993ASPC...43..291A}, OGLE,
 which also started in 1992, \citep{1992AcA....42..253U}, and DUO 
\citep{1995Msngr..80...31A,1996IAUS..173..215A}. \cite{1991ApJ...366..412G} proposed that objects responsible
 for gravitational microlensing be called Massive Astrophysical Compact Halo
 Objects (MACHOs). This term is used to refer to lensing by non-luminous objects.

The name OGLE stands for `Optical Gravitational Lensing Experiment'. The first phase of the OGLE experiment,`OGLE-I' began in 1992 and observations were continued for the next four seasons until the end of the 1995 season, using the $1$m Swope telescope at the Las Campanas Observatory, Chile. Due to limited telescope availability, 
observations were limited to the Galactic bulge, and the area covered was relatively small. However, the scientific achievements were still considerable and included
the first determination of the microlensing optical depth towards the Galactic bulge \citep{1994AcA....44..165U} (which was measured as $\tau =  (3.3 \pm 1.2)\times10^{-6}$.

EROS stands for `Exp\'{e}rience pour la Recherche d'Objets Sombres', and its main aim is research into dark massive objects, otherwise known as `MACHOs' which are gravitationally bound to our Galaxy. This has included observations of the Galactic bulge, spiral arms, and both the large and small Magellanic Clouds.
Early determinations by EROS of the optical depth to the Galactic Centre \citep{2003A&A...404..145A} gave a value of $\tau = (0.94 \pm 0.29)\times10^{ - 6}$ which was not consistent with the values calculated by the MACHO and OGLE groups. Later analysis \citep{2006A&A...454..185H} with 
better statistics and a much longer temporal baseline resulted in a new value of $\tau = (1.62 \pm 0.23)\times10^{ - 6}$
$\rm{e}^{[-\rm{a}(|\rm{b}|-\rm{3} \rm{~deg})]}$
with $a = (0.43 \pm0.16)\rm{~deg}^{-1}$ which was more consistent with the latest results at that time from MACHO and OGLE-II. 

In \cite{2007A&A...469..387T}, optical depth toward the Large Magellanic Cloud was calculated to be $\tau<0.36\times10^{-7}$ (95\%CL), which would correspond to a halo mass fraction of less than $8$\%, which disagreed strongly with the MACHO result (which is described below).

The MACHO project \citep{1993ASPC...43..291A} is a joint US/Australian collaboration whose main stated aim is to test the hypothesis that a significant fraction of the matter in our Galaxy 
(and hence by implication, many other galaxies) is composed of low mass, non-luminous objects such as brown dwarfs and planets,
 which could be described as MACHOs. The project first started taking data in $1993$ \citep{1993BAAS...25..859A} and began by monitoring $6$ million stars in the LMC and Galactic bulge.
 Results using $5.7$ years of observations towards the LMC were published in \cite{1999AAS...195.4802B} and \cite{2000ApJ...542..281A} and in this second paper
 an optical depth of $\tau = 1.2^{+0.4}_{-0.3}\times10^{-7}$ was calculated using events with timescales between $2$ and $400$ days, based on $13$-$17$ events.
This was a significantly larger number than the $\sim$ $2$ to $4$ events expected from lensing by known stellar populations, which led to a predicted MACHO halo fraction of $20$\%.
Using seven years of MACHO data, \cite{2005ApJ...631..879P} were able to use a subset of $42$ out of the total of $62$ selected microlensing candidates which had sources which were ``clump giants'' 
to estimate the optical depth to microlensing towards the Galactic bulge as being $\tau = 2.17^{+0.47}_{-0.38}\times10^{-6} $ at $ (l,b)=(+1.50 $ deg,$-2.68 $ deg). Giant source stars were chosen for 
this estimate in order to reduce the problems caused by blending of flux from stars other than the true source star.

Despite the huge amount of effort put into surveys towards both the Milky Way and its satellites, the discrepancies between the results of the different major collaborations have still not entirely disappeared. When observing towards the Galactic Centre, however, it has become clear that the observed differences in optical depth are less due to
differences between survey methodologies and more to do with a clear correlation observed between the measured optical depth and the angular positions on the sky of the fields integrated over. A clear gradient in optical depth was observed in MACHO data \citep{2005ApJ...631..879P} in the sense that MACHO fields closer to the Galactic Centre have higher optical depth. This is clearly consistent with the model predictions of increasing mass density towards the center of galaxies.
This gradient has been modelled as a linear fit \citep{2005ApJ...631..879P} and, when the best fit is removed from data from all surveys, results are consistent with one another. 
    The most obvious remaining discrepancy between surveys lies in the results of
 analysis of data from the LMC and SMC.
 As mentioned above, the MACHO project announced the discovery of $13$-$17$ microlensing events
 towards the LMC in \cite{2000ApJ...542..281A}. This is significantly more than
 would be expected from known stellar populations and hence they deduce that there must be a significant fraction of dark matter in the LMC. EROS, however, only claim to
 have discovered one microlensing event towards \emph{both} Magellanic Clouds
 \citep{2007A&A...469..387T}.
 OGLE have published only two microlensing events towards the LMC
 \citep{2009MNRAS.397.1228W}, which is clearly more consistent with the EROS
 result than with MACHO.
 The optical depths to microlensing towards the LMC calculated by the EROS and
 OGLE collaborations remain incompatible with that calculated by the MACHO
 collaboration within published experimental
 errors.
 Much effort has already gone into attempting to understand where the
 inconsistency between these results lies, but no clear answer has been arrived
 at. One possible clue may be found in the fact that the MACHO fields are more 
 compactly arranged toward the centre of the LMC, whereas the EROS fields cover a
 much larger area. This might have had some influence on the relative optical
 depths expected, but the predictions with which the results are compared are arrived at by the different
 collaborations using extensive modelling of their individual circumstances, and
 should therefore also produce consistent answers.
   The responsibility for further investigation of the reasons for the clear
 discrepancies between the results of different collaborations towards the 
 Magellanic Clouds will have to be passed to currently running next-generation
 surveys such as SuperMACHO \citep{2002AAS...201.7807S}.    
    
 Another difficulty, when the statistics are low as they still are in measuring lensing towards the LMC and especially the SMC, is how to be certain whether the lens that caused a particular event resided in the Milky Way halo or that of the Cloud, especially as the two halos may even be continuous. Observations toward the Galactic Centre have to deal with high values of extinction and reddening due to dust in the plane of the Galactic disk, which has made making definite conclusions about the distribution of matter in the Galaxy much harder.

The Andromeda Galaxy has several major advantages over the Milky Way with respect to the difficulties mentioned above \citep{1992ApJ...399L..43C}. It lies out of the disk plane of our Galaxy, thus significantly reducing the Galactic extinction problem. We can explore the inner halo, rather than the outer halo of the Milky Way that can be observed in the direction of the LMC, for example. Due to Andromeda having a favourable geometry of its own in that it is almost, but not quite, edge on to us the modelled lensing probability has a clear variation with respect to galactocentric position, meaning that the microlensing origin of events can be proved
using statistical analysis. Also, both its greater outer rotation speed and the Tully-Fisher relation \citep{1977A&A....54..661T}, for example, imply that Andromeda has a greater mass \citep{2009MNRAS.397.1990B}, and hence a larger halo than the Milky Way, giving it a relatively increased lensing cross section. Of course it is also advantageous to be able to observe the halo as a whole from outside, rather than being immersed within it.
Observations in Andromeda could allow resolution of the disagreements over methodology
which may have contributed to the inconsistent results found in the LMC/SMC observations. Therefore, collaborations began to contemplate making observations towards M31.

\subsection{Microlensing Surveys of Andromeda}

The microlensing surveys conducted towards the Andromeda Galaxy began with
 AGAPE, who reported the first microlensing candidate in the 
direction of M31, detected in 1995, which had a $t_{\rm{FWHM}}$ of $5.3\pm0.2$ days \citep{1999A&A...344L..49A}. SLOTT-AGAPE gave a partial
 analysis of their early (1998-1999) data, using images which were stacked 
nightly in \cite{2002A&A...381..848C},
reporting five candidate events. The analysis of the full data set was concluded in \cite{2003A&A...405..851C} with twice as long a baseline, which resulted in all the five previous candidates being rejected due to additional clear variations in the lightcurve being found, but an additional three candidates being put forward. These new candidates even the authors stated were ``inconclusive''. This highlights one of the major issues faced by microlensing surveys in Andromeda which is that due to the blending of many stars in the same pixel, some of which are likely to be variable, variation of the microlensing baseline due to variable stars is a major problem and needs to be seriously considered. Further discussion of this issue will be given in Chapter \ref{Chapter_4}.
 WeCaPP, which began in 1997, described their method and gave some sample lightcurves of
variable stars in 
\cite{2001A&A...379..362R}. The MEGA survey began in 1998, \citep{2004A&A...417..461D}, followed by the NainiTal \citep{2005A&A...433..787J}
 and POINT-AGAPE \citep{2005A&A...443..911C}, \citep{2005MNRAS.357...17B} surveys. POINT-AGAPE and MEGA used the same data, from the Wide Field Camera on the Isaac Newton telescope,
 but the two surveys used independent data analysis techniques.
 POINT-AGAPE and MEGA have both
 published results, finding several microlensing candidates.

\subsubsection{The POINT-AGAPE Survey}

The POINT-AGAPE collaboration have so far published two papers giving independent and different analyses of their data. These were \cite{2005MNRAS.357...17B} and \cite{2005A&A...443..911C}. In \cite{2005A&A...443..911C}, 6 microlensing 
candidates were found in their analysis of 3 years of data whose positions in M31, taken together with Monte Carlo simulations led them to conclude
that they detected at least a $20\%$ MACHO contribution to the halo of M31. \cite{2005MNRAS.357...17B}, however, using different cuts, only found three top level events. The most important difference between these two analyses was that \cite{2005A&A...443..911C} limited the timescales of their events to less than 25 days, whereas \cite{2005MNRAS.357...17B} applied no limit on timescale, but still found fewer events. \cite{2005MNRAS.357...17B} did not tend to find short timescale events whose lightcurves are contaminated with flux from blended variable stars (which seems to be due to the strictness of their first cut, detailed in Chapter \ref{Chapter_4}), whereas \cite{2005A&A...443..911C} did. More details about some of the cuts applied by these two surveys will be made in Chapter \ref{Chapter_4} in relation to the cuts applied in the selection procedure used in the work performed for this Thesis.

The many variable stars which are inevitably detected by such surveys were analysed in detail by \cite{2004MNRAS.351.1071A}. Knowledge of the spatial distribution
of such stars has great relevance to the information that can be extracted from surveys looking specifically for microlensing events since variable stars are the major contaminant for measuring the spatial asymmetry of the microlensing signal produced by the high inclination of M31 towards us. The facts that on some lines of sight the disk
is in front of the densest part of the bulge and in others the bulge is in front of the disk, along with the asymmetry on the sky caused by the proposed barred bulge \citep{1956StoAn..19....2L,1958ApJ...128..465D,1977ApJ...213..368S}, lead to
likely source and lens populations, and hence the microlensing optical depth, varying over the face of the galaxy in a systematic way that can be modelled.
It has been found that the variable stars share some of this signature, which makes extracting an unambiguous signal from microlensing candidate spatial distributions much more difficult. \cite{2004MNRAS.351.1071A} suggested that one way this could be done is to look for an East-West asymmetry, rather than the near-far asymmetry previously proposed in \cite{1992ApJ...399L..43C}, because both the resolved stars and the various groups of variables are more or less symmetric about an axis oriented North-South.

The Classical Novae  discovered by the POINT-AGAPE survey were reported in 
\cite{2004MNRAS.353..571D}. $20$ novae were reported, of widely varying speed classes and lightcurve morphologies. These were detected using an automated pipeline which minimised the required amount of inspection of the lightcurves by eye. This was followed by a more in depth analysis of the nova rate and statistical characteristics of the nova population in 
\cite{2006MNRAS.369..257D}.

\subsubsection{The MEGA Survey}
MEGA reported $14$ microlensing candidates, although they acknowledged that their cuts were less 
stringent than those used by the 
POINT-AGAPE collaboration, which may explain the difference in numbers. In addition, the pixel lensing techniques used by the two teams were different. Specifically, in \cite{2004A&A...417..461D}, it is stated that the difference in the number of candidate microlensing events found by the MEGA and POINT-AGAPE projects was due to the ``severe cuts'' used by POINT-AGAPE, examples of which include the specification that $t_{1/2} < 25$ days and the ``much higher signal to noise'' specified. This second claim is not simple to verify, however, as direct comparison of the methods used by the different groups is not easy. The MEGA collaboration simply specified that
$\chi_{\rm{const}}^2 - \chi_{\rm{pac}}^2 > 100$, where $\chi_{\rm{const}}^2$ is the $\chi^2$ of the best constant fit to the event lightcurve and $\chi_{\rm{pac}}^2$ is the $\chi^2$ of the best reduced Paczy\'nski fit. \cite{2004A&A...417..461D} were evidently comparing themselves to the method of \cite{2003A&A...405...15P} which found $4$ candidates and who, in order to select high signal to noise candidates, calculated the probability of the bump related to the proposed microlensing event being caused by chance, and demanded that $-\ln(P) > 100$ in r band and $-\ln(P) > 20$ in at least one other filter. The four candidates finally selected, however, had total signal to noise, (defined as
$S/N = \sum_{1}^{n} \frac{{\rm{Flux}}_i - {\rm{Flux}}_{\rm{baseline}}}{\rm{error}_{i}}$
 where $60<$ $S/N$ $<1600$, and $n$ is the number of points defined as 
``in the peak''), which was apparently high compared to typical
 events in the database.
Contrary (in some circumstances) to this, however, \cite{2004A&A...417..461D} used a $\chi_{\rm{pac}}^2$/d.o.f. cut which could be as low as $<$ $\sim1.5$ in r band and $<$ $\sim2.0$ in i band but would usually be looser for higher signal to noise events (depending on the total significance, or ``signal to noise'' as defined above, of points in the peak) according to the recipe
$\chi_{\rm{pac}}^2$/d.o.f.$ < 1.5 + 0.1(\frac{S/N}{n} - 1)$, where $S/N$ is as defined above, and $n$ is again the number of points in the peak.
So, for example, if there were only $5$ points in the peak, with $S/N$ contributions of $(3,3,3,5,5)$ (i.e. $3$ with $3\sigma$ above the baseline and $2$ with $5\sigma$), then $S/N$ would be $19$, and the $\chi_{\rm{pac}}^2$/d.o.f. limit would be $1.5 + 1.8 = 3.3$. (Of course, this formula can break down if the bump goes below the baseline on average, but clearly only positive bumps will be selected).
 Whereas the $\chi_{\rm{pac}}^2$/d.o.f. cut described above 
began at a lower level, \cite{2003A&A...405...15P} used a flat cut of $\chi_{\rm{pac}}^2$/d.o.f.$ < 5$
, and, later, another POINT-AGAPE analysis \citep{2005A&A...443..911C} used only $\chi_{\rm{pac}}^2$/d.o.f.$ < 10$.
The reason that direct comparison is difficult is not just because cuts with a particular purpose are constructed differently. It is also that even if one cut is indeed tighter in one project's analysis, this can be more than compensated for by another cut (or cuts) being less strict, meaning that the total effect on the number of selected candidate events is difficult to predict.
However, as pointed out in \cite{2000ApJ...542..281A}, there is always a strong element of subjectivity to setting the types and levels of cuts, and every group seems to do this in subtly (or unsubtly) different ways. However, if the following (usually Monte Carlo) selection efficiency calculations are done properly, the corresponding efficiencies calculated should compensate for the number of events selected, and the resulting optical depth to microlensing $\tau$ should be the same.

\subsection{Development of difference imaging}
\label{dia}
Most microlensing surveys in recent years have used a technique called Difference Image Analysis (DIA), otherwise called ``image subtraction''.

The feasibility of this technique was demonstrated in \cite{1994AAS...185.1701T}, first proposed for use in microlensing surveys by \cite{1996AJ....112.2872T} and thereafter utilised in the Columbia/VATT survey, which was the pilot study for MEGA. It was first developed for a M31 survey because the crowding problem for variable sources is much greater there than in the Milky Way.
Later, from 1998 onwards \citep{1998AAS...19310804D} it was first employed in Milky Way by the MACHO survey of the Milky Way bulge.

 The basis of this method is that
 if it is possible to scale and subtract two images of the same region of sky taken 
at different times, the resulting image will show only the objects whose flux has changed between 
the two epochs. Since bright non-variable stars will effectively disappear during this process, this method enables weakly varying sources to be detected with far greater sensitivity.

The difference imaging technique was first automated by \cite{1998ApJ...503..325A} in their ``Optimal Image Subtraction" method \citep{1998ApJ...503..325A}. In this method a reference image, which is usually a good image with as small and circular point spread function (PSF) as possible, is convolved with a kernel ($K$) which is a function which describes how to transform the shape of the PSF of the reference image ($R$) to that of any other chosen image ($I$).
A smooth function is used to model the background galaxy surface brightness ($B$) in the resulting subtracted image and then the residuals to the fit are minimised using the linear least-squares method in order to approximate the true difference image. This process can be summarised by Equation \ref{dif_im_min}. 

\begin{equation}
\sum_{i}^{}D(x_{i},y_{i}) = min\{{\sum_{i}^{}[(R(x_{i},y_{i}) \otimes K(u,v)) - I(x_{i},y_{i}) -B(x_{i},y_{i})]^2}\}
\label{dif_im_min}
\end{equation}

where $x_{i},y_{i}$ are the coordinates of each pixel $i$, $(u,v)$ are coordinates
centred on the middle of the kernel, and ``$\otimes$'' represents a convolution.

To make their code more computationally efficient \cite{1998ApJ...503..325A}
decided to decompose their kernel function into the sum of several Gaussian basis functions which are also scaled by a polynomial function of position.
This process can be summarised by Equation \ref{kernel_decomp}

\begin{equation}
K(u,v) = \sum_{(k\le{N},ij) \left({i\le n_k,(i+j)\le n_k}\right)}^{}a_{m}{exp({-({u^2+v^2})/{2{\sigma_{k}}^2}})}u^{i}v^{j}
\label{kernel_decomp}
\end{equation}

where $K$ is composed of $N$ Gaussian-like components each with width given by $\sigma_{k}$, and the range of the two indices $i,j$ is given by the degree $n_k$ of the polynomial associated with each component. The scaling factors $a_{m}$ are individual to each summed component, so the values of $m$ are implied by the summations over the other indices.

The least squares fitting is often performed within sub-regions of the image, known as ``stamps'' due to their square shape, which can drastically reduce computing time. This requires an assumption that the background and PSF do not change significantly on the spatial scale of the stamp, which may not always be true.

We are also 
using this method in the Angstrom project, using a modified version of the ``ISIS'' code of \cite{1998ApJ...503..325A}.
We have modified this code substantially, extracting the particular subroutines which perform the difference imaging and integrating them with the ADAP.
This code now uses the standard cfitsi/o routines, and is able to utilise image masking, unlike ISIS.
 The sequence of processes performed in the complete Difference Imaging Pipeline is described below, in Section \ref{descDIA}.

\subsection{The study of variable stars from large area surveys}

As described in \cite{1996AAS...188.6503C}, among others, the time series made available by the high cadence monitoring performed by microlensing surveys such as MACHO, OGLE and EROS, have revolutionised the level of knowledge
of variable stars in, for example, the Magellanic 
Clouds \citep{Cook_Alcock_et_al}.
 Virtually complete catalogues have 
been constructed for many of the more common types of variable star, such as RR Lyrae, Cepheids and long period variables.
The DIRECT project, in \cite{2003AJ....126..175B} used difference imaging to look for variable stars in M31, in order to attempt to directly measure the distances to this galaxy and also M33.
This search covered a much wider area than Angstrom, being $22^{\prime}\times22^{\prime}$, and thus did not concentrate on the M31 bulge.
In \cite{2003ApJ...591L.111B}, a difference imaging process was also used, to make observations in M83, looking for variable stars, particularly Cepheids.

\section{The Angstrom Project\label{angstrom_project}}
\subsection{Rationale}

\begin{figure}[!ht]
\vspace*{9.5cm}
  \includegraphics{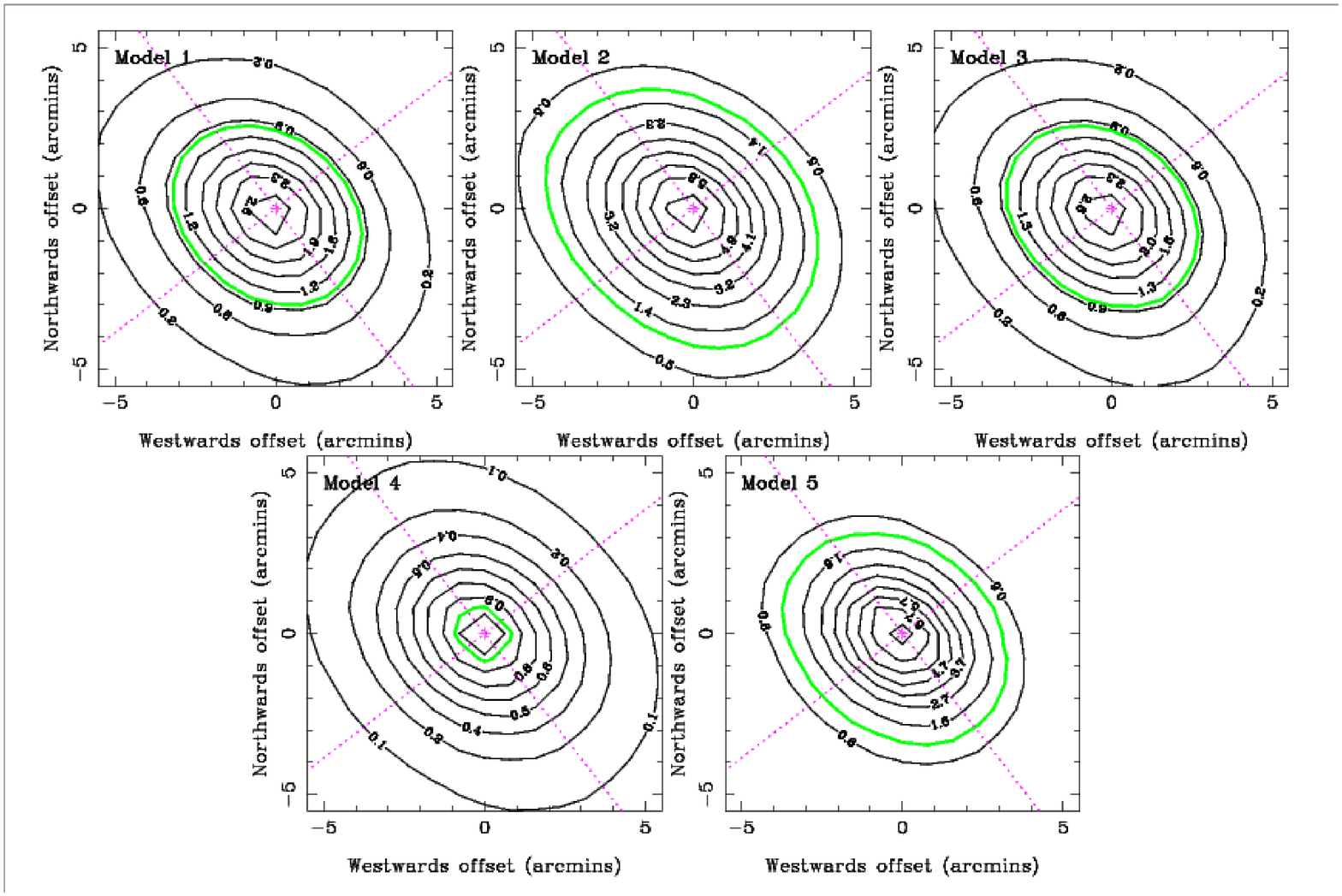}
\caption[The pixel lensing spatial distributions for galaxy models 1-5 of \cite{2006MNRAS.365.1099K}.]{The pixel lensing spatial distributions for galaxy models 1-5 of \cite{2006MNRAS.365.1099K}, including events with visibility timescales $t_{v}$ = 1-100 days.
The origins of the plots are the centre of M31 and the pink cross indicates the major and minor axes of the M31 disk. The major axis is the line going top left to bottom right. Event rate contours are shown in black and are labelled in events per year per arcminute$^{2}$. For comparison, the green contour indicates a rate of 1 in those units.
Model 1: ``Light'' disk, ``Light'' exponential bulge. Model 2: ``Heavy'' disk, ``Heavy'' exponential bulge. Model 3: ``Light'' exponential bulge, ``Heavy'' disk. Model 4: To explore the effect of the luminosity function, this model has a disk luminosity function and ``Heavy'' disk mass function for both the bulge and disk populations. Model 5: ``Light'' disk, power law bulge.
}
\label{rate_predictions}
\end{figure}

 The Angstrom (Andromeda Galaxy Stellar Robotic Microlensing) project is using five telescopes to conduct a high cadence microlensing survey
of the bulge of M31. The collaboration is led from the United Kingdom and has members in Korea and the United States.
M31 is considered a good target for a lensing survey for the reasons
already given in Section \ref{milky_way_surveys}. In addition, the bulge can be covered in a single telescope field. The main aim of this survey is to detect microlensing 
events with durations above 1 day due to low mass stars and brown dwarfs in M31.
 Previous M31 surveys have not observed as frequently as we are doing and so were not
 sensitive to these short timescale events and hence to lens masses as low as we 
are sensitive to. We are performing difference analysis in ``real time", i.e. analysing one day's data in less than a day.
With sufficient statistics we should
be able to provide valuable information about the low mass end of the stellar
 Initial Mass Function (IMF) in Andromeda \citep{2001MNRAS.323L..31B}, and about the distribution of mass in the bulge.
 The data will also help to constrain the parameters of the galactic bar in the
 central
 regions of M31 \citep{1987A&AS...69..311W,1994ApJ...426L..31S}
 The most recent investigation of the central regions of M31, a photometric analysis of infra-red survey data from ``Two Micron All Sky Survey" (2MASS) 6X program was \cite{2005AAS...20713503B} and then
 \cite{2007ApJ...658L..91B}, which found that the axis of the ``boxy'' bulge is
 oriented at approximately $10^{\circ}$ to the major axis of the outer disk, and
 confirmed that M31 is a barred spiral, like the Milky Way.
In \cite{2006MNRAS.365.1099K}, the stellar bar is modelled as a triaxial elliptical distribution
with one axis having a greater scale length than the other two by a factor $1/0.6$.
Since this long axis is predicted to be misaligned with the major axis of the disk,
this should enhance the event rate distribution in two of the four quadrants defined
with respect to the disk orientation. This has been modelled in \cite{2006MNRAS.365.1099K} for different combinations of disk and bulge models, to produce spatial distributions of the predicted event rate for each, shown in Figure \ref{rate_predictions}.
 The nearly edge-on
orientation of Andromeda to us ($77^{\circ}$) should accentuate any
 asymmetry between the near and far sides of the galaxy in event rate
 or in the positions of lenses and sources. The far side would be predicted to
produce more far-disk sources and bulge lenses, whereas the near side would be predicted to
produce more bulge sources and lenses in the near disk due to the varying lines of sight through the galaxy. Due in part to the differences in mass/luminosity functions between the bulge and disk, the angle at which the galaxy is viewed, and the probable existence and particular orientation of a bar, the number of events expected on the far side of the galaxy is higher. These effects are enhanced over a given angular field of view, due to the foreshortening effect of the almost edge-on galaxy. There is also an additional effect due to absorption of the light from the far side of the galaxy by dust, which may reduce the number of detected events from the far side, or even change their distribution.

Our calculations \citep{2006MNRAS.365.1099K} estimate that
approximately $27$ events per season would be detected by the Angstrom Project
over the field of view of the BOAO telescope, which is $11$ arcmin$^{2}$, if an IMF which is truncated 
at the hydrogen burning limit is assumed. (This corresponds to the ``light'' or ``standard'' stellar IMF model used in \cite{2006MNRAS.365.1099K}).
Of these events, about half last less than $10$ days. However, if the
 IMF contains a substantial brown dwarf population, as in the ``heavy'' IMF models, the predicted 
yield rises to about $64$ events per season. Of these, about $80\%$ have timescales of 
less than $10$ days.

\subsection{Angstrom Telescope Network}
\label{Angstrom_Telescope_Network}
\begin{figure}[!ht]
\vspace*{9cm}
  \includegraphics{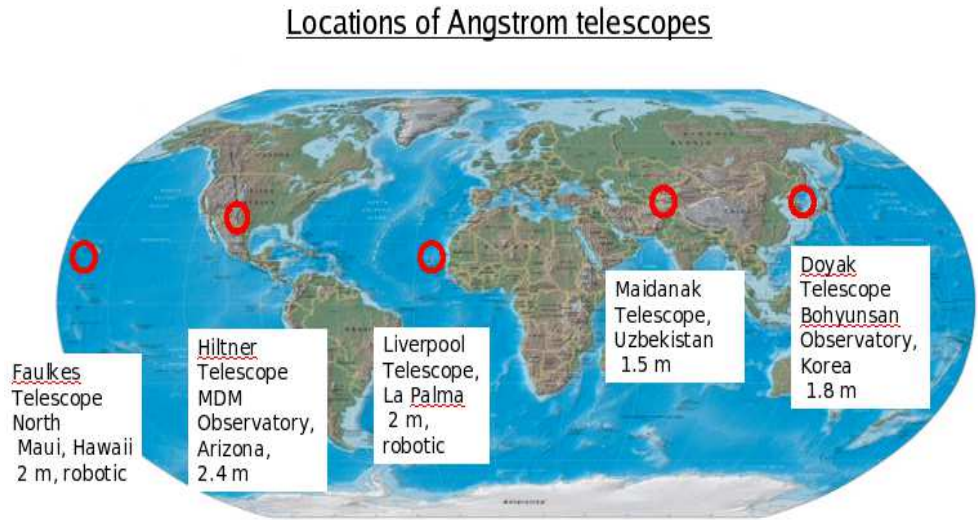}
\caption[A Map showing the locations around the Earth's circumference of the Angstrom telescopes.]{A Map showing the locations around the Earth's circumference of the Angstrom telescopes.
}
\label{telescope_locations}
\end{figure}

 Angstrom has used five telescopes in the Northern Hemisphere, 
fairly well spaced in longitude to provide good time coverage. These are:
 the Liverpool Telescope (henceforth ``LT''), on La Palma in the Canary Islands, the Faulkes 
Telescope North (FTN) on Maui in Hawaii, the Hiltner telescope on Kitt Peak in
 Arizona, the Doyak telescope at the Bohyunsan Observatory in Korea and the Maidanak
telescope in Uzbekistan. 
The LT and FTN telescopes are both 2 metre diameter 
robotic telescopes, the Hiltner is 2.4 m, the Doyak is 1.8 m and the Maidanak is 1.5 m.
 As can be seen from Figure \ref{telescope_locations}, the telescope locations are reasonably evenly spaced around the circumference of the Earth,
 which gives the best chance to get evenly spaced data points several times over the course of a day.

 Data were only obtained from the Maidanak telescope in Uzbekistan during the fourth season, but due
 to the excellent seeing properties of Mt. Maidanak, on which the telescope is sited, the data are
 of high quality.
Data from the non-robotic telescopes are pre-processed and difference imaging is performed on them in Korea, while data from the robotic (LT and FTN) telescopes are processed in the U.K, where all the data from the different telescopes are combined into lightcurves. In addition to new data from the telescopes described above, Angstrom has access to the POINT-AGAPE project image data. These have been re-processed using the Angstrom data analysis pipeline. The use of these data is of great benefit to us as it significantly increases the length of the temporal baseline of our data, which in turn helps to distinguish true lensing events, which usually do not repeat, from variable stars, which repeat. The fields observed by the POINT-AGAPE and MEGA collaborations partially overlap with the fields
 which the Angstrom collaboration are currently observing. 
 The Angstrom fields are more centrally concentrated than the
POINT-AGAPE fields which have a gap in the very centre.
However in the areas which do overlap the benefits to the analysis of the data are
 very significant, albeit with sparser time sampling than Angstrom. 

 A graphical summary of the Angstrom collaboration pipeline is given below in Figure \ref{angstrom_pipeline_flow}.
Already preprocessed data from all five telescopes are sent to Liverpool for processing. The stages performed include PSF quality control, removal of cosmic rays, defringing, alignment and stacking. Next the difference imaging is performed on both the Angstrom and PA data, and variable sources are detected in the difference images.
The lightcurves of variable objects which are produced are then fed into the candidate selection pipeline developed for this thesis, which selects the microlensing (and variable star) candidates. The same lightcurves educate the Angstrom
Project Alert System \citep{2007ApJ...661L..45D} (see Section \ref{APAS}). Any promising alerts can then be distributed to the wider astronomical community who may make follow up observations of interesting transient targets.

\begin{figure}[!ht]
\vspace*{20cm}
   \includegraphics{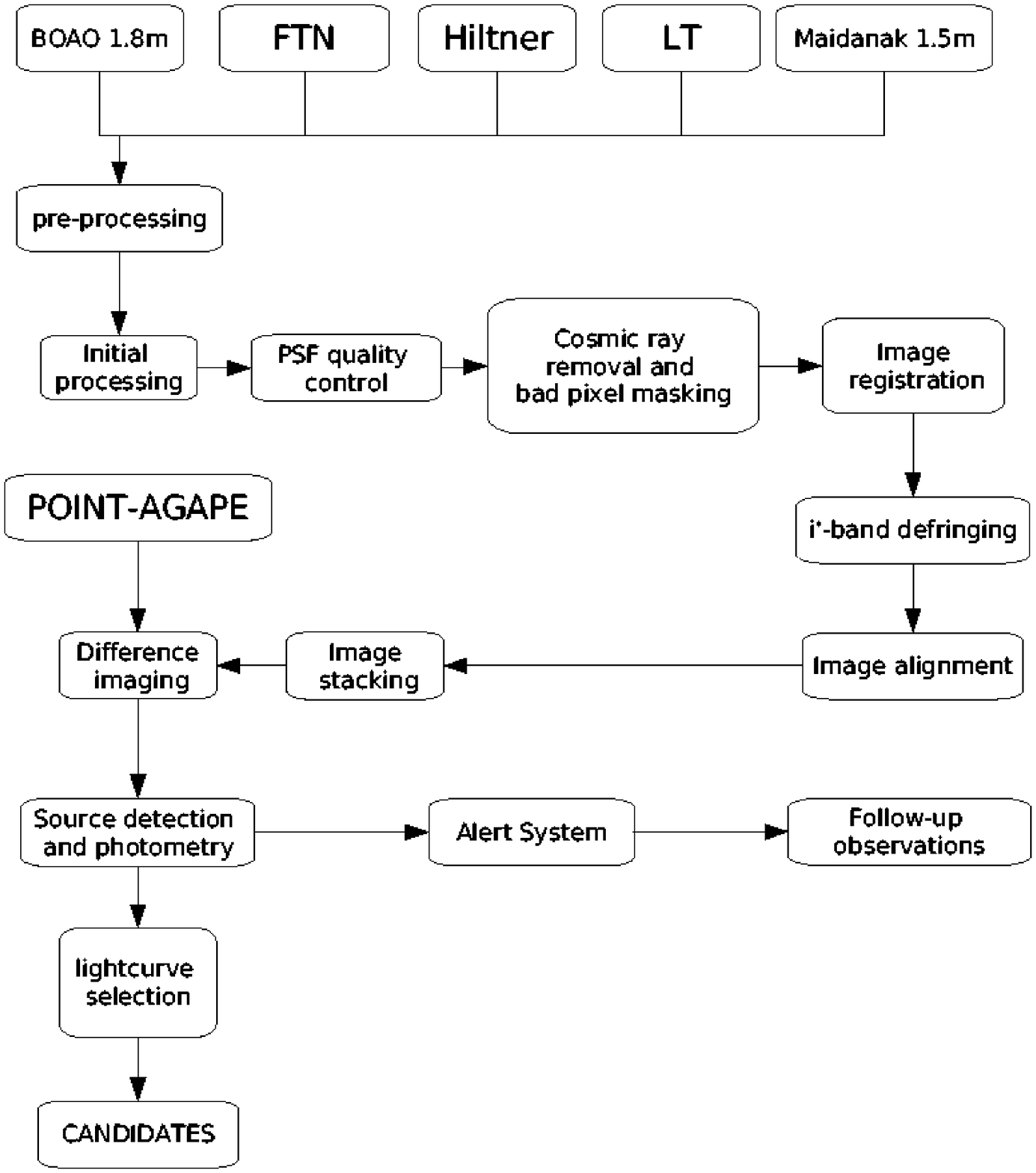}
\caption[A figurative representation of the Angstrom collaboration pipeline.]{A figurative representation of the Angstrom collaboration pipeline. This plot is
described more fully in Section \ref{Angstrom_Telescope_Network}.
}
\label{angstrom_pipeline_flow}
\end{figure}

\subsection{Observing Strategy}

As mentioned above, many of the predicted microlensing events have
 timescales of 10 days or less. Therefore we are taking data at the rate 
of several ($\sim 2-3$) epochs each night when possible. This should enable
 us to characterise events with a timescale of only a few days. 
Observations were initially made in multiple wavelengths in order to
 assist with the identification of events. This was because microlensing events
 are achromatic whilst non-microlensing events are likely 
to be chromatic. The LT observed in Sloan i' and r' filters, the 
FTN in Sloan i' and Bessel R. The Doyak telescope is using Johnson-Cousins 
I and the Hiltner uses Gunn-Thuan i. The Maidanak telescope data are also Johnson-Cousins I band.
 The differing sizes of the CCDs used on the Angstrom telescopes are shown in Table \ref{angstrom_ccd_sizes}.
The Maidanak telescope has a $4096$x$4096$ pixel Fairchild CCD camera, which gives a pixel size of $0.26^{\prime\prime}$.

\begin{table}
\caption[Table showing the various sizes of the CCD chips used in the Angstrom Survey.]{Table showing the various sizes of the CCD chips used in the Angstrom Survey.}
\center{\begin{tabular}{|l|l|l|}
\hline
\hline
 Telescope & Mean CCD size (arcmin) & aspect ratio \\
\hline
 LT, FTN & $4.6$ & $1$    \\
 Doyak     & $11$  & $1$ \\
 Maidanak  & $18$  & $1$ \\
 Hiltner   & $4.5$ & $1.52$ \\
\hline
\end{tabular}}
\label{angstrom_ccd_sizes}
\end{table}

Each exposure in the sequence is kept short (no more than $200$ seconds) to avoid saturating the M31 bulge. In the pilot season, the total exposure time of each epoch was $10$ minutes, but in subsequent seasons this was increased to $30$ minutes in order to increase our sensitivity to microlensing anomalies.
Unfortunately, the LT r and FTN R band data proved at an early stage 
to be of lower quality, with fewer variable objects visible.
For a fixed allocation of telescope time, i' band data give a greater sensitivity to
variable stars, thus allowing better exclusion of false positives than by using a 
sparser time sampling spread between two bands (for e.g. (i' + r) or (i' + R))
Unfortunately this meant that doing a comprehensive check as part of the candidate selection
process for achromaticity was not always possible, but the pipeline was written in such a way that
if the data existed, the achromaticity condition could be utilised, perhaps in this project in future, or in future surveys.

\subsection{The Angstrom Project Alert System (APAS)}
\label{APAS}
As part of the Angstrom Project, a real time alert system has been developed \citep{2007ApJ...661L..45D} to enable the alerting of interesting developing events to the astronomical community so that they may be followed up with other telescopes, as has been done for several years by the Milky Way surveys such as OGLE and MOA. The successful operation of this
kind of system requires the Angstrom data to be processed in ``real time''. This has been achieved and in the observing season 2007/2008 real alerts were issued on promising looking events \citep{2007ATel.1192....1D}. A more detailed description of the
operation of the APAS will be given in Section \ref{angstrom_alert}.

\section{Aims of thesis}
\label{aims_of_thesis}
 In this Thesis I will describe the Angstrom Project, the work performed by myself as part of the Angstrom Project
and present the scientific results which have been produced as a result of this work so far.
The main aim of this Thesis is to develop a Candidate Selection Pipeline for the Angstrom Project which can select, from the totality of variable object lightcurves produced by the ADAP, the candidates which are most likely to be caused by microlensing, while simultaneously
selecting as few objects as possible which are more likely to be due to
variable stars, classical novae, supernovae or other non-microlensing phenomena. 
If this primary aim is achieved, then whatever statistical analysis that may be possible of any lensing candidates will be performed. For example, it would be interesting to examine the spatial distribution of candidates within M31 and the distribution of lensing timescales. Any 
particularly interesting lensing candidates will be investigated in more detail.
It is expected that a large fraction of the variable objects found will be due to variable stars, and so any examination of the statistical properties of these objects which is possible would be useful. Again, the spatial distribution of variable star candidates across the LT field would be interesting to investigate, for its own sake, and because of its relevance to the spatial distribution of microlensing events, which is predicted to be detectably different, given enough detected events.


\chapter{Liverpool Telescope Pilot Season}
\label{Chapter_2}
\section{Introduction}
This Chapter describes the first ``pilot'' season of Angstrom using the
Liverpool Telescope (henceforth ``LT'') data
 and the analyses that were performed using them. The robotic operation of the LT is also described.

\section{Robotic Telescopes}

The LT and FTN telescopes are both ``robotic''. This means that no human operator is required on the mountain at the telescope. Robotic telescope data comprised 74.9\% of Angstrom
unstacked image frames up to the end of Season 3 (2006/7). The LT and FTN are built to identical designs. The LT is sited at the international Observatorio del Roque de los Muchachos on the summit of the island of La Palma, in the Canary Islands.
The Faulkes Telescope North is located on the mountain of Haleakala, on the Hawaiian island of Maui. The altitude is 10,000 feet above sea level meaning that the telescope sits above the cloud layer for the majority of the time.
 The mean seeing at this site is around one arcsecond.
Both telescopes are enclosed within a protective dome which is designed 
 to open fully rather than having a narrow slit through which to observe.
 This innovative dome design is referred to as ``clamshell''.
When the dome is fully open, it is possible to move the telescope rapidly from one 
side of the sky to the other down to an artificial horizon level of 20$^\circ$ above the horizontal
 with a slew rate of at least 2 degrees 
per second. This ability to slew rapidly is very important in 
 relatively new areas of astronomy where 
a rapid response is required.
 Some of the areas of research in which robotic telescopes are making an important
 contribution are long term monitoring programs, fast transients such as novae/supernovae and gamma 
 ray bursts.

Robotic telescopes do not require an operator to constantly be at 
the telescope deciding which target to observe and for how long 
\citep{1995_BodeMF}. The telescopes are designed to operate autonomously, 
as if there were an operator present \citep{2001_SteeleIA}.
 This leads to considerable cost and convenience benefits,
 as pointed out by \cite{1997_BodeIAU}.
 However, it is also possible for an observer to take 
direct control of one of the telescopes remotely over the internet in order to direct 
it to observe particular targets. This capability is used by users in the United Kingdom 
such as
schools, through the ``National Schools Observatory'' to learn about telescopes and 
observational astronomy. A significant fraction of the observing time of the LT is
 allocated for these educational purposes.

\subsection{Instruments}

On the LT there are currently four operational instruments. These are:
 ``RATCam",  an optical CCD camera, 
 ``SupIRCam", an infrared array camera,
 ``RINGO", an optical polarimeter, and 
 ``the Meaburn Spectrometer".

The Angstrom project uses RATcam, which has a $2048$x$2048$ CCD which gives a pixel size of $13.5$ microns, which corresponds to $0.14^{\prime\prime}$.
Most of our observations are made in Sloan i' band, which has an approximate wavelength range of 
$6930$ - $8670$ \AA , but observations have also been taken in Sloan r' which spans $5560$ - $6890$ \AA .

\subsection{Pre-processing at the LT}

 As well as the observations being scheduled and carried out 
autonomously, the data are preprocessed at the telescope and then
placed on a website, where it may be accessed remotely by Principal Investigators
 \citep{2000_SteeleIA}. In order to do this, regular flat field observations
 are scheduled automatically.
The operations which are performed automatically at the telescope are:
 bias subtraction, overscan trimming, and flat fielding.

From operational experience, it is known that RATcam does exhibit a small
 gradient down each column of the image,
and a first order fit to this is required to remove it.
 Overscan areas are trimmed from the image leaving a $2048$x$2048$ pixel image.
 Cosmic ray rejection is not performed, nor bad pixel masks applied at the telescope.
 Users of the telescope who are attempting to produce accurate photometry
will need to know exactly what masking has been applied, so it is considered better 
that these functions are performed by individual projects.
However, the Angstrom project has generated bad pixel masks as a necessary part of our pipeline 
and have made them available for general use.
At the moment dark subtraction is not part of the LT preprocessing pipeline,
 (although the facility does exist in the reduction 
pipeline to add it at a future date), as it has been determined experimentally that at the temperatures the telescope experiences in normal operation dark current is not significant.
The optical beams on the LT and FTN are vignetted to a small degree by the filter wheel. Each filter has a slightly different effect.
This vignetting is usually well removed by flat fielding.
De-fringing is currently not performed at the telescope, although master fringe 
frames, which have been produced by stacking many deep integrations, are provided to
enable users to perform their own de-fringing.

\subsection{Scheduling}

 A sophisticated automated scheduling system takes all the information
about the projects which have been accepted for observation, for e.g.
 order of priority and observing requirements and decides which target
 should be observed \citep{1998_SteeleIA_ProcSPIE}, given the current 
observing conditions and 
 astronomical activity (e.g. the occurrence of gamma ray bursts). This 
 leads to a more efficient use of telescope time, and the possibility  
 of quickly turning the telescope to observe targets of opportunity as 
 required.

\section{Introduction to Preprocessing}

At the very start of the author's work on this project some time was spent becoming 
familiar with the processes which were necessary to perform preprocessing on the raw
unstacked fits images and to combine them into stacked images. This also allowed
familiarity to be developed with some of the relevant IRAF \citep{1986SPIE..627..733T}
\footnote[1]{IRAF is distributed by the National Optical Astronomy Observatories,which are operated by the Association of Universities for Research in Astronomy, Inc., under cooperative agreement with the National Science Foundation.}
routines and to learn some
 of the basics of C-Shell programming. This was to prove very useful for the rest of 
the PhD project.

\section{Preprocessing}

What might be called ``initial'' preprocessing had already been performed before 
the author began working on the FITS images. This meant that processes such as 
de-biasing, dark current subtraction and flat fielding had already been 
performed.

The first few months of study were spent learning to use basic IRAF routines
such as ``IMREPLACE'',``DAOFIND'',``XYXYMATCH'',``GEOMAP'' and ``IMCOMBINE'',
 which were necessary
to perform the preprocessing on the LT
images of M31. The processes performed by the eventual script were:
bad pixel masking, bright object finding, calculation of image relative translations 
and rotations (and possibly scalings) and image stacking.
 Some quite extensive experimentation was required to find 
suitable parameter ranges for these routines which would work for all images
 and produce useful output which could then be used by the next step in the
chain. 
 The actual operations performed on the images were as follows:

 IMREPLACE was used to replace the values of groups of individual pixels with
 the number -1. This was chosen as a minimum value of 0 was set later, in 
IMCOMBINE, resulting in values less than 0 being ignored. The groups of pixels
 that it was necessary to replace were a strip of 5 pixels along the bottom of
 each image which were bad due to an error in the previous processing, and a
 vertical strip 1 pixel wide, starting at pixel coordinate (753,1) and ending
 at (753,2048), which appeared to be a bad column on the CCD.

Next, DAOFIND was used to fit Gaussian peaks to areas in the images that were 
more than $3.5 \sigma$ brighter than their surroundings. This was done to find the
 positions of bright stars in the image that could be matched to the 
equivalent stars in other images. The value of $\sigma$ was one of the variable 
parameters which had to be found by experimentation. Since some of the images
 in the data set had longer exposure times than others, the average brightness
 of the images, and hence the value of $\sigma$ which produced the best results,
 was not constant for the whole data set. This had to be borne in mind when 
processing the data automatically. The first images taken of M31 were 
generally longer and hence fewer each night, but due to telescope tracking
 difficulties, a large proportion of these were deemed unusable due to the
 telescope moving during the exposure. To counteract this, by reducing the 
number of unusable images, the exposure time of later images was reduced.
 Therefore, the ``early-long" and ``late-brief" images had to be processed 
separately using different values of $\sigma$. The values of $\sigma$ were chosen so
 as to produce between $15$ and $40$ hits. By visual examination of the output
 from DAOFIND in the form of lists of the coordinates of the ``stars" found, 
it was possible to see that DAOFIND was finding relatively large numbers of 
hits in the areas of the images where the core of the galaxy is. These are 
assumed to be in most cases merely random fluctuations in the background 
brightness, and not stars. In an attempt to ignore these, a parameter in the
 function XYXYMATCH which sets the minimum separation allowable for ``stars''
 was set to
 $150$ pixels. This eliminated almost all the spurious hits as they tended to
 be clustered together quite tightly, and hence much closer together than the
 real stars, which were separated by significantly more than 150 pixels in
 most cases. In any case it is not necessary to find \emph{all} the stars in 
the image, just enough to be able to match them to other images. i.e., a 
minimum of one to find translations, and two to find any rotation between
 images. A few more stars than this are useful, however, to provide error
 checking and to increase the accuracy of the matching.

XYXYMATCH was then used to take the lists of stars produced by DAOFIND and to 
find matches between the stars in each of the images and thereby calculate the 
transformation between each star in an image relative to a reference image in 
each coordinate direction.

The output from XYXYMATCH could then be input into the GEOMAP function, which
 uses the coordinate transformations in a polynomial 
 fit to calculate a mean translation in each coordinate direction and also a 
rotation of each image relative to the reference image. 

Originally, the rotation of the images was fixed first 
as a separate step using the IRAF function ``rotate'', but in the final pipeline, XYXYMATCH is used to calculate the required rotations
 (and translations) and then ``GEOMAP'' is used to perform all the transformations simultaneously.

 The translations produced as output of GEOMAP were originally used as input to the 
IMCOMBINE function. This takes all the images in an epoch and lays them on top
 of each other, shifting them horizontally and vertically by the translations 
calculated previously. The values of all stacked pixels were averaged, and the
 brightness of the image was scaled by the median pixel value in a square area
 of the image centred on the middle, the bottom left corner being at pixel
 ($724$,$724$) and the top right being at ($1324$,$1324$). The function 
``crreject'' within IMCOMBINE was used to reject pixels that are more than $3$
 times the expected CCD readout noise ($\sigma$) different than the area 
immediately around them. This is done in an attempt to remove most cosmic 
rays.

In order to facilitate the automatic processing of large numbers of images,
IRAF commands from within UNIX were inserted within a Unix script to perform 
the running of IRAF functions automatically. This enabled the whole 
directory of images to be processed in one long computing run, rather than 
doing each epoch semi-manually.

 When this automated script had been perfected, it was passed over to 
Dr Andrew Newsam (Liverpool JMU), who incorporated the principles and processes contained 
within it into his own more extensive script, called ``stackEpochLT.csh'' 
which tied together pre-processing and photometry of the pilot season data.

\section{Investigating the Common Overlap Region\label{commonoverlap}}

During the LT pilot season there were major problems with the pointing accuracy of the LT relative to the intended
 (specified) field. Epoch images had considerable translational and rotational errors
 relative to one another (which could be of the order of hundreds of arcseconds and tens of degrees) which had the effect of severely reducing the common overlap between them 
which could be used to produce time-series of difference imaging data for the purposes of finding 
time varying sources.
The reason that this investigation became necessary was that the original ISIS image subtraction
 package was unable to handle image masks. Therefore the region on which ISIS was able to produce
lightcurves was the region in which every image in the stack overlapped, which was reduced every time a further misaligned image was added to the stack. At that time, the
question of what was the best available compromise between overlap area and number of time points (equivalent to the number of images in the stack) in each lightcurve was one that needed to be clarified. This compromise would be the one in which the total amount of usable data was optimised.

 In order to investigate this issue, again using IRAF software, an image was produced for one example reference image 
which showed the pattern of overlaps, and the relative ``image density'' of various regions of the 
combined image (i.e. how many images were
coincident at that point of the super-image). Visual inspection of this image implied that the number
 of overlapping images fell off very rapidly as one moved away from the region in which all the images
 overlapped, implying that only a small gain in coverage area might be made if one relaxed the 
constraint on the number of overlapping images. This was investigated further more analytically, as
 described below.

In order to discover whether it was possible to make a gain in the area covered by all images 
considered, by removing some (hopefully small) number of images that were outliers in position 
and/or rotation, and thereby the cause of a disproportionate reduction in overlap area, an analytical
 study was made of the distribution of images with respect to one another in displacement and angle. 
Several scripts were developed with the aim of deciding firstly whether there were any images which 
were having a particularly deleterious effect on the overlap area, and if so, which ones and how many
 should be discarded. The ``Quality Factor'' which needed to be optimised relates to the total amount 
of microlensing information available in the images, and is equal to the number of images used 
multiplied by the overlap area remaining. i.e. Quality Factor $QF = {N}\cdot{A}$. The idea was to
 discover which of the images were particularly bad outliers, and then remove these one by one and
 calculate the $QF$ at each stage to discover whether it had a maximum and if so, at which point.
 It was important to sort the images in order of ``centrality'' since the remaining area after $n$ 
images depends sensitively on the order in which these are added to the stack. To maximise the 
remaining area and hence the quality factor, the best images must be added first and the worst
 images last.

\begin{figure}[!ht]
\vspace*{12cm}
   \includegraphics{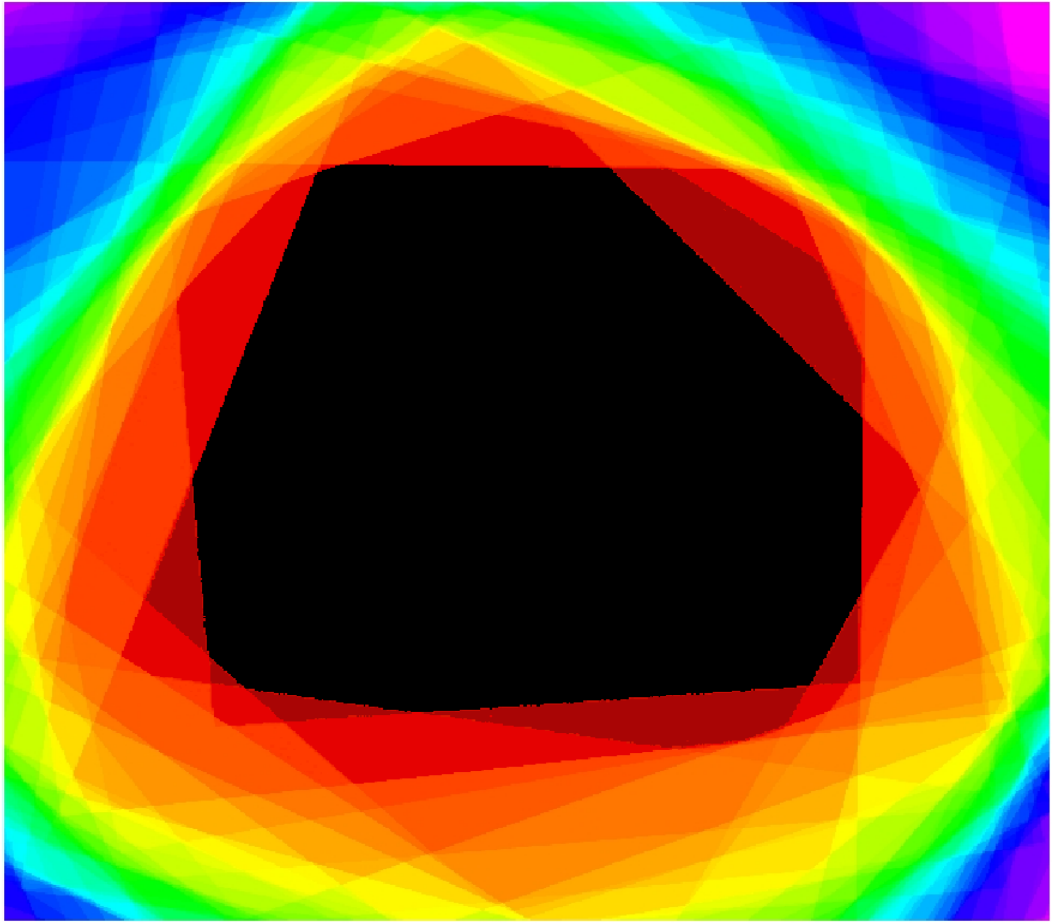}
\caption[The overlap image from August $2004$ (all $49$ images collected up to that point included) for LT i-band reference image LT\_i\_$20040816$\_$001$.]{The overlap image (all $49$ images collected up to that point included) for the LT i-band reference image LT\_i\_$20040816$\_$001$, showing the variation in density of images over the overlap region. The colours encode the numbers of images overlapping in each area, which generally increase towards the centre, in which all $49$ images overlap in the black region. In the magenta regions (for e.g. the top right corner, only the reference image exists.)
}

\label{overlapimage}
\end{figure}

  The most accurate way (i.e. requiring no approximations) to do this would be the ``brute force'' method- 
trying every combination of reference image (the image upon which the overlap area of all other images
 is calculated) and other images. A script was written to attempt to accomplish this, starting with
 the object (star coordinate) files for each image previously found by starfind, and calculating 
from first principles the rotation, the $x$ and $y$ shift, and hence the overlap area.
However it was found that the computing time using this method was too great,
so it was clear that an alternate method would be better. Therefore, a method was developed 
whereby for each reference image a weighting factor would be calculated for every other image based 
on that image's displacement in the two coordinate directions and angular rotation relative to the 
reference image. The total weighting $W_{T}$ was calculated by calculating geometrically the area of ``overhang'' 
i.e. the non-overlap of the image considered with the reference image due to the displacements in the
 three directions independently, and then multiplying these together as in Equation \ref{common_overlap_weights}.

\begin{equation}
W_{T} = W_{x}W_{y}W_{\theta}
\label{common_overlap_weights}
\end{equation}

where the individual weights in the three directions $W_{x}$,$W_{y}$ and $W_{\theta}$ were defined by the Equations \ref{common_overlap_weights_x}, \ref{common_overlap_weights_y} and \ref{common_overlap_weights_theta} respectively.

\begin{equation}
W_{x} = |{\frac{\Delta{x}}{x_{im}A_{\hat{\phi}} }}|
\label{common_overlap_weights_x}
\end{equation}

\begin{equation}
W_{y} = |{\frac{\Delta{y}}{y_{im}A_{\hat{\phi}} }}|
\label{common_overlap_weights_y}
\end{equation}

where $x_{im}$ and $y_{im}$ are the dimension of one side of an image in whatever units are used (here $x_{im}$ = $y_{im}$ = $2048$ pixels)

\begin{equation}
W_{\theta} = 2 \sin (|{\Delta{\phi}}|) \cos(\Delta{\phi})  / ( \sin(|{\Delta{\phi}}|) + \cos(\Delta{\phi}) + 1.0 ) ^2
\label{common_overlap_weights_theta}
\end{equation}

where $\Delta{\phi}$ is the rotation angle in radians minus $\pi/2$.
 In order to make the three 
terms have the same maximum possible magnitude, the $x$ and $y$ terms were scaled by a factor $A_{\hat{\phi}} = 2/(2 + \sqrt(2))^2$ so that when 
$\frac{\Delta x}{x} = A_{\hat\phi}$, (where $A_{\hat\phi}$ is the maximum possible overhang 
area due to rotation which occurs at $45$ degrees if the sides of the original square image are aligned
 with the coordinate axes), the weight $W_{x}=1$. In other words the $x$ and $y$ weights could correctly have a larger
 effect since the overhang areas can theoretically be equal to the total area of the image, whereas the $\phi$ weight cannot.
     Once the single iteration of the original overlap area script
had been run, it was realised that once the translations/rotations between one reference image and 
all the other images were known, the translations/rotations between any other pair of images could 
be calculated by geometry. i.e. to go from image B to image C with image A as a reference, the reverse of the AB transformation is used followed by the AC transformation. This method was then used to
 calculate the transformation for all other combinations of images in order that the weightings for 
each image pair could be calculated. Once these weightings had been calculated, the median weighting
 for each reference image was calculated. This gave information about the geometrical relationship 
of the reference image to all the other images. A high median weighting in this case implied that the
 particular reference image is widely separated from the centroid of all the other images. In addition the 
distribution of these medians was plotted to give some idea whether there were a few outliers which 
were far from the bulk of the images or whether all the images were close together in median and hence
 no great improvement could be expected by removing the worst images. By calculating the medians the 
images could be ordered in (estimated) overlap area. 
   Then the images were stacked in the order of ``goodness'', and the remaining overlap which included
 ALL the stacked images was calculated, along with the quality factor.
 Then the images with the worst median weightings (i.e. the images that were, on average most separated
 from the bulk of the rest of the images) were removed from the stack one by one in order of ``badness'',
 and the overlap area and quality factor re-calculated. 
 When this was done for different reference images, the same two images were found to have the worst 
positions. These were the observation from 22.10.2004 followed by the one from 15.11.2004. For the 
``best'' reference image, which was taken on 21.08.2004, removal of these two images resulted in an 
increase in the overlap area of all the remaining images from $0.409$ to $0.448$ of the original image 
area, a fractional increase of $9.5$\%. For this reference image, the quality factor increased by $7.5$\%.
 Removing more than these two worst images from the original stack of $56$
 (i\_band) epochs resulted in 
a consistently lower quality factor, as expected.
The plot of overlapping images for the stack of images is shown in Figure \ref{overlapimage}, showing the kind of overlap region which existed at this time.
 Due to it having been shown above that the overlap region of all images was rather small (less than $50$\%)
and therefore much information would otherwise be lost in the edges of the image, the ISIS code was modified by us. It was decided that a more efficient method of utilising more of the
 information in the images would be to make a ``super image'' which was a
 rectangular frame of dimensions which were large enough to include the furthest out pixel on the rotated images so
 that all the image pixels from the original images were utilised and no
 information was discarded at this early stage. Later on in the process, in the early stages of the candidate selection pipeline,
cuts were used to discard lightcurves (equivalent to image stacks of
 one pixel) containing less than a certain number of data points (images).

    \subsection{LT/FTN difference image assessment+masking\newline after the second season}
\label{image_flaws}
In the first two seasons of LT and FTN images there were many problems,
 particularly with the FTN, which caused
several different kinds of degradation of the resulting difference images.
In some cases the flaws on the difference images were localised and could therefore be masked out, and in other cases they were more distributed across the whole image. In these cases, a decision had to be made on whether the image was usable at all.
After the end of the second season of data taking, i.e. in mid-2006, all the LT and FTN images to date were examined by eye, to assess the quality of the images and to assess their usability or to discover any flaws in the difference images. These images numbered $174$ LT and $167$ FTN, subdivided into LT i'-band ($130$), LT r-band ($44$), FTN i'-band ($85$) and FTN R-band ($82$). The problems/flaws with the images could basically be grouped into
ten categories, which are listed below.

$\cdot$ \textbf{Large Bipolar galactic centre}\newline
$\cdot$ \textbf{PSF mismatch}\newline
$\cdot$ \textbf{Dark and/or light blobs}\newline
$\cdot$ \textbf{Edge effects}\newline
$\cdot$ \textbf{Corner shading/lightening}\newline
$\cdot$ \textbf{Fringing}\newline
$\cdot$ \textbf{Stripes/lines}\newline
$\cdot$ \textbf{Smearing}\newline
$\cdot$ \textbf{Oil-Speckling}\newline
$\cdot$ \textbf{General Noisiness}\newline
$\cdot$ \textbf{Large PSF}\newline

For each difference image, these flaws were noted, where present, and in addition
the image was given a general classification as ``C'', ``P'', or ``U'', where these are defined as:
C  =  ``Completely or almost completely usable"
P  =  ``Partially usable-some significant fraction of the image can be used"
U  =  ``completely Unusable at that moment, although some may have been correctable".
In total over all images, $87$ received a ``P'' classification and only $6$ a ``U''. However, the
``P'' classifications were clearly clustered in the FTN R-band category- $46$ out of the total of $82$ images had this classification. Many of these were due to some extent of ``Speckling/stippling'', which was mostly concentrated in the brighter regions of the image, i.e. around the galactic centre.
The cause of this effect was thought to be an oil leak which occurred on the FTN which unfortunately sprayed the surface of the CCD chip, which could not be immediately replaced.

\begin{figure}[!ht]
\vspace*{10cm}
$\begin{array}{c}
\vspace*{11cm}
   \leavevmode
 \includegraphics{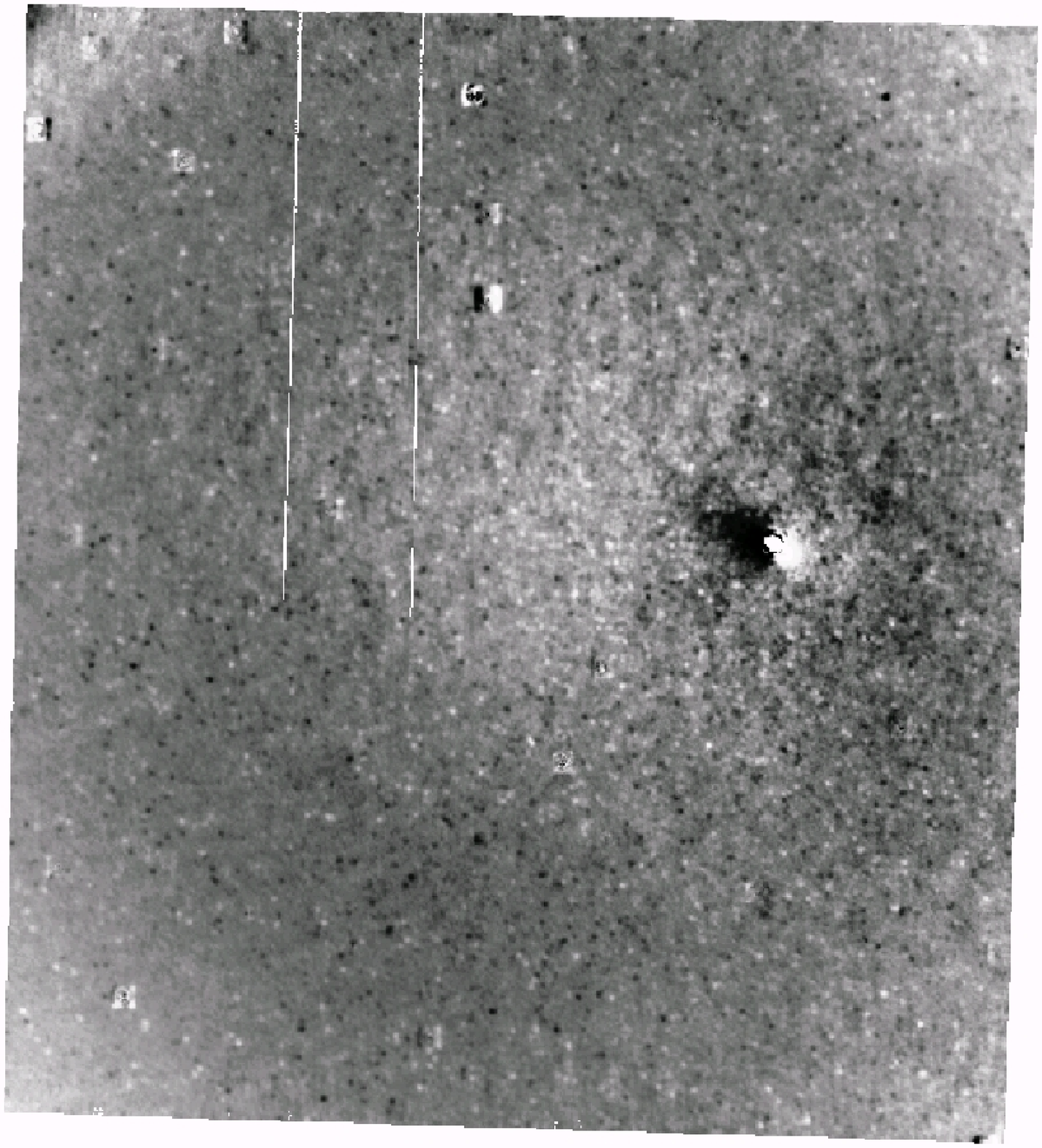} \\
\vspace*{1cm}
   \leavevmode
 \includegraphics{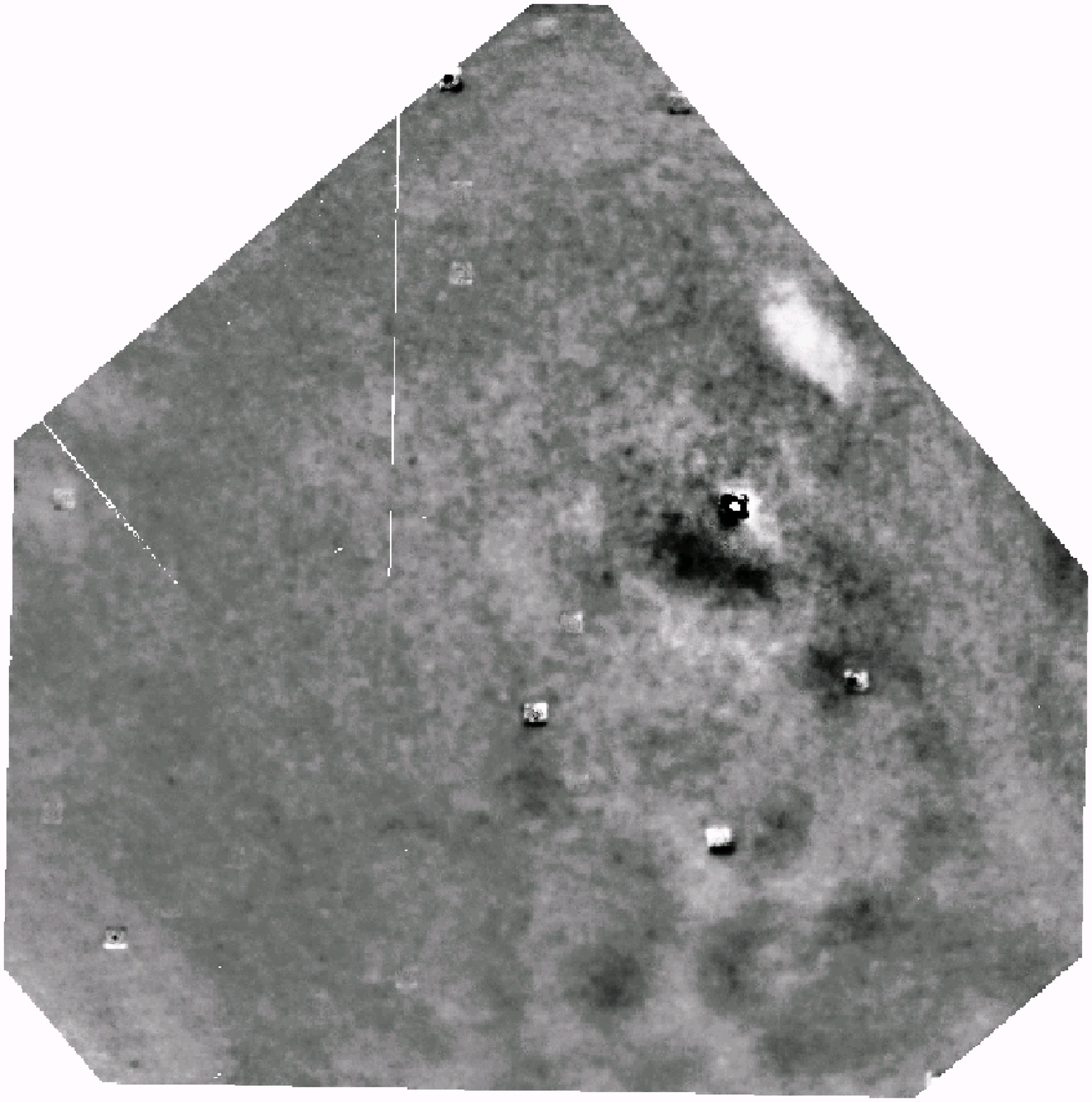} \\
\end{array}$
\caption[Examples of faults in the difference imaging (1).]{Examples of faults in the difference imaging (1). Top) This is a more serious example of a particularly common phenomenon in the difference imaging. A minor scattered light problem has lead to a poor fit to the overall 
galaxy model image. Bottom) The cause or causes of the noisy galactic centre to this image are not clear but may possibly be due to a poor flat field.}
 \label{quadrupole_bg_general_noise}
\end{figure}

\begin{figure}[!ht]
\vspace*{6cm}
$\begin{array}{c}
\vspace*{7cm}
   \leavevmode
 \includegraphics{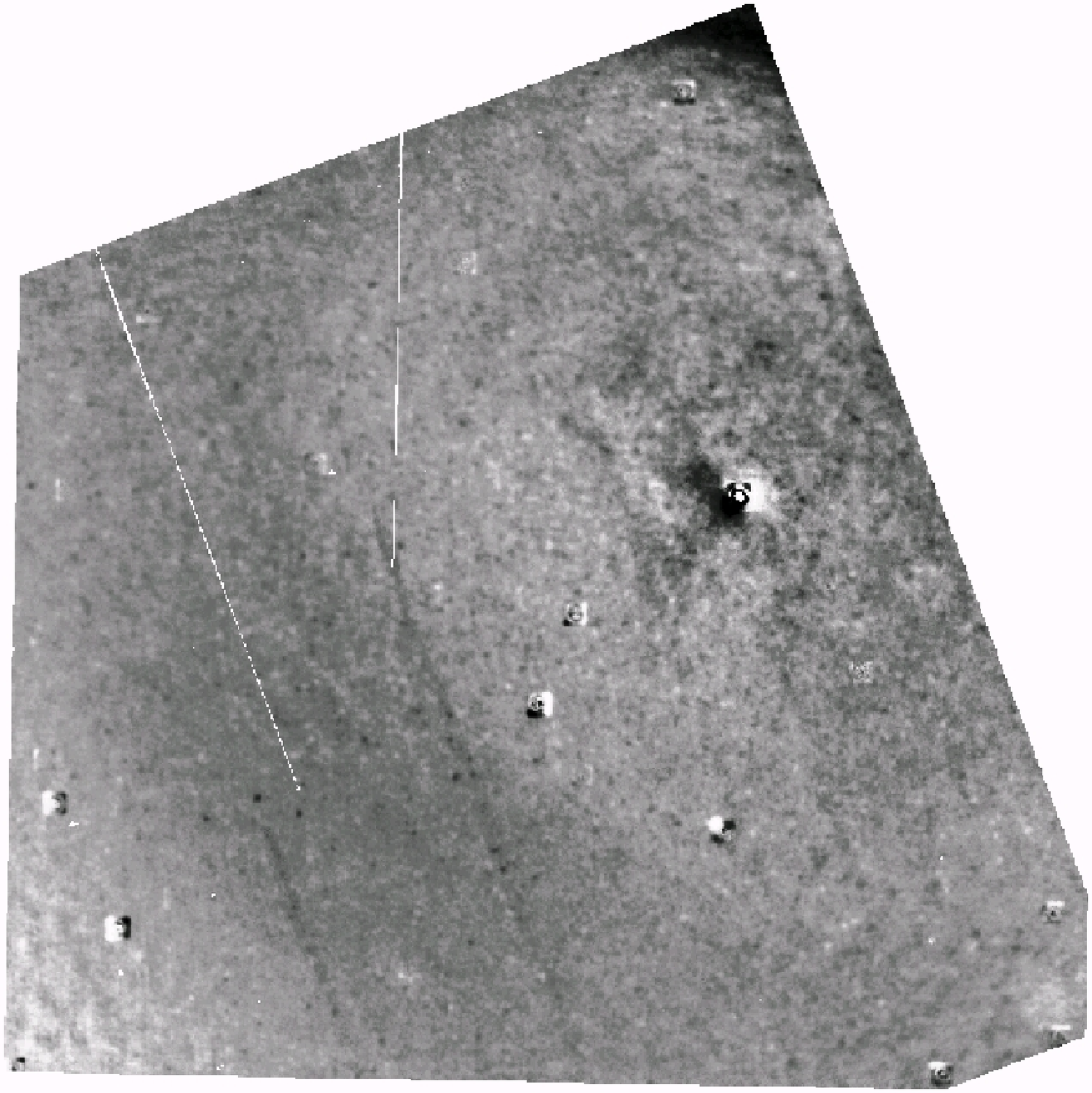} \\
\vspace*{7cm}
   \leavevmode
 \includegraphics{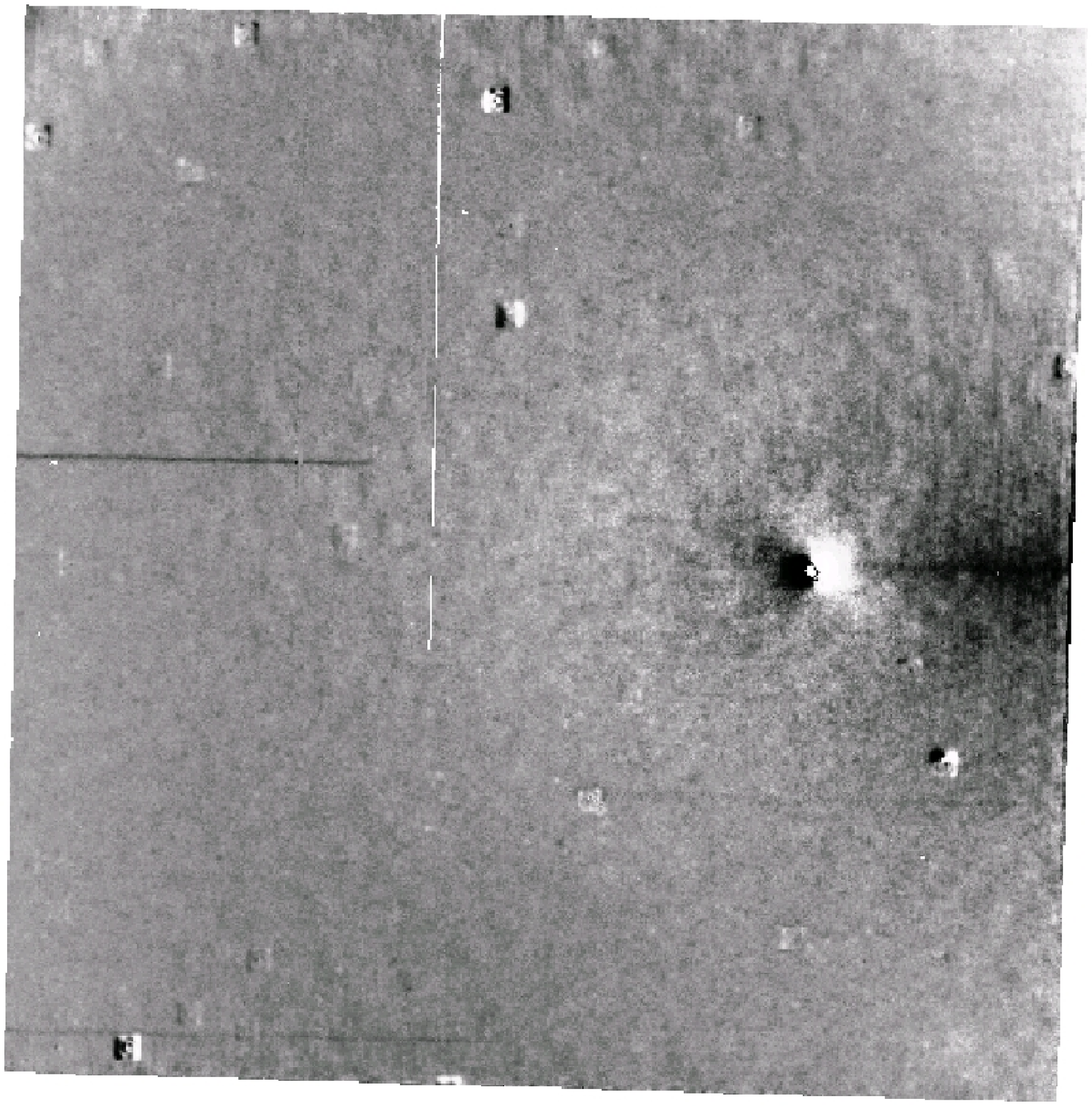} \\
\vspace*{1cm}
   \leavevmode
 \includegraphics{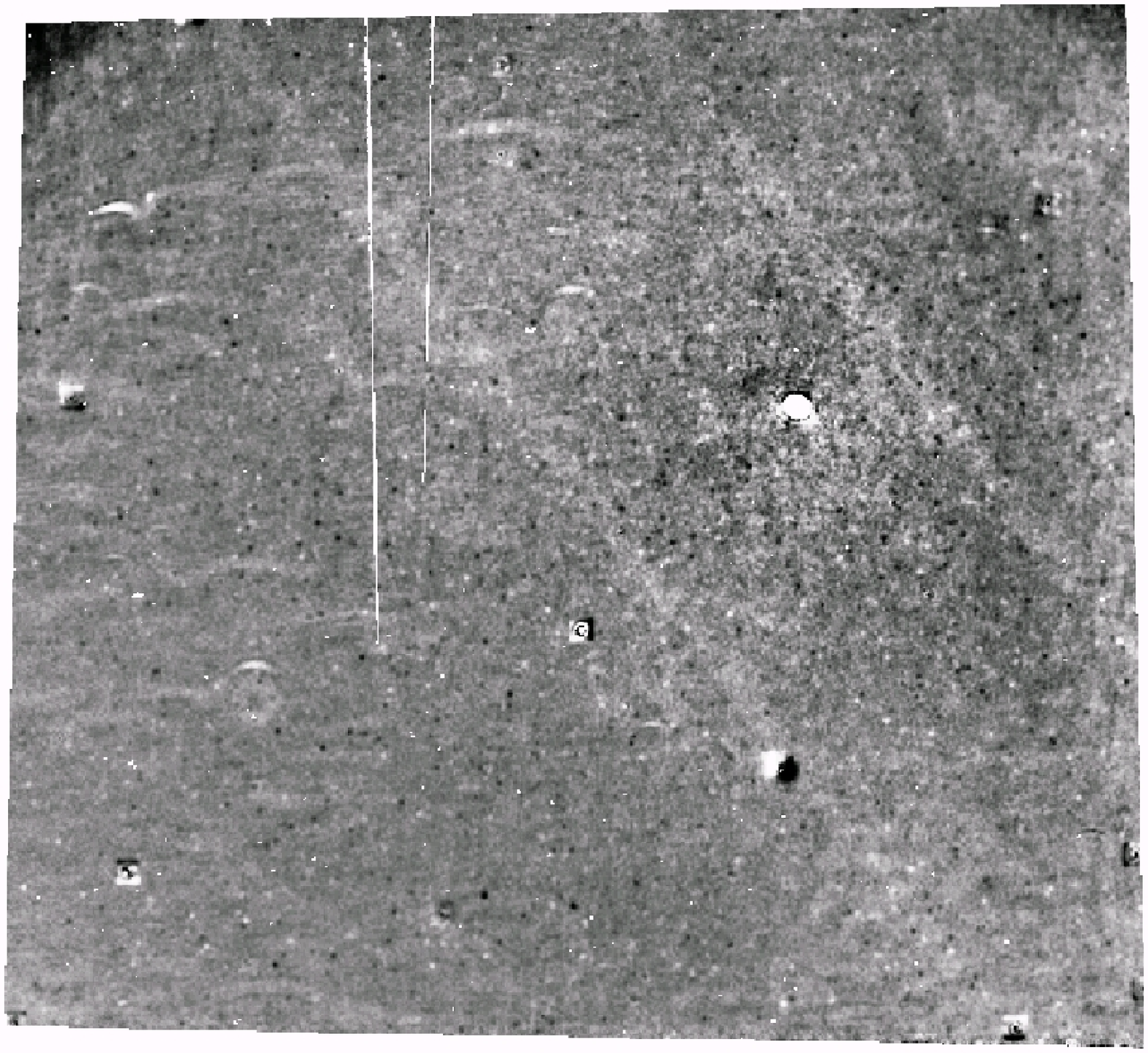} \\
\end{array}$
\caption[Examples of faults in the difference imaging (2).]{Examples of faults in the difference imaging (2). Top) The darker stripes in this image were caused by scattered moonlight. This problem was fixed on the LT by the installation of new baffles in $2007$. After this modification a very significant improvement in the data quality from the LT was seen. Middle) The dark horizontal lines are probably due a slight misalignment of the difference imaging. Bottom) The cause of this problem is currently unknown.}
 \label{dark_and_light_stripes_1_2_3}
\end{figure}

\begin{figure}[!ht]
\vspace*{10cm}
$\begin{array}{c}
\vspace*{11cm}
   \leavevmode
 \includegraphics{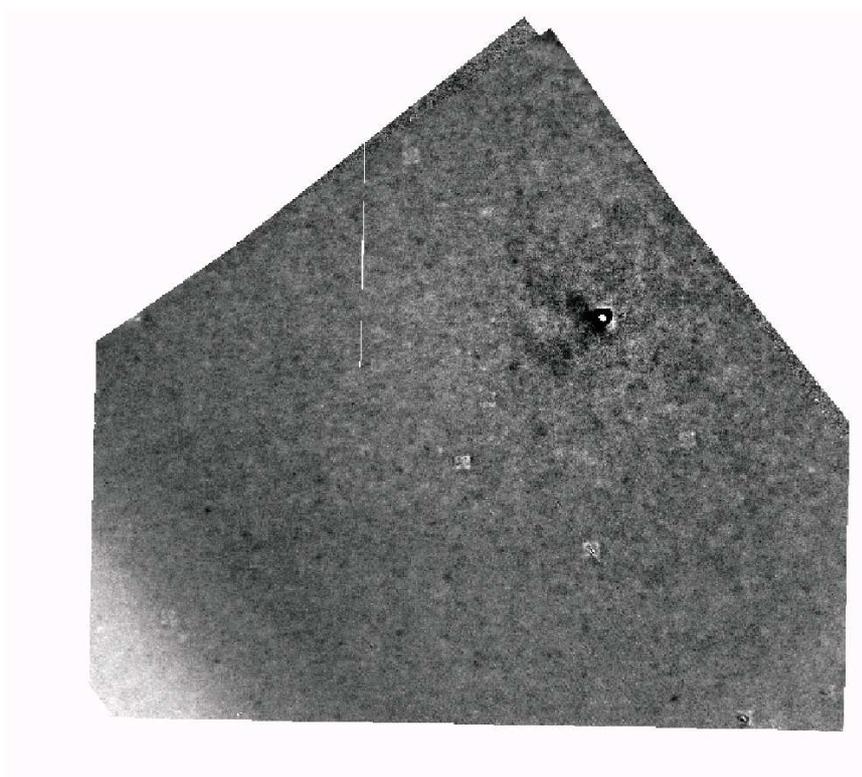} \\
\vspace*{0cm}
   \leavevmode
 \includegraphics{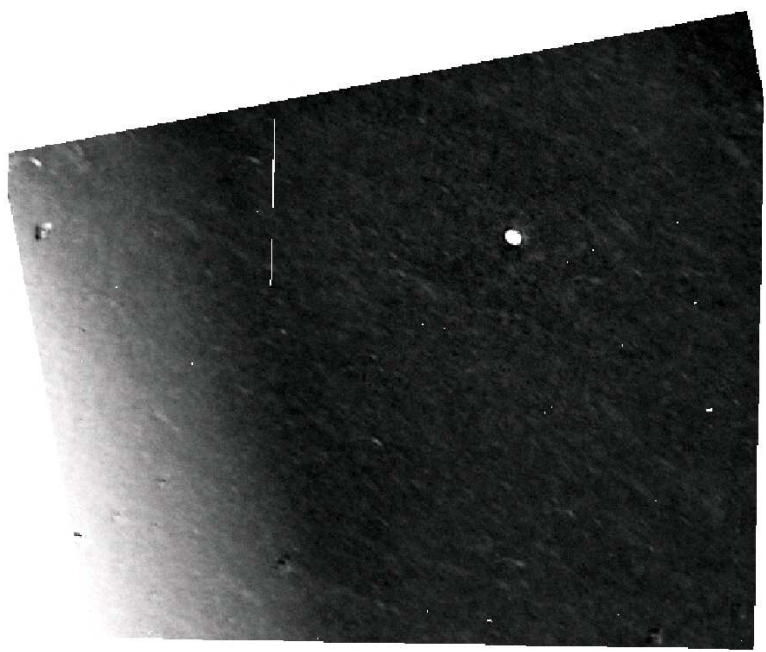} \\
\end{array}$
\caption[Examples of faults in the difference imaging (3).]{Examples of faults in the difference imaging (3). Top) An image with one side lighter than the rest of the image, possibly caused by scattered moonlight. Bottom) This image also has one side much lighter than the other. This may be caused either by a serious scattered light problem or perhaps a total misalignment in the difference imaging pipeline.}
 \label{light_side_1_2}
\end{figure}

\begin{figure}[!ht]
\vspace*{6.5cm}
$\begin{array}{c}
\vspace*{7cm} 
   \leavevmode
 \includegraphics{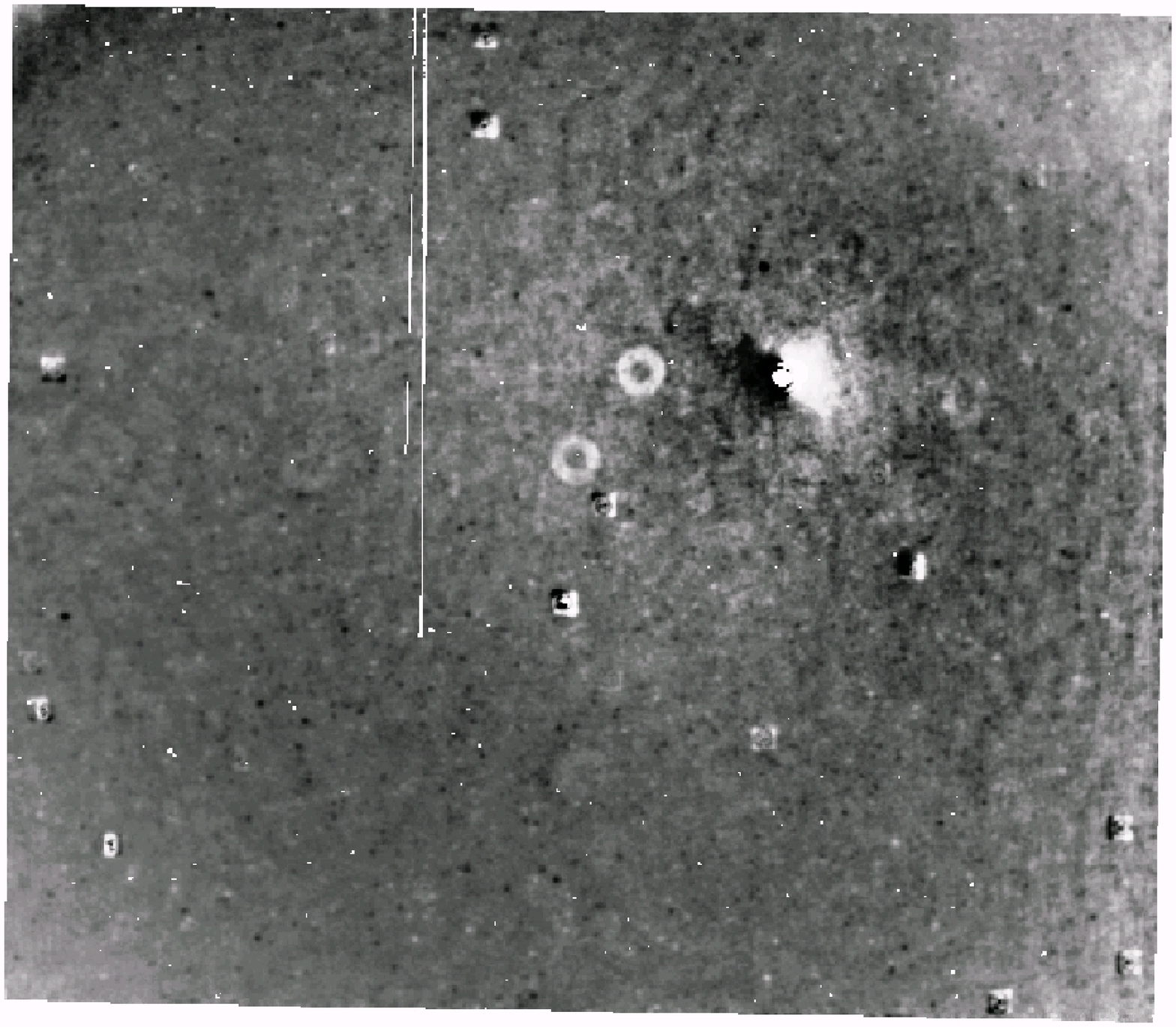} \\
\vspace*{7cm}
   \leavevmode
 \includegraphics{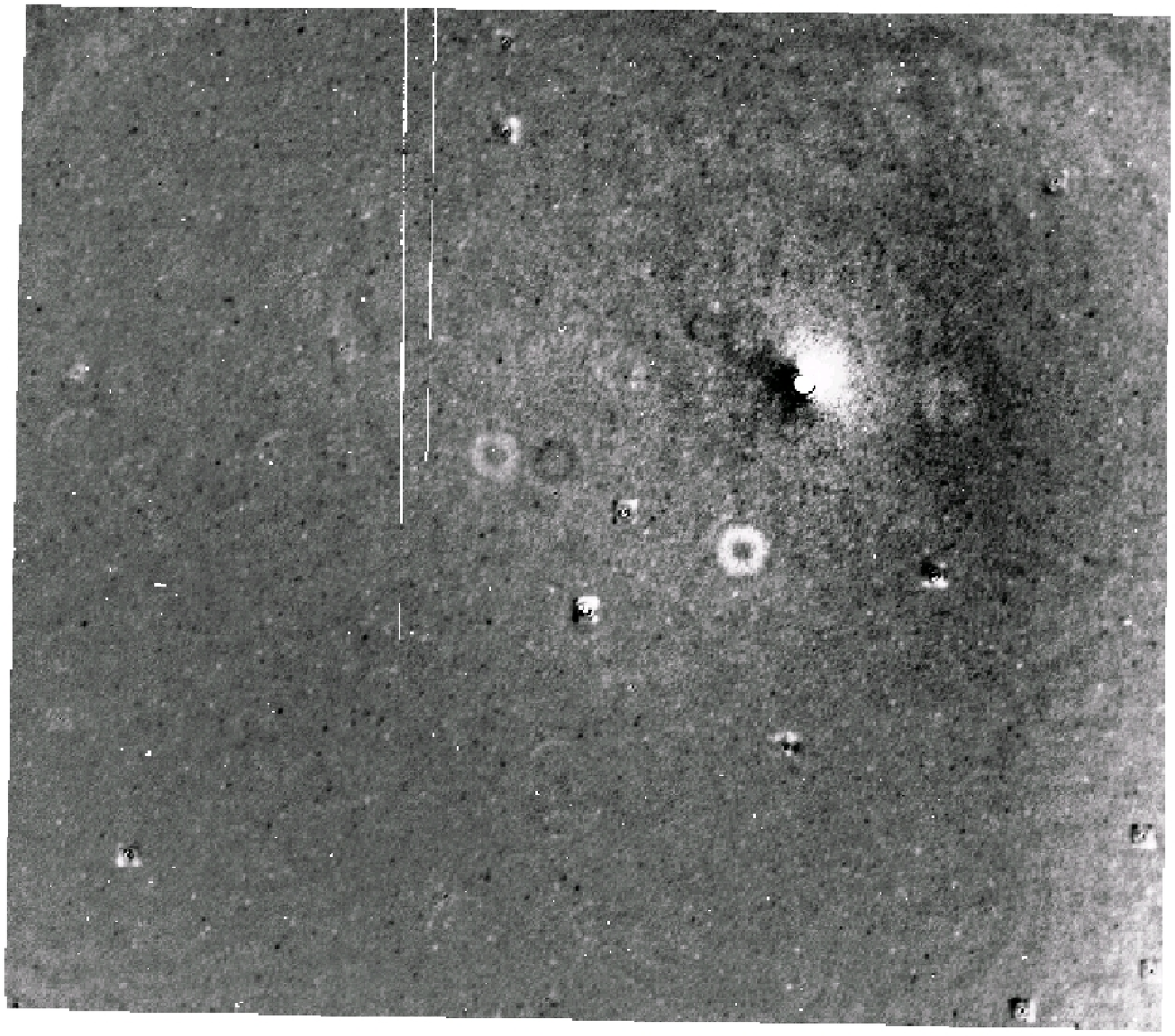} \\
\vspace*{1cm}
   \leavevmode
 \includegraphics{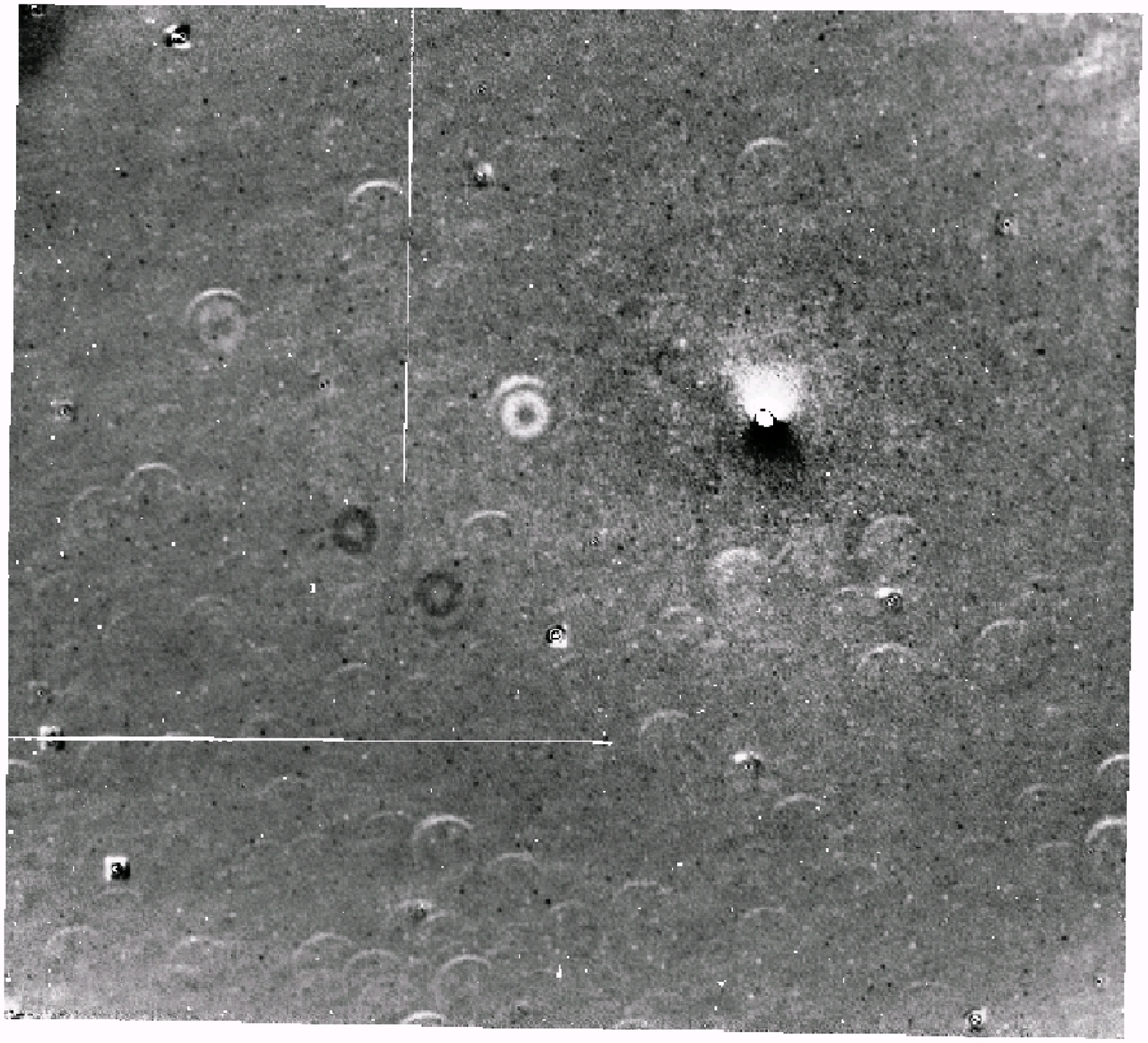} \\
\end{array}$
\caption[Examples of faults in the difference imaging (4).]{Examples of faults in the difference imaging (4). Top), Middle) and Bottom) All three images are likely to be examples of a poor flat field. The doughnut shapes are caused by the out of focus images of specks of dust}
 \label{doughnuts_1_2_3}
\end{figure}

\begin{figure}[!ht]
\vspace*{15cm}
   \includegraphics{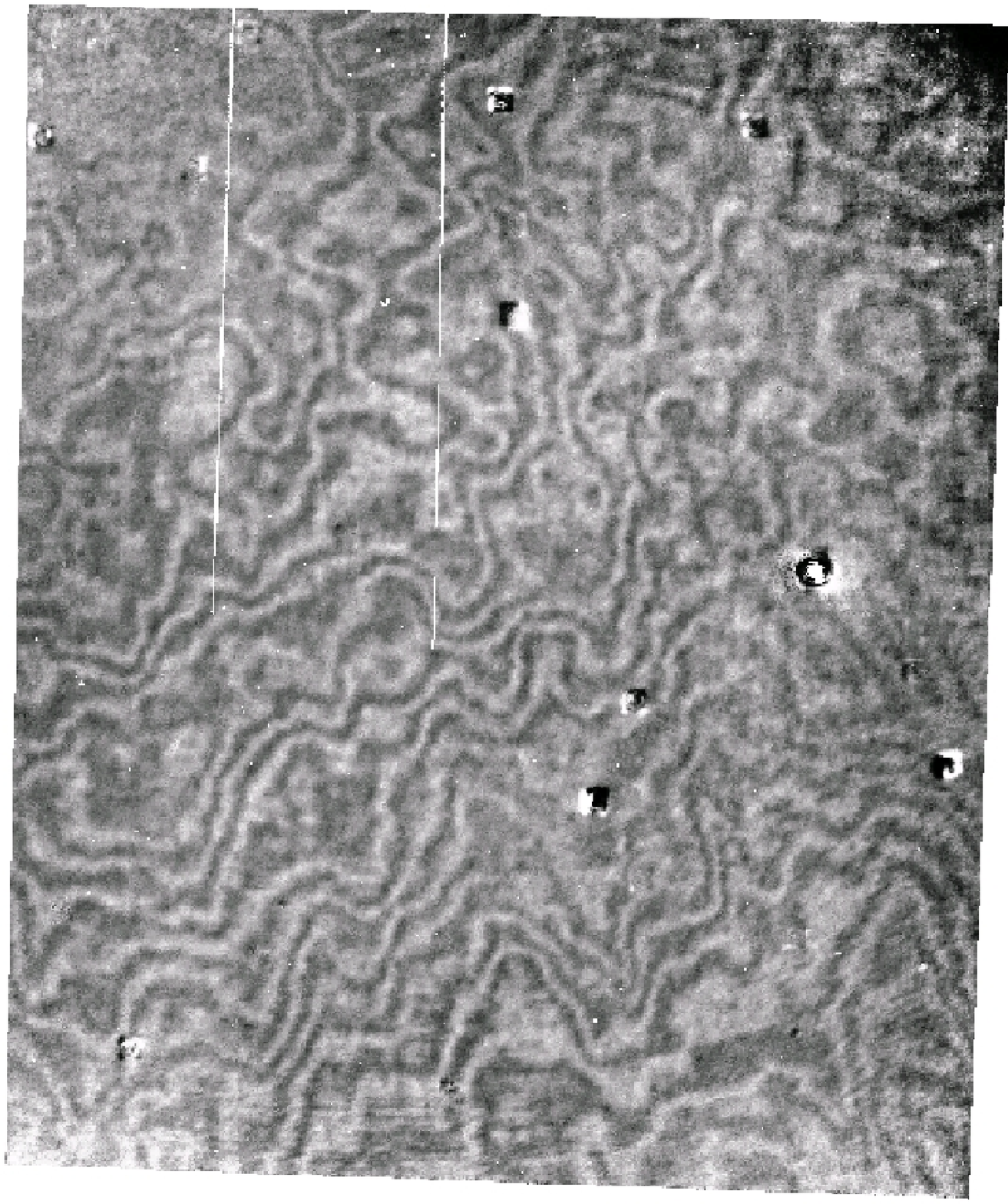}
\caption[Examples of faults in the difference imaging (5).]{Examples of faults in the difference imaging (5). This image clearly shows a good example of fringing which has not been successfully removed.
}
\label{big_fringing}
\end{figure}

Examples of difference images containing some of the categories of flaw described above are shown in the images below, Figures \ref{quadrupole_bg_general_noise} to \ref{big_fringing}, which are all selected from LT i'-band data.

In the top image of Figure \ref{quadrupole_bg_general_noise}, a  minor scattered light problem has lead to a poor fit to the overall 
galaxy model image, whereas in the  
bottom image of Figure \ref{quadrupole_bg_general_noise}, the cause or causes of the noisy galactic centre to this image are not clear but may possibly be due to a poor flat field.
The top image of Figure \ref{dark_and_light_stripes_1_2_3} shows darker stripes which, it is thought, were caused by scattered moonlight. This problem was fixed on the LT by the installation of new baffles in $2007$. After this modification a very significant improvement in the data quality from the LT was seen.
The middle image of Figure \ref{dark_and_light_stripes_1_2_3} shows dark horizontal lines which are probably due a slight misalignment occurring in the difference imaging process.
The bottom image of Figure \ref{dark_and_light_stripes_1_2_3} shows both a ``crescent moon''-shaped smearing of the background which is likely to be caused by scattered moonlight and also two clear lighter lines, the cause of which is unknown.
The two images in Figure \ref{light_side_1_2} may both involve scattered light as a cause. The top image has one side lighter than the rest of the image, possibly caused by scattered moonlight. The left hand side of the bottom image in Figure \ref{light_side_1_2} is also much lighter than the rest of the image. This may be caused either by a serious scattered light problem or perhaps a total misalignment in the difference imaging pipeline.
All three images in Figure \ref{doughnuts_1_2_3} contain ``doughnut-shaped'' dark and light artefacts, which are caused by out of focus dust particles.
Figure \ref{big_fringing} shows a good example of a failed defringing process.

 \clearpage
\newpage

 \clearpage
\newpage

 \clearpage
\newpage

 In the cases where the flaws with the images were localised and could therefore be masked 
 out, for example, ``Dark and/or light blobs''
``Edge effects'' or ``Corner shading'', this was done by creating masks using the facility in the DS9 \citep{2003ASPC..295..489J} program for 
creating geometrical ``region'' (.reg) files. These contain data about the shapes and sizes of whichever geometrical shapes were used to compile the desired mask, and are in a very similar format to that read in by the IRAF ``mskregions'' task. A brief script was used to convert the .reg files into IRAF readable ones, and then IRAF was used to create image masks which consisted of mainly zeros, with ones where the masking was to occur.
The position of each variable object from the variable 
object database was then checked against the masked areas in each image in the lightcurve in turn, and data points which fell inside a masked region were removed. Each time a new difference imaging photometry pipeline run (with different parameters/algorithm or a new reference image) was performed, this masking had to be repeated
 as the positions of the variable objects found by the DIA would be different.

 During the Third Season, it was discovered fortuitously that one of the 
causes of some of 
the poor difference images was that the reference image was slightly
mis-aligned from the rest of the data. This had previously had the effect of producing a noticeable ``bipolar'' effect on the difference images, which caused bright sources, particularly the galactic centre, but also variable objects to be darker than average on one side of the source position and lighter on the other. This caused the flux contained in the object to be
spatially smeared out, and hence lowered the signal to noise of our lightcurves.
When fixed, the lightcurves appeared noticeably better by eye. The dark edge pixels had also been fixed by this time in the improved pipeline. Therefore, the decision was made to not
go through the tedious process of re-creating the masks from scratch (which would then have been for many more images than before) as it was not felt that the potential gains would justify the extra work, given the now-improved quality of the difference images.

\chapter{Angstrom Pipelines and Data}
\label{Chapter_3}
\section{Introduction}

In this Chapter, the operation of the Angstrom Project Alert System is briefly described.
Then, the operation of the data analysis pipeline is summarised.
Finally, the data set upon which the forthcoming analysis is based is 
described.

\section{The Angstrom Project Alert System (APAS)}
\label{angstrom_alert}

The normal mode of operation of the Angstrom Data Analysis Pipeline (ADAP) is to process the data collected over one observing season offline at the end of that season. However, the ADAP is also used to process $1024$x$1024$
re-binned copies of the $2048$x$2048$ robotic telescope images which are small enough to be processed in real time \citep{2007ApJ...661L..45D}. 
Many of the stages of this processing are the same as used by the main offline Pipeline, and so a brief summary is given below. 
 An initial quality control check is made on the images based on
 the size and shape of the PSF using bright
 foreground stars. Images that fail this check are not used in the
 following Difference Image Analysis as to do so would result in lowering
 the quality of the lightcurves produced.
On the images which pass this test, the APAS performs defect masking, cosmic ray rejection and image alignment. Next it defringes each image, using an iterative method, described below in \ref{descDIA}.
AngstromISIS is used to PSF match and then stack groups of images taken
sequentially in one run of data-taking at the same epoch. One image stack, usually the one with the smallest PSF, is selected as the reference image. This is convolved with, and then subtracted from, the other images to create a sequence of difference images. From each stack of images, a likelihood map of variable sources is constructed by using an indicator of statistical significance which is similar to that of Cash \citep{1979ApJ...228..939C} shown in Equation \ref{Cash_mod} below. From these maps, discrete sources are identified and positionally matched with previous epochs. A master list of
object positions is maintained and updated with new objects if they have positions which have not been seen before. Photometry is performed on the difference images using PSF fitting. The interval between observation and
adding new photometric points to the database is about $2$ hours.
The legacy POINT-AGAPE data are also analysed by the same pipeline, off-line, in those regions that overlap with the LT/FTN field.

\subsection{Candidate Selection Criteria of the APAS}

There follows a brief description of the criteria used by the APAS to decide whether a particular object should be (currently) classed as ``flat", ``variable", ``followed", ``alert", or ``old".

If the lightcurve of an object has no detected peaks, then it is classified as ``flat". A ``peak'' is defined, basically, as a cluster of five points which are at least $5\sigma$ above the baseline for that lightcurve. The baseline estimate is determined by the median lightcurve flux. Up to three non-significant points are allowed within a peak, with the provisos that these points occur during a period of bad seeing and they are followed by at least one more significant point within the same observing season. This classification can of course change with the addition of new data points. If the lightcurve contains no peaks within the current season, but does have at least one peak in one or more previous seasons, then it is classed as ``variable" and is also subject to re-analysis when there are new data.
If an object lightcurve does have a peak in the current season, and also has a peak or peaks in a previous season or seasons, then it must still be analysed by the alert system because of the high probability in the central regions of M31 that blending exists in the lightcurves of transient and microlensing events due to the crowding of variables, leading to 
variations in the baselines of these events in a similar way, for e.g., to microlensing candidate PA-99-N1, as described in \cite{2003A&A...405...15P}.
Thus objects with both a variation in the current season and a previous season are not necessarily automatically classed as ``variable". If the most recent variation has an amplitude above a given threshold 
 in the reference image and is at least $50\%$ greater than the highest previous variation then it is still classified as a transient, otherwise it is classified as a ``variable''. All surviving objects are considered to be transients and are therefore flagged as ``followed".
The FTN and POINT-AGAPE lightcurves (where these exist) are then scaled to the reference image of the LT data, using the previously calculated flux scalings described in Section \ref{fitting_functions}. Next, the lightcurves are fitted with a reduced
Paczy\'nski curve (see Equation \ref{red_pac_curve}). It is required that the time of the peak flux of the fitted curve does not occur before the first observation of the current season and also not after the anticipated start of the next season. Also, the duration of the event, as measured by the full width half maximum time $t_{\rm{FWHM}}$, is required to be within the range $1 \leq t_{\rm{FWHM}} \leq 365$ days. The last criterion is that the amplitude of the event is positive as all physical phenomena that might
be of interest (novae, microlensing) are \emph{positive} deviations from a baseline. If an object passes all the above criteria it is classed as an ``alert". Any object failing any of the criteria is still classed as ``followed", unless it was previously classed as ``alert", in which case it is re-classified as ``old". The results of the APAS analysis are displayed on a web interface for a human observer to inspect. The final decision as to whether to alert a particular object is currently left to this observer. The APAS also displays lightcurves for all variable objects within $3^{\prime\prime}$ of the object of interest, which assists in assessing whether an object might be a real alert but contaminated with light from a nearby variable object.

The first alert ever given by the system, from the $2005$/$6$ observing season, shows special promise as a possible microlensing candidate. This appears to be a high signal to noise, very short timescale ($t_{\rm{FWHM}}\approx1$ day) microlensing event.
This event has been the subject of a study by the author, the details of which are reported in Chapter (\ref{Chapter_5}). The security of the identification of this event as lensing has been increased by the
addition off-line of some reduced data from the Maidanak telescope, which confirm the gradual smooth rise of the lightcurve, greatly reducing the chances that this could be an extremely short duration classical nova.

\section{The Angstrom Data Analysis Pipeline}
\label{angstrom_dap}

\subsection{Data Collection}

\begin{figure}[!ht]
\vspace*{14.5cm}
   \includegraphics{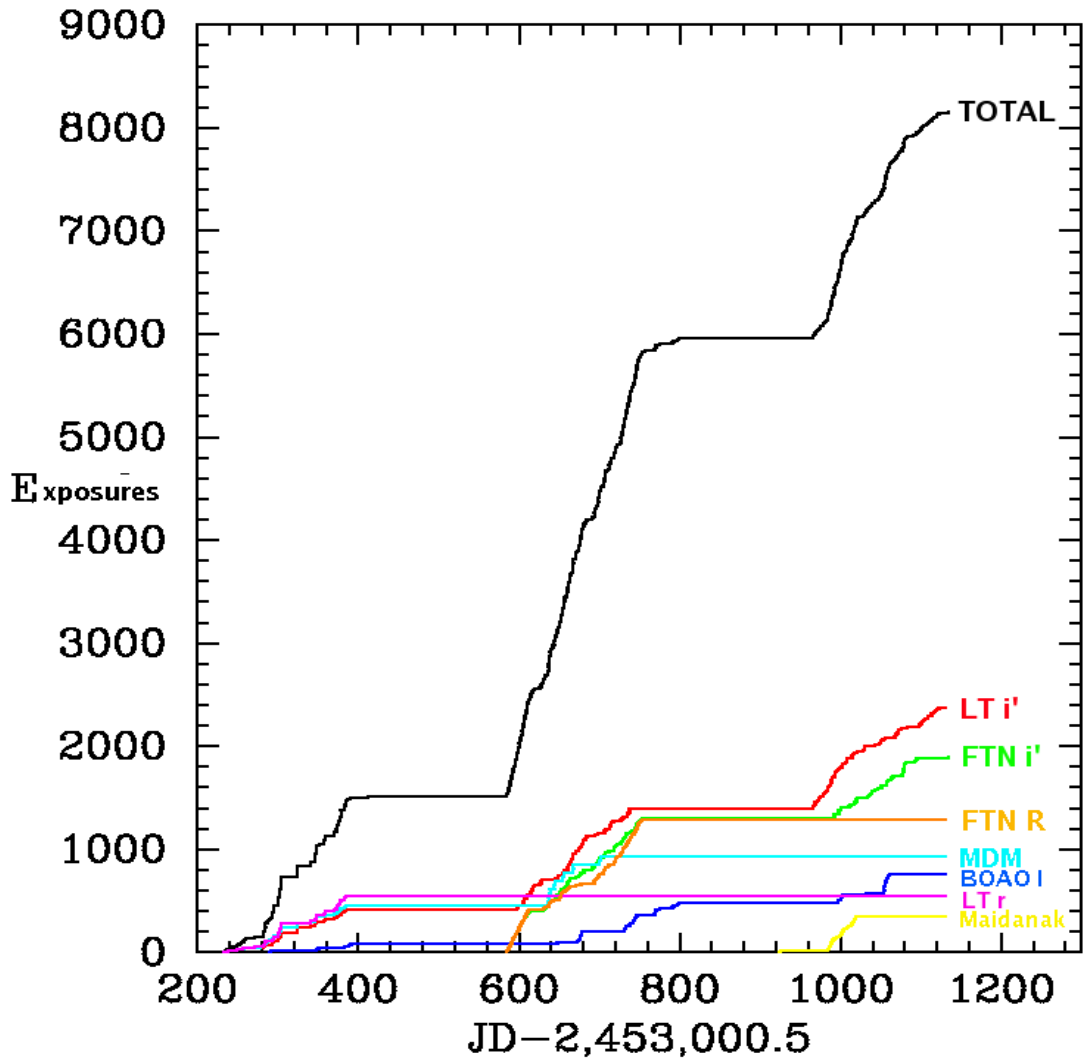}
\caption[Plot showing a summary of the accumulation of Angstrom data during the first three seasons.]{Plot showing a summary of the accumulation of Angstrom data during the first three seasons.}
\label{angstrom_data}
\end{figure}

 During the first ``pilot'' season, only the LT, Doyak and Hiltner telescopes
 were available. $86$ frames of data were obtained from the BOAO $1.8$m telescope, $418$ frames from the LT in i' band 
 plus $550$ in r band along with $453$ frames from the Hiltner telescope through the MDM observatory in Arizona.
 However, in the second season, the FTN also became
 available through RoboNet open time, and hence the rate of data collection 
 increased markedly as compared with the pilot season. 
 The FTN contributed $2592$ frames out of the total of $4447$ for the season, $1303$ in i' band and $1289$ in R band. The next largest contribution was from the LT in i' band, with $984$ frames, and then the Hiltner ($481$) and Doyak ($390$).
In the Third Season, the FTN R band observations were discontinued, but a new telescope, the Maidanak in Uzbekistan was added to our collaboration, contributing $352$ frames of data. The LT (i') obtained $973$ frames, the FTN (i') $600$, and the Doyak took $286$.
Also very important to the project were $591$ stacked epochs from the POINT-AGAPE project, which span the four seasons
$1998$/$9$, $1999$/$2000$, $2000$/$2001$ and $2001$/$2$. These data were from the two POINT-AGAPE (henceforth, ``PA'') fields $2$ and $3$ which overlap substantially with the top and bottom part of the Angstrom field.
The above information is summarised in 
Figure \ref{angstrom_data} and in Table \ref{data_summary_table}. It can also be seen from Figure \ref{angstrom_data}
 that the majority of the data collected were from LT and FTN together, 
with relatively smaller contributions from the other three non-robotic telescopes. It should be noted that no attempt has been made to scale the plot according to exposure length, so this graph does not accurately reflect the amount of real information contributed by each telescope, just the number of epochs.
 At the end of the Third Angstrom Season in February $2007$, $8165$ ``raw''
 (unstacked) frames of data had been collected by the collaboration.

\newpage

\begin{sidewaystable}
\begin{tabular}{|c|c|c|c|c|c|c|c|c|c|c|}
\hline
\hline
     & \multicolumn{9}{c}{Telescopes}    &  Seasons totals \\
\hline
     & Maidanak  & Doyak &\multicolumn{2}{c|}{LT}    
    &\multicolumn{2}{c|}{FTN}  & Hiltner  & PA 2 & PA 3 &   \\
\hline
                 &   &   & i' &   r   &  i'   &  R  &  i  &  i  &  i  &   \\   
\hline
Pre-Angstrom     &           &          &           &          &           &           &           & 
/306  & /285 & /591  \\
\hline
Season 1 (pilot) &           & 86/20  & 418/54  & 550/44 &           &           & 453/55  &           &    &  1507/173 \\
Season 2         & 352/28  & 390/32 & 984/82  &          & 1303/83 &           & 481/13  &           &    &  4447/280          \\
Season 3         &           & 286/12 & 973/121  &          & 600/- & 1289/70 &           &           &    &  2211/ 161   \\
\hline
Telescope Totals &  352/28   & 762/64      & 2375/257      &   550/44       & 1903/ 83      & 1289/70      &  934/68       &
 591 &     & 8165/614    \\
\hline
\hline
\end{tabular}
\caption [Table summarising the data collected by Angstrom over the first three seasons.]{Table summarising the data collected by Angstrom over the first three seasons. In each line and column, the data is displayed as ``unstacked exposures/stacked epochs''.}
\label{data_summary_table}
\end{sidewaystable}

\newpage

\label{descDIA}
 The structure of the Angstrom Difference Imaging Pipeline can be
 summarised in eight sections as follows:

 $\cdot$\textbf{Quality control}

 $\cdot$\textbf{Defringing}

 $\cdot$\textbf{Alignment}

 $\cdot$\textbf{DIA}

 $\cdot$\textbf{Source Detection}

 $\cdot$\textbf{Merge lists of newly detected and previously detected objects}

 $\cdot$\textbf{Perform photometry of new objects in all observations}

 $\cdot$\textbf{Perform photometry of all objects in the current image}

 These eight operations are performed on each epoch of data
 in ``real time'' i.e. the epoch is analysed before the next epoch arrives.
 Currently, the Pipeline is operating on a reduced data set consisting 
 of ``binned'' images which contain $1024$x$1024$ pixels, as opposed to the 
 original $2048$x$2048$ pixels of the original LT/FTN CCDs. It was 
 decided to do this so that the Pipeline would be 
 capable of analysing one night's data in less than $24$ hours.

The most complicated of the sections above and most costly in terms of computing time are the first three, ``Quality Control, Defringing and Alignment''.

                      \subsubsection{Stacking}

The first thing that is done is to apply the mask of known bad pixels
for each camera. The routine then attempts to ``fix'' the values of the rejected pixels by interpolating across them using all good data in surrounding pixels. 
A mask is then made consisting of the bad pixels. All subsequent operations in the Pipeline are also performed on these mask images as well as on the ``good data'' images, to allow the relative contribution of bad pixels to every other pixel after all other operations have been performed to be assessed.
    Next, an initial guess is made at the PSF of the epoch, using the 
 IRAF task ``starfind''. This routine searches for ``extended point-like''
 objects that have a peak flux that is above a given threshold value. The
 background is estimated locally by starfind, which is an important
 requirement for our observations. 
It then fits an elliptical Gaussian profile to these pixels and hence 
estimates their FWHM and ellipticity.
 The average of these two properties from the majority of the starlike 
objects found in each image (some extreme outliers are rejected) is taken as the initial estimate of the PSF size and shape.
 On the basis of this, individual observations are classified as ``Good''
``Bad'' or ``Ugly'' in terms of the size and ellipticity of the PSF.
The criteria used for this categorisation for the reduction of the data investigated in this thesis are:

 If the ``Roundness'' or ellipticity, which is defined as in Equation \ref{roundness} where $\sigma_x$ and $\sigma_y$ are the widths of the fitted 
Gaussian profile along its major ($x$) and minor ($y$) axes respectively, is larger than $0.6$ or the FWHM of the PSF is more than $11.0$ pixels then the observation is classified as ``Ugly''.
 Otherwise, if the ``Roundness" is between $0.4$ and $0.6$ or FWHM is between $8.75$ and $11.0$ pixels then the observation is classified as ``Bad''.
 In all other cases the observation is classified as ``Good''.

 ``Good'' and ``Bad'' classifications are treated in the same way, but ``Ugly''
 observations are rejected entirely and not used further.

\begin{equation}
e=\sqrt{1-\frac{\sigma_{y}^{2}}{\sigma_{x}^{2}}}
\label{roundness}
\end{equation}

 The next task that is performed is cosmic ray removal which utilised 
another IRAF task; ``craverage''. This task requires the estimate of the PSF calculated above as a parameter in order to decide what is a cosmic ray (which are on the whole very compact, with little energy spreading to surrounding pixels) and what might be a ``real'' feature of the image.

 Finally, a thin border, currently $4$ pixels in width for binned data, is masked out 
around the edge of the images to remove edge effects (over-bright or over-dark pixels) which occur in the outer few lines.

                     \subsubsection{Alignment-part I}

The first thing that is done is to align the centroids of point-like sources in the images.

 Point-like sources in the images are found using the IRAF task
``starfind'', using the previously calculated estimate of the PSF as a guide to the size of object the routine should initially expect to find.
 If the number of sources found is insufficient (currently defined as less than 8) then the 
image is not used in any following calculations.
 If sufficient objects are found, then the list of found objects is 
compared to the list of stars from the astrometric reference image (the DIA reference image could be different but is aligned to the astrometric reference image). The above process is repeated for
all cameras being used. Pairs of objects in the two lists are matched up using the IRAF task ``XYXYmatch''. Again, if there are insufficient matched pairs then the image is dropped. If sufficient pairs are found then the transformation between the current image and the reference image is calculated using the IRAF ``geomap'' task, initially allowing three degrees of transformation; a translation, a rotation and a scaling factor.
 As a sanity check, the calculated scaling factor should be very close to 1.
 If the scaling factor is not ~1, the limits being $0.95 \leq$ magnification $\leq 1.05$, then the image is dropped.
All geometrical transformations on images within each epoch should be similar (this is checked).
 If this is true then the object matching stage is repeated,
 this time giving the average of all calculated transformations as a 
``first guess''. This may occasionally find a few more pairs of objects than before and consequently improve the calculated transformations. On the second iteration, the transformations are calculated allowing only a shift and a rotation. At the same time the inverse transformation is calculated.

                      \subsubsection{Defringing}

A subset of Angstrom i' band observations contain fringe patterns which are clearly visible by eye. For an example, see Figure \ref{big_fringing}. 
These are thin-film interference patterns which originate in multiple reflection from the top and bottom surfaces of the CCD. The effect of these fringes may be increased in the difference images when the fringe patterns are misaligned, causing a significant modulation in the background flux level.

Fortunately, the fringe pattern does not change much over time.
Therefore, a master fringe frame, which may be labelled ${\bf F}$, can be built up by adding a number of images, and then scaled and subtracted from
a target image to minimise fringing in that image. However, in the case of
the signal of interest being of the same order of magnitude as the level of fringing and much lower in magnitude than the background flux level, as in our data, this can be computationally intensive.
  Fortunately, we are able to take advantage of the problems experienced 
in the pilot season with large pointing misalignments and rotator problems (see Section \ref{commonoverlap}) on the LT.
By adding together all the aligned frames, the now-randomly-oriented fringe
patterns largely cancel out, giving a first order fringe free template, ${\bf M}$. This gives a good first estimate of the background galaxy light distribution in our images. The fringe level in the individual images can
 be computed and then removed by minimising the sum:

\begin{equation}
\label{fringe_mini}
{{\chi}^2}(m,f,s) = \displaystyle\sum_{i=1}^N {[{\bf I}_i - m{\bf M}_i - f{\bf F}_i -s]}^2/{\bf I}_i
\end{equation}

over all pixels $i$, where ${\bf I}$ is the target image to be defringed,
and $m$, $f$ and $s$ are the scalings and offset respectively to be computed.
    By using the stack of first order defringed images as a second order 
fringe-free model, and repeating the process above, the level of fringing can be further reduced. No obvious evidence of residual fringing is found in the images after this second level of defringing, so the iterative process does not need to be continued any further.

Due to the similarity between the LT and FTN, which includes CCD cameras and filters, it is possible to use the LT fringe-free frame as the first order FTN image. Unfortunately, it has not yet been possible to produce reliable fringe frames for the FTN, so to date the FTN has perforce been processed without defringing.

For the Maidanak telescope, a first order fringe-free image is produced by taking a series of dithered observations of the Maidanak M31 field.
These observations are combined in the same way as for the LT data, and the above algorithm used to remove the fringes.

                      \subsubsection{Alignment-part II}

Using the previously calculated geometric transformations, the re-alignment
of the current image into the frame of the reference image is performed.
The same re-alignment is also performed on the bad pixel masks.
After the re-alignment has been completed, the ``starfind'' routine is once again used to make a new determination of the PSF. This will usually be the same as when it was previously calculated.

                      \subsubsection{Stacking}

The images in each epoch are stacked. The bad pixels, as given by the masks, are not included. In order to calculate the new stacked pixel values, if there are less than five images in the stack then the mean value is taken but if there are $5$ or more images then the median value is taken. A new mask is then created giving information about how many images contribute to each pixel.

                      \subsubsection{Difference Imaging}
\label{Angstrom_DIA}
\textbf{Angstrom Project DIA}

The difference imaging (see Section \ref{dia}) is performed using the ``Angstrom version'' of the ISIS \citep{1998ApJ...503..325A} image subtraction program, ``AngstromISIS''.
The kind of reference image used has changed along with the development of the DIA pipeline. For the $2007$ photometry, which was used for the analysis of variable stars contained in Chapter \ref{Chapter_5}, a single image was used. This was chosen as one of the best DIA images, with a PSF which was as small and round as possible.
This does mean, however, that when the reference image is subtracted from itself the
result should be very close to zero, and this means it is not possible to correctly estimate the error on the point. For this photometry, therefore, the flux point corresponding to the reference image was not used as its error was untrustworthy (see Section \ref{ref_im}).
For the 2008 photometry, used for the microlensing analysis (also in Chapter \ref{Chapter_5}) a reference image was constructed from a stack of good images, which removed the problem with error estimation and hence the need to remove the corresponding flux point from the fitted photometry.

 Parameters used for the DIA include:
 \newline
Number of stamps along the X and Y axes: $20$

Degree of polynomial to fit the background variations: $2$

Sigma of first Gaussian component = $0.7$

Sigma of second Gaussian component = $2.0$

Sigma of third Gaussian component = $6.0$

Degree associated with 1st Gaussian = $6$

Degree associated with 2nd Gaussian = $4$

Degree associated with 3rd Gaussian = $3$

Degree of the fit of the spatial variations of the Kernel = $3$

The best values for these parameters were found by experimentation, using our image data.

             \subsubsection{Source Likelihood Map Generation}

For each pixel in each difference image, an elliptical Gaussian of known widths is fitted. Only the peak flux is allowed to vary. From the fitting, the best fitting peak flux and a statistic which is similar, but not identical to ${\chi^2}$ are obtained. This statistic is a modified version of the Cash statistic \citep{1979ApJ...228..939C}. The original Cash statistic was designed for use with low photon counts, and has been modified for use with high photon
counts (when approximating the Poisson noise as being drawn from a Gaussian distribution is reasonable).
 At a given point (or pixel) in a difference image, the modified Cash
 statistic ($C_{mod}$) is determined by fitting a simple model for the PSF, 
 and minimising the sum

\begin{equation}
\label{Cash_mod}
{C_{mod}}^2 = \displaystyle{\sum_{i}{\left({\frac{(d_i - b_i) - m_i(f)}{\sigma_i}}\right)}^2}
\end{equation}

over a suitably large region around the nominal position. In Equation \ref{Cash_mod}, $d_i$ are the pixel counts at pixel $i$, $b_i$ is the local background, $m_i(f)$ is the value for the PSF model, centred on pixel $i$ and scaled by flux $f$, and $\sigma_i$ is the estimated photon noise for pixel $i$. Only the flux $f$ is varied in the fit.

Two more maps are produced: One containing the fitted fluxes and another with the ``likelihoods'', ${C_{mod}}^2$. At this stage, known bright (foreground) stars and the centre of the galaxy are masked out. These masked areas are also scaled in size with the changes in the PSF.
 Source detection is performed on the likelihood map by considering all 
pixels above a given threshold (of value $2.0$). For sources found in these pixels (which are defined as peaks in the value of the likelihood distribution), the peak likelihood value and centroid of the pixel values is found, which is assumed to be a good estimate of the position of the source. A list of new sources found in the current epoch is compiled.
     The ``new'' and ``old'' source lists are merged, using the following 
algorithm: For each object in the ``new'' object list, the program goes through the list of old objects and finds the closest match to it in position. The separation (measured in positional error bars) is calculated. Using two threshold constants ``$n_1$'' and ``$n_2$'', whose values are $2.0$ and 
$6.0$,
if the position is greater than $n_2\sigma$ away from any other object, then it is considered to be a new object, and is added to the list of objects. If the position is between $n_2\sigma$ and $n_1\sigma$ then the object is probably an old object but this is not certain enough to use the position to improve the average position. If the position is closer than $n_1\sigma$, then the object is considered to be a match with the old object, and the new position is combined with the previous average position of the old object and hence improves the knowledge of its position by reducing its error.

                  \subsubsection{Photometry}

The pixel corresponding to the position of each object is found, and the
sum of the flux in the PSF of the extended object surrounding that pixel is read. In order to calculate the error on this flux, the values of pixels taken from around the central pixel (the pixels used have to be independent, so they are more than the $\sigma$ of the PSF apart and in rings centred on the nominal pixel) are read and their scatter is used as an estimate of the likely error on the central pixel.
Photometry is then performed on all objects in the merged (old + new) object list in the current difference image.

            \subsubsection{Spatial Distribution of variable objects}

\begin{figure}[!ht]
\vspace*{5cm}
   \includegraphics{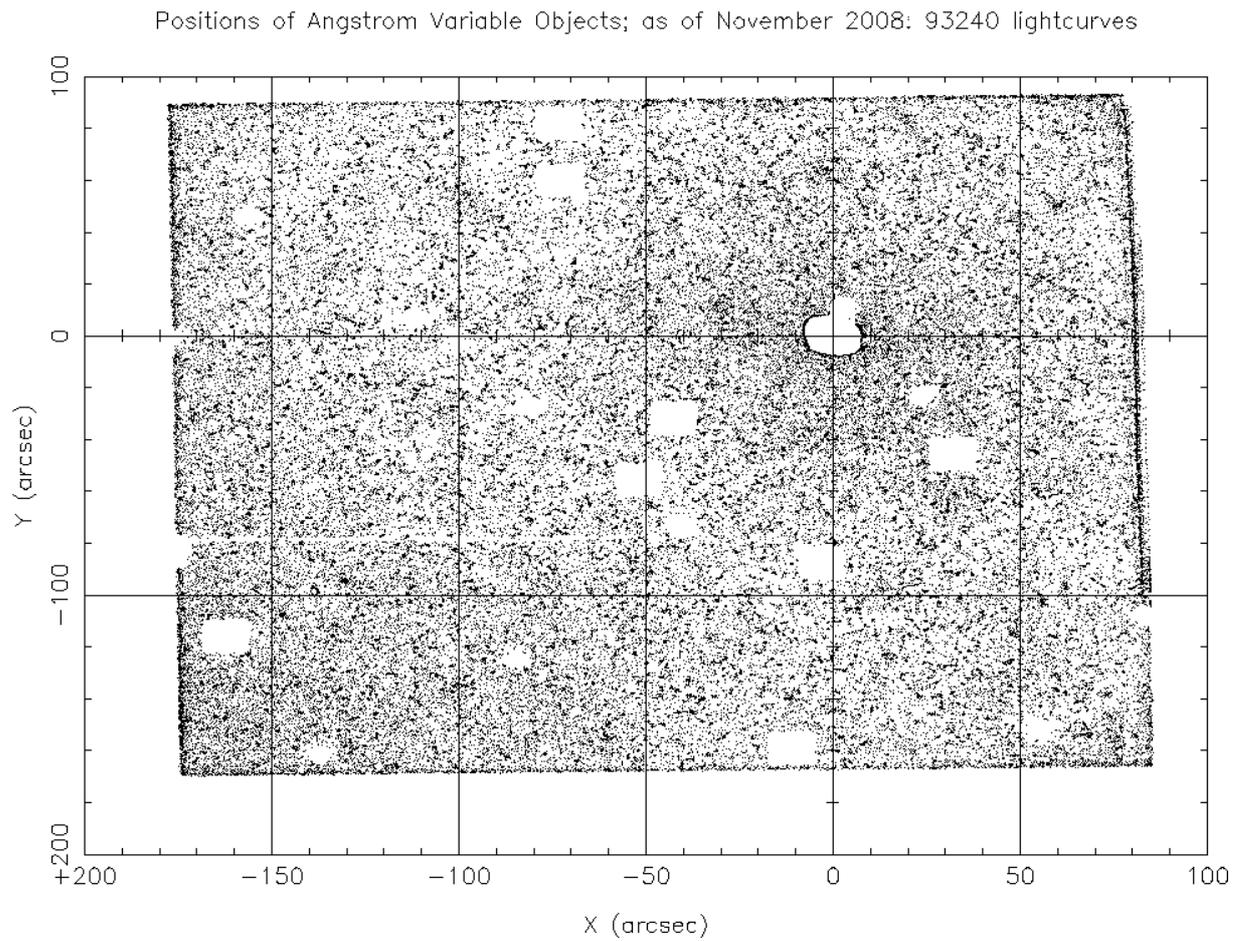}
\caption[Diagram showing the spatial distribution of the 93240 objects in the Angstrom Variable Object Database at November 2008.]{The spatial distribution of the 93240 objects in the Angstrom Variable Object Database at 27th November 2008. Units of X and Y are arcsecs on a square image: (the LT frame is $4.5^{\prime}$ (= $270^{\prime\prime}$) on a side). The M31 galactic centre is at (0,0). Square gaps are due to masking of resolved bright stars.
}
\label{variable_object_dist}
\end{figure}

The spatial distribution of the final set of all $93240$ variable objects
from which the objects selected in this thesis and reported in Section \ref{lensing_candidates_2008} were selected, is shown in Figure \ref{variable_object_dist}. The M31 centre is
defined as being at (0,0). The blank area around this and various other patches is due to these areas being masked out in the difference images. This was done for areas around bright resolved stars and the centre of the galaxy because image artefacts caused by the difference imaging process combined with Poisson noise due to the the large central surface brightness of the centre of the galaxy make detecting real variable objects in these regions impossible in the difference images.

\chapter{The Candidate Selection Pipeline}
\label{Chapter_4}
\section{Introduction}
In this Chapter the operation of the candidate selection
pipeline is
described and explained.
Some aspects of the development process are detailed.
The selection process used to select microlensing candidates
after the main pipeline has been run is described.
Finally, the investigation of the ratio of flux amplitudes between the
POINT-AGAPE data and the other telescopes, using variable star
candidate lightcurves is also described.

    \subsection{Modelling periodic variable star lightcurves}
\label{skewsin_model}
It was known from previous microlensing surveys in Andromeda that in all likelihood
large numbers of periodic variable stars such as Mira variables would be found in the
 Angstrom catalogue of variable objects. Many previous surveys have automatically
 rejected any lightcurve that shows any evidence of periodic variability as 
a suitable candidate
 to claim as a microlensing candidate. One analysis of the PA survey data \citep{2005A&A...443..911C} does
include a similar Paczy\'nski + sinusoid fit in their analysis, and, as their sixth and final cut, reject a lightcurve as a probable variable star if both
the Johnson-Cousins R band ($\Delta\phi$) flux difference between the modelled microlensing bump and another bump in the lightcurve which is due to the variable component is smaller
than $1$ magnitude and the time width of the sinusoidal
part is compatible with that of the lensing bump within a factor of $2$.  
Their sinusoid function is only a simple non-skew one, however. The other PA analysis, by \cite{2005MNRAS.357...17B}, starts by defining annuli of radius 6, 3, or 1.5 pixels around resolved stars (depending on their magnitude) which are then masked out.
Then, as their first selection criterion, they considered the improvements in $\chi^2$ of fit of a
simple sinusoid over a constant baseline, and also the improvement in $\chi^2$ for a Paczy\'nski fit over the same constant baseline. The actual cut then was to require that  $\Delta\chi_{\rm{var}}^2 < 0.75\Delta\chi_{\rm{micro}}^2$.
 
 However, in \cite{2005MNRAS.357...17B} a joint variable plus lensing fit was not explicitly employed.
 Many other variable stars were also removed by another cut in the two
 colour pseudo-magnitude plane \citep{2004MNRAS.351.1071A}.

\begin{figure}[!ht]
\vspace*{4cm}
   \includegraphics{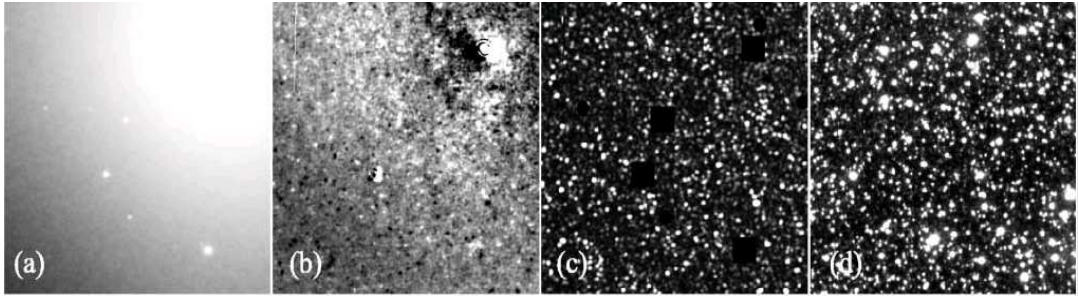}
\caption[Images illustrating the density of variable objects in the Angstrom field compared to the density of all stars for microlensing surveys in the Milky Way.]{Images illustrating the density of variable objects in the Angstrom field compared to the density of all stars for microlensing surveys in the Milky Way. Image taken from \cite{2007ApJ...661L..45D}.
From left to right: a) LT i'-band image of a $2^{\prime}$ region of the Angstrom field. b) The difference image corresponding to a), showing variable objects as black and white spots. c) The corresponding significance map, showing variable sources visible above background. Blending of variable sources is visible across the image. Blackened squares are masked out resolved stars. d) For comparison with c), an image from an OGLE-III Milky Way bulge field (OGLE image is from the OGLE
Early Warning System \citep{2003AcA....53..291U} for alert OGLE-2005-BLG-172). All of the objects in c) are variable, whereas only a handful of the stars in d) will be variable, and yet the density of objects is similar.}
\label{variable_objects_comparison}
\end{figure}

Since the density of variable objects in the Angstrom field is so high \citep{2007ApJ...661L..45D},
(see Figure \ref{variable_objects_comparison} for an example) it is considered essential to attempt to include the modelling
of variable stars in the microlensing selection procedure so that it is possible
 to find the significant fraction of the microlensing events that occur which are blended with light from 
a variable star. Of course, multiple blended variables are also possible and even probable,
but modelling the sum of an arbitrary number of variables is considered too complex to attempt at this stage.

It was clear that certain classes of variable star lightcurves could be modelled as some kind of sinusoid function. In many known cases, the rise time of the lightcurve is shorter than the fall time. Therefore some kind of skew sinusoid was required.

   \section{Operation of the pipeline}
\label{pipeline_operation}

    \subsection{Introduction}

 The main aim of the candidate selection pipeline was clearly to select promising
microlensing candidates from the large number of variable lightcurves
produced by the DIA pipeline. However, a secondary aim was to investigate the population of variable stars in M31, and to this end a sinusoidal function was
fitted to the data to select those lightcurves which vary periodically and in a regular way. This process was complicated greatly by the number of lightcurves which clearly vary in an obvious way, but are not necessarily a good fit to a single sinusoidally varying component. These might
be examples of blended or multiply blended variable stars, or variable stars blended
with microlensing candidates, or classical novae blended with variable stars, or some other combination of phenomena. Previous authors describing microlensing searches in M31 have, to a large degree, tried to exclude all periodically variable baselines from their selection of microlensing candidates \citep{2004A&A...417..461D,2005MNRAS.357...17B,2005A&A...443..911C}. It was decided at an early stage to attempt to include the variable star model in the selection process for microlensing candidates, in order to allow the detection of microlensing events closer to the core, where the probability of blending is higher.

Therefore, a ``combined'' or ``mixed'' fitting function which consisted of a reduced Paczy\'nski curve added to the periodic variable function was included as one of the options the code could choose to fit to the data, as well as the variable star function or the reduced Paczy\'nski curve separately.
During the analysis of each lightcurve, a five digit code was given to each one which described what was the general classification of the lightcurve (pure lensing, periodic variable, mixed, or flat),
which of the cuts had been passed or failed, and what were the calculated signal to noise categories of both the variable and microlensing components.
These data along with the calculated parameters for each of the cuts were output to enable post-analysis to be performed. A separate code
 was written which utilised this stored data to facilitate the relatively fast
 iteration and experimentation with values of cut parameters (without having to
 repeatedly run the much slower main code each time a cut threshold value was changed) and to roughly order the finally selected events in terms of how ``well''
 the cuts had been passed collectively and hence how ``good'' each candidate was.
 This code performed exactly the same cuts in the same order as the main code but
 only used as input the fitting output data. Therefore its advantage was that no
 fitting had to be performed, which was the slowest operation performed by the main
 pipeline. However, if any changes to the fitting routines, or the way they were
 used, were made then the main code had to be run again. To ensure that the necessary
 data were always available to allow later experimentation
with the cut values in a wide range, the values of the cuts used in the main pipeline were set to be rather loose. Certain cuts were then tightened significantly after the second stage, when experimentation with their values had been completed.

\begin{figure}[!ht]
\vspace*{22cm}
   \includegraphics{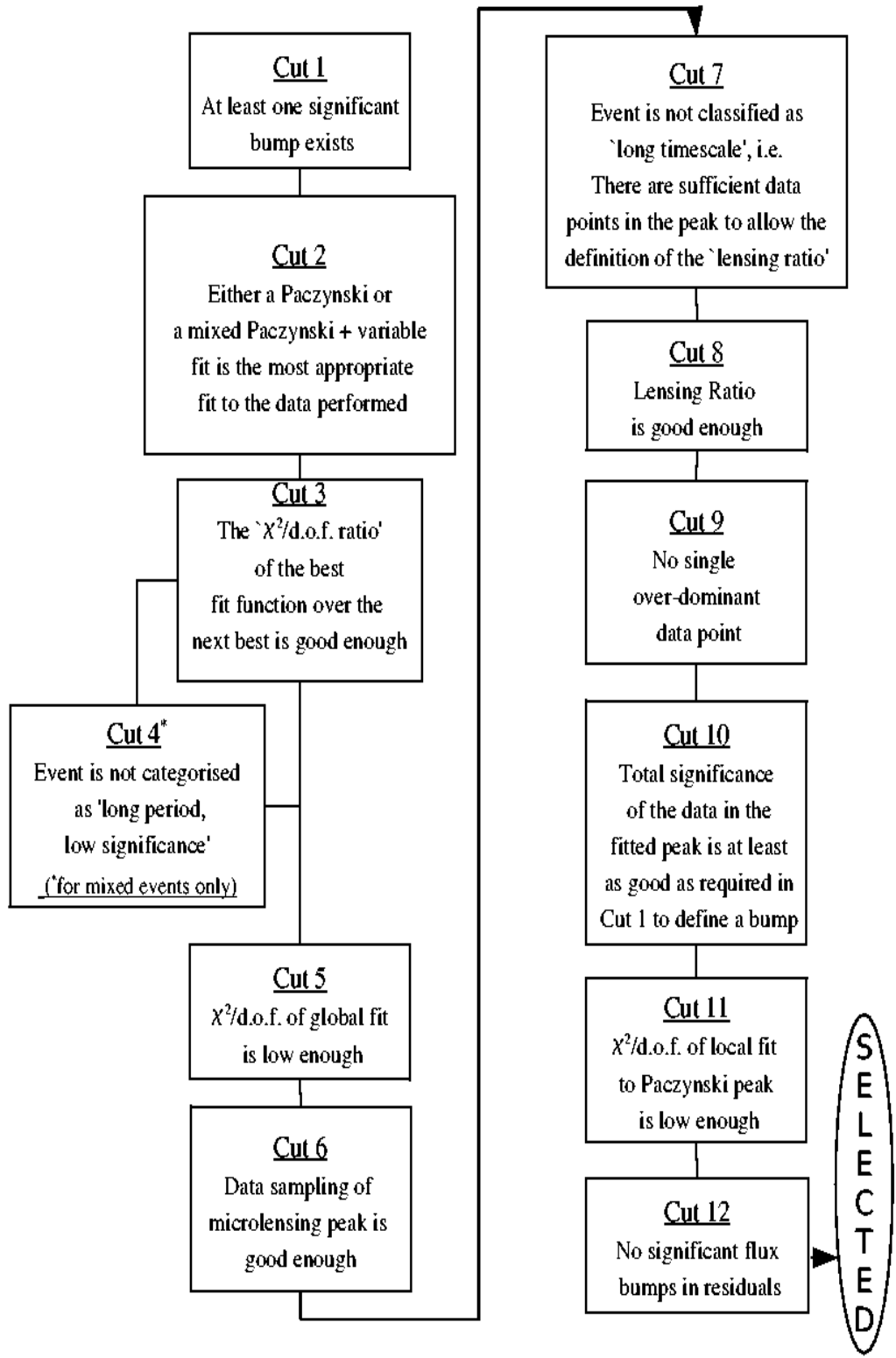}
\caption[A graphical summary of the main cuts applied to the lightcurves.]{A graphical summary of the main cuts applied to the lightcurves in order to select likely microlensing events.
}
\label{Cuts_summary}
\end{figure}
\clearpage

\subsection{Summary of main actions of the main pipeline}

The main actions taken by the pipeline are summarised below in the order
they are performed:
In order of action:

1) a) Reading in data; masking of reference image
   b) Sigma clipping of extremely small or large errors or fluxes

2) Bump finding

3) Bump rationalisation between telescope bands, if bands overlap in time

4) Lomb-Scargle periodogram calculation

5) Classification into ``one significant bump'' and ``many bumps'' and ``periodic''
 or ``non-periodic''

6) Classification of the ``many bumped'' into ``one bump significantly higher than
 the rest'' or ``all bumps similar in size''

7) Calculation of best constant fit

8) Depending on results of 5) and 6), functions fitted are either
 a) reduced Paczy\'nski
 b) skew cosinusoid
 c) sum of the two functions above, so called ``mixed'' Paczy\'nski + cosinusoid function

9) Simultaneous calculation of ``$\chi^2$ ratio'' = ($\chi^2$(A)-$\chi^2$(B))/$\chi^2$(A) (where A and B represent fits of two different functions to a lightcurve to be compared) for all combinations of fits performed on that lightcurve

10) Lightcurve loosely classified as whatever the lowest $\chi^2$ fit is

11) Calculation of $\chi^2$ per degree of freedom for best chosen fit

12) F-test to weed out those lightcurves where the improvements in $\chi^2$ do not justify the number of additional
parameters. These are downgraded to the next lowest $\chi^2$ fit.

13) Exclusion of fits with extreme flux ratios between bands

14) Calculation of signal to noise estimators both for lensing and for variables. Classification into sub-categories for each
class based on these estimators.

15) For classifications where microlensing is involved, the same criterion for number and significance of points in the finally fitted bump as was previously required in 2) 
to define a bump was re-applied. This was necessary
as sometimes a Paczy\'nski bump was fitted by the ($\chi^2$ minimising) fitting routines to what was basically one high point with a small error, surrounded by noise, rather than the intended bump found by the bump finding routines. In general there was no guarantee that the fitting will find the same bump as preferred by the bump finding routine.

16) Calculation of local $\chi^2$ per degree of freedom solely in the region around 
the Paczy\'nski
peak. This was useful because the previously calculated reduced $\chi^2$ for the whole lightcurve may be dominated by the variations in the
background or by the poor fit to the variable part of a mixed fit.

17) Cut on ``$\chi^2$ ratio'' was applied

18) For ``mixed'' events: Assess whether lightcurve is ``long period, low contrast''.
(see Section \ref{long_period_low_contrast})

19) Temporal sampling of Paczy\'nski bumps calculated.

20) Cut on signal to noise and significance of points in the main bump applied.

21) Cut on ``local $\chi^2$/d.o.f.'' is applied.

22) Cut on Paczy\'nski bump sampling is applied.

23) Cut looking for significant remaining bumps in the residuals to the best fit is applied.

24) Selected events and useful data written out to file

25) Plots of interesting lightcurves are created, for both variable and microlensing candidates.

The main cuts are summarised in Figure \ref{Cuts_summary} and will be described in greater detail from Section \ref{insufficient_points} onwards.
 The numerical labelling of particular cuts used in Figure \ref{Cuts_summary} is maintained throughout the sections below.

\subsection{Reading in data}

 It is possible to request the reading of any subset of the available
 telescopes or wavebands of data through the use of a parameter array. In 
other words, analysis can be performed on any number of 
 telescope wavebands greater than one and less than an arbitrary maximum 
 of ten corresponding to the largest expected (number of telescopes multiplied by 
number of wavebands per telescope) for our project.
 In the original data files generated by the ADAP, (see Section \ref{angstrom_dap}), each data time-point is classified as being either
 `CLEAN', `CONTAMINATED', or `NOERROR'. `CONTAMINATED' data occurs when,
 for example, there is a bad pixel or a cosmic ray on the particular
image being considered, within the PSF of
the variable object for which the photometry is being calculated, which may make the flux calculated less reliable in an unquantifiable way. `NOERROR' means that it is not possible to estimate the error for that particular time-point. Both the `GOOD' and `CONTAMINATED' points are read in by the Pipeline, but only `GOOD' points are thereafter used in the calculations.
 The numbers of `GOOD' points in each band that were
 read in by the data-reading routine are counted and used throughout the 
 rest of the pipeline. 

\subsubsection{Reference Images}
\label{ref_im}

 For the reason stated in Section \ref{Angstrom_DIA} the data points associated with
 the reference image were removed from the $2007$ photometry lightcurves before the
 pipeline analysed the data. For this
photometry, the LT i'-band reference image was ``LT\_i\_20060131\_012.fits'', image taken on 31/01/2006. The LT r-band reference image was ``LT\_r\_20050114\_037'', and the equivalent FTN i' and r band images were ``FTN\_i\_20051216\_006'' and ``FTN\_r\_20051123\_001''.

Reference image masking was not required for the $2008$ photometry data
which used the combination of several good images as the reference.

    \subsection{Using ``Sigma Clipping'' to remove outliers}

During the development of the Candidate Selection Pipeline, it had been noticed that
 there exist data points which have flux values and/or error values much greater than the average values
By examining the plots of the best fits to the data it has also been noticed 
that for fits with a moderately low value of $\chi^2$, the Paczy\'nski peak was placed at a 
particular $t_{0}$ because around that time point there were one or two data points with higher than
average flux but smaller than average magnitude of error bar. Hence the $\chi^2$ was being minimised by making the fitted function pass as close as possible to these 
``small error'' points. 

The estimation of errors in the D.I.A. pipeline is a complicated process and it is difficult to prevent the occasional error bar being mis-estimated due to unlikely but possible occurrences in the pipeline.
 Therefore it was considered reasonable to apply sigma clipping not only 
to the flux values, but also to the error values.
A subset of the images corresponding to the data points rejected were examined at the coordinates of the variable objects to which the lightcurves belonged in order to investigate whether any clear reasons existed for the ``bad'' data points.
The majority of those examined were points close to the kind of major localised image flaws shown in the images in Section \ref{image_flaws}, or just in very noisy difference images, but there were also in some images what appeared to be
un-masked cosmic rays, as is almost unavoidable.
 
\subsubsection{Sigma clipping on flux values}

The discrepant flux point which occurs most frequently is most often numbered $86$th in the order of
data from POINT-AGAPE field number $2$, although this can vary minus a few if previous time points are missing.
This data point occurs at $(JD -2400000.5) = 51803.25$. This point usually has fluxes of the
order of thousands of units, where the mean flux is of order tens of units. In the $2007$ photometry, for example, almost all the data at this time index is marked by the DIA pipeline as ``CONTAMINATED'', but still in a few lightcurves some data is classified as ``CLEAN'' and some of these points have either fluxes or error bars that are very large, as described above.
It is found that setting
a first cut on the flux at flux mean$\pm10\sigma_{(\rm{original})}$ for standard deviation ``$\sigma_{(\rm{original})}$''
 in each data band successfully removes this point and others like it, 
without impacting on the bulk of the data which is not discrepant. Once this first iteration has been performed,
and discrepant points removed from the list of data,
the standard deviation is re-calculated and the same cut performed using the new value of $\sigma$.
 This iteration is repeated until the total number of points remaining 
becomes constant.

\subsubsection{Sigma clipping on error values}

Since errors on flux points are always positive in sign, a cut such as performed on the flux points which is symmetrical around the calculated mean will not work, as for some values of the multiplicative constant parameter (which has the value $10$ in the flux cut above) the lower cut would be at a negative value. The errors are expected to have a Poissonian distribution which has a most commonly occurring value near the mean value and tails on either side of this which extend down to zero error and upwards to infinity. In order to allow a simple cut to be made both on errors which are much smaller and much larger than the mean value, but using only one multiplicative constant parameter $n$, a data point was cut if the error value either exceeded $n\times\rm{Mean Error}$ or if it was less than $\rm{Mean Error}$/$n$. For example, if the constant was chosen to be $10$, the cuts would be made in the two regions Error $< \rm{Mean Error}/10$ and Error $> 10\times\rm{Mean Error}$.

Data about each point which was rejected by the sigma clipping algorithm,
 either due to a flux or error discrepancy, were written to a file
along with the reason it was rejected and whether the limit exceeded was the upper or lower one.

 The parameter space of $n$ was sampled as far out as $n=200$, and even at the upper end of this range,
 there were still significant numbers of points which exceeded both the upper and lower error bounds.
This seems to imply that the geometric nature of the use of the cut parameter $n$ does not lead to
too great an imbalance in the numbers of low and high points rejected. The statistics of the total 
number of points and their breakdown into ``too low'' and ``too high'' were analysed for both flux and error
 clipping, using only the first $2000$ lightcurves (to speed up the process somewhat),
 and are presented in Table \ref{clipping_parameters}: 

\begin{table*}
\caption[Table showing the numbers of data points rejected by the sigma clipping routine for a range of different values of the error parameter $n$.]{Table showing the numbers of data points rejected by the sigma clipping routine for a range of different values of the error parameter $n$. (The flux clipping parameter was always set at $10\sigma$).}
\begin{footnotesize}
\begin{center}
\begin{tabular}{|c|c|c|c|c|c|c|c|c|}
\hline
\hline
 \footnotesize{Error clipping}      &\multicolumn{3}{c|}{\footnotesize{Number Rejected}}    &\multicolumn{3}{c|}{\footnotesize{Number Rejected}} &    \multicolumn{2}{c|}{\footnotesize{Total Number of}}  \\
\footnotesize{Parameter n}    &\multicolumn{3}{c|}{\footnotesize{for flux}}           &\multicolumn{3}{c|}{\footnotesize{for error}}       &  \multicolumn{2}{c|}{\footnotesize{Data Points Rejected}} \\
\hline
                     &   \footnotesize{Low}      &    \footnotesize{ High}     &  \footnotesize{Flux}  &     \footnotesize{Low}    &   \footnotesize{High}    &  \footnotesize{Error}   & \footnotesize{Total} & \footnotesize{Total}\\
                     &       &       &  \footnotesize{Total}  &        &       &  \footnotesize{Total}   & \footnotesize{(num.)} & \footnotesize{(\%)}\\
\hline
 $20$                & $397$  & $17$  & $414$  & $2081$ & $429$ & $2510$ & $2924$ &  $0.182$     \\
 $40$                & $416$  & $17$  & $433$  & $351$  & $323$ & $674$  & $1107$ &  $0.069$     \\
 $60$                & $426$  & $18$  & $444$  & $126$  & $214$ & $340$  & $784$  &  $0.049$     \\
 $100$               & $435$  & $18$  & $453$  & $30$   & $85$  & $115$  & $568$  &  $0.035$     \\
 $150$               & $444$  & $18$  & $462$  & $8$    & $29$  & $37$   & $499$  &  $0.031$     \\
 $200$               & $444$  & $18$  & $462$  & $5$    & $6$   & $11$   & $473$  &  $0.029$     \\
\hline
\end{tabular}
\end{center}
\end{footnotesize}
\label{clipping_parameters}
\end{table*}

It can be seen that the numbers of points rejected for flux discrepancies varies with the value of $n$,
even though the equivalent parameter for flux clipping was constant at $10$. This is because if, for example,
the error cut parameter $n$ is set higher, then the number of data points rejected in the first iteration is
 lower, which means that more data points remain to be considered for the flux cut in the second iteration.
In many cases, if a point has an overly large error it also has a proportionately large flux value, and so is
rejected on the basis of flux in the second iteration whereas previously it was rejected on the basis of error
in the first iteration. The number of points rejected on the basis of flux became constant for larger error
clipping parameters than $150$, which indicates that there are no more points with high flux magnitudes which have 
errors which are more than $150$ times the mean error. It is clear that the final value for the number of points 
rejected will converge to $462$ if $n$ is increased much further as almost no points are rejected on the basis of
error when $n$ is as large as $200$. The total number of data points contained in the $2000$ lightcurves was $1,605,674$, which is divided into the total number of points discarded to give the percentages of the whole in the final column.

\subsection{Rejection of lightcurves with insufficient data}
\label{insufficient_points}
  The number of ``good'' data points in each individual telescope waveband are 
totalled together. If this number is greater than a threshold value, set at $25$, then the pipeline proceeds with the analysis, else it ``fails'' the lightcurve and moves on to the next.

\subsection{Iteration over residuals}

The pipeline is designed to find the best fit to the data, using functions representing constant flux, periodic variable stars, point lens point source microlensing or an additive combination of lensing and variability. When this fit has been decided upon, the residual of the fit is calculated. The pipeline then iterates by examining the residual in a very similar way as it did the original flux data,
up to a maximum number of degrees of residual, currently set to be three.
 Iteration over residuals for a particular lightcurve is cut short when:

1) The $\chi^2$ of the best fit function increases from the value found by 
 the previous iteration, or if

2) The $\chi^2$ goes below the value of the parameter for the cut on
 global reduced $\chi^2$ (Cut 5) (see below, Section \ref{reducedchisqcut}). In other 
 words a sufficiently good fit has been found, or if

3) A constant is the best fit to the lightcurve, or if

4) No bumps (as defined below, in Section \ref{bumpfinding}) were
 found in the current iteration's residual data.

\subsection{Bump Finding\label{bumpfinding}}

           \subsubsection{First iteration (flux data)}

The first process performed on the data is always to search for ``bumps'' in the data. That is, coherent areas in time within the data which show a positive deviation with respect to the baseline of the lightcurve. On the first iteration the data are always examined using a routine referred to hereafter as ``bump\_find'', which uses a point-by-point counting algorithm. This is done on each telescope band in turn, and a count is kept of the number of bumps found in each band.
 The first action performed is to estimate a baseline for the 
lightcurve. This is achieved by calculating the five point moving average for the lightcurve and then taking the minimum value of that function.
Specifically five points were chosen as being a reasonable compromise between being sufficiently localised to react to short term downward fluctuations, but not so localised as to prevent some smoothing over noise in the lightcurve.
 Next, the lightcurve is examined point by point looking for positive 
deviations above this baseline value.
A parameter $f_{dif} = (F_{i} - B)/(E_{i})$, where $F_{i}$ is a flux point, B is the calculated baseline and $E_{i}$ is the error bar on the flux point, is calculated for each point $i$. If this number goes above $3$, i.e. $3\sigma_{i}$ for any point then the count of points $n_{p}$ in the possible bump begins.
If $f_{dif}$ goes above $5$ then a separate count of ``high signal to noise'' points $n_{h}$ is begun. While a bump continues, the total ``significance'' of the bump is calculated by summing $f_{dif}$ quadratically if $f_{dif}$ is positive.
 If a point drops below $f_{dif}$ = $3$ then a count of 
non-significant 
points $n_m$ is started. If this number reaches $2$ when $n_p > 5$ \emph{and} the $n_{h} > 3$ and the $n_{h}$ points are consecutive then the bump is simultaneously aborted at that point and counted as a ``real'' bump. This combination of
significances guarantees that any bump found is at least $8.77 \sigma$ above the baseline.
     An additional prescription is introduced largely to mitigate a
 perceived inconsistency in the logic caused by the ``$n_m = 2$ to end a 
bump" condition. It is possible for a bump to be begun and then for a single point to have such a large negative $f_{dif}$ that the totalled $f_{dif}$ for the bump becomes less than zero, but without triggering the $n_m = 2$ condition.
If the new point still has $0 < f_{dif} < 3$, i.e. is positive but not large enough to be considered ``significant enough'' then nothing unusual happens. Its $f_{dif}$ is added quadratically as before. Hence even points which are considered ``not high enough to be part of the bump'' can still contribute in a small way to the total significance.
If the new point has a negative $f_{dif}$ and subtracting the $f_{dif}^2$ of the new point number $i$ from the current total of $f_{dif}^2$ of the bump so far would not make the total $\Sigma{f_{dif,i}}^2$ up to the $i$ th point in the bump negative then this is done, by 

\begin{equation}
\mbox{Running total of bump significance} = \Sigma{f_{dif,i}}^2 = \sqrt{\Sigma{f_{dif,(i-1)}}^2 - {f_{dif,i}}^2 }
\end{equation}

Thus the negative $f_{dif}$ data point reduces the significance of the bump but does not end it.
However, if calculating $f_{dif,(i-1)}^2 - f_{dif,i}^2$ results in a negative number and there have been insufficient points with positive $f_{dif}$ then this is classed as a ``false start'' and the bump is aborted without being counted as ``real''.
 The start and end of the bump are set to the first and last points 
where $3 < f_{dif}$. The routine also returns the total $\chi^2$ of each bump, and an estimate of the central time of the bump calculated by 
a mean weighted by $f_{dif,i}^2$ for all the time points $i$ in each bump.

           \subsubsection{Subsequent iterations (flux residuals)}

In situations where a ``good'' fit has been found between the applied model and the data, it is expected that any remaining significant coherent deviations from zero will be much less obvious than those in the original data.
``good'' here means a fitted $\chi^2$/d.o.f. which is reasonably low, but not low enough to pass Cut 5 (see Section \ref{reducedchisqcut}).
 Therefore on subsequent iterations, which attempt to find remaining variations in 
 the residuals to the first fit, a different routine is
 used. This is designed to detect \emph{the one most significant} deviation (bump)
 in the lightcurve. This is mainly aimed towards detecting microlensing events which
 may have been overwhelmed on the first iteration by a more obvious variation
 (likely to be stellar variability), but which can be detected after this first
 order variation has been subtracted.
 This algorithm works by calculating the $f_{dif,i}$ of each point $i$ in 
the lightcurve and ordering them with a sort routine so that the most
 significant deviation is at the top of the list. Next it looks for a 
cluster of five or more points at the top of the list which are consecutive in time order, with the exception that there can be one ``outlier'' interposed within this group. If such a group is found, the search is extended to the next point on the list, which is either added to the bump if it is contiguous with it or not if it is not. The bump is
extended until no more points can be added. The $f_{dif,i}$ is used as the determinant of significance because if $f_{dif,i}^2$ were used a large negative deviation from zero residual would be treated in the same way as
a large positive one, meaning that roughly half the time it would subsequently be attempted to fit a negative dip in the lightcurve with a positive Paczy\'nski curve. In addition, residual lightcurves with a lot of ``noiselike'' variations where consecutive time-points are often alternately positive and then negative would commonly be detected as having bumps in $f_{dif,i}^2$ when this in reality only shows high variability, not a positive signal.
 The central peak time is calculated using a
$f_{dif,i}$-weighted mean of all the time points in the group. The width of the hypothetical 
Paczy\'nski peak is estimated from a $f_{dif,i}$-weighted mean difference from the central peak time calculated from all the points in the bump group.

\subsection{Bump Combining}

\begin{figure}[!ht]
\vspace*{10cm}
   \includegraphics{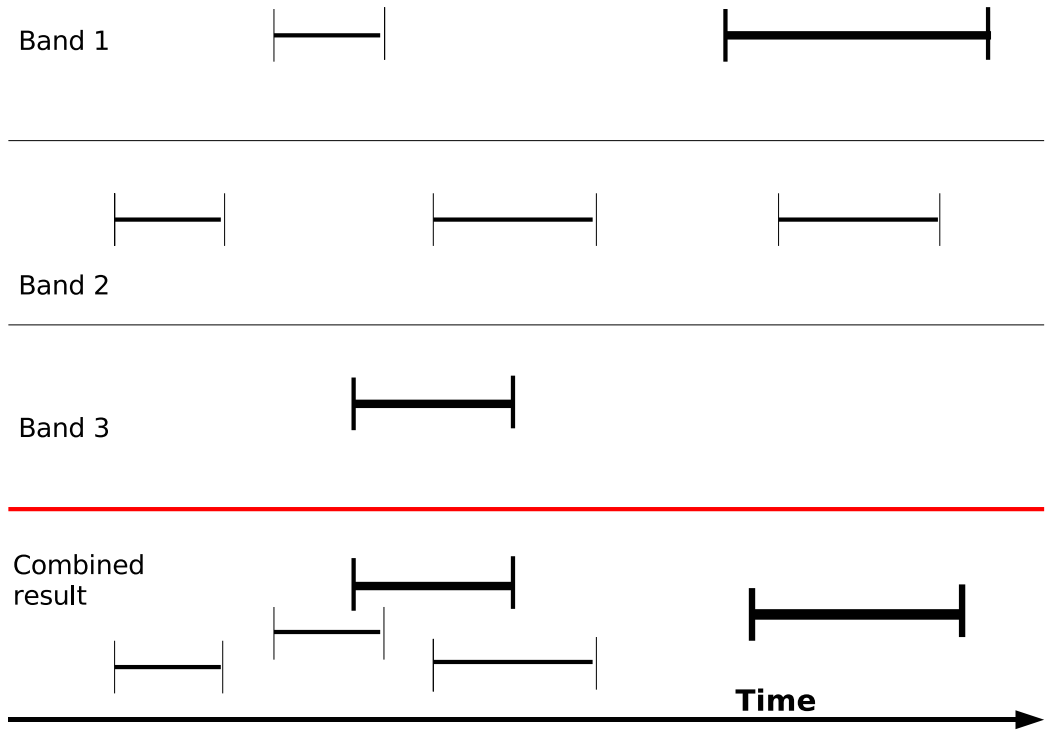}
\caption[An example of how the bump-combining process works.]{An example of how the bump-combining process works. The three 
bands of data shown in the top three panels are analysed to find any
pairs of bumps which lie completely inside one another. These are then combined into one by adding the $\Sigma{f_{dif,i}^2}$ values and finding the $\Sigma{f_{dif,i}^2}$ -weighted average of each of the end points. The boldness of the lines represents the relative significance of each bump.
}
\label{bump_combining}
\end{figure}

Because the data analysed are contained within several separate telescope wavebands, which could not be directly combined due to lack of good knowledge about the relative flux scaling factors, it is possible that one physical bump in the lightcurve may have been detected in several different bands
at once. Therefore in order to correctly assess whether the lightcurve can be characterised as having ``one bump'' or should be better classed as ``many bumps'', it is necessary to decide how to combine bumps which overlap in time in different bands. There are several ways of doing this, but the more conservative convention adopted for this work was that two bumps in different bands should only be combined if the boundaries of one bump are completely contained within the boundaries of the other.
Two bumps which overlap partially at one end are allowed to remain as separate entities. A hypothetical example of this is shown in Figure \ref{bump_combining}.
Three bands of data are shown with an initial total of six detected bumps. The start and end of each bump are shown by the extent of the horizontal brackets. The significance of each bump is illustrated by the thickness of the horizontal line, more significant being thicker.
It can be seen that after the combination process has been completed, the two bumps in Bands $1$ and $2$ which were latest in time have been combined into one larger bump in the bottom panel. The pipeline simply adds together (quadratically) the $\Sigma{f_{dif,i}^2}$ values of two combined bumps to show that there is more evidence for the combined bump due to it having been detected in two bands. The end points of the combined bump are calculated by finding the average of the two initial bumps' end points, weighted by their initial $\Sigma{f_{dif,i}^2}$ values. Hence the boundaries of the final bump will be closer to those of the more significant original bump. These bump width values are only used as starting points for the MINUIT (CERN, 2006) minimisation to follow and so need only be approximate. The combination of bumps occurs serially,
so that bumps considered later are compared with the new combined end points of any bumps already found to overlap, rather than trying to find ``nests'' of bumps which \emph{all} lie inside one another from the totality of bumps and combining them all at once.

It has been considered that this algorithm might have problems if the beginning of a physical bump is detected in the data of one band, which then has no more observations, and the end of the same bump is then detected in observations beginning in a second band. In this case, if the data in the two bands does not overlap or overlaps only partially, one physical bump might be classified as two. However, since LT and FTN data are collected simultaneously during the observing season, this situation should only occur at the boundary between PA data and Angstrom data, or alternatively
for very short timescale bumps which insert themselves between clusters of data in different bands.
Even in the instance of this occurring, the most serious consequence for the pipeline would be that a mixed (Skew-Cosine + Paczy\'nski) fitting function would be used instead of a simple Paczy\'nski fit (see Section \ref{choosingfits}). This should not be too serious, as either the Paczy\'nski or the ``Cosinusoid'' function, or both, are free to vary their amplitudes down as low as zero and so, practically, can regain the fit that should ideally have been found.

\subsection{Periodogram Analysis}
\label{periodogram}
After the number of bumps in the lightcurve has been established,
A period-finding routine which implements the Lomb-Scargle Periodogram
algorithm \citep{1975MNRAS.172..639L,1972MNRAS.156..181S,1972MNRAS.156..165S,1982ApJ...263..835S}
 is used on each band of data separately to find significant periods. Currently, $1$ million 
frequency divisions
are used which was a number merely chosen large enough (since this was a computationally fast algorithm) to ensure the smallest step in period was sufficiently small to be unimportant, especially since the periods found are only used as ``initial seed'' values for the fitting. From the periodogram, periods are first selected which are local maxima 
and then, the most significant periods are chosen using a cut of
$\rm{Periodogram(frequency)} > 30$.

 The period range searched by the periodogram
began at $0.5$ days for the analysis of the $2007$ data, but was reduced to $0.01$ days for the $2008$ data. The highest period searched is $1.5\times(t_{\rm{max}}-t_{\rm{min}})$ where $t_{\rm{max}}$ is the latest data point for the particular lightcurve and $t_{\rm{min}}$ is the earliest data point. A multiple of $1.5$ is chosen based on the logic that for a period as large as twice the range of the data $\Delta{t} = (t_{\rm{max}}-t_{\rm{min}})$, a variable star might not be reliably distinguished from a Paczy\'nski curve, whereas a slightly shorter period of $1.5\times\Delta{t}$ should have a noticeable rise at one end or the other above the hypothetical Paczy\'nski baseline, if there is sufficient data coverage.

The number of significant periods found, including when zero periods are found, along with the number of bumps found, is used to make further decisions about what fitting operations to perform on the lightcurve. This is further described in Section \ref{choosingfits}.

\subsection{Fitting functions used}
\label{fitting_functions}
 The CERN MINUIT (CERN, 2006) (originally written by \cite{1975compj..10..343})
software was used to minimise the $\chi^2$ of the fit.
Where possible the allowed ranges of parameters were left unconstrained as this increases the efficiency of the search algorithm, but since MINUIT is a ``local minimisation algorithm'' and not designed to find ``global'' minima, it was also found to be important to pre-calculate guesses for the starting values of parameters as well as possible.
The particular fitting function used in all fitting routines was ``MINI''. This in turn calls the MIGRAD routine, but switches to the SIMPLEX routine if MIGRAD fails. MIGRAD uses variable-metric (also known as ``quasi Newton'') algorithm, based upon the algorithm of \cite{1970compj..13..317-322}, with a stable metric updating scheme, and checks for positive-definiteness of the calculated covariance matrix, because the search directions determined by MIGRAD are guaranteed to be downhill only if the covariance matrix is positive-definite. This algorithm estimates the matrix of second derivatives (the ``Hessian'') of the defined ``quality function'' with respect to the various parameters and uses the inverse of this matrix to iterate the parameter values towards the minimum of the quality function. Thus
the computing cost will rise as the size of the matrix that needs to be inverted rises. The matrix involved is of size n$\times$n, where n is the number of fitted parameters. The stated main weakness of this algorithm is that it depends heavily on knowledge of the first derivatives, and fails if they are very inaccurate.
The SIMPLEX routine is based upon the simplex method of \cite{1965compj..7..308-313}.

 Attempts were made to reduce the number of fitted parameters to a minimum, in order to increase the efficiency and reduce the required computing time of the
 fitting, and consequently the overall running speed of
 the program. To this end it was possible to link the fitted amplitudes of LT i'-band
 and FTN i'-band difference image data and also those of LT r-band and FTN R-band, thus
 reducing the number of parameters by one when each of these pairs of data sets exist and are used. The average ratios between these telescope bands were found by comparing many small equivalent areas of an image for each telescope. The ratios found were

 $\rm{Flux}_{\rm{LT i'}} = $Flux ratio$_{\rm{(LTi'/FTNi')}}\times\rm{Flux}_{\rm{FTN i'}}$

 and
 $\rm{Flux}_{\rm{LT r}} = $Flux ratio$_{\rm{(LTr/FTNr)}}\times\rm{Flux}_{\rm{FTN r}}$

where the constant ratios between the difference fluxes in bands assumed were

Flux ratio$_{\rm{(LTi'/FTNi')}} = 1.081 $ and Flux ratio$_{\rm{(LTr/FTNr)}} = 0.305$

The fitting code can either be used with or without maintaining a fixed scaling factor between the flux amplitudes in the PA and (LT and/or FTN) bands.
 The first links the PA flux amplitude to the LT and/or FTN flux amplitudes using
 $\rm{Flux}_{\rm{PA i'}} = \frac{\rm{Flux~ratio}_{\rm{(PAi/FTNi)}}}{\rm{Flux~ ratio}_{\rm{LTi/FTNi}}}\times\rm{Flux}_{\rm{LT i'}}$
and hence
$\rm{Flux}_{\rm{PA i'}} = $Flux ratio$_{\rm{(PAi/FTNi)}}\times\rm{Flux}_{\rm{FTN i'}}$.
 The second
uses the flux amplitudes in PA data as variable fitting parameters and does not assume any knowledge of the value of the flux amplitude ratios between PA and LT or FTN.
 The advantage of linking the PA flux amplitude to the other bands used is that
fewer fitting parameters were required, which simplifies the parameter space used, meaning that the code can run faster and more efficiently.
However, for initial investigations of the flux ratio between PA and LT and/or FTN data the ``unlinked'' version had to be used, for obvious reasons.
Using the ``unlinked'' version also has other disadvantages in that the fitting routines are free to find any values of flux amplitude which have the effect of reducing the
 overall $\chi^2$, which can occasionally lead to unphysical ratios between waveband flux ratios or strange combinations of Paczy\'nski and Sinusoid parameter values. 
These solutions may improve the fit to a particular data set, as measured by $\chi^2$, but at the expense of flux amplitudes and hence flux ratios between bands which could never occur in reality.
 Thus additional prescriptions and safeguards had to be introduced (which are described below) to minimise the number of these ``wild'' fits which pass through the selection cut system.

   \subsubsection{The reduced Paczy\'nski function for microlensing fits}

A reduced Paczy\'nski curve is fitted
in which the flux varies with time as,

\begin{equation}
F\rm{(t)}=\rm{B}+\Delta\rm{F}/\sqrt{1+12[(t-t_{0})/t_{\rm{FWHM}}]^2}
\label{red_pac}
\end{equation}

where $t$ and $t_{0}$ are the epochs of observation and
maximum brightness, $B$ is the baseline flux, $\Delta{F}=F{(t_0)}-B$ is the
maximum flux deviation and $t_{\rm{FWHM}}$
is the duration of the full-width at half-maximum.

   The initial conditions for the fit i.e. the $t_0$ and the 
estimation of the order of magnitude of $t_{\rm{FWHM}}$ are primed by the output obtained by bump\_find from the width and position of the (largest or only) bump.
In order to exert some control on Paczy\'nski peaks with huge amplitudes being fitted with $t_0$'s in gaps in the data, (attempting to minimise the $\chi^2$ to an upwardly deviant point near the edge of the data immediately before or after the gap), the Paczy\'nski amplitude was limited to $100\times(\rm{Flux}_{\rm{max}} - \rm{Flux}_{\rm{min}})$.

   \subsubsection{The Skew Cosinusoid Function for variable stars}
\label{skew_cos_function}

The skew cosinusoid function chosen, which was based on $\phi \propto t^S$,
 was designed to be capable of being adjusted by the fitting
 routine continuously from zero skewness (a simple cosine curve) to high 
skewness where the rise time is much shorter than the fall time.

In this model, the phase rises from zero initially as $t^S$, where $S$ is the skewness parameter.
After the middle of the period, at $P/2$, it rises the rest of the way to $\phi =  2\pi$ as 
the double reflection around 
$\phi = \pi$ and $t = P/2$, i.e. $\phi(t)$ initially has an increasing gradient, but after half 
the period has a decreasing gradient back down to a gradient of zero at $t = P$, $\phi = 2\pi$.
For a skewness of $1$, this function reduces to $\phi = t$ and has smooth gradient changes over
the whole period, as required. Since the rise of this function steepens rapidly with increasing $S$,
 in order that the function be continuous at $\phi = \pi$ and $t = P/2$, the function must be normalised to equal $\pi$ at this point by dividing by $\frac{(P/2)^S}{\pi}$

For a practical reason, (since the zero phase of a Cosine function occurs at the peak of the
 function and bumps in lightcurves are being searched for) a Cosinusoid was actually used 
instead of a Sinusoid. To model a variable star lightcurve,
 parameters required are : the amplitude of the cosinusoid, $F_{\rm{max}}$, the baseline flux $B$, the zero phase of the cosine $\phi_{0}$ where
 phase $\phi = \frac{t}{P}2\pi$ and the phase function $A$. The temporal variation of the flux $F$ is represented by: 

\begin{equation}
\label{skewcos1}
       F = {F_{var}}\cos({A - \phi_{0}}) + B
\end{equation}

  The phase function $A$ is defined by:

\begin{equation}
\label{skewcos2}
  A = \left\{ \begin{array}{ll} 
 \pi^{1-S}\phi^S -\frac{\pi}{2}& \mbox { if } 0 \leq \phi \leq \pi \\ 
2\pi - \pi^{1-S}{\left(2\pi-\phi\right)}^S -\frac{\pi}{2} & \mbox { if } \pi \leq \phi \leq 2\pi \\
  \end{array}
   \right\}  
\end{equation}

 \textbf{The skewness parameter, ``$S$''}

The cosinusoid function was originally conceived varying in the range $1<S$. In this range,
 for relatively moderate values of the skewness parameter,
  a point of inflection, occurring at $\phi = \pi$,
 begins to be visually obvious around $1.5 < S < 2$. By choosing examples of 
lightcurves that were deemed ``acceptable'' or ``unacceptable'', the upper limit for $S$ in the fitting routines was reset to be $S < 1.574$ which corresponds to the rise time being a fraction of one period of $0.356$.
This was the limit used for all fitting on the 2007 data, including the investigation of variable stars (Section \ref{variables}). During later development it was found that $S$ could also be allowed to be less than one and still produce a physically useful skew cosinusoid function. In order to get a skew cosinusoid which has 
a rising portion which is shorter in time than the falling portion (as in many variable stars), when using $S>1$, 
 the sign of $F_{\rm{var}}$ in Equation \ref{skewcos1} must be negative, but if $S<1$ then the sign of $F_{\rm{var}}$ must be positive to achieve the same end.
When $S<1$ the skew function is smoother and has a different shape than when $S>1$ and does not suffer from the function becoming no longer smooth as the value of the skewness parameter becomes lower.
 In the fitting algorithms
these conditions are imposed by formally limiting the ranges of the fitting variables used in MINUIT. The maximum skewness used in the fitting
corresponds to the rise time being $>5\%$ of the total period, which is about the most extreme seen in variable stars.
The equivalent limit in the skewness when $S<1$ is $0.02314<S<1$. With the above function, the fraction of the period taken by the rise time is given by
 Equation \ref{skew_rise_time}. Mathematically this second $S$ limit is equivalent to using $\frac{t_{\rm{rise}}}{P} = 0.95$ in Equation
 (\ref{skew_rise_time}), as when the function is reflected when the flux amplitude changes sign, the rise time then becomes the fall time and vice versa.
Both possible regimes of the function, i.e.

 $0.02314<S<1$, $F_{\rm{var}} > 0$ and
 $1<S<1.574$ , $F_{\rm{var}} < 0$ were utilised in the fitting to the $2008$ data.

\begin{equation}
\label{skew_rise_time}
 \frac{t_{\rm{rise}}}{P} = 1-\sqrt[S]{\frac{1}{2}}
\end{equation}

\begin{figure}[!ht]
\vspace*{10cm}
   \includegraphics{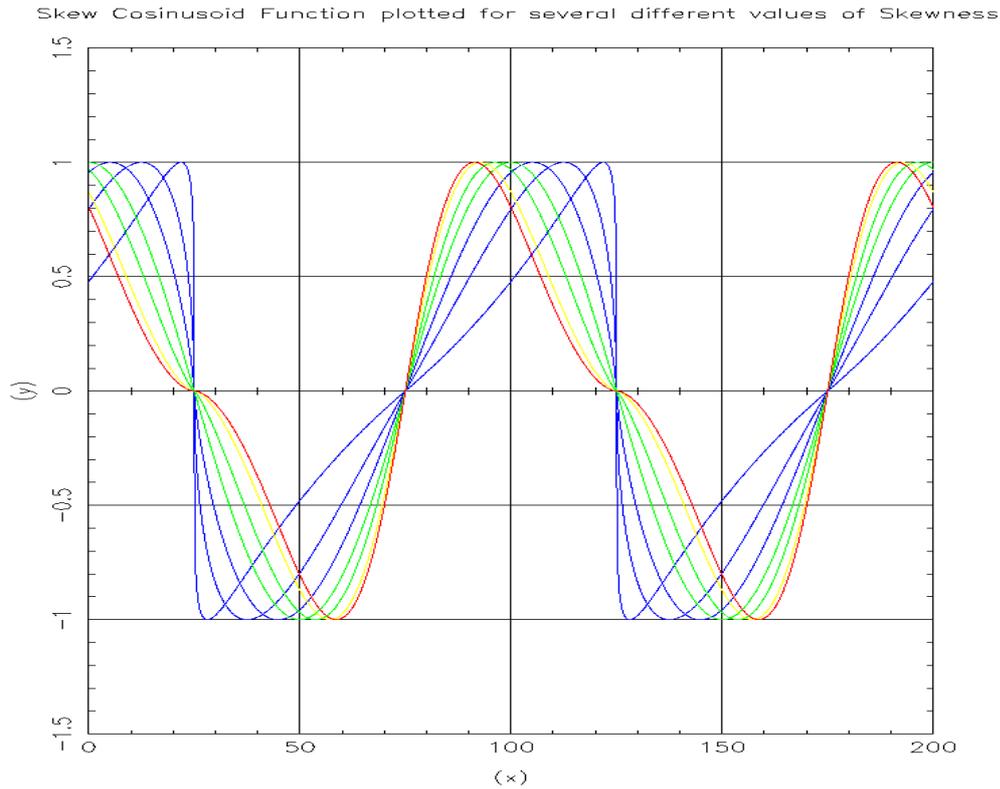}
\caption[Plot showing the skew cosinusoid function used in the candidate selection pipeline for varying skewness parameter.]{Plot showing examples of the skew cosinusoid function used in the candidate selection pipeline for varying skewness parameter. The skewnesses shown are $0.2314$,$0.50$,$0.75$,$1.00$,$1.25$,$1.574$ and $1.75$. The first three, which have $S<1$, are shown in blue- in practice these would be effectively time-reversed by using a negative amplitude. The next two (1.0,1.25) are shown in green. The upper limiting value of $1.574$ is shown in amber, and a value of $1.75$ which is considered to have too extreme a point of inflexion to be usable is shown in red. 
}
\label{skewcos_varying_skewness}
\end{figure}

Some examples of the skew cosinusoid function, plotted for varying values of
skewness parameter, are shown in Figure \ref{skewcos_varying_skewness}

 The $t_0$ of the largest bump in the lightcurve, found from the bump-finding routine is 
 used to define the starting point for the fitting of the
zero phase of the cosine function.

 If the periodogram has been calculated and produces any significant 
periods, then these are used as initial conditions for the Skew Cosine part of the fits. Otherwise when either the periodogram is not used, or if it is used and does not find any periods for a particular lightcurve, the period range between $0.5$ to $1000$ days is equally sliced in $\log(P)$.
 In both this function and the joint fit below, MINUIT limits on 
the fitted period are only defined (to give the size of the required P slice) if P is sliced, but allowed to vary freely if the periodogram is used.
For the $2007$ photometry, $52$ overlapping slices are used to avoid having touching slice boundaries and to give $50$ slices covering the actual region of interest, but this is reduced to $6$ for the analysis of the $2008$ photometry in order to reduce computation time.

   \subsubsection{The Joint Paczy\'nski + Skew Cosinusoid Fit Function}

This consists of a simple sum of the reduced Paczy\'nski and Skew Cosinusoid Fit functions described above. It was necessary to include this kind of function because
Angstrom is observing close to the M31 core, where stellar crowding is
a serious problem. Therefore a large fraction of variable objects contained some kind of detectable contribution from one or more variable stars. If a ``traditional'' cut specifying a requirement for a flat baseline outside the lensing peak had been used,
many events in our fields would be discarded.

\subsection{Selecting which functions to fit}
\label{choosingfits}

 Based on the output of bump\_find (or chi\_bump) and the 
Periodogram, the Pipeline decides which functions are most appropriate to attempt to fit to the 
lightcurve.

\begin{figure}[!ht]
\vspace*{10cm}
   \includegraphics{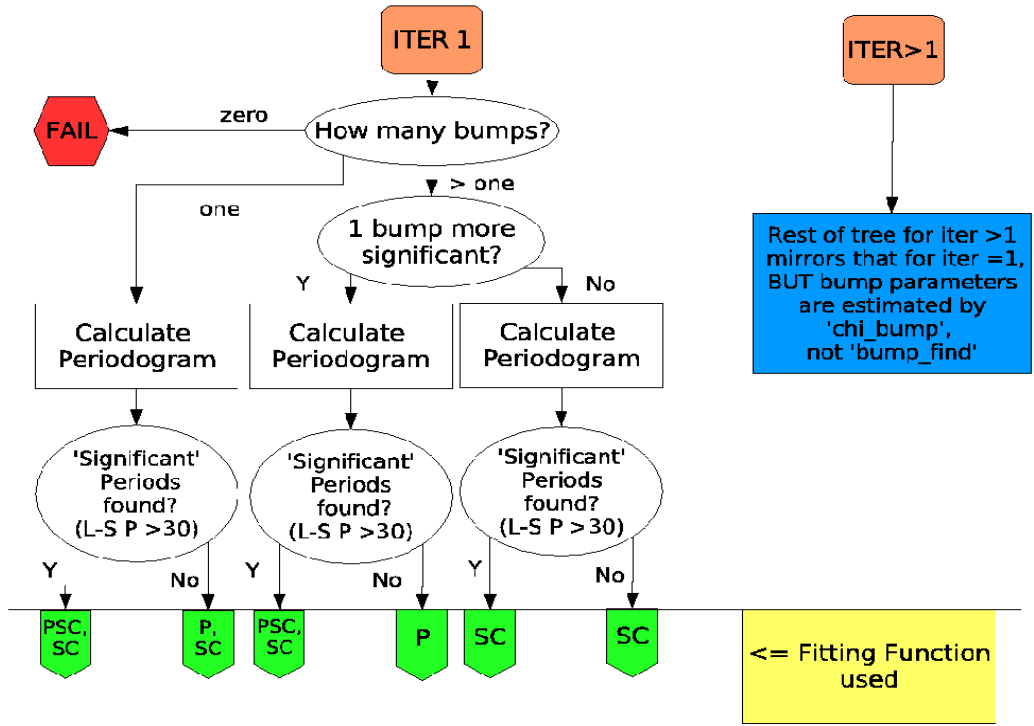}
\caption[Flow-Diagram showing which fitting functions are used given 
particular attributes of the lightcurve.]{Flow-Diagram showing which fitting functions are used given 
particular attributes of the lightcurve such as a) number of bumps
 found (if any), b) dominance of one bump over any others, c)
 periodicity or lack of it. On the bottom (green) level the fit functions 
used are designated as P = Paczy\'nski, SC = Skew Cosine, PSC = Paczy\'nski + Skew Cosine. On the second level from the bottom, ``L-S P''
refers to the value of the Lomb-Scargle Periodogram at the period of interest.
}
\label{decision_tree}
\end{figure}

 In all cases the best fit constant value is calculated along with its 
$\chi^2$. This is then compared to the more sophisticated
fits used. There are several layers of filters in the logic tree, and this is more easily shown in its entirety by Figure \ref{decision_tree}.
However, the basic philosophy is that if the lightcurve has only one bump
then a Paczy\'nski fit is performed, along with the Skew Cosine function for comparison. If the lightcurve has more than one bump, then it is likely that a Skew Cosine function is a possible fit, so this fit is performed.
 However, if one of the bumps is significantly ``more prominent'' than the rest,
 then the possibility of microlensing superimposed on a variable star is considered,
 and a joint (Paczy\'nski + Skew Cosine function) fit is also performed.

For the definition of ``more prominent'' currently chosen 
see Cut 2 in Section \ref{the_cuts} below.

\section{The Cuts}
\label{the_cuts}
\textbf{Cut `0' Rejection of lightcurves with insufficient data }

  The number of ``good'' data points in each individual telescope waveband 
are totalled together. If this number is greater than a threshold value, set at $25$, then the pipeline proceeds with the analysis, else it ``fails'' the lightcurve and moves on to the next.

\textbf{Cut 1 At least one significant bump exists}

In order to be a possible candidate, a lightcurve must have a coherent
area of the lightcurve in which the difference flux is significantly higher than either the rest of the lightcurve in the case of a sufficiently flat baseline, or at least, if the variability is sufficiently high, its immediate neighbourhood in time. Therefore an algorithm was constructed (see Section \ref{bumpfinding} above) which identifies ``bumps'' in the lightcurve.
If no such bumps are identified, the lightcurve is classed as ``constant'', and no further fitting is performed, meaning that it cannot be either a lensing or variable star candidate.
 
\textbf{Cut 2 Microlensing is required for the best fit to the lightcurve }

When the $\chi^2$/d.o.f. of whichever fits were performed on a particular lightcurve are compared, to be selected as a lensing candidate
:
a) Either a Paczy\'nski or a mixed fit must have been fitted to the lightcurve and
b) One of these must have the lowest $\chi^2$/d.o.f. fit of all fits performed.
c) An ``F-Test'' (see Section \ref{F-test}) must decide that either a Paczy\'nski or mixed fit is the most appropriate fit, given the numbers of fitting parameters.

For condition a) to occur a criterion must be satisfied
that is designed to look for lightcurves with one bump which is more prominent than the others, and which would therefore be more likely to be well fitted by a joint fit.
 
 The most prominent bump is compared to the second most prominent bump in two 
separate statistics, namely the significance of the bump as defined by the $f_{dif}$ values (Section \ref{bumpfinding}), normalised by the number of points in the bump $n_p$, and the largest $f_{dif}$ component due to a single data point in each bump. (This looks more for brief but sharp flux spikes).
The final criterion was:
If the number of bumps found was more than $1$ and either 

    $\sqrt{\Sigma{f_{dif,i,1}^2}}$/$n_p > 1.5\sqrt{\Sigma{f_{dif,i,2}^2}}$/$n_p$,
    
  where subscripts 1 and 2 refer to the most significant and second most significant bumps respectively,  
 or,

${{\hat{f}_{dif,i,1}}} > 1.5$ ${{\hat{f}_{dif,i,2}}}$,

where $``{\hat{f}_{dif,i}}"$ is the largest value of ${f_{dif,i}}$ within the bump for a single data point i,
then a mixed fit was fitted in a) rather than a Paczy\'nski fit. 

If there was only one significant bump in the lightcurve, then a
simple reduced Paczy\'nski curve was fitted.

\textbf{Cut 3 Selection on $\chi^2$ difference ratios }

In order to be selected as a candidate, either as a microlensing event or as a variable star, a lightcurve must pass another cut, which is designed to ensure that the chosen best fit is \emph{clearly} better than the next best. The best fit function is only selected as a candidate if the percentage difference between $\chi^2$/d.o.f. of the best fitting function and the second best fitting function is greater than some threshold value. Equation \ref{chisq_ratio} shows a 10\% reduction in $\chi^2$/d.o.f.
 The quantity which must be evaluated to assess this is referred to as 
the ``$\chi^2$ difference ratio'', or ``CDR'' and it is
defined between competing fits A and B,

\begin{equation}
\label{chisq_ratio}
\rm{CDR} = \frac{\chi^2_{B}/\rm{d.o.f}-\chi^2_{A}/\rm{d.o.f}}{\chi^2_{A}/\rm{d.o.f}} > 0.1
\end{equation}

The threshold chosen for the final selection in this work was $0.4$.

This cut is
 similar to the first cut used by \cite{2005MNRAS.357...17B} except that in my
 analysis the joint fit is also performed and possible lensing events can
 be classified either as ``mixed'' or ``Paczy\'nski'', depending on which fit 
 has the lowest $\chi^2/\rm{d.o.f}$.
 
\textbf{Cut 4 ``Long period, low contrast'' mixed events}
\label{long_period_low_contrast}

 For ``mixed'' fits, it is harder to be confident in the fit to the lensing part of the ``mixed'' fit
if in turn we are less confident about the variable part. One circumstance which may reduce our confidence in the variable part of the fit is when the period of the variable is very long compared with the time span of the data for a particular lightcurve. Then the fitted variable function
may not repeat sufficiently, and it is not possible to say for certain whether
the behaviour of the lightcurve would actually remain the same outside this region, if the data existed.
Therefore, it is demanded in the case that the period is shorter than half the time span of the data (the variable does not repeat fully at least once) that the ratio of the fitted
Paczy\'nski amplitude and the fitted variable amplitude is greater than $5$. In the case of the period being longer than this criterion, a secondary condition is applied on the amplitude ratio, described below.

The area of parameter space in which the Paczy\'nski peak and the peak of the variable component may most easily be confused is when the time span of the peaks is of the same order of magnitude. This occurs roughly when $\rm{period}/2 \sim t_{\rm{FWHM}}$. Hence in this region a higher ratio is demanded between the fitted Paczy\'nski and variable amplitudes
in order to avoid classification as ``long period low contrast''. Outside this region, a lower criterion on this amplitude ratio is demanded.
The central region surrounding the above condition is chosen as $t_{\rm{FWHM}} < \rm{period} < 4$ $t_{\rm{FWHM}}$, and the resulting dual conditions which must be satisfied for a lightcurve to
pass the cut are:
in the region $t_{\rm{FWHM}} < \rm{period} < 4$ $t_{\rm{FWHM}}$ :
$F_{\rm{pac}}/F_{\rm{var}} > 5.0$,

 where $F_{\rm{pac}}$ is the flux amplitude of the fitted Paczy\'nski peak and $F_{\rm{var}}$ is again the flux amplitude of the variable component, 
 
 and in the surrounding regions,
 $\rm{period} \leq t_{\rm{FWHM}}$ 
or
$\rm{period} \geq 4$ $t_{\rm{FWHM}}$,

$F_{\rm{pac}}/F_{\rm{var}} > 0.5$.

Lightcurves failing any of these conditions are labelled as ``long period, low contrast'' events and are saved in a separate file.
The original motivation for developing this cut is described in greater detail, with an example lightcurve, below in Section \ref{mixed_events_with_long_periods}.
 This cut has a similar effect to the sixth and final cut of \cite{2005A&A...443..911C}.

\textbf{Cut 5 Selection on reduced $\chi^2$}
\label{reducedchisqcut}

 To ensure that only good fits to the data are passed, a cut is made on 
the reduced $\chi^2$ of the data relative to the best fit function. If all the physical variation in the lightcurve is entirely modelled by the fitted function, this number would be expected to be of the order of 1. However, experience gained modelling a particular lightcurve found by the Angstrom Project Alert System (APAS) \citep{2007ApJ...661L..45D}, showed that in our data there is much variation of the ``background'' to the objects which cannot always be modelled by simple one or two object fits such as described here. This may be caused by the high density of stars in the M31 bulge causing additional variable objects to be unavoidably blended in with the PSF of our studied objects. Therefore a preliminary loose cut is imposed in the main pipeline of a value for the reduced $\chi^2 < 15$. Tighter cuts on reduced $\chi^2$ are imposed in the subsequent second part of the analysis using the post-run analysis program.

\textbf{Cut 6 Sufficient points in central peak}
\label{bump_sampling}

For lightcurves fitted with either a reduced Paczy\'nski curve or a combined Paczy\'nski + Skew Cosine fit, a cut is imposed to ensure only candidates which have sufficient data points in the central region of the peak are allowed through. Without this cut, many Paczy\'nski peaks would be fitted with peaks sitting in gaps in the data. Also, it was found that this cut could be used as one of the strongest
ways of preventing classical novae being selected as (particularly short timescale) lensing events. One of the lightcurve examples which informed the evolution is described below in Section \ref{pipeline_development}. A difference was made in the treatment of Paczy\'nski peaks with values of $t_{\rm{FWHM}}$ above and below $2$ days. This stemmed from the desire to only select fast-rising lightcurves if there was enough data in the early and late rising and falling parts respectively of the curve (i.e. the ``wings'' of the curve) to assist in distinguishing between fast novae and microlensing, which can look quite similar given sparse sampling around the time of peak flux. Hence a stricter cut was applied for shorter timescale lightcurves in the wing regions (defined as between $t_{\rm{FWHM}}/2$ and $2t_{\rm{FWHM}}$ away from the central $t_0$).

For lightcurves with $t_{\rm{FWHM}} \leq 2$ days:

 Within the time range $(t_0 \pm \frac{t_{\rm {FWHM}}}{4})$ number of data points $\geq 1$.\newline
 Within the range $(t_0 \pm \frac{t_{\rm {FWHM}}}{2})$ number of data points $\geq 3$.
 
 Within the range $(t_0 + \frac{t_{\rm{FWHM}}}{2}) \leq t < (t_0 + 2t_{\rm{FWHM}})$ number of data points $\geq 2$.
 Within the range  $(t_0 - 2t_{\rm{FWHM}}) < t \leq (t_0 - \frac{t_{\rm{FWHM}}}{2})$ number of data points $\geq 2$.

For lightcurves with $t_{\rm{FWHM}} > 2$ days:

  Within the time range $(t_0 \pm \frac{t_{\rm {FWHM}}}{4})$ number of data points $\geq 1$.\newline
  Within the range $(t_0 \pm \frac{t_{\rm{FWHM}}}{2})$ number of data points $\geq 3$.\newline
 Within the ranges $(t_0 + \frac{t_{\rm{FWHM}}}{2}) \leq t < (t_0 + 2t_{\rm{FWHM}})$\newline OR $(t_0 - 2t_{\rm{FWHM}}) < t \leq (t_0 - \frac{t_{\rm {FWHM}}}{2})$ number of data points $\geq 2$.

So to summarise: for timescales less than or equal to $2$ days, a minimum of $(1+3+2+2) = 8$ data points are required inside the region $(t_0 \pm 2t_{\rm {FWHM}})$. In addition, there must be $4$ points in the central region, and at least $2$ points \emph{in each} of the ``wings''.

 For timescales greater than $2$ days, a minimum of $(1+3+2) = 6$ data points are
 required inside the region $(t_0 \pm 2t_{\rm{FWHM}})$. In addition, there must be
 $4$ points in the central region, and at least $2$ points somewhere in one or both
 of the ``wings''.

The specification in both cases that there must be one point in the very central region $(t_0 \pm \frac{t_{\rm{FWHM}}}{4})$ as well as $3$ in the less central region $(t_0 \pm \frac{t_{\rm{FWHM}}}{2})$ is an attempt to reduce the instances of clumps of three points being very close together in time followed by a data gap precisely where the peak is. Data tend to be clumped in this way as, for example, during good weather conditions many images can be taken.

\textbf{Cut 7 The lensing timescale }

Definition of the signal to noise parameter as defined for lensing candidates is discussed in more detail below in Section \ref{lensing_signal_to_noise}, but in summary, any lightcurve where there are fewer than $10$ data points outside the peak region $t_0 - 2t_{\rm FWHM} > t > t_0 + 2t_{\rm FWHM}$ cannot have the signal to noise of the peak assessed by the method used for the majority of lightcurves. Therefore these lightcurves are not included in the main group for selection, although with the addition of more data points after the peak, they may yet be included in future. The lightcurves which fail this cut will be henceforth referred to as ``long timescale'', as in the majority of cases when there are insufficient data in the baseline to define a standard deviation this is because the width of the $t_{\rm FWHM}$ found in the fitting is comparable to or greater than the time spread of the data.
Conversely, those that are \emph{not} rejected at this stage are referred to below as ``normal'' candidates.

\textbf{Cut 8 Sufficient signal to noise }

 For each lightcurve which has been selected as a lensing candidate by passing all the cuts
described above, and also for each lightcurve classified as a variable candidate
 a ``signal to noise factor'' is calculated that quantifies how clearly the desired
 signal is seen above the noise. Different routines are used for variables and for
 lensing candidates, and these are described below.

 \subsection{Variables}

\label{variable_signal}

 For variable candidate lightcurves, first the mean error of all the points is calculated,
along with the mean flux in each band and the standard deviation of the fluxes in each band from this mean value. Next, if the number of points in each band is greater than $10$ (to ensure that the standard deviation value is reliable), then the mean value of the ratio $R$ given by

\begin{equation}
\label{variable_ratio}
R = \frac{\sigma_{\rm{flux,band}}}{\overset{-}{\rm{E}}_{\rm{band}}}
\end{equation}

 is calculated for each band, where $\overset{-}{\rm{E}}_{\rm{band}}$ is 
the mean error in each band, and $\sigma_{\rm{flux,band}}$ is the flux standard deviation in each band.
Then, the weighted mean of the ratio values for each band (in which a standard deviation could be calculated) is calculated, to find a mean value of the ratio for all data points, as if all fluxes had the same flux scale. Hence, variability in bands with more data points have a greater weight in the final result.

The final quantity calculated can be represented by 

\begin{equation}
\label{variable_signal_to_noise}
S_{\rm{var}} = \frac{\sum_{i=1}^{k}{n_{i}R_{i}}}{\sum_{i=1}^{k}n_{i}}
\end{equation}

where $k$ is the number of data bands with a number of points $n_{i}$ greater than $10$

 This quantity $S_{\rm{var}}$, referred to as the ``variable signal to noise ratio''
gives a value to the size of the variable 
signal relative to the expected noise as quantified by the error bars. The reasons for this definition are described in more detail later, in Section \ref{variable_flux_ratios}.

For a lightcurve which was completely random, and hence distributed in a Gaussian manner, it would be expected that on average $68.2\%$ of points should lie within $\overset{-}{\rm{E}}$ of the mean, where $\overset{-}{\rm{E}}$ as above equals the mean error of all the points. Hence the ratio of
the standard deviation of the points to the mean error should be $\sim1$. Therefore, any non random lightcurve i.e. one which has, on average, variations greater than the size of its error bars will have
a ratio greater than $1$. Hence a value $>1$ is required by the pipeline for a lightcurve to be selected as having a detectable amount of variation above the noise. Lightcurves are graded from $0$ to $9$ on the basis of this quantity, the top of the ninth category corresponding to a signal to noise parameter of $30$, and the category divisions being linear and equal. It would be possible to impose a lower cut at a higher value than $1$, but the investigation of the statistics of lower signal to noise variables is included in this work. It should be remembered also that a variable lightcurve is only fitted to the data if at least one significant period has been found in the data by the Lomb-Scargle periodogram algorithm. Thus truly random lightcurves are eliminated from the variable sample and a higher percentage of true signals would be expected.

 \subsection{Lensing Candidates \label{lensing_signal_to_noise}}

 For lensing candidates, one of the desired qualities which must be 
quantified is the effective signal 
to noise of the lensing peak measured above the fitted baseline flux. Obviously it would be impossible to estimate the noise level by using all data points in the lightcurve, as these would include the candidate signal. Hence a region is defined outside the main peak, where
the modelled lensing signal is insignificant, in which to estimate the background noise in the data. This region is chosen as being  $t_0 - 2t_{\rm {FWHM}} > t > t_0 + 2t_{\rm {FWHM}}$. Initially, each band has its fitted baseline offset or variable component removed, so that the baseline is zero. For the lightcurves which are modelled as a joint Cosine + Paczy\'nski fit, the Cosinusoidal part of the fitted model is first subtracted to leave only the lensing peak, before the above calculation is performed. The variable signal to noise ratio of the Cosine part of the fit is then also calculated.
The parameter quantifying the signal to noise of the event is then calculated by finding the mean value of the quantity $f_{dif}$ (see Section \ref{bumpfinding}) using all data points \emph{within} the central Paczy\'nski peak region $t_0 - 2t_{\rm {FWHM}} < t < t_0 + 2t_{\rm {FWHM}}$. The ratio between this mean $f_{dif}$ level (the ``signal'') and the standard deviation of all data points \emph{outside} the peak region, where the lightcurve should be almost flat, (the ``noise''), is then taken. This quantity, referred to as the ``lensing signal to noise ratio'', or ``lensing ratio'' (or LR) for short, shows how many ``sigma'' above the baseline is the mean measured flux in the peak. If there were no lensing signal at all, this quantity would be zero. Objects were initially defined as having sufficient signal to noise for the initial selection using the most relaxed cuts if the lensing ratio $> 2$, which allowed examination of a greater selection of lightcurve fits, but in the final selection a higher value is required.
 Of course, for the same reasons as in the calculation of the variable signal to noise ratio, the standard 
deviation of the non-peak region is not calculated using low numbers of data points, so if the number of points in this region is $<10$ then the lensing signal to noise ratio is not calculated. This situation will be more likely for candidate events with large $t_{\rm FWHM}$, as there will be fewer data points in the lightcurve which are not part of the peak.
 Thus only allowing through events with a defined value of lensing signal to noise 
ratio would remove some perfectly valid long timescale events and hence bias the final sample.
 Therefore, lightcurves for which this quantity cannot be defined were not entirely thrown away. This class of lightcurves have therefore been subjected to one fewer cut than all the others and so are kept in a separate output file. In this group there may be a small number of valid lensing candidates, but there are also a large number which have fitted amplitudes and hence signal to noise which are too low. To prevent lightcurves in which a long period variable star could mimic a long timescale lensing event from being allowed through as easily, the level of Cut 3 which is applied to these lightcurves is higher than the initial loose cut applied to all others; a value of $0.1$ instead of $0.01$. The lightcurves with ``defined but too low'' lensing signal to noise ratio \emph{or} undefined lensing signal to noise ratio and too low $\chi^2$ difference ratio are cut.

\textbf{Cut 9 No single over-dominant data point }

Several lightcurves were found which had one or perhaps two data points in the fitted bumps with very large difference flux values and small errors, giving them a very highly significant deviation above the average of other points in the lightcurve. Since it is not felt to be a good thing to rely on a
very few points for all the significance of an event (even if sufficient other data points exist to pass Cut 6), a check is made that of the data points in the fitted Paczy\'nski peak in the region $(t_{0}-2t_{\rm{FWHM}})<t_{0}<(t_{0}+2t_{\rm{FWHM}})$
the numbers of significant points pass the original specifications for the numbers of significant points in a bump, i.e. if
 $n_p < n_{\rm{pmin}}$ or $n_h < n_{\rm{hmin}}$ in the final fitted bump then a 
 lightcurve is rejected. If the numbers of data points within this region are fewer
 than $n_{\rm{pmin}}$ then the region 
is widened until the number of data points is equal to that number. This means that if there are only $n_{\rm{pmin}}$ data points in the bump then all of them must be
more than $3\sigma$ above the baseline in order for this cut to be passed.

\textbf{Cut 10 Significance of fitted peak}

Since the fitting routine is not currently forced to place the Paczy\'nski peak within the bounds of the most significant bump in the lightcurve found in the bump finding section, it is possible that a lower $\chi^2$ solution could be found by placing $t_0$ elsewhere, especially when combined with a cosinusoid function. Hence it is necessary to check that the points within the fitted peak have at least the same significance above baseline as was required to define a bump in the first place. The required value of the geometrical mean value of $f_{dif}$ in the peak region is therefore also set as $> 8.77$.

\textbf{Cut 11 $\chi_{\rm{local}}^2$/d.o.f. of the Paczy\'nski peak}

All data points within the region $(t_0 - 2.0$ $t_{\rm{FWHM}})< t < (t_0 + 2.0$ $t_{\rm{FWHM}})$ were considered to be inside the fitted bump, and, using these points only, a ``local'' $\chi_{\rm{local}}^2$/d.o.f. parameter was constructed by using the formula

\begin{equation}
\label{local_chisq}
 \chi_{\rm{local}}^2/\rm{d.o.f.} = \frac{{\chi_{\rm{bump}}^2} n_{\rm{total}}}{{(n_{\rm{total}} - N)} n_{\rm{bump}}}
\end{equation}

where $\chi_{\rm{bump}}^2 = \sum_{i=0}^{n_{\rm{bump}}}({F_{i}-F_{\rm{fit}}})/{\rm{error}_{i}}$,
$n_{\rm{total}}$ is the total number of points in the lightcurve,
$n_{\rm{bump}}$ is the number of points inside the bump, and $N$ is the
number of fit parameters. This is the same as finding the $\chi^2$/d.o.f. using the contributing points in the bump only, then scaling the result up as if the whole lightcurve had contributed. The cut on $\chi_{\rm{local}}^2$/d.o.f. is set at $\chi_{\rm{local}}^2$/d.o.f.$ < 5$.

\textbf{Cut 12 No significant coherent residual bumps}

By ``significant coherent residual bumps'' is meant groupings of consecutive data points all showing a significant deviation from the best-fitted model curve.
 Both positive and negative bumps
are searched for, since a large positive bump or bumps in a lightcurve in addition to that fitted decreases significantly the confidence in the microlensing model and increases the likelihood that the flux variations are due to other repeating variable phenomena. Negative bumps decrease the confidence in the variable baseline model fit and hence, by implication, the confidence in the lensing component fit, which relies on the subtraction of a satisfactory baseline fit.
 It must be emphasised that this cut is applied to lightcurves which have already passed the two
$\chi^2$/d.o.f. cuts, both of the global (Cut 5) and local (Cut 11) kind, so the fact that it became clear that this cut is required demonstrates that relying only on $\chi^2$/d.o.f. to decide if a fit to a model was good enough is insufficient for our purposes.
If the deviation/s, while clearly significant in themselves, are not too large, their excess $\chi^2$ may be averaged out over the rest of lightcurve and the $\chi^2$/d.o.f. may still stay below a reasonable cut
level. This can be possible if part of the lightcurve is an especially good fit to a part of the model (for e.g., the lensing peak) and has a below average $\chi^2$/d.o.f., thus leaving more ``free'' $\chi^2$ to be shared out among the other data points, including the deviation.
 Simply put, relying on $\chi^2$/d.o.f. requires that the model used explains (and 
hence removes from the residuals) any non-random flux variations. In our case, there exist many different combinations of physical phenomena which could produce lightcurves more complex than our most complex model, and hence non-random flux excursions can and do remain in the residuals.

Coherent flux deviations in the residuals are searched for using two variants on the routine used
to make Cut 1. For positive bumps, as in the original bump finding routine, $n_p$ is required to be $5$ and $n_h$ is $2$, using the same two significances of $3$ and $5\sigma$ respectively as used in Cut 1. $n_{m}$, the number of points allowed below threshold before a bump ends, is set to be more sensitive than for the original bump finding, with $n_{m} = 1$, as opposed to $2$ previously.
Negative bumps are allowed to be broader with a lower peak significance, since they are not expected to look anything like a lensing lightcurve, requiring $n_{p} \ge 5$, $n_{h} = 0$, $n_{m} = 2$.
In addition, groups of data with no points above the $n_{h}$ threshold are allowed to accumulate using more data points, as long as the geometrical mean value of $f_{dif}$ is as great as it would have been for the criteria above.

    \section{Development of the Pipeline}
\label{pipeline_development}

\subsubsection{Checking that no useful data are being thrown away}

 The microlensing candidate lightcurves which are most likely to have the most extreme flux variations and hence the most extreme 
departures from their mean fluxes are the three lightcurves associated with the position of the ``short event'' described in Chapter \ref{Chapter_5}, which in the $2007$ photometry are the three events $4427$, $6389$ and $12147$. These lightcurves were used to check
 whether the $10\sigma$ flux cut caused any of the points from ``real''
 microlensing candidate events to be
unintentionally thrown away by the sigma clipping routine. For the event $6389$, the peak flux value magnitude is $137.59$ ADU/s whereas the standard deviation of the whole lightcurve is $23.78$. Thus, $10\sigma$ above the 
mean (which is close to zero) is $\sim237.8$ which is clearly larger than $137.59$ with a reasonable margin.
 One point in this lightcurve 
\emph{is} clipped, but this is one of the ``typically'' bad single points in the POINT-AGAPE (PA) data described above, and is 
$16.38\sigma$ above the mean flux for the band. It is therefore known that the $10\sigma$ threshold is reasonable
 both on the high and the low side, being roughly between the ``normal'' deviant flux values and the most extreme known
``real'' flux values.

     \subsection{Using the F-Test to decide the most useful model fit}
\label{F-test}
 In the early stages of the development of the Candidate Selection Pipeline, the model (i.e., Paczy\'nski only, mixed Paczy\'nski + Variable, pure Variable, 
or Constant) which was chosen to represent the classification of the lightcurve
was decided by choosing the model with the lowest $\chi^2$. However, it was quickly realised that since the models compared had differing numbers of degrees of freedom,
that like was not being compared with like, and that in some cases it might be better
to choose a simpler model with fewer degrees of freedom, even though it might have a
slightly higher $\chi^2$. The statistical test which can decide whether a model with
more parameters but a lower $\chi^2$ makes a big enough improvement in the $\chi^2$ 
as compared to a simpler model, to justify the use of the extra fitting parameters, is the ``F-Test''.

For the current application, the ``F-ratio'' which must be calculated
in order to perform the test can be written as in Equation \ref{F-Ratio_chisq_2}

\begin{equation}
\label{F-Ratio_chisq_2}
\mbox{F-Ratio} =\frac{\frac{({\chi^2}_1 - {\chi^2}_2)}{(p_1 - p_2)}}{\frac{{\chi^2}_2}{n_{\rm{total}} - p_2}}
\end{equation}

(where $n_{\rm{total}} = \sum_{k=1}^{n_{\rm{bands}}} n_k$), to express the F-ratio in terms 
of the ``observables'' in the
selection pipeline, the ${\chi^2}$'s of each model, numbers of parameters $p_1$ 
and $p_2$, and the 
number of data points $n_{\rm{total}}$.

 A confidence level of 
$0.05$, which is the standard level used for F-tests, was used. This means that the probability
 that the \emph{more} complicated model is NOT better is 5\%.

In all cases the model which had been found to have the lowest ${\chi^2}$ fit was only
 compared, using the F-test, to models with lower numbers of parameters than itself,
 as if a model was more complicated, \emph{and} had a worse ${\chi^2}$, then it would
 definitely be a worse model anyway.

If the status of the lowest ${\chi^2}$ model was ``downgraded'' because the F-test showed that it
 is not the best model to use, then the ``classification'' of the lightcurve was changed to that
 of the simpler model which displaced it. Then the new ``best model'' was compared in turn
 with any remaining simpler models to see if it was in fact the ``simplest and best'' of all models
fitted. The classification remained fixed only if a model ``wins the contest'' with the next simplest
 model.

    \subsection{Extreme flux ratios}

In the version of the code where the flux amplitude ratios between PA and LT i'-band data are not fixed, it is sometimes possible for the fitting routines to 
return extreme flux ratios, especially in the ``mixed'' fits.
There are several ways that the fitting routine can combine
a Paczy\'nski function ``creatively'' with a skew cosinusoid to slightly reduce the $\chi^2$ of a basically flat or simple periodic variable
lightcurve. Therefore, extreme fitted
flux ratios are not retained as classified lensing but downgraded to the
classification with next lowest $\chi^2$. For example, in the case of the flux ratio between the amplitudes of the lensing components in PA and LT data the status of the candidate is only changed if two conditions apply:

a) The flux amplitude ratio of the fitted Paczy\'nski components is more than $3$ times what is expected, i.e. more than $3$ times the ``normal''
amplitude ratio which is taken as $8.53$ for these purposes
and
b) The $t_0$ of the fitted Paczy\'nski peak is close to the boundary of the two data sets, which occurs at around $(JD - 2400000.5) = 53000$.
For these purposes, ``close'' is defined as 
i) within ($t_0 + t_{\rm{FWHM}} > 53000$) for a $t_0$ in the PA data and
ii) within ($t_0 - t_{\rm{FWHM}} < 53000$) for a $t_0$ in the LT data.
 In this area, there is a chance of some interaction between the Paczy\'nski
 peak in both bands, whereas if the central peak is only covered in one band, the
 fitted amplitude ratio would be expected to be almost random anyway, without this
 being something to cause undue concern.

    \subsection{The Quality Factor of selected events \label{quality_factor}}

As a tool to assist in deciding the final values of the cuts and showing the most interesting lightcurves to examine further, the second part of the selection pipeline
was expanded to include a numerical tool which attempted to put the selected lightcurves in order of their ``quality'' with respect to confidence in the microlensing interpretation of the data.
 This was done according to the value of a ``quality factor'' constructed from a 
product of five factors, one for each major cut, quantifying how ``well'' (i.e. by how far) the lightcurve had passed that particular cut.

Four of the five factors were those already used as cut parameters and hence defined above in Section \ref{the_cuts}. The fifth, called the ``bumpsample quality factor'', or ``BQF'', quantifies how well the fitted bump is temporally sampled.

The value of the ``bumpsample quality factor'', ``BQF'', is given by the product of the mean of the number of data points
sampling each of the four quadrants defined below with a scaling factor representing the fraction of the total number of these four quadrants which contain at least one data point. The four quadrants used were four of those used in the definition of the bump sampling cut, i.e. if 

 the number of data points within $t_{0} < t < \left({t_{0} + {\frac{ t_{\rm{FWHM}}}{2}}}\right)$ is $n^{+}_{\frac{1}{2}}$,

 the number of data points within $t_{0} > t > \left({t_{0} - {\frac{ t_{\rm{FWHM}}}{2}}}\right)$ is $n^{-}_{\frac{1}{2}}$,and the number of data points within $\left({t_{0} + {\frac{ t_{\rm{FWHM}}}{2}} }\right)< t <\left({t_{0} + 2t_{\rm{FWHM}}}\right) $ is $n^{+}_{\frac{1}{2},2}$,

 the number of data points within $\left({t_{0} - {\frac{ t_{\rm{FWHM}}}{2}}}\right) > t > \left({t_{0} - 2t_{\rm{FWHM}}}\right) $ is $n^{-}_{\frac{1}{2},2}$

and the
number of quadrants with at least one data point is $n_{\rm{>0}}$ 

then the value of the bumpsample quality factor is given in Equation \ref{bumpsample_quality_factor}

\begin{equation}
\label{bumpsample_quality_factor}
\rm{BQF} = \log{[\frac{[n^+_{\frac{1}{2}} n^-_{\frac{1}{2}} n^+_{(\frac{1}{2},2)} n^-_{(\frac{1}{2},2})]}{4} \frac{n_{\rm{>0}}}{4}]}
\end{equation}

The reasoning behind this definition was that it was not only the number of data points in the fitted flux peak which was important for modelling the lightcurve, but also that they were evenly distributed in time. Therefore even if one lightcurve had many data points, but they were all closely clustered in time, this should not be valued as highly as a lightcurve with the same or slightly fewer data points well spaced over the peak. Also, for modelling microlensing, good data coverage of both the central peak and the wings are required. The logarithm was taken to reduce the relative weighting of the BQF in the overall quality factor for larger values of the BQF, where the data points are so close together compared to the timescale of the Paczy\'nski peak that each extra point adds less to the information known about the peak than for lower values.

The expression for the value of the quality factor ``QF'' is given in Equation \ref{quality_factor_def} in which the lensing signal to noise is represented by ``LSN'' the $\chi^2$ difference ratio by ``CDR'', as previously, and the ``bumpsample quality factor'' by ``BQF''.

\begin{equation}
\label{quality_factor_def}
\rm{QF} = (\chi^2_{\rm{local}}/\rm{d.o.f.})(\chi^2_{\rm{global}}/\rm{d.o.f.})\rm{(LSN)}\rm{(CDR)}\rm{(BQF)}
\end{equation} 

\section{Optimising the levels of the cuts}
\label{optimising_cuts}

It became clear that running the complete candidate selection pipeline each time the level of one of the
five major cut parameters was changed, to investigate the effect of that change, would be an impractical proposition, given the number of possible combinations of the levels of these
cuts that might need to be investigated and the time the selection pipeline took to run.
Therefore, a C-Shell script was written which mirrored the operation of the main cuts of the first part of the pipeline
precisely, producing the same selected events for any level of cut. This was achieved by ensuring that the calculated values of any parameters that were used to cut on were printed to file by the first (``main'') part of the selection pipeline, along with the coded description of how the lightcurve had been classified.
When the main pipeline was run, values of the cuts were chosen
 which were very ``relaxed'' (in other words, they allowed through more
 lightcurves as microlensing candidates than could be conceivable for the
 final cut levels). However, the cut levels were chosen so that
 the events selected would still show a reasonable correspondence with
the desired microlensing form, even if the list produced contained candidates that were of lower quality than would be expected from those in the final selection. The cuts were chosen to be relaxed for two reasons; firstly so that many examples of the kinds of lightcurves which are being considered were produced, to allow errors in the selection procedure to be more easily spotted and diagnosed, and secondly to ensure that the most detailed analysis required to produce the calculated values of the cut parameters (which are written to file and later read by the second stage pipeline script) is performed on a larger number of lightcurves that are likely to be selected when the cuts are later tightened.
 This in turn ensures that the second component of the pipeline 
always has the necessary data to work with. If the cuts had been chosen to be too tight for the main program run, some parts of the program might never be accessed because they are ``inside'' a cut condition which was failed, and hence the data required for the script to reproduce the results would never be calculated or written out.
The initial values for the cuts which were chosen for this ``general cut parameter calculation'' run were as follows: 
global $\chi^2$/d.o.f. = local $\chi^2$/d.o.f. = $15.0$,
CDR = $0.1$, CDR (for ``long timescale events'') = $0.01$, ``lensing ratio'' = $1.5$.
The global $\chi^2$/d.o.f. and local $\chi^2$/d.o.f. cut values were set initially at $15$ as a value above which it was not believed any real confidence could be placed in the fits, since generally fits to data having values of $\chi^2$/d.o.f.$> 10$ are not considered reliable. The value of CDR must be a positive number, and when set at $0.1$ represents a $10\%$ improvement in $\chi^2$/d.o.f. by using the best fitting model used.
This value was chosen to at a level which was lower than that at which it was anticipated the final cut would be set.
Lensing ratio must also be a positive number and from theory should have had a value greater than at least $2$ since this would represent approximately a bump which was on average $2\sigma$ above the background flux level, so this cut level was set lower than this at $1.5$ to allow it to be progressively tightened across the interesting range. All of the above cuts were initially set at levels which it was felt would be slightly too lax to be used in the final selection, for the same reason.  

To get a feel for the limiting practical values of the cut parameters for the data set being used, the second stage pipeline script was run for $35$ different sets of cut parameters. In each case all except one parameter were given the starting values above, and the last was varied steadily in the direction required to produce a stricter cut until no lightcurves were allowed through. The required range of the parameter being varied was found by experimentation; the cut level was made progressively more tight until no lightcurves remained selected. Several runs of the program were required just to discover the practical limit of each parameter.

For the run of the pipeline
(using the third season ($2007$ photometry) data, and the version of the pipeline which does NOT link the magnitudes of PA and (LT or FTN) flux amplitudes) the parameter values used and the resulting numbers of lightcurves selected, both in the ``normal'' and ``long timescale'' categories, are shown in Table \ref{varying_cut_parameters} below.
Clearly, at least while the pipeline was being developed and checked, the values of the cut parameters each had to be less strict than the limiting values which did not allow any lightcurves through, as when used in combination, all the cuts together with their real values allow through fewer lightcurves than only having one ``tight'' cut as in Table \ref{varying_cut_parameters} with all the others at their most relaxed base values.

\begin{table*}
\caption[Table showing the number of lightcurves passed by the selection pipeline for varying values of the main cut parameters.]{Table showing the number of lightcurves passed by the selection pipeline for varying values of the main cut parameters.}
\begin{footnotesize}
\begin{center}
\begin{tabular}{|c|c|c|c|c|c|c|c|}
\hline
\hline
   &   &   & \multicolumn{2}{c|}{$\chi^2_{max}$} & & &  \\
\hline
         & $\chi^2$    & $\chi^2$       &          &         &          &        &   \\ 
  Label  & diff. ratio &  diff. ratio   & (global) & (local) &  Lensing & Normal & Long \\ 
         &             &(long timescale)&          &         &   Ratio  &        & Timescale \\
\hline
 -  & 0.01 & 0.1 & 15.0 &  15.0 &1.5 & 142  &  50    \\
 a  &    " &   " & 12.5 &   "  & "  & 138  &  50    \\
 b  &    " &   " &  8   &   "  & "  & 120  &  43    \\
 c  &    " &   " &  5   &   "  & "  &  72  &  38    \\
 d  &    " &   " & 2.5  &   "  & "  &  12  &  11    \\
 da &    " &   " & 1.75 &   "  & "  &   3  &   5    \\
daa &    " &   " & 1.0  &   "  & "  &   2  &   0    \\
 db &    " &   " & 0.5  &   "  & "  &   0  &   0    \\
\hline
 e  & 0.01 &   " & 15.0 & 15.0 & 2.0 &  96  &  50    \\
 f  &   "  &   " &  "   &   "  & 3.0 &  46  &   "    \\
 g  &   "  &   " &  "   &   "  & 4.0 &  21  &   "    \\
 h  &   "  &   " &  "   &   "  & 6.0 &   8  &   "    \\
 ha &   "  &   " &  "   &   "  & 8.0 &   2  &   "    \\
\hline
 i  & 0.2  &   " & 15.0 & 15.0 & 1.5 & 125 &  50    \\
 j  & 0.4  &   " &   "  &   "  & "  &  92  &   "    \\
 k  & 0.6  &   " &   "  &   "  & "  &  63  &   "    \\
 l  & 0.8  &   " &   "  &   "  & "  &  43  &   "    \\
 m  & 1.0  &   " &   "  &   "  & "  &  35  &   "    \\
 n  & 2.0  &   " &   "  &   "  & "  &   5  &   "    \\
 o  & 3.0  &   " &   "  &   "  & "  &   2  &   "    \\
 oa & 4.0  &   " &   "  &   "  & "  &   0  &   "    \\
\hline
 q  &   "  & 0.2 & 15.0 & 15.0 & 1.5 & 142 &  37    \\
 r  &   "  & 0.3 &   "  &  "   & "  &   "  &  26    \\
 s  &   "  & 0.4 &   "  &  "   & "  &   "  &  17    \\
 t  &   "  & 0.6 &   "  &  "   & "  &   "  &  10    \\
 u  &   "  & 0.8 &   "  &  "   & "  &   "  &   6    \\
 v  &   "  & 1.0 &   "  &  "   & "  &   "  &   6    \\
 w  &   "  & 2.0 &   "  &  "   & "  &   "  &   3    \\
 wa &   "  & 2.5 &   "  &  "   & "  &   "  &   1    \\
\hline
 x  & 0.01 & 0.1 & 15.0 & 12.5 & 1.5 & 128 &  50    \\
 y  & "    &   " &   "  &  8.0 & "  &  88  &  43    \\
 z  & "    &   " &   "  &  5.0 & "  &  51  &  38    \\
 za & "    &   " &   "  &  2.5 & "  &  12  &  10    \\
zaa & "    &   " &   "  & 1.75 & "  &   5  &   5    \\
zaaa& "    &   " &   "  &  1.0 & "  &   1  &   0    \\
\hline
\end{tabular}
\end{center}
\end{footnotesize}
\label{varying_cut_parameters}
\end{table*}

At this stage of the candidate selection investigation, the cut 
levels used by the pipeline emulator script
were made significantly stricter, to make the selected lightcurves 
more likely to be good candidates.
The cuts used at this stage were:
global $\chi^2$/d.o.f. = local $\chi^2$/d.o.f. = $10.0$,
CDR = $0.1$,
CDR (for ``long timescale events'') = $0.1$,
``lensing ratio'' = $2.0$.
Using these cut levels, $62$ ``normal'' lightcurves were selected,
 along with $42$ in the ``long timescale''
category. A quality factor was also calculated for the long timescale 
lightcurves in an analogous way to the normal ones with the exception
that there could only be four factors multiplied together as the lensing signal to noise
factor is undefined for the long timescale class, by definition.

It is interesting to note that the numerical values for the 
quality factor were of the same order of magnitude for the majority of lightcurves in both groups,
despite the long timescale group having lost a multiplicative factor
 (although only usually of order $1$) from their quality factor calculation. This
was apparently due to the quality factor associated with the data sampling of the bump being on average  
larger for the long timescale group, as flux deviations which are longer in
duration will on average contain more data points.
Only two lightcurves (numbers $1033$ and $2675$) at this stage were selected on the second iteration.

     \subsection{Data sampling of the Paczy\'nski peak}

  A cut was introduced to ensure that microlensing candidates had good data sampling 
of the area around the
 peak of the lightcurve in the region of $t_0$. The fitting routines have
a tendency to insert Paczy\'nski peaks into gaps 
in the time coverage of the data, in order to reduce the overall $\chi^2$.
 This will usually happen when one or more data points on either side of
 the gap in the data are significantly higher in flux than those immediately
outside of them with respect to the $t_0$ of the peak. This is usually a 
caused by random variations in the data. In a small number of instances, this
kind of flux variation may actually be caused by a true microlensing event 
which just happens to unfortunately fall in a gap in the data. There is 
nothing that can be done in this case to distinguish whether the changes
 in flux are real or random and so it cannot be established whether a real 
microlensing event is occurring. Therefore, a good data coverage of the peak
 must be demanded to increase the confidence that the modelled event could be real.

In the first instance, a routine was written which counted the number of data points
within the ranges $t_{0}\pm{ t_{\rm{FWHM}}}$ and $t_0\pm{\frac{ t_{\rm{FWHM}}}{2}}$.
The routine also calculated the normalised distances 
$\Delta t_{\rm{norm},n} = \frac{t-t_0}{t_{\rm{FWHM}}}$ at which
 the $n$th data point exists, where $n$ was calculated in the range $1$ to $10$. 
This was done to enable cuts over different time ranges 
than just $t_0\pm{ t_{\rm{FWHM}}}$ or $t_0\pm{\frac{ t_{\rm{FWHM}}}{2}}$ to be used.
For example, if it was required that at least $5$ data points exist within
 $t_0\pm{3 t_{\rm{FWHM}}}$ then the cut would be ``$\Delta t_{\rm{norm},5} < 3$''

The minimum number of points which were required to exist in this time range was
 specified as a cut; initially, $3$ points within $t_{0}\pm{t_{\rm{FWHM}}}$ was chosen.

\begin{figure}[!ht]
\vspace*{6cm}
$\begin{array}{c}
\vspace*{6cm}
   \leavevmode
 \includegraphics{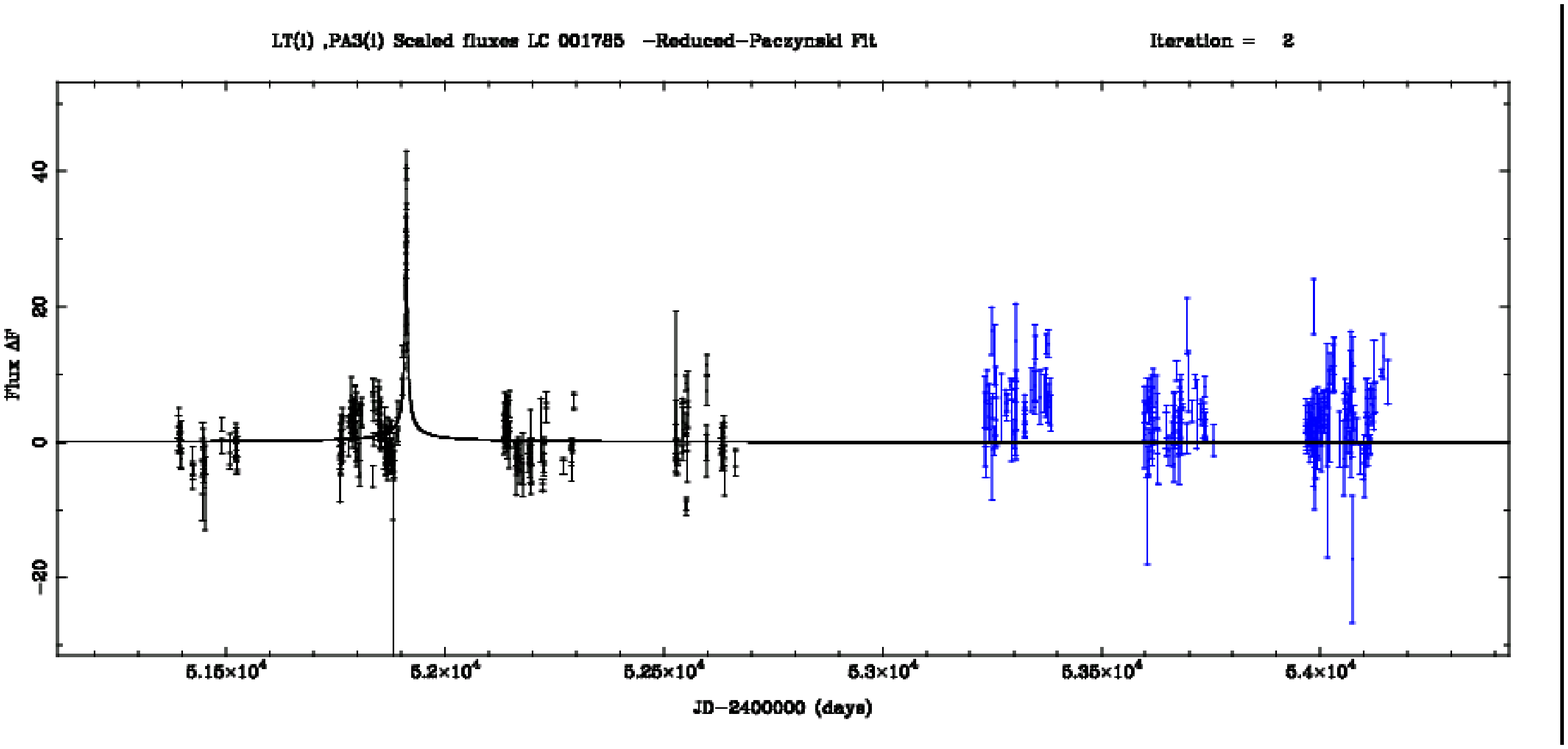} \\
\vspace*{0cm}
   \leavevmode
 \includegraphics{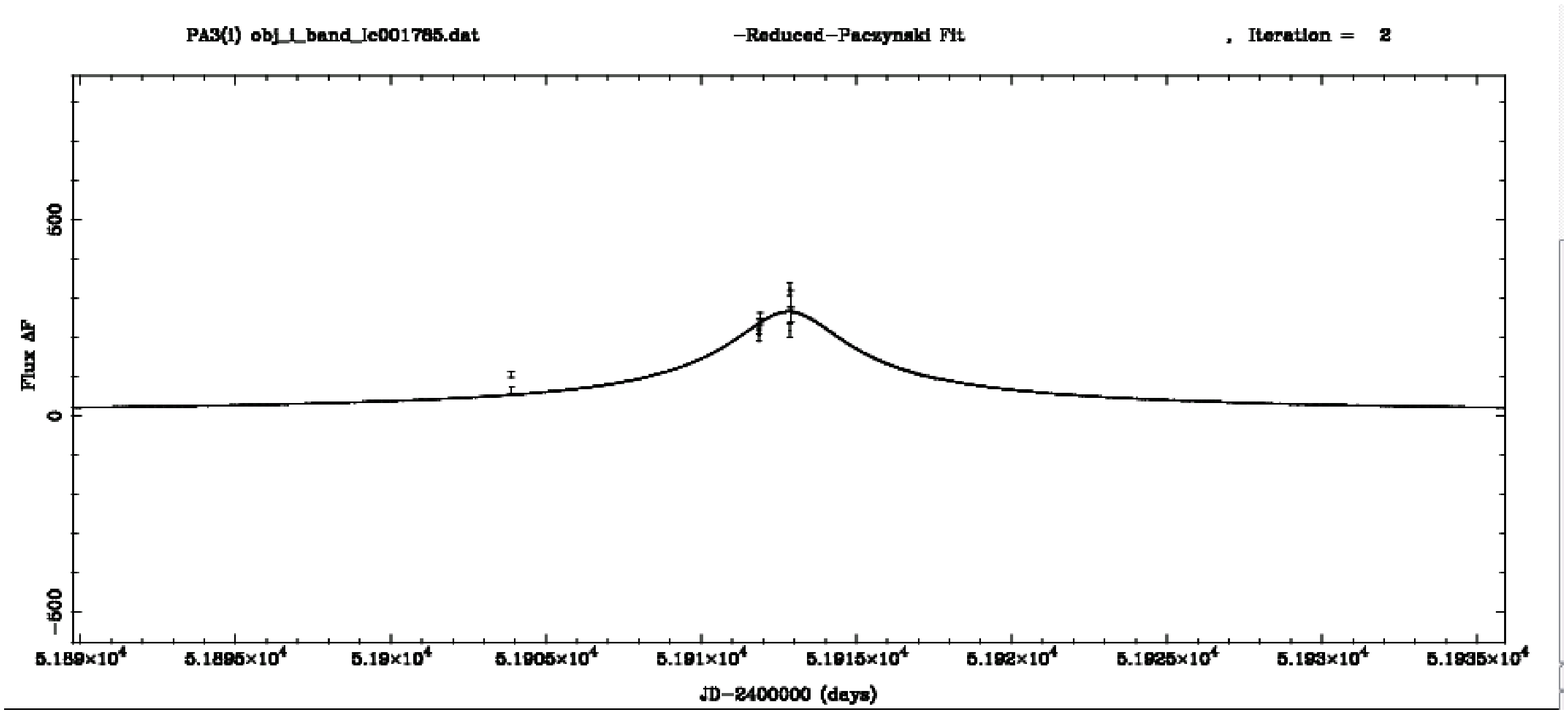} \\
\end{array}$
\caption[Two plots of lightcurve 1785 in the 2007 photometry.]{Two plots of Lightcurve 1785 in the 2007 photometry. a) The whole lightcurve b) The peak region (within $t_0\pm2t_{\rm{FWHM}}$) showing more clearly the time sampling of the flux peak, insufficient to pass the selection cut.}
 \label{LC1785_and_peak}
\end{figure}

This worked reasonably well, but one example in particular forced a reassessment of
whether this was an over-simple way to perform this cut.
In one case, which was numbered lightcurve (LC) $1785$, which happened also to be one of the 
few lightcurves in which some evidence of microlensing-like behaviour was found 
on the \emph{second} iteration 
of the pipeline after the subtraction of variable behaviour on the first iteration,
(i.e. found using the ``$f_{dif}$ clustering'' routine), the lightcurve was selected
 (not rejected) by the pipeline, and a $\chi^2$/d.o.f.$= 8.05$ fit
 was found to the data using a Paczy\'nski fit. 
However, on examination of the fits, it was clear that this event, although a good fit to
a very-short-timescale Paczy\'nski peak of $t_{\rm{FWHM}} = 6.41$ days 
was much more likely to be a classical nova candidate (from the very steep rise in flux). This lightcurve is shown in Figure \ref{LC1785_and_peak}.

The reason why doubt exists about the explanation for this lightcurve is that the data coverage
ends very close to the modelled $t_0$ and data only exist on the low-t side of the event.
This event demonstrated that if data only exist on one side of the flux peak it can be difficult or even impossible to characterise the lightcurve as either microlensing or, for example, a nova. 

With this event in mind, therefore, the cut on data coverage across the peak was changed so that
a given number of points must exist on \emph{both} sides of the peak, within symmetrical time ranges, as described in Section \ref{bump_sampling} above. The time regions $(t_{0} + \frac{t_{\rm{FWHM}}}{2}) \geq t < t_{0} + 2t_{\rm{FWHM}})$ and its equivalent on the negative side of $t_0$ were chosen because these are the approximate regions in which a microlensing curve differs the most from a classical nova lightcurve; where the flux is not at baseline but not yet at peak. This is especially true for the flux rise, where novae rise
much more steeply than a lensing event of the equivalent overall best fit timescale.
It was also specified that data should exist close to the time of modelled peak flux, as this region is very important in modelling the magnitude of the maximum flux rise from baseline. Hence at least one data point was also required within the range $(t_{0} - \frac{t_{\rm{FWHM}}}{4}) \geq t < t_{0} + \frac{t_{\rm{FWHM}}}{4})$.

\subsection{Long-period variation in the baseline of mixed microlensing candidates}
\label{mixed_events_with_long_periods}

\begin{figure}[!ht]
\vspace*{6.1cm}
$\begin{array}{c}
\vspace*{5.6cm}
   \leavevmode
 \includegraphics{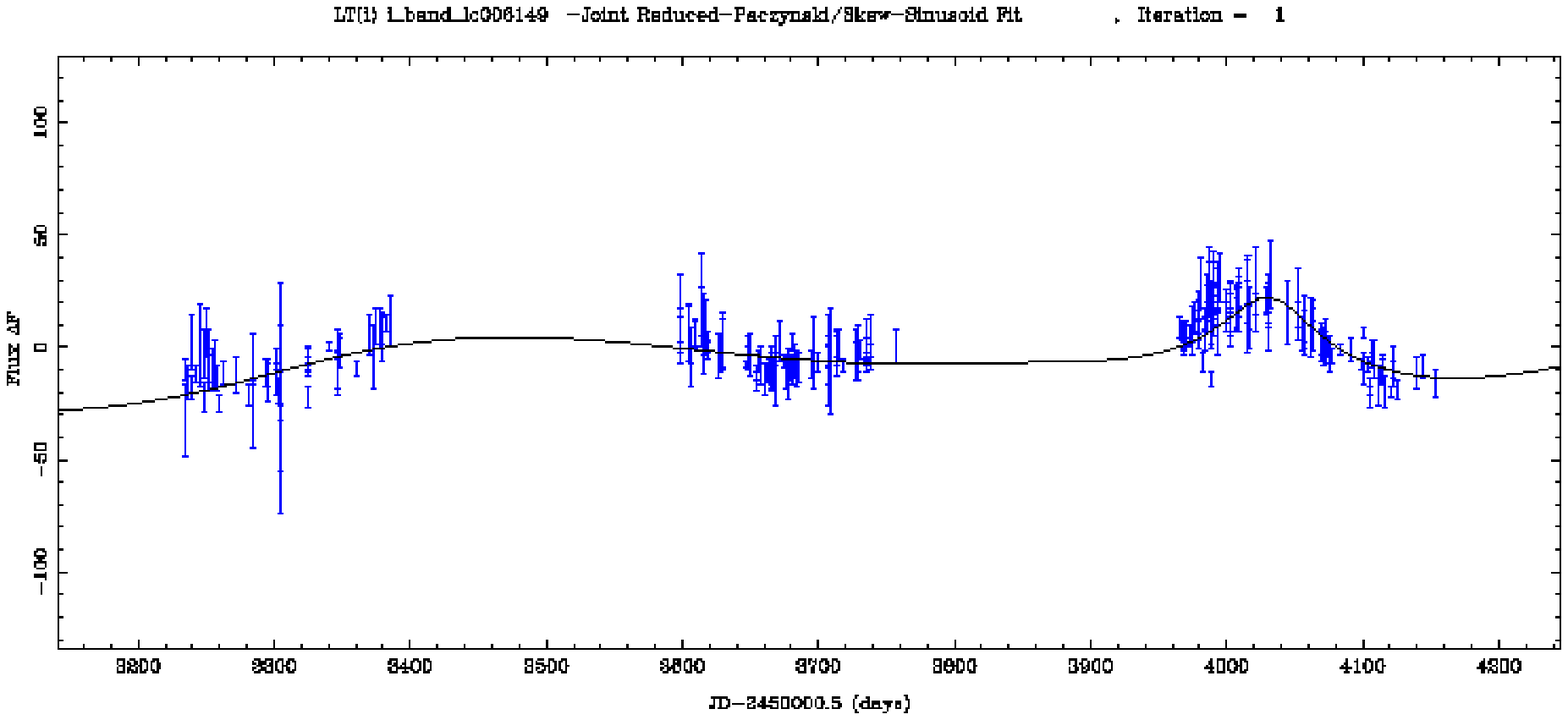} \\
\vspace*{0cm}
   \leavevmode
 \includegraphics{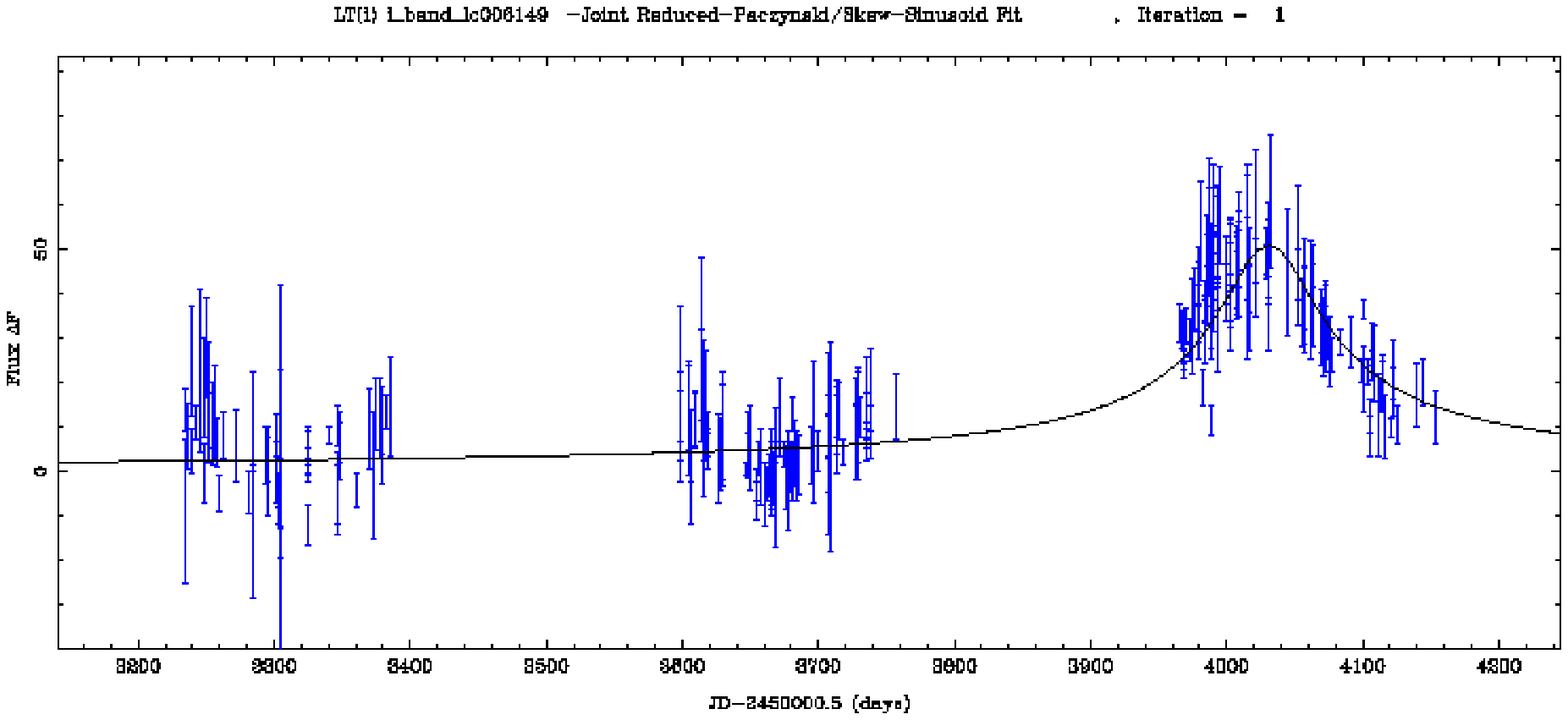} \\
\end{array}$
\caption[Two plots of lightcurve $6149$ in the $2007$ photometry.]{Two plots of lightcurve $6149$ in the $2007$ photometry: a) With its joint skew cosinusoid + Paczy\'nski fit, showing the long period variation in the baseline with comparable amplitude to that of the Paczy\'nski peak. b) With the variable component of the fit subtracted, plotted with the Paczy\'nski part of the joint fit.}
 \label{LC6149_two_plots}
\end{figure}

At one point in the development process of the candidate selection pipeline,
after the quality factor classification had been introduced (see Section \ref{quality_factor}), the lightcurves with the highest quality factor were being examined to ensure that only lightcurves that were convincing candidates had been selected. It was noted that lightcurve $6149$ had the second highest value of quality factor at that time. When the lightcurve was examined, however, although all the fits (shown in the first panel of Figure \ref{LC6149_two_plots}) were correct, and the fit to the microlensing component (shown in the second panel of Figure \ref{LC6149_two_plots}) seemed to be exactly as would be expected for a lightcurve with a high quality factor, the overall lightcurve seemed insufficiently convincing to rate it as the second highest quality factor. After some thought about the possible reasons for this, it was decided that the amplitude of the microlensing component was not much larger than that of that of the variable component, coupled with a second factor that the period of the variable component was long compared with the time-span of the data. Only LT data were available for this object, giving a timespan of only $3$ seasons, and the fitted period of the variable was $966$ days; comparable to the span of the data. This meant that less than a whole period of the variable component could be within the data. This in turn meant that, although the joint skew cosinusoid + Paczy\'nski fit does fit the data well, this interpretation was not guaranteed to be the correct one since the behaviour of the variable part had not been properly established. An alternative interpretation might have been that there could be one or several variable star(s) which happened to brighten just when the Paczy\'nski peak occurred.
Therefore, since confidence in the appropriateness of the joint fit depends to a large extent on the believability of the fit to the baseline, it was decided that this sort of lightcurve should not be allowed in the final selection, although with a longer time series of data the behaviour of the baseline could become better established, allowing the fitted Paczy\'nski peak to become a selected microlensing candidate.

By analogy with previous works which demanded microlensing candidates have a flat baseline
outside the Paczy\'nski peak, it was therefore demanded that the regular periodic nature of the modelled ``baseline'' variability be firmly established before the exceptional nature of the Paczy\'nski peak could be confidently claimed. To this end, a cut was devised
that had two conditions either of which had to be passed if the lightcurve was to get through. They were:

1) The period of the variable component must be less than half the time span of the data (i.e. at least two periods must be contained in the data) 
OR

2) The peak magnitude of the Paczy\'nski peak must be $> 5$ times the amplitude of the sinusoid.

After this cut was first applied, the number of joint (mixed) fit lightcurves which passed through the main pipeline decreased from $213$ to $169$. The above two conditions were eventually modified to form Cut 4.

\section{Investigations of variable lightcurves and telescope flux ratios}
\label{variable_investigations}

As part of the process of developing the pipeline and understanding of the data, several investigations have been carried out into lightcurves which were classified by the selection pipeline as ``variable''.
 These initially included any kind of variability, not only 
periodic variables. At the start of these investigations, the lightcurves being worked with were from an earlier DIA run which had used a different reference image to the one used to create the
data on which much of this thesis is based. The reference image relevant to the data used in this thesis was chosen, and the DIA pipeline re-run, in August 2007. Therefore all lightcurve numbers 
applicable at earlier times than this would not correspond to those after this date and so this relatively early stage of investigation will be reported in general terms without specific lightcurve examples. 
The selection pipeline used to classify the variables was also at a much earlier stage of development, and so if the experiments performed were repeated using the current pipeline, the results would differ.
 It is not thought, though, that the differences would change any of the conclusions reported below, because the routines used to fit the variable function, and the branch of the pipeline to which variable-type
 lightcurves were directed were not significantly changed. Most of the development work had been concentrated on the lensing branch.

 The selection pipeline was run with the following cut parameters:

 Only LT i'-band and FTN i'-band fluxes (or LT r-band and FTN R-band if used) were linked in magnitude:  
$\chi^2$/d.o.f. $= 10$, $\chi^2$ ratio$ = 0.3$, bump data sampling $= 3$ points within $t_{0}\pm t_{\rm{FWHM}}$. This produced $326$ lightcurves selected as microlensing candidates.

 Next the pipeline was run again, with tighter cut parameters:
$\chi^2$/d.o.f. $= 10$, $\chi^2$ ratio$ = 0.5$, bump data sampling $= 3$ points within $t_{0}\pm \frac{t_{\rm{FWHM}}}{2}$. This selected $143$ lightcurves as microlensing candidates.

The skewness limits were set so as to preserve a simple monotonic rise and fall of the function and avoid the more complicated behaviour that
emerges for higher skewness values.
The limiting value was chosen by finding two lightcurves, one of which appeared to be near the borderline of acceptability, but on the right side of the border, and another which was slightly too skew to be acceptable.
 The values of skewness of these two lightcurves were $1.574$ and $3.151$ respectively. Therefore the skewness parameter was limited from this point onwards at $1.574$. This investigation was performed using only 
skewness parameters $S>1$ (see Section \ref{skew_cos_function}) as at the time this was the only range being used.

     \subsection{PA-LT/FTN flux amplitude ratios}
\label{variable_flux_ratios}

Previous to this point, the only knowledge that had been gained of the magnitude of the ratio between the PA flux measurements and those of LT and hence FTN had been from the exploration of the short event 
(see Chapter \ref{Chapter_5}), where, over a long baseline, good evidence of at least one 
variable star was found in the baseline, allowing the fitting of the flux ratio between the PA i' band variable flux amplitude and the FTN i' band flux amplitude as equal to $8.53$, making the PA/LT ratio $= 8.53/1.081 = 7.89$.
 It was felt that relying on this one example and assuming that the same 
ratios applied for all lightcurves was not safe, and so efforts were made to find out with greater certainty what the relationships between flux and flux ratios actually were. This aim was of high importance as the ability to fix
the flux ratio between two bands has several benefits. One fewer fitting parameter would be required, reducing the time spent by the fitting routine and reducing the multi-dimensional parameter space by one dimension, which makes
 a considerable difference to how easily the lowest $\chi^2$ solutions may be found. Also, since the time indices of data in the LT and FTN data do not overlap with those in the PA data, the only way the flux amplitudes in different
 telescopes or wavebands can be related is by using some long-term, repeating phenomenon which can be detected in both telescopes (such as variable stars). However, a microlensing event will only rarely be of sufficiently long timescale
 and have $t_0$ close enough to the boundary between the data groups to have enough of a detectable change in flux that it can be accurately characterised in both bands. Usually, data around the peak would be required to know the microlensing
 flux amplitude, which almost always will occur in only one band. Therefore, if knowledge of the average relationships between
telescopes/wavebands is not available, then the peak fluxes of microlensing events must be fitted individually with independent fitting parameters.
Because only the amplitude of the fit in the band containing $t_0$ can usually be believed, this makes appropriately scaling data to be able to display all of the data points from all bands on one time axis very complicated, especially
 in the case where a mixed fit has been applied, and the lensing peak has a similar amplitude to the variable amplitude.
 
For example, a microlensing candidate with $t_0$ in the PA data might be fitted in LT i'-band with a very large amplitude (much larger than would be expected from
the average expected flux ratio between PA and LT). The reason for this might have been that the data at the beginning of the LT lightcurve appear to have some random positive trend.

 In this case, simply subtracting this component to leave the variable 
component does not give the correct best fitting variable component as the two components were fitted jointly, each with ``knowledge'' of the existence of the other. Hence only one constant flux offset parameter would have been used,
 which in this case would be large and negative, to almost cancel out the magnitude of the large lensing peak, to just leave the small positive random swing visible in the LT data. 

Reducing the incidence of these ``extreme flux ratio'' fits was the motivation behind limiting
the acceptable range of flux ratios for mixed fits, and regrading those with extreme flux ratio values to simply be variable star candidates.
 Except when it is known that the telescopes and filters are
 truly identical, ideally it should not not be automatically 
 assumed that the flux ratios that
 should be used for the variable component and for the Paczy\'nski
 component of a mixed fit are identical. Although the PA i'-band and LT/FTN i'-band filters
 should be very close, they will not be identical, and since the source star
 for a microlensing event will almost always not be the same star as the
 variable component, therefore usually having different colours and hence flux ratios, there may
 still be small differences in the flux ratios for microlensing components and
 for variable star components. Moreover, although the flux ratio between wavebands is expected to be
 constant for microlensing events as these are achromatic, in general
 the flux ratio for variable stars would be expected to change slightly, as the temperature of the
 star varies with time as it expands and contracts. Since both bands used here are i'-band, in this 
case these differences will be negligible.
 For the purposes of applying any
 consistent flux ratio relationship that might be found, it was necessary to assume
 that the same ratio be applied to both components, because, as described
 above, Angstrom does not have any data which overlap with the PA to allow
 the flux ratios between PA data and other bands, for microlensing components,
 to be investigated individually, or overall, in a more direct fashion.

Therefore, since ideally, if a simple relationship did exist between
the fluxes in the PA and LT/FTN data it would be useful to know it, to avoid 
all the complications above and make the fitting faster and more accurate.
This was attempted by examining fits to lightcurves with purely periodic variability i.e. no suspected lensing components.
The flux ratios in this section were all 
calculated and are quoted as if they were between PA i' and FTN i' bands, which is because the scripts were written when three data bands were being used.

This investigation was begun by selecting lightcurves from the previous run of the pipeline where $\chi^2$/d.o.f. = $10$, $\chi^2$ ratio$ = 0.5$, bump data sampling $= 3$ points within $\pm \frac{t_{\rm{FWHM}}}{2}$, in other words, ``good variable candidates''. This selected $341$ lightcurves.

A further selection was made from this list of lightcurves which might be expected to have the best chance of having a well defined flux ratio
between PA i'-band and either LT and FTN i'-band fluxes, based on the number of points in each band. These criteria were later improved upon and described in more detail below. 

In an attempt to further increase the sample,
 the original lightcurve plots were re-examined, looking for
 ``mixed'' events which had clearly variable baselines.
  $13$ were selected as suitable and these were added to the original
 $341$ lightcurves. 

$18$ lightcurves were selected with sufficient data and non-zero fitting errors.
The calculated mean flux ratio of these $18$ lightcurves was $8.53$, with a standard deviation of $5.69$. The contribution to the overall uncertainty in the mean from the fitting errors was only $^{+0.18}_{-0.17}$ and so negligible compared to the scatter.

Although the magnitude of the amplitude seemed reassuringly close to what was expected from investigations of the short event, the large spread was worrying.
Closer examination by eye of the selected lightcurves showed that on the whole
they were low signal to noise variables. This could be explained because variables with large signal to noise would often have other detectable variations in addition to the simple sinusoid modelled by the fitting code and so would fail the $\chi^2$/d.o.f. cut.

 Therefore, a method seemed to be required to classify variable 
lightcurves according to the ``amount'' of variability present. In other words, some kind of of variable signal to noise quantifier was required
so that the more ``interesting'' variables could be preferentially selected and examined more closely without having to go through all of the thousands of lightcurves which were classified as ``variable''.

One method which would have been possible to implement would have been to take the amplitude of the best fitting sinusoid curve and to compare this with
a measure of scatter, such as the standard deviation, of data points about this curve. A lightcurve which is purely noise would have a large
scatter but the amplitude of the sinusoid would be very low, whereas
a lightcurve which is a very good fit to a sinusoid curve would have a finite sinusoid amplitude but a low standard deviation around that curve.
This method was not chosen for several reasons. Firstly, through examining a random sample of lightcurves classified as ``variable'', it had been noticed that a few of them had extremely large, apparently random flux variations, but were not well fitted by a sinusoid curve.

 In these cases the scatter is huge, but the amplitude is very low, so they would not be picked out by the method described, and in fact would be actively ``suppressed'' in their variability classification. These lightcurves are interesting to know about, so that they can be either
removed from the list and/or investigated, and so incorrectly classifying their variability would have been undesirable. Secondly, the method relies on the assumption that all variable stars that might be interesting are a good fit to the chosen sinusoid model, because otherwise a lightcurve might be a very interesting periodic variable which just happens not to match a simple sinusoid. This would make the scatter from the sinusoid large (but not random; rather, coherently clustered in time), and thereby reduce the significance of the signal to noise classification.

The method which was preferred made a comparison of the scatter of
data points in the lightcurve from the mean flux with the mean error over the whole lightcurve. (The mathematical form of this quantity is shown above in Equation \ref{variable_signal_to_noise}). This had the advantage that it should pick out the large signal to noise random variables, but also works for regular periodic variables.

In addition, a cut was made on the fraction of data points above a given multiple of the mean error away from the mean flux. Initially, a multiple of $10$ times the mean error, with $20\%$ of points being above this level was demanded. This resulted in $21$ lightcurves being selected.
This was tested by running the main pipeline for this list of lightcurves and examining the plots. $18$ of the $21$ lightcurves were classified as variables and $3$ as some kind of microlensing candidate, either mixed or pure Paczy\'nski. The method seemed to work well and so it was incorporated into the main selection pipeline.

The number of points in both PA and the total of all LT and FTN data present were required to have a minimum number of data points to ensure that
each band would have
sufficient points to properly define the flux amplitude in that band

 1) total number of points $> 250$
 2) PA points $> 80$
 3) total of other bands $> 80$

 An additional constraint was introduced that the MINUIT fitting errors were
 not allowed to be zero (which is often an indication that a fit has not fully
 converged).

The two variability cuts described above were also applied, but with slightly relaxed levels of

 a) Fraction of points above $n(\overset{-}{\rm{E}})$ 
away from the mean = $0.15$
 b) where $n = 7.5$ and $\overset{-}{\rm{E}}$ is the mean error on the data points.

With these cuts, $62$ lightcurves were selected:
These were examined by eye, selecting those which had a reasonably good fit to the skew cosinusoid function and in which variations can clearly be seen in both PA AND in LT or FTN.

After this step, $33$ lightcurves remained.
When lightcurves with zero fitting errors were rejected as before,
only $10$ remained.

The mean flux ratio obtained from these lightcurves was $25.72$, and the standard deviation was $13.77$ with fitting errors $^{+0.15}_{-0.15}$. Clearly this is not very compatible at $\pm1 \sigma$ with the earlier
result. The largest value was $74.4$, from LC $2004$ and the smallest value was $6.89$ from LC $10685$ - a very large range indeed.
The plots of the model fits to these two extreme ratio lightcurves
were examined by eye and were reasonable fits to the data and consistent with the above ratios.

The next experiment tried was to attempt to use larger numbers of
lightcurves (better statistics) to see visually whether any correlation
existed between, for example, PA flux amplitude and flux amplitude ratio. Clearly, if the ratio was a constant, this correlation would be a horizontal straight line at the value of the ratio. Any non-constant correlation would imply a more complicated relationship was present.
The classifications given to lightcurves by the selection pipeline were used to select lightcurves that were classified as variables.
Instead of requiring a given value of $\chi^2$/d.o.f. and a low $\chi^2$ fit to a skewsinusoid fit, however, neither of these conditions were applied. This allowed $11341$ lightcurves to get through the first stage. These consisted purely of lightcurves for which a variable fit was the lowest $\chi^2$ of the ones fitted.
 $5818$ of these passed the above cuts on the number of points in each 
band, and $1185$ of these had non-zero fitting errors.

 The flux ratios of these $1185$ lightcurves were plotted in a scatter 
plot with their PA variable flux amplitude on the x-axis.
The resulting plot is shown in Figure \ref{1185_lightcurves_13_08_07_flux_ratios_scatter} below. 

\begin{figure}[!ht]
\vspace*{13cm}
 \includegraphics{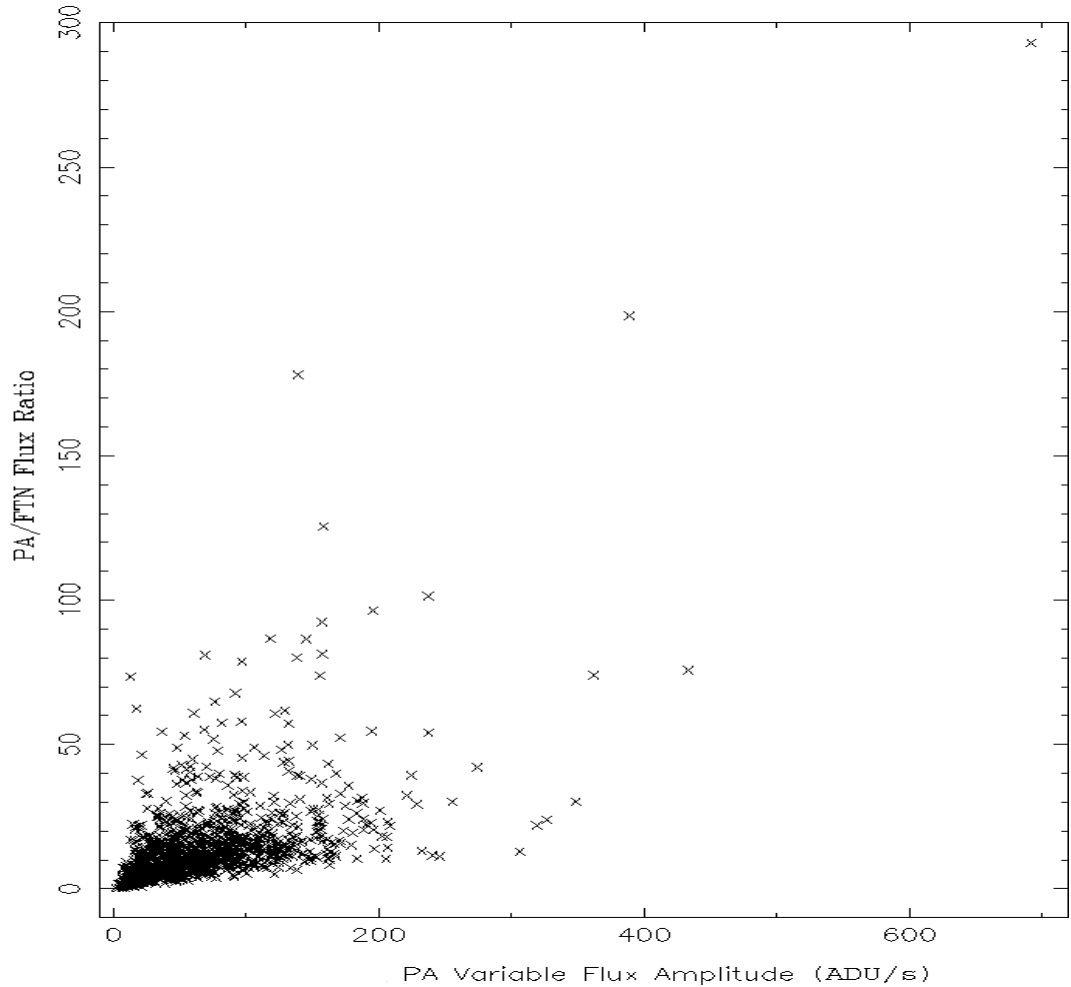}
\caption[Scatter plot of PA variable flux amplitude versus flux amplitude ratio for $1185$ variable lightcurves.]{Scatter plot of PA variable flux amplitude versus flux amplitude ratio for $1185$ variable lightcurves.
}
\label{1185_lightcurves_13_08_07_flux_ratios_scatter}
\end{figure}

It could clearly be seen in this figure that there was some kind of correlation between PA flux amplitude and flux ratio which was not described by a constant function. However, the correlation was very wide, and had much scatter, which was not really a surprise considering the deliberately relaxed criteria used to select lightcurves. It appeared that the correlation
passed through, or close to, ($0,0$) and was possibly approximately linear.
The densest part of the main mass of points was indeed around the $5-15$ range that had been repeatedly occurring in the above experiments. The wide range of values obtained so far was consistent with the wide range of the main mass of points on the plot.

At a later date the issue of flux ratios was returned to, in an attempt to gain a better understanding of these issues. In order to be more certain whether the wide spread in flux ratios found in the first two investigations above was due to the method preferentially selecting
low signal to noise lightcurves, a similar investigation was carried out,
but using an additional cut on the variable signal to noise parameter (as defined in Equation \ref{variable_signal_to_noise}) of $S_{\rm{var}} > 2$. This resulted in $80$ lightcurves being selected, which was reduced to only $5$ when the cuts above on the number of points in bands and non-zero fitting errors had been applied. The mean flux ratio
from these $5$ lightcurves was found to be $9.14$ with a standard deviation of $3.31$ and fitting errors $^{+0.24}_{-0.24}$

The process was repeated using as a base group the list of $355$ lightcurves with any value of variable signal to noise parameter, i.e. not necessarily $S_{\rm{var}} > 2$. The value for the ratio now obtained was $10.42$ with standard deviation $5.74$ and combined fitting errors
$^{+0.26}_{-0.26}$.

The several investigations conducted so far, although not seeming to converge on one value, did seem to be consistently indicating a value for the flux ratio which was $<20$.

 The above method was repeated later,
 after small changes had been made to the pipeline. This time, $6$
 lightcurves were selected. However, only $3$ of them coincided with the previous $5$ above. The
 discrepancies were all checked and were due to the parameter changes in the pipeline. The flux
 ratio from these $6$ was found to be $13.12$ with standard deviation $7.67$ and fitting errors
 $^{+0.32}_{-0.32}$.
 The three overlapping lightcurves had identical fit parameters, so
 it was possible simply to combine the lists of $5$ and $6$
 lightcurves into one list of $8$. The fits to these lightcurves were
checked by eye. Mostly they were reasonable, but it was clear that
their larger values of variable signal to noise parameter were mostly caused by possessing one or a few very discrepant data points, rather than the whole lightcurve having a larger signal to noise overall.

The possibility of the existence of a coherent spatial variation in the flux ratio between LT and PA data was checked for by plotting points colour-graded by amplitude in an X-Y scatter plot of the $1185$ data points earlier selected 
for Figure \ref{1185_lightcurves_13_08_07_flux_ratios_scatter} above.

\begin{figure}[!ht]
\vspace*{13cm}
   \includegraphics{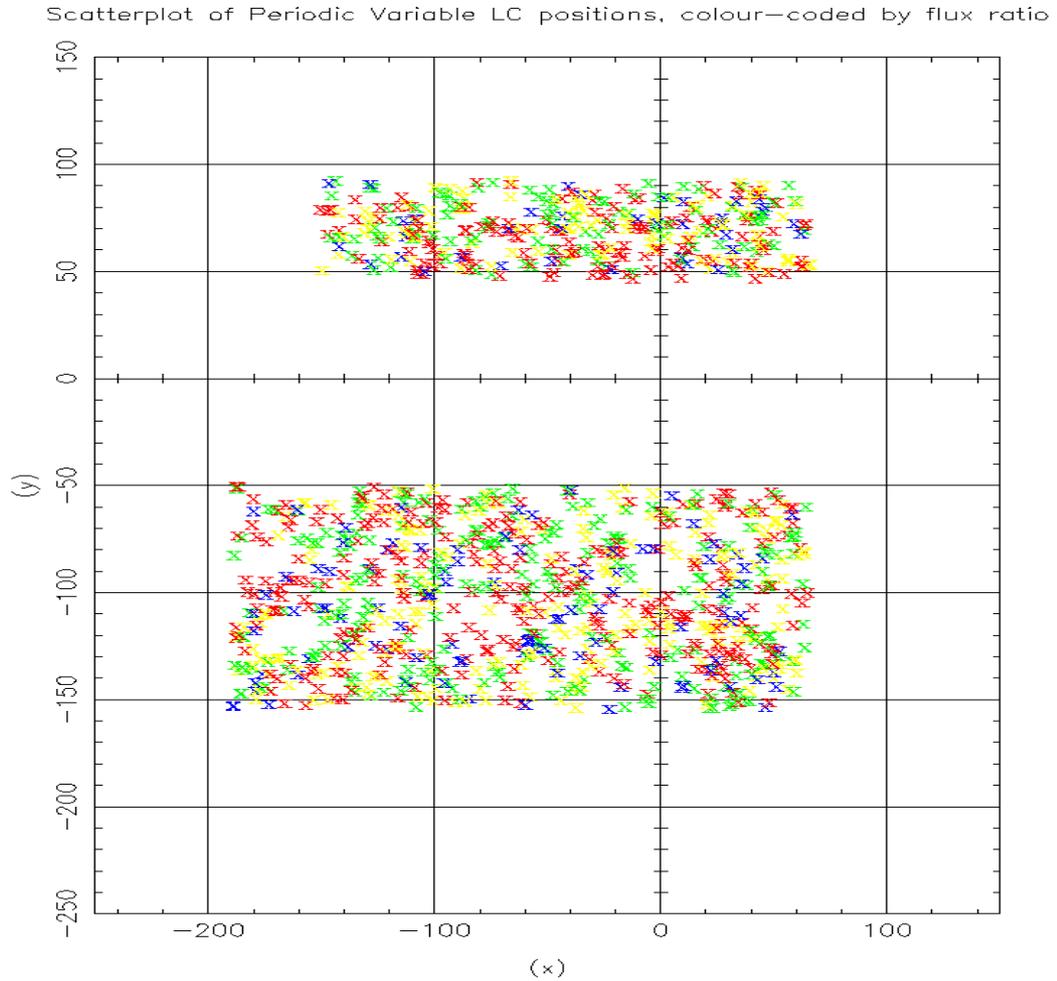}
\caption[Scatter plot of the spatial distribution of the $1185$ lightcurves previously selected for Figure \ref{1185_lightcurves_13_08_07_flux_ratios_scatter}, colour coded by the flux ratio between the PA and LT variable flux amplitudes.]{Scatter plot of the spatial distribution of the $1185$ lightcurves previously selected for Figure \ref{1185_lightcurves_13_08_07_flux_ratios_scatter}, colour coded by the flux ratio between the PA and LT variable flux amplitudes.
The flux ratios corresponding to the colour bands are: Blue: 0-5, Green: 5-10, Yellow: 10-15 and Red: 15 and over.
}
\label{colour_coded_flux_ratio_spatial_correlation_1185}
\end{figure}

The overlap regions between the two PA image fields which overlap with the LT and FTN fields were clearly delineated in Figure \ref{colour_coded_flux_ratio_spatial_correlation_1185}, but no pattern or correlation could be seen in the colours. This strongly implied that
a spatial correlation was unlikely to be found. Nevertheless, the equivalent plot was also made using the eight lightcurves selected above. No spatial pattern could be seen in the low number of data points.

Instead of selecting and classifying lightcurves based upon 
the signal to noise parameter, which after all was known to be sensitive to non-periodic variations and discrepant points, it was wondered whether a better selection criterion might be the PA flux amplitude itself, which should perhaps be more closely correlated with the ``true'' variables.

By this stage, only two bands of data were being used for the fitting;
LT i'-band and PA i'-band.
The script used to select variable lightcurves was modified so that
it first selected lightcurves classified as variable that had $\chi^2$/d.o.f. $< 15$, and then applied the cuts on the number of data points

 1) total number of points $> 250$
 2) PA points $> 80$

Next the lightcurves were grouped according to PA flux amplitude allowing the possibility that the fitting errors might be zero. The flux ratio  
was again calculated and is quoted as if it were between PA and FTN bands, which is because the scripts were written when three data bands were being used.  
The categories were defined as  int$(\frac{\rm{PA~flux~amplitude}}{10})$, e.g.
category $0$ contained PA flux amplitudes $0-10$, category $1$ contained
PA flux amplitudes $10-20$, etc. On inspection of the flux ratio
and amplitude information, $5$ lightcurves were discarded because some of the data were zero, implying a bad convergence of the fitting. The numbers of lightcurves found to be in each amplitude category are displayed in Table \ref{PA_flux_amp_variable_categories}.

\begin{table}
\caption{Table showing number of variable lightcurves found to be in each
PA flux amplitude bin, where each bin is of width $10$ ADU/s.
}
\begin{center}
\begin{tabular}{|c|c|}
\hline
\hline
PA flux amplitude &  number of   \\
 Category         &  lightcurves \\
\hline
        0          &      20       \\
        1          &      52       \\
        2          &      44       \\
        3          &      26       \\
        4          &       9       \\
        5          &       8       \\
        6          &       9       \\
        7          &       8       \\
        8          &       8       \\
        9          &       7       \\
        10         &       3       \\
        11         &       3       \\
        12         &       3       \\
        13         &       2       \\
        14         &       3       \\
        15         &       0       \\
        16         &       1       \\
        17         &       2       \\
        18         &       0       \\
        19         &       0       \\
        20         &       0       \\
        21         &       2       \\
\hline
\end{tabular}
\end{center}
\label{PA_flux_amp_variable_categories}
\end{table}

\begin{figure}[!ht]
\vspace*{11cm}
   \includegraphics{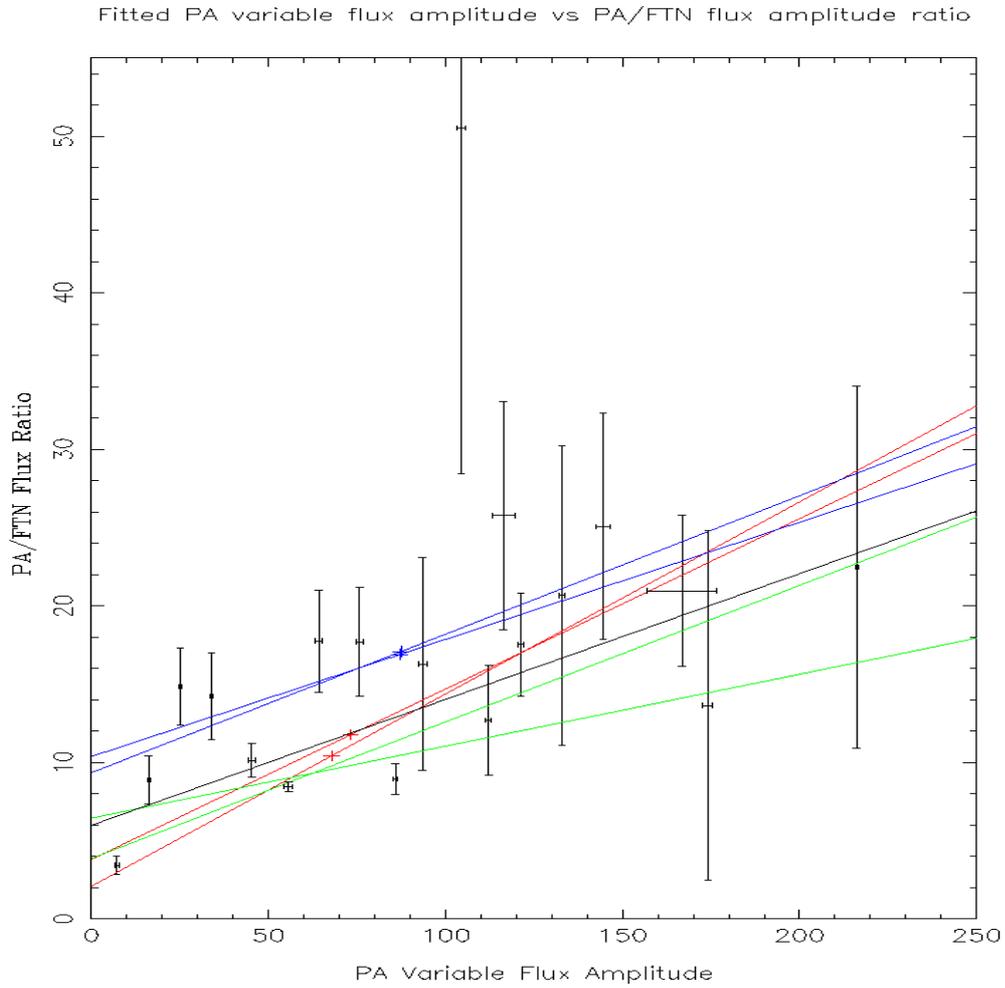}
\vspace*{2cm}
\caption[Scatter plot showing the mean PA/FTN flux amplitude ratios found for each PA flux amplitude bin, together with the various linear fit lines calculated.]{Scatter plot showing the mean PA/FTN flux amplitude ratios found for each PA flux amplitude bin, together with the various linear fit lines calculated. The two blue lines were found by linear regression based on the
unweighted mean (assuming throughout that all errors are equal).
The red lines were also found by linear regression, but forced to go through the mean weighted by the flux ratio errors. The third, green, set of lines show the best fit lines obtained by a full $\chi^2$ minimisation fit, using the flux ratio errors only. There are always two lines of each colour because the calculations were performed both with and without the first (Category $0$) data point.
}
\label{mean_flux_ratio_with_fits}
\end{figure}

It was interesting to note that the numbers of lightcurves in each category rose with falling category number, until about
category $1$, when they began to fall. It would be expected that fainter
stars would always be more common than brighter stars, at least down to the detection limit, so it appeared from this table that the decline in numbers of variable candidates with lower amplitude variability
could best be explained by the lower signal to noise categories being too faint to be identified consistently. Since the flux ratio between PA and LT was known to be of order $10$, any lightcurves classified as having PA flux amplitude $<10$ will have LT flux amplitude $\sim1$ ADU/s, which is certainly extremely hard to detect, given the known noisy LT data.
Next, the mean
flux ratio of all categories containing more than $1$ lightcurve was calculated, along with the standard error on the mean, given by the standard deviation about the mean divided by the square root of the number of lightcurves. Due to the low statistics, the errors on the mean were large in most cases, but for almost all data points the error was less than the value of the mean. In addition, in an attempt to improve the flux ratio errors (at the expense of the PA flux errors), categories $10$-$13$ were grouped together, and categories $14$-$21$ were grouped together, the two groups therefore containing $11$ and $8$ lightcurves, respectively. The mean PA flux values of each data point were calculated using the actual values, and in an analogous way to the flux ratio, the standard errors on the mean were calculated. It seemed clear that these would be under-estimates of the true flux spread, since each data point comes from a pre-selected non-random group with a range limited to $10$ ADU/s.
  The now $19$ data points were plotted in a PA flux vs mean flux ratio
 scatter plot (comparable
 with Figure \ref{1185_lightcurves_13_08_07_flux_ratios_scatter}).
 These are the data shown in Figure \ref{mean_flux_ratio_with_fits}.

It had been observed that the estimated error in the PA flux direction of the last data point (the $21$st flux category) was much smaller than for any other category. Indeed this was found to be because the PA flux amplitudes for the two specific lightcurves involved ($5289$ and $7081$) 
were extremely similar, although their flux ratios were very different.
On checking the positions on the sky of these two lightcurves, it was found that their positions on the LT field were both close to ($-17$,$57$). Close examination of the two lightcurves showed that the PA data were practically identical, whereas the LT data were different, although similar in form, implying that two different (but adjacent) LT objects had been matched with the same PA
object at different times. This introduced some doubt over the estimate of the errors in both directions, as the two lightcurves are only semi-independent data points, although the magnitude of the flux ratios derived still look sensible on average. To be safe, this data point was not included in the following work.

 A clear correlation can be seen by eye in Figure
 \ref{mean_flux_ratio_with_fits}, although finding a line of best fit
which satisfactorily describes the data was made harder by the way all the points with small flux ratio spreads are clustered at the bottom left of the plot. This drags any fit which includes the errors downwards,
so that the majority of the data are above the trend line.
In Figure \ref{mean_flux_ratio_with_fits}, the various fits that were performed
are shown. In summary, the blue lines were found by linear regression based on the
unweighted mean (assuming throughout that all errors are equal).
The red lines were again found by linear regression, only again forced to go through the mean weighted by the flux ratio errors. The third set of lines, which are green, show the best fit lines obtained by a full $\chi^2$ minimisation fit, using the flux ratio errors only. There are always two lines of each colour because the calculations were performed both with and without the first (Category $0$) data point, which also has some doubt attached to it, for the reasons described above. As would be expected, not including this first point allows the gradient of the lines to move away from where the data point once was on the left hand side, meaning that the fits performed without the first data point have \emph{lower} gradients and \emph{higher} offsets.

Possible fitting functions other than simple straight lines were considered, for example a logarithmic curve, but it was felt that not including the first point removed any motivation for trying this as it was the point which had the flux ratio which was the lowest relative to the others.

It was felt that the range of the six curves described above nicely
bracketed the range of fit lines which might be described as ``possible''.
The seventh line, which is coloured black, has the mean gradient and offset of the previous six, that is: 
Gradient $= 0.08038$, offset $=5.98$

The equivalent parameters for the $\chi^2$ minimisation fit are:

1) Including Category $0$ point:

Gradient $= 0.087\pm{0.017}$ offset $=3.86\pm{0.56}$

$\chi^2 = 59.54$, $\chi^2$/d.o.f. $= 3.72$ 

2) Not including Category $0$ point:

Gradient $= 0.046\pm{0.010}$ offset $= 6.4\pm{1.0}$

$\chi^2 = 50.90$, $\chi^2$/d.o.f. $= 3.39$

It is not clear whether the underlying behaviour of the flux ratio with PA flux does flatten out as PA flux increases beyond $200$ ADU/s, but all
proposed linear fits stay within the $1\sigma$ flux ratio error of the final unused data point, which provides some confidence, even though this data point is not a perfect fit, that continuing the linear fit is not
wildly wrong.

 However, this question must remain unanswered for the moment, as variables with larger flux ratios (and good fits to a skew cosinusoid) apparently do not exist in the data, and other kinds of variables are not currently detected by the selection pipeline.

\chapter{First Results and Analysis of the Angstrom Project Dataset}
\label{Chapter_5}
\section{Introduction}
In this Chapter the main results of the research described in this Thesis
are described. The results of investigations into one particularly interesting transient event, of high signal to noise and short timescale are presented.
Some lightcurves which are highlighted by the pipeline as having a relatively high signal to noise bump, but which do not fit the point-source point-lens lightcurve
model well enough are shown. These may include classical novae.
Investigations into the properties of variable star candidates are described,
and the information about their spatial distribution that has so far been extracted is also described.
The lensing candidates selected by the pipeline are shown, along with summaries of their properties.
Finally, a test of the pipeline, conducted using ``fake'' microlensing event lightcurves, is described.

\section{Comparisons with previous surveys}

 Since we use the POINT-AGAPE data to extend the Angstrom survey baseline
 it is interesting to ask questions like: ``Do any of the published POINT-AGAPE 
 events lie in the Angstrom field?" and if so, ``Are they detected as variable 
objects independently by Angstrom?".
 If any of these microlensing candidates had either repeated, or had showed some 
clear high signal to noise variation since the end
of the POINT-AGAPE survey, this would be very interesting, as the first possibility
would either most likely eliminate the event as a possible lensing event, or be an exceedingly rare repeating lensing event (possibly a binary event as investigated by \cite{2008AcA....58..345J}, \cite{2009MNRAS.393..999S}), and the second possibility would cast some doubt on whether the original event was due to microlensing.
 The POINT-AGAPE papers \cite{2003A&A...405...15P}, \cite{2005MNRAS.357...17B} and 
\cite{2005A&A...443..911C} have been checked and the positions of the lensing candidates reported therein have been converted to arcseconds and compared with the
limits of the Angstrom field illustrated by Figure \ref{variable_object_dist}.
The POINT-AGAPE field is considerably larger than the Angstrom field, and none of their events have positions within it.

    Hence, the LT or FTN data have no power to confirm (or
 otherwise) the possible nature of the POINT-AGAPE events - however it is possible
 that the BOAO or Maidanak data, with their larger fields or view, may be able to
 assist in this quest. However, from the point of view of the Angstrom Candidate 
Selection Pipeline, object detection is only performed in the areas overlapping with the LT and FTN reference images, so, none of our events can
correspond to these previously detected events.
 Obviously, this means that the selection pipeline is not able to re-detect these 
 events.

\section{The very short microlensing event              ANG-06B-M31-01\label{short_event}}

\subsection{The Angstrom Project Alert System (APAS)}

The event described in this paper was detected in the first instance by the Angstrom Project Alert System (APAS) \citep{2007ApJ...661L..45D}. This is a pipeline which analyses
 the data in ``real time'' (see Section \ref{angstrom_alert}). It is analogous to those used by Milky Way searches such as OGLE \citep{2006ApJ...636..240S} and MOA \citep{2003ApJ...591..204S}. Each day, 
the APAS presents to the observer a limited list of typically no more than a few dozen 
lightcurves, via a web interface. These are the most interesting transient signals 
found in the database of lightcurves produced by the difference imaging. Currently this 
database contains of the order of $90000$ lightcurves and during the observing season is 
updated each day. The observer
is then able to review the list and decide which, if any, are worthy of further study. 
In order to allow the assessment of possible lightcurve contamination from close-by objects, the APAS also records the positions of all variable signals in the immediate neighbourhood of the lightcurves in the list.

\subsection{Data and Analysis}

\subsubsection{Data}

In addition to almost 3 seasons of LT i'-band and one season of
each of LT r'-band, FTN i'-band and FTN R-band Angstrom data for the position
 of this particular lightcurve, archive i'-band data from the
POINT-AGAPE dark matter microlensing survey of M31
\citep{2003A&A...405...15P} were available.
 POINT-AGAPE used the wide-field camera of the $2.5$m
Isaac Newton Telescope on La~Palma to survey a $0.6$~deg$^2$ area of the
M31 disk and bulge. Around $65\%$ of the LT/FTN reference field overlaps with
the PA data. All Sloan
i'-band data from the PA survey obtained between $1999$ and
$2001$ have been reprocessed using the Angstrom Data Analysis Pipeline (Section \ref{angstrom_dap}).
These data are very useful to us as they provide an extended data baseline
 for the subset of light curves lying in both Angstrom and PA survey
 regions to enable better discrimination between variable stars and
 microlensing events.

\subsubsection{Analysis}

 Visual inspection of the data confirmed the presence of a clear short
 timescale spike in the $2006$/$7$ season of LT i' data. There is also an obvious
 periodic variation of the baseline of the event which is also present in
 the PA data. These features can be seen in Figure \ref{short_event_2008_photom}.

\begin{figure}
  \begin{tabular}{cccc}
\includegraphics[height=43mm]{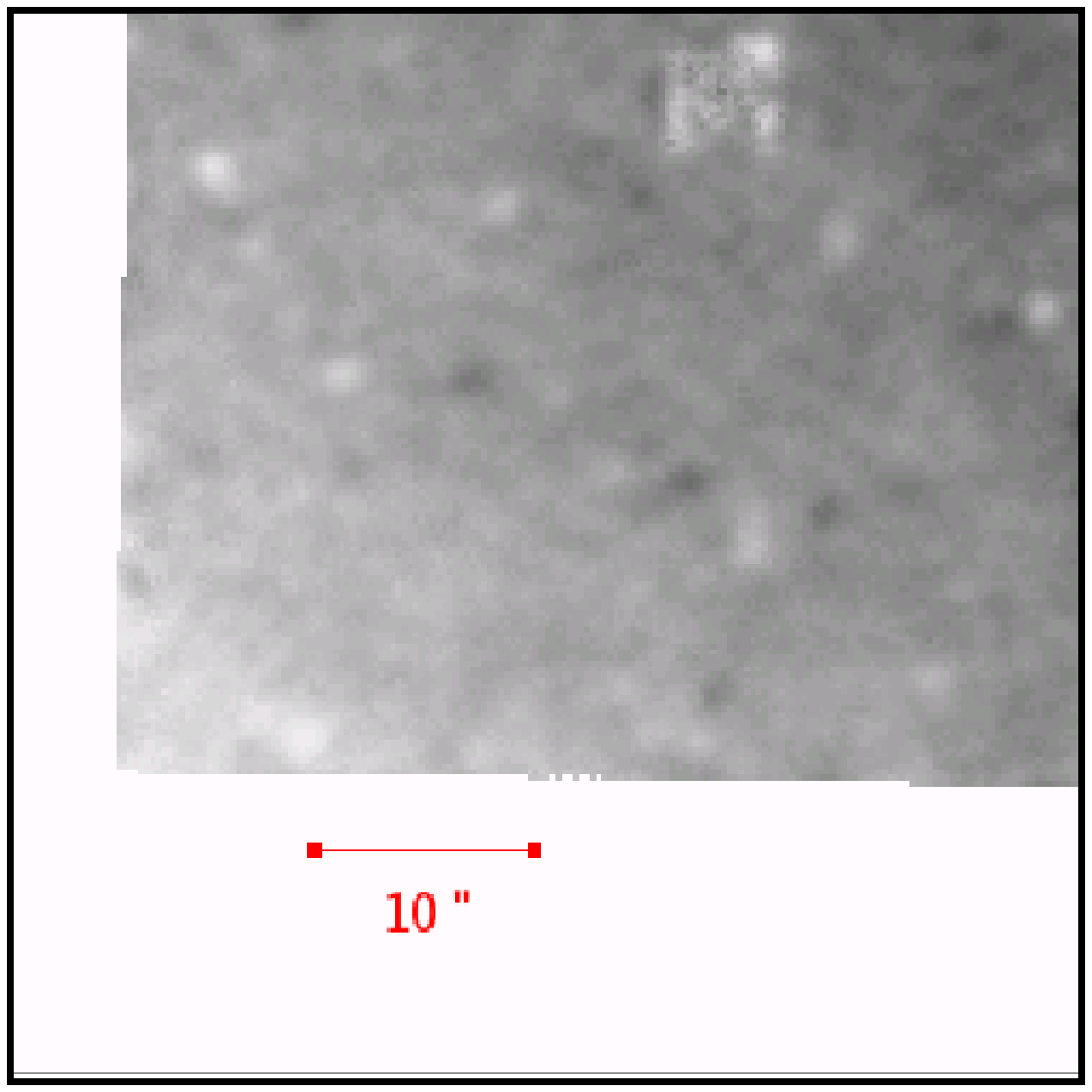} \\
\includegraphics[height=43mm]{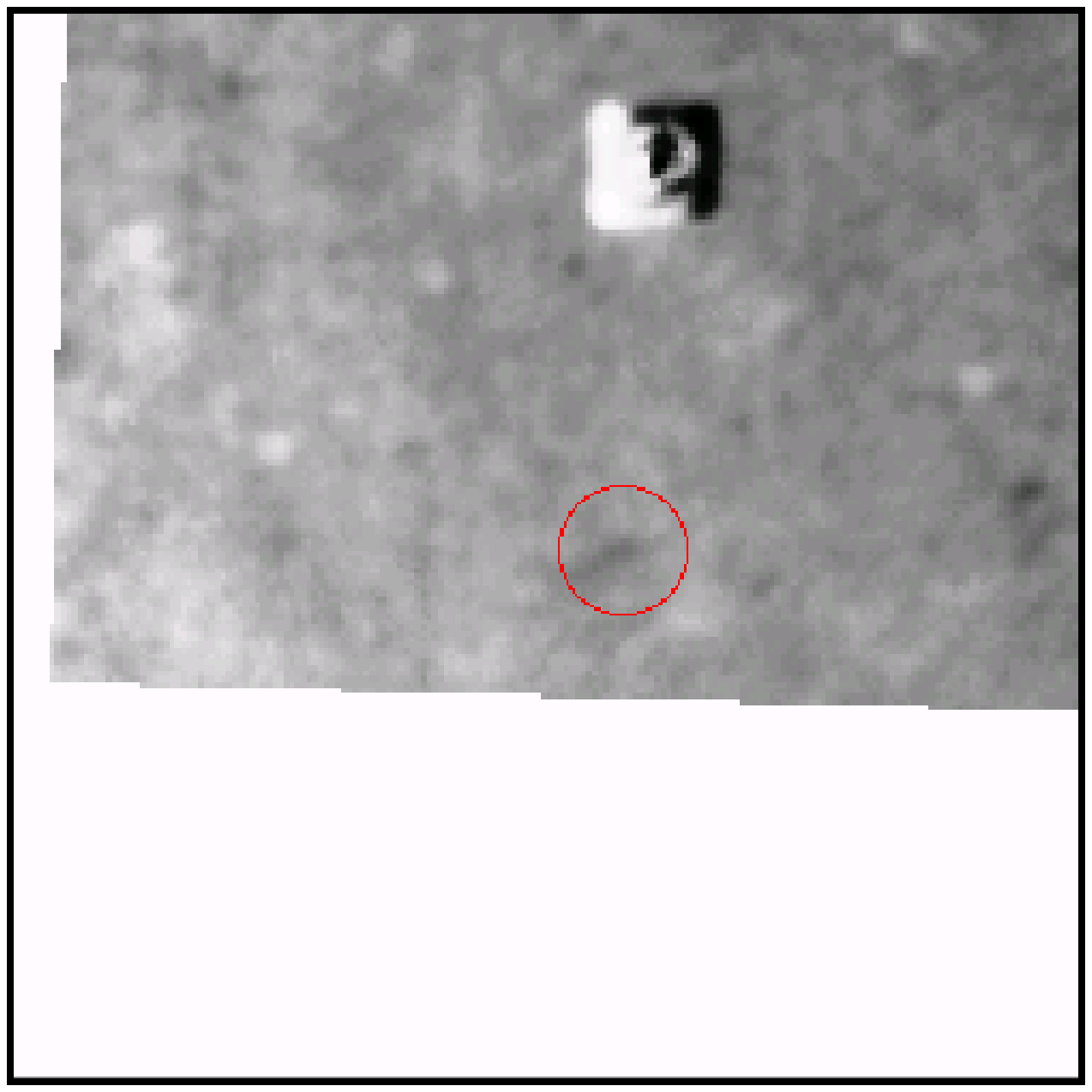} \\ 
 \includegraphics[height=43mm]{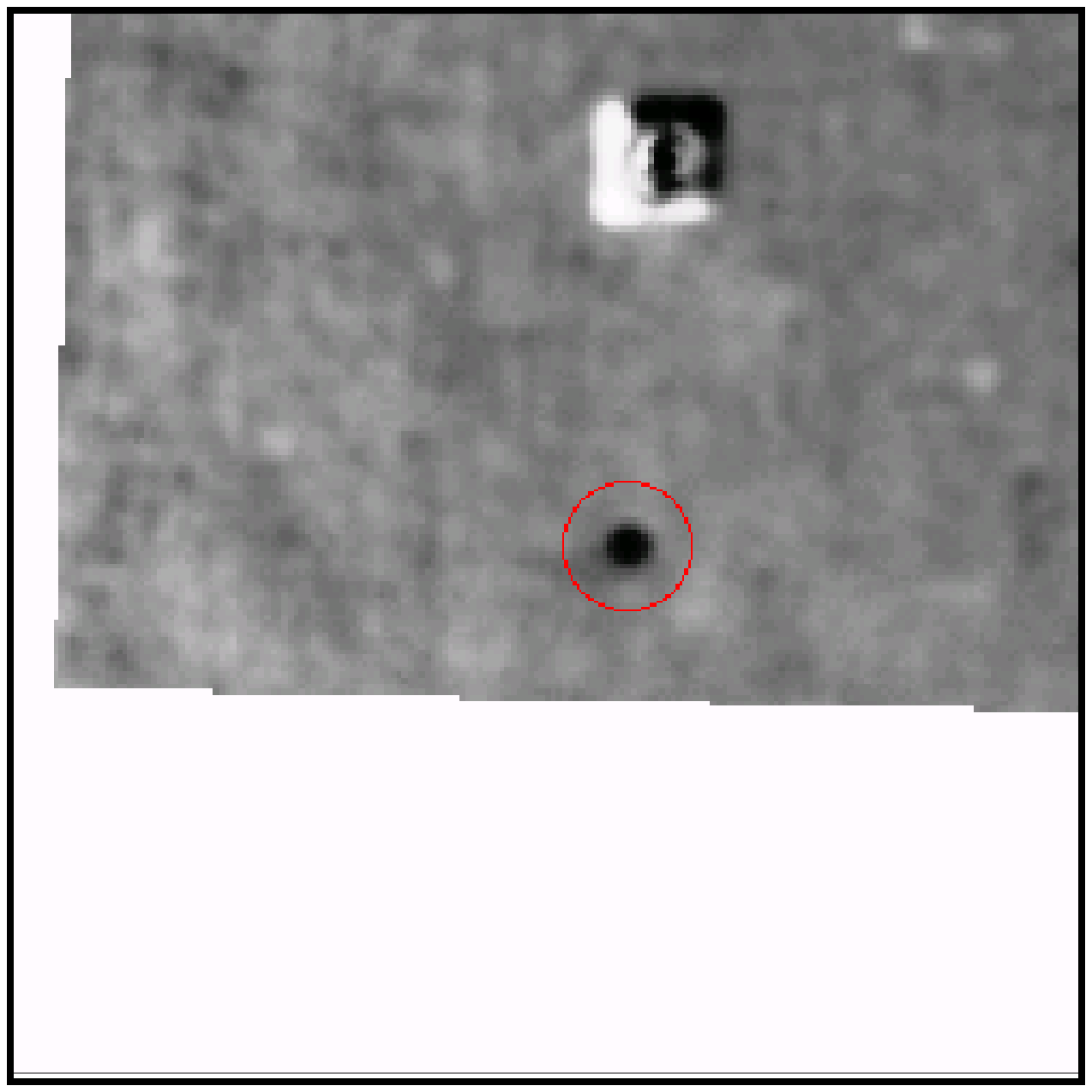} \\
\includegraphics[height=43mm]{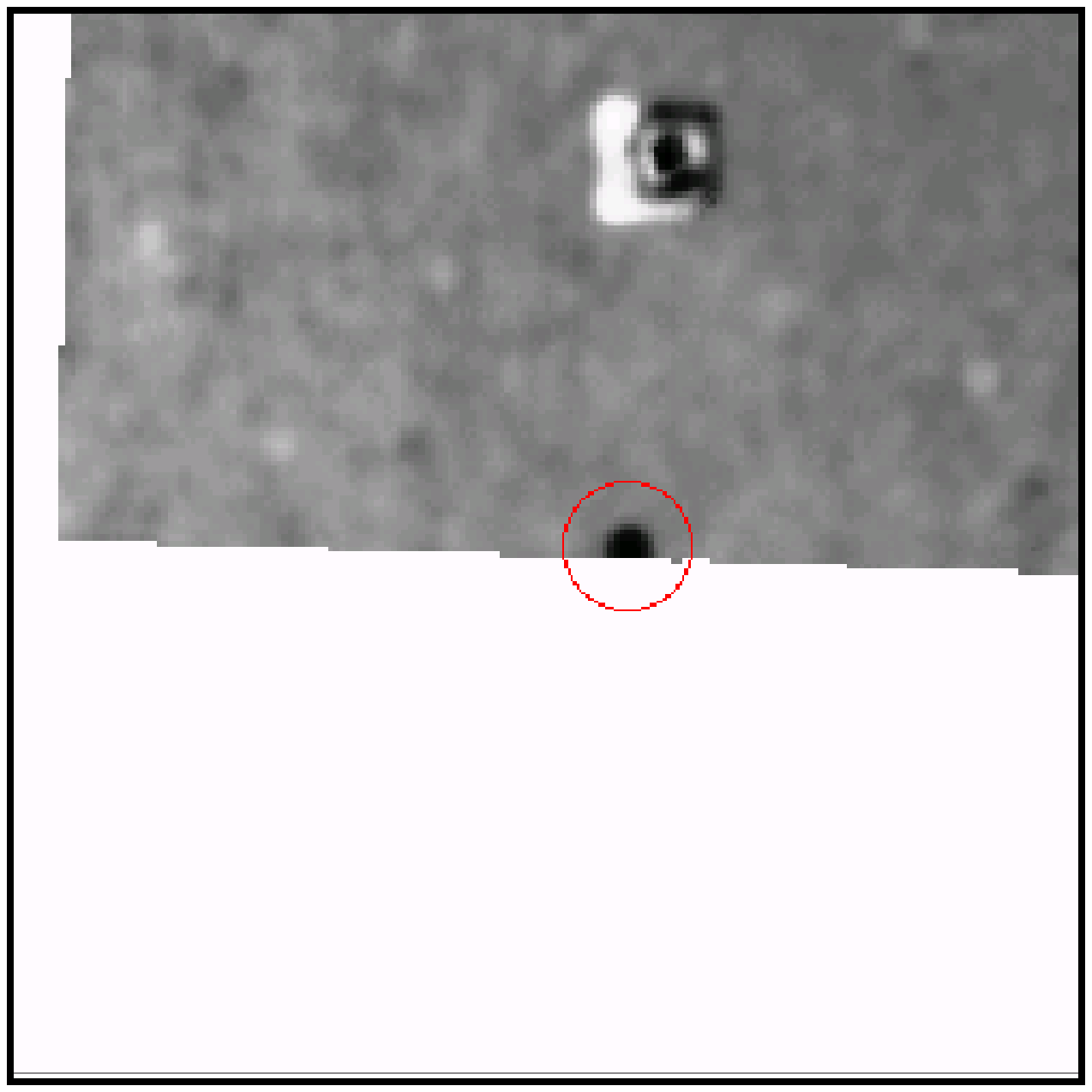} \\
  \end{tabular}
  \begin{tabular}{cccc}
   \includegraphics[height=43mm]{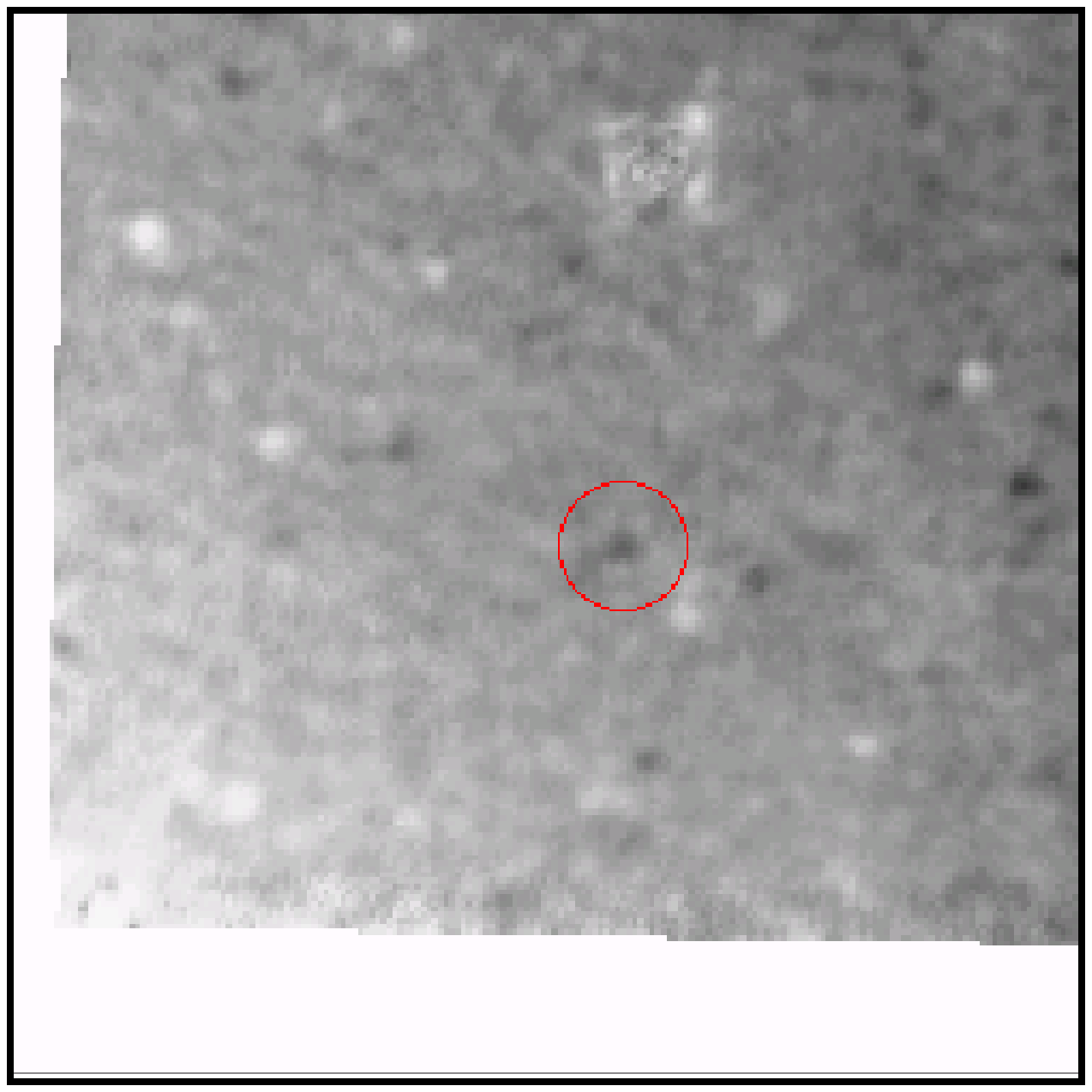} \\
   \includegraphics[height=43mm]{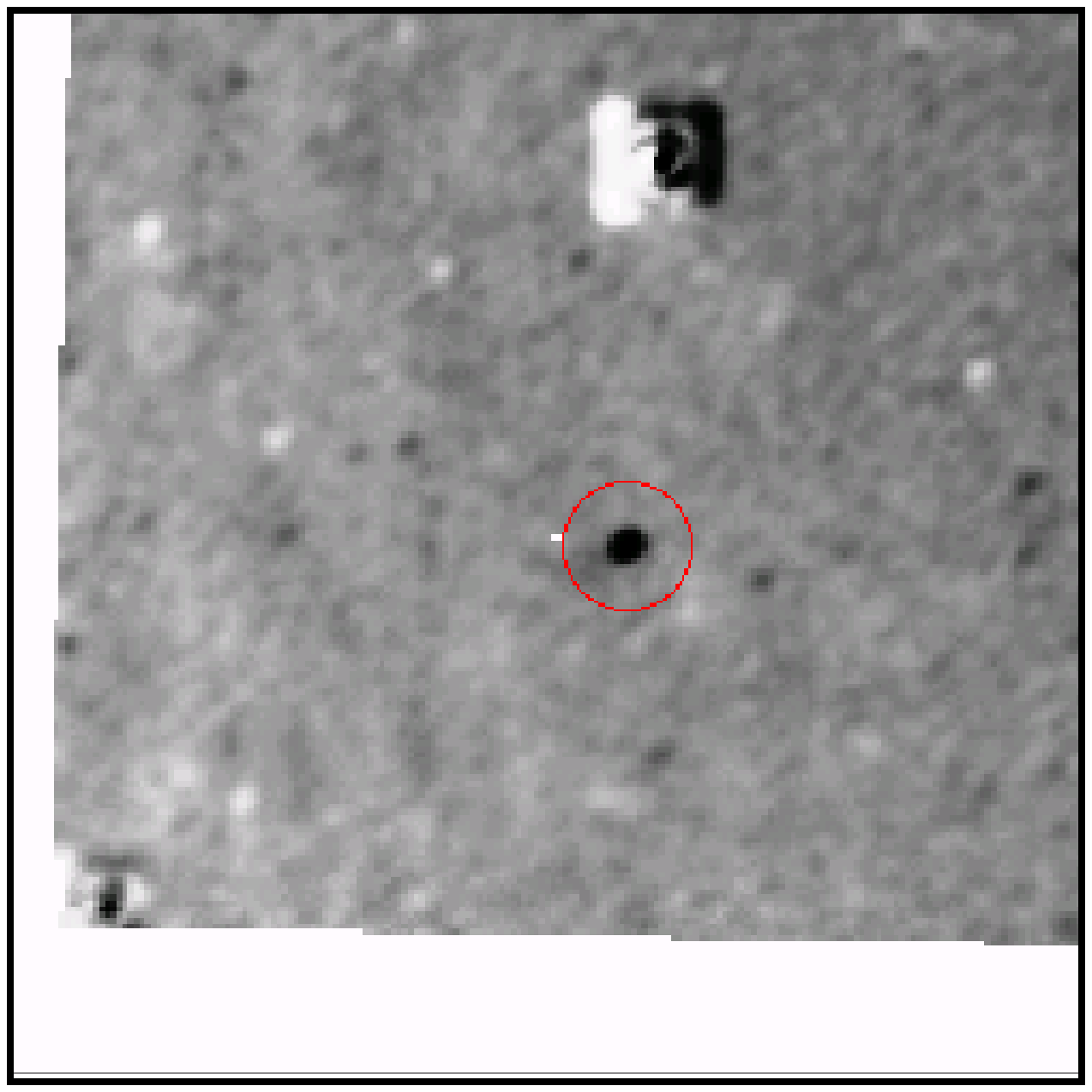} \\
   \includegraphics[height=43mm]{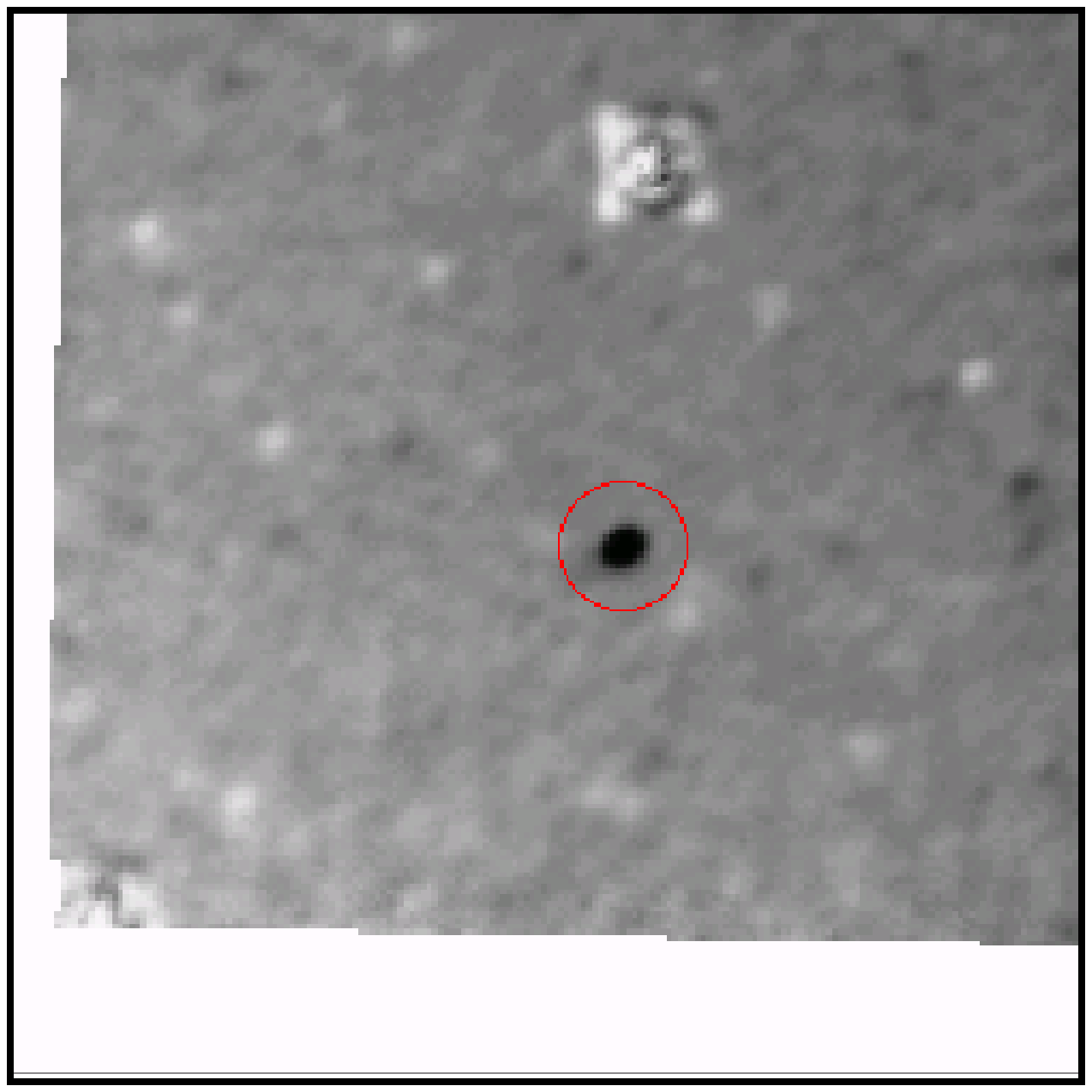} \\
   \includegraphics[height=43mm]{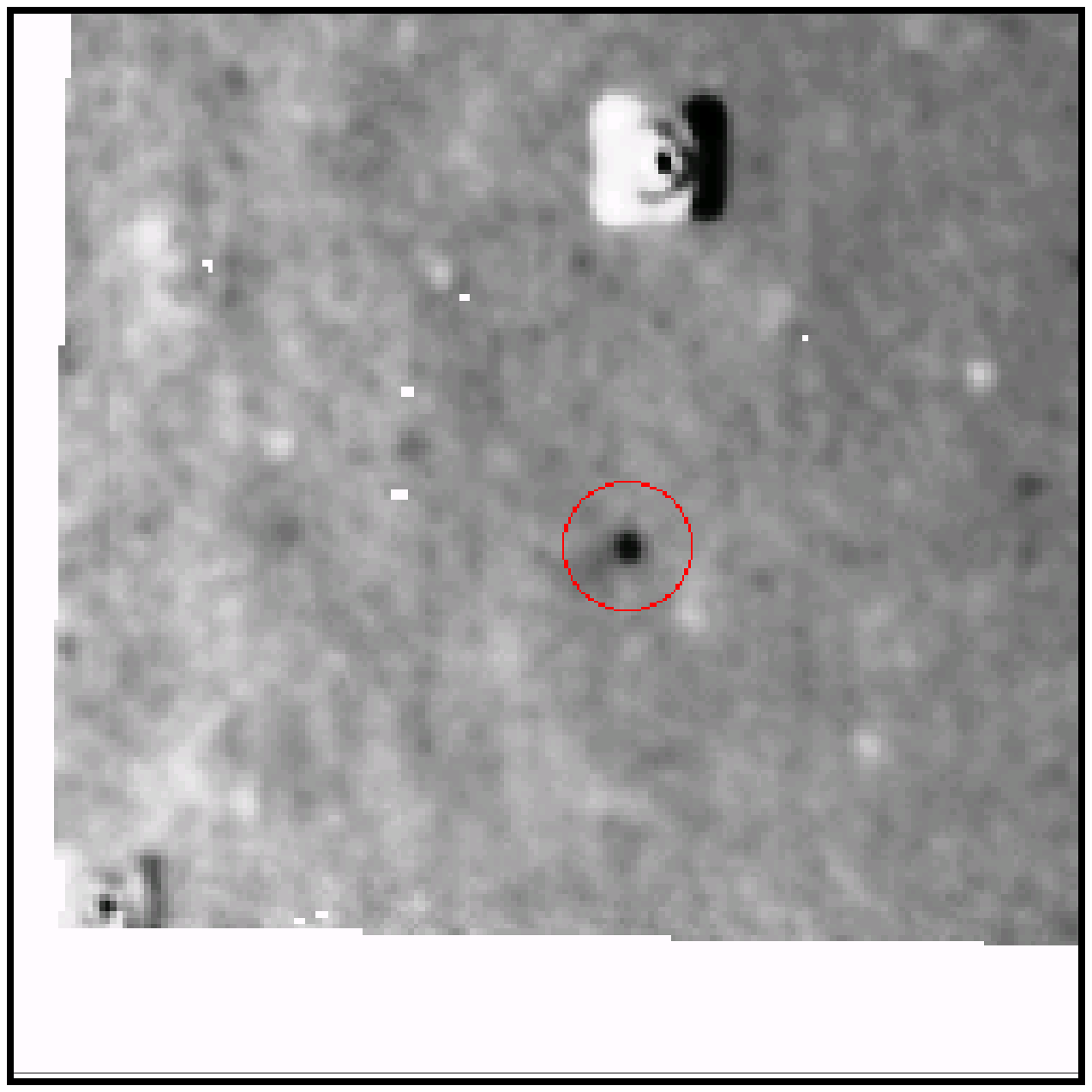} \\
  \end{tabular}
  \begin{tabular}{cccc}
   \includegraphics[height=43mm]{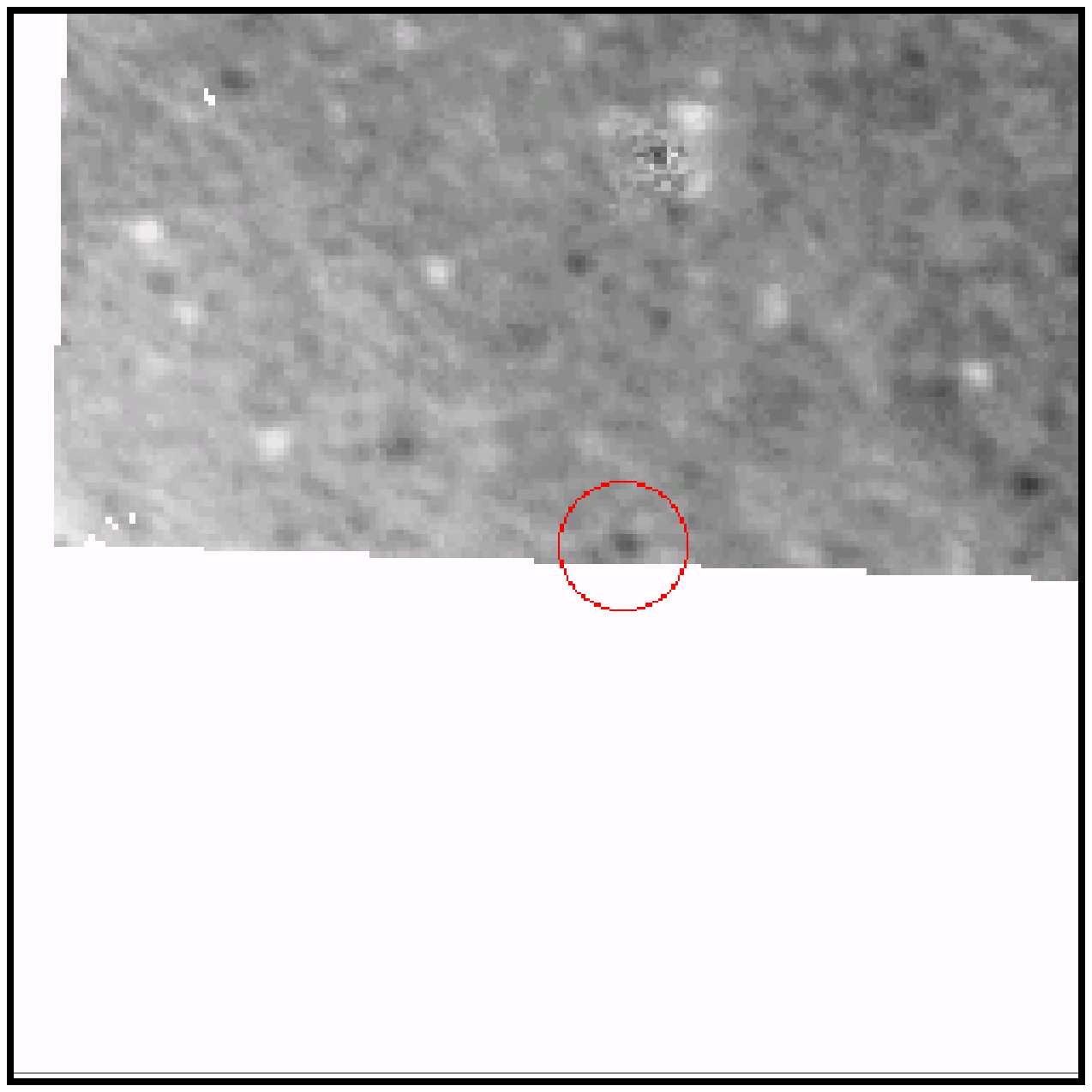} \\
   \includegraphics[height=43mm]{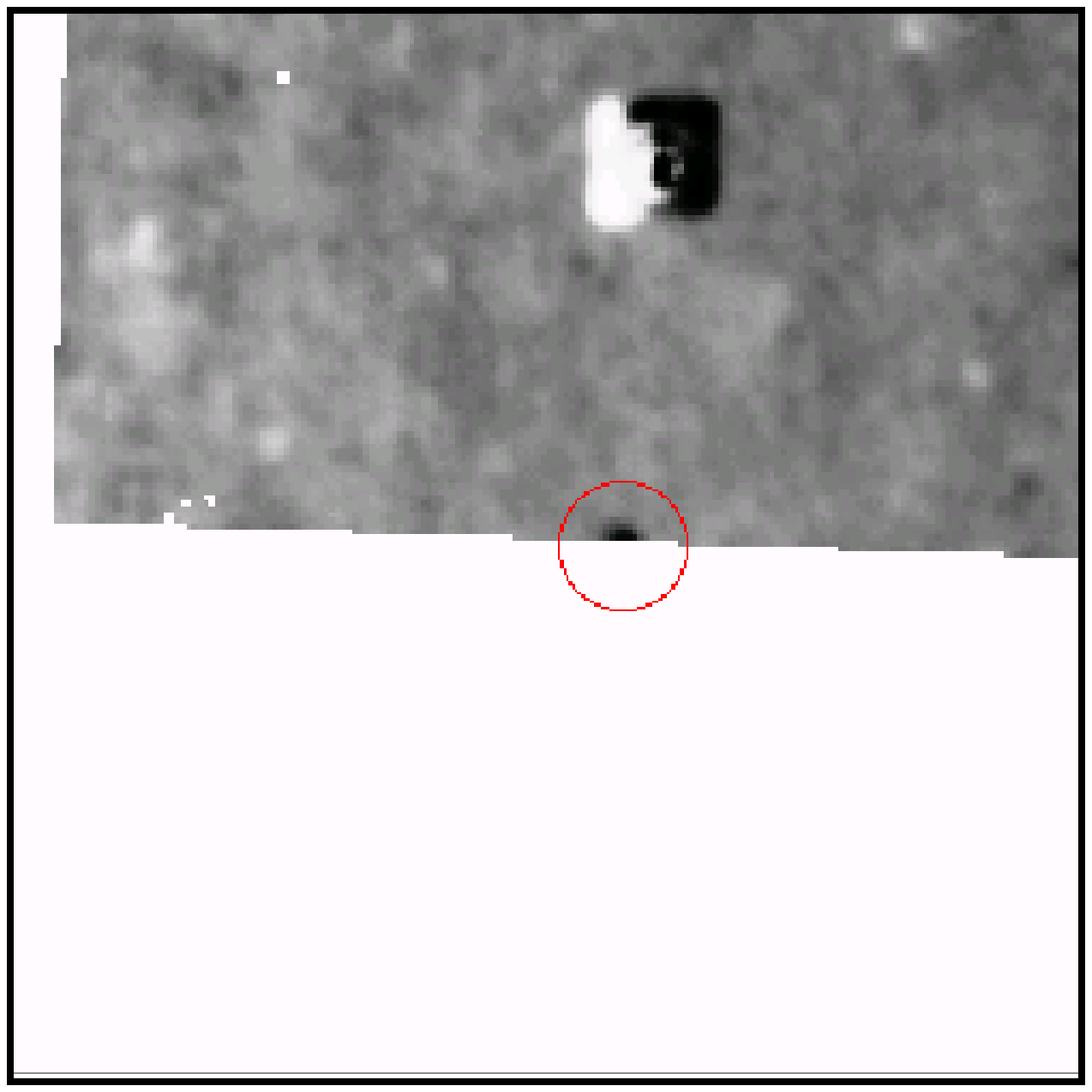} \\
   \includegraphics[height=43mm]{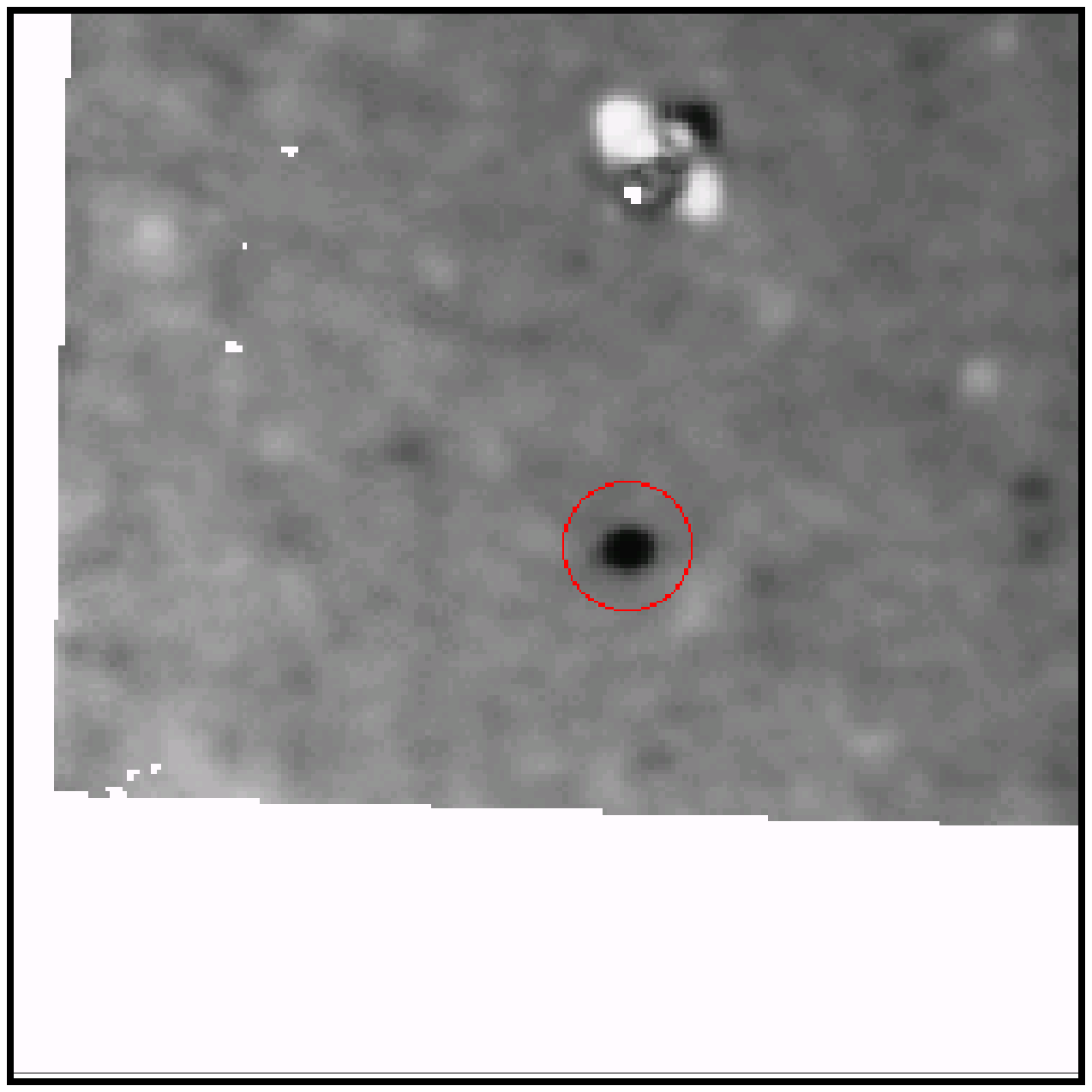} \\
   \includegraphics[height=43mm]{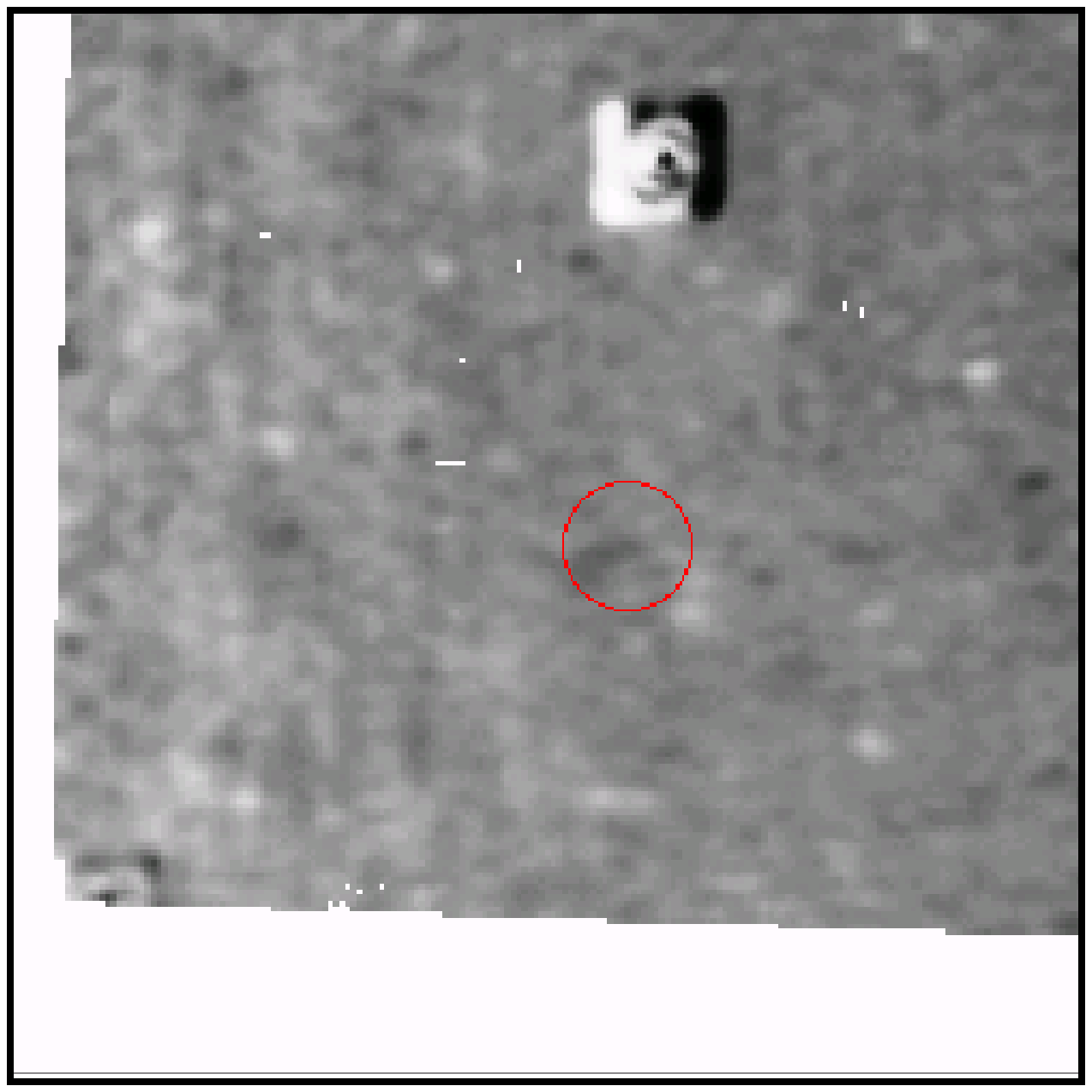} \\
  \end{tabular}
 \caption[The ``short event'': Twelve i-band images covering the main flux spike]{Portions of the Angstrom difference images covering the
 main flux spike. The first image in the series is 11th September $2006$;
 the last is 20th September $2006$. The image sequence is left to right,
 top to bottom. The seven central points where the
 source was brightest span 14th-16th September. The scale of the images is given in the first image:
 (each sub-image is approximately $1/12$ of the area of an LT field). The red line in the
 top left image is roughly $10^{\prime\prime}$ in length. The white area at the bottom and left of each image is the corner/edge of the LT field difference images.}
 \label{09388_peak_dia}
\end{figure}

 It is clear from inspecting the difference images that this is a real 
astronomical event and is not caused by some exotic data processing effect. 
The relevant images bracketing the flux spike, spanning the dates 11th - 20th 
September $2006$ are shown in Figure \ref{09388_peak_dia}

 Therefore, the event is modelled by fitting the flux with the
 sum of a reduced Paczy\'nski curve
(with flux varying with time as given in Equation \ref{red_pac}
where $t$ and $t_{0}$ are the epochs of observation and
maximum brightness, $B$ is the baseline flux, $\Delta{F} = F(t_0)-B$ is the
maximum flux deviation and $t_{\rm{FWHM}}$
is the duration of the full-width at half-maximum), and a skew cosinusoid function.
 The CERN MINUIT\footnote[1]{http://wwwasd.web.cern.ch/wwwasd/cernlib/} software 
is used to minimise the $\chi^2$ of the fit.
The skew cosinusoid function used is designed to be capable of being adjusted by the fitting routine continuously from zero skewness (sine curve) to high skewness. The function chosen is defined by Equations (\ref{skewcos1}) and (\ref{skewcos2}), in Chapter \ref{Chapter_4}.

\subsection{Fitting Details}

This event was first fitted in $2006$, and was subsequently re-investigated each time
the Angstrom pipeline was significantly improved and the lightcurves re-generated.

The following describes sequentially the work that has been done to investigate the behaviour of this lightcurve, concentrating to a greater extent on the analysis of more recent data which were of noticeably better quality than the earlier data.

\subsubsection{Lightcurve copies}
\label{lightcurve_copies}
In the DIA pipeline, new objects are defined as soon as a localised variation in flux appears which is not consistent in position with
any previously detected object position. In many cases, there exist more
than one defined varying object within the same PSF for a given epoch.
As the fluxes of these objects vary independently, the centroid of their combined PSF moves back and forth so that it is nearer to the object which happens to be the brightest at that moment. Sometimes, the centroid of a PSF moves so far from a previously defined object that it can no longer be considered consistent, to within the positional estimation errors, with being part of the same object. At this point a new variable object is defined. However, the fluxes for the (now two) objects are drawn from the
integrated flux of the whole PSF. Hence, although the objects may be positionally independent, photometrically they are almost identical as their individual PSFs overlap almost entirely. Therefore, in many cases
the lightcurves of objects which are very close in position to one another are very similar, or sometimes identical.

\subsection{Investigating the variable component}

Firstly, an eight parameter skew cosinusoid fit was made to the above three band
 data, with the region ($53982 < t < 54006$) (JD - $2400000.5$) around the possible Paczy\'nski spike masked out. The best fit parameters for this are shown in Table \ref{skewsinfit_params}.
This fit had a $\chi^2$ of $2237.85$, which corresponds to a $\chi^2$/d.o.f. of $4.61$,
 as only $494$ data points were included. All these parameters are consistent with the
 equivalent ones found using the earlier $11$ parameter fit, which has not been quoted here. 
 
\begin{table}
\caption{Fitting parameters for best skew cosinusoid fit to ``short event'' data in the $2007$ photometry, not including possible Paczy\'nski spike.}
\begin{center}
\begin{tabular}{|l|l|l|l|}
\hline
\hline
Parameter &  Value & MINUIT   &  MINOS errors \\
  Name    &        & fitting error & (if calculated)\\
\hline
 $\phi_{\rm{0}}$ & $53210.92$ & $0.70$ & $+0.69 -0.71$\\
   $P$      & $245.29$   & $0.14$ & $\pm0.14$\\
   $S$      & $1.155$   & $0.053$ & $+0.021 -0.016$ \\
 $B_{\rm{LT}}$   & $9.01$ & $0.17$ & $\pm0.17$\\
 $B_{\rm{FTN}}$ & $2.03$ & $0.19$ & $\pm0.19$\\
 $\Delta F_{\rm{FTN,sin}}$ & $-7.55$ & $0.20$ & $\pm0.20$\\
 $B_{\rm{PA}}$ & $-18.62$ & $0.91$ & $\pm0.91$\\
 $\Delta F_{\rm{PA,sin}}$ & $-88.6$ & $1.6$ & $\pm1.6$\\
\hline
\end{tabular}
\end{center}
\label{skewsinfit_params}
\end{table}

\subsubsection{Periodogram Analysis}
\begin{figure}
\vspace*{9.5cm}
$\begin{array}{c}
\vspace*{0cm}
   \leavevmode
 \includegraphics{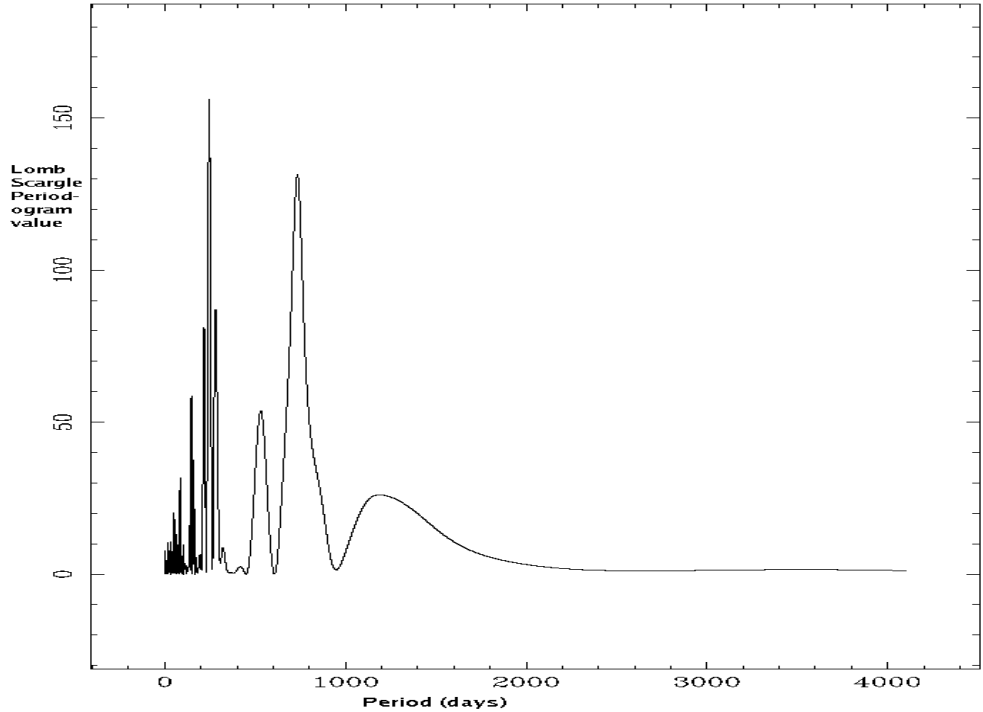} \\
\end{array}$
\caption{Periodogram of scaled combined variable signal for the event ANG-06B-M31-01.}
\label{all_data_periodogram}
\end{figure}

\begin{figure}
\vspace*{10cm}
$\begin{array}{c}
\vspace*{0cm}
   \leavevmode
 \includegraphics{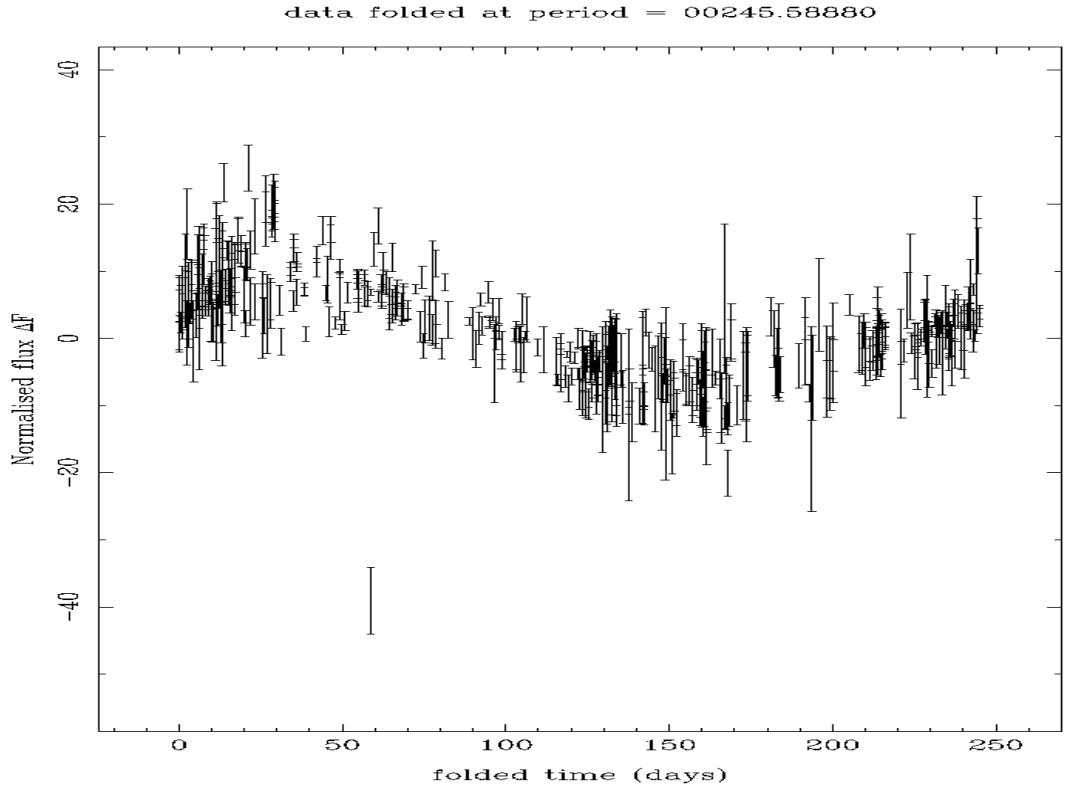} \\
\end{array}$
\caption{Plot showing the variable star component of the microlensing candidate ANG-06B-M31-01. All data with flux spike subtracted; each band scaled by the amplitude of the skewcosinusoid fit,
 interleaved in time and folded at P = $245.8$ days}
\label{variable_part_folded_at_245d}
\end{figure}

 Due to the way the annual gaps in the data occur, it is quite easy to see a period in the data
 (especially in the PA data)
 equivalent to roughly $750$ days. This effect is apparently caused by aliasing between the above
 periods and the annual observing cycle as several obvious resonances can be calculated. To
 provide further supporting evidence for the best fit period found, the data were subjected to a
 Lomb-Scargle periodogram analysis
\citep{1975MNRAS.172..639L,1972MNRAS.156..181S,1972MNRAS.156..165S,1982ApJ...263..835S}
which can cope with
 non-regularly spaced data such as ours, unlike Fourier Analysis. The ratios found above between $\Delta{F}_{\rm{LT,sin}}$, $\Delta{F}_{\rm{FTN,sin}}$
 and $\Delta{F}_{\rm{PA,sin}}$ were used to scale the measured variable fluxes (with the Paczy\'nski peak
 still masked out) to a consistent level, and then the data from all three bands were concatenated
and ordered in date order to form one combined data set. The periodograms of these combined data, 
plus the individual bands, were taken. Also, the above skew cosinusoid fit was subtracted from 
the original raw data, and the residuals were scaled and combined in the same way, and periodograms again taken.
  The periodograms of the scaled combined data confirmed that there were
 significant periods in the data at $P = 245.8$ and $P = 736.1$ days as well as
 some other less significant periods. This is shown in figure
 \ref{all_data_periodogram}. The data were folded at all significant
 periods found as another check on the periods indicated by the
 periodogram. The light curve folded at a period of $P = 245.8$ days is
 shown in Figure \ref{variable_part_folded_at_245d}.

\begin{figure}
\vspace*{9cm}
$\begin{array}{c}
\vspace*{0cm}
   \leavevmode
 \includegraphics{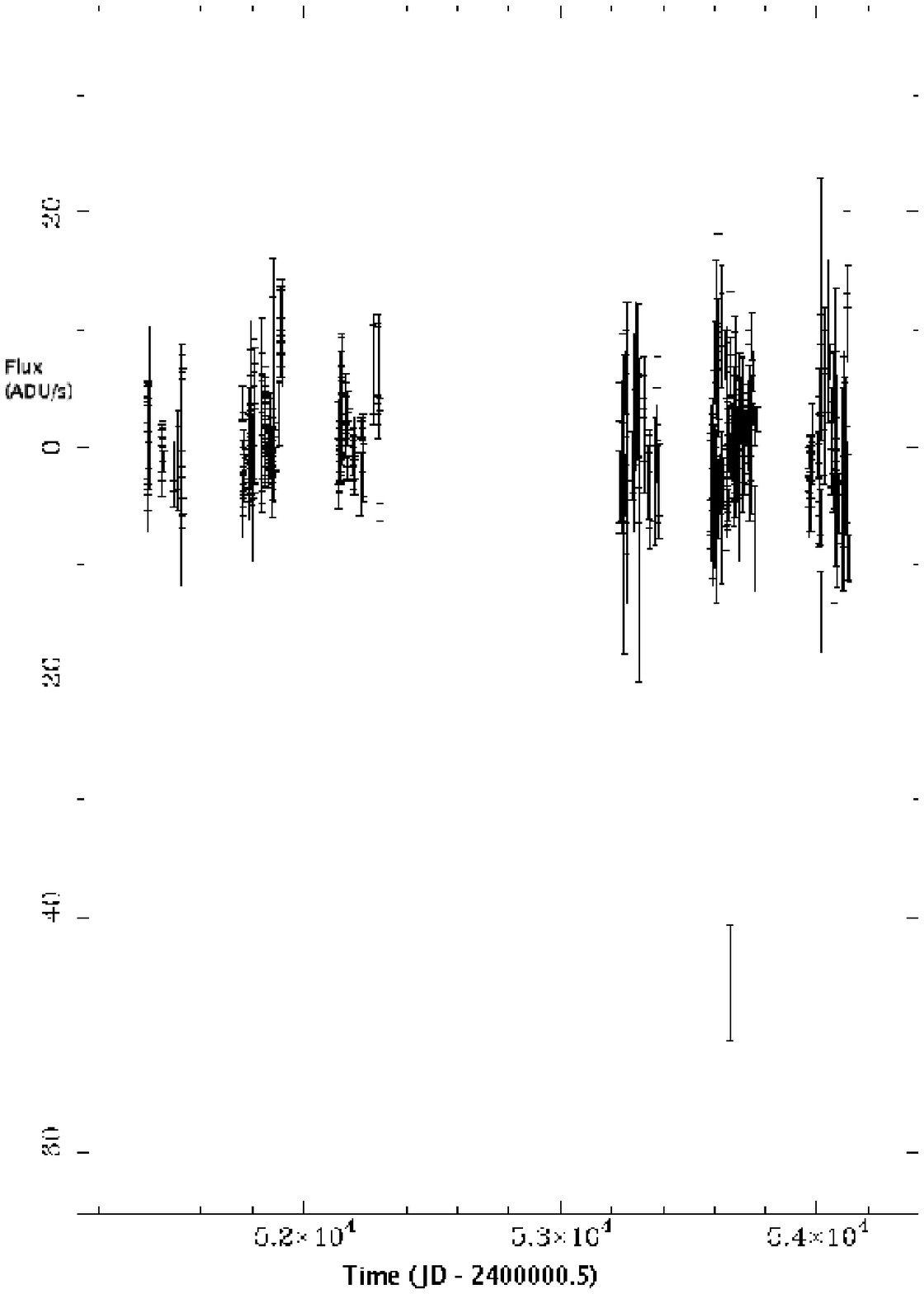} \\
\end{array}$
\caption{The residuals after removal of the variable fit to the baseline of ANG-06B-M31-01}
\label{residual_data_after_skew_sin_subtraction}
\end{figure}

Folding the data at the most significant periods contained in the periodogram 
of the residuals to the best skew cosinusoid fit showed no evidence for a second period in the data.

   When the best fit skew cosinusoid has been subtracted from the original
 data, the baseline of the Paczy\'nski peak does indeed look flat within
expected statistical variation. This can be seen in Figure \ref{residual_data_after_skew_sin_subtraction}.

The periodogram of the residuals shows that no significant periods remain in the data.

\subsection{Investigating the Paczy\'nski peak}

The remaining spike in the data after subtraction of the variable component was then 
modelled individually using a full Paczy\'nski fit in order to investigate whether it
 was possible to extract all the lensing parameters.
 An eight parameter fit was attempted, using $t_{\rm{0}}$, $t_E$, the Einstein crossing
time, $\beta$, the minimum impact parameter, $B_{\rm{LT,pac}}$, $F_{\rm{FTN,pac}}$,
 $B_{\rm{FTN,pac}}$, $F_{\rm{PA,pac}}$ and $B_{\rm{PA,pac}}$. The Paczy\'nski amplitude in the PA band was allowed to vary freely this time. Obviously, if the subtraction of the variable had been
 done correctly we would have expected the baseline parameters to be approximately $0$, and 
indeed they were all much less than $1$ ADU/s.
     The data could be fitted fairly well with a full Paczy\'nski profile, with a 
best fit $\chi^2$ of $3221.34$, corresponding to a $\chi^2$/d.o.f. of $6.29$. There 
was insufficient information in the lightcurve to find a unique solution, however, with solutions having equally good $\chi^2$ being found over a wide range of beta values.
All the fits to the short event lightcurve presented in this section may be seen to have $\chi^2$/d.o.f. values which are considerably higher (i.e. $\chi^2$/d.o.f.$\gg 1$) than would normally be expected from a model fit which correctly explains the data. We believe that this may be explained by the facts discovered by the investigation soon to be described in \cite{Kerins09etalinprep}. This work has found that the intrinsic noise in the data after the difference imaging process is $40\%$ higher than pure photon noise. This in turn implies that all $\chi^2$ and $\chi^2$/d.o.f. described within this thesis as fits to Angstrom data may effectively be divided by a factor of $1.4^2$, i.e. $1.96$. In the degenerate region, $2\sqrt{3}\beta t_E = 7.69\pm0.03$.
As $\beta$ changes, so does 
the lensing amplification, and hence the original flux required by the fit to reproduce the rise in flux $\delta F$ observed. 

\subsection{Adding in the Maidanak data}
\label{adding_Maidanak}
Early in 2008 the data from the Maidanak telescope were processed.
We had been aware for some time that these data fortuitously spanned the
event time of the ``short event'', and so being able to utilise these data
to inform the interpretation of this event and hopefully cut down the degeneracy of the model solutions described above was very important.

Unfortunately, the Maidanak data have only $40$ data points, spanning a little under $11$ days so it proved not possible to do a full combined Pac + Cosinusoid fit for all the data simultaneously, since the amplitude of the variable component related to the Maidanak data was hardly constrained at all by the data.

\begin{figure}[!ht]
\vspace*{11cm}
   \leavevmode
   \includegraphics{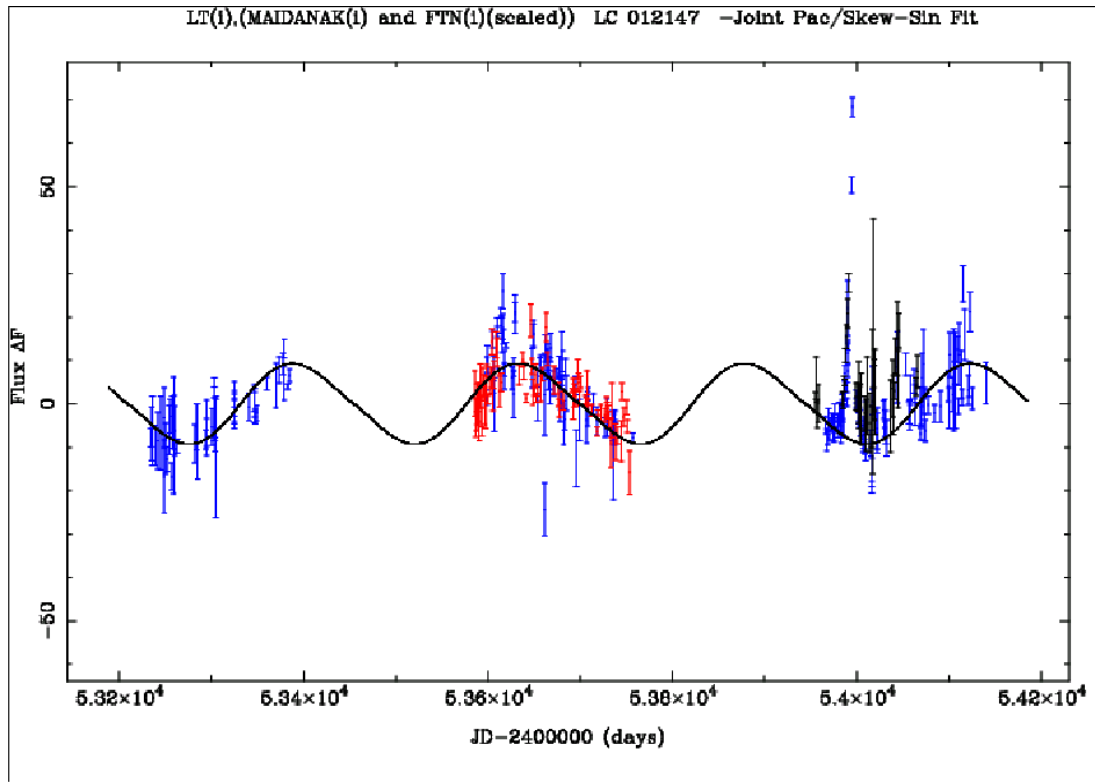}
\caption[The lightcurve of the short event using LT, FTN and Maidanak data.]{The lightcurve of the short event, using LT (Blue), FTN (Red) and Maidanak (Black) data. These are the raw data as they were before the process of scaling the Maidanak data to the other data had begun.
}
\label{short_event_three_bands_plus_variable_fit}
\end{figure}

Therefore, a different method was required.
First, the previously found (see description of earlier work above, and Table \ref{LT_FTN_variable_flux_parameters}) variable contribution to the LT and FTN data was subtracted from those two bands, to leave only the modelled microlensing components. As can be seen from Figure~\ref{short_event_three_bands_plus_variable_fit} this function was a good fit to the baseline variability. Nothing was subtracted from the Maidanak data at this stage for the reason given above.

\begin{table*}
\caption{Table showing the parameters of the modelled 
Variable component which was subtracted from the data before the lensing peak was modelled.}
\begin{center}
\begin{tabular}{|c|c|}
\hline
\hline
  Parameter    &  Parameter  \\
   Description &   Value  \\                     
\hline
  Phase $\phi_0$               &  $53148.02$\\
  Period $P$                   &  $244.99$\\
  Skewness $S$                 &  $-0.875$\\
  LT flux amplitude $A_1$      &  $-9.26$\\
  LT flux offset $\Delta{F_1}$ &  $7.44$\\
 FTN flux amplitude $A_2$      &  $-8.57$\\
 FTN flux offset $\Delta{F_2}$ &  $2.10$\\
           
\hline
\end{tabular}
\end{center}
\label{LT_FTN_variable_flux_parameters}
\end{table*}

\begin{figure}[!ht]
\vspace*{10cm}
   \leavevmode
   \includegraphics{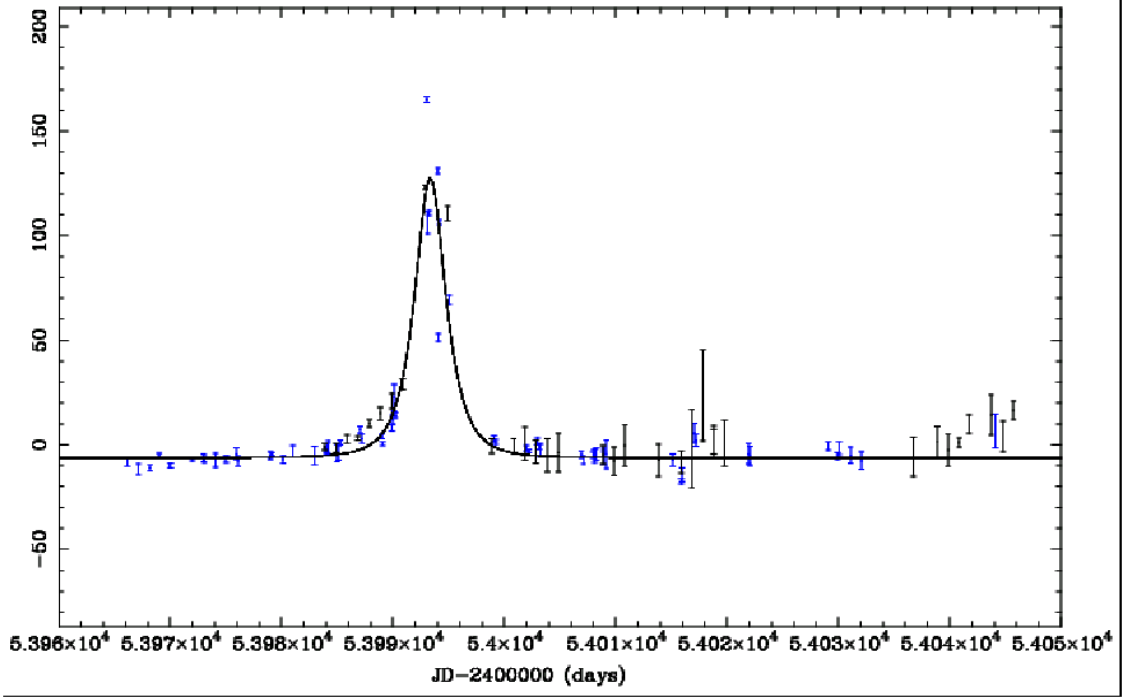}
\caption[A moderate zoom in on the flux peak region of the lightcurve of the short event, using using LT, FTN and Maidanak data.]{A moderate zoom in on the flux peak region of the lightcurve of the short event, using using LT, FTN and Maidanak data.
These are the raw data as they were before the process of scaling the Maidanak data to the other data had begun.
}
\label{short_event_mid_zoom_peak_before_fit}
\end{figure}

\begin{figure}[!ht]
\vspace*{10cm}
   \leavevmode
   \includegraphics{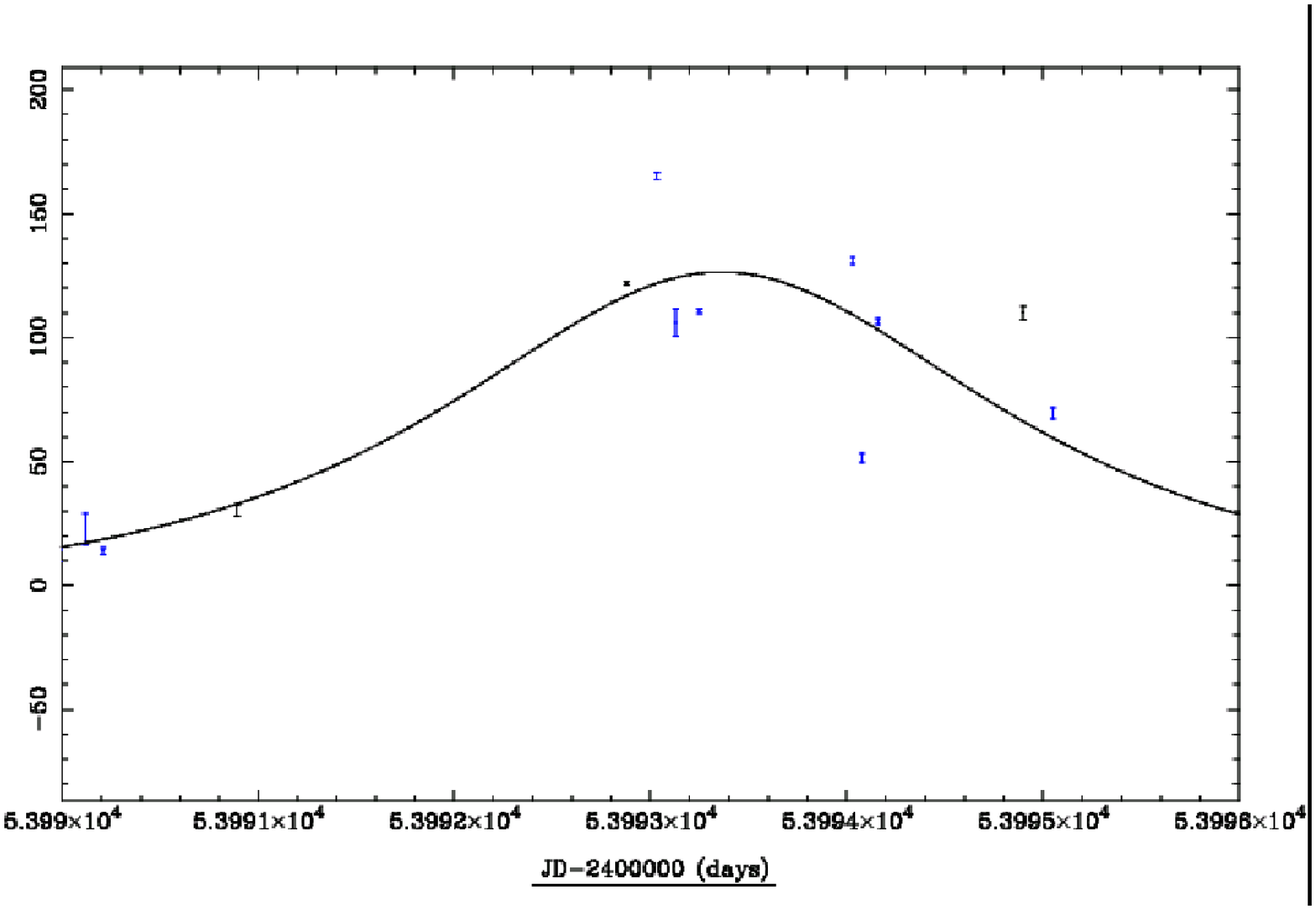}
\caption[Plot showing a close zoom in on the peak of the short event lightcurve using LT, FTN and Maidanak data, before the scaling factor or offset between the bands had been adjusted.]{Plot showing a close zoom in on the peak of the short event lightcurve, using LT, FTN and Maidanak data, before the scaling factor or offset between the bands had been adjusted.
}
\label{short_event_close_zoom_of_peak}
\end{figure}

Then the data were fitted with a full Paczy\'nski curve. The resulting amplitudes of the Paczy\'nski curves of the best fit were taken as an indication of the flux ratios between the bands.
As can be seen from Figure~\ref{short_event_mid_zoom_peak_before_fit} there was a clear excess flux on the left hand side of the Maidanak flux peak which was fitted in the first iteration, corresponding fairly convincingly to the downward slope with time of the currently un-subtracted variable component, as can be seen in Figure~\ref{short_event_three_bands_plus_variable_fit}. A closer zoom in on the central peak area is shown in Figure~\ref{short_event_close_zoom_of_peak}, which highlights the large scatter ($\gg$ the magnitude of the error bars) on the data points in and after the peak.
 This will be discussed further later in this Chapter.

  In order to make a first
order estimate of the unknown variable contribution in the Maidanak data, despite the short time span of those data, a constant fraction of the variable component function from the LT fit was subtracted from the Maidanak data, and the resulting flux re-fitted with the full Paczy\'nski curve. The constant fraction was iterated manually, attempting to minimise the value of $\chi^2$ of the fit. The minimum value of $\chi^2$ was found when the fraction was equal to $0.45$. No flux offset of the variable component was introduced as this would only introduce an offset 
into the Maidanak flux which would then be fitted by the Paczy\'nski curve fitting.
By eye from Figure~\ref{short_event_three_bands_plus_variable_fit}, though, any offset anyway appeared to be small, as the Maidanak and LT points in the steep rise lay almost on top of each other.
 The FTN data were also fitted to assist with the characterisation of the baseline flux.
As in the work above, the LT and FTN fluxes were linked, reducing the number of parameters by one.
 
The $\chi^2$ does not change very quickly with the chosen fraction and this is probably to be expected given the short timescale in the Maidanak data over which any change in the flux gradient present due to the variable component can act. Hence this must be considered a very preliminary result which would almost certainly change given a longer Maidanak time series.

Once the amount of variable component to subtract had been optimised, the resulting best fit parameters (with fitting errors) were as shown in Table \ref{pac_fit_with_maidanak_params}. However, when the errors in the LT data were scaled so that
$\chi^2$ for the fit to the whole lightcurve = $1.0$ 
the errors for the fitting parameters $\beta$ and $t_{E}$ became undefined, implying that there is insufficient information in the lightcurve to distinguish separate values for these two parameters. Therefore it seems we must satisfy ourselves with a value for $t_{\rm{FWHM}}$.
However, the new fitted values were similar to those found with the unscaled errors, namely, $t_{E}$ = $2.34$, $\beta$ = $0.748$.
If the combination 
$2\sqrt{3}\beta t_{E}$ is then formed from these values, then the
result is $t_{\rm{FWHM}}$ = $6.06$ days,
compared to the fit using unscaled errors which gave a value of $5.672$ days, and the $t_{\rm{FWHM}}$ found for the earliest joint fit without Maidanak data which was $0.711$ days. The later values found using the Maidanak data 
should be preferred over the fit to the earliest data, as adding the Maidanak data to the peak greatly increases the number of data points and average data quality over the flux peak, and should thus have improved the quality of the fit which was achieved.
Although the full Paczy\'nski fit above does find values for $t_{E}$ and $\beta$ separately which when combined give a reasonable value for $t_{\rm{FWHM}}$, although not completely consistent with the directly fitted $t_{\rm{FWHM}}$ values, it must still be strongly suspected that, in reality, these two parameters are degenerate, since the relatively high value of $\beta$ corresponds to an amplification of only $1.34$ in the high magnification regime, and there can be no stars in M31 with unlensed magnitude bright enough to have the measured peak magnitude of this event with only this magnification (see Section \ref{short_event_source_stars}).
Therefore it must be borne in mind that the errors quoted in Table \ref{pac_fit_with_maidanak_params} are merely the fitting errors calculated by MINUIT for the particular data points provided to it in the lightcurve, and do not show the true degeneracy and thus uncertainty of the $t_E$ and $\beta$ parameters. It might be expected that if it were possible to perform a Monte Carlo study in which many realisations of the statistical noise on each point in the lightcurve were selected and the fitting re-performed, that the individual parameters $t_E$ and $\beta$ would individually vary to a much greater extent than their fitting errors would suggest, but that their product would remain much more constant, as this is the actual well-determined physical quantity. 

\begin{table}
\caption{Table showing the fitting parameters for a Paczy\'nski fit for the ``short event'' using the Maidanak data.}
\begin{center}
\begin{tabular}{|l|l|l|l|}
\hline
\hline
Parameter &  Value & MINUIT   &  MINOS errors \\
  Name    &        & fitting error &               \\
\hline
  $t_{0}$   & $53993.346$ & $0.015$ & $\pm0.015$\\
  $t_E$     & $2.240$ & $^{+0.030}_{-0.024}$ & $\pm0.024$\\
  $\beta$   & $0.7310$ & $^{+0.012}_{-0.0024}$ & $\pm0.0024$\\ 
 $B_{LT}$   & $-6.24$ & $0.014$ & $\pm0.014$ \\
 $B_{FTN}$ & $-1.78$ & $0.17$ & $\pm0.17$\\
 $B_{Maid}$ & $+1.31$ & $^{+0.45}_{-0.46}$ & $\pm0.46$\\
 $\Delta F_{FTN,pac}$ & $196.4060$ & $^{-}_{5.71}$ & $\pm0.0037$\\
 $\Delta F_{Maid,pac}$ & $191.4$ & $^{+2.3}_{-5.7}$ & $\pm2.3$\\
\hline
\end{tabular}
\end{center}
\label{pac_fit_with_maidanak_params}
\end{table}

\begin{figure}[!ht]
\vspace*{10cm}
   \leavevmode
   \includegraphics{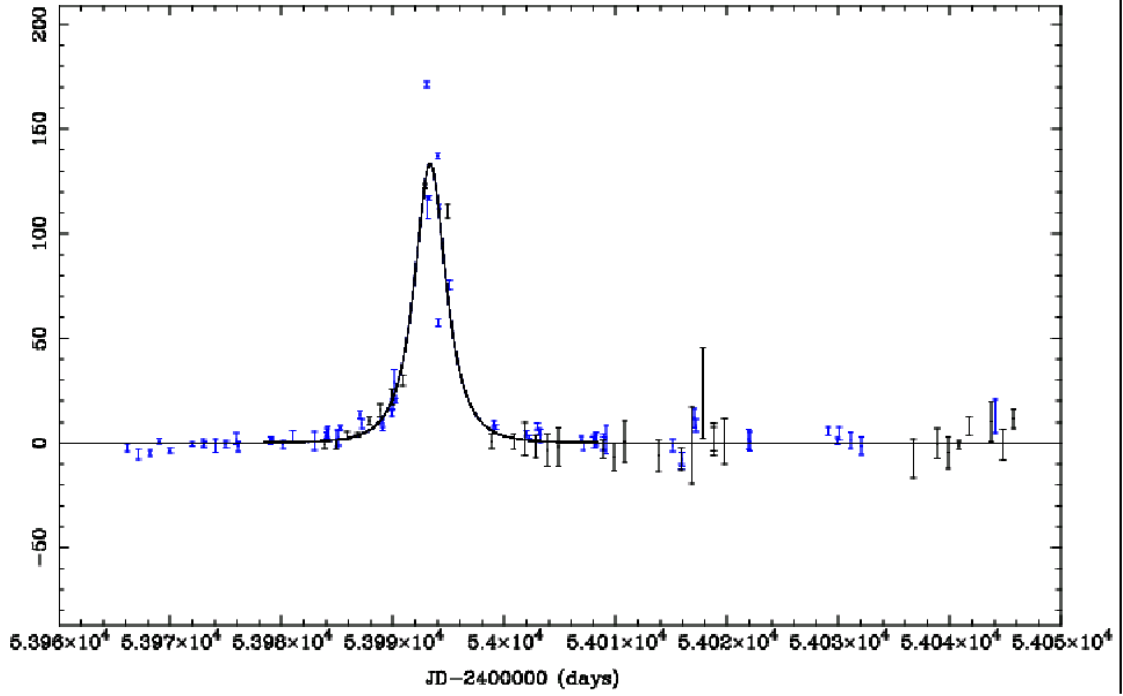}
\caption[Plot showing the peak region of the best full Paczy\'nski fit to LT plus Maidanak data, for the ``short event''.]{
Plot showing peak region of the best full Paczy\'nski fit to LT plus Maidanak data, for the ``short event''.}
\label{short_event_wide_rescaled_best_pac_fit}
\end{figure}

In order to display the Maidanak and LT data in a way that allows direct comparison of their fits to the model, the flux data were normalised to the LT flux
amplitude by subtracting the
(small) fitted offsets and multiplying the Maidanak data by the peak Paczy\'nski flux amplitude ratio between the Maidanak and LT, which
was equal to $1.11$. The re-scaled two band fit to the Paczy\'nski curve is shown in Figure~\ref{short_event_wide_rescaled_best_pac_fit}. A close-up of the peak region is shown in Figure~\ref{short_event_zoom_rescaled_best_pac_fit}.
The best fit to the Maidanak data only is shown below in Figure~\ref{maid_final_fit}

\begin{figure}[!ht]
\vspace*{10cm}
   \leavevmode
   \includegraphics{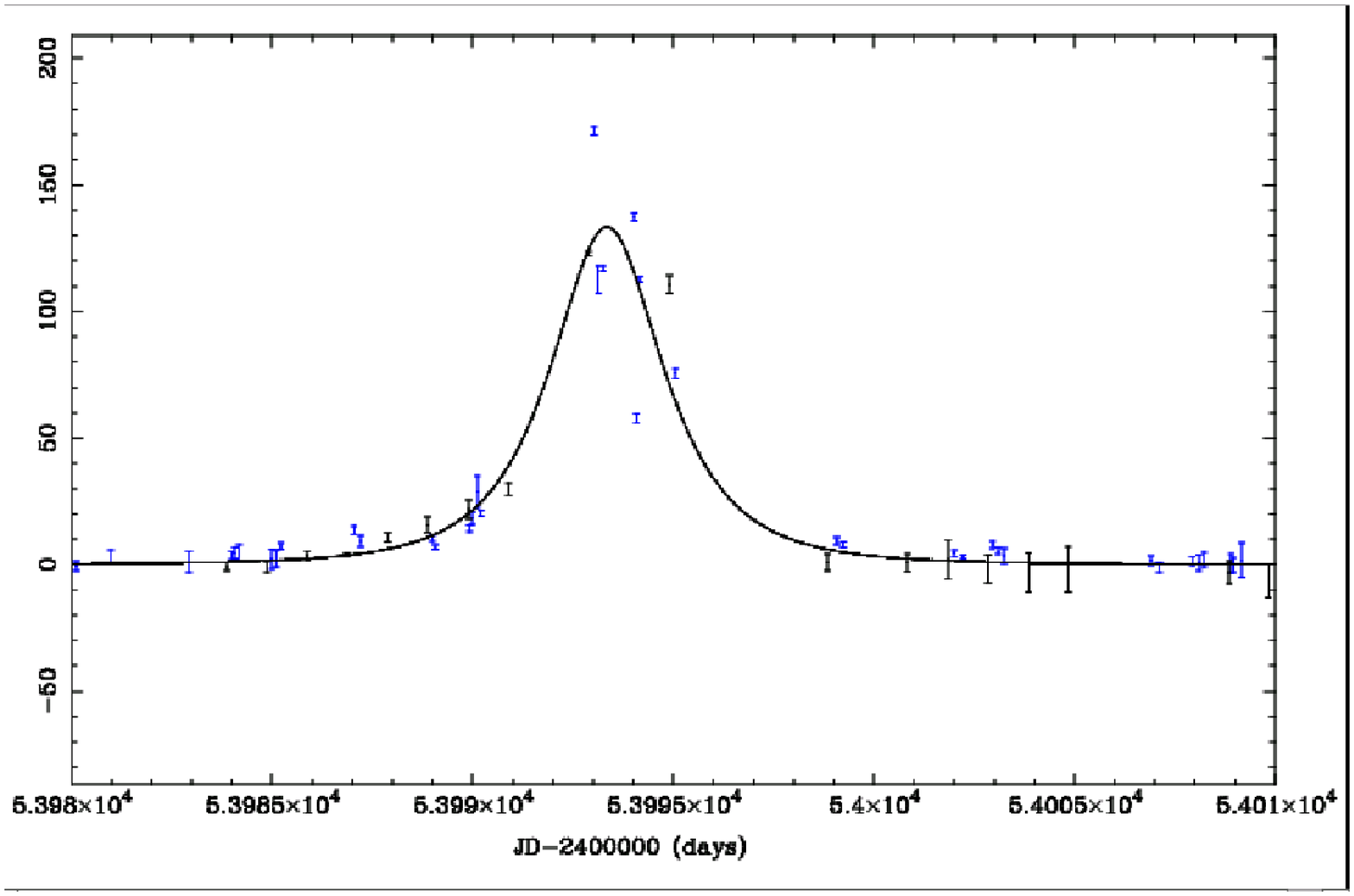}
\caption[Plot showing the central peak region of the best full Paczy\'nski fit to LT plus Maidanak data, for the ``short event''.]{Plot showing the central peak region of the best full Paczy\'nski fit to LT plus Maidanak data, for the ``short event''.
}
\label{short_event_zoom_rescaled_best_pac_fit}
\end{figure}

\begin{figure}[!ht]
\vspace*{11cm}
   \leavevmode
   \includegraphics{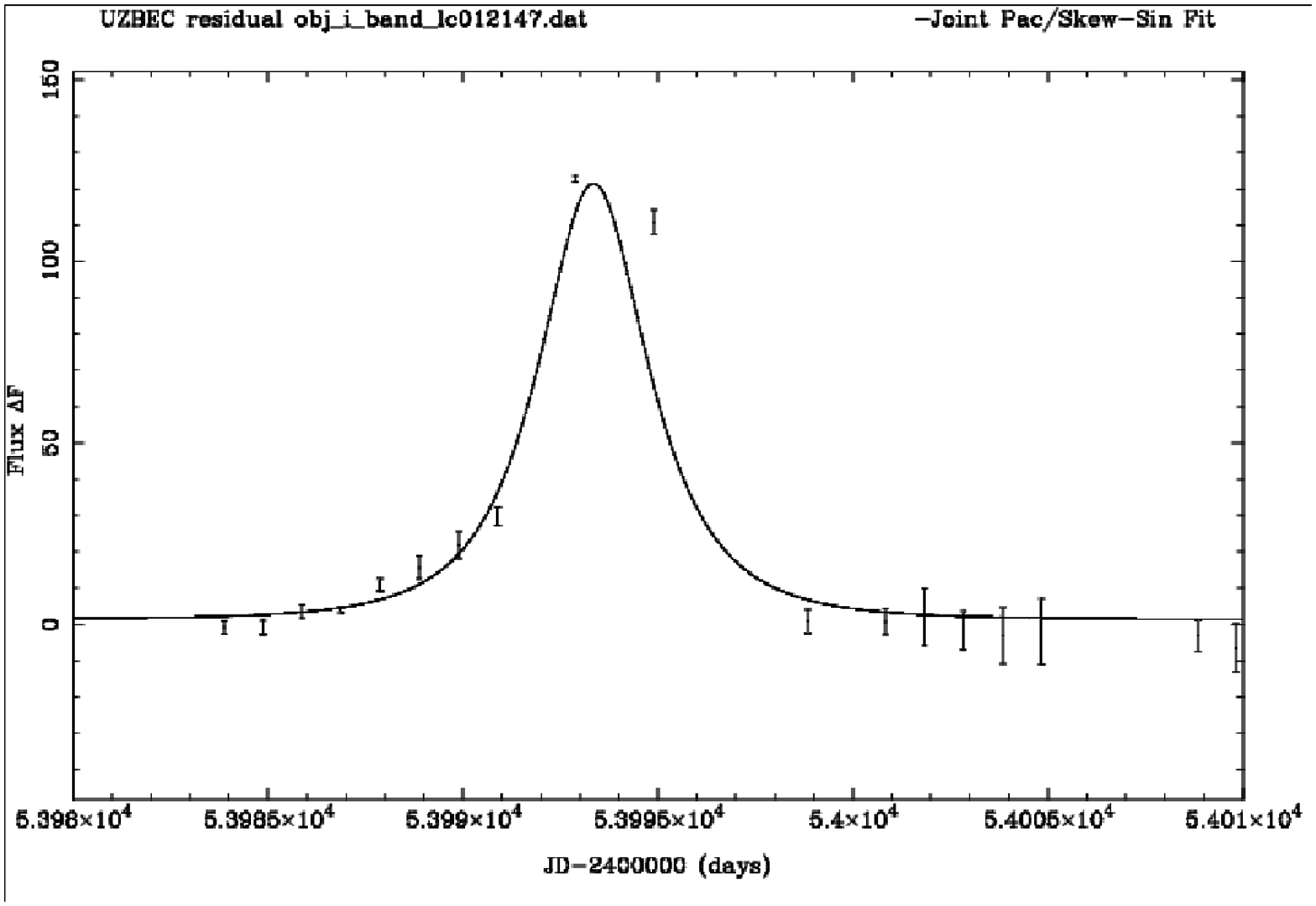}
\caption[Plot showing the best full Paczy\'nski fit to the Maidanak data only, for the ``short event''.]{
Plot showing the best full Paczy\'nski fit to the Maidanak data only, for the ``short event''.}
\label{maid_final_fit}
\end{figure}

\subsection{Rediscovery of the event in the $2008$ photometry}
\label{rediscovery_in_2008}

By searching for objects in the variable object database with the same position as the previously identified lightcurves, the equivalent lightcurves to the short event were re-found in the most recent photometry.

These lightcurves had not been selected by the candidate selection pipeline, and
more detailed investigation revealed that these lightcurves had failed the pipeline primarily on Cut 6 (bump sampling) due to not having any data points in the central region $t_0 \pm t_{\rm{FWHM}}/4$.
The global $\chi^2$/d.o.f. was found to be $3.39$, considerably better than before, and the local $\chi^2$/d.o.f. was found to be $5.40$, which, although still worse than the global fit is also much better than before.
As can be noted from the $\chi^2$/d.o.f. values above, which are clearly significantly larger than $\sim1$ and which were even larger in the data from earlier
iterations of the DIA pipeline, there still does seem to be an issue with either unexplained scatter in the data or incorrect estimation of the error bars. We believe there may be at least three possible explanations for this. Firstly, there could still be some poorly understood problem with the DIA process; for example a badly fitting kernel. Secondly, there really could be low level variations going on from
physically varying objects in the galaxy which have the effect of enhancing the ``noise'' to our fits. Thirdly, problems can be caused by poor background fits to the galaxy surface brightness. Interestingly, we are not the first microlensing survey to
have experienced stubborn $\chi^2$/d.o.f.$>1$. The OGLE early warning system has produced data which have $\chi^2$/d.o.f.$\sim2$ for bright events, which, although not as high as our data, is still significant given the amount of time they have had to develop and test their pipeline.
 This issue of high $\chi^2$ which we experience, which is also seen in my sample of 
variable stars, as shown by Figure \ref{reduced_chisq_dist_3160_LCs}, has been investigated more fully and the results will be described in \cite{Kerins09etalinprep}. It has been found that the intrinsic noise in the data after the difference imaging process is $40\%$ higher than pure photon noise. This implies that all $\chi^2$ and $\chi^2$/d.o.f. described within this thesis as fits to Angstrom data may effectively be divided by a factor of $1.4^2$, i.e. $1.96$. This is a very nice independent confirmation of the information separately and previously shown by Figure \ref{reduced_chisq_dist_3160_LCs}, whose peak was found to occur at $1.97\pm{0.19}$. Now that this is known, it can be seen that the local and global
$\chi^2$/d.o.f. above of $3.39$ and $5.40$ respectively, actually correspond to 
$\chi^2$/d.o.f. of about $1.72$ and $2.76$ which look much more like what what be expected for good fits to data. This issue has also affected the $\chi^2$/d.o.f.
cut levels used by the candidate selection pipeline, set at $5$ and $7$, which are now known to effectively correspond to $2.55$ and $3.57$ respectively.

\begin{figure}[!ht]
\vspace*{9cm}
$\begin{array}{c}
\vspace*{10cm}
   \leavevmode
 \includegraphics{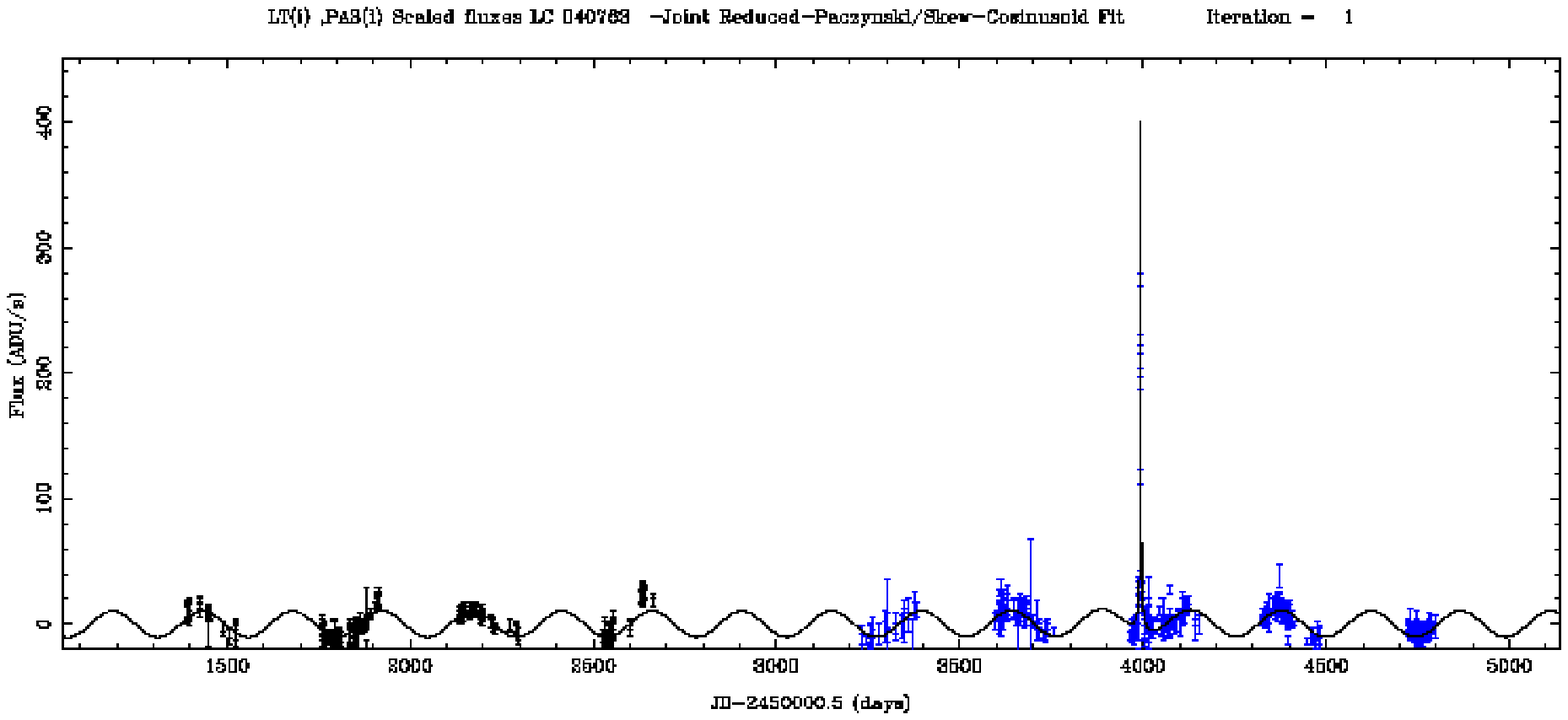} \\
\vspace*{0cm}
   \leavevmode
 \includegraphics{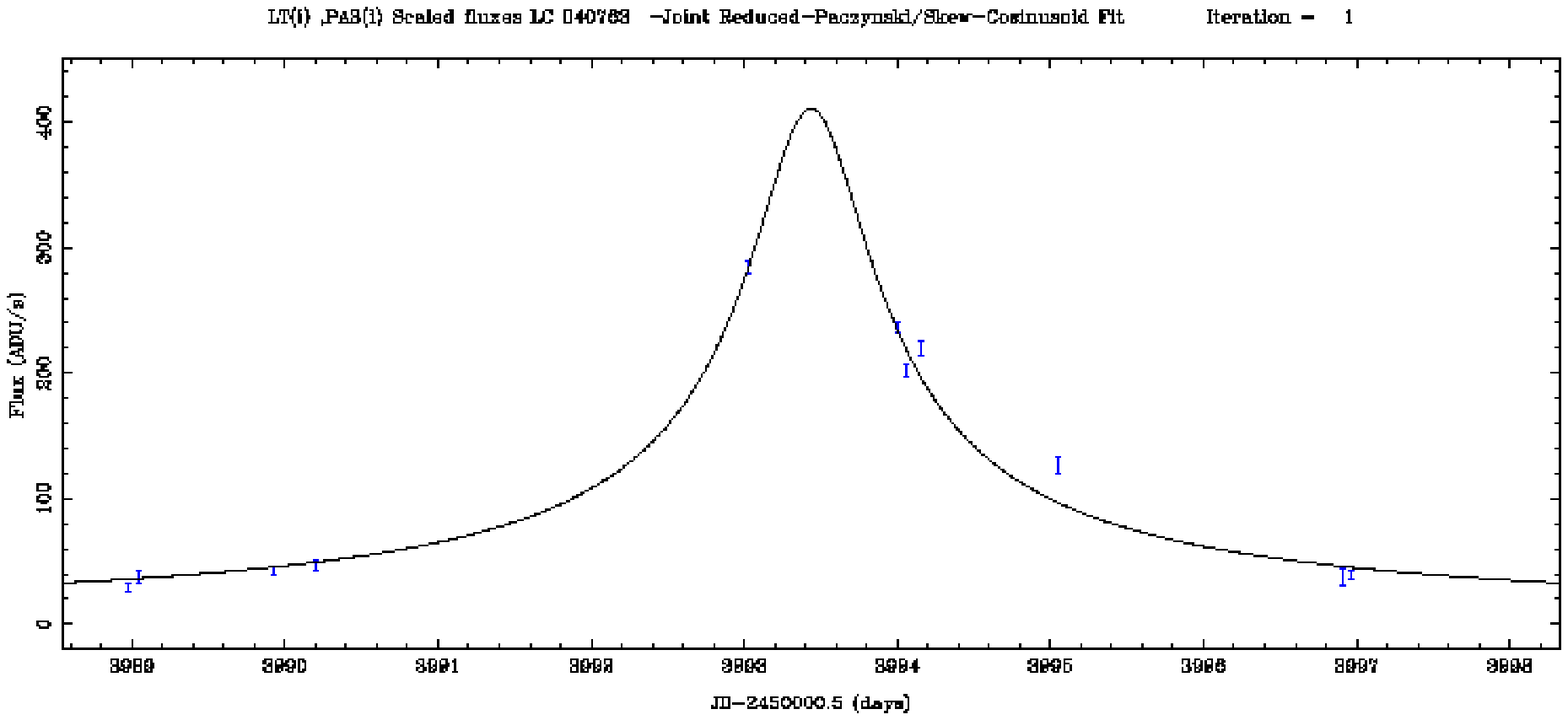} \\
\end{array}$
\caption[Plots showing the microlensing candidate ANG-06B-M31-01, rediscovered in the 2008 photometry.]{Plots showing the microlensing candidate ANG-06B-M31-01, rediscovered in the 2008 photometry as objects $40763$ and $45097$ a) The whole lightcurve, spanning $9 \frac{1}{2}$ years, with $8 \frac{1}{2}$ seasons of data, plotted with the full mixed lensing + variable fit b) The central peak region, showing the fit to the lensing part of the fit, after the variable component has been subtracted.}
 \label{short_event_2008_photom}
\end{figure}

The lightcurve shown with the best fits is shown in Figure \ref{short_event_2008_photom}. It can be seen from the second panel that the majority of the data points are well fitted by the model, but the two worst fitted points are consecutive and both on the right hand side of the peak.

The best fit parameters for the fit shown in Figure \ref{short_event_2008_photom} are shown in Table \ref{joint_fit_params_2008}. The amplitude of the Paczy\'nski component of the fit in the PA data is so large, and the errors undefined, as the fit is almost entirely unconstrained by the PA data.

It can be seen from Table \ref{joint_fit_params_2008} that the width $t_{\rm FWHM}$ of the peak is still found to be very narrow, being only $1.36$ days, compared to $0.71$ days with the earliest $2007$ data.
 This timescale would make this event the shortest known microlensing 
event found in M31 to date, the next shortest being $\sim1.8$ days reported by \cite{2001ApJ...553L.137A}.

\newpage
\clearpage
\clearpage
\newpage

 \begin{table}
\caption{Table giving the fitting parameters for the best joint (Paczy\'nski + skew cosinusoid) fit, for the ``short event'' in the $2008$ photometry.}
\begin{center}
\begin{tabular}{|l|l|l|l|}
\hline
\hline
Parameter &  Value & MINUIT   &  MINOS errors \\
  Name    &        & fitting error & (if calculated)\\
\hline
 $\phi_{0}$          & $54131.11$   & $0.99$ & $\pm0.99$\\
   $P$               & $245.15$   & $0.14$ & $\pm$\\
   $S$               & $1.00000008$   & - & $\pm4.35e^{-9}$\\
  $t_{0}$            & $53993.436$   & $0.01$ & $\pm0.01$\\
$t_{\rm FWHM}$       & $1.36$   & $^{+0.10}_{-0.11}$ & $\pm0.11$\\
 $\Delta F_{LT,sin}$ & $-10.42$   & $0.22$ & $\pm0.22$\\
 $B_{LT}$            & $-8.40$   & $0.22$ & $\pm0.22$\\
 $\Delta F_{LT,pac}$ & $411$   & $^{+24.0}_{-19.7}$ & $\pm21.7$\\
 $\Delta F_{PA,sin}$ & $-71.5$   & $1.39$ & $\pm1.39$\\
 $B_{PA}$            & $-30.6$   & $1.15$ & $\pm1.15$ \\
 $\Delta F_{PA,pac}$ & $37811$   & - & - \\
\hline
\end{tabular}
\end{center}
\label{joint_fit_params_2008}
\end{table}

\newpage

From the first panel of Figure \ref{short_event_2008_photom} it can be seen that the behaviour of the lightcurve which existed before the
sudden peak has returned in the period after it. The previously found variable fit is still consistent with the lightcurve. This return to the
previous baseline, although not ``flat'', along with the improved fit to a lensing lightcurve model must be taken as strengthened evidence in support of the microlensing interpretation of this event.

\subsection{The lens mass of ANG-06B-M31-01}

The radial velocities and in some cases velocity dispersion of different physical components of M31 such as planetary nebulae \citep{MerretH}, globular clusters \citep{Perret_et_al_2002}, or HII regions \citep{1970ApJ...159..379R} have been measured by several authors \citep{1999JRASC..93..175V}. The rotational velocity outside a radial distance of about $20$ kpc has been measured as $\sim260$ km s$^{-1}$, but considering the range of the LT field, which at $4.5^{\prime}$ corresponds almost exactly to $1$ kpc square at the distance of M31, most authors, for e.g. \citep{1970ApJ...159..379R,1980ApJ...235...30R,1984A&A...141..195B,1995A&A...301...68L},
show a consistent picture of the rotational velocity being $\sim \pm200$ km s$^{-1}$ at a radial distance of $1$ kpc, and then rapidly falling to zero in a linear fashion at the centre. The velocity dispersion, which may
increase the relative velocity of lens and source stars, has been measured for globular clusters as about $138\pm3 $ km s$^{-1}$ \citep{Perret_et_al_2002}. The actual relative velocity of the lens and source stars involved in this particular event cannot, of course, be precisely known, but it is interesting to consider, using average quantities, what
would be the expected mass range of the lens star, given the measured timescale above. The maximum possible expected relative velocity can, however, be estimated, as the rotation curve is flat out to large radial distance and so twice the maximum value of $\sim260$ km s$^{-1}$, plus a contribution from the velocity dispersion above, gives a value of the order of $650$ km s$^{-1}$. This should only be possible for a disk source on the far side of the galaxy and a disk lens on the near side. Due to the small angle of ($\sim13^{\circ}$) between the disk and our line of sight it is possible for objects which are both close to the centre of the galaxy, but have radii large enough to have the maximum rotational velocity, to have this large a relative velocity difference. 
For example, for a line of sight through the galactic centre, assuming radii for both source and lens of greater than $1.25$ kpc as assumed below (but in opposite directions), the height of both objects above the disk centre would have to be greater than $0.28$ kpc, compared to the disk scale height assumed below of $0.4$ kpc. Lines of sight to objects over most of the area of the galaxy will, however, have rotational velocities of source and lens in one direction only, and so will have a much smaller velocity difference.

Equation $15$ of \cite{1996ARA&A..34..419P} gives a relation, in practical units, between the relative velocity of source and lens stars and the mass of the lens, if the Einstein time is known.
It is (slightly rearranged for practical reasons):

\begin{equation}
\label{velocity_lens_mass}
t_{E} = 0.214 (\rm{yrs})\left(\frac{M}{M_{\odot}}\right)^{\frac{1}{2}}\left(\frac{D_l}{10 \rm{kpc}}\right)^{\frac{1}{2}}\left(\frac{D_s-D_l}{D_s}\right)^{\frac{1}{2}}\left(\frac{200 \rm{km s}^{-1}}{v_t}\right)
\end{equation}

where $D_l$ is the distance from the observer to the deflector (lens), $D_s$
is the distance from the observer to the source, $v_t$ is the relative transverse velocity 
of source and lens, and $t_E$ is the Einstein crossing time measured in years.

Equation \ref{velocity_lens_mass} can be easily rearranged to get the predicted lens mass in terms of the Einstein time in days. This results in the following
equation:

\begin{equation}
\label{velocity_lens_mass_rearranged}
\displaystyle{\frac{M}{M_{\odot}} =
\frac{{t_{E} (\rm{ days})}^2}{{0.214 \rm{ (years)}}^2} \frac{{v_t}^2}{({200\rm{ km} \rm{s}^{-1}})^2} {\left(\frac{1}{365.2422}\right)}^2\left(\frac{10 \rm{ kpc}}{D_l}\right){\frac{D_s}{D_s-D_l}}
}
\end{equation}

It is also necessary to make a connection between the value of $t_{E}$ which is required by the formula above, and $t_{\rm{FWHM}}$, which is fitted by the selection pipeline.
 It is almost certainly justified
to assume that this event is in the high magnification regime, and hence Equation \ref{high_mag_approx_2} applies.

The bulge was simulated both as in the ``exponential bulge'' and ``power law bulge'' model, as in \cite{2006MNRAS.365.1099K}. The bulge profile was therefore given by Equation \ref{exponential_bulge} for an exponential bulge,

\begin{equation}
\rho_{b} = \rho_{b,0}e^{-{[{\frac{x_b}{a}}^2 + {\frac{y_b}{qa}}^2 + {\frac{z_b}{qa}}^2]}^s}
\label{exponential_bulge}
\end{equation}

where $q$ parametrises the elongation of the bar and has the value $0.6$, and $a$ is the bulge scale length with the value $1$ kpc. The power law index, $s$, has the value $0.75$.
For the power law bulge model, the disk mass profile is given by Equation \ref{power_law_bulge}.
The power law index, $s$, has the value $3.0$. 

\begin{equation}
\rho_{b} = \rho_{b,0}{(1 + {{\frac{x_b}{a}}^2 + [{\frac{y_b}{qa}}^2 + {\frac{z_b}{qa}}^2}]^\frac{s}{2}})
\label{power_law_bulge}
\end{equation}

 The mass distribution of the disk was assumed to be the normal double exponential profile,

\begin{equation}
\rho_{d} = \rho_{d,0}e^{-|(Z/H)|}e^{(-R/\rm{h})}
\label{disk_mass_profile}
\end{equation}

where $H$ is the disk thickness scale length taken to be $0.4$ kpc and $h$ is the disk radial scale length, taken as $5.8$ kpc.

The distance to M31 was assumed to be $784$ kpc, as measured by \cite{1998ApJ...503L.131S}.

\subsubsection{Velocities perpendicular to the line of sight}

Rotational velocities of lenses and sources in the disk were assumed to be constant at $235$ km s$^{-1}$
outside a radius of $1.25$ parsec. Inside this radius, they are assumed to act as a solid body rotator, so the rotational velocity decreases in proportion to radius, reaching zero at the centre.
In the galactic frame, the coordinates of the ANG-06B-M31-01 in kpc are ($-0.025,-0.805$). Therefore, the line of sight penetrates the centre of the disk at a galactic radius of roughly $3.6$ kpc, which means that the orbital velocity takes its maximum value of $235$ km s$^{-1}$. In the plane of the sky, this also translates to a rotational speed of almost identical magnitude.

The magnitudes of the one dimensional velocity dispersions assumed for the M31 bulge and disk respectively are identical to those used in \cite{2006MNRAS.365.1099K}.
That is, where a ``heavy'' disk model was used, (Models 1,3 and 5),
$\sigma_{\rm{disk},1d} = 60$ km/s. For the ``light'' disk models 2,4 and 6, $\sigma_{\rm{disk},1d}$ was $\sqrt{3}$ lower.
Where a ``light'' exponential bulge model was used, (Models 3 and 4), this was taken as $\sigma_{\rm{bulge},1d} = 90$ km/s. For the ``heavy'' exponential bulge (Models 1 and 2), $\sigma$ was $\sqrt{3}$ higher, and for the power law bulge (Models 5 and 6) it was $\sqrt{2}$ higher. In all cases below, a $\sigma_{2d} = \sqrt{2}\sigma_{1d}$, i.e. an isotropic velocity dispersion is assumed.

For a disk lens and source, the rotational velocities of lens and source are assumed to cancel, leaving only the velocity spread of the disk, $\sigma_{\rm{disk}}$. The overall motion was
modelled using a summed velocity dispersion = $\sqrt{2}{\sigma_{\rm{disk},2d}}$.

For a bulge lens and source, there is no bulk rotation- only the random motion
described by the bulge, $\sigma_{\rm{bulge}}$. The overall motion was
modelled using a summed velocity dispersion = $\sqrt{2}{\sigma_{\rm{bulge},2d}}$

The magnitude of the typical transverse velocity and velocity spread are calculated by assuming an isotropic Maxwellian velocity distribution (such as that given in Equation $6$ of \cite{2006MNRAS.365.1099K}) and integrating over both the line of sight and radial angle directions, leaving only the dependence on the transverse speed $v_t$.

This dependence is shown in Equation \ref{maxwellian_speed_distribution}, where $\rho$
is the mass density.

\begin{equation}
 F(\rho,v_t) = v_t(\frac{\rho}{\sigma^2})exp{-\frac{{v_t}^2}{2{\sigma^2}}}
\label{maxwellian_speed_distribution}
\end{equation}

The typical magnitude of the velocity is taken as the mean speed of this distribution and the spread as its HWHM (half-width half maximum).

When a mix of disk lens and bulge source is modelled, or vice versa, both rotational and random motions exist and so these must be combined. This was done by adding a rotational velocity vector and a randomly oriented dispersion vector and then integrating the sum over the full range of angles of the dispersion vector to find the average total length. This magnitude can be written as $(v_{\rm{rot}}^2. + ((2\sigma_{\rm{total}})/\pi)^2)^{0.5}$. In this case the 2d velocity dispersions from both the bulge and the disk are combined to form the overall dispersion by $\sigma_{\rm{total}} = \sqrt({\sigma_{\rm{disk},2d}^2 + \sigma_{\rm{bulge},2d}^2})$.

Only velocity vectors are used (and not proper motion vectors) in the above calculation, as both lenses and sources are at approximately the same distance from us, and so the normal correction may be neglected.

The calculated typical speeds and spreads, which are used as the uncertainties in the speeds, are given in Table \ref{typical_speeds}. 

\begin{table}
\begin{center}
\begin{tabular}{|c|c|c|c|c|c|c|}
\hline
\hline
   &\multicolumn{6}{c|}{galaxy model} \\
\hline
   &    1   &    2   &    3    &   4   &    5   &    6 \\
\hline
DD & \small{$282\pm192$} & \small{$163\pm111$} & \small{$282\pm192$} & \small{$163\pm111$} & \small{$282\pm192$} & \small{$163\pm111$} \\
BD & \small{$294\pm278$} & \small{$281\pm241$} & \small{$271\pm212$} & \small{$257\pm161$} & \small{$283\pm247$} & \small{$269\pm205$} \\
DB & \small{$294\pm278$} & \small{$281\pm241$} & \small{$271\pm212$} & \small{$257\pm161$} & \small{$283\pm247$} & \small{$269\pm205$} \\
BB & \small{$366\pm250$} & \small{$366\pm250$} & \small{$211\pm144$} & \small{$211\pm144$} & \small{$299\pm204$} & \small{$299\pm204$} \\
\hline
\end{tabular}
\end{center}
\caption[Table giving typical relative lens-source speeds in the plane of the sky along with their estimated uncertainties. The values are calculated for 6 galaxy models and the 4 combinations of disk or bulge sources or lenses.]{Table giving typical relative lens-source speeds in the plane of the sky along with their estimated uncertainties. The values are calculated for 6 galaxy models and the 4 combinations of disk or bulge sources or lenses. Speeds are given in km/s.}
\label{typical_speeds}
\end{table}

\subsubsection{Source-lens separations}

The typical separation of lenses and sources $(D_{s} - D_{l})$ is estimated by calculating the mean lensing rate-weighted separation over the source and lens density distributions. This is done for all six galaxy models and the four combinations of lens and source locations; Disk-Disk, Disk-Bulge, Bulge-Disk and Bulge-Bulge.
Uncertainties in these values are taken as equal to the values themselves.

The values found are given in Table \ref{source_lens_distances}.

\begin{table}
\begin{center}
\begin{tabular}{|c|c|c|c|c|c|c|}
\hline
\hline
   &\multicolumn{6}{c|}{galaxy model} \\
\hline
   &    1   &    2   &    3    &   4   &    5   &    6 \\
\hline
DD & $1.5330$ & $1.5330$ & $1.5330$ & $1.5330$ & $1.5330$ & $1.5330$ \\
BD & $1.3648$ & $1.3648$ & $1.3648$ & $1.3648$ & $1.3809$ & $1.3809$ \\
DB & $1.3295$ & $1.3295$ & $1.3295$ & $1.3295$ & $1.3488$ & $1.3488$ \\
BB & $0.7666$ & $0.7666$ & $0.7666$ & $0.7666$ & $0.8533$ & $0.8533$ \\
\hline
\end{tabular}
\end{center}
\caption[Table giving the lensing rate-weighted mean values of the source-lens separation in the M31 bulge (kpc).]{Table giving the lensing rate-weighted mean values of the source-lens separation in the M31 bulge (kpc). The values are calculated for 6 galaxy models and the 4 combinations of disk or bulge sources or lenses.}
\label{source_lens_distances}
\end{table}

The values given in Table \ref{source_lens_distances} are equal for galaxy models 1-4 as these all use the exponential bulge model, with various scaling factors. Since the integral is effectively normalised by the scaling factors, only a change in the \emph{shape} of the bulge distribution has any effect on the final value. Hence the two power law bulge models 5 and 6 have slightly different values for those combinations involving the bulge.

\subsubsection{The magnitude of the event at peak}
\label{short_event_peak_magnitude}

Although there are no LT data points very close to the peak flux of this event,
it is possible to estimate the magnitude of the event for seven of the data points around the peak which are the only ones in which the object is a resolved source.

In order to estimate and calibrate these magnitudes, photometry of resolved sources within the Angstrom data was carried out on background subtracted data. The background of each image was estimated using the IRAF task {\tt rmedian} with an inner radius of $25$ pixels and an outer radius of $49$ pixels. Instrumental magnitudes were computed using the SExtractor \citep{1996A&AS..117..393B} package from within the STARLINK GAIA \cite{GAIA} software suite. Relative photometry was calculated between observations and was then calibrated against standard stars from \cite{1992AJ....104..340L}.
 By using the difference flux values of the same points to calibrate the fitted peak
 Paczy\'nski Flux, the peak magnitude of the event is estimated as $i = +18.07\pm0.06$.

\subsubsection{The range of likely source stars}
\label{short_event_source_stars}
In order to estimate the range of the parameter $\beta$ which is required to produce the magnification necessary to reproduce the measured brightness of this event, the
range of possible source stars was first considered.
 Examination of the colour-magnitude diagrams which form Figure 3 of
 \cite{2006MNRAS.365.1099K} shows that the upper end of the Main Sequence (MS) in M31
 has been modelled by us as ending at an R-band absolute magnitude of about $-1$,
 whereas the Red Giant Branch (RGB) ends at about a magnitude of $-2$. Correcting for
 the appropriate values of (R-I) colour, and for the distance to M31 leads to I-band
 magnitudes for the MS and RGB respectively of $m_{I,MS} = +23.47$,
$m_{I,RGB} = +21.87$. We choose this (as the highest value) as the upper limit.
Similar values may be obtained by examining the Colour-Magnitude (CM) diagrams which are Figure 6a of \cite{1995A&A...304...69P}, although the plotted $H_p$ (Hipparcos) broad band ($375$-$750$nm) filter magnitudes are only partially compatible with Sloan i' band, being closer to r' or g'.
At the lower stellar mass and magnitude end of the CM diagram, the stars which have a significant probability of being sources are limited both by the decreasing probability of the required high magnification event and also by peak magnifications beginning to be suppressed by the finite source effect, although the number of stellar sources increases markedly as the mass decreases. The finite source effect begins to become significant when the radius of the source star $R_s\sim\beta{R_E}$. Therefore, $\beta_{min}\sim R_s/R_E$, which is minimised for low $R_s$ and high $R_E$. For main sequence stars, $R_s\propto{M}^{0.8}$, and for low mass stars this changes to $R_s\sim{0.1 R_{\odot}}$, i.e. $R_s \cong \rm{constant}$. This means that $R_s$ does not get much below $0.1 R_{\odot}$.
Since the size of the Einstein ring is governed mainly by the lens mass and the source-lens separation (see Equation \ref{einsteinring}), if we assume a low mass star of mass $0.5 M_{\odot}$ and radius $0.1 R_{\odot}$ with a typical source-lens separation of $\sim1$ kpc, then the value of $\beta$ which gives the maximum magnification can be calculated to be $0.00023$. Since peak magnification $A \sim 1/\beta$ in the high magnification regime, this means that the maximum magnification for this star is $\sim4300$. One could of course get higher magnifications by raising the lens mass to increase $R_E$, but since the lensing rate scales as  $\beta{(m^{0.5})}\phi(m) \propto \beta{m^{-1.8}}$, where $\phi$ is the lens mass function, which decreases steeply towards higher masses. This means that if the lens mass is increased by a factor $100$, for example, the amplification only increases by a factor of $10$, but the chances of this event happening decrease by a factor of $\sim40000$.

 Therefore the lowest values of $\beta_{min}$ and hence the largest values of the magnification can be achieved somewhere near the bottom of the main sequence. Hence a reasonable value to choose for the lowest mass star to consider might be $M\sim0.5M_{\odot}$.
Knowing that the absolute magnitude of the sun is $M_I = +4.08$, (see Table 2.1, page 53, \cite{1998_Binney_and_Merrifield}), enables us to calculate that the magnitude of a solar mass star in M31 would be $M_I = 28.55$. Using the relationship $L\propto M^{3.5}$ which is approximately valid for main sequence stars enables us to extrapolate to the luminosity of a star of mass $0.5 M_{\odot}$, which is $L = 0.0884 L_{\odot}$. This equates to a magnitude difference of $2.634$, making its magnitude 
$M_I = +31.18$.
The magnifications required to achieve the peak magnitude of the event for the upper and lower limits of the above-described source star range are given by 
$A_{min} = 10^{0.4{-3.8}} = 33.1$ and $A_{max} =  10^{0.4{-13.11}} = 176035$.

Because of the argument detailed above, this value of $A_{max}$ never occurs in reality as finite source effects would dominate.
If a maximum value of $4336$ is assumed, then, reversing the above reasoning, the maximum magnitude difference that could occur would be $-9.09$. This in turn would give an unmagnified magnitude for the source star of $27.16$, slightly brighter than a solar mass star, in fact, equivalent to a star of $1.44 M_{\odot}$. This assumed lower limit on the magnification means that the minimum likely value of $\beta$ is $\sim0.00023$.

Using the inverse of Equation \ref{magnification}, which is, for the peak flux, given by Equation \ref{magnification_inverse} for high magnifications,

\begin{equation}
\label{magnification_inverse}
\beta = \sqrt{2}{\left[A\left(A^{2} -1 \right)^{-\frac{1}{2}} - 1  \right]}^\frac{1}{2}
 \simeq 1/A
\end{equation} 

the upper limit of $\beta$ is calculated using the lower limit on the magnification given above. $\beta$ is found to range from 
roughly $0.0001$ to $0.0302$, and so the expected value is simply assumed to be
$0.015 \pm 0.015$, although the actual number may not in fact be central to its range.

Due to the remaining uncertainty in the fitted timescale of the event $t_{\rm{FWHM}}$,
its value was taken as equal to $2.1\pm0.8$ days. Combined with the central value of $\beta$,
this gives $t_E = 41\pm{+44}$ days.
This is
formally consistent with zero, although we know the Einstein time must be a
positive quantity. Therefore we can only conclude that, given the uncertainties in the method we have been forced to adopt, we cannot make any quantitative statement about the Einstein time, other than it is likely to be $t_E < 85$ days.

\subsubsection{Estimated uncertainties}

The fractional error in the mass is calculated by propagating the errors in the individual variables, and by including the variations in calculated masses between galaxy models.

\begin{equation}
\displaystyle{\frac{\Delta M}{M} \cong \sqrt{ {2\left(\frac{\Delta t_{E}}{t_{E}}\right)}^2 + {\left(\frac{\Delta D_d}{D_d}\right)}^2 +{\left(\frac{\Delta D_s}{D_s}\right)}^2 +{\left(\frac{\Delta (D_s - D_l)}{(D_s - D_l)}\right)}^2 + {2\left(\frac{\Delta v}{v}\right)}^2 } }
\label{velocity_lens_mass_uncertainty_estimate} 
\end{equation}

where

\begin{equation}
\displaystyle{\frac{\Delta t_{E}}{t_{E}} \cong \sqrt{ {\left(\frac{\Delta{t_{\rm{FWHM}}}}{t_{\rm{FWHM}}}\right)}^2 + {\left(\frac{\Delta {\beta}}{\beta}\right)}^2
} }
\label{t_E_uncertainty_estimate}
\end{equation}

\subsubsection{Mass estimate results}

The results of the mass estimate described above are summarised in Table
\ref{mass_estimate_results}.
 The fractional errors are calculated using only the first order approximation and so should only be taken as indicative of the magnitude of uncertainty, but generally range between $2.6$ and $2.9$.

\begin{table}
\begin{center}
\begin{tabular}{|c|c|c|c|c|c|c|c|c|}
\hline
\hline
   &\multicolumn{6}{|c|}{galaxy model}& & \\
\hline
   &    1   &    2   &    3    &   4   &    5   &    6  & Mean Mass & S.D. \\
\hline
DD & $3.50$ & $1.25$ & $3.50$ & $1.25$ & $3.50$ & $1.25$ & $2.30$ & $1.25$ \\
BD & $4.25$ & $3.75$ & $3.50$ & $3.25$ & $3.75$ & $3.50$ & $3.73$ & $0.35$ \\
DB & $4.25$ & $4.00$ & $3.75$ & $3.25$ & $4.00$ & $3.50$ & $3.80$ & $0.38$ \\
BB & $11.75$ & $11.75$ & $4.00$ & $4.00$ & $7.00$ & $7.00$ & $7.50$ & $3.50$ \\
\hline
\end{tabular}
\end{center}
\caption[Table giving the estimates of lens mass for ANG-06B-M31-01.]{Table giving the estimates of lens mass for ANG-06B-M31-01 in units of the solar mass. The results are given for the 4 combinations of source/lens location and for 6 galaxy models.}
\label{mass_estimate_results}
\end{table}

As can be seen from Table \ref{mass_estimate_results}, the predicted masses are generally in the giant star range, but with a large uncertainty. The largest contribution to the uncertainty comes from the uncertainty in $\beta$, which in turn comes from the large range of possible source stars, given our lack of colour information for this event.
Utilising the calculations of lensing rate distributions plotted in Figure 5 of \cite{2006MNRAS.365.1099K} as a guide as to which of the mass estimates above are more likely in practice, it may be seen that the innermost contour for disk lenses is $0.2$ events per year per square arcminute, whereas for bulge lenses it is $2.4$. A similar ratio would be expected if the rates were divided by the location of the source. Therefore the bulge lens, bulge source estimate is the most probable, followed by the mixed estimates, and lastly the disk - disk estimate. This also increases the chances of the mass of the lens being at the top end of the estimates in Table \ref{mass_estimate_results}. Therefore, if the variations in the results for different galaxy models are folded in with the errors in the individual mass estimates, the most likely mass estimate is $7.5\pm{21} M_{\odot}$. This is formally consistent with zero, but the mass cannot be less than zero. So to first order, we can only conclude that 
the lens mass is likely to be less than $28.5 M_{\odot}$. Since almost all stars have masses less than this, this is a very weak constraint. The uncertainty of the lens mass estimate is mostly due to the large uncertainties in both the true values of $\beta$ and the relative lens-source velocity. 

     \section{Some interesting ``near misses'' in the 2007 data\label{near_misses}}

\begin{figure}[!ht]
\vspace*{7.3cm}
   \leavevmode
   \includegraphics{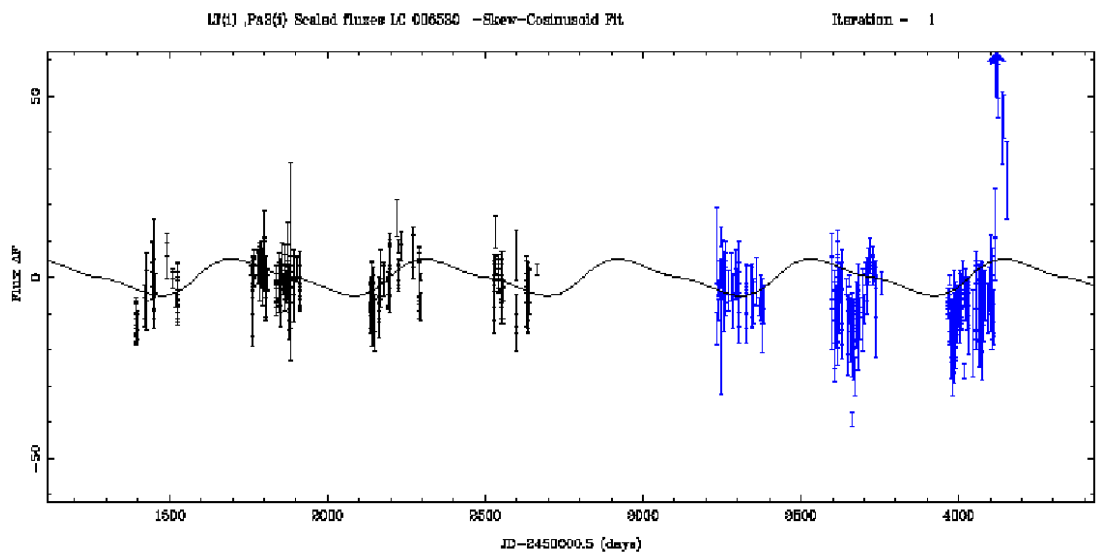}
\caption[Plot showing the first iteration periodic variable fit to lightcurve 6530.]{Plot showing the first iteration periodic variable fit to lightcurve 6530.
}
\label{LC_6530_variable_baseline_fit}
\end{figure}

 While scanning through files looking for periodic variables which could be used
 to estimate the flux ratio between the PA and LT data (see Chapter \ref{Chapter_4})
 two lightcurves, close together in lightcurve number, were noticed. Both had clear
 coherent ``spike''-like deviations in their lightcurves, and did not appear to be
 the normal type of lightcurve expected to be classified as ``variables''. The first
 of these was LC 6530 (see Figure \ref{LC_6530_variable_baseline_fit}).
 The selection pipeline merely classified this lightcurve as a 
``variable'', with no mention in any of the first iteration performed of even attempting to fit a Paczy\'nski or mixed Paczy\'nski + Variable model. This seemed strange, given how clear was the short duration flux variation at the end of the third season of LT data. Therefore, further investigations were carried out, to establish the reasons for this.
 In the first iteration, no mixed fit was performed, due to the lightcurve failing the criterion required by Cut 2. This was because 
 the long low bump, containing many data points,
spanning time points $3684.97$ to $3974.06$ (JD-$50000$) had a lower, (but not much lower) accumulated $\chi^2$ than the bump which is the most obvious in flux, spanning $4106.96$ to $4153.86$ (JD-$50000$). A major contributing factor to this was that the earlier wide bump contained $34$ data points, while the later sharp bump only contained $13$.
 This fact, and the observation that it would clearly be
desirable to attempt to fit a mixed fit to lightcurves containing this kind of obvious spike and a variable baseline, meant that the criterion to decide whether the biggest bump was significantly larger than the next was changed to use the quantity $\chi^2_{\rm{bump}}/n_p$ where $n_p$ is the number of data points in the bump, as defined by the ordinary bumpfind routine. Thus a long low bump which has the same integral underneath it as a short high bump would no longer be ``equivalent'' in the eyes of this criterion. Always bumps which accumulate $\chi^2$ at a greater rate would be preferred. Using this new criterion, a mixed fit was indeed fitted by the pipeline, and produced $\chi^2$ $\sim7900$ compared to that of the pure variable fit of $\sim8200$.

\begin{figure}[!ht]
\vspace*{5cm}
$\begin{array}{c}
\vspace*{7cm}
   \leavevmode
 \includegraphics{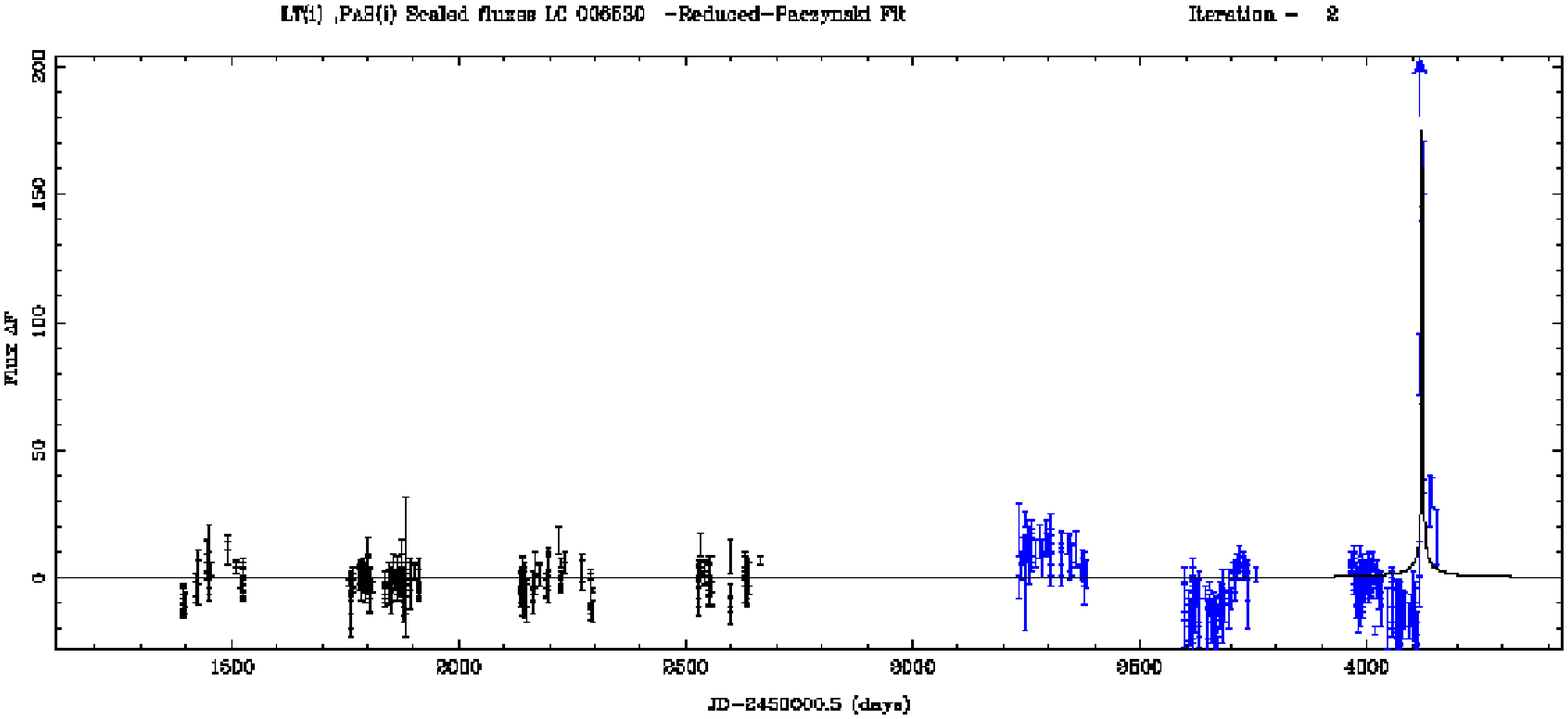} \\
\vspace*{0cm}
   \leavevmode
 \includegraphics{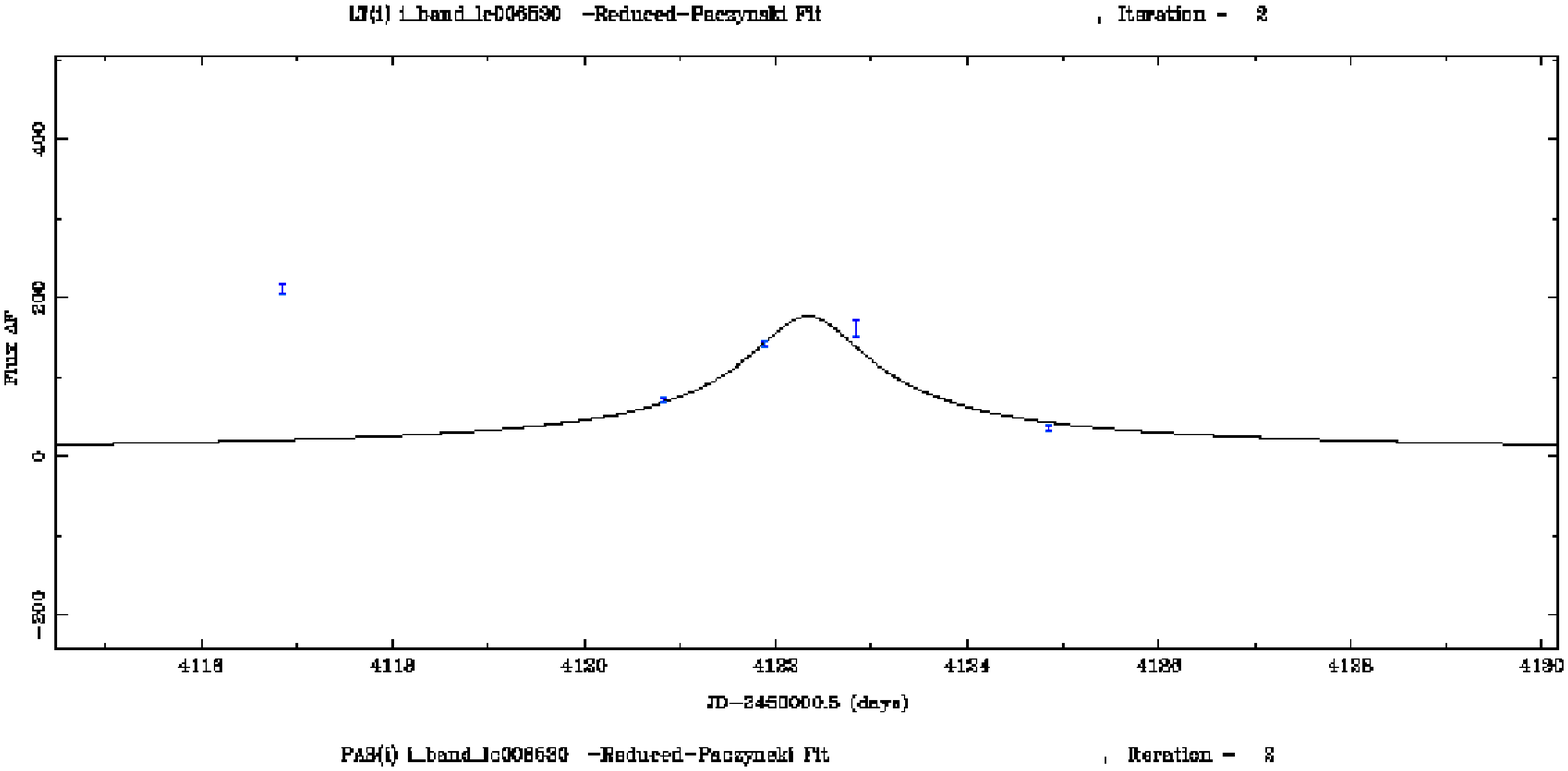} \\
\end{array}$
\caption[Plot showing two different aspects of the lightcurve $6530$ in the 2007 photometry.]{Plot showing two different aspects of the lightcurve $6530$ in the 2007 photometry: a) the best second iteration reduced Paczy\'nski fit to lightcurve 6530, after the
first variable component has been subtracted in the first iteration b) the central Paczy\'nski peak region of the second iteration fit.}
\label{two_06530_plots}
\end{figure}
\newpage

However, the final classification was still as a variable.
 Further investigation showed that the $t_0$ of the best fit Paczy\'nski bump was being placed somewhere in the PA data, not where naively expected in the flux bump at the end of the LT data. It was presumed that this is because a single period sinusoid is not sufficient to correctly model the baseline variation of this lightcurve, as indeed can be seen from Figure~\ref{LC_6530_variable_baseline_fit}, especially in the LT data. In this situation, the $\chi^2$ minimisation routine will still try do its work and
has found that a Paczy\'nski bump can more effectively be used to match to one of the remaining variable bumps and hence reduce the size of its residual, rather than
reducing the peak in the LT. In confirmation of the observation that a single sinusoid is insufficient to model this lightcurve, the output of the periodogram was noted to have detected two clear periods; of $150.6$ days and $688.2$ days.
 As a knock-on effect of this unexpected placement of the Paczy\'nski component in 
time, the fitted ratio between the PA and LT variable component flux amplitudes was
found to be only $1.68$, which caused the lightcurve to fail the ``unrealistic flux ratio'' condition which in this case had a lower limit of $2.63$, and therefore be downgraded in classification to a ``mere'' variable.
Given that the baseline characteristics of this lightcurve are more complicated than the most complex model available in the pipeline to attempt to fit it, it was concluded that the behaviour of the pipeline in the first iteration was not in fact incorrect and was to be expected. To model this lightcurve correctly, it would be necessary to use the ``normal'' bumpfind routine, which is able to find more than one bump, on second and subsequent iterations of the pipeline. This is because the
``large-$f_{dif}$ cluster'' routine which by default is used to evaluate the criterion for Cut 1 after the first iteration
is only capable of finding the one largest deviation. Therefore is it normally not possible to find a variable on the first iteration and a mixed fit on the second,
as this requires more than one bump to be found on the second iteration. It would be better if it were possible to find a mixed fit on the first iteration and then
deal with any remaining variability on the second iteration, but this is apparently not possible within the current framework without intervening to force the Paczy\'nski peak to be within the range of the short flux spike, even if it meant that the resulting solution was not the lowest $\chi^2$ one that could be found. This has not been implemented, as it has been a principle in constructing this pipeline to allow the fitting to find its own fits as much as possible and not to attempt to influence the results through more complicated methods which might not themselves always work in every situation. Fitting only the sinusoidal variable on the first iteration, as was finally chosen, is not ideal either, as it can be seen from 
Figure~\ref{LC_6530_variable_baseline_fit} that this fit has been skewed, especially in the LT, by the influence of the high flux spike.
 The option to use the first iteration bumpfind routine instead has been included as an option in the pipeline, but is not used by default as it was felt that a combination of two variable components and a Paczy\'nski component would be too
 flexible and could fit many different lightcurves, without necessarily being the
 correct model. As a tool to be used in individual cases, (such as this) where
 inspection by eye can be used as a guide, then perhaps this technique could be of
 some use.

  On the second iteration, having removed some of the baseline variation by 
subtracting the first variable component, the pipeline did indeed manage to find the expected 
simple Paczy\'nski fit to the flux bump at the end of the LT data. This had a global (whole lightcurve) $\chi^2$/d.o.f. of $8.66$, but a local (just the central peak) $\chi^2$ of $1200.8$! As can be seen from the second panel of Figure~\ref{two_06530_plots}, most of the points in the central region of the peak are a surprisingly good fit to the model with the exception of the point at time $4116.85$ which is $31.4$ times its own error bar above the model fit, and therefore contributes $986.7$ out of the total
$\chi^2$ of $1200.8$. In addition, if one looks at the first panel of Figure~\ref{two_06530_plots}, there are several points to the right of the main peak which are clearly above the model. 
Taken together, the three points of evidence that 1) There is a highly discrepant
point which is earlier than the fitted peak, 2) The flux appears, from the few data points available to have at least two bumps and 3) The part of the declining
lightcurve covered by the end of the data is clearly above the model, point towards the best explanation for this lightcurve being a classical nova. Clearly this would have been easier to determine if the event had not been so close to the end of the season and there had been more points in the tail of the lightcurve.
 However, if this event was indeed a nova, then it certainly declined very rapidly 
indeed. Some idea of the timescale can be gained by the best fitting Paczy\'nski curve FWHM timescale $t_{\rm{FWHM}} = 2.17$ days. 
   For completeness, the second iteration Paczy\'nski fit also failed two more of 
the pipeline cuts in addition to the local $\chi^2$ one.
Firstly, it failed the bump time sampling cut as only $4$ data points are within
$\pm2t_{\rm{FWHM}}$ of $t_0$, and, for related reasons, failed 
Cut 9, as the significance of the peak was shared by too few ($<5$) data points.

\section{Results of Pipeline} 
\label{results}

\subsection{Variable stars}
\label{variables}

All the analysis of variable star candidates in this section was made using the data from the $2007$ photometry.
As mentioned previously, two versions of the pipeline were maintained, one in which a constant ratio was maintained between the flux amplitudes of the cosinusoidal function in the PA and in the LT and FTN data fits, and another in which this ratio was allowed to vary freely as the PA and LT and/or FTN cosinusoid flux amplitudes were all independently fitted variables.

 Experimentation during the writing of these codes had confirmed that the use of a greater number of fitting variables, and/or more data points caused the fitting routine to run slower. Also, due to the larger dimensional space which must be covered by the minimisation routine, more parameters lead to it being more difficult for the routine to find the global minimum in $\chi^2$ space. In other words, the efficiency of the fitting routines at finding the best solutions decreases as the number of fitting parameters increases.
  When using the freely varying flux amplitude version the solution found, if it is the global minimum, will represent the true lowest $\chi^2$ and hence the true best fitted flux ratio as opposed to fixed flux ratios which will usually only find solutions which are close to the best.
 So, both versions of the code have their own advantages.

Therefore, to maximise both the number of variables found and the accuracy of the minimum $\chi^2$ solution for a given lightcurve and hence the accuracy of the fitted flux ratio,
several runs of the pipeline were performed using both versions of the code, and several different fixed flux ratios. In total, four runs were used to produce the results described in this section, as well as the results for microlensing for the analysis of the $2007$ photometry data. Three of these were fixed flux ratio runs, using values of the ratio designed to span the most probable range of this quantity found from the work in Section \ref{variable_flux_ratios} and plotted in Figure \ref{mean_flux_ratio_with_fits}. After the runs of the main pipeline had been made,
 the lightcurves which were classified as either ``variable'' or ``mixed'' were selected and 
the fitted flux ratio calculated. Information about the other fitted parameters such as period and skewness was collated from the output files of the main pipeline runs and then the data were outputted to files, binned according to LT flux amplitude, with the width of each bin being equivalent to 1 ADU/s, in a similar way to Section \ref{variable_flux_ratios}.
Some details of the exact operation of the variable lightcurve selection scripts will
now be described.
The first selection performed on lightcurves by the codes was on their classification as either ``variable'' or ``mixed'' by the main pipeline.
It is an important point to note that for ``mixed'' events it would not have been desirable while specifically investigating the properties of variable stars to have accepted lightcurves which had a significant modelled microlensing component as this might have skewed the modelling of the variable component parameters. Therefore only ``mixed'' events which had failed the main pipeline cut on lensing signal to noise were chosen for this part of the work. Therefore any contribution
from the microlensing component to the fit was small.
No specification was made at this stage on the signal to noise of the variable component, as one of the things to be investigated was the distribution of fitted flux amplitudes. Therefore even lightcurves with very small fitted amplitudes (which had a higher chance of being spurious) were admitted. As will be seen later, these very low signal to noise lightcurves turned out to be rare, (as also seen in Section \ref{variable_flux_ratios}), so their contribution to distributions of other fitting parameters was minor.
The next selection cut made was to require the reduced $\chi^2$ of the fit to be less than $15$ as in the main pipeline (Cut 5). This was deliberately
chosen to be a loose cut, as a good range for the later investigation of the distribution of reduced $\chi^2$ among the selected lightcurves
would be required. It would always be possible to impose a stricter reduced $\chi^2$ cut at a later stage, if required. 
As in Section \ref{variable_flux_ratios}, a cut was made on the number of data points in each data band, also using the same cut values as previously, namely:
 1) total number of points $> 250$, 2) PA points $> 80$.
In the case of freely fitted flux amplitudes (and hence flux ratios between PA data and other bands), a cut was made that the flux ratio
should be $\leq 100.0$, which was designed to cut out the very small number of
attempted fits to extremely noisy data, or where the lensing and variable component amplitudes had derived abnormally large values
(perhaps largely cancelling one another out in magnitude to fit a particular lightcurve).

 Events rejected by the main pipeline as having too long a period and too low a lensing signal to noise were allowed through into the sample of ``mixed'' events to allow fair comparison between ``variable'' and ``mixed'' parameter distributions. It was good to note from the point of view of lensing selection that several of the newly allowed lightcurves did indeed form part of the extreme outlier tail, for example in their extremely large values of period and flux amplitude, although others did lie in the body of the main distribution.
As shown in Table \ref{first_level_groupings} the numbers involved in this change turned out to be small.

 Four versions of this script were also made, each designed to select out a different set of
lightcurves, namely those where only LT or FTN data existed, or where PA data also existed, or lightcurves classified as ``mixed'' or ``variable''. Hence $16$ (=$4$x$4$) separate groups of lightcurves were initially maintained to allow the similarities or differences in the statistics of the fitted
parameters in these different groups to be investigated. The reason for keeping the lightcurves
with differing combinations of data bands separate was to investigate whether, and if so how, the extra baseline length gained by adding PA data changed the periods that were found.

The numbers of lightcurves selected in each of the $16$ groups are shown below in Table \ref{sixteen_groups}. The numbers do not vary with changing fixed flux ratio when there are only LT data, as in this case the flux ratio between PA and LT (for example) is irrelevant to the fitting.

\begin{table}
\caption{Table showing the number of lightcurves classed as either ``variable'' or 
``mixed-(low lensing component)'' for the four selection pipeline runs, subdivided into
telescope data source groups.
}
\small
\label{sixteen_groups}
\begin{center}
\begin{tabular}{|c|c|c|c|c|c|}
\hline
\hline
   & Ratio  &\multicolumn{2}{c|}{\footnotesize{LT and/or FTN data}}    &\multicolumn{2}{c|}{\footnotesize{PA+LT and/or FTN data}}  \\
\hline
   & \footnotesize{PA/LT,PA/FTN}  &     Variable     &      Mixed     &      Variable     &      Mixed     \\
\hline
\multirow{3}{*}{\footnotesize{Fixed Flux ratio}}      & $7.89,8.53$  & $792$ & $372$ & $303$ & $700$ \\
                                       & $11.10,12.00$ & $792$ & $372$ & $303$ & $736$ \\
                                       & $14.80,16.00$ & $792$ & $372$ & $304$ & $728$ \\
\hline
\multirow{2}{*}{\footnotesize{Variable Flux ratio}}     & $-$         & $651$ & $258$ & $187$ & $444$ \\
                                        &             &       &       &       &       \\
\hline
\end{tabular}
\end{center}
\end{table}
\normalsize

To allow visual comparison between the $16$ sub-divisions, ((LT+FTN) data only or (LT+FTN+PA) data, Variable or Mixed fit, Fixed flux amplitude ratio low, medium, high or unfixed) and to establish that combining the groups did not destroy information or add two or more inconsistent populations, the distributions of period (plotted as $\log{(P)}$) and skewness parameter, $S$ were plotted. In each of the summary plots
the three vertical columns also first divide the data into two parts, for low (0-5 ADU/s) flux amplitudes and higher flux amplitude ratios and then in the third column show the undivided total distributions. This division is made because the most obvious differences in the plotted distributions usually occur between the two groups illustrated; it might be more accurate to say that the low amplitude variables differed from the rest, but they also dominate hugely in number/ADU.
The equivalent distributions 
to the skewness distributions were also plotted as fractional rise-time for each variable candidate, (where rise-time is defined as the fraction of one period taken up by the time taken to rise from the lowest flux point to the highest). 
It should be remembered that the skewness parameter can be converted into a fractional rise-time using
Equation \ref{skew_rise_time}. However, for the particular skewness distributions found the rise-time plots gave no additional information so are not shown here.

\begin{figure}
\vspace*{4.5cm}
  \begin{tabular}{ccc}
\vspace*{5cm}
   \leavevmode
 \includegraphics{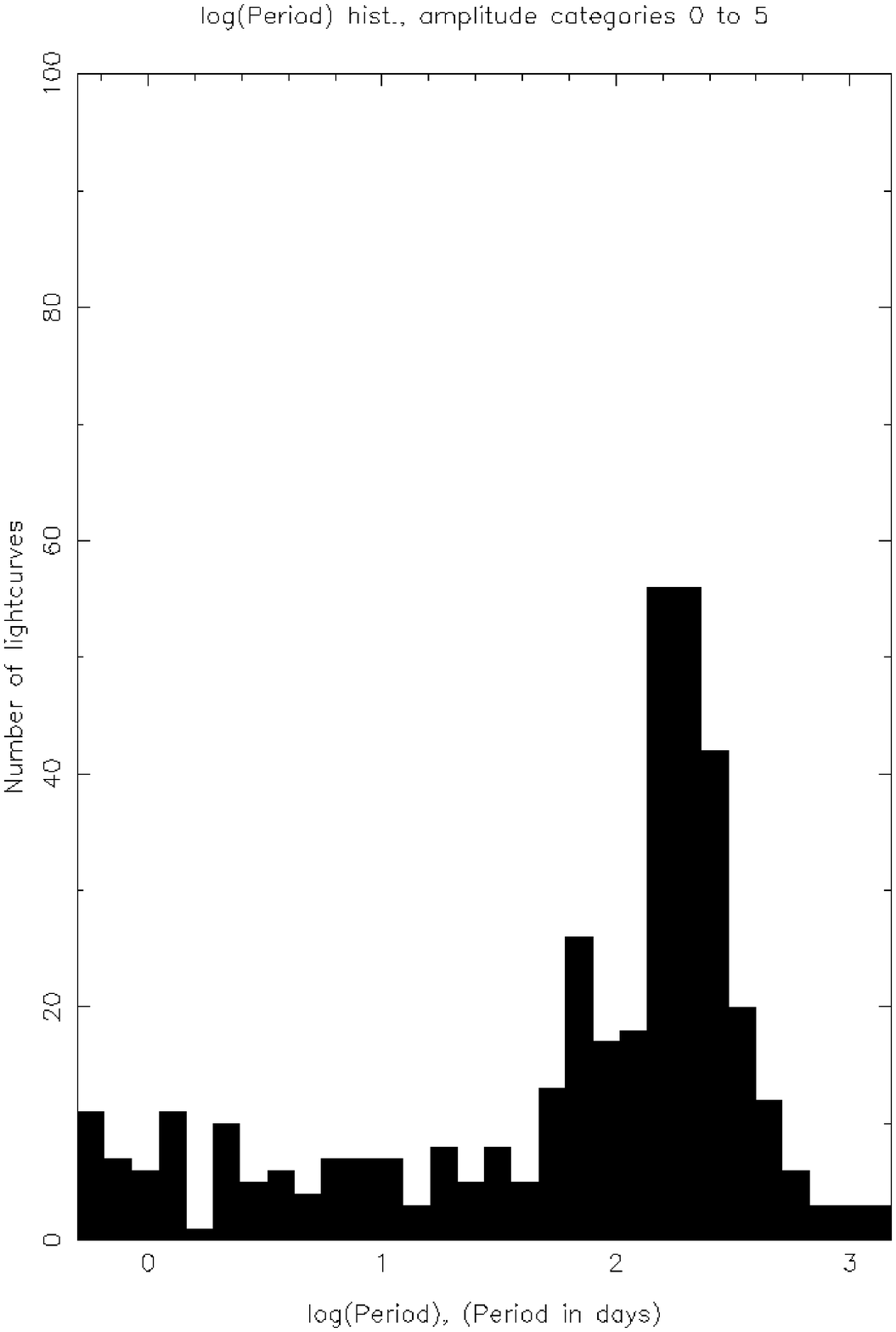} \\
\vspace*{5cm}
   \leavevmode
 \includegraphics{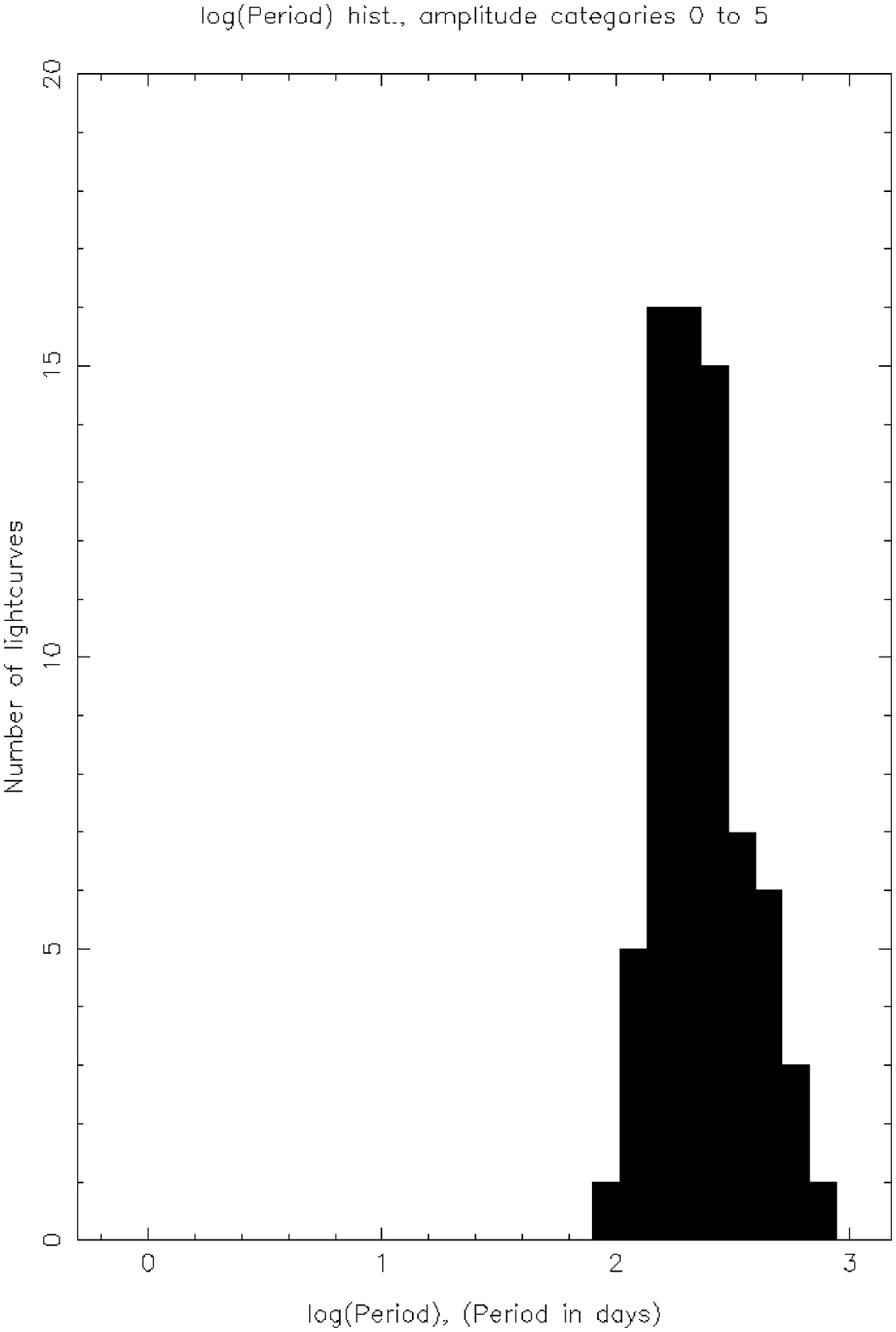} \\
\vspace*{5cm}
   \leavevmode
 \includegraphics{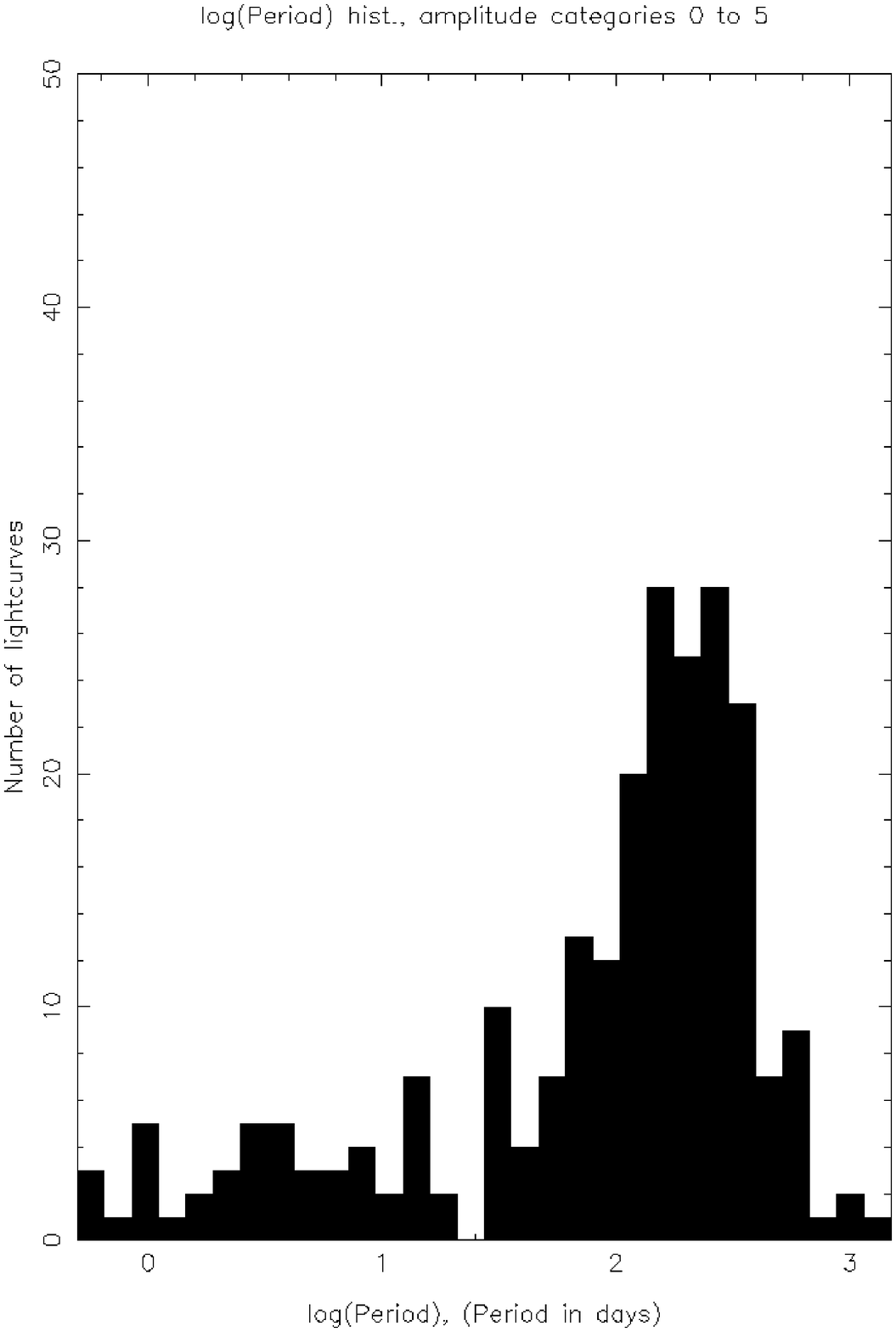} \\
\vspace*{0cm}
   \leavevmode
 \includegraphics{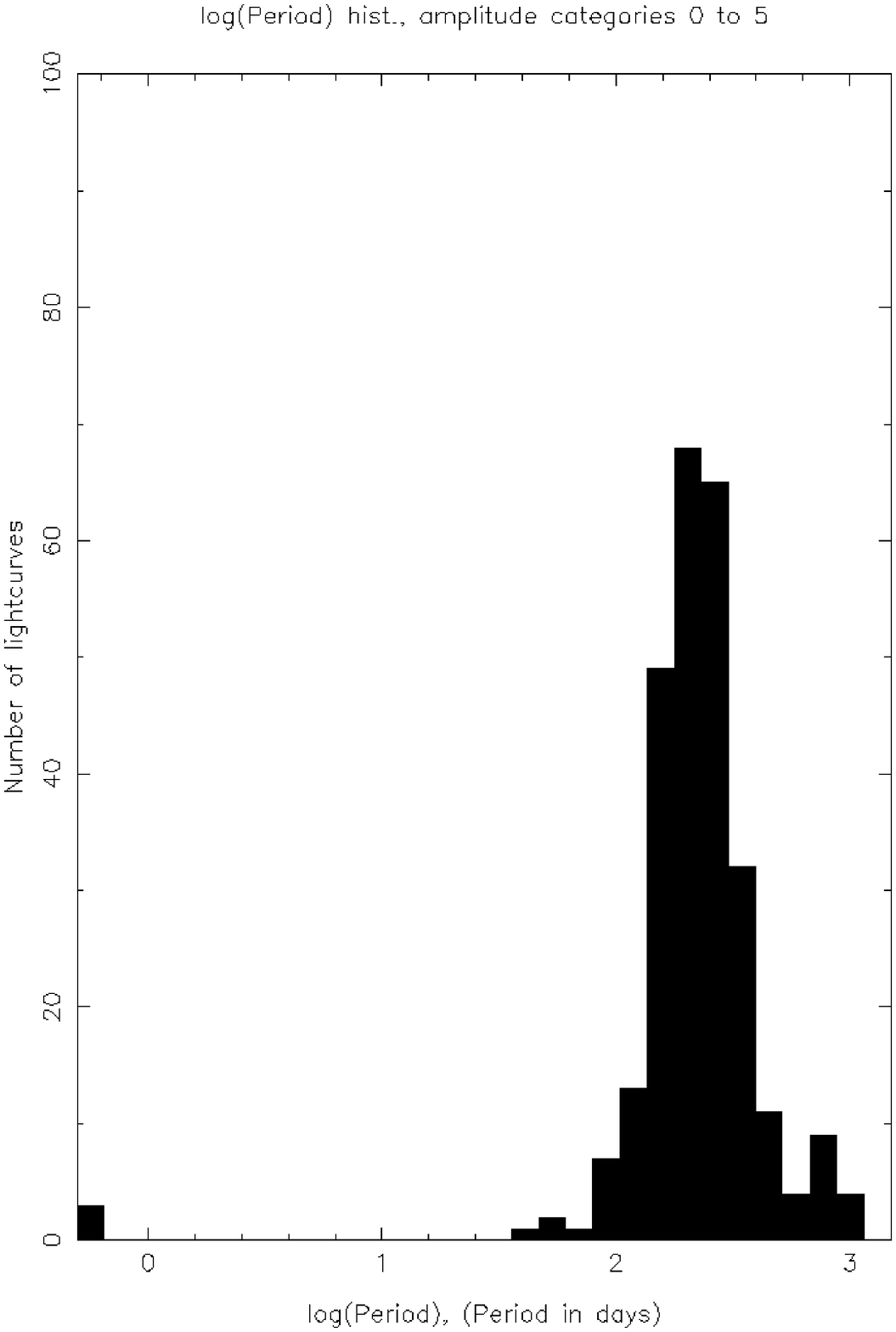} \\
  \end{tabular}
\hspace*{4cm}
  \begin{tabular}{ccc}
\vspace*{5cm}
   \leavevmode
 \includegraphics{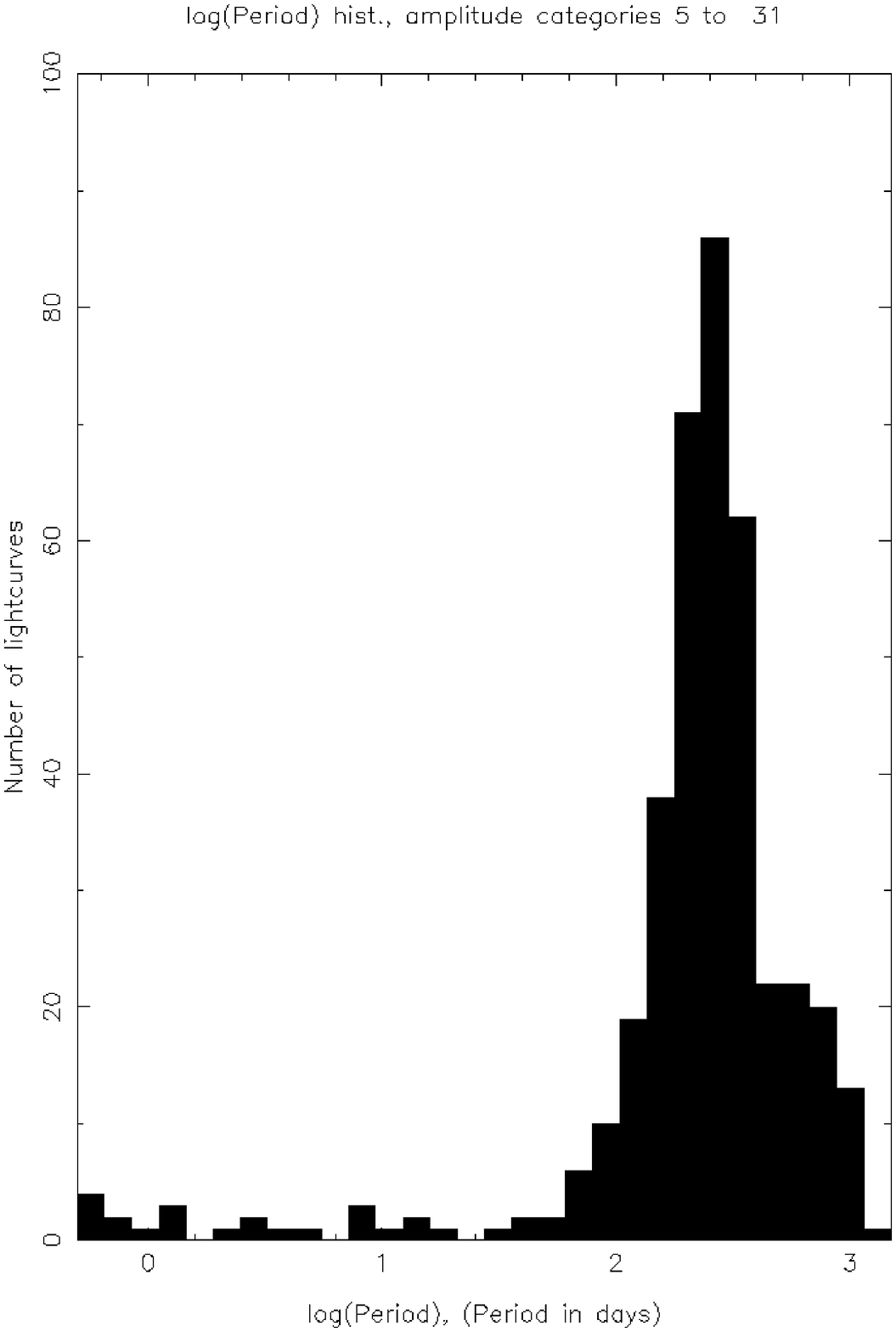} \\
 \vspace*{5cm}
   \leavevmode
\includegraphics{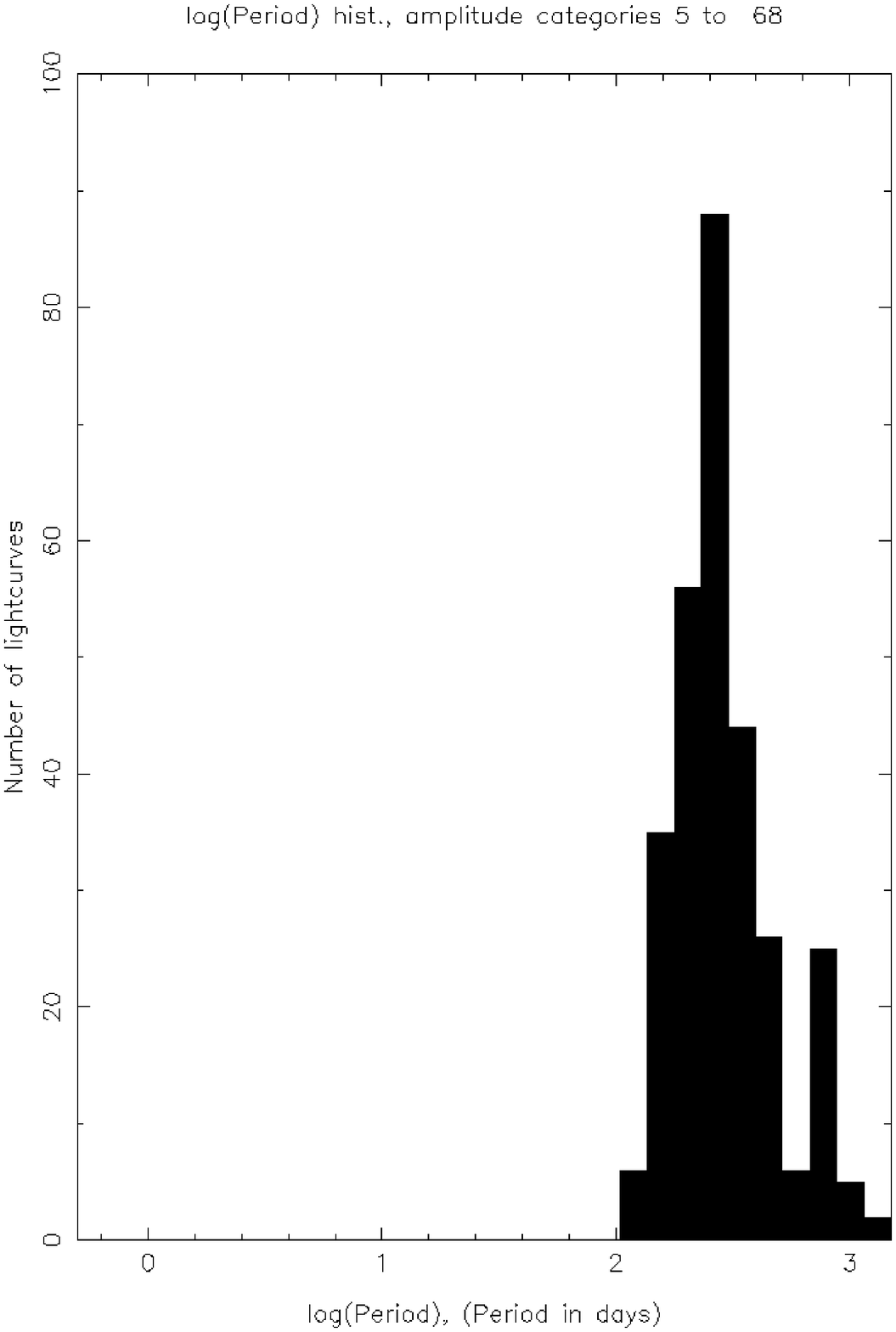} \\
\vspace*{5cm}
   \leavevmode
 \includegraphics{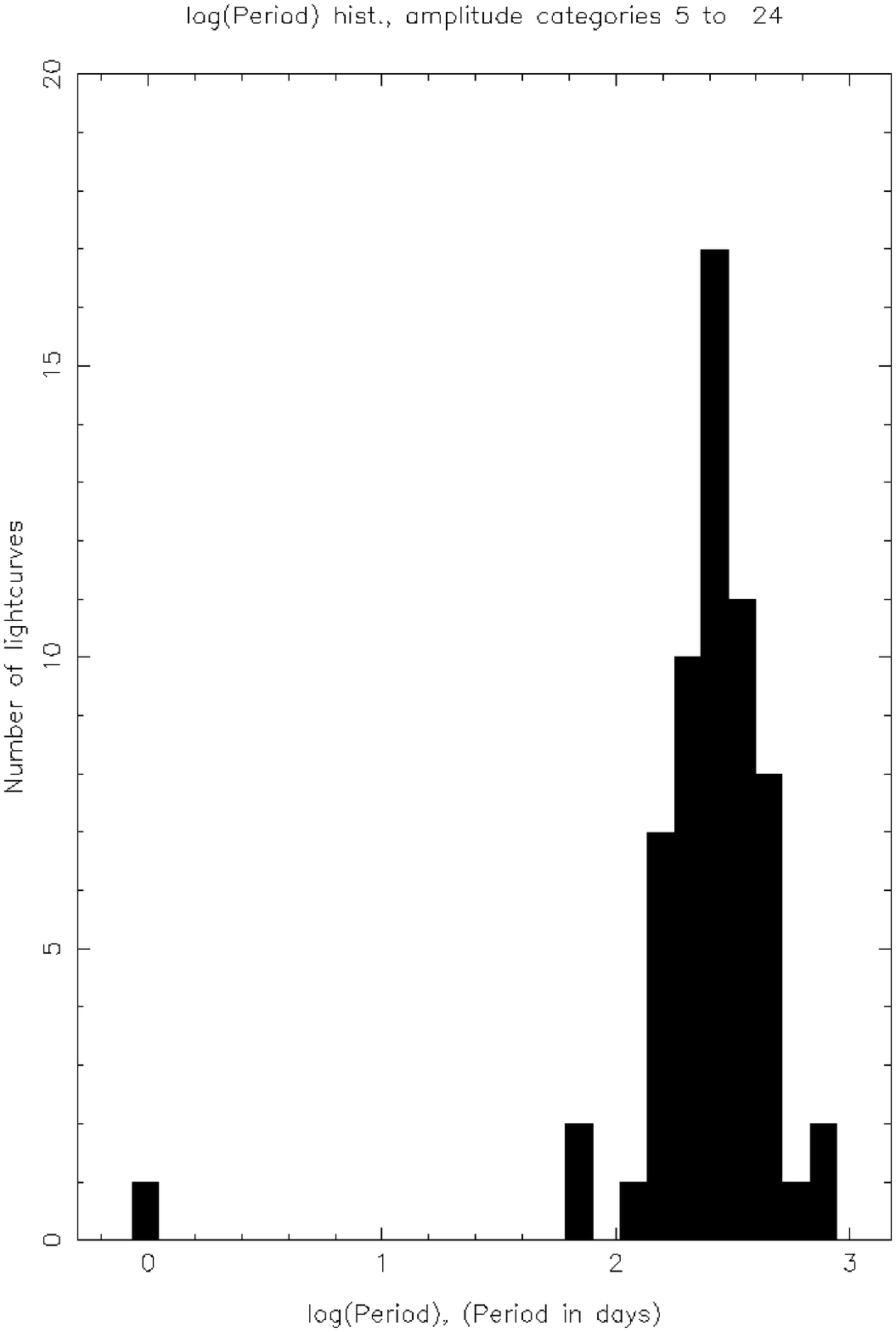} \\
\vspace*{0cm}
   \leavevmode
 \includegraphics{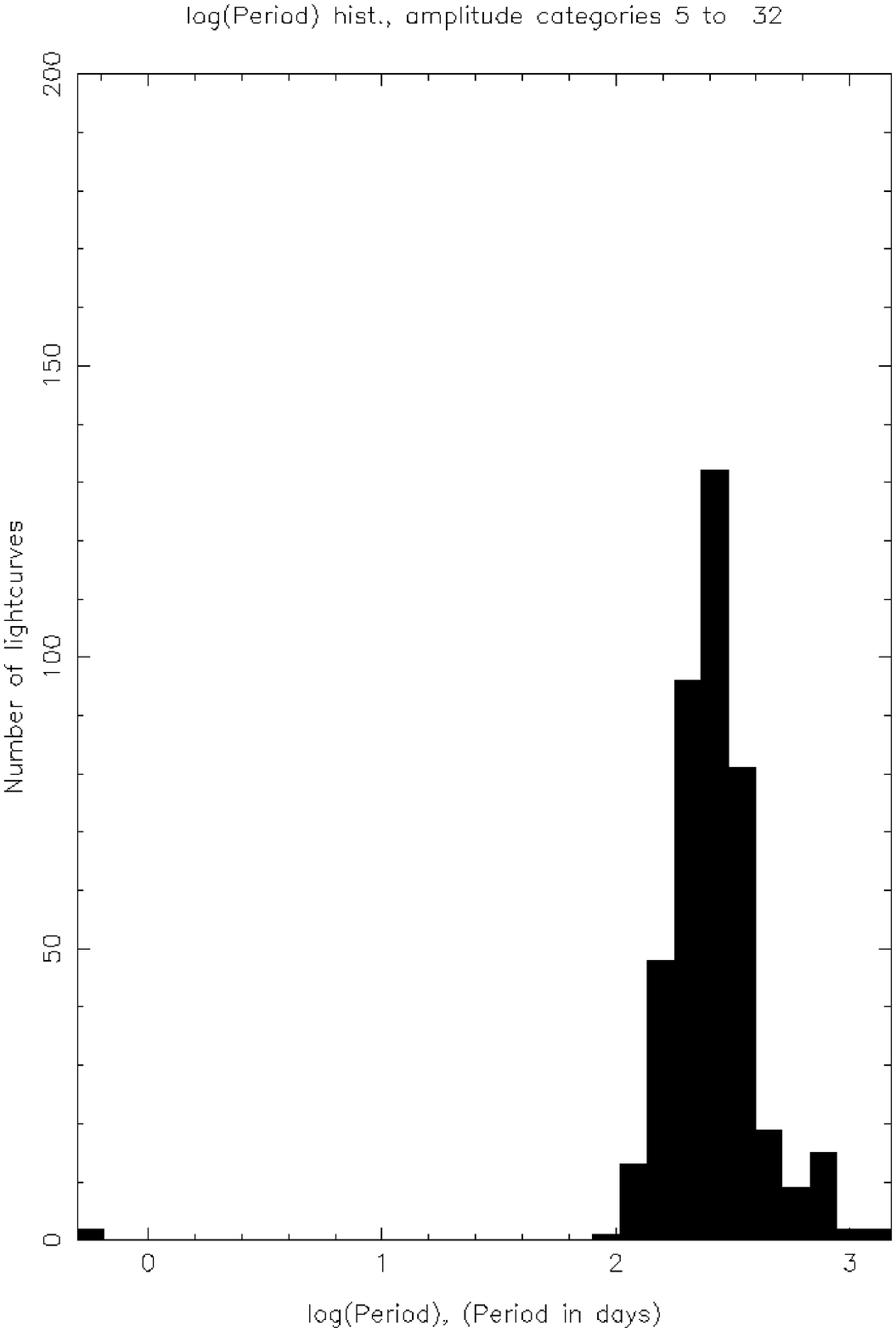} \\
  \end{tabular}
\hspace*{4cm}
  \begin{tabular}{ccc}
\vspace*{5cm}
   \leavevmode
 \includegraphics{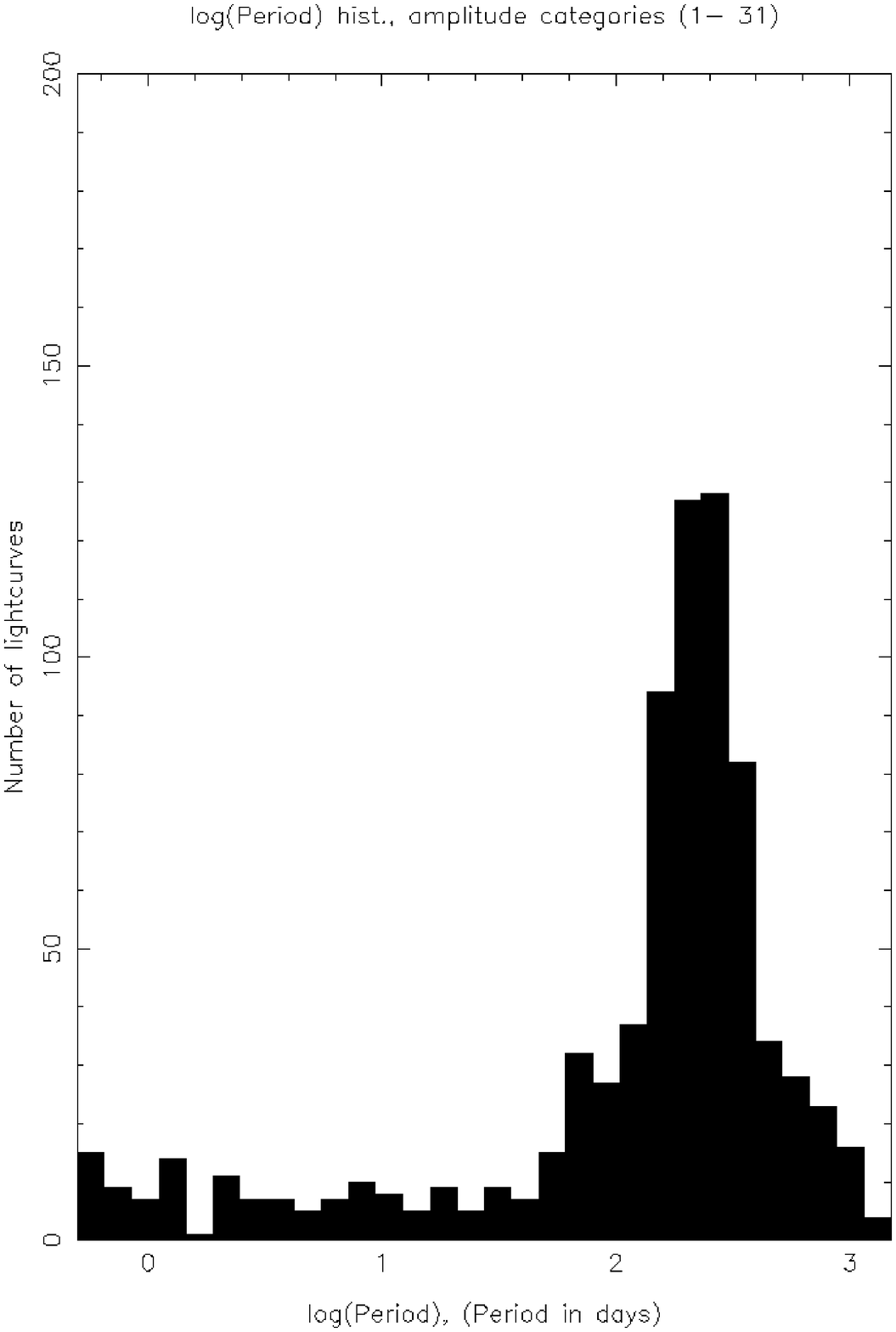} \\
\vspace*{5cm}
   \leavevmode
 \includegraphics{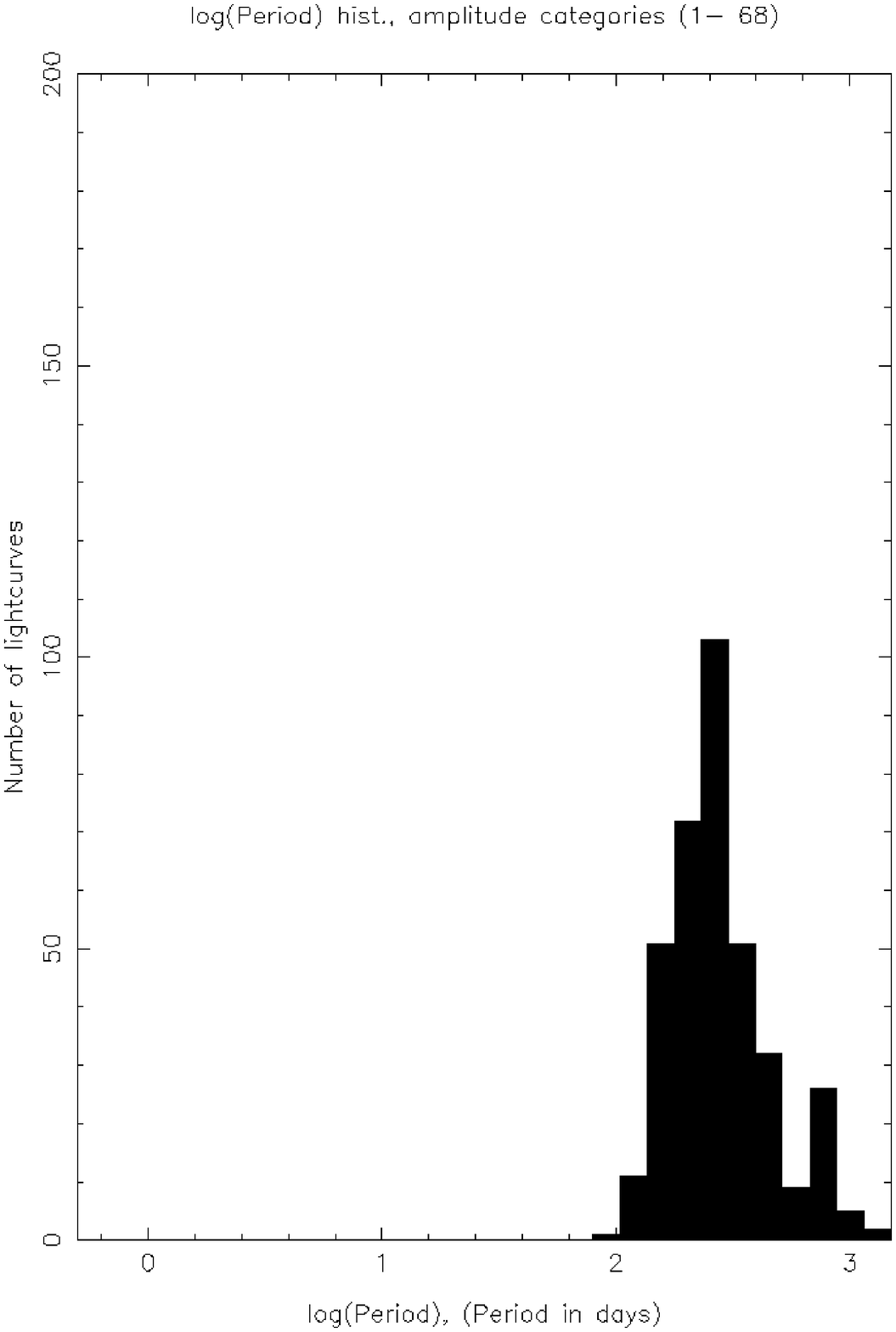} \\
\vspace*{5cm}
   \leavevmode
 \includegraphics{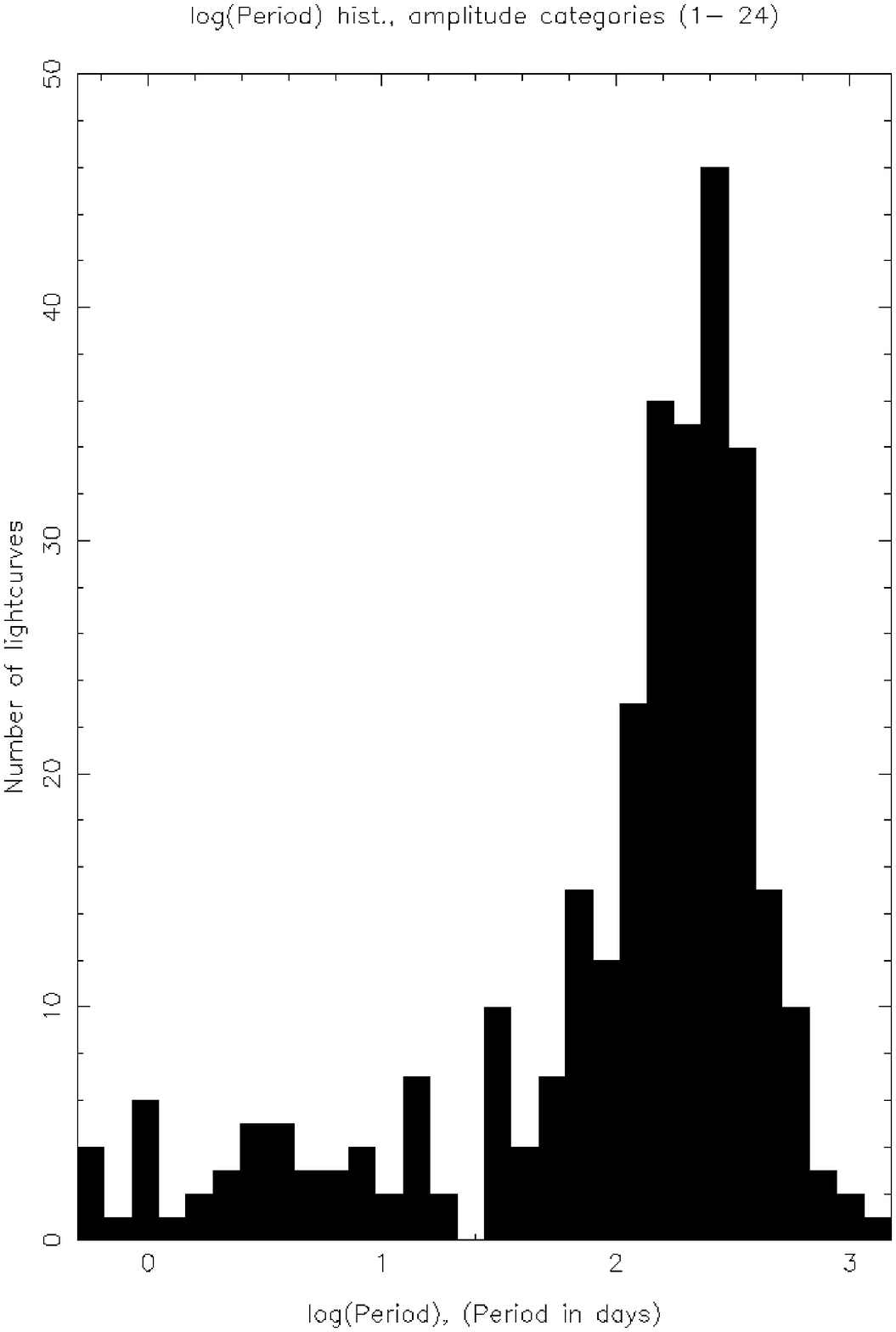} \\
\vspace*{0cm}
   \leavevmode
 \includegraphics{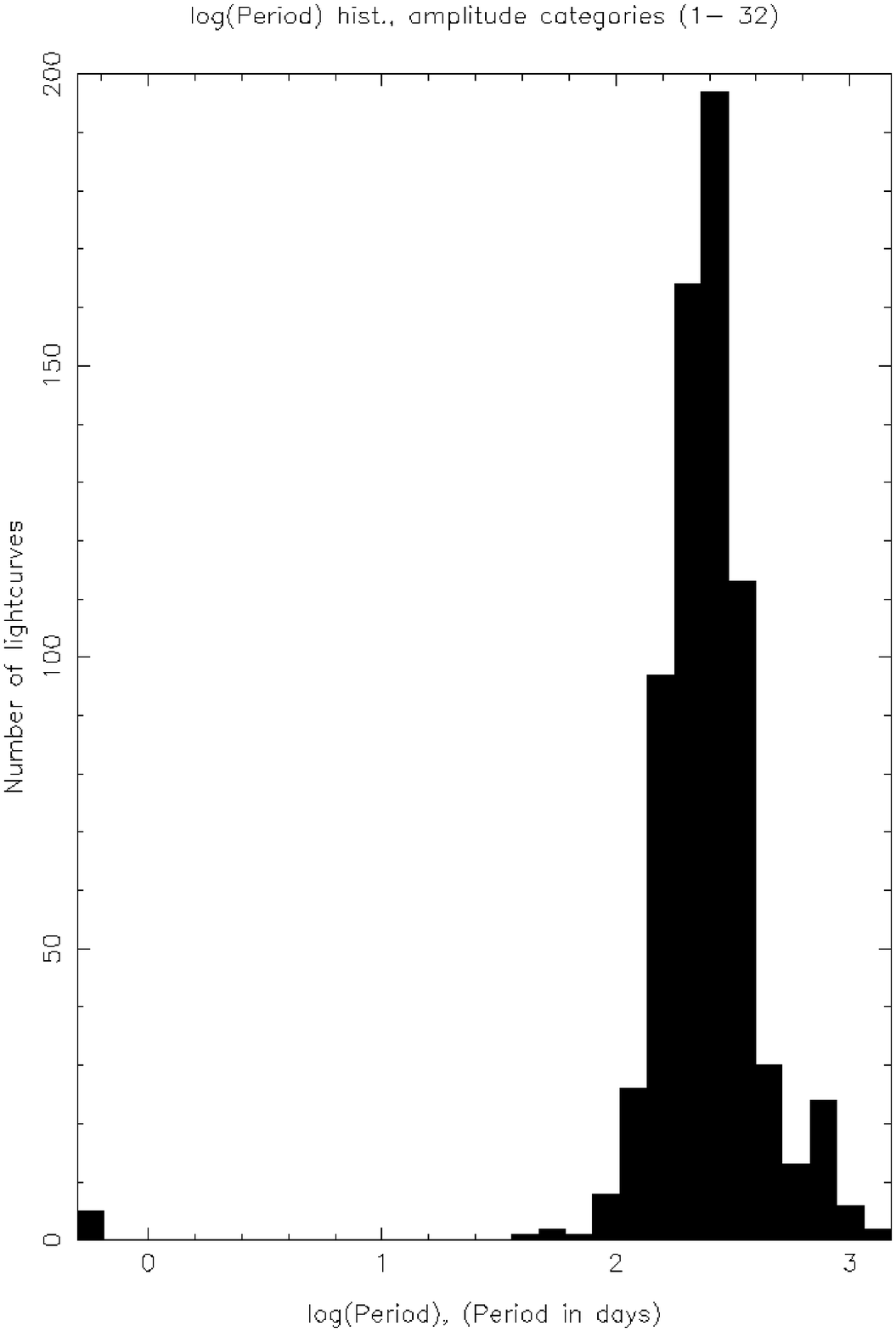} \\
  \end{tabular}
\caption[Histograms of variable lightcurve $\log{(\rm{period})}$ distributions: PA/LT flux ratio = 7.89.]{Histograms of variable
 lightcurve $\log{(\rm{period})}$ distributions: PA/LT flux ratio = 7.89; Rows, top to bottom:
1)`variable' lightcurves containing (LT) or (LT + FTN) data only,
2) ``mixed'' lightcurves containing (LT) or (LT + FTN) data only,
3)`variable' lightcurves containing (PA + LT) or (PA + LT + FTN) data,
4) ``mixed'' lightcurves containing (PA + LT) or (PA + LT + FTN) data.
Columns, left to right: Flux amplitudes, 1) 0-5, 2) 5-`MAX' 3) Sum of all categories
`MAX' varies with row as: 1) 31 2) 68 3) 24 4) 32 .}
 \label{Log(P)_all_categories_low_ratio}
\end{figure}

\begin{figure}
\vspace*{4.5cm}
  \begin{tabular}{ccc}
\vspace*{5cm}
   \leavevmode
 \includegraphics{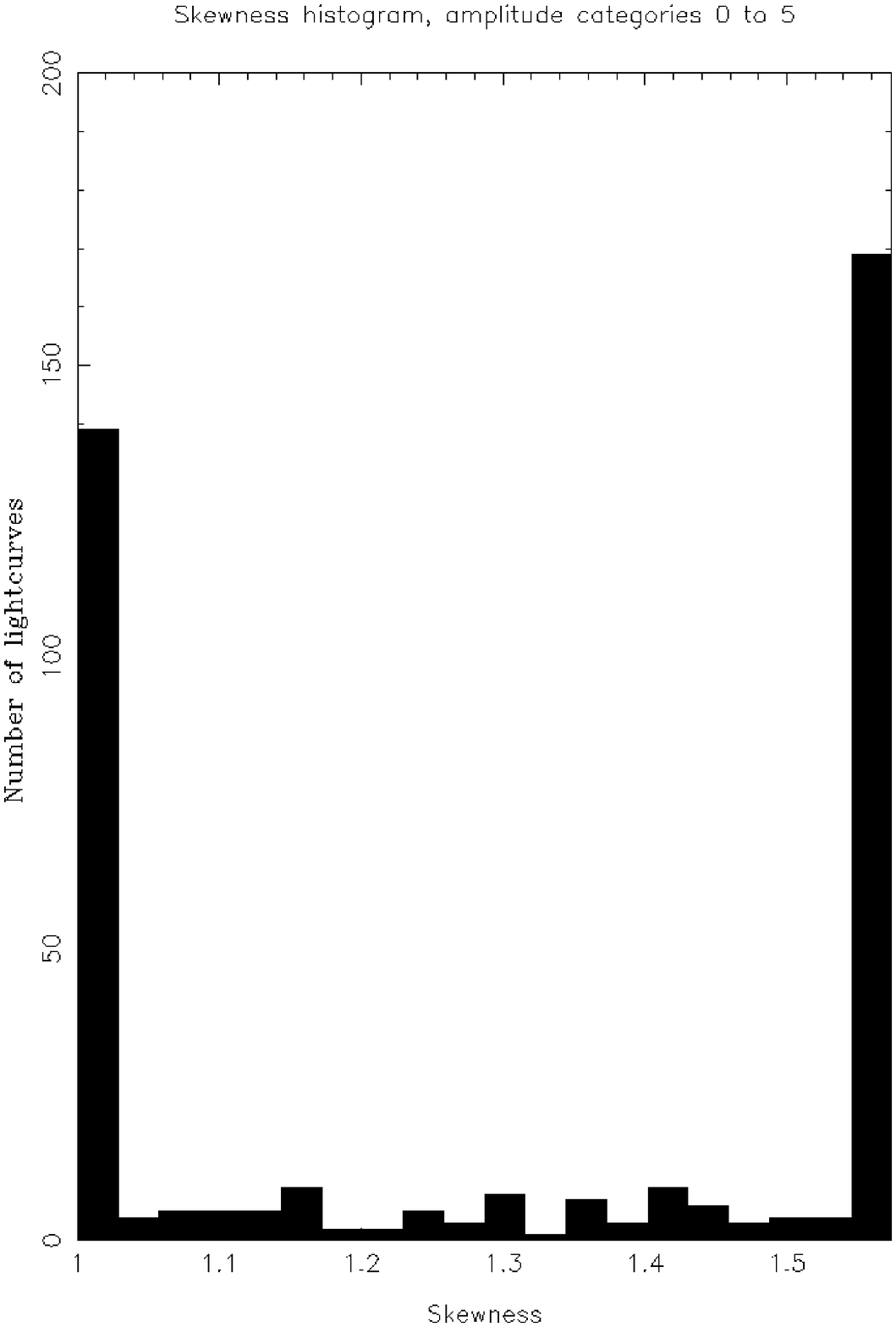} \\
\vspace*{5cm}
   \leavevmode
 \includegraphics{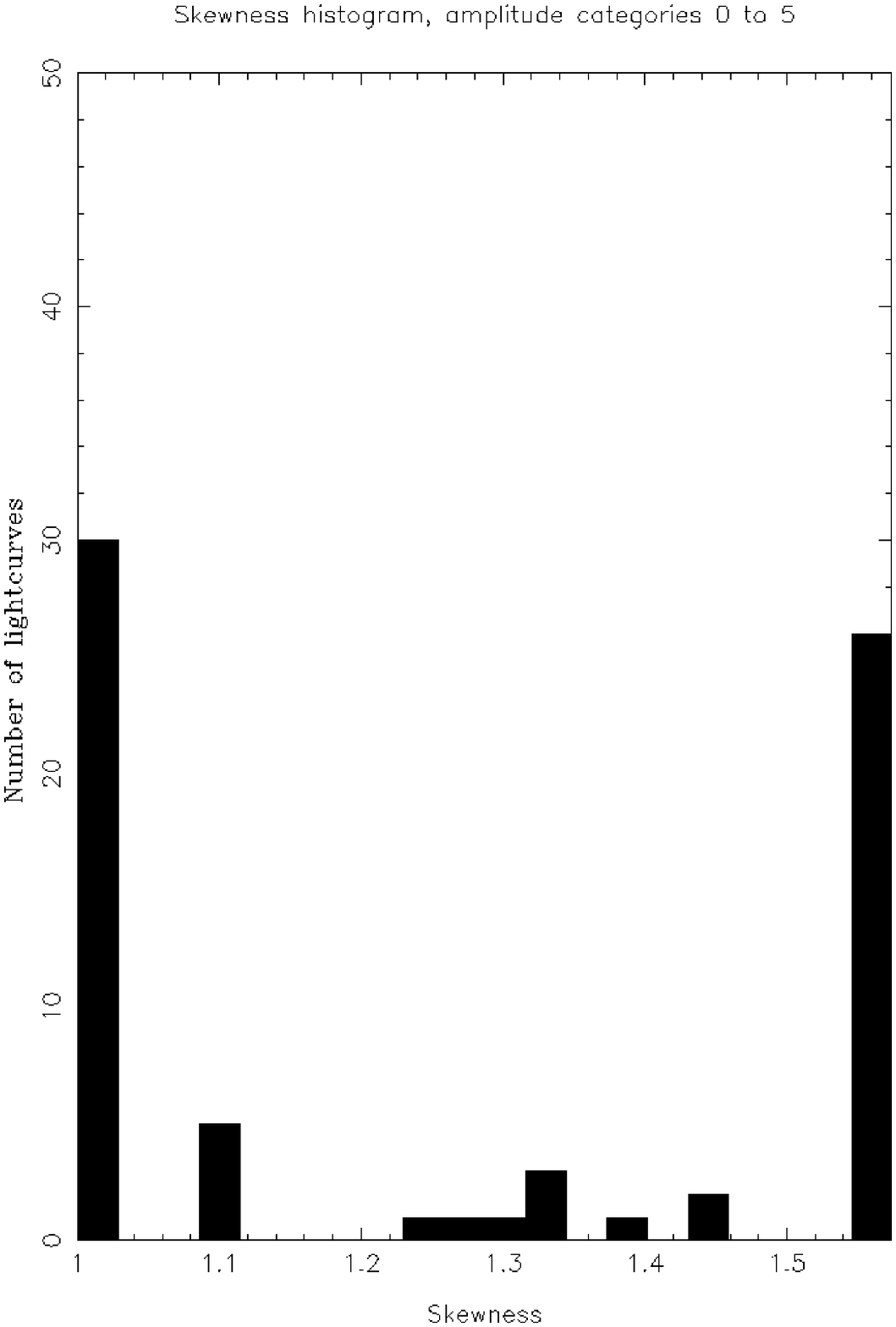} \\
\vspace*{5cm}
   \leavevmode
 \includegraphics{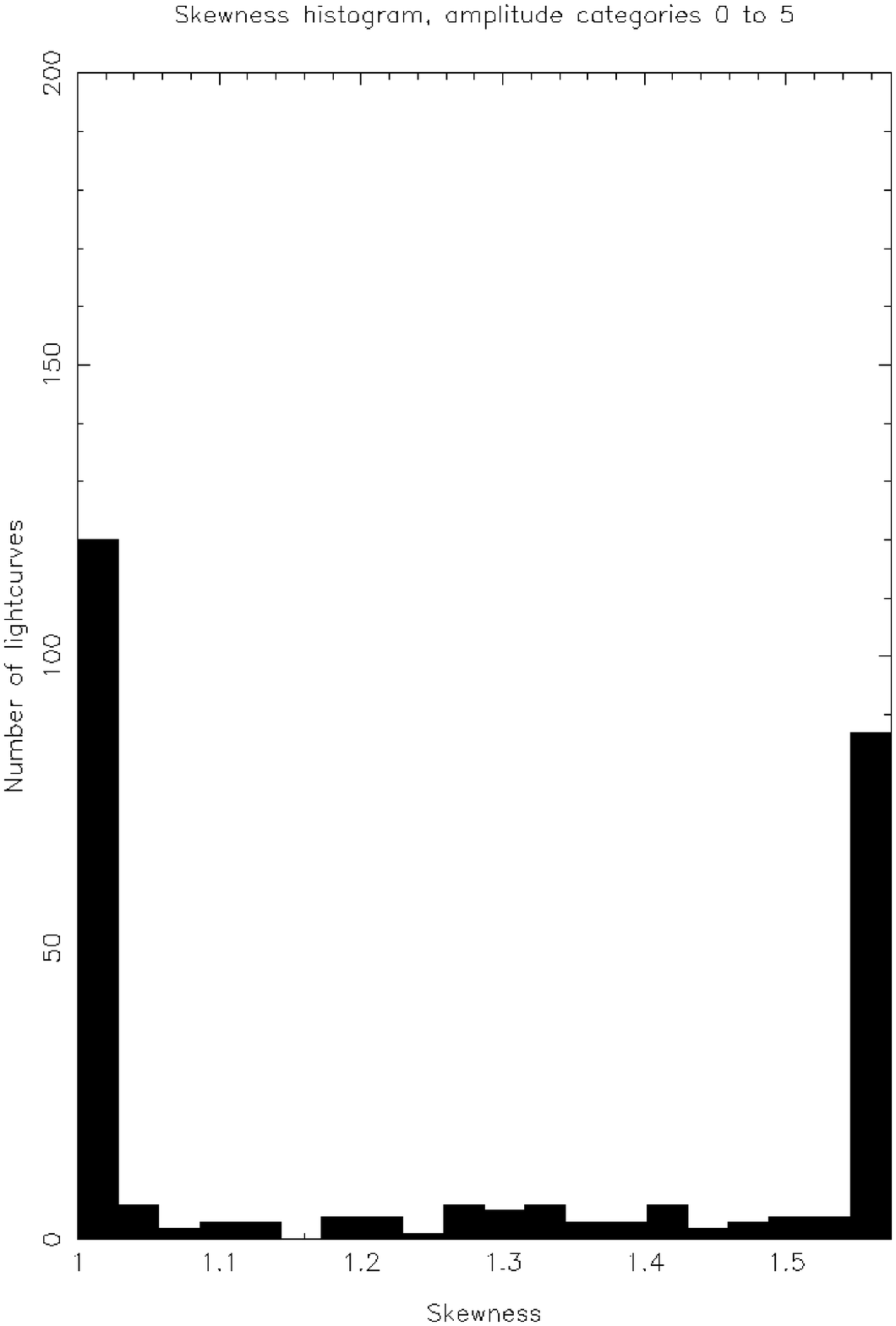} \\
\vspace*{0cm}
   \leavevmode
 \includegraphics{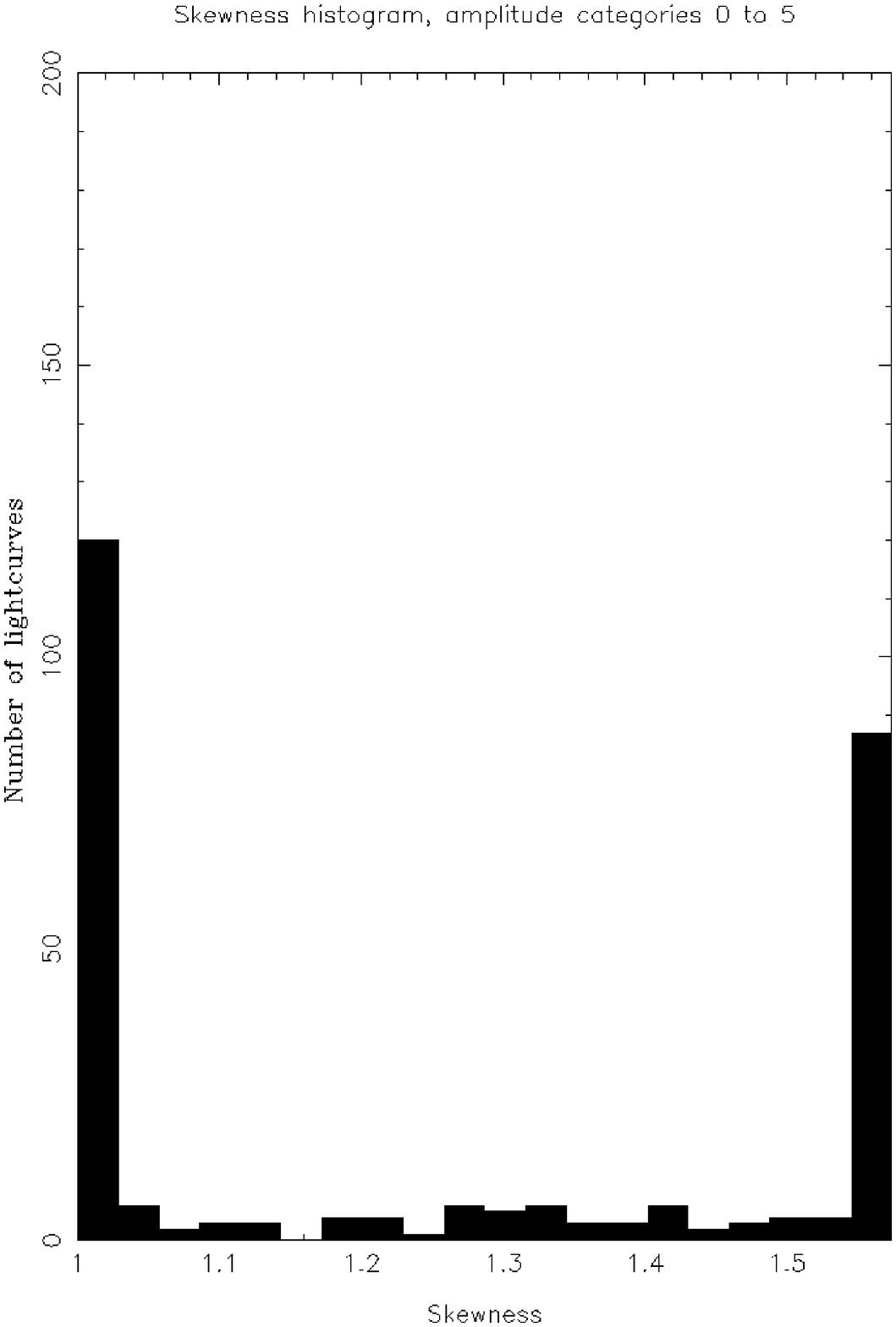} \\
  \end{tabular}
\hspace*{4cm}
  \begin{tabular}{ccc}
\vspace*{5cm}
   \leavevmode
 \includegraphics{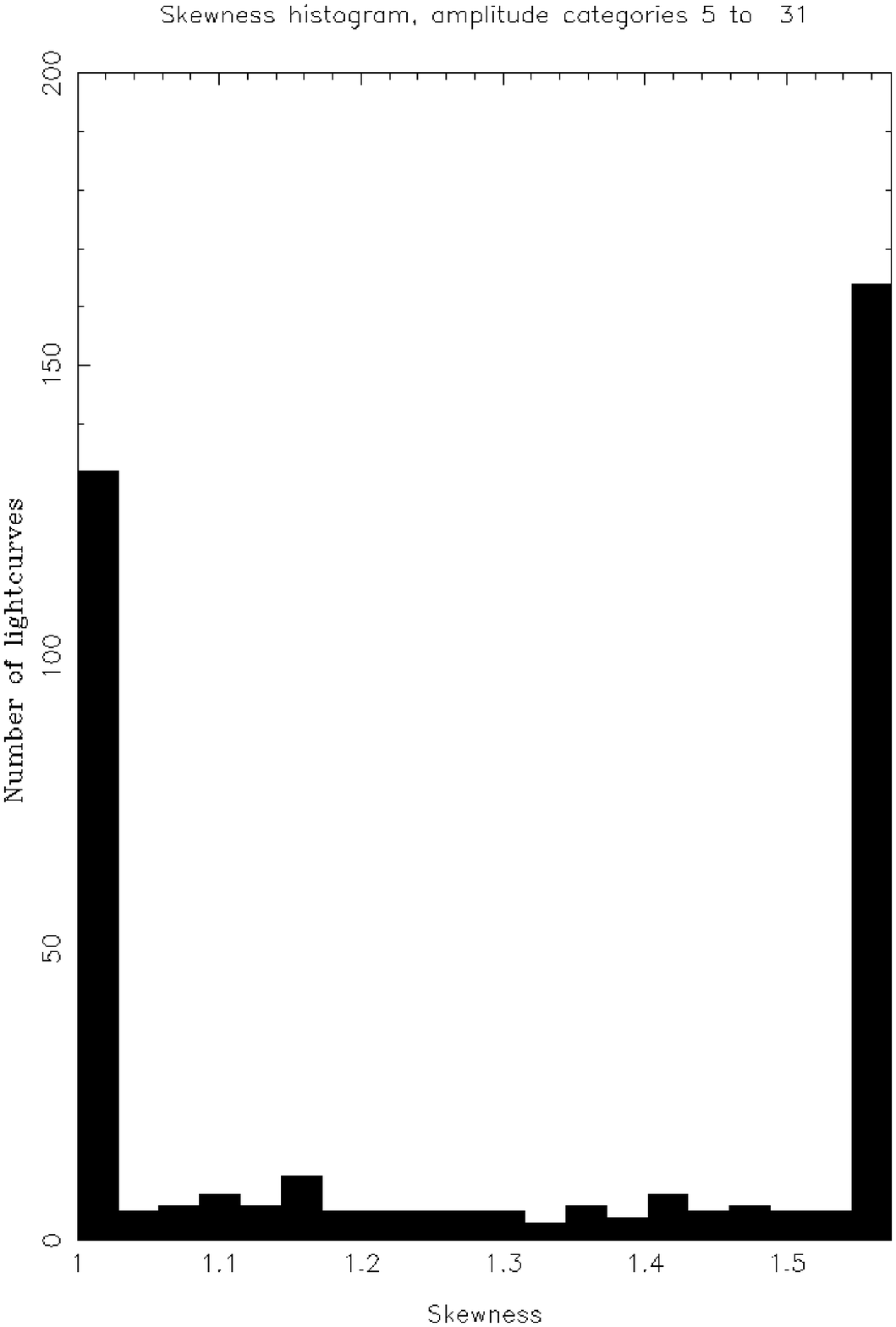} \\
 \vspace*{5cm}
   \leavevmode
\includegraphics{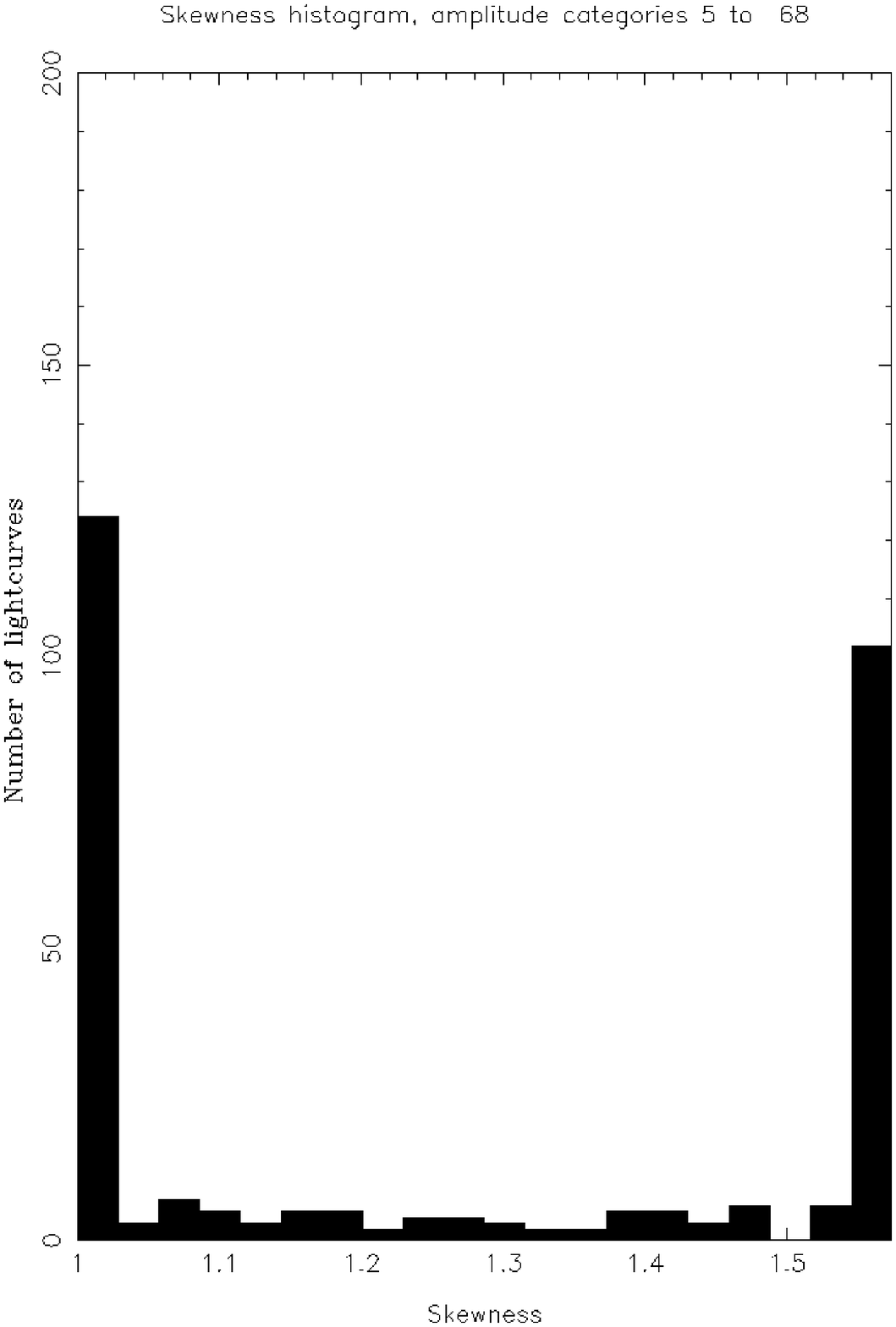} \\
\vspace*{5cm}
   \leavevmode
 \includegraphics{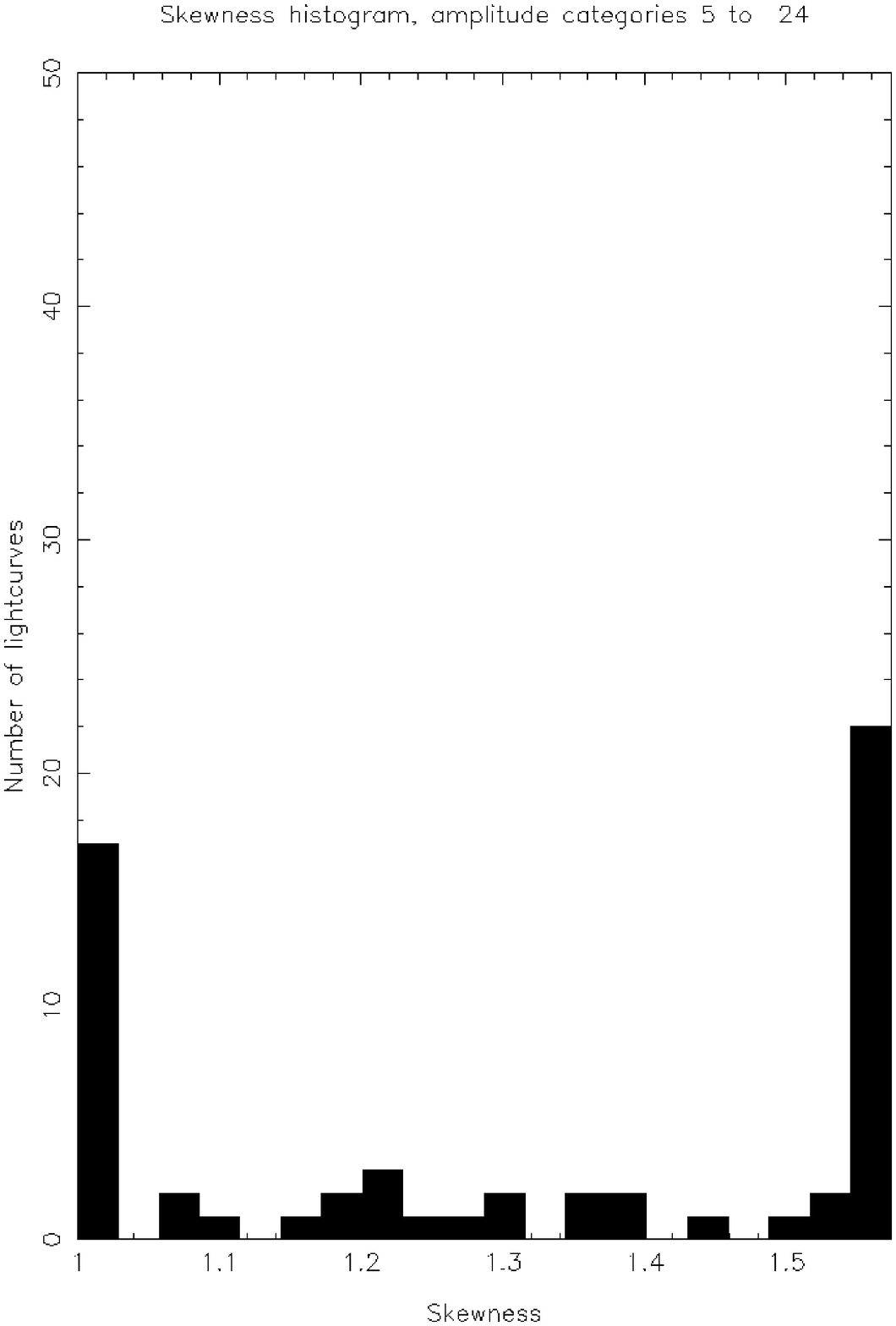} \\
\vspace*{0cm}
   \leavevmode
 \includegraphics{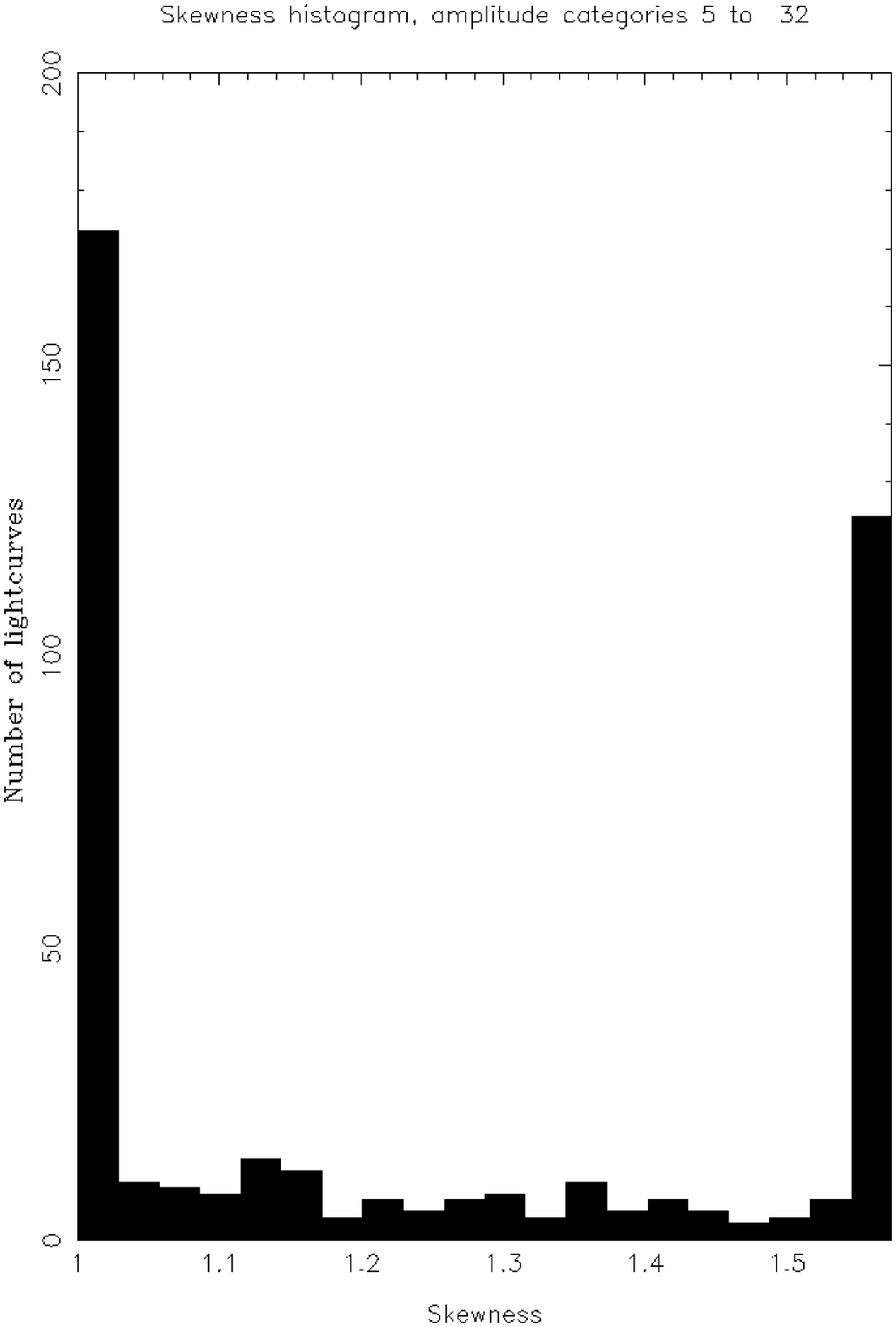} \\
  \end{tabular}
\hspace*{4cm}
  \begin{tabular}{ccc}
\vspace*{5cm}
   \leavevmode
 \includegraphics{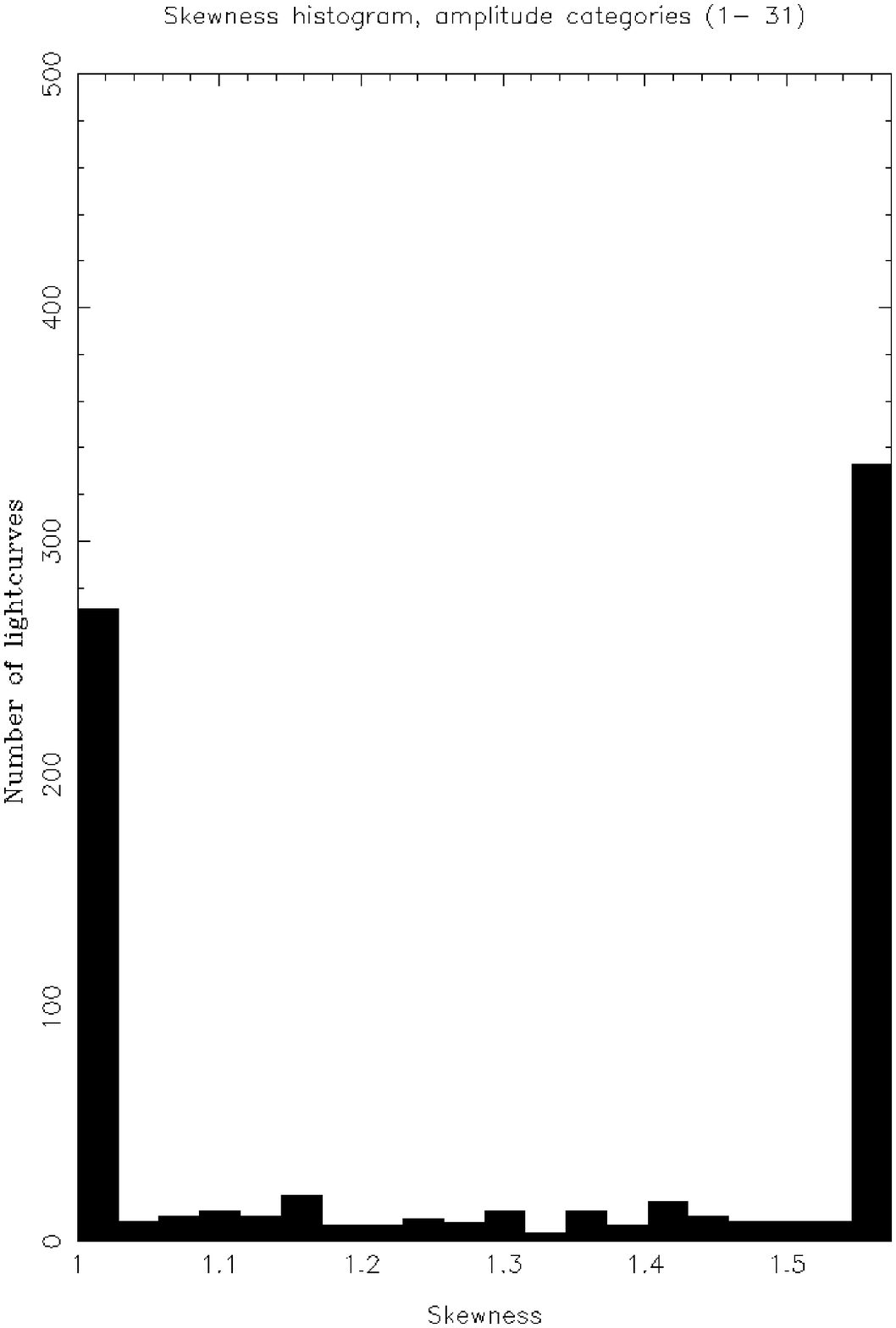} \\
\vspace*{5cm}
   \leavevmode
 \includegraphics{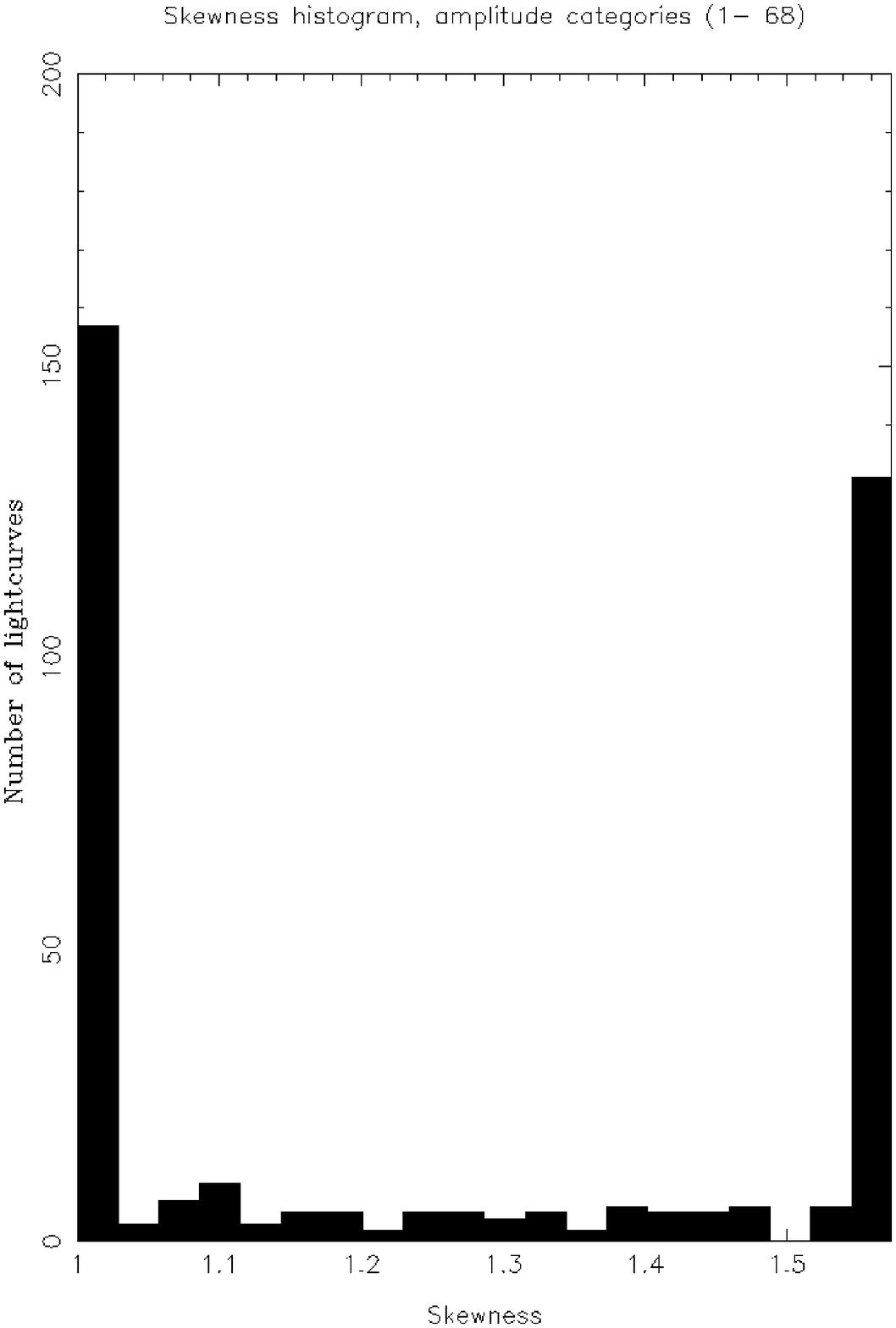} \\
\vspace*{5cm}
   \leavevmode
 \includegraphics{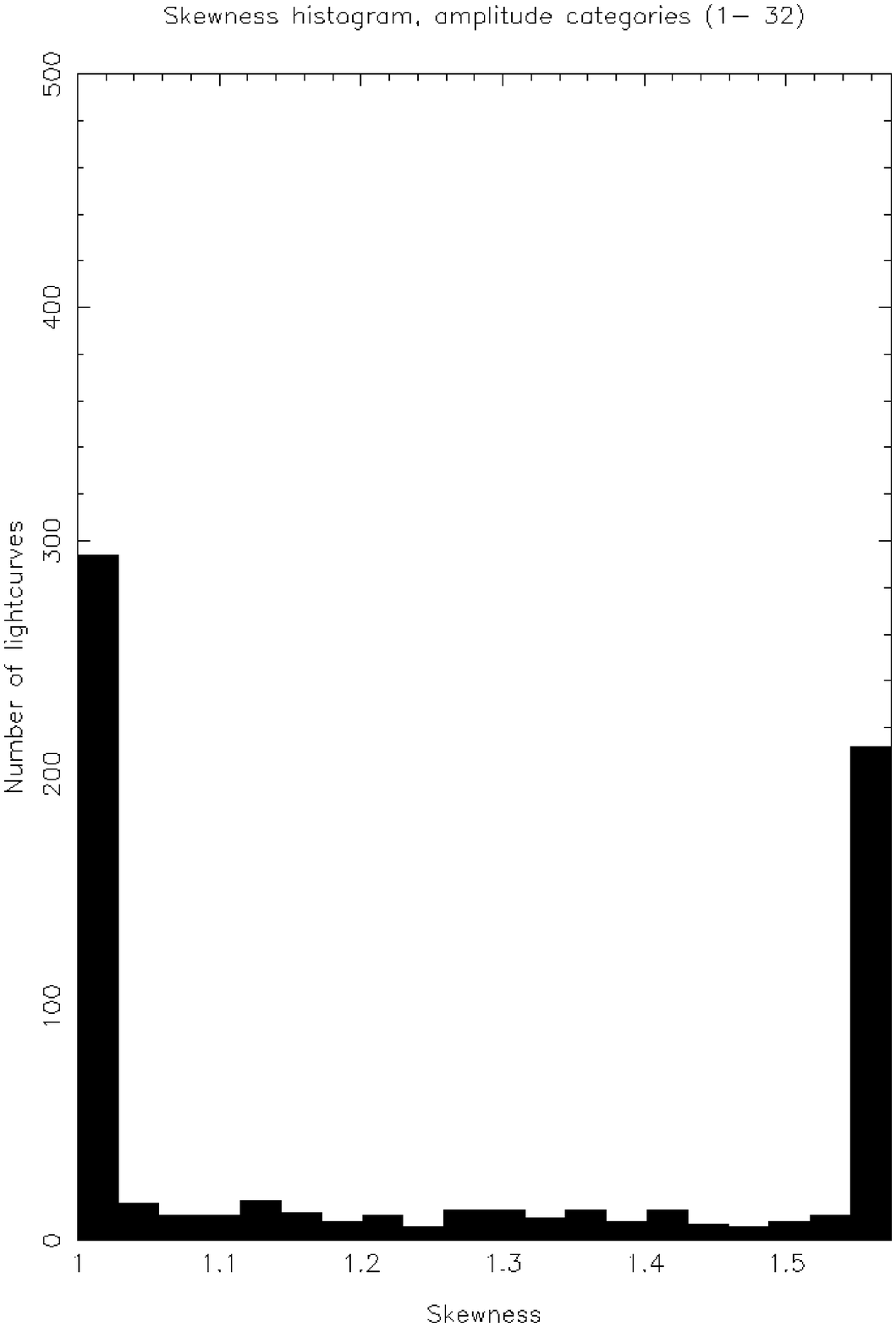} \\
\vspace*{0cm}
   \leavevmode
 \includegraphics{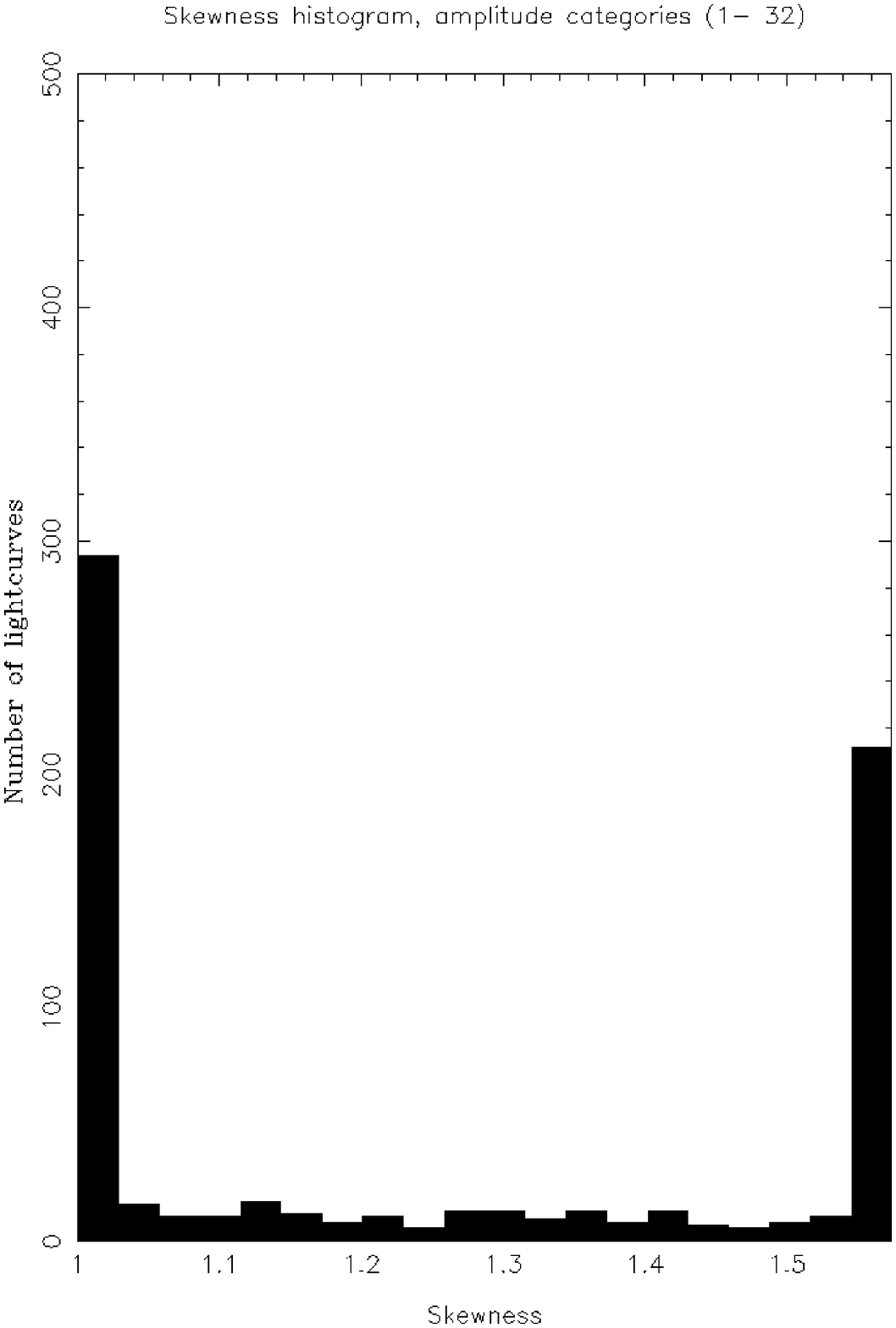} \\
  \end{tabular}
\caption[Histograms of variable lightcurve Skewness distributions: PA/LT flux ratio = 7.89. ]{Histograms of variable
 lightcurve Skewness distributions: PA/LT flux ratio = 7.89; Rows, top to bottom:
1)`variable' lightcurves containing (LT) or (LT + FTN) data only,
2) ``mixed'' lightcurves containing (LT) or (LT + FTN) data only,
3)`variable' lightcurves containing (PA + LT) or (PA + LT + FTN) data,
4) ``mixed'' lightcurves containing (PA + LT) or (PA + LT + FTN) data.
Columns, left to right: Flux amplitudes, 1) 0-5, 2) 5-`MAX' 3) Sum of all categories
`MAX' varies with row as: 1) 31 2) 68 3) 24 4) 32 .}
 \label{Skewness_all_categories_low_ratio}
\end{figure}

An example for one only of the $4$ flux amplitude ratio categories of summary plots for each of skewness and $\log{(P)}$ variables are shown in Figures \ref{Log(P)_all_categories_low_ratio} and \ref{Skewness_all_categories_low_ratio}.

\newpage

\subsubsection{Combining variable star categories}

Having established that distributions of period and skewness were reasonably consistent
between different fixed flux ratio pipeline runs and also between fixed and unfixed
flux ratio pipeline runs, the sixteen categories detailed above were combined in several different combinations to enable further investigations of consistency to be performed (for example between ``variable'' and ``mixed'' and ``fixed flux ratio'' and ``unfixed flux ratio'' classifications) and also to increase the number of lightcurves in each grouping to improve
statistics. Six obvious combinations of the $16$ sub-categories were formed as follows:

1) all fixed ratio ``variables'' (only LT data present)

2) all fixed ratio ``variables'' (at least LT and PA data present)

3) all fixed ratio ``mixed'' (only LT data present)

4) all fixed ratio ``mixed'' (at least LT and PA data present)

5) all ``variables'' (all four flux ratios and both LT only and (LT+PA) data)

6) all ``mixed'' (all four flux ratios and both LT only and (LT+PA) data)

Having concatenated members of the group of sixteen sub groupings appropriately to form the
above groups, the six groups above were processed to resolve the occasions where 
lightcurves were duplicated from different pipeline runs within the same group. There existed a substantial overlap between each pipeline run of the lightcurves selected and categorised,
as would be expected due to the wide reduced $\chi^2$ window used, but clearly it was required to select the best fit from the maximum of four possible fits performed. Therefore the groupings above were consolidated by selecting only the
pipeline run for each lightcurve which had the lowest reduced $\chi^2$. In Table \ref{first_level_groupings} below
the starting numbers of lightcurves in each of the above groups and the numbers remaining after consolidation by selection on lowest reduced $\chi^2$ are detailed.
This step had also previously been performed when the ``long period low contrast'' lightcurves
had not been included and so those numbers are also detailed in the table to give an idea of the relatively low numbers of lightcurves added by including these. Only the lines involving ``mixed'' events change because the ``long period low contrast'' cut is only applicable to mixed
lensing candidates.

\begin{table}
\caption{Table showing the numbers of lightcurves in each of $8$ categories after consolidation according to lowest reduced $\chi^2$. The equivalent numbers if the ``long period low amplitude'' lightcurves are also given for comparison.}
\small
\begin{center}
\begin{tabular}{|c|c|c|c|c|}
\hline
\hline
      &\multicolumn{2}{c|}{NOT incl. ``long period''}    &\multicolumn{2}{c|}{incl. ``long period''}     \\
\hline
   Grouping   &  Before  $\chi^2$ &  After  $\chi^2$ &  Before  $\chi^2$  &  After $\chi^2$ \\
              & \tiny{Consolidation}     &   \tiny{Consolidation}  &  \tiny{Consolidation}     &   \tiny{Consolidation}\\
\hline
  1. \footnotesize{Fixed ratio, Vars, LT}      & $792$ & $792$ & $792$ &  $792$    \\
  2. \footnotesize{Fixed ratio, Vars, +PA}     & $910$ & $403$ & $910$ & $403$     \\
  3. \footnotesize{Fixed ratio, Mixed, LT}     & $315$ & $315$ & $372$ & $372$     \\
  4. \footnotesize{Fixed ratio, Mixed, +PA}    & $2148$ & $968$ & $2164$ & $974$   \\
  5. \footnotesize{All ratios, Vars, LT, +PA}   & $2540$ & $1522$ & $2540$ & $1522$ \\
  6. \footnotesize{All ratios, Mixed,LT, +PA}  & $3117$ & $1548$ & $3238$ & $1638$ \\
\hline
\end{tabular}
\end{center}
\label{first_level_groupings}
\end{table}
\normalsize

Adding together groups 1 and 3 and also adding groups 2 and 4 to give purely ``LT only'' and ``(LT + PA) only'' data sets respectively, gave fractions of completely maximal allowed skewness of $34.3\%$ and $30.7\%$ respectively for the LT and (LT + PA) groups.
Examining each of groups 5 and 6 (representing all ``variable'' and all ``mixed'' lightcurves in that order) for the maximal skewness value produced proportions of $36.9\%$ and $27.0\%$ respectively. The number of lightcurves involved in all of the four groups above was over $1000$, so the statistics were fairly good. Therefore it seemed that the proportion of high skewness lightcurves was more strongly correlated with the categorisation into ``variable'' or ``mixed'' than with the length of the baseline of data. One speculation as to the reason for this might be that the ``mixed'' lightcurves, although selected to have low lensing contribution, might nevertheless allow the fitting routine to make use of it to
better fit any steeper rising portion of the lightcurve, leaving less ``work'' for the variable component to do. It has not yet been ascertained whether this suggestion is correct.

Observations: There are several things which may be observed from examination of the histograms
which are plotted, (Figures \ref{Log(P)_all_categories_low_ratio} and \ref{Skewness_all_categories_low_ratio} being examples).

It is interesting to see the perhaps unexpectedly large proportion of the lightcurves which
were best fitted with the maximal allowed skewness value of $1.574$. 

 Since the maximum fractional rise-time explored using the cosinusoid function being employed for this pipeline run was only $0.356$, this peak probably represents the integral of all lightcurves which would have been better fitted with a higher skewness, and there seems plenty of room between fractional rise-times of $\sim0.05$ (roughly the most extreme known) and $0.356$ for this to be possible. Two examples of variable stars, (specifically in this case Cepheid variables) which have fairly simple lightcurves and fractional rise-times less than $0.356$ can be found in \cite{1955RA......3..257O}.

The distribution of the reduced $\chi^2$ of all fitted cosinusoids to the above $3160$
lightcurves, this number being the sum of the consolidated totals of groups 5 and 6 in Table \ref{first_level_groupings}, was plotted in Figure \ref{reduced_chisq_dist_3160_LCs} below. The distribution appeared highly Poissonian
with a peak which was close to $2$. If the centre of the peak is taken to be in the centre of the $11$th bin then the peak occurs at $1.97\pm{0.19}$, assuming an approximate reduced $\chi^2$ error of $\pm1$ bin width.
If the error estimates for these lightcurves had been made correctly then it would be expected that the peak of this distribution would be at $\sim 1$, so this seemed to imply that re-scaling of the errors by multiplying by a number $\sim \sqrt{2}$ might be justified, if the assumption was made that the model used completely explained the variations in the data.
Alternatively, if the errors have been correctly estimated then
there must be additional variations in the data which are not explained by the model.

\begin{figure}[!ht]
\vspace*{9cm}
   \leavevmode
   \includegraphics{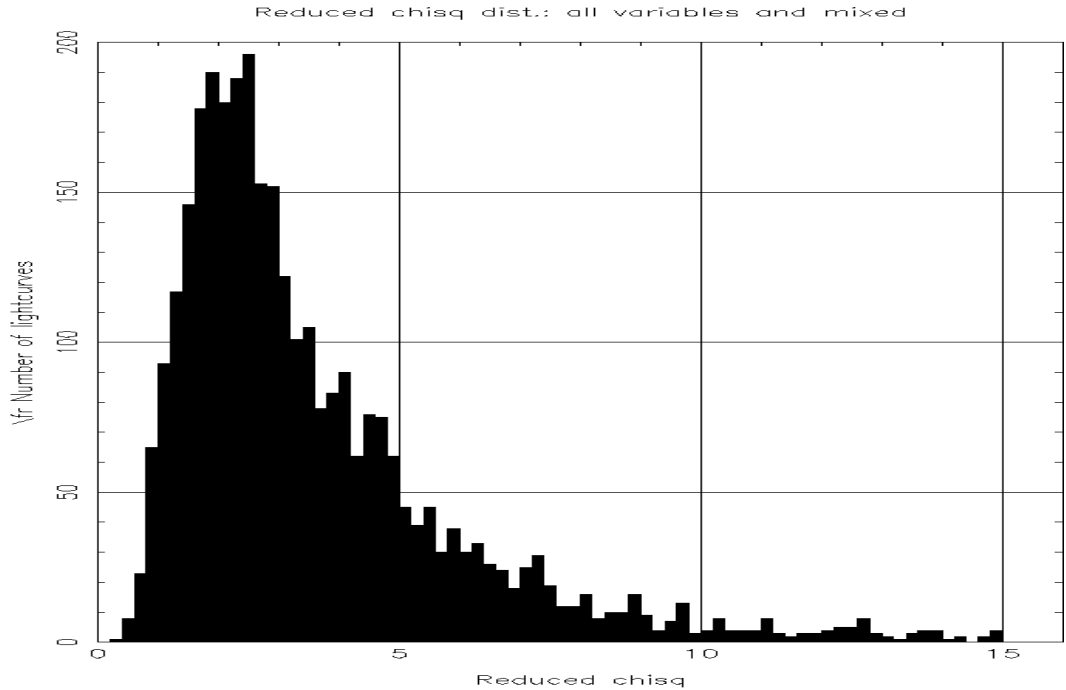}
\caption[Reduced $\chi^2$ distribution for all $3160$ variable and mixed (with low lensing contribution) classified lightcurves.]{Reduced $\chi^2$ distribution for all $3160$ variable and mixed (with low lensing contribution) classified lightcurves.
}
\label{reduced_chisq_dist_3160_LCs}
\end{figure}

When a tighter cut on reduced $\chi^2$ of the fits was applied (using $\chi^2$/d.o.f.$< 5.0$), the number of lightcurves
retained in the ``all flux ratios, all data, all fitting functions'' group was reduced from $3160$ to $2546$, ($80.6$\% of the original number, which seemed a reassuringly high fraction of the whole, implying that there were not many ``bad'' fits).

\subsubsection{Relationships and correlations between fitted parameters}

\begin{figure}[!ht]
\vspace*{8.5cm}
   \leavevmode
   \includegraphics{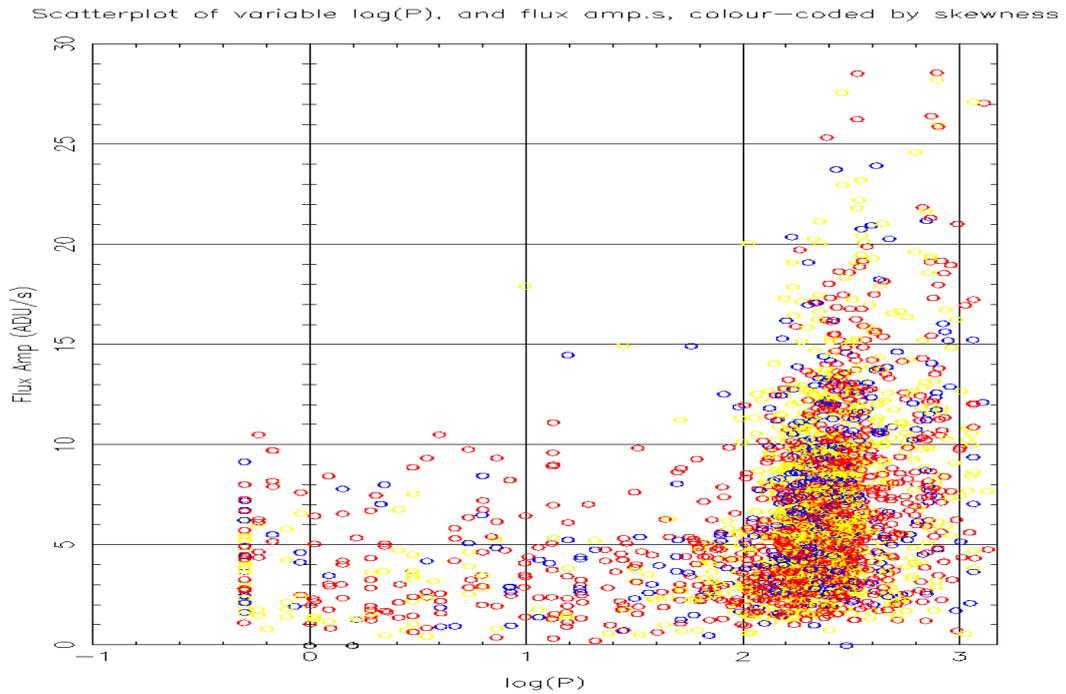}
\caption[Scatter plot showing the relationship of $\log{\rm{(period)}}$ and LT flux amplitude for $2546$
variable star candidates.]{Scatter plot showing the relationship of $\log{\rm{(period)}}$ and LT flux amplitude for $2546$
variable star candidates selected to have reduced $\chi^2 < 5$. Colours represent the skewness of the cosinusoid used to fit the lightcurve. Yellow = low skewness: $1.0 < S < 1.1$, blue = medium skewness: $1.1 < S < 1.474$, red = highest skewness: $1.474 < S < 1.574$.
}
\label{LTflux_v_logP}
\end{figure}

In Figure \ref{LTflux_v_logP}, a clear correlation can be seen, for the periodic variables with periods between about $100$ and $500$d, between period and flux amplitude. Specifically, as the flux amplitude increases, so does the mean period of the distribution. This is indeed in the direction that the correlation would be expected to be since larger stars have longer (fundamental) periods on average, and are also brighter on average, and hence have larger mean variability amplitudes.
There have been several other examples of this sort of correlation being found. Some examples are:
 \cite{2003ApJ...591L.111B} plotted the relationship between
the period and flux amplitude of Cepheids as $\log(P)$ versus $\log(F_{\rm{amp}})$. A correlation was found to exist, but it was not very tight.
\cite{Cook_Alcock_et_al}
also plotted the $P$-$F_{\rm{amp}}$ relationship 
and described a ``suggestion'' of rising flux amplitude with rising period for bright, long period variables from MACHO survey of the LMC, SMC and Milky Way. This should be the same kind of star sampled by the stars in the main group in Figure \ref{LTflux_v_logP}. They also state that the RR Lyrae stars found in their survey form groups at $\log{(P/{\rm{days}})} = -0.3$ ($P = 0.5 $d) for RRab stars and $\log{(P/{\rm{days}})} = -0.5$ ($P = 0.3 $d) for the RRc stars.
These stars are the class of variables with the shortest currently known periods, although the shortest currently known Cepheid variable has a period of only $1.24$ days \citep{1991JApA...12..119S}.

It could also be seen, more clearly in Figure \ref{logP_F_dist_after_transformation}, that the main group is
elongated towards higher $\log{P}$, especially for lower flux amplitudes, and that there was a hint of a separate very weak second grouping at periods above about $525$ days.
 This possibility was investigated further, below.
In an attempt to make the relationship between $P$ and $F_{\rm{amp,LT}}$ in our data clearer, the data were plotted as a scatter plot, using $\log{(P)}$ (instead of $P$) as the $x$ coordinate. This is shown below in Figure \ref{LTflux_v_logP}.

The main group of data between about $\log{(P/{\rm{days}})} = 1.6$ and $\log{(P/{\rm{days}})} = 3.5$ appeared to follow a linear relationship with the flux amplitude, so the gradient of this strong correlation was calculated using linear regression. The equation of a line which is a good fit to all data between $\log{(P)/{\rm{days}}} = 1.9$ and $\log{(P/{\rm{days}})} = 2.72$ was found to be

\begin{equation}
F_{\rm{amp,LT}} = G \log{(P)} + K
\label{logP_LT_flux_amp_relationship}
\end{equation}

 where the values of $G$ and $K$ were $11.92^{+90.76}_{-5.63}$ and $-21.16^{-213.90}_{+13.27}$ respectively.

\begin{figure}[!ht]
\vspace*{9cm}
   \leavevmode
   \includegraphics{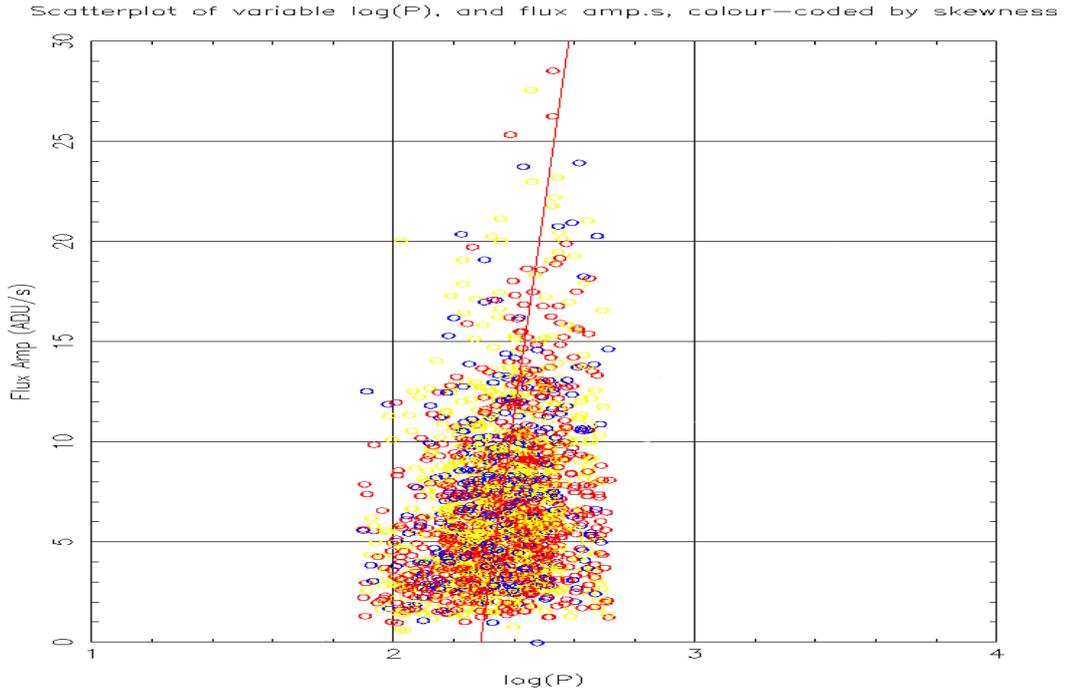}
\caption[Scatter plot showing the relationship of $\log{\rm{(period)}}$ and LT flux amplitude for the main group of variable star candidates with $\log{\rm{(period)}}$ $> 1.9$ and the best fit lines.]{Scatter plot showing the relationship of $\log{\rm{(period)}}$ and LT flux amplitude for the main
group of variable star candidates with $\log{\rm{(period)}}$ $> 1.9$ and the best fit line to the strong linear correlation. Colours represent the skewness of the cosinusoid used to fit the lightcurve. Yellow = near minimum, blue = intermediate, red = near maximum.}
\label{LTflux_v_logP>1.9}
\end{figure}

\begin{figure}[!ht]
\vspace*{9cm}
   \leavevmode
   \includegraphics{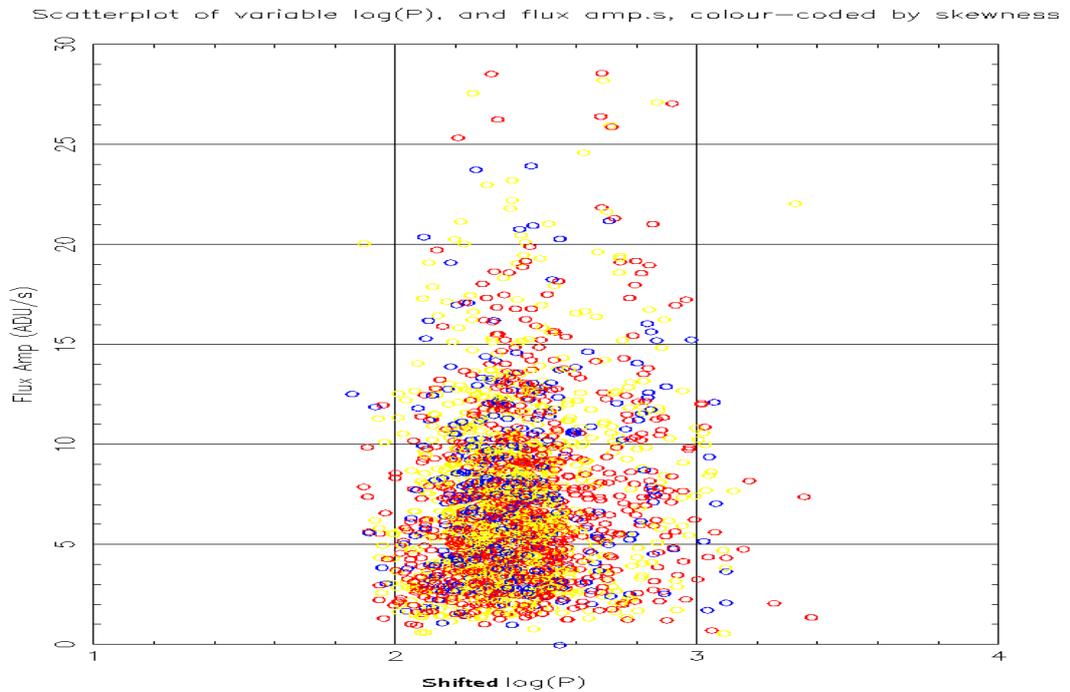}
\caption[Scatter plot of ``shifted'' $\log(P)$ versus variable flux amplitude after the correlation of $\log(P)$ and flux amplitude has been removed by transforming $\log(P)$.]{Scatter plot of ``shifted'' $\log(P)$ versus variable flux amplitude after the correlation of $\log(P)$ and flux amplitude has been removed by transforming $\log(P)$.
}
\label{logP_F_dist_after_transformation}
\end{figure}

The scatter plot of all data with $\log(P/{\rm{days}})>1.9$,
 are shown in Figure \ref{LTflux_v_logP>1.9}. 
As can be seen in the Figure, the line which fitted best the trend formed by the major grouping of points was
the red line corresponding to regression minimisation of the $\log{(P)}$ variations. This had equation $F = 102.68 \log{(P/{\rm{days}})} -235.06$. 
Equation \ref{logP_LT_flux_amp_relationship} can also be written in the form

\begin{equation}
P =  10^{(F_{amp,LT}-K)/G}
\label{P_LT_flux_amp_relationship}
\end{equation}

which allows the expectation value of the period to be estimated given the flux amplitude of the variable.
The calculation of the fit line through the main body of points was performed using
only those lightcurves with $1.9 < \log(P/{\rm{days}}) < 2.72$. These two limits correspond to the
lower edge of the main group of points and the point which is roughly the upper boundary
between the main group and the other possible smaller group of lightcurves which lie at periods 
above $P \sim 525$ days. This meant that $302$ lightcurves were not used because they were 
below the lower
limit and $272$ because they were above the higher limit. This left $1972$ lightcurves to be used in the fitting.

When the fit given above had been found, all the original $2546$ data points were
transformed to remove the fitted dependency on flux amplitude, so that at every value of amplitude the distribution in Period was centred on the same value of $\log(P)$. This was done because in Figure \ref{LTflux_v_logP>1.9} the
distribution in cross section in the $\log(P)$ direction seems to be of a consistent width, and so could be more fundamental than after it has been ``smeared out'' in $\log(P)$ space by the correlation with flux amplitude.

This was done not with a rotation (for example, around the origin), but rather a simple horizontal translation of the data points, as it was not required or desired to change the coordinate value in the flux amplitude (y) direction. However, it was also not desired to 
move the mean value of the group of points, so this was corrected for afterwards.

The translation was done by an initial amount which ``pivoted'' around the best fit line's point of contact with the $\log(P)$ (x) axis. Then the difference between the $\log(P)$ value of this intercept and the original mean value of $\log(P)$ was added back to bring the (now vertical) distribution back to approximately its original position in $\log(P)$.
 Hence, in summary, if $\phi$ is the angle between the line of best fit and the 
vertical (i.e. $\phi = \frac{\pi}{2} - \theta$ where $\theta$ is the angle corresponding to the fitted gradient above) the correction applied is shown in Equation \ref{log_P_transformation}:

\begin{eqnarray}
\log{(P/{\rm{days}})}' &=& \log{(P/{\rm{days}})} - F \tan(\phi) + \bar F \tan(\phi),\\ \nonumber F' &=& F
\label{log_P_transformation}
\end{eqnarray}

This is equivalent to rotating by an angle $\phi$ around the mean of both $\log{(P/{\rm{days}})}$ and $F$ 
but maintaining the $F$ values.

The scatter plot of ``shifted'' $\log(P)$ versus F after the dependence of $\log(P)$ on flux amplitude has been removed is shown in
Figure \ref{logP_F_dist_after_transformation}.

\begin{figure}[!ht]
\vspace*{9cm}
   \leavevmode
   \includegraphics{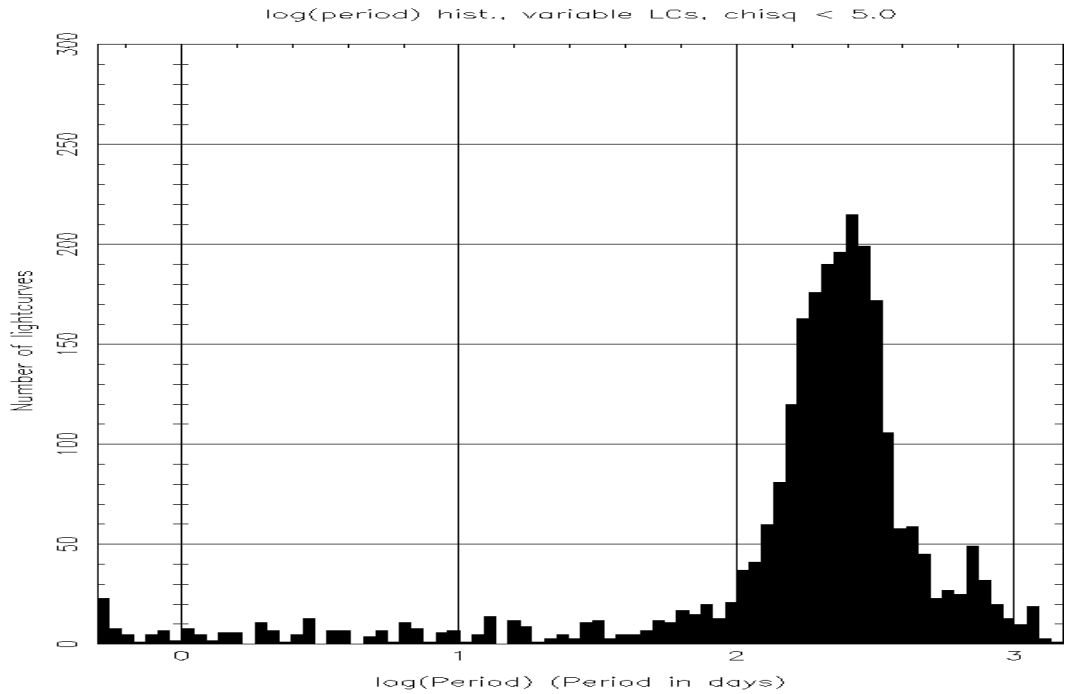}
\caption[Histogram of $\log(P)$ \emph{before} the main group of lightcurves has been transformed
in $\log(P)$ to remove the dependency on flux amplitude.]{Histogram of $\log(P)$ \emph{before} the main group of lightcurves has been transformed
in $\log(P)$ to remove the dependency on flux amplitude.
}
\label{logPhist_chisq_lt_5_no_trans}
\end{figure}

\begin{figure}[!ht]
\vspace*{9cm}
   \leavevmode
   \includegraphics{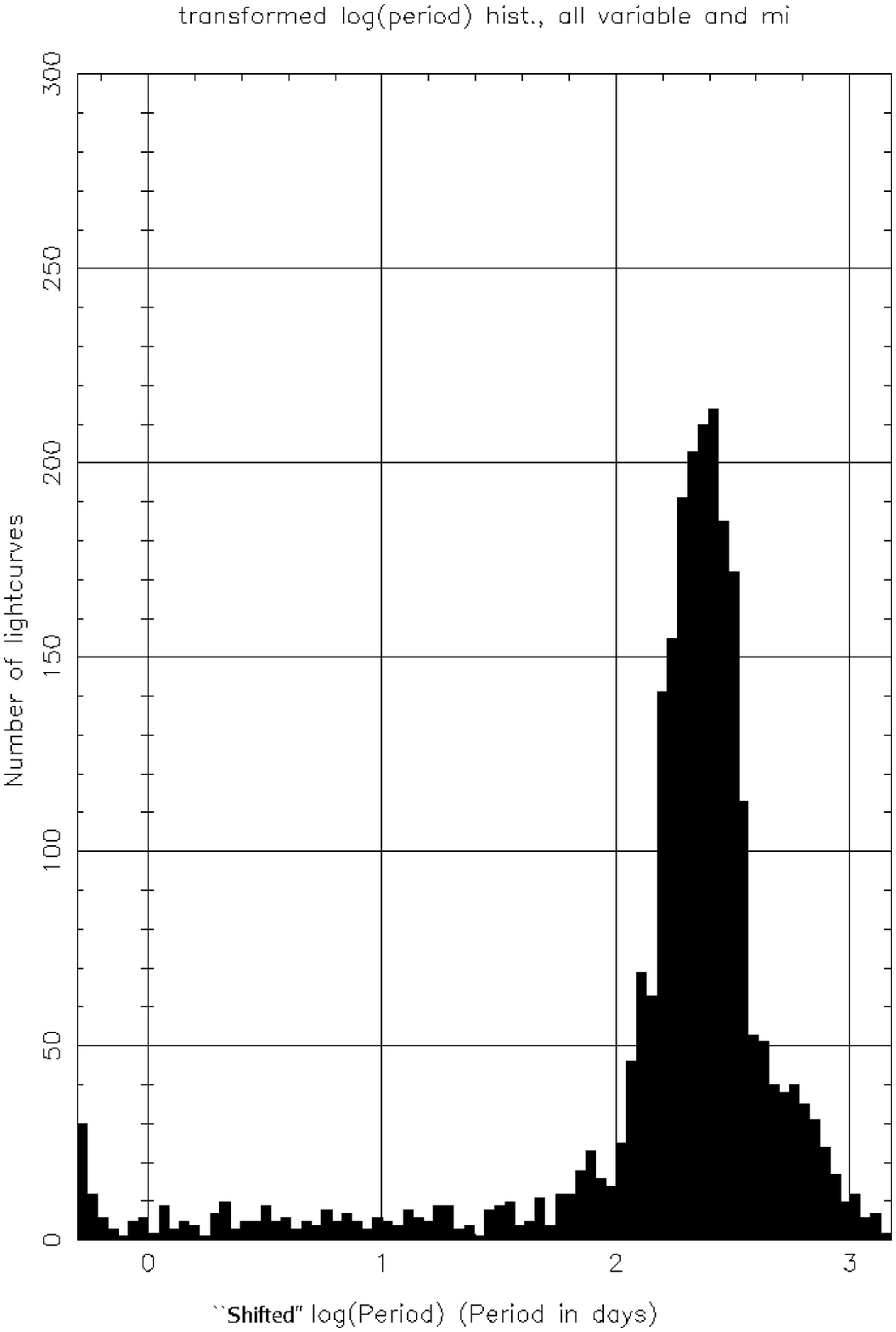}
\caption[Histogram of ``shifted'' $\log(P)$ \emph{after} the main group of lightcurves has been transformed
in $\log(P)$ to remove the dependency on flux amplitude.]{Histogram of ``shifted'' $\log(P)$ \emph{after} the main group of lightcurves has been transformed in $\log(P)$ to remove the dependency on flux amplitude.
}
\label{logPhist_chisq_lt_5_after_trans}
\end{figure}

After this transformation had been applied, the histogram of the ``shifted'' $\log(P)$ distribution
was re-plotted. This is shown in Figure \ref{logPhist_chisq_lt_5_after_trans} with the
equivalent figure before the transformation given in Figure \ref{logPhist_chisq_lt_5_no_trans}
for comparison. Since the transformed angle is not large in $\log(P)$-$F$ space, the difference is not huge, but the peak of the distribution has clearly and predictably shifted downwards in the shifted $\log(P)$ direction and become more rounded. The first result that this process has provided is the ``true'' distribution of
$\log(P)$ at any constant value of $F$; in other words the horizontal cross sectional distribution through the
main group of points. The second motivation for doing all this was to investigate whether the second group of points which occupy the area with $P>\sim525$ days are a second true grouping of variables or 
a grouping caused by some feature of the pipeline program. It was anticipated that if the first option were true, then their distribution would become ``sharper'' when the same transformation
was applied to them as to the main group, since they should follow the same relationship as that group, and would now not be smeared out as much in $\log(P)$ space. However, it can be seen in Figure \ref{logPhist_chisq_lt_5_after_trans} that in fact the opposite has happened, and the previously seen peak has become more smeared out, rather than less. This is apparently 
indicative that the original group with $P>\sim525$ days did not exhibit any clear trend with $F$ and
were on average at a constant value of $\log(P)$. This seems to make it more likely that they are
a grouping which is caused or accentuated by the pipeline program rather than being grouped for some physical reason. This does not mean that these lightcurves, which after all comprise $8.6\%$ of the selected lightcurves, are not valid variable stars, but that the $\log(P)$ distribution in this area should not be over-interpreted. Also, the statistics are still low, and
it may still be possible for features in the distribution of $\log(P)$ versus $F$ to appear, change or disappear given more or better data.

The relationship between period and skewness was also investigated, as this seemed an interesting area to look for possible new correlations. No significant correlations were
found in any of the above $16$ smaller groups, or in the combined sample, so the plots are not presented here.
The only visible systematic correlation was that already investigated above, in which
lower variable amplitude lightcurves are found on average to have lower periods, which is also shown by Figure \ref{LTflux_v_logP} above.

\begin{figure}[!ht]
\vspace*{14cm}
   \leavevmode
   \includegraphics{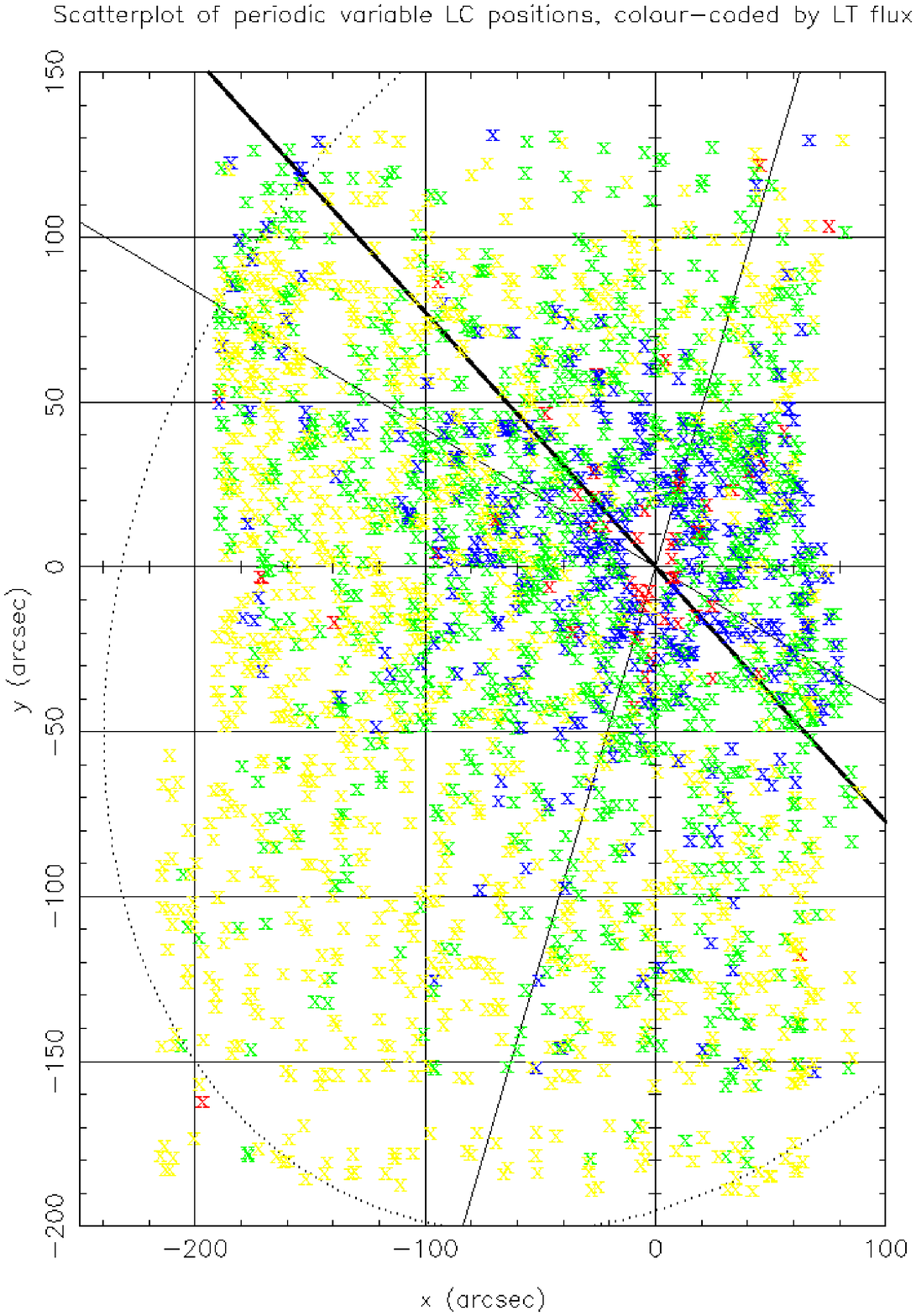}
\caption[Spatial distribution of $2546$ variable and mixed lightcurves selected to have reduced $\chi^2 < 5.0$.]{Spatial distribution of all $2546$ variable and mixed lightcurves selected to have
reduced $\chi^2 < 5.0$.
}
\label{var_spat_dist_all_lcs}
\end{figure}

The spatial distribution of all selected ``variable'' and ``mixed'' objects with 
cosinusoid fits having reduced $\chi^2 < 5$ were plotted in Figure \ref{var_spat_dist_all_lcs}.

An additional dimension for investigation was made available by using the variable flux amplitude.
 The spatial plot was coarsely divided into four flux amplitude bins (represented by the different 
colours in Figure \ref{var_spat_dist_all_lcs}) in order to investigate whether the spatial distribution changed significantly with variable amplitude.
 Consistently in all the spatial distribution plots presented below, the colour codes 
are defined as follows: 
LT variable flux amplitude, F : 
\newline Yellow: $ F < 5.0$ ADU/s,
\newline Green: $5.0 \leq F < 10.0$ ADU/s,
\newline Blue: $10.0 \leq F < 20.0$ ADU/s,
\newline Red: $20.0 < F $ ADU/s.

The dark line, which is inclined at $37.7^\circ$ to the y axis, represents the orientation or ``position angle'' of the M31 disk, as measured by \cite{1958ApJ...128..465D}, and the smaller ellipse represents one possible model for the orientation 
of the M31 bulge, represented as a triaxial ellipsoid being $0.955$ kpc along the long axis of the ellipsoid. The particular model illustrated is taken from \cite{1977ApJ...213..368S}, namely the model with $\phi = 44^\circ$. Although the original model had the length of the long axis of the ellipsoid
equal to $2.76$ kpc, this has been halved to fit on the scale of the LT field. The orientation of the ellipse describing the outside surface of this ellipsoid, as seen by us, is $10^\circ$ below (on the left) the M31 disk, consistent with \cite{1956StoAn..19....2L}. The pair of less dark lines
represent the major and minor axes of an ellipsoidal bulge which appears to the observer to be oriented $10^\circ$ below (on the left) the M31 disk.

 Several features can immediately be observed in Figure \ref{var_spat_dist_all_lcs}.
 Firstly, the density of points in general increases markedly towards the centre (0,0)
 of the galaxy. This is clearly consistent with the increasing density of stars towards
 the centre of the galaxy. Secondly, the central concentration of variables can be
 seen to increase as the magnitude of the flux amplitude increased. This is consistent
 with variable stars with greater variability being comparably more easily observed near to the
 centre of the galaxy, where the photon noise from the subtracted galaxy background is
 the greatest. The converse of this statement is to notice that there is a relative lack of yellow (relatively low amplitude) variables observed in the central region.
This may be explained by their variable signal being swamped by the high photon noise in the centre.
Thirdly, there appears to be a clear horizontally-oriented gap in the data at a y value of roughly $-160$ to $-170^{\prime\prime}$. The reason for this is not clear, as any dust lane would be expected to be oriented at the same angle as the disc, i.e. about $37.7$ degrees to the vertical, but, if the position is compared with the bottom edge of the objects which contain PA data seen in Figure \ref{colour_coded_flux_ratio_spatial_correlation_1185} the y coordinate appears similar. The reason the overlap of LT+PA data as shown by the coloured points in Figure \ref{colour_coded_flux_ratio_spatial_correlation_1185} is limited on the bottom edge is because of the edge of the LT data, not the PA, so it is possible that the gap mentioned above coincides with the most frequent position of the edge of the LT frame, which is masked out due to image defects
on the edges of images.

Finally, the general distribution (looking, for example, at the blue points) does seem to be elliptical, as expected (remembering the x and y scales on the plot are different!). The apparent ratio of y radius / x radius appears to be $\sim2.0$, e.g. $\sim50^{\prime\prime}$ in y compared to $\sim100^{\prime\prime}$ in x, as shown by Figure \ref{gridded_contour_plot}. This ellipse does not, by eye, appear to be oriented in the same direction as the M31 disk line, and does seem to be lower than it on the left of the centre, as expected, but to a greater degree than might be expected. This is investigated further below in Section \ref{bulge_angle}.

To further investigate point one above, the variables are grouped according to their flux amplitude, in bins of size $1$ ADU/s, up to a maximum of $16-17$ ADU/s.
Then the spatial distributions are plotted for each amplitude bin.
The boundary between the distribution being concentrated around the edge of the area plotted (with a noticeable central hole) and it being centrally concentrated on the galaxy (with no central hole) appeared to be at around $4-5$ ADU/s. To illustrate this, the three groups spanning this flux amplitude are presented below in Figure \ref{var_spat_dist_F_3_to_6}. From the full range of these $16$ plots there are several examples of non-random-looking distributions involving clusters of points, implying that perhaps these represent real transverse distributions which vary in detail from the global average ``symmetrical, centrally concentrated elliptical'' distribution.

\begin{figure}[!ht]
\vspace*{7cm}
$\begin{array}{c}
\vspace*{7cm} 
   \leavevmode
 \includegraphics{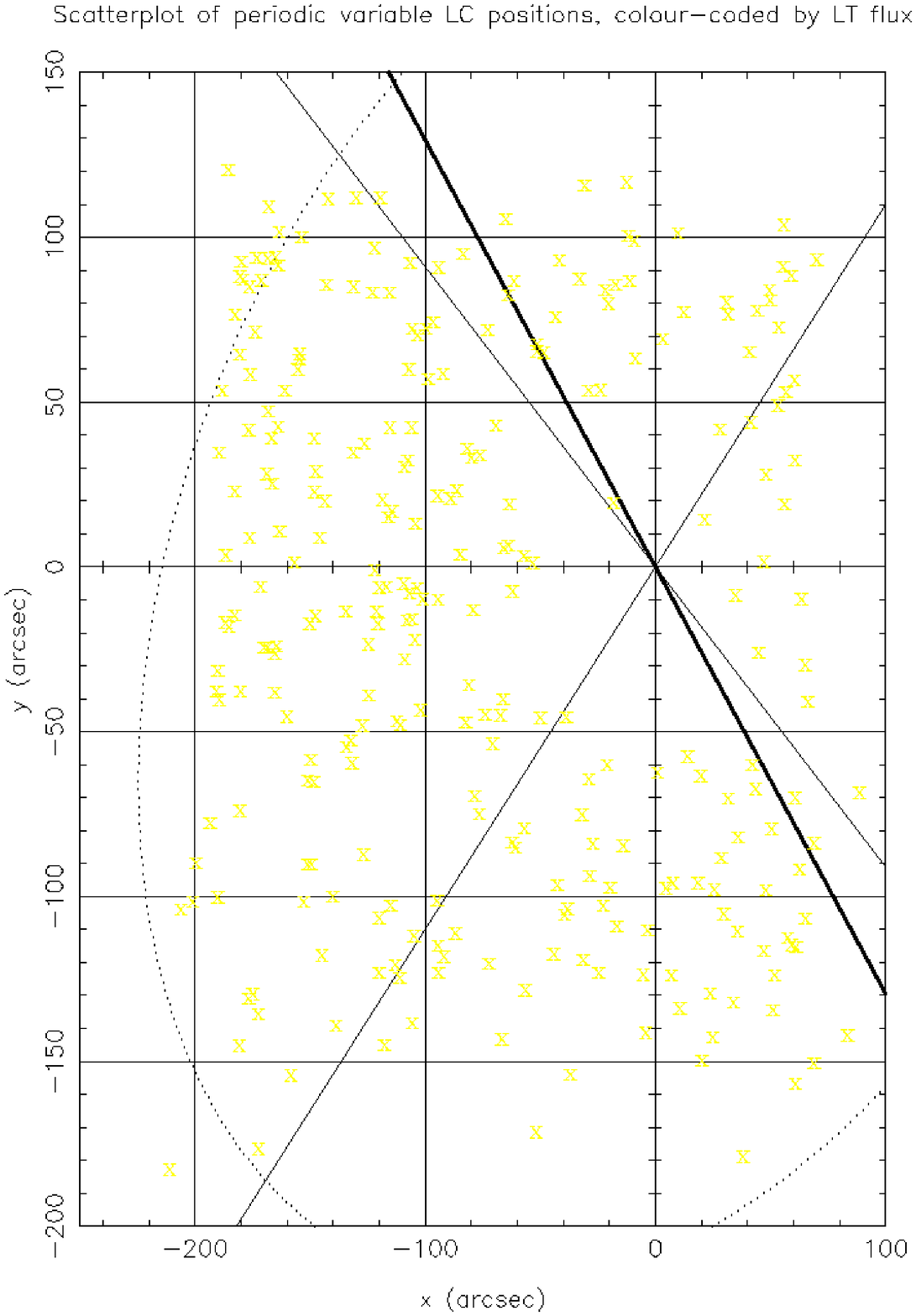} \\
\vspace*{7cm}
   \leavevmode
 \includegraphics{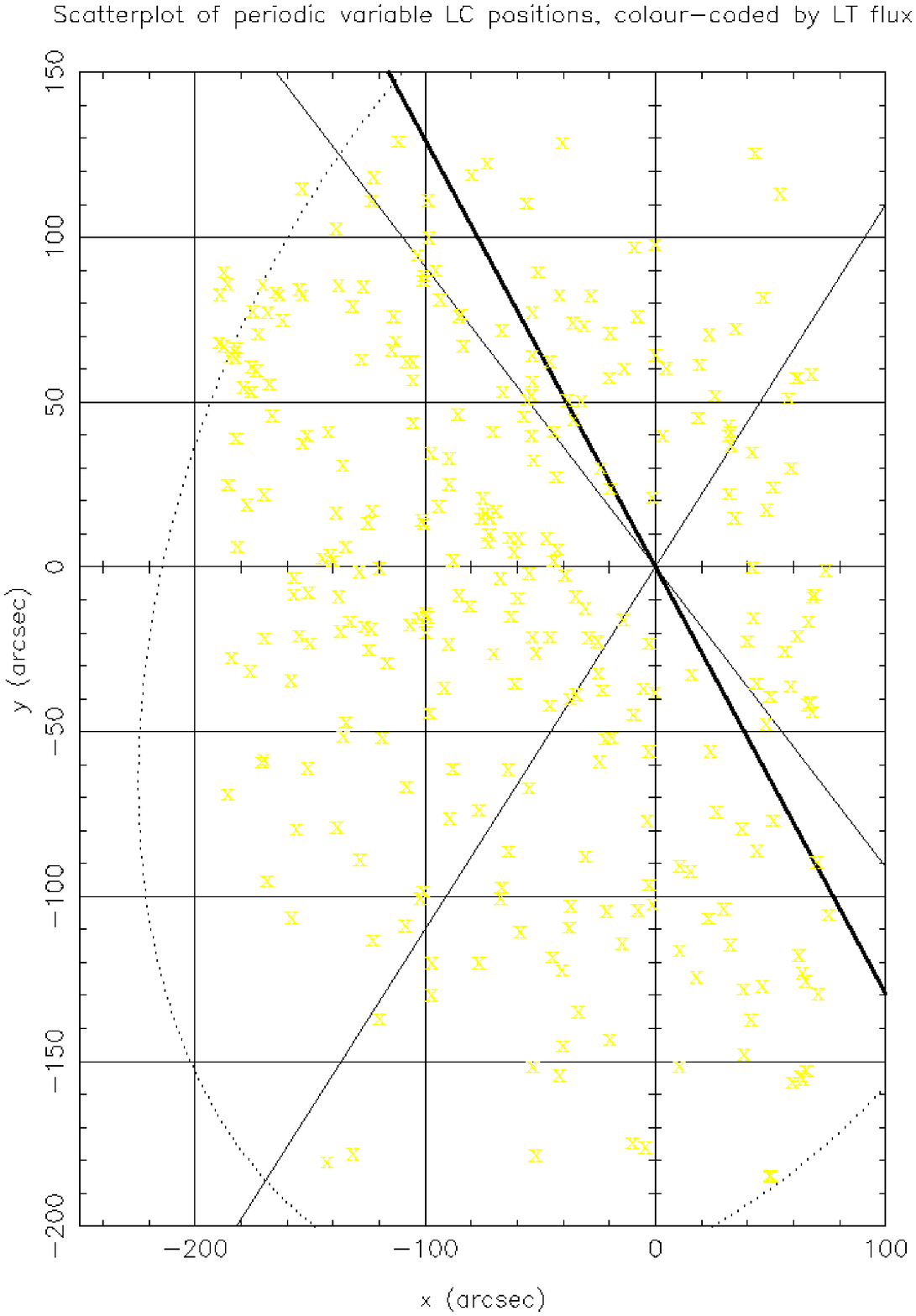} \\
\vspace*{1cm}
   \leavevmode
 \includegraphics{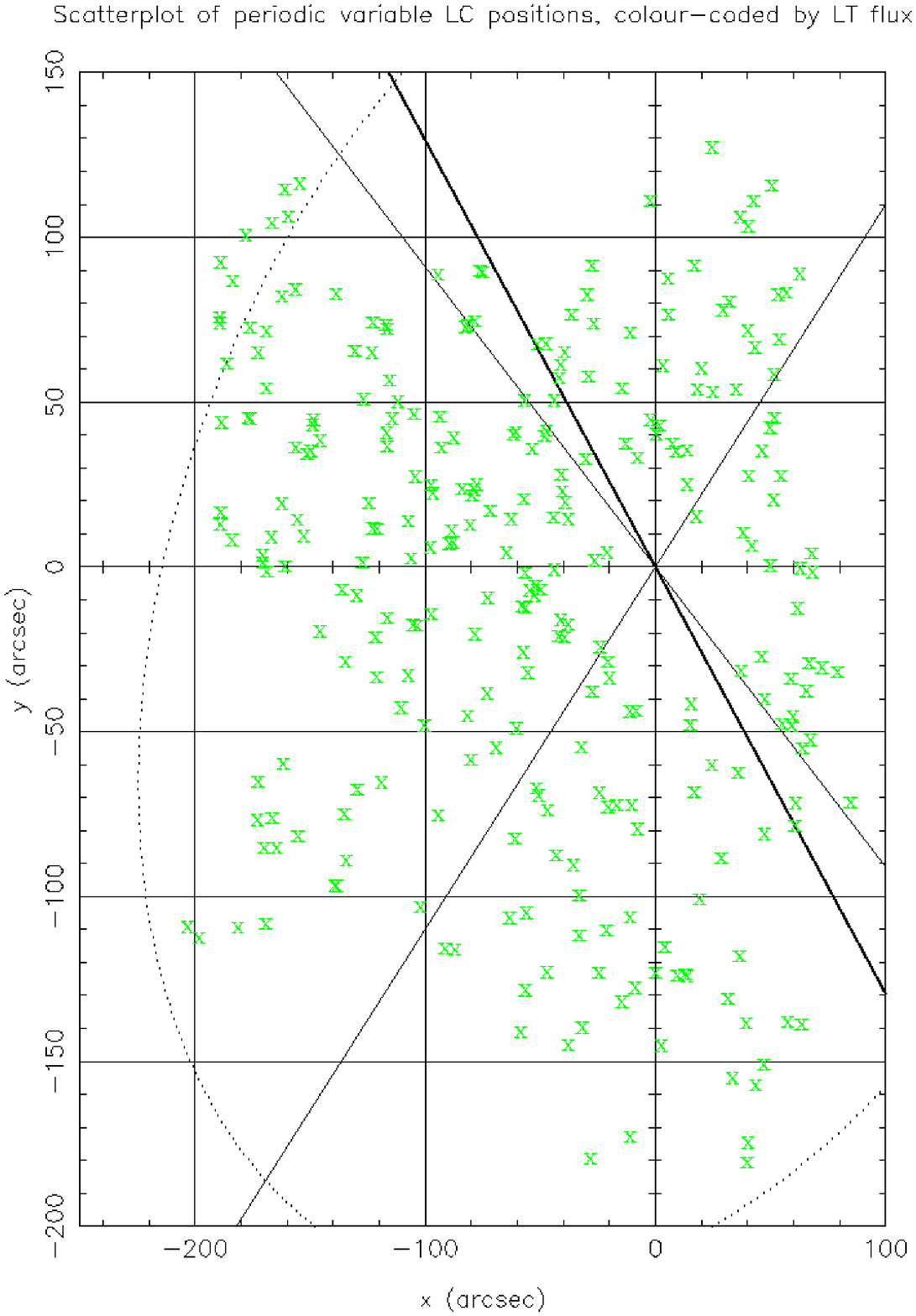} \\
\end{array}$
\caption[Spatial distribution of variable stars with flux amplitudes in the ranges 3-4, 4-5 and 5-6 ADU/s.]{Spatial distribution of variable stars with flux amplitudes in the ranges a) 3-4, b) 4-5 and c) 5-6 ADU/s, showing the cross-over region between central dominated and central sparse distributions.}
 \label{var_spat_dist_F_3_to_6}
\end{figure}

\begin{figure}
  \begin{tabular}{cc}
   \includegraphics[height=95mm]{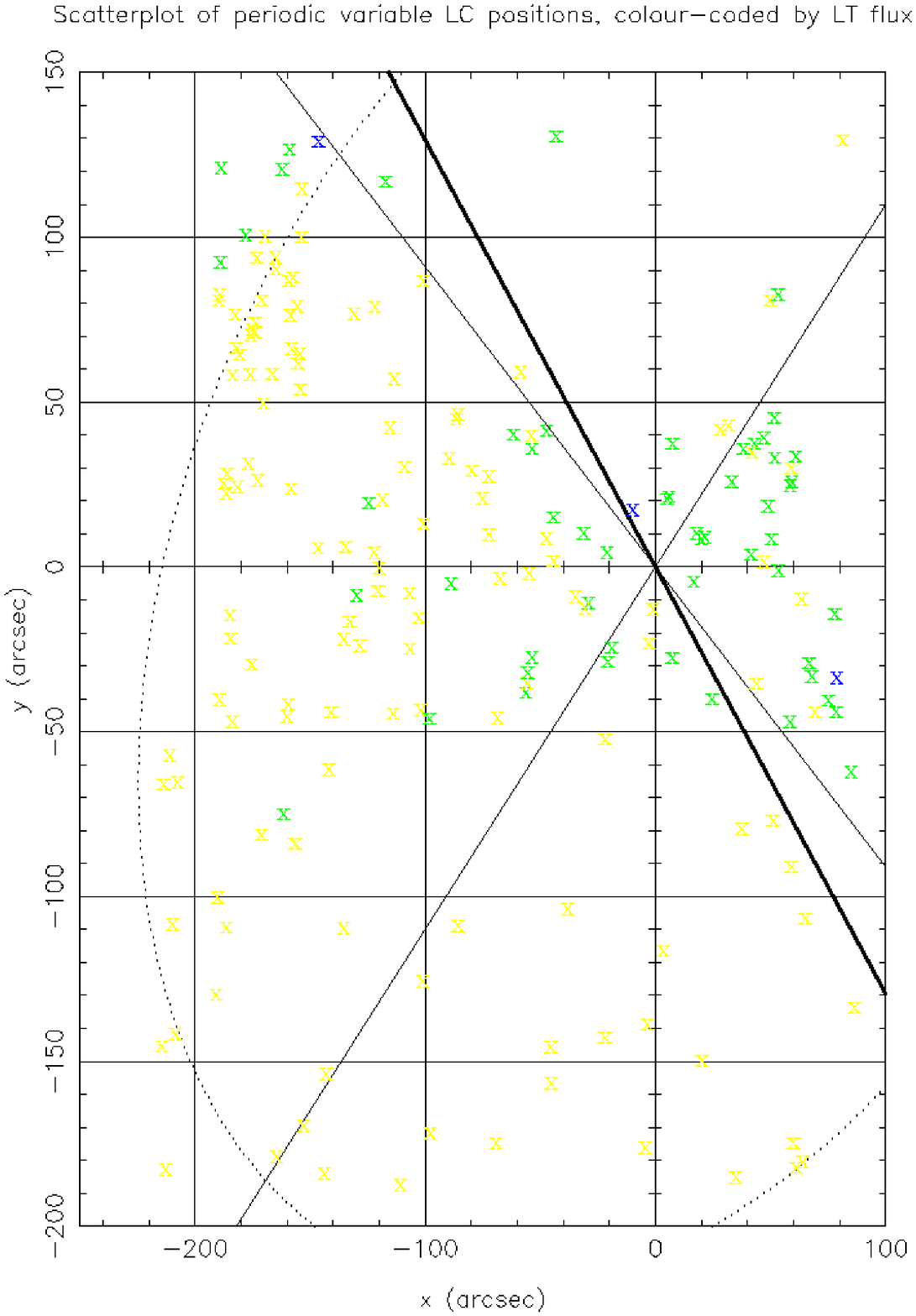} \\
   \includegraphics[height=95mm]{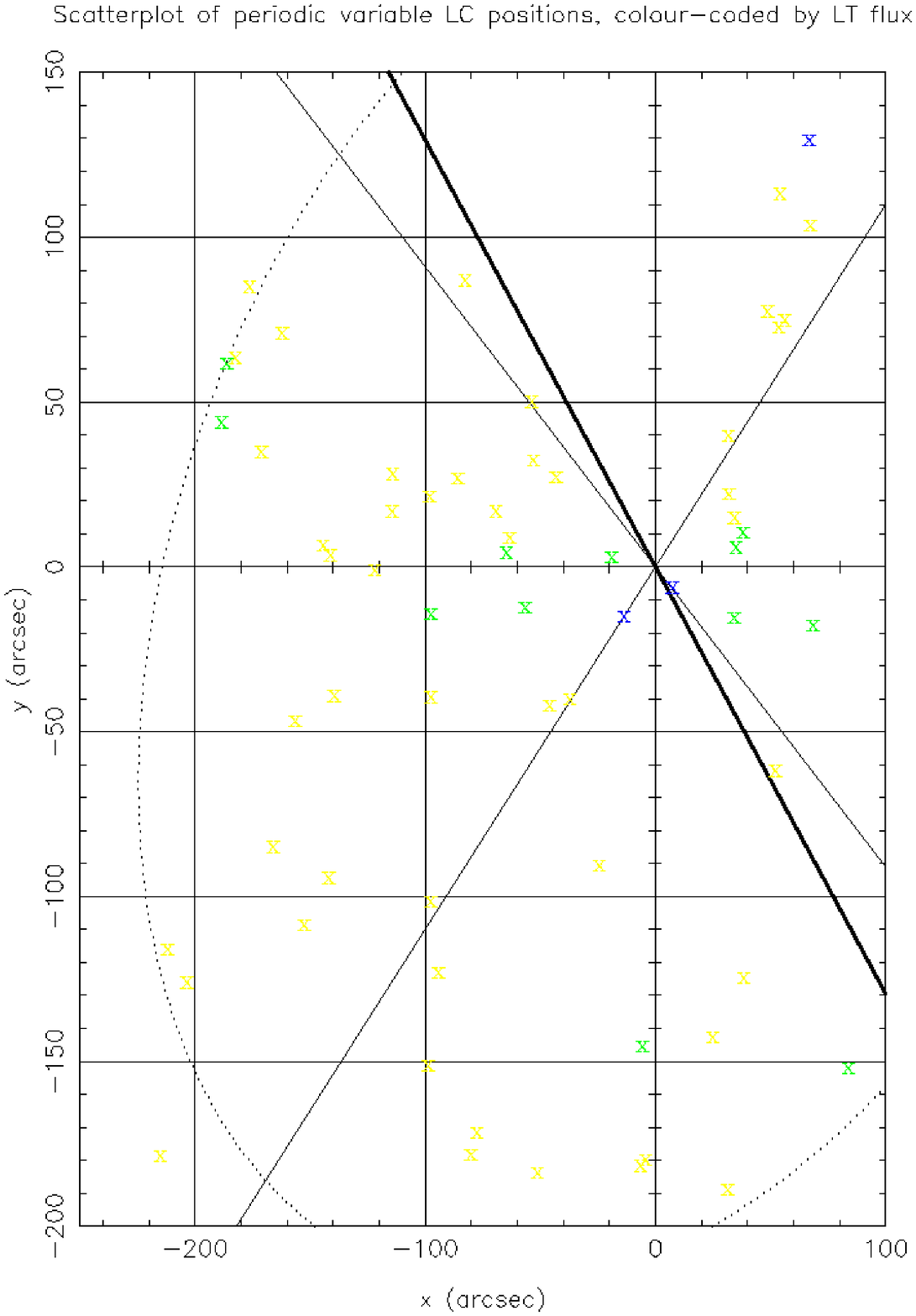} \\
  \end{tabular}
  \begin{tabular}{cc}
   \includegraphics[height=95mm]{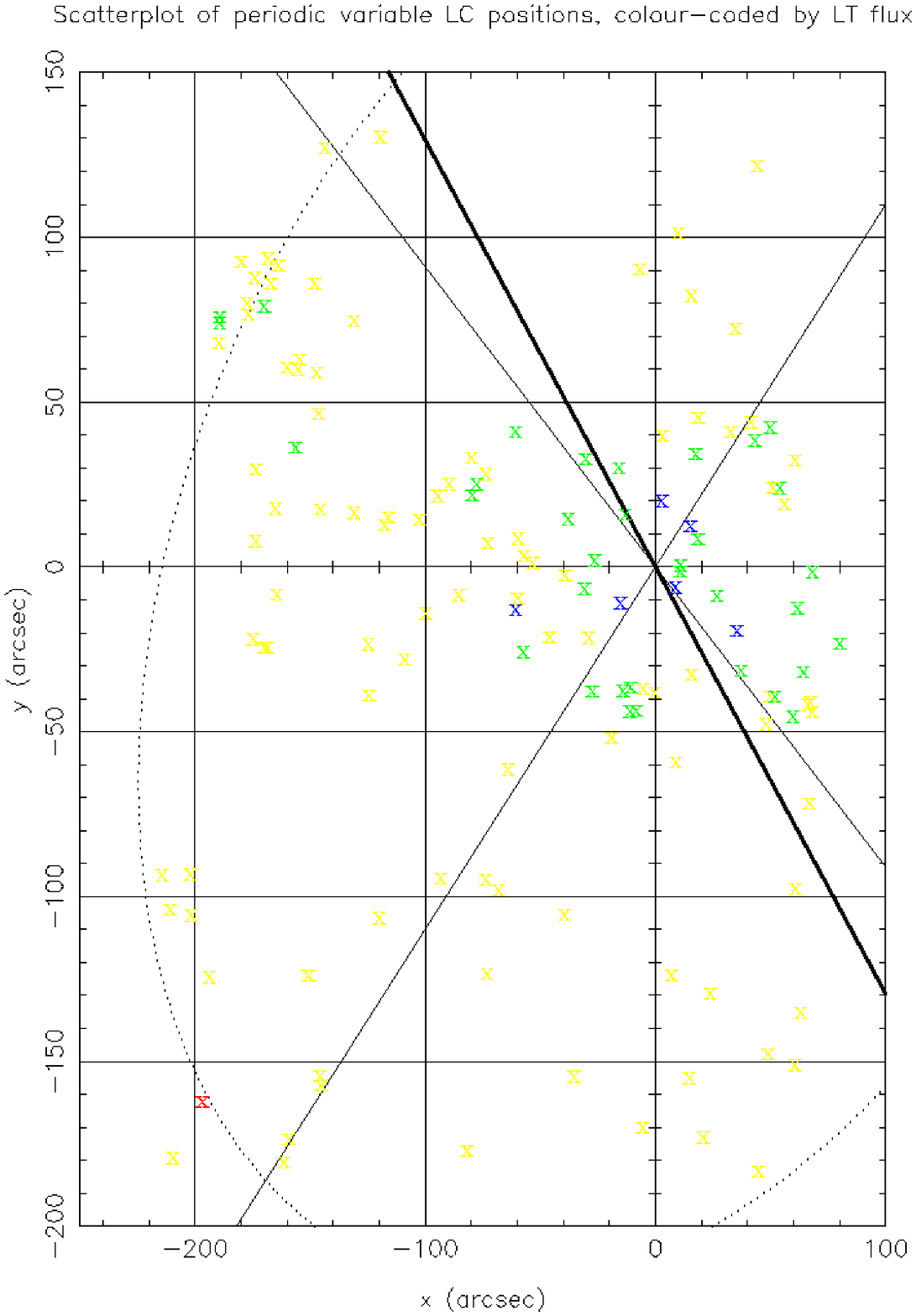} \\
   \includegraphics[height=95mm]{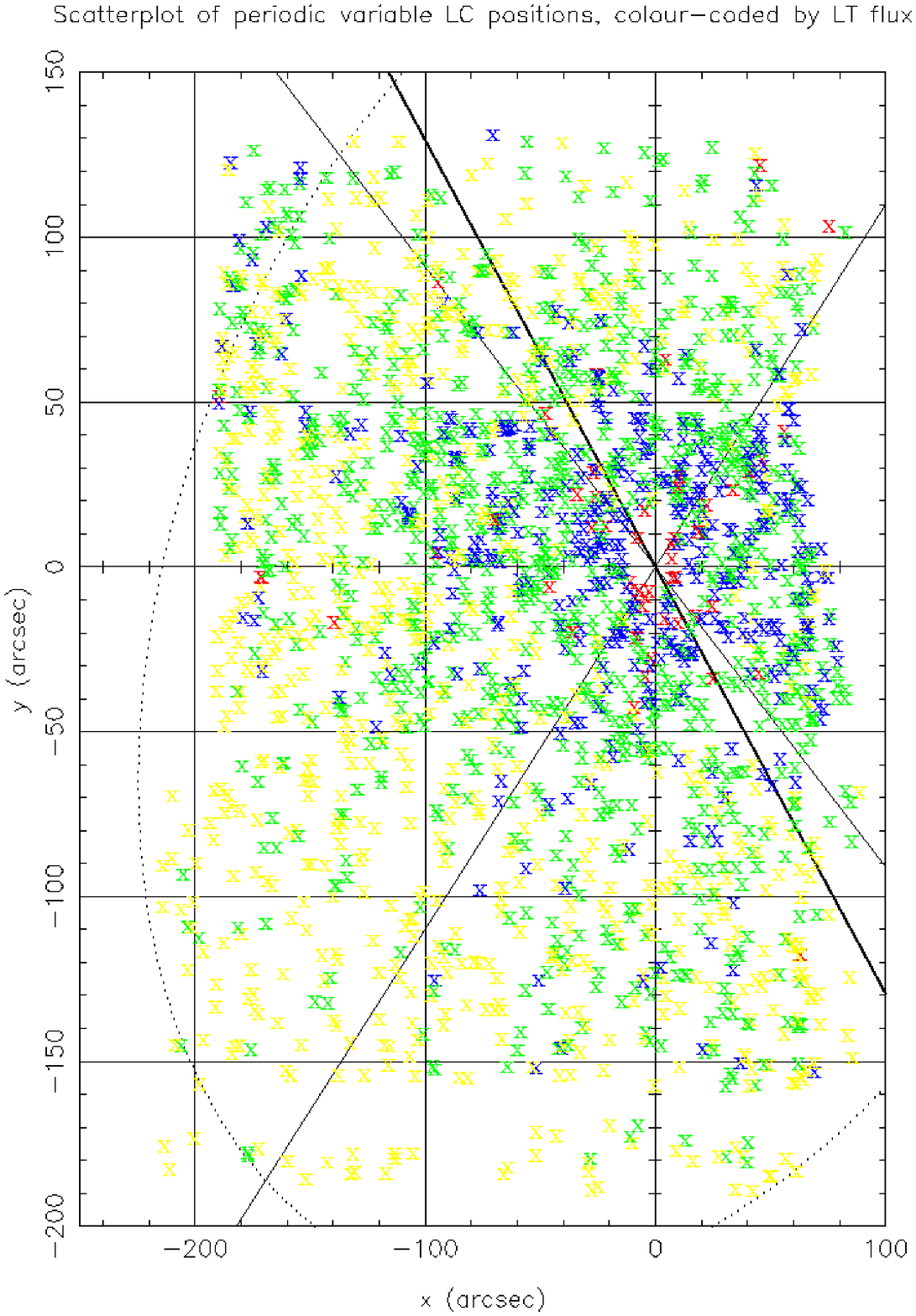} \\
  \end{tabular}
 \caption[Spatial distribution of variable stars with fitted periods in four different
 ranges.]{Spatial distribution of variable stars with fitted periods in four different
 ranges- (left to right and top to bottom): a) below 10 days b) between 10 and 30 days c) between 30 and 100 days d) greater than 100 days.}
 \label{spatial_dist_various_period_slices}
\end{figure}

It is also interesting to plot the spatial distributions of variable stars grouped according to period, to see whether the obvious groups on the Period-Amplitude plots varied in their spatial distributions and so might or might not be separate physical populations.
Four groupings according to period are shown in Figure \ref{spatial_dist_various_period_slices}, being $P \leq 10$ days, $10 \leq P < 30$ days, $30 \leq P < 100$ days and $100$ days $< P$. These divisions were guided by the apparent concentrations of lightcurves in period that appeared in Period-Amplitude plots such as Figure \ref{logP_F_dist_after_transformation}.

\clearpage

\subsubsection{Orientation and Ellipticity of the distribution of variable objects}
\label{bulge_angle}

One of the earliest things that was done in attempting to investigate the orientation of the M31 bulge was to attempt to better visualise the data.
To this end, contour plots of the density of variable candidates in the LT field were constructed, by binning the data in square bins. 
With a quite coarse bin size of $40^{\prime\prime}$,
some idea of the underlying structure could be gained. The resulting contour plot is shown below in Figure \ref{gridded_contour_plot}. The edges of the data affected the contours nearest to the edge, but there was sufficient central data to show a clear central quasi-elliptical peak (elliptical within the limits of the coarse binning), which appeared to be oriented in an almost horizontal direction with respect to the edges of the field, surrounded by (especially in the lower left) other contours which seem to lie in a similar direction as is expected for the M31 disk (shown by the diagonal line on the plot). Examining the three innermost contours surrounding the central peak,
the outermost of these, at the $50$ contour level, is roughly $220^{\prime\prime}$ in diameter in the x direction and $110^{\prime\prime}$ in the y direction. The next innermost, at the $65$ contour level, is roughly $180^{\prime\prime}$ in x and $70^{\prime\prime}$ in y. The two ratios of elliptical major axes divided by minor axes (a/b) derived from these two contours are $2.0$ and $2.6$. 
It may also be observed from Figure \ref{gridded_contour_plot} that there appear to be two central peaks, and both of them are centred about $10-15^{\prime\prime}$ above the zero in the y direction. 
 The gap between the two peaks appears to be consistent with the 
expected incompleteness of the survey in the very central regions of the field. This is investigated in more detail below in Section \ref{radial_distribution}.

\begin{figure}[!ht]
\vspace*{13cm}
   \leavevmode
   \includegraphics{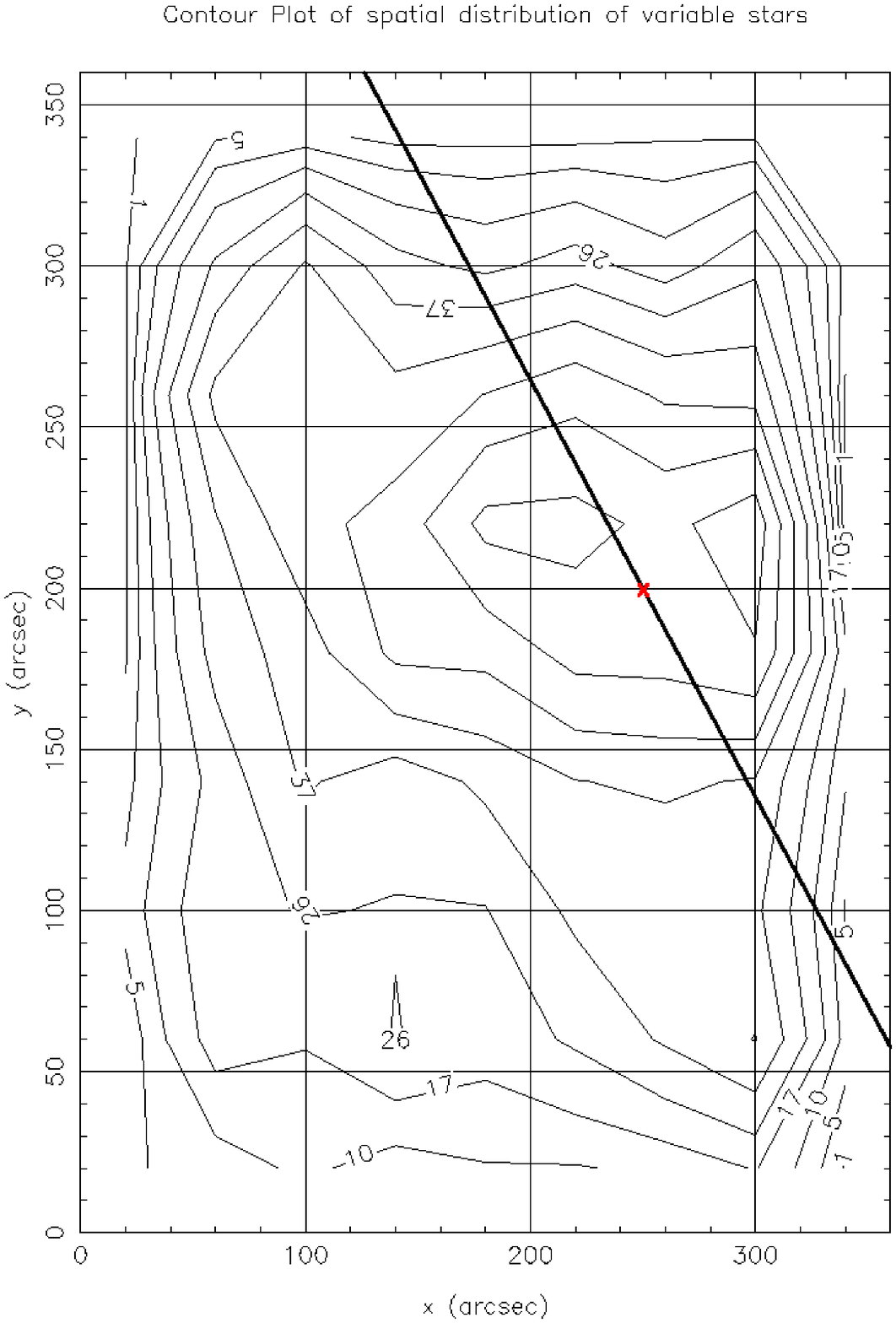}
\caption[Contour plot showing the way the density of variable stars changes over the LT field.]{Contour plot showing the way the density of variable stars changes over the LT field. The data were binned using $9$ x $40^{\prime\prime}$ bins in each direction. The diagonal line represents the orientation of the M31 disk, as measured by de Vaucouleurs \citep{1958ApJ...128..465D}.
}
\label{gridded_contour_plot}
\end{figure}

\textbf{Investigating the azimuthal distribution}

 The second investigation that was performed was to divide the field into circular annuli and then to divide each annulus into azimuthal sectors and to bin the objects in that direction also.
The aim was to see if any coherent trend could be detected in the angle of the azimuthal sector which had the most objects in. If so, then the distribution of objects could be confirmed as non-random, and could therefore contain potentially useful information.
If the assumption is made that the distribution of objects on the plane of the sky may be modelled by a two dimensional elliptical Gaussian distribution with some unknown vertical (number) radial dependence, and this distribution is integrated around circular annuli, it would be expected that each annulus would possess two diametrically opposite peaks corresponding to the major axis of the elliptical distribution, along which the higher contours project relatively further out into a particular circular annulus than on the minor axis, where the contours would be more radially compressed. This hypothesis was tested by dividing the field into circular annuli.

The maximum radius of the outermost ring was set at $200^{\prime\prime}$, close to the left hand edge of the data, but well inside the data point with the largest radius, which lay at $281^{\prime\prime}$.
Different combinations of numbers of sectors and annuli were experimented with
to find the best compromise between angular and radial resolution and statistics in each bin, with the aim of getting the maximum amount of information out of the available data, without straying into the regime of noise. Combinations which produced reasonable results were
(16 sectors, 4 rings), (32 sectors, 4 rings), (32 sectors, 8 rings) and (64 sectors, 4 rings). With either more rings or more sectors than this, the statistics of each bin were such that a clear peak could not easily be defined due to the $\sqrt{N}$ noise dominating any signal. 
The criterion used to decide whether a particular ring had good enough statistics to be selected as a data point was in two parts. ``peaks'' in the data were first defined as an angular region having continuously falling number density on either side of a local maximum.
Then the numerical difference between whichever was the bin with the highest number of objects in it and the bin corresponding to the highest bin of the second highest peak had to be greater than the sum of the square roots of the two numbers involved. 
In other words, the largest peak had to be clearly highest by the sum of the two Poissonian errors of the two highest peaks. Secondly, to avoid ruling out any ring which happened to detect the diametrically opposite
peak that would be expected on the other side of the galaxy centre at an almost equal level to the first, secondary peaks were only considered if they lay within $\pm 90^{\circ}$ of the largest peak.
    Many rings contained data which were clearly highly correlated with rings 
around them, which increased confidence that the peaks found were non-random (i.e. physical) phenomena. Applying the above two criteria strictly to individual rings with the above four sets of sectors and rings led to $9$ sufficiently clear data points (out of a possible $20$). Due to the incomplete angular coverage of the data (due in turn to the edges of the rectangular LT field) not all angular bins contained points and those which intersected the edge of the data contained fewer points than they would have done in an infinite field. Rings with smaller radii had better angular coverage, being further from the edges of the data. 
In order to make best use of data not selected in the first selection process, and to gain as much information as possible on the inner regions of the galaxy, some rings adjacent to one another which failed the above tests individually were co-added to increase the statistics in the resulting wider ring. The errors on the mean radius were, of course, increased in proportion. These summed sets of bins were then also subjected to the above tests. Only $4$ further data points could be derived in this way, making a total of $13$. It should be noted, that because binning the data in these various different ways does not invent new data, the $13$ points are not completely independent, and should be expected to be significantly correlated.
However, because using a greater number of angular bins narrowed the angular error (usually at the expense of the best attainable radial error) and vice versa, using different combinations of bin sizes in the two directions better defined the limits of the useful information which could be extracted from this data set.
The final thirteen data points, with their limits, are given below in Table \ref{azimuthal_angle_data}.

\begin{table}
\caption{Table listing the data points derived for the investigation of
the azimuthal angle of the peak of the azimuthal number density distribution of variable star candidates.}
\begin{center}
\begin{tabular}{|c|c|c|c|c|c|c|c|}
\hline
\hline
 Number  & Number of &  Rings  &  Peak  &  Peak & Angle &  Mean  & Radius \\
of Rings &  Sectors  &  Used   & Sector & Angle & Error & Radius &  Error \\
\hline
 $16$ & $4$ & $1$   &   $9$   & $163.550$ & $11.2500$ & $175.0$ & $25.0$ \\
 $16$ & $4$ & $2$   &   $9$   & $163.550$ & $11.2500$ & $125.0$ & $25.0$ \\
 $16$ & $4$ & $3$   &   $9$   & $163.550$ & $11.2500$ & $75.0$  & $25.0$ \\
 $16$ & $4$ & $3+4$ &   $9$   & $163.550$ & $11.2500$ & $50.0$  & $50.0$ \\
 $32$ & $4$ & $1$   &   $17$  & $157.925$ &  $5.6250$ & $175.0$ & $25.0$ \\
 $32$ & $4$ & $3$   &   $18$  & $169.175$ &  $5.6250$ & $75.0$  & $25.0$ \\
 $32$ & $4$ & $1+2$ & $17,18$ & $163.550$ & $11.2500$ & $150.0$ & $50.0$ \\
 $32$ & $4$ & $2+3$ &   $18$  & $169.175$ &  $5.6250$ & $100.0$ & $50.0$ \\
 $32$ & $4$ & $3+4$ &   $18$  & $169.175$ &  $5.6250$ & $50.0$  & $50.0$ \\
 $32$ & $8$ & $1$   &   $17$  & $157.925$ &  $5.6250$ & $187.5$ & $12.5$ \\
 $32$ & $8$ & $2$   &   $18$  & $169.175$ &  $5.6250$ & $162.5$ & $12.5$ \\
 $32$ & $8$ & $5$   &   $17$  & $157.925$ &  $5.6250$ & $87.5$  & $12.5$ \\
 $64$ & $4$ & $1$   &   $33$  & $155.125$ &  $2.8125$ & $175.0$ & $25.0$ \\
\hline
\end{tabular}
\end{center}
\label{azimuthal_angle_data}
\end{table}

 It can be seen that there is a tight correlation in the angles in each ring
 which contained the most data points.
 The data clearly cluster around $\sim161^{\circ}$, measured (as in all this
 work) from the positive x axis, which corresponds in turn to $\sim33^{\circ}$ 
 \emph{above} (on the right) the angle of the M31 disc measured by \cite{1958ApJ...128..465D}.
 The best fit constant value for the angle corresponding to the peak number
 density was found to be $161.3^{\circ}$($\pm1.61^{\circ}$ fitting error).

\begin{figure}[!ht]
\vspace*{20cm}
   \leavevmode
   \includegraphics{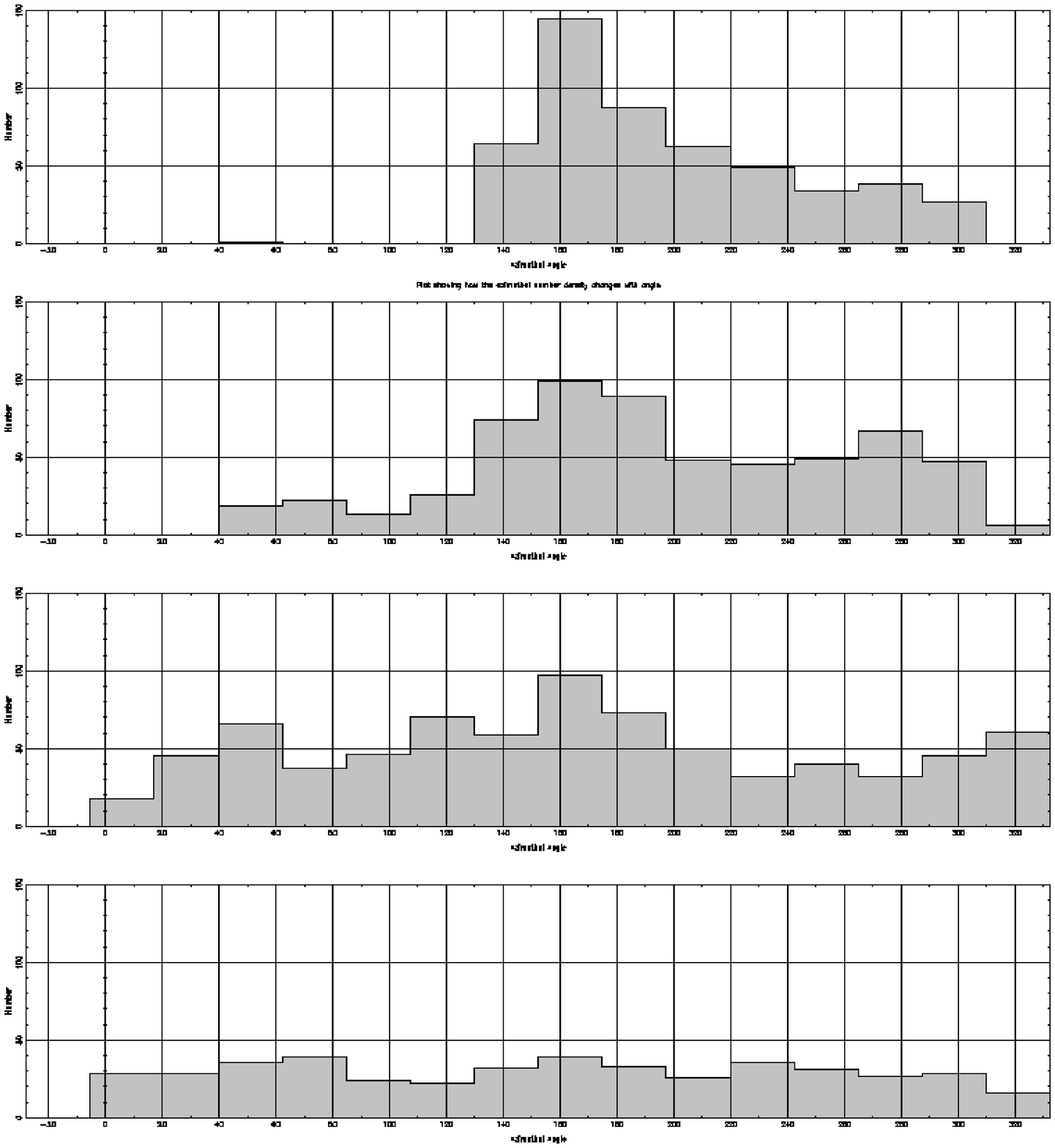}
\caption[Histogram showing the variation in the angularly binned (with $16$ equal bins) numbers of variable candidates, divided also into four radial annuli.]{Histogram showing the variation in the angularly binned numbers of variable candidates, divided also into four radial bins, each of radial width $50^{\prime\prime}$. The major division in azimuthal angle (horizontally), is $20$ degrees, and in number, $50$ objects.)
}
\label{azimuthal_histogram_16_4}
\end{figure}

\begin{figure}[!ht]
\vspace*{20cm}
   \leavevmode
   \includegraphics{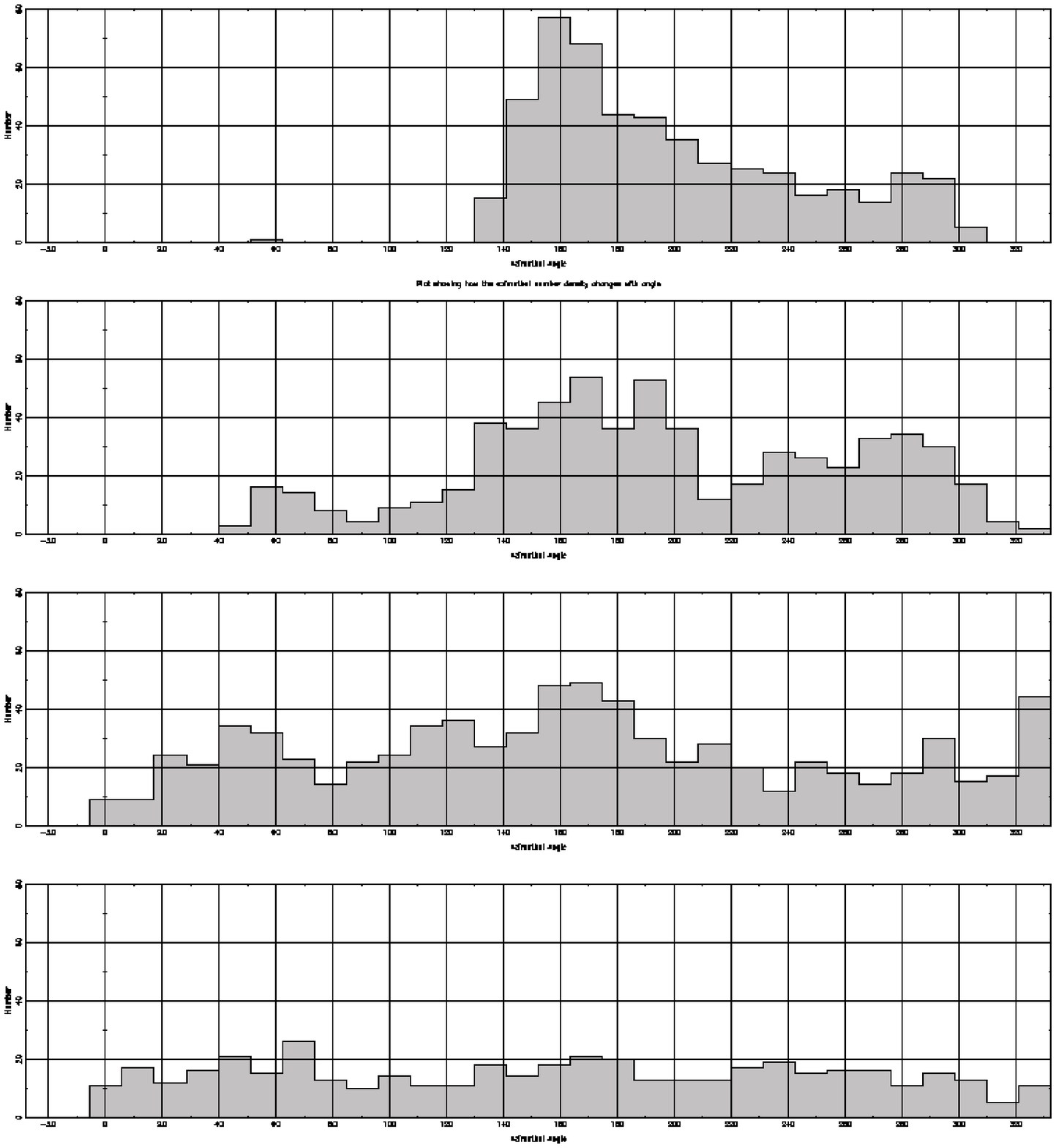}
\caption[Histogram showing the variation in the angularly binned (with $32$ equal bins) numbers of variable candidates, divided also into four radial annuli.]{Histogram showing the variation in the angularly binned (with $32$ equal bins) numbers of variable candidates, divided also into four radial annuli, each of radial width $50^{\prime\prime}$. The major division in azimuthal angle (horizontally) is $20$ degrees, and in number, (vertically), $20$ objects.
}
\label{azimuthal_histogram_32_4}
\end{figure}

Two examples of azimuthal distributions, as investigated above, plotted as histograms for differing
radial bin size and angular bin size, are shown in Figures \ref{azimuthal_histogram_16_4} and \ref{azimuthal_histogram_32_4}.
The main peak can clearly be seen to appear in the three outer radial annuli in both plots.

In order to correctly interpret the information gathered above and apply it to the investigation of the bulge of M31 it would be necessary to know the completeness of our survey for the types of variable objects selected here. At this stage, this information is not known. If it were, and it could be proved that the variable objects used here are indeed variable stars and they are located in the bulge, then the information contained in the spatial distributions of objects described in this section might be used as a proxy for the more general distribution of matter in the inner M31 bulge. Given what is currently known, however, it is only possible to conclude that the distribution of variable objects selected by the candidate selection pipeline is centrally concentrated, with what appears to be a central hole probably due to the incompleteness of our survey, and that the peak of its radial distribution does not align either with the angle of the outer disk measured by \cite{1958ApJ...128..465D} or with the angle of the bulge measured by e.g. \cite{1956StoAn..19....2L}. We cannot conclude that the actual distribution of variable stars is inconsistent with either of these orientations, due to our lack of knowledge of the completeness of our survey.
The radial distribution of variable objects, and specifically the form of the central hole in the data, is investigated in the next section, to gain what knowledge can be gained at this stage about the central incompleteness.

\subsubsection{Radial distribution of variables and central completeness}
\label{radial_distribution}

\begin{figure}[!ht]
\vspace*{10cm} 
   \leavevmode
   \includegraphics{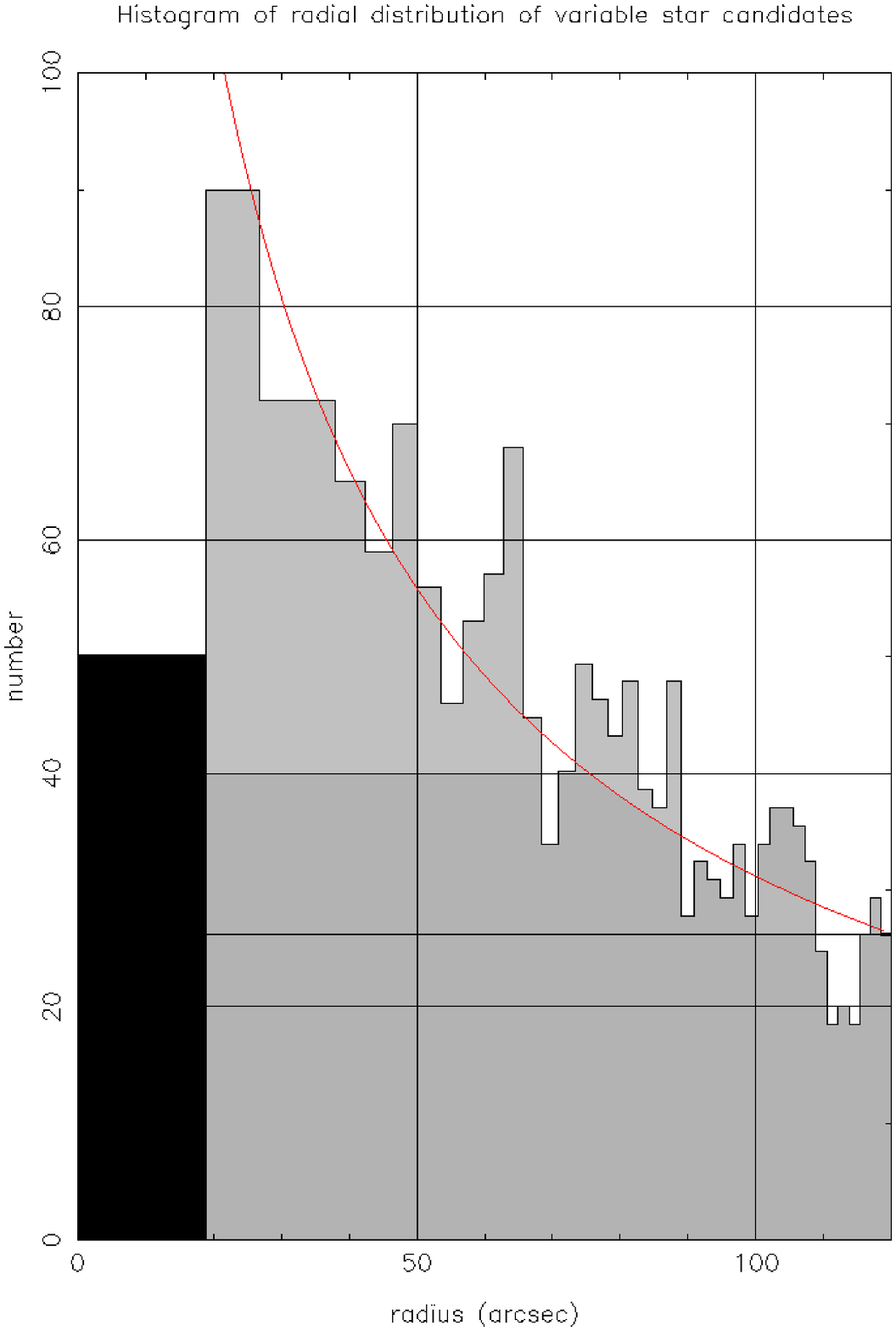}
\caption[Histogram showing the radial distribution of variable star candidates in the inner $120^{\prime\prime}$ of the M31 bulge along with the best fit de Vaucouleurs profile.]{Histogram showing the radial distribution of variable star candidates in the inner $120^{\prime\prime}$ of the M31 bulge along with the best fit de Vaucouleurs profile. This plot used $220$ circular annuli of equal area between $r=280^{\prime\prime}$ and $r=0$.
}
\label{radial_dist_and_de_V_fit}
\end{figure}

\begin{figure}[!ht]
\vspace*{10cm}
   \leavevmode
   \includegraphics{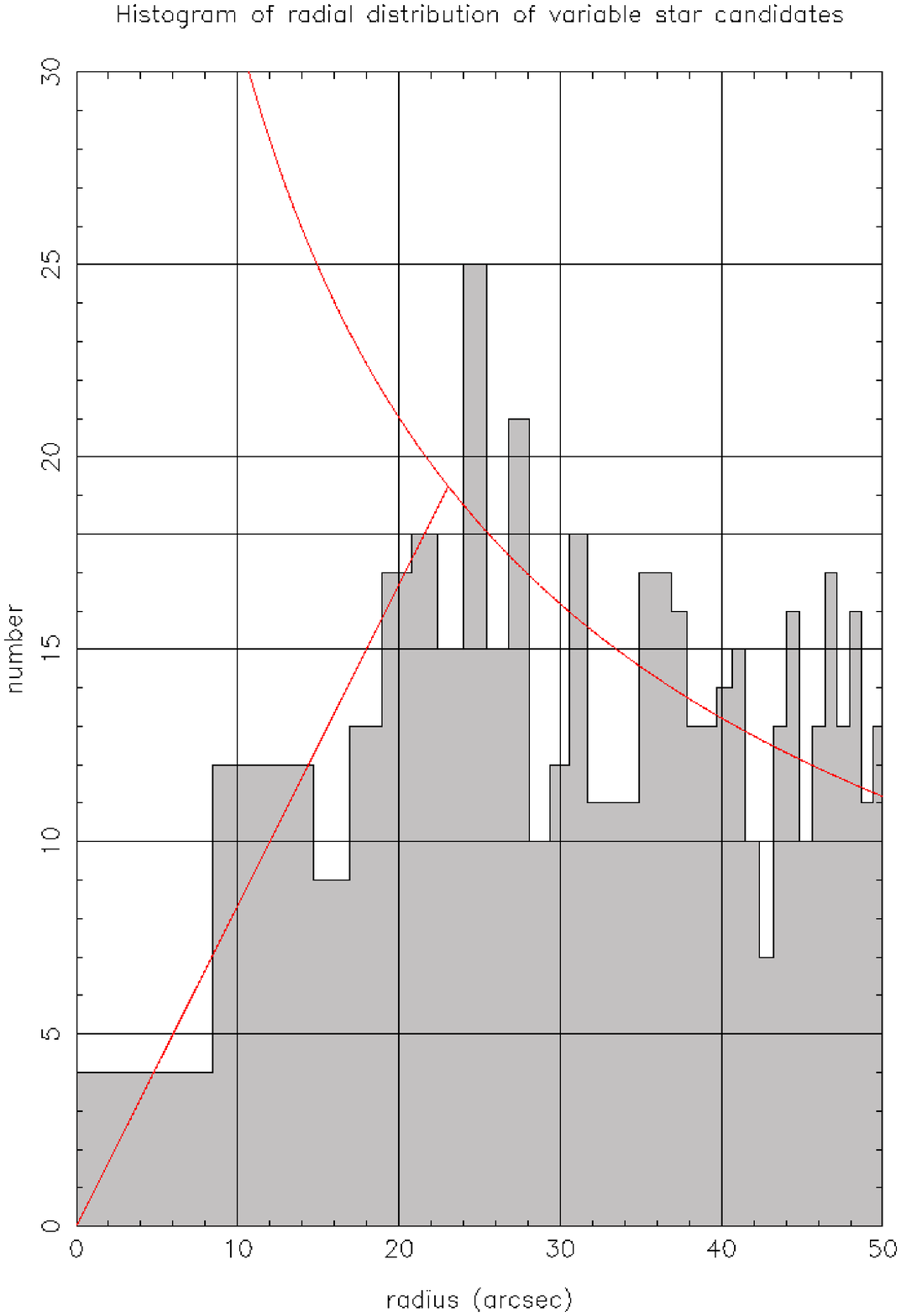}
\caption[Histogram showing the radial distribution of variable star candidates in the inner $50^{\prime\prime}$, focusing particularly on the inner $20^{\prime\prime}$.]{ Histogram showing the radial distribution of variable star candidates in the inner $50^{\prime\prime}$, focusing particularly on the inner $20^{\prime\prime}$. This plot used 1100 circular annuli of equal area between $r=280^{\prime\prime}$ and $r=0$.
}
\label{inner_radial_dist_and_linear_fit}
\end{figure}

As can be seen from Figure \ref{var_spat_dist_all_lcs}, the centre of the galaxy, and hence that of the spatial distribution of variable candidates, is not central to the mean image frame position. This made investigating the radial distribution a little more complicated than would otherwise be the case. It was decided to divide the space into concentric circular annuli, this time of equal area,
since then the numbers of objects within each ring could be fairly compared.
It was noted that out to a radius of about $60^{\prime\prime}$ the distribution is not affected by the edges of the data, and further out in radius to about $120^{\prime\prime}$ the circular annuli are only crossed by the edge of the data on the right hand side. At greater radii than this, corrections to the number distribution would become much more complicated due to the uneven shape of the data distribution, and by intersecting the corners.
 Noting the above two groups of $r < 60^{\prime\prime}$ and $60^{\prime\prime} < r < 120^{\prime\prime}$, the outer 
radius of the largest circle and the number of circles were carefully selected
so as to make one circular annulus have its outer rim at $60^{\prime\prime}$ and another at $120^{\prime\prime}$. (These values were $r_{\rm{max}} = 280^{\prime\prime}$ and $n_{\rm{ring}} = 220$). Once this had been set, the number of annuli could be adjusted by multiplying or dividing by a whole number in order to adjust the scaling of the data in each bin and hence the spatial resolution and corresponding statistical noise of each bin. The data inside $r=60^{\prime\prime}$ were taken unchanged, while the data in the second, outer, area were selected to have their angle $\theta$ measured from the positive x axis NOT in the sector $\arctan(-2)\geq\theta\leq \arctan(2) $ which corresponds to being outside two lines drawn between the origin and coordinates (in arcsecs) of $(60,120)$ and $(60,-120)$ respectively.
A small amount of data had therefore to be discarded, but not a significant amount.
The remaining data were scaled by the angular factor
$\frac{360}{360 - (2 \arctan(2))}$ required to correct for the lost sector.
In this way a non-biased sample was obtained out to $r = 120^{\prime\prime}$.
It was clear, both from the resulting histogram and from the original scatter plot (Figure \ref{var_spat_dist_all_lcs}) that the data were seriously incomplete in the very central region. The numbers in each bin continued rising, however, at 
a consistent and smooth rate moving inwards from the outer regions until inside roughly 
($r < 20^{\prime\prime}$) where a sudden fall-off began. Since no clear rounding off of the rise in numbers could be detected (with these statistics) outside $r = 20^{\prime\prime}$ it seemed reasonable to attempt to fit an appropriate function to the data outside this radius in order to quantify the way the inner completeness changed with radius.
The appropriate function chosen was de Vaucouleurs ``$r^{\frac{1}{4}}$'' law \citep{1948AnAp...11..247D}, generalised to indices other than $4$ by Sersic \citep{1963BAAA....6...41S} which has been applied both to elliptical galaxies and to the bulges of spiral galaxies. This function can be written:

\begin{equation}
\label{de_V_r_1_4}
I(r) = I_{0} e^{-b({\frac{r}{r_e}}^{\frac{1}{n}} - 1)}
\end{equation}

where $I(r)$ represents the surface brightness at radius $r$, the constant $b$ is chosen so that half the light falls within the scale
length $r_e$ and $I_{0}$ is the surface brightness scaling constant (i.e. the surface brightness at $r_e$). $n$ in this case is the ``Sersic index'', where the value $n=4$ was chosen by de Vaucouleurs. For values of $n > 1$, $b \approx 1.999n -0.327$. Therefore, with $n=4$, $b$ was chosen to be $7.67$.
The choice of this function is not meant to imply that the distribution of variable objects is being assumed to be a proxy for the shape of the bulge, but only that it is a reasonable choice of a centrally rising function with which to fit the rising portion of the histogram presented in
Figures \ref{radial_dist_and_de_V_fit} and \ref{inner_radial_dist_and_linear_fit}.

 Since the incompleteness in the central regions was unknown and was the
current object of investigation, it was not thought that the data justified
 varying the Sersic index $n$ from $4$, since the central incompleteness and
 the index affect one other circularly; only one can be chosen as the
unknown.

Making the necessary assumption that the data outside $r \approx 20$ are sufficiently close to being complete (although it was already known from plots such as, for example, the upper panel of Figure \ref{var_spat_dist_F_3_to_6} that this assumption is not true for low variable amplitudes), all data points outside
$r = 20$ were fitted with Equation \ref{de_V_r_1_4}, using only two variable parameters, the scale length $r_{e}$ and the scaling factor $I_{0}$.
To obtain a fit, errors in the numbers $N$ in each bin were taken to be $\sqrt{N}$. Using these errors, the best fit value of $I_{0}$ was $0.57^{+0.13}_{-0.11}$ and the best fit scale length $r_{e}$ was found to be $1900^{+570\prime\prime}_{-410}$, (or $32^{\prime} $).
For the reasons stated above it is not very surprising that these figures are not physically meaningful.
The histogram of radially binned data and the best fitting function above are plotted in Figure \ref{radial_dist_and_de_V_fit}.

In order to investigate the region inside $r = 20^{\prime\prime}$ in more detail, the number of annuli was increased to $1100$, which was the largest number found to produce sufficiently numerically stable results in this region. This many divisions is too many for the outer regions, but was necessary to provide sufficient resolution in the central region of interest. It was found that the numbers of objects declined linearly, within the noise, to zero with decreasing radius. This decline was modelled by a linear function which passed through the centre of the seventh bin. This had gradient $0.83391$ arcsec$^{-1}$.
The point at which this line met the previously fitted de Vaucouleurs function (re-scaled appropriately downwards by a factor of five) was found by iteration to be $r = 23.065^{\prime\prime}$.
The new histogram with the modelled fits in the two regions is shown in Figure \ref{inner_radial_dist_and_linear_fit}. 

\begin{figure}[!ht]
\vspace*{10cm}
   \leavevmode
   \includegraphics{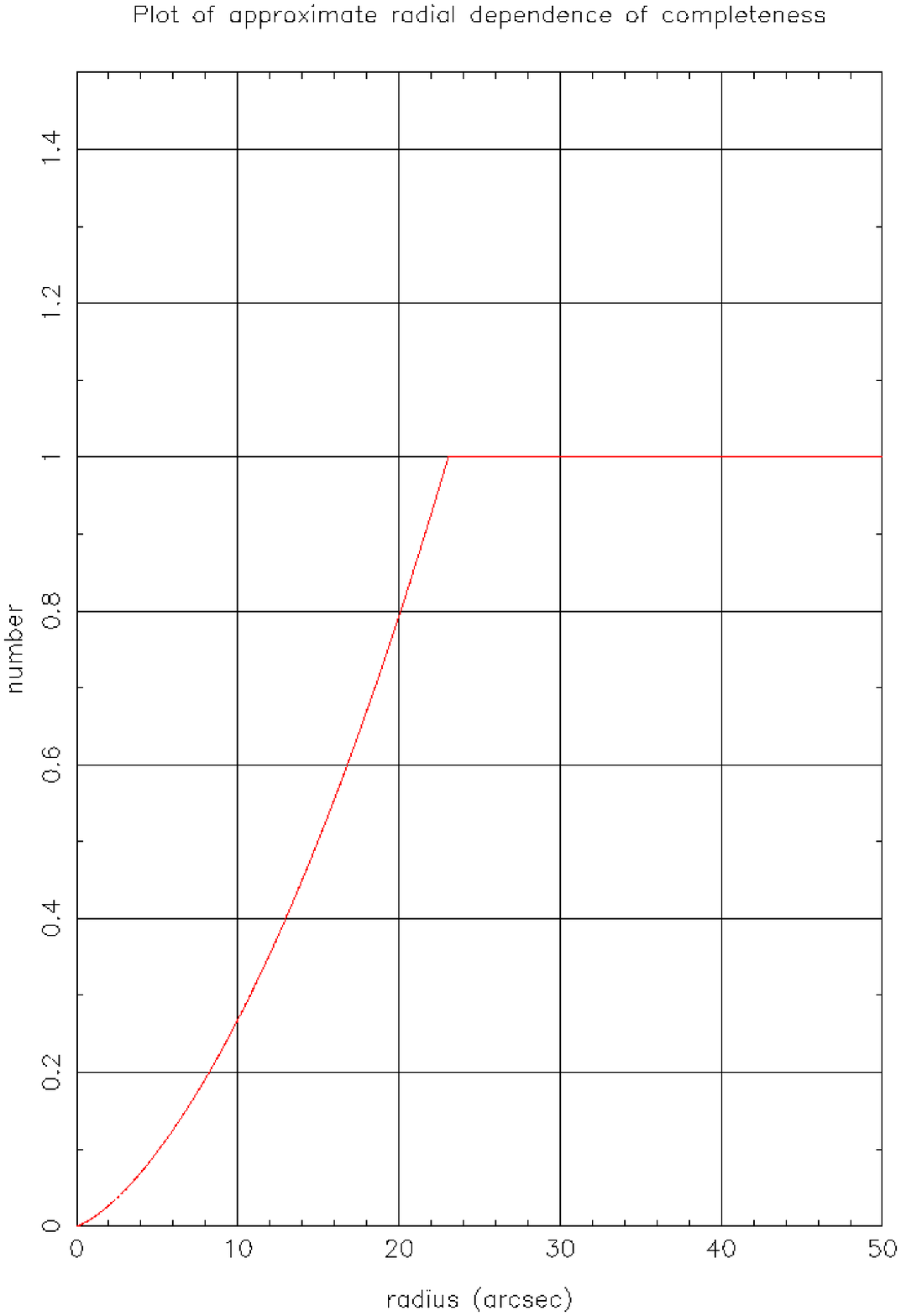}
\caption[Plot of the radial dependence of the estimated spatial completeness function in the inner $50^{\prime\prime}$ of the bulge.]{Plot of the radial dependence of the estimated spatial completeness function in the inner $50^{\prime\prime}$ of the bulge, assuming that the completeness is $1$ at $r=120^{\prime\prime}$, and using a de Vaucouleurs profile fit to the real data for $r > 23.065^{\prime\prime}$ and a linear fit inside that.
}
\label{spatial_completeness_fit}
\end{figure}

Assuming that, in the region $r < 23.065^{\prime\prime}$, the original number of objects with no decline in completeness would have risen in the same way as the de Vaucouleurs fit, then the radial variation of the completeness could be modelled as the ratio of the linear and de Vaucouleurs fit functions, which can be written as:

\begin{equation}
\label{modelled_completeness}
C(r<23^{\prime\prime}) = m r N_{0} (220/n) e^{-7.67({\frac{r}{r_e}}^{\frac{1}{n}} - 1)}
\end{equation}

 where $m$ is the gradient of the linearly declining portion = $0.83391$arcsec$^{-1}$,
 $N(0)$ is the fitted number scaling factor $= 0.5713$, $n$ = number of rings used
 and $r_{e} = 1905.6$. This function is shown plotted in Figure \ref{spatial_completeness_fit} for the inner $50^{\prime\prime}$.

It should be clearly stated that this model is only an approximation based on
several major assumptions and that a calculation of the true completeness function (including its variation with the amplitude of the variable) would require a full Monte Carlo analysis, the scope of which is beyond that of this thesis. The function above, in reality, would almost certainly not equal $1$ at $r = 23^{\prime\prime}$ as it has been previously shown above that there is some incompleteness affecting larger radii than this for lower amplitude variables.
Also, the details of the central completeness ``hole'' modelled here will vary
depending on the details of the photometry pipeline used, for example whether
the mask over the centre of the galaxy varies in size with each image, or is taken to be a constant.
 The data here selected as ``variable stars'' will not have detected all those 
objects in that category due to the algorithm used, which will not be perfect.
   However, using the currently selected data it is the best approximation that can be made to the true radial variation of the completeness for amplitudes of
 variable in this data set greater than about $5$ ADU/s. This function should
 not be assumed to be applicable to microlensing events due to their
 differing amplitude distribution and selection criteria.

\subsection{Microlensing Candidates}
\label{lensing_candidates}
\subsubsection{Results using 2007 photometry run lightcurves}
\label{lensing_candidates_2007}
Once the four runs of the candidate selection pipeline (using three varying PA/LT flux ratios and also no fixed ratio) had been completed, a further script was used to do the selection using tighter values of the cut parameters. Doing the selection in two parts like this
had the advantage that one still has all the necessary information on a wider range of lightcurves if it is required to go back and change the cut values for any reason. As shown in Figure \ref{reduced_chisq_dist_3160_LCs}, it was discovered after analysing the variable stars that apparently the sizes of the error bars on the flux points had apparently been underestimated by a factor of approximately $\sqrt(1.97)$. Therefore, using
values of $\chi^2$/d.o.f. as dictated by normal theory (originally chosen to be global $\chi^2$/d.o.f. $< 5.0$, local $\chi^2$/d.o.f. $< 2.0$) would have resulted in cuts that were much too strict and rejected many candidates that were in fact good enough. Therefore, the selection process was repeated using more relaxed values (by a factor $1.97$, making them now  global $\chi^2$/d.o.f. $< 9.85$, local $\chi^2$/d.o.f. $< 3.94$) for the two cuts which relied on $\chi^2$/d.o.f.
 During the course of this work it became apparent that in some lightcurves
 more than one variable period is clearly visible. This would be expected
 from experience as the probability that the PSFs of variable objects in the
 Angstrom variable object database overlap is non-negligible, given the
 density of variable objects, as can be seen in Figure 
\ref{variable_objects_comparison}. Ideally all of these periods would be
 subtracted before attempting to find microlensing.
However, in this work the ``mixed'' fit only contains the possibility of modelling the periodic baseline using a single period, and the pure reduced Paczy\'nski clearly none, and hence there will often be residual variations left in the baselines, which always act to increase the $\chi^2$ of a fit if the model does not include them. These may or may not be fitted well by subsequent residual iterations, but if the mixed fit was conducted in an iteration where there existed more than one periodic variation in the baseline, then the applied $\chi^2$/d.o.f. cuts will still be too strict even after the error scaling described above. Since the global $\chi^2$/d.o.f. fit was maintained at a higher figure than the local one anyway, to cater for the larger portion of baseline included in this figure, it was felt that the major un-catered-for effect would be on the varying baseline causing the fit to the Paczy\'nski component to be degraded. This would obviously affect fits with longer $t_{\rm{FWHM}}$ to a greater extent. To cater for this possibility, the selection was also performed with a more relaxed local $\chi^2$/d.o.f. of value $5$, referred to in Table \ref{2007_results_numbers_summary} below as the ``intermediate'' cut levels.

In Table \ref{2007_results_numbers_summary}, all the results for the four selection runs and for the three sets of cut values described above, as well as for the same ``loose'' $\chi^2$/d.o.f. cuts employed by the main selection pipeline, are summarised. The other cut values employed were always constant at: ``lensing ratio'' $> 2$, roughly equivalent to requiring at least a $2\sigma$ detection, $\chi^2$ difference ratio $>0.1$ and the peak sampling condition as described in Section \ref{optimising_cuts} (for both long period and ``normal'' candidates). The value of the $\chi^2$ difference ratio $>0.1$ was chosen to be relatively low, after some experimentation, to allow non-zero numbers of candidates to pass even with the stricter values of $\chi^2$/d.o.f..
 The numbers of lightcurves categorised as ``long period, low contrast'', or ``l.p.l.c.'' have also been
 included. The final column describes the final result after the results from the four flux ratio runs
 were concatenated, ordered by lightcurve number, and then the fit with the highest quality factor
 chosen in each case where the same lightcurve had been found by more than one of the four runs. Since
 some lightcurves are only selected by one run and not by others, these combined figures would usually 
be expected to be larger than all of the individual runs.

\begin{tiny}
\begin{table}
\caption[Table summarising the numbers of microlensing candidates 
selected for three varying cut levels and four PA/LT flux ratio regimes, using the 2007 photometry.]
{Table summarising the numbers of microlensing candidates selected 
for three varying cut levels and four PA/LT flux ratio regimes, using the 2007 photometry.}
\begin{center}
\begin{tabular}{|c||c|c|c|c|c|c|c|c|c|c|}
\hline
\hline
  &\multicolumn{2}{c|}{\footnotesize{Free Ratio}} & \multicolumn{2}{c|}{``low''} & \multicolumn{2}{c|}{\footnotesize{``medium''}}&
 \multicolumn{2}{c|}{``high''} & \multicolumn{2}{c|}{\footnotesize{``combined''}} \\
  &\multicolumn{2}{c|}{} & \multicolumn{2}{c|}{ } & \multicolumn{2}{c|}{ }&
 \multicolumn{2}{c|}{} & \multicolumn{2}{c|}{} \\
\hline
 \footnotesize{$\chi^2$ Cuts}        & \multicolumn{2}{c|}{\footnotesize{$F_{PA}/F_{LT}$ =}  }&\multicolumn{2}{c|}{7.89}&\multicolumn{2}{c|}{11.1}& \multicolumn{2}{c|}{14.8}&\multicolumn{2}{c|}{}\\
\hline
 values   & \footnotesize{norm.} & \footnotesize{lplc} & \footnotesize{norm.} & \footnotesize{lplc} & \footnotesize{norm.} & \footnotesize{lplc} & \footnotesize{norm.} & \footnotesize{lplc} & \footnotesize{norm.} & \footnotesize{lplc}   \\
 (global,local)  &  &  &  &  &  &  &  &  & &    \\
\hline
 \footnotesize{orig.$(5,2)$}         & \footnotesize{$2$}  &  \footnotesize{$8$}  & \footnotesize{$2$}  & \footnotesize{$8$}   & \footnotesize{$1$}  & \footnotesize{$7$}    & \footnotesize{$2$}  &  \footnotesize{$7$}    & \footnotesize{$4$}  & \footnotesize{$17$} \\

 \footnotesize{scaled}\tiny{$(9.85,3.94)$} & \footnotesize{$20$} &  \footnotesize{$28$}  & \footnotesize{$9$}  & \footnotesize{$47$}   & \footnotesize{$9$}  & \footnotesize{$49$}   & \footnotesize{$16$} &  \footnotesize{$46$}   & \footnotesize{$33$}(\tiny{$-1$)}  & \footnotesize{$93$} \\

 \footnotesize{``inter.''$(9.85,5)$}   & \footnotesize{$29$} &  \footnotesize{$40$}  & \footnotesize{$20$} & \footnotesize{$66$}   & \footnotesize{$16$} & \footnotesize{$68$}   & \footnotesize{$24$} &  \footnotesize{$64$}   & \footnotesize{$49$}\tiny{($-1$)}    & \footnotesize{$121$} \\
 \footnotesize{``loose''$(15,15)$}     & \footnotesize{$63$} &  \footnotesize{$49$} & \footnotesize{$55$} & \footnotesize{$103$}  & \footnotesize{$54$} & \footnotesize{$107$}  & \footnotesize{$60$} &  \footnotesize{$103$}  & \footnotesize{$115$}\tiny{($-3$)}   & \footnotesize{$362$} \\
\hline
\end{tabular}
\end{center}
\label{2007_results_numbers_summary}
\end{table}
\end{tiny}

The ``combined'' results of the four PA/LT flux ratio regimes for the events selected by the ``intermediate'' cuts are presented in Tables \ref{2007_results_fitting_data_summary_part_1} and \ref{2007_results_fitting_data_summary_part_2}. Included are the values of the parameters used to make the cuts
and the resulting quality factor values. The table has been ordered by quality factor. Where the candidate was selected by more than one flux ratio run, the letters N,L,M,H have been used to indicate that it was found by the ``No fixed flux ratio'', ``Low'', ``Medium'' or ``High'' flux ratios respectively. Within each lightcurve where there are multiple flux ratio fits the fits were ordered and the one included in Tables \ref{2007_results_fitting_data_summary_part_1} and \ref{2007_results_fitting_data_summary_part_2} chosen by the lowest $\chi^{2}_{\rm{\rm{glob}}}$/d.o.f.. The values of PA/LT flux amplitude for the three fixed ratio runs were $7.89$, $11.1$ and $14.8$ respectively. The type of fit performed, i.e. reduced Paczy\'nski or ``mixed'' is shown under ``Fit''.
In cases where exactly the same fit was found by several of the runs (for example when only LT data existed and hence the flux ratio with PA data was irrelevant), the above letters are separated using an ``=''. `BSQF' represents the ``bump sample quality factor'', and ``LSN'' represents the ``lensing signal to noise'' parameter. It will also be noticed that there is one instance in Table \ref{2007_results_fitting_data_summary_part_1} of two lightcurves with identical parameters (161,7319). This was due to a duplication (or near duplication) of lightcurves which could occur in the DIA pipeline,
as explained above in Section \ref{lightcurve_copies}. The combined figures for ``normal'' and ``long period'' events in Table \ref{2007_results_numbers_summary} have been corrected for this effect, and the number of independent \emph{objects} are shown in the table (as opposed to lightcurves), with the number of duplicates removed shown in brackets, if any existed. The four candidates retained even with the strictest $\chi^2$/d.o.f. cuts are highlighted with asterisks on the left hand side. (``CDR'' again stands for ``$\chi^2$ difference ratio'')

\begin{table}
\caption[Table (part I) summarising the selection information of microlensing candidates 
selected for global $\chi^2$/d.o.f. $< 9.85$, local $\chi^2$/d.o.f.$ < 5.0$, lensing ratio $> 2.0$ and the combined results of the four PA/LT flux ratio regimes, using the 2007 photometry.]
{Table (part I) summarising the selection information of microlensing candidates 
selected for global $\chi^2$/d.o.f.$ < 9.85$, local $\chi^2$/d.o.f.$ < 5.0$, lensing ratio $> 2.0$ and the combined results of the four PA/LT flux ratio regimes, using the 2007 photometry.                                                                   }
\begin{center}
\begin{tabular}{|c||c|c|c|c|c|c|c|c|c|c|}
\hline
\hline
LC & QF & Run/s & iter & Fit &$\chi^{2}_{\rm{loc}}$& LSN & CDR &$\chi^{2}_{\rm{\rm{glob}}}$& \footnotesize{BSQF} & $t_{0}$ in\\
\hline
 \footnotesize{3315} & \footnotesize{6.063} &  \footnotesize{N,H,M,L} & 1 & M & \footnotesize{2.398} & 5.924 & 0.808 & \footnotesize{2.821} & 1.477 &  PA\\
 \footnotesize{13776} & \footnotesize{6.007} &  N & 1 & M & \footnotesize{2.565} & 2.321 & 3.840 & \footnotesize{1.860} & 1.183 &  PA\\
 \footnotesize{6535} & \footnotesize{3.543} &  \footnotesize{N,H,M,L} & 1 & M & \footnotesize{3.973} & 6.686 & 0.550 & \footnotesize{2.603} & 1.124 &  PA\\
 \footnotesize{2064} &\footnotesize{2.595} &  H,L & 1 & M & \footnotesize{3.306} & 2.778 & 1.144 & \footnotesize{3.046} & 1.384 &  PA\\
 \footnotesize{3282} & \footnotesize{2.563} &  \footnotesize{L=M=H,N} & 1 & M & \footnotesize{3.738} & 2.221 & 1.739 & \footnotesize{2.346} & 1.282 & \footnotesize{ANG}\\
 \footnotesize{8679} & \footnotesize{2.510} &  H,L & 1 & M & \footnotesize{3.799} & 4.274 & 0.188 & \footnotesize{3.554} & 1.818 &  PA\\
 \footnotesize{9086} & \footnotesize{2.168} &  \footnotesize{M,L,H} & 1 & M & \footnotesize{0.599} & 2.091 & 0.468 & \footnotesize{2.102} & 0.954 &  PA\\
 \footnotesize{7319} & \footnotesize{2.061} & N & 1 & M & \footnotesize{2.749} & 2.047 & 1.074 & \footnotesize{3.468} & 1.509 &  PA\\
 \footnotesize{161} & \footnotesize{2.061} & N & 1 & M & \footnotesize{2.749} & 2.047 & 1.074 & \footnotesize{3.468} & 1.509 &  PA\\
 \footnotesize{5947} & \footnotesize{1.860} & \footnotesize{N,L,H} & 1 & M & \footnotesize{4.006} & 2.163 & 2.708 & \footnotesize{3.633} & 0.886 &  PA\\
 \footnotesize{7874} & \footnotesize{1.841} & \footnotesize{N,L,M,H} & 1 & M & \footnotesize{4.767} & 2.797 & 1.820 & \footnotesize{3.000} & 0.906 &  PA\\
 \footnotesize{7201} & \footnotesize{1.725} & M & 1 & M & \footnotesize{2.781} & 2.114 & 0.314 & \footnotesize{3.176} & 1.850 &  PA\\
 \footnotesize{762} & \footnotesize{1.607} & \footnotesize{H,M,N,L} & 1 & M & \footnotesize{2.630} & 3.098 & 0.308 & \footnotesize{4.185} & 1.352 &  PA\\
 \footnotesize{9799} & \footnotesize{1.536} & H,L & 1 & M & \footnotesize{0.209} & 5.584 & 0.544 & \footnotesize{3.746} & 0.352 &  PA\\
 \footnotesize{9920} & \footnotesize{1.504} & L & 1 & M & \footnotesize{4.247} & 3.693 & 0.693 & \footnotesize{7.192} & 1.376 &  PA\\
 \footnotesize{16055} & \footnotesize{1.456} & N & 1 & M & \footnotesize{4.578} & 3.691 & 0.842 & \footnotesize{3.593} & 0.875 &  PA\\
 \footnotesize{4853} & \footnotesize{1.452} & N & 1 & M & \footnotesize{1.903} & 2.510 & 0.412 & \footnotesize{1.806} & 0.760 &  \footnotesize{ANG}\\
 \footnotesize{6056} & \footnotesize{1.432} & N & 1 & M & \footnotesize{3.265} & 2.165 & 0.739 & \footnotesize{4.282} & 1.435 &  PA\\
 \footnotesize{1643} & \footnotesize{1.418} & H & 1 & M & \footnotesize{4.387} & 2.203 & 0.661 & \footnotesize{4.507} & 1.723 &  PA\\
 \footnotesize{5791} & \footnotesize{1.310} & \footnotesize{L,N,M,H} & 1 & M & \footnotesize{3.471} & 2.570 & 1.289 & \footnotesize{4.486} & 0.886 &  PA\\
 \footnotesize{9189} & \footnotesize{1.278} & N & 1 & M & \footnotesize{2.910} & 2.121 & 0.493 & \footnotesize{4.243} & 1.443 &  \footnotesize{ANG} \\
 \footnotesize{1732} & \footnotesize{1.252} & H & 1 & M & \footnotesize{4.760} & 3.644 & 0.184 & \footnotesize{4.729} & 1.377 &  PA\\
 \footnotesize{5383} & \footnotesize{1.244} & L,N & 1 & M & \footnotesize{4.838} & 2.607 & 0.619 & \footnotesize{3.193} & 1.183 &  PA\\
 \footnotesize{2335} & \footnotesize{1.221} & N & 1 & P & \footnotesize{1.713} & 2.041 & 0.775 & \footnotesize{1.325} & 0.512 &  \footnotesize{ANG}\\
 \footnotesize{5381} & \footnotesize{1.199} & \footnotesize{N,M,H} & 1 & M & \footnotesize{4.119} & 2.178 & 0.191 & \footnotesize{3.881} & 1.850 &  PA\\
\hline
\end{tabular}
\end{center}
\label{2007_results_fitting_data_summary_part_1}
\end{table}

\begin{table}
\caption[Table (part II) summarising the selection information of microlensing candidates 
selected for global $\chi^2$/d.o.f. $< 9.85$, local $\chi^2$/d.o.f.$ < 5.0$, lensing ratio $> 2.0$ and the combined results of the four PA/LT flux ratio regimes, using the 2007 photometry.]
{Table (part II) summarising the selection information of microlensing candidates 
selected for global $\chi^2$/d.o.f.$ < 9.85$, local $\chi^2$/d.o.f.$ < 5.0$, lensing ratio $> 2.0$ and the combined results of the four PA/LT flux ratio regimes, using the 2007 photometry.                                               }
\begin{center}
\begin{tabular}{|c||c|c|c|c|c|c|c|c|c|c|}
\hline
\hline
LC & QF & Run/s & iter & Fit &$\chi^{2}_{\rm{loc}}$& LSN & CDR &$\chi^{2}_{\rm{glob}}$& \footnotesize{BSQF} & $t_{0}$ in\\
\hline
 \footnotesize{18106} & \footnotesize{1.114} & H & 1 & M & \footnotesize{3.423} & 2.033 & 0.173 & \footnotesize{3.725} & 1.670 &  PA\\
 \footnotesize{7662} & \footnotesize{1.026} & H & 1 & M & \footnotesize{3.652} & 3.424 & 0.396 & \footnotesize{5.566} & 0.989 &  PA\\
 \footnotesize{11057} & \footnotesize{0.995} & N,L & 1 & M & \footnotesize{3.661} & 2.085 & 0.504 & \footnotesize{3.934} & 1.204 &  PA\\
 \footnotesize{6736} & \footnotesize{0.990} & N & 1 & M & \footnotesize{4.470} & 2.969 & 0.574 & \footnotesize{2.180} & 0.704 &  PA\\
 \footnotesize{2075} & \footnotesize{0.987} & H,M,L & 1 & M & \footnotesize{3.865} & 2.602 & 0.312 & \footnotesize{6.354} & 1.477 &  PA\\
 \footnotesize{2675} & \footnotesize{0.978} & L & 2 & M & \footnotesize{4.085} & 14.215 & 1.491 & \footnotesize{2.933} & 0.097 &  \footnotesize{ANG}\\
 \footnotesize{1033} & \footnotesize{0.976} & H & 2 & M & \footnotesize{3.349} & 3.212 & 0.382 & \footnotesize{4.613} & 0.875 &  PA\\
 \footnotesize{12648} & \footnotesize{0.946} & \footnotesize{L,M,H} & 1 & M & \footnotesize{4.604} & 2.881 & 1.036 & \footnotesize{2.862} & 0.602 &  PA\\
 \footnotesize{4654} & \footnotesize{0.905} & N & 1 & M & \footnotesize{4.842} & 3.117 & 0.643 & \footnotesize{3.807} & 0.764 &  \footnotesize{ANG}\\
 \footnotesize{3328} & \footnotesize{0.894} & L,M & 1 & M & \footnotesize{3.877} & 2.065 & 0.933 & \footnotesize{9.280} & 1.473 &  PA\\
 \footnotesize{15888} & \footnotesize{0.833} & M,H & 1 & M & \footnotesize{4.279} & 2.558 & 0.306 & \footnotesize{6.793} & 1.380 &  PA\\
 \footnotesize{13817} & \footnotesize{0.704} & \footnotesize{L=M=H} & 1 & M & \footnotesize{3.717} & 2.145 & 0.889 & \footnotesize{4.570} & 0.720 &  PA\\
 \footnotesize{3911} & \footnotesize{0.617} & M & 1 & M & \footnotesize{4.035} & 2.481 & 0.183 & \footnotesize{6.048} & 1.061 &  PA\\
 \footnotesize{7714} & \footnotesize{0.616} & N & 1 & M & \footnotesize{1.789} & 2.747 & 1.042 & \footnotesize{8.114} & 0.544 &  PA\\
 \footnotesize{15294} & \footnotesize{0.577} & N & 1 & M & \footnotesize{3.854} & 3.174 & 0.364 & \footnotesize{6.304} & 0.677 &  PA\\
 \footnotesize{13966} & \footnotesize{0.573} & N & 1 & M & \footnotesize{3.406} & 5.454 & 1.188 & \footnotesize{6.061} & 0.227 &  PA\\
 \footnotesize{9092} & \footnotesize{0.528} & N & 1 & M & \footnotesize{1.989} & 3.303 & 0.525 & \footnotesize{5.402} & 0.387 &  PA\\
 \footnotesize{6163} & \footnotesize{0.491} & H & 1 & M & \footnotesize{2.673} & 2.724 & 0.172 & \footnotesize{6.698} & 0.720 &  PA\\
 \footnotesize{13318} & \footnotesize{0.446} & H & 1 & M & \footnotesize{1.576} & 2.290 & 0.121 & \footnotesize{6.705} & 0.720 &  PA\\
 \footnotesize{7256} & \footnotesize{0.389} & \footnotesize{N,H,M} & 1 & M & \footnotesize{2.656} & 2.317 & 0.431 & \footnotesize{8.897} & 0.677 &  PA\\
 \footnotesize{6354} & \footnotesize{0.322} & N & 1 & M & \footnotesize{3.609} & 3.300 & 1.467 & \footnotesize{7.864} & 0.227 &  PA\\
 \footnotesize{16823} & \footnotesize{0.309} & N & 1 & M & \footnotesize{3.078} & 4.765 & 0.862 & \footnotesize{2.482} & 0.097 &  PA\\
 \footnotesize{4286} & \footnotesize{0.289} & L & 1 & M & \footnotesize{4.881} & 7.786 & 0.745 & \footnotesize{4.222} & 0.097 &  PA\\
 \footnotesize{1237} &\footnotesize{0.157} & N & 1 & M & \footnotesize{3.705} & 2.717 & 0.251 & \footnotesize{5.436} & 0.211 &  PA\\
 \footnotesize{8239} & \footnotesize{0.055} & N & 1 & P & \footnotesize{4.493} & 2.049 & 0.215 & \footnotesize{4.257} & 0.097 &  PA\\
\hline
\end{tabular}
\end{center}
\label{2007_results_fitting_data_summary_part_2}
\end{table}

If the threshold for Cut 8 (lensing ratio) was raised to $3.0$, which would be a
more normal value to choose, being effectively a $3\sigma$ detection as opposed to a $2\sigma$ one, then the number of selected candidates was reduced to $18$. These are 
shown in Table \ref{2007_results_fitting_data_summary_lensing_ratio_>3} with their
fitted values of $t_{\rm{FWHM}}$. Clearly, as this threshold is raised, the number of ``false'' detections should decrease, but there were still several notable candidates within the 
larger list in Tables \ref{2007_results_fitting_data_summary_part_1} and \ref{2007_results_fitting_data_summary_part_2} that might still be considered ``good'' candidates, despite their lower signal to noise, since they have very low $\chi^2$ values. Obviously one cannot take this process to its limit (lensing ratio$\rightarrow 0$, $\chi^2\rightarrow 1$) due to likely confusion with variable stars but the boundary is necessarily ``fuzzy''. Two examples from Tables \ref{2007_results_fitting_data_summary_part_1} and \ref{2007_results_fitting_data_summary_part_2} which appeared by eye to at least be worthy of further consideration were lightcurves (161,7319) and 2335, and these are shown in Figures \ref{third_season_new_ref_candidate_0161} and \ref{third_season_new_ref_candidate_2335} below. (In all lightcurve plots below this point, blue points = LT data, red = FTN data and black = PA data).

\begin{figure}[!ht]
\vspace*{8cm}
$\begin{array}{c}
\vspace*{4.4cm} 
   \leavevmode
 \includegraphics{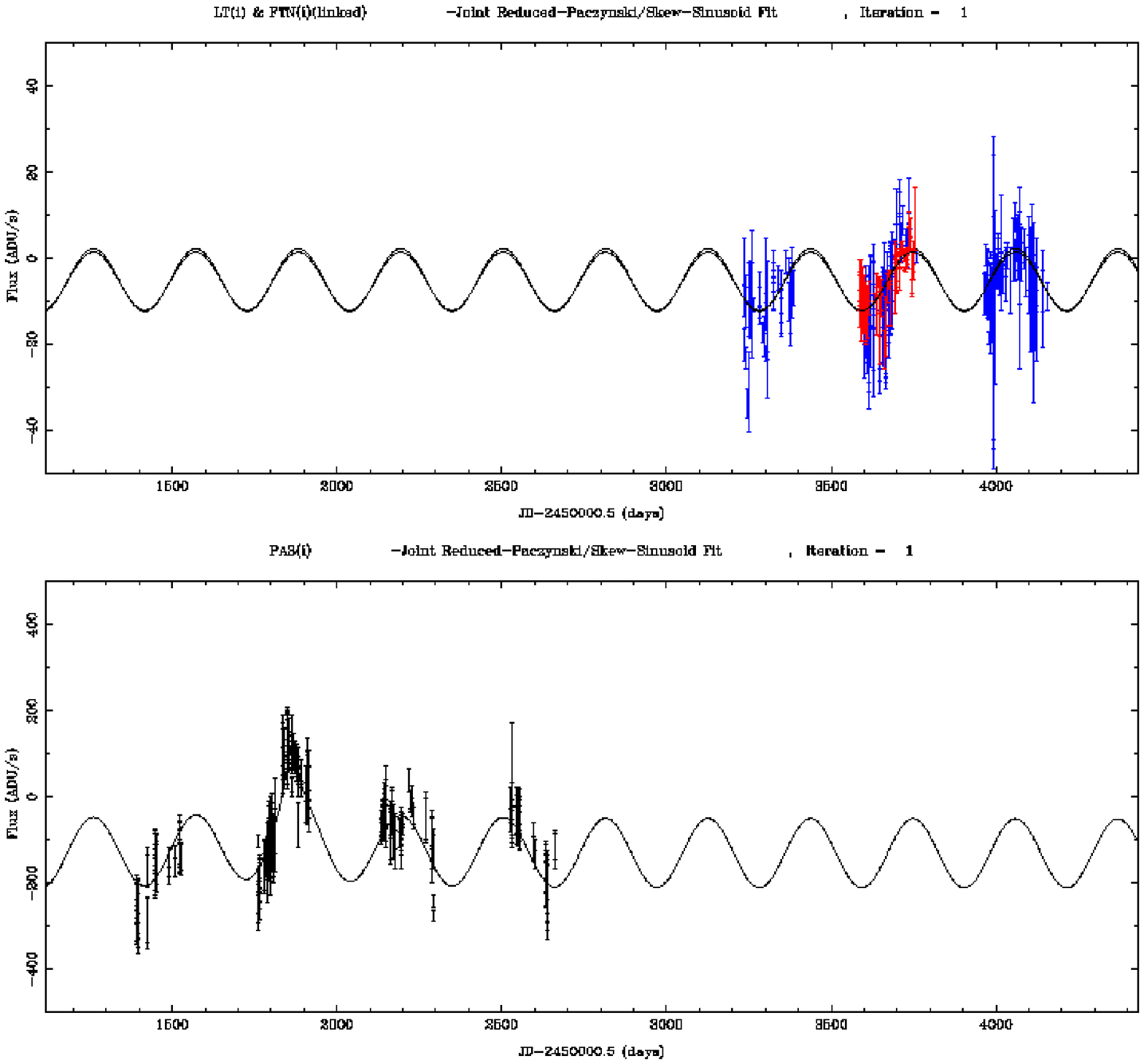} \\
\vspace*{4.4cm}
   \leavevmode
 \includegraphics{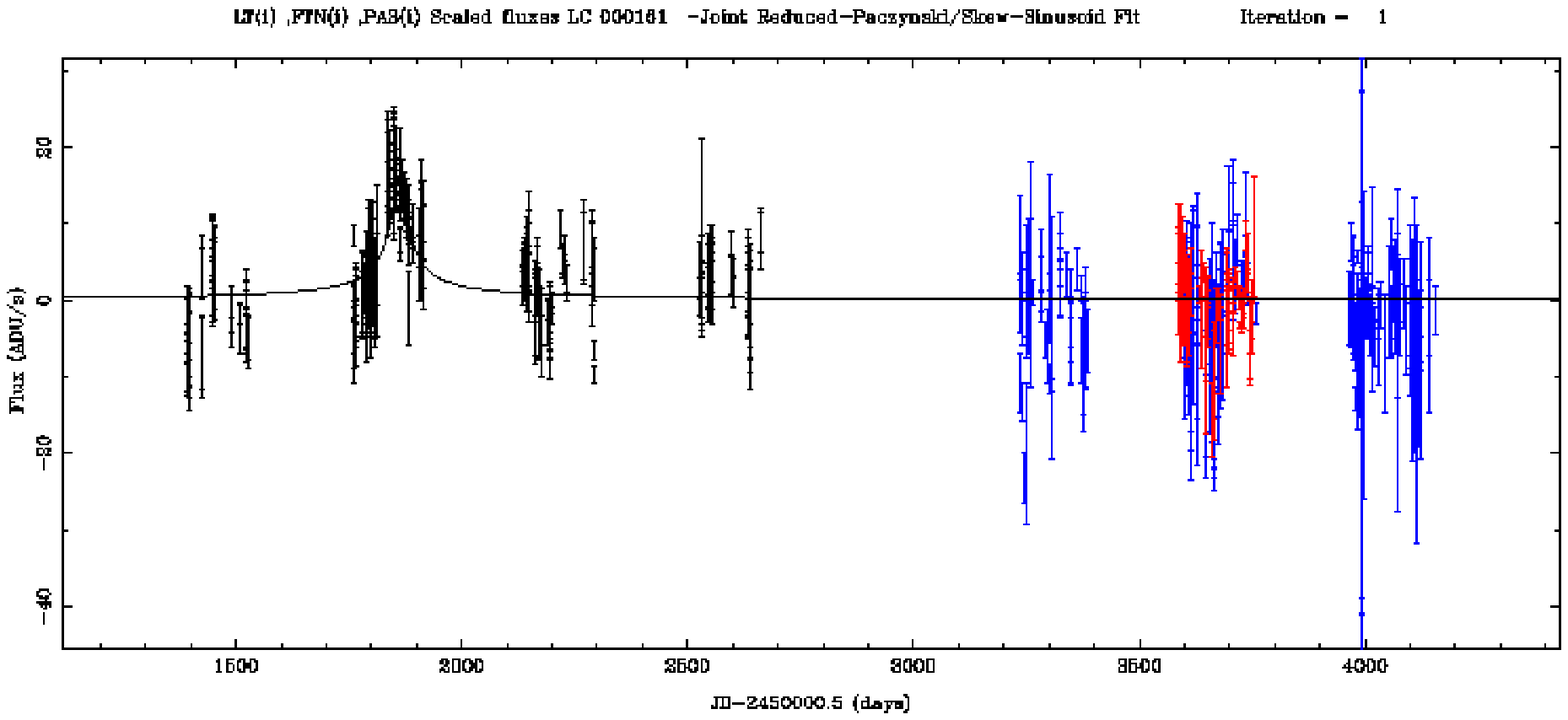} \\
\vspace*{4.1cm}
   \leavevmode
 \includegraphics{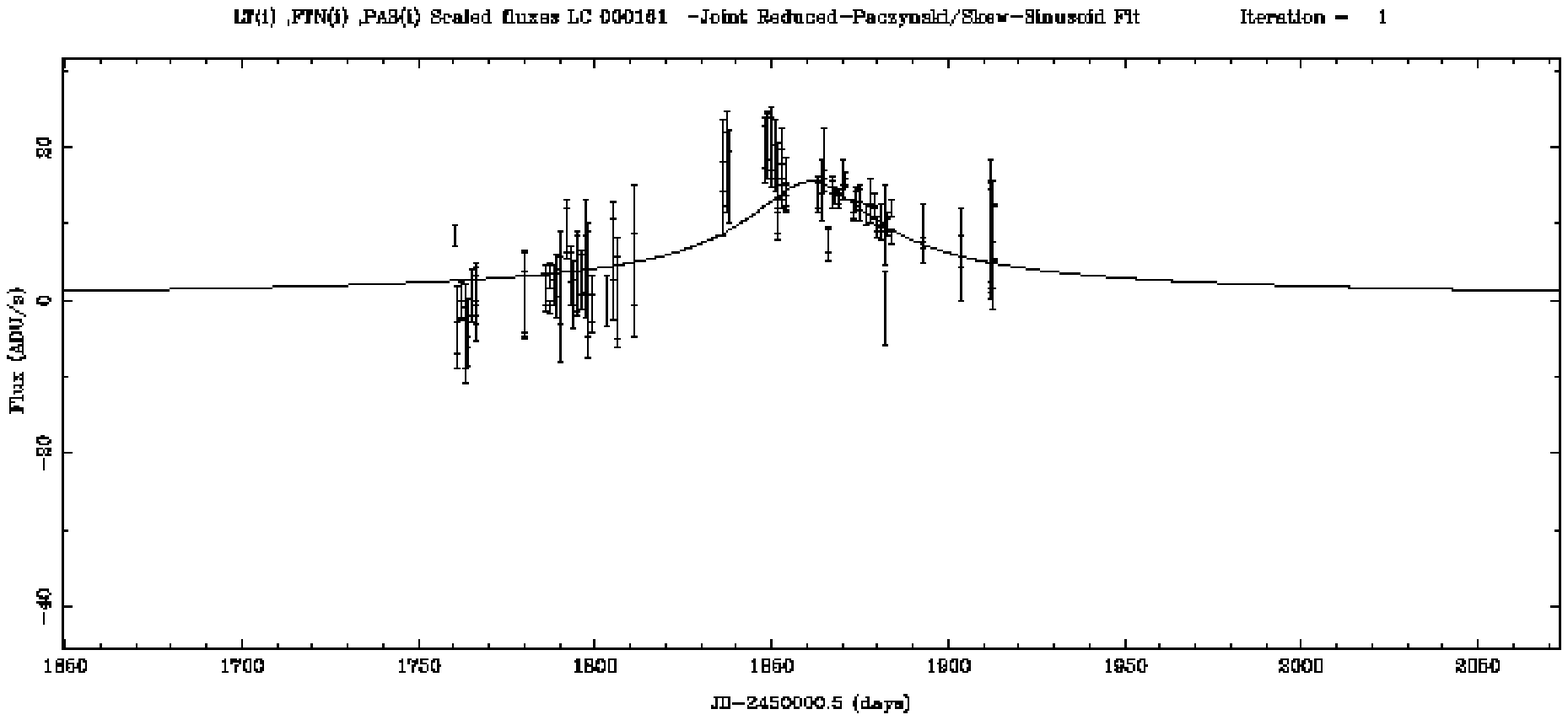} \\
\vspace*{0cm}
   \leavevmode
 \includegraphics{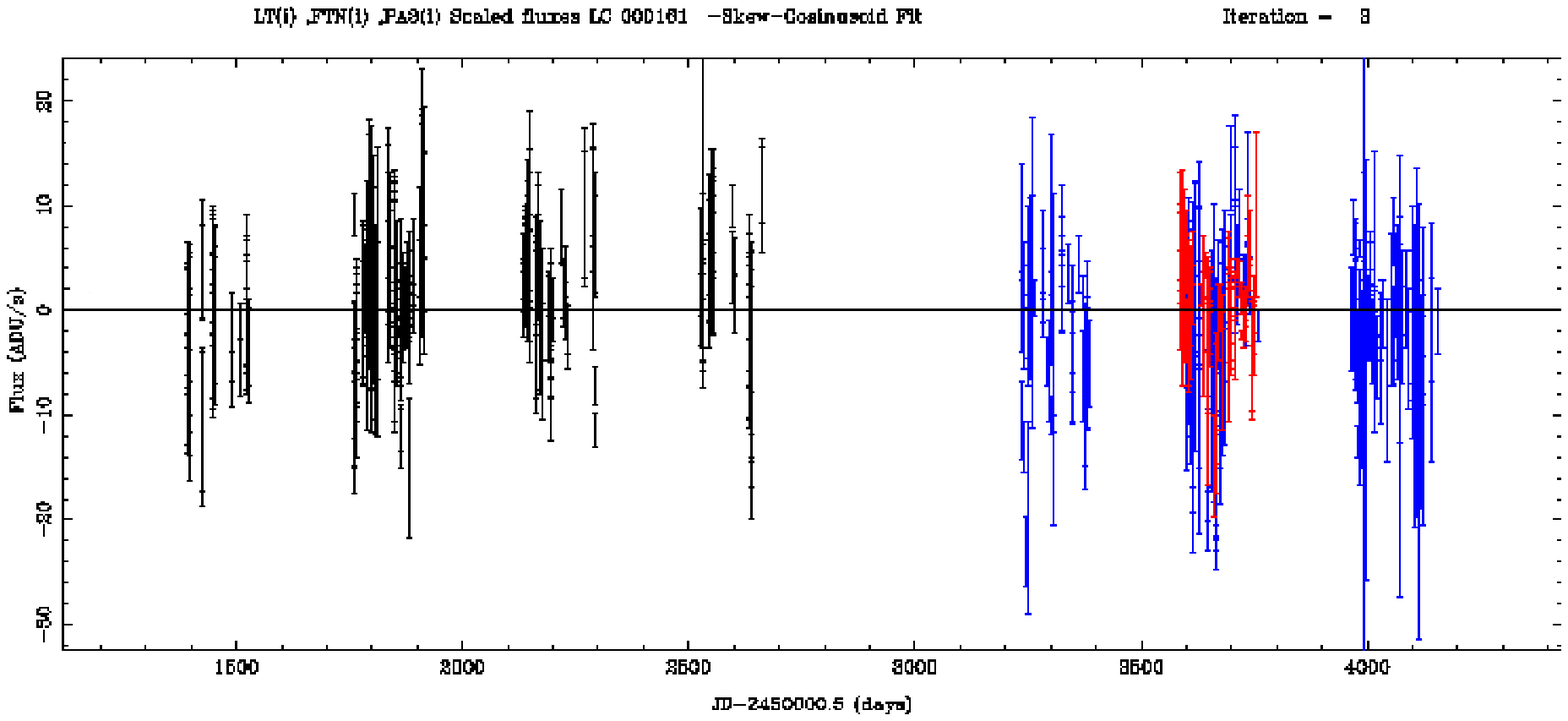} \\
\end{array}$
\caption[Plots of the lightcurve $161$/$7319$ in the LT Third Season photometry.]{Plots of the lightcurve $161$/$7319$ in the 2007 photometry. The panels are: 1) \& 2) The original lightcurve in three bands with the full mixed fit. 3) The full lightcurve when the first iteration variable component has been subtracted and the data have been scaled and plotted on the same time axis. 4) A zoom in on the peak region of panel 3). 5) The residuals remaining after subtraction of the mixed fit shown in panels 1) \& 2).}
 \label{third_season_new_ref_candidate_0161}
\end{figure}

 \clearpage
\begin{figure}[!ht]
\vspace*{6cm}
$\begin{array}{c}
\vspace*{6.5cm} 
   \leavevmode
 \includegraphics{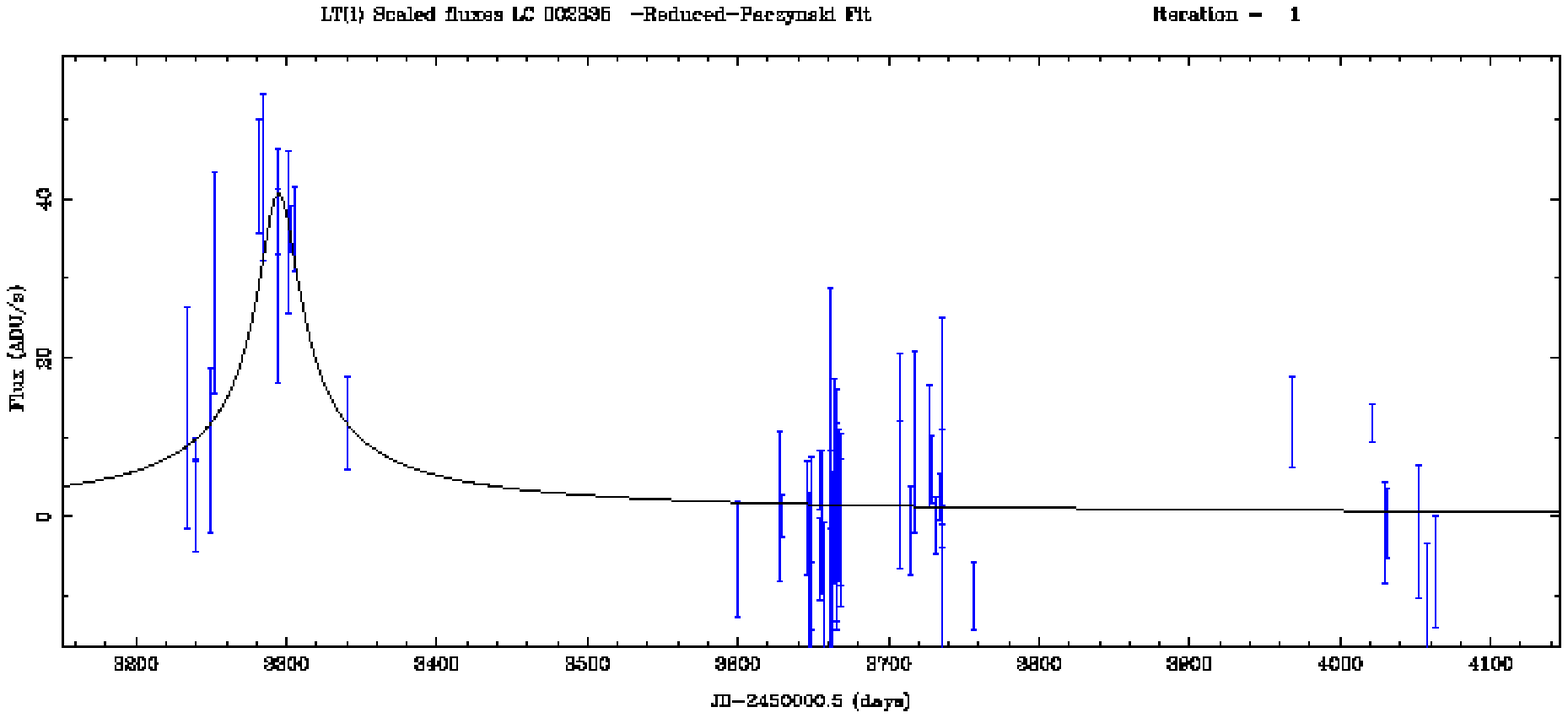} \\
\vspace*{6.5cm}
   \leavevmode
 \includegraphics{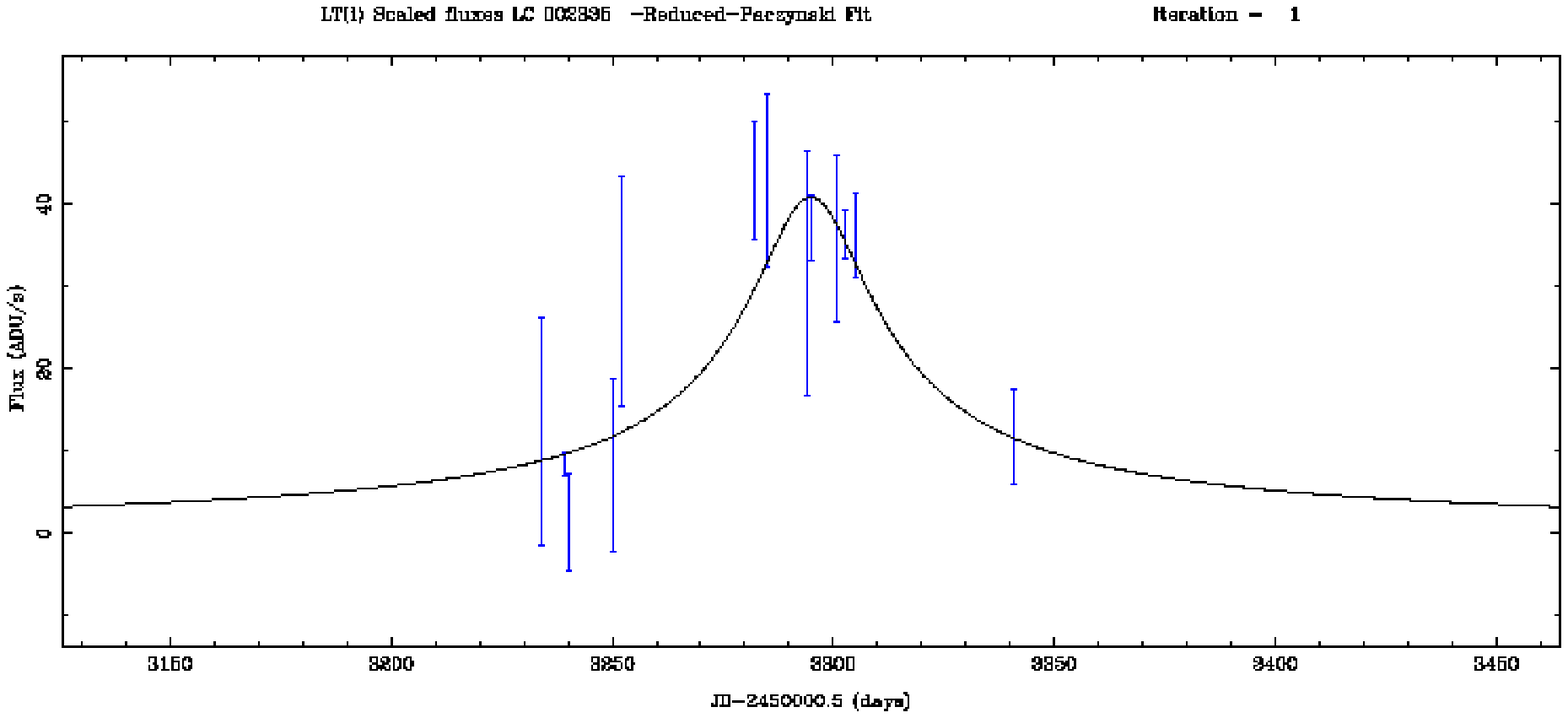} \\
\vspace*{2cm}
   \leavevmode
 \includegraphics{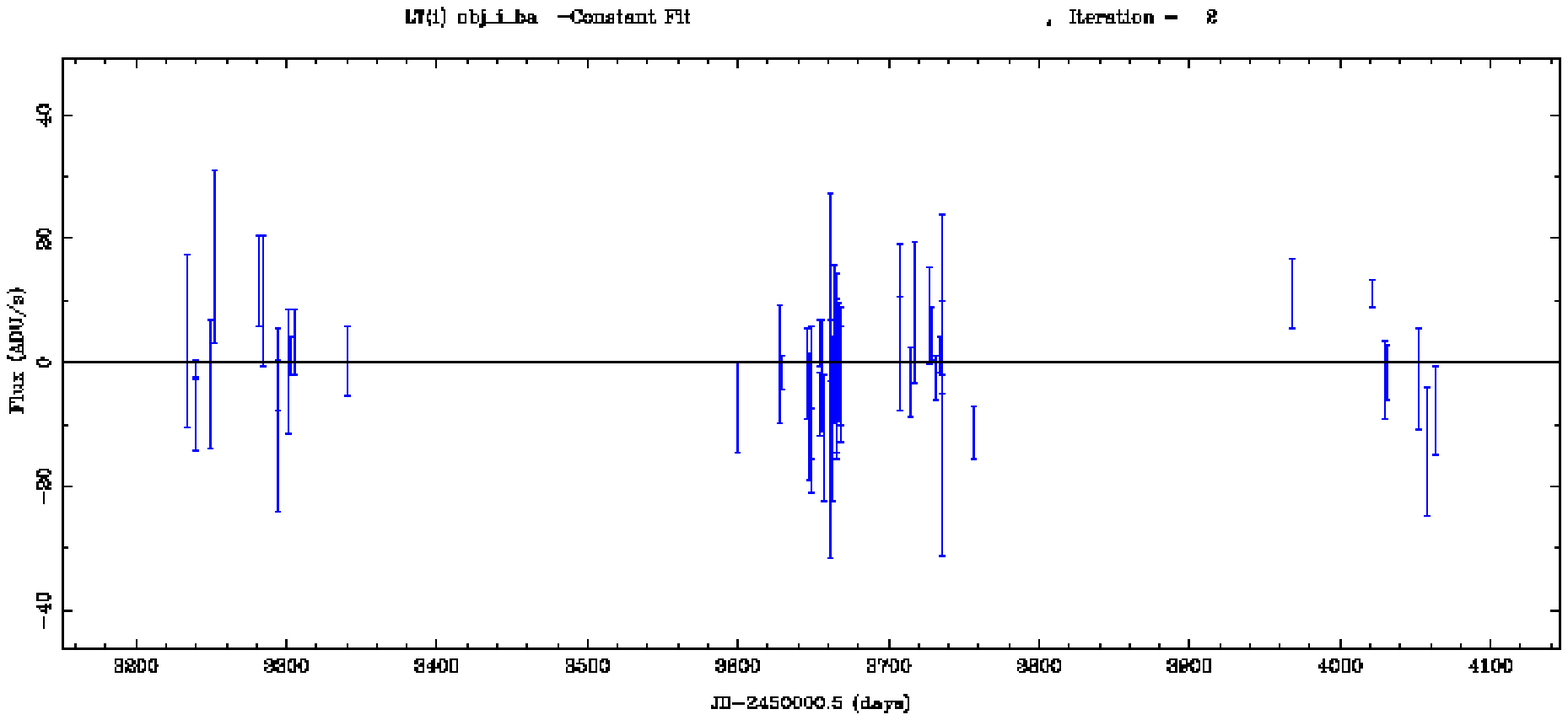} \\
\end{array}$
 \caption[Plots of the lightcurve $2335$ in the Third Season photometry.]{Plots of the lightcurve $2335$ in the 2007 photometry. Panels are: 1) The LT lightcurve (no PA or FT data existed for this lightcurve) with the reduced Paczy\'nski fit. 2) Zoom in on the peak region of 1). 3) Residuals after subtraction of the reduced Paczy\'nski fit.}
 \label{third_season_new_ref_candidate_2335}
\end{figure}

\begin{tiny}
\begin{table}
\caption[Table summarising the information about the fitted peak of microlensing candidates 
selected for global $\chi^2$/d.o.f. $< 9.85$, local $\chi^2$/d.o.f. $< 5.0$, lensing ratio $> 3.0$ and the combined results of the four PA/LT flux ratio regimes, using the 2007 photometry.]
{Table summarising the information about the fitted peak of microlensing candidates 
selected for global $\chi^2$/d.o.f. $< 9.85$, local $\chi^2$/d.o.f. $< 5.0$ and the combined results of the four PA/LT flux ratio regimes, using the 2007 photometry.}
\begin{center}
\begin{tabular}{|c||c|c|c|c|c|c|}
\hline
\hline
LC & QF & Run/s & iter & Fit & $t_{0}$ in & $t_{\rm{FWHM}}$\\
\hline
3315 & 6.063 & N,H,M,L &  1 & M & PA  & 82.018 \\
6535 & 3.543 & N,H,M,L &  1 & M & PA  & 130.399 \\
8679 & 2.510 & H,L &  1 & M & PA & 388.182 \\
762 & 1.607 & H,M,N,L &  1 & M & PA & 84.963 \\
9799 & 1.536 & H,L &  1 & M & PA & 6.536 \\
9920 & 1.504 & L &  1 & M & PA & 40.901 \\
16055 & 1.456 & N &  1 & M & PA & 19.428 \\
1732 & 1.252 & H &  1 & M & PA & 76.637 \\
7662 & 1.026 & H &  1 & M & PA & 83.242 \\
2675 & 0.978 & L &  2 & M & LT  & 2.881 \\
1033 & 0.976 & H &  2 & M & PA & 6.195 \\
4654 & 0.905 & N &  1 & M & LT & 24.966 \\
15294 & 0.577 & N &  1 & M & PA & 21.057 \\
13966 & 0.573 & N &  1 & M & PA & 2.958 \\
9092 & 0.528 & N &  1 & M & PA & 7.838 \\
6354 & 0.322 & N &  1 & M & PA & 3.081 \\
16823 & 0.309 & N &  1 & M & PA & 7.312 \\
4286 & 0.289 & L &  1 & M & PA & 8.917 \\
\hline
\end{tabular}
\end{center}
\label{2007_results_fitting_data_summary_lensing_ratio_>3}
\end{table}
\end{tiny}

As can be seen from Tables \ref{2007_results_fitting_data_summary_part_1} and \ref{2007_results_fitting_data_summary_part_2}, the large majority of the candidates selected were ``Mixed'' with $t_0$ situated within the PA data. The next largest category was ``Mixed,Angstrom'' and there were only one each of (Pac,PA) and (Pac,Angstrom). The great majority of ``mixed'' events was not too surprising, given the degree of blending expected in the core of M31, but the size of the majority of candidates with $t_0$ in the PA data \emph{was} surprising. A majority of some kind would have been predicted simply from the number of data points in each band, of which the PA data at the time had more, with a greater number of complete seasons (4) to the LT's two (+ the pilot season). Another reason could be a phenomenon noticed during the development of the selection routine, namely the prevalence of very short, almost vertical ``spikes'' within the PA data, which would naturally be attractors to a Paczy\'nski fit. It is not yet known whether these photometric
phenomena are caused by physical events or due to some unusual data processing problem, and this will require further investigation.

The third and least pleasant explanation might be that the quality of the photometry from the version of the DIA pipeline used to generate the flux points used in this investigation was not as good as it might have been. From the investigation of the short high signal to noise event presented in this Chapter in Section \ref{short_event}, and similar investigations into other high signal to noise variations highlighted by the APAS, certain phenomena such as apparent photometric oscillations after particularly bright flux points
had been noted. Changes were made (among others) to the way the position of the objects were calculated in order to ensure that photometry was calculated closer to the position corresponding to the brightest fluxes (after these had occurred).
 All the improvements to the DIA pipeline were complete before the generation of the 
photometry for the $2008$ candidate selection run, reported below.
As later discovered (see Section \ref{rediscovery_in_2008}), given that a ``perfect'' event in the $2008$ photometry would have $\chi^2$/d.o.f. $= 1.96$
on average, the sizes of the best $\chi^2$/d.o.f. values in Tables \ref{2007_results_fitting_data_summary_part_1} and \ref{2007_results_fitting_data_summary_part_2} are about what would be expected, with only $3$ out of the $50$ selected events having $\chi^2$/d.o.f. values less than $1.96$.
For illustration, lightcurves for the first five selected events in Table \ref{2007_results_fitting_data_summary_part_1} are presented below in Figures (\ref{LC_03315_2007} to \ref{LC_03282_2007}).

\begin{figure}[!ht]
\vspace*{10cm}
$\begin{array}{c}
\vspace*{10cm}
   \leavevmode
 \includegraphics{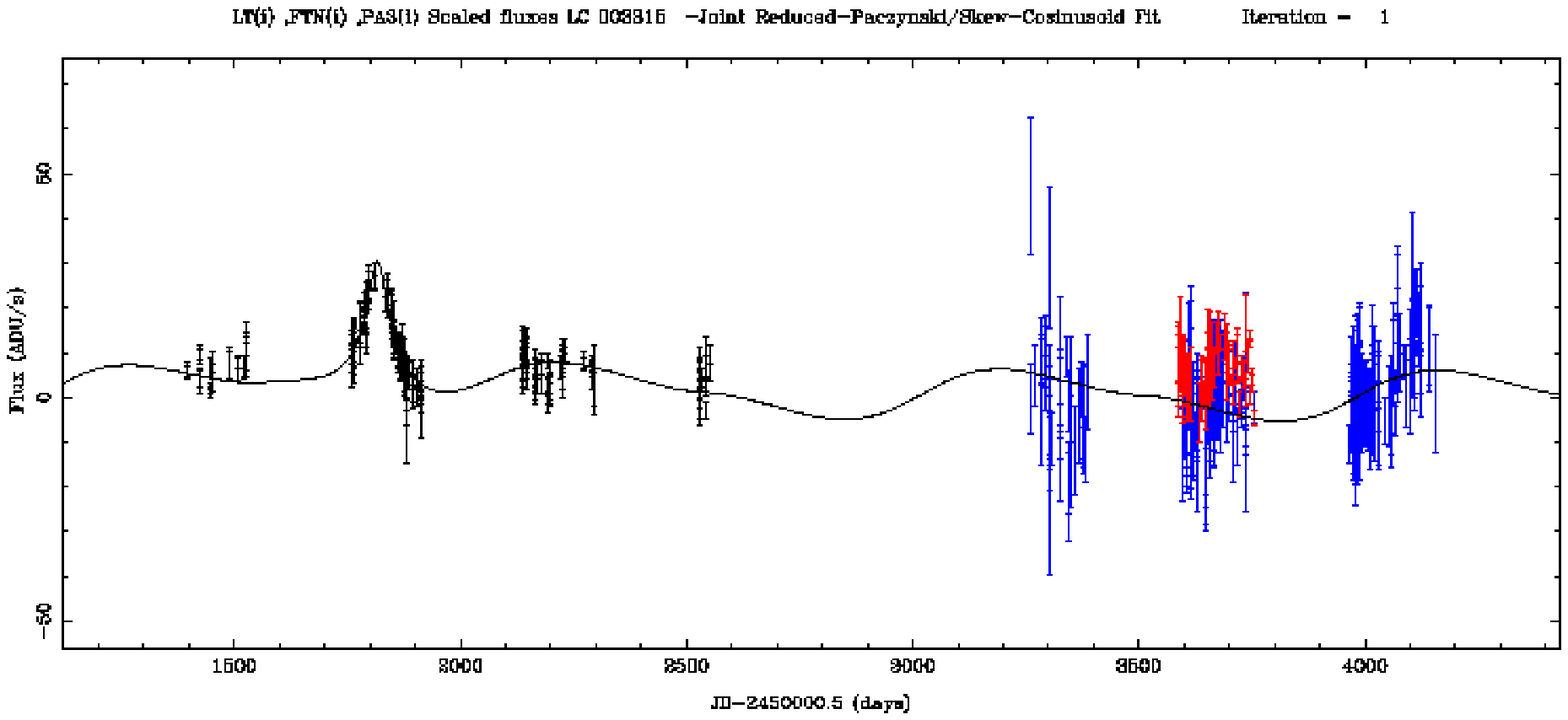} \\
\vspace*{2cm}
   \leavevmode
 \includegraphics{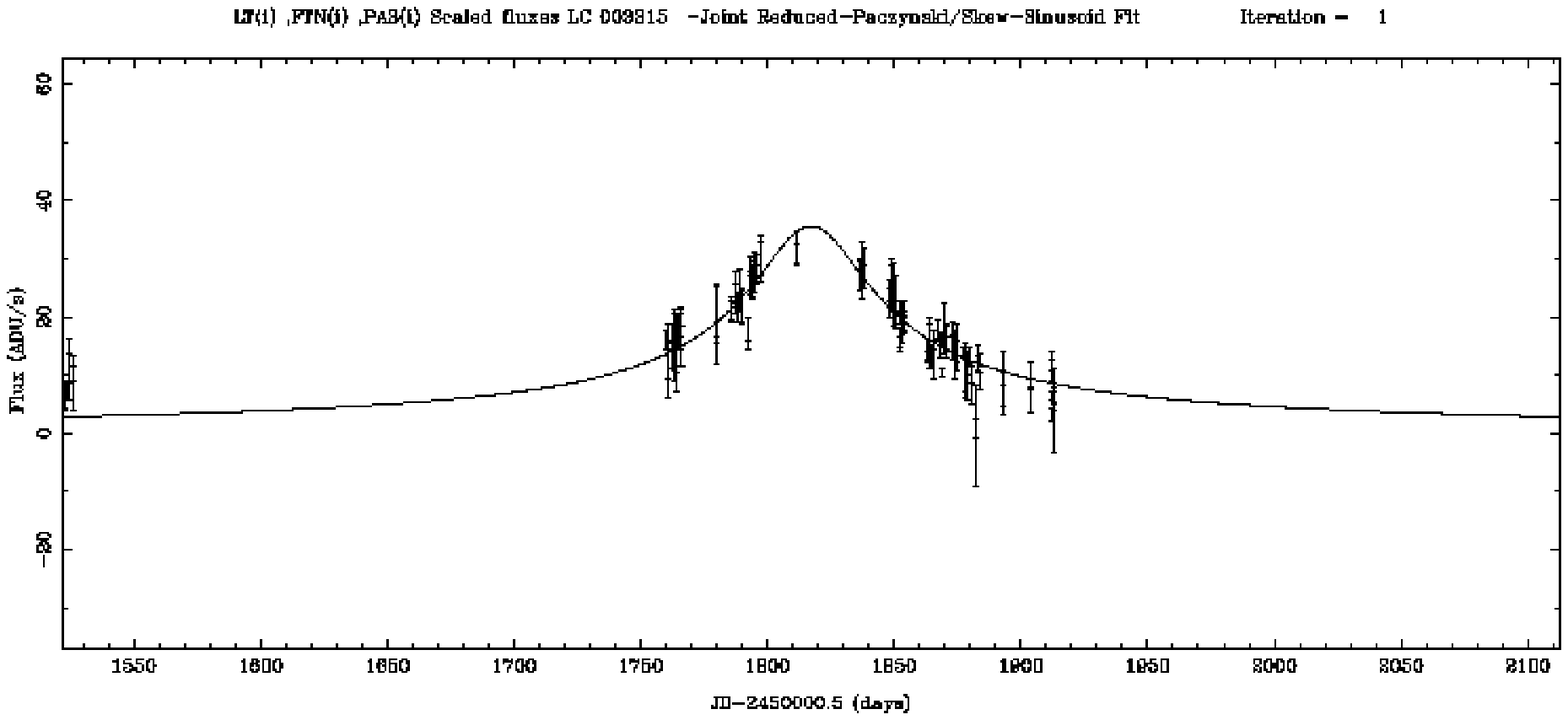} \\
\end{array}$
\caption[Lightcurve of object number $3315$ in the 2007 photometry, showing a) the original mixed fit and
b) the lensing component only, after the variable component has been subtracted.]{Lightcurve of object number $3315$ in the 2007 photometry, showing a) the original mixed fit and b) the peak region of the lensing component only, after the variable component has been subtracted.}
\label{LC_03315_2007}
\end{figure}

\newpage

\clearpage

\begin{figure}[!ht]
\vspace*{10cm}
$\begin{array}{c}
\vspace*{10cm}
   \leavevmode
 \includegraphics{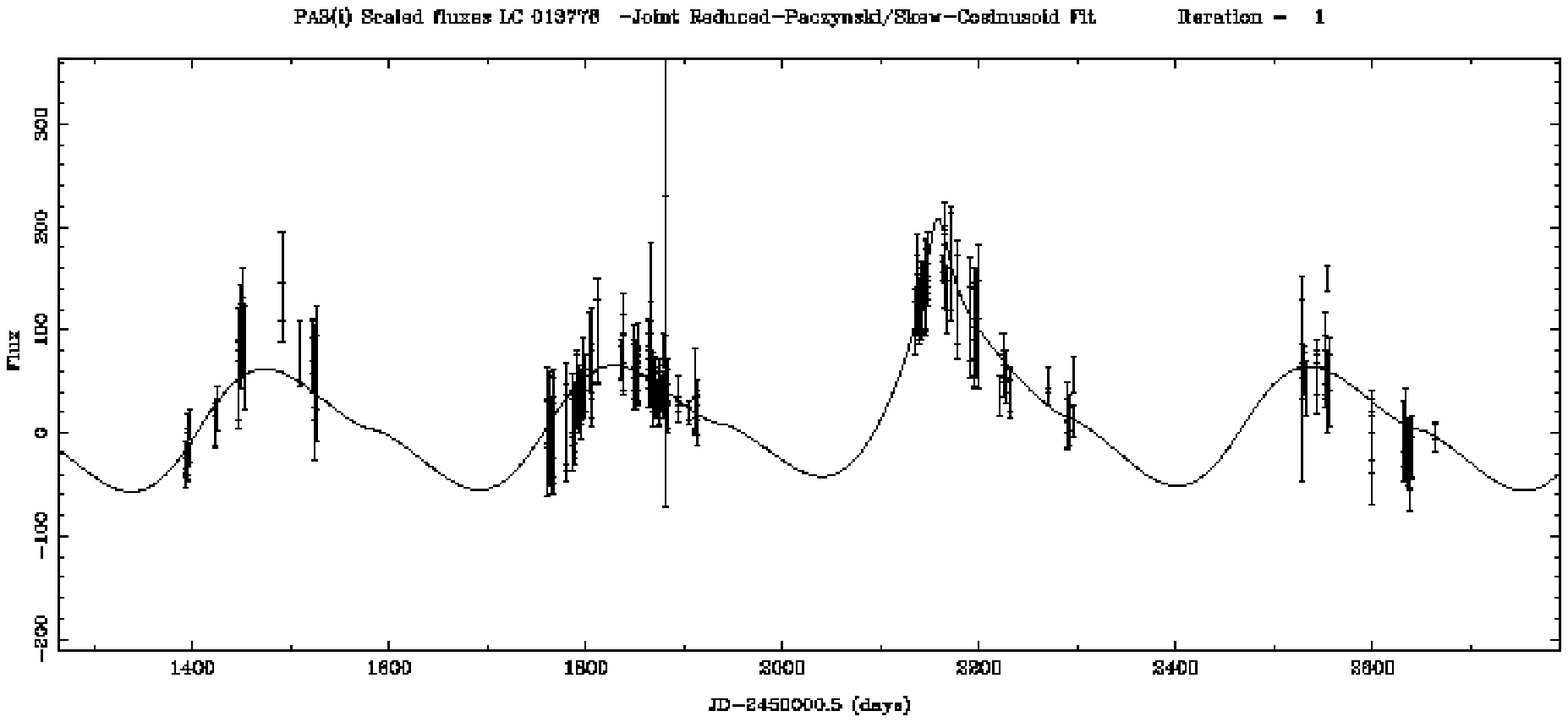} \\
\vspace*{2cm}
   \leavevmode
 \includegraphics{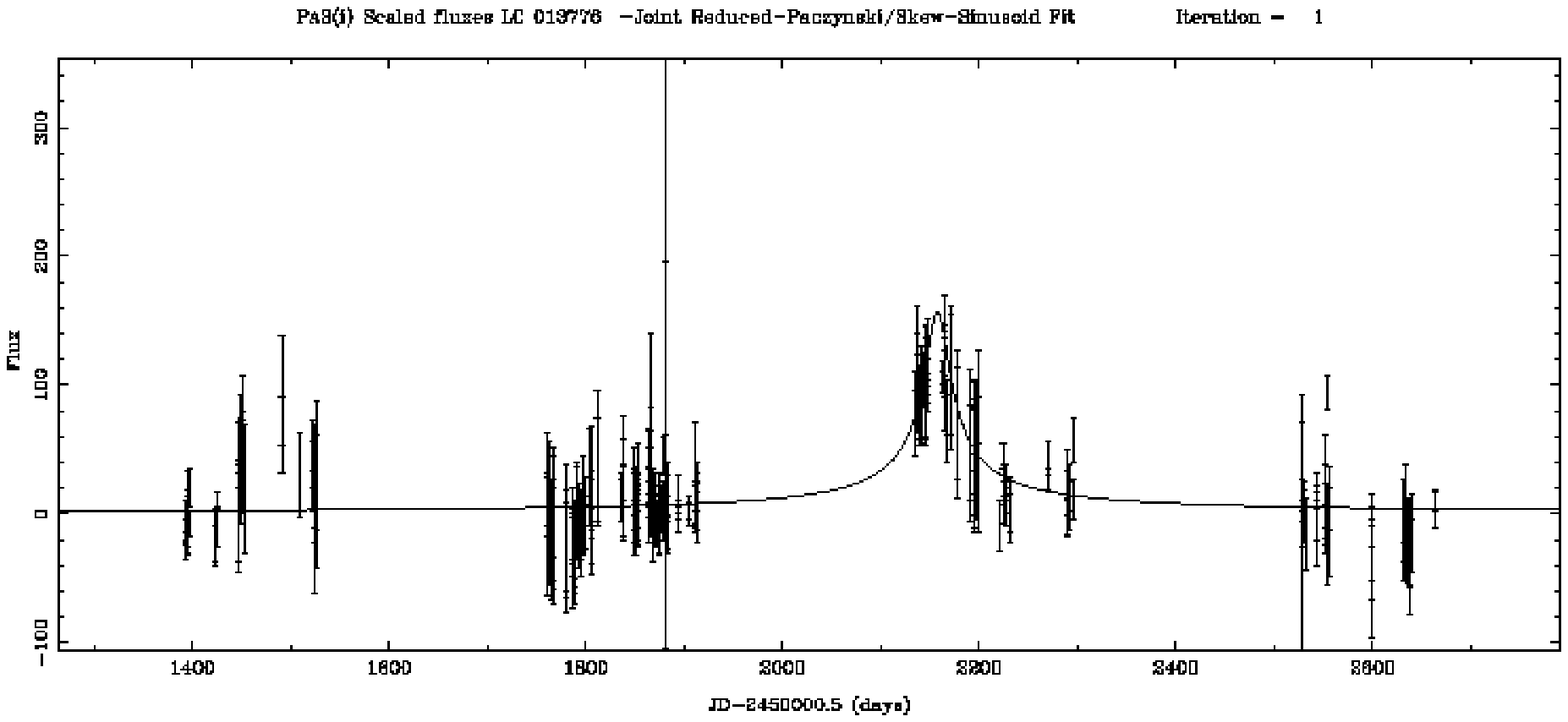} \\
\end{array}$
\caption[Lightcurve of object number $13776$ in the 2007 photometry, showing 1) the original mixed fit and
2) the lensing component only, after the variable component has been subtracted.]{Lightcurve of object number $13776$ in the 2007 photometry, showing 1) the original mixed fit and 2) the peak region of the lensing component only, after the variable component has been subtracted.}
\end{figure}

\clearpage

\begin{figure}[!ht]
\vspace*{10cm}
$\begin{array}{c}
\vspace*{10cm}
   \leavevmode
 \includegraphics{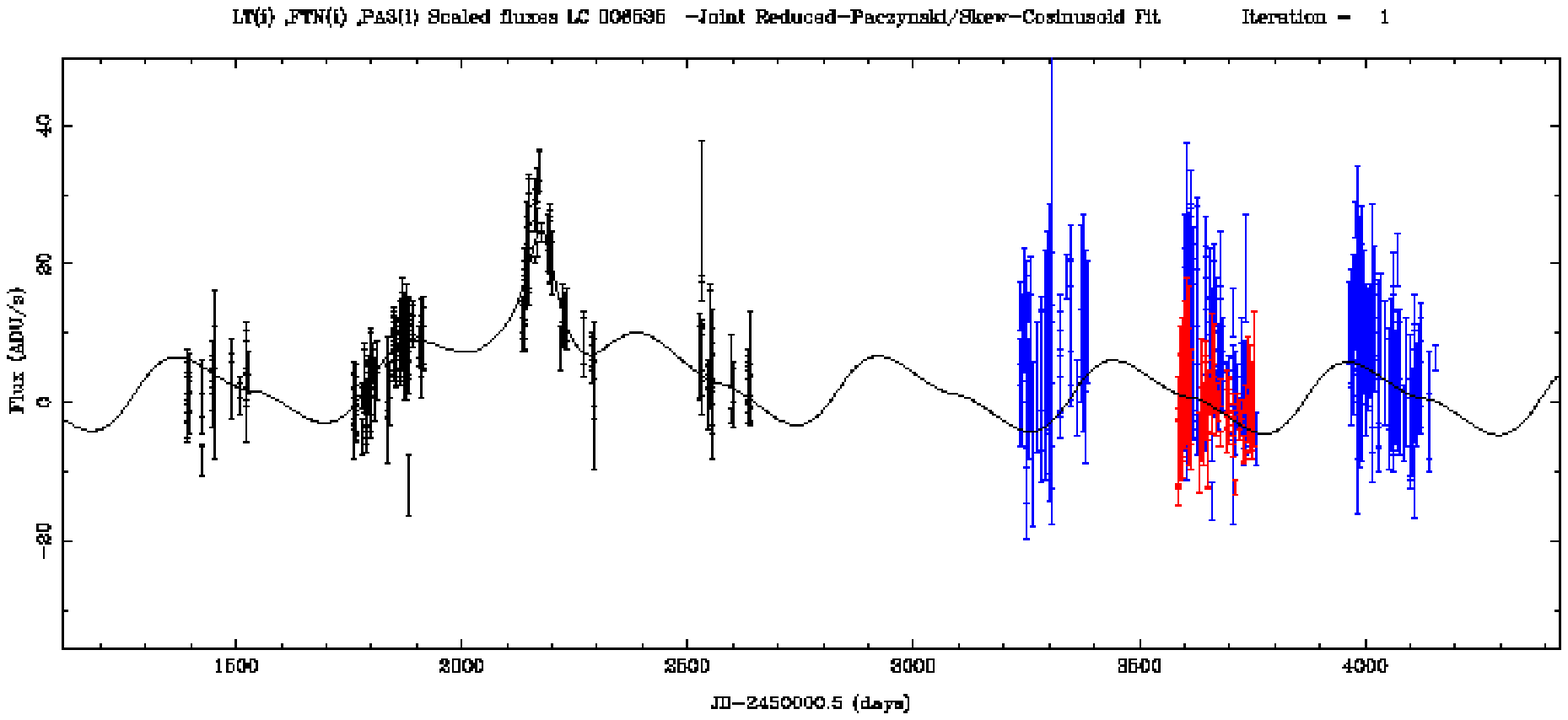} \\
\vspace*{2cm}
   \leavevmode
 \includegraphics{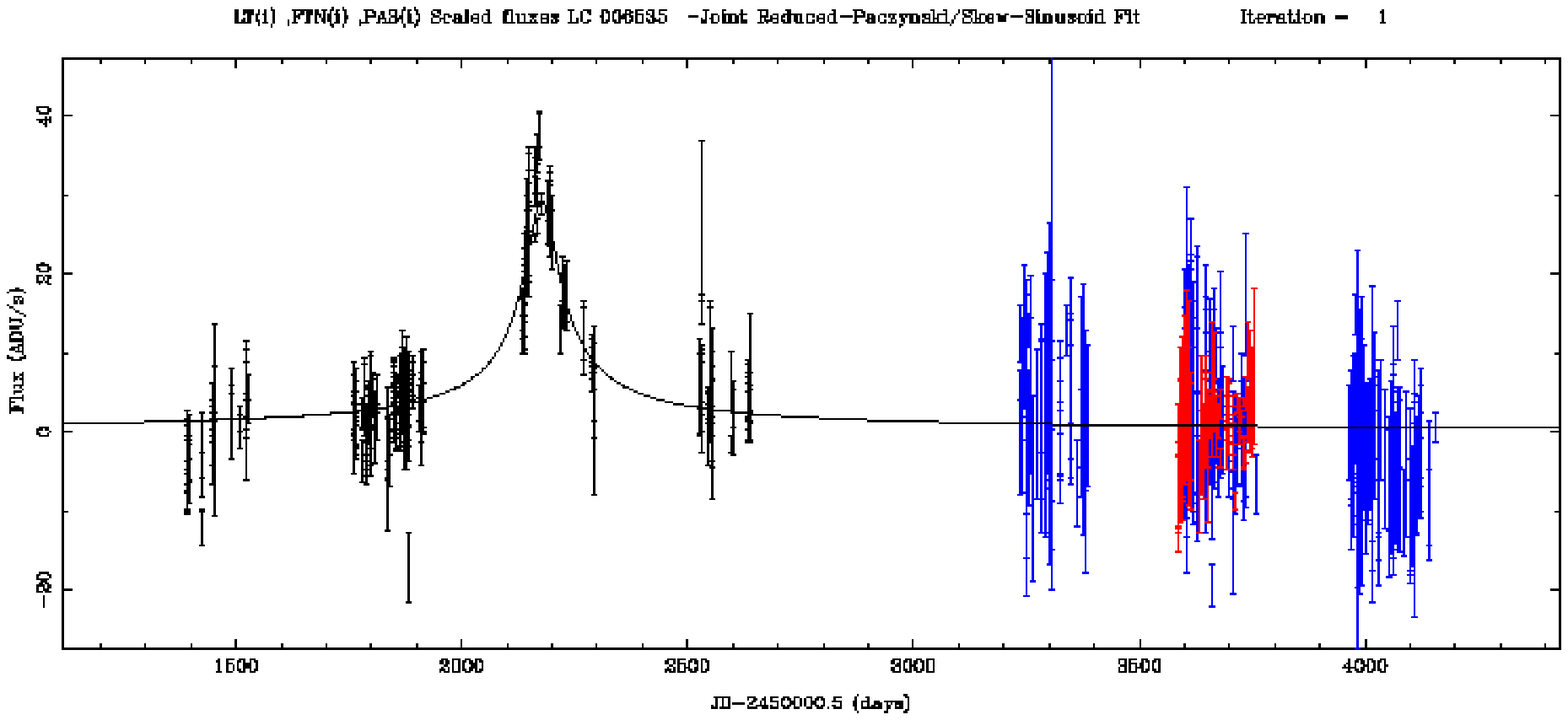} \\
\end{array}$
\caption[Lightcurve of object number $6535$ in the 2007 photometry, showing a) the original mixed fit and
b) the lensing component only, after the variable component has been subtracted.]{Lightcurve of object number $6535$ in the 2007 photometry, showing a) the original mixed fit and b) the peak region of the lensing component only, after the variable component has been subtracted.}
\end{figure}

\newpage

\clearpage

\begin{figure}[!ht]
\vspace*{10cm}
$\begin{array}{c}
\vspace*{10cm}
   \leavevmode
 \includegraphics{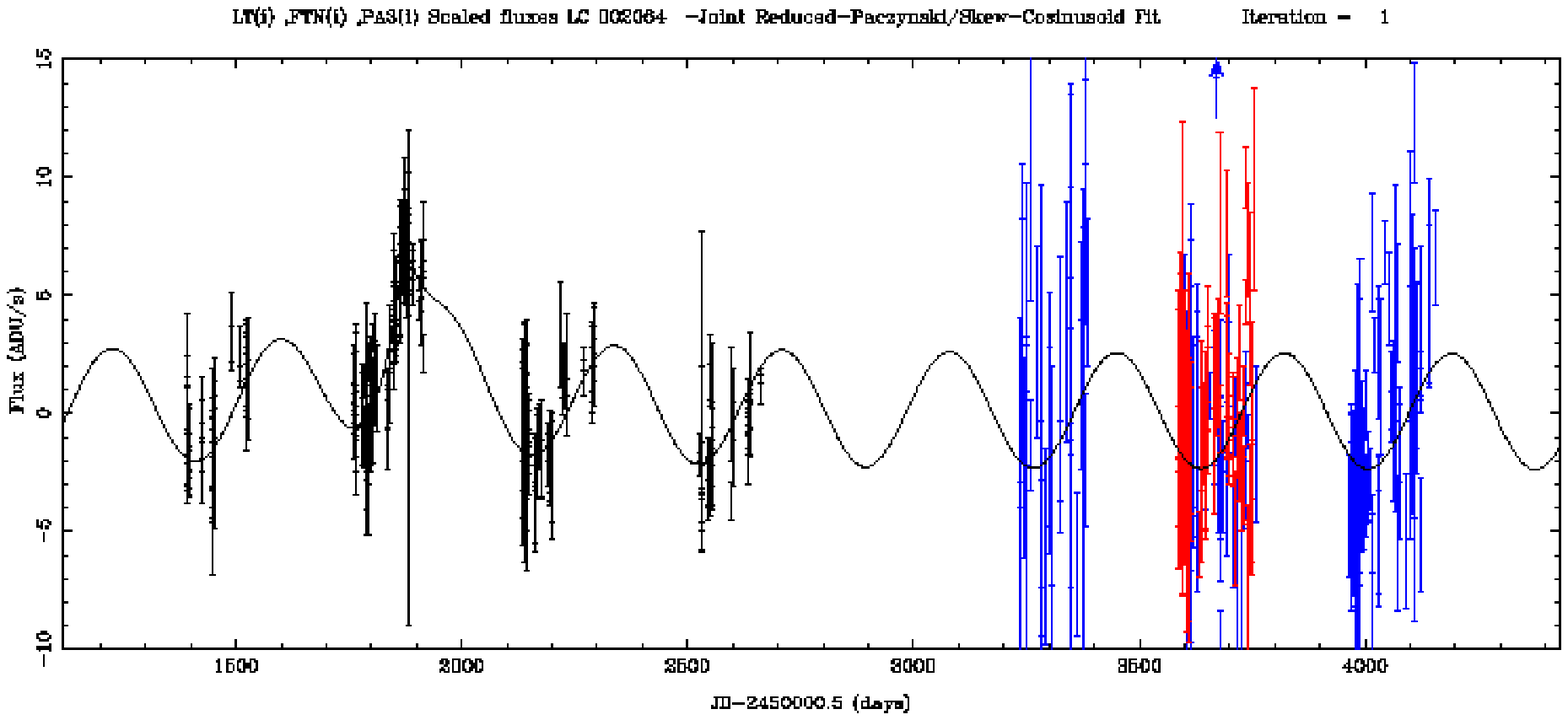} \\
\vspace*{2cm}
   \leavevmode
 \includegraphics{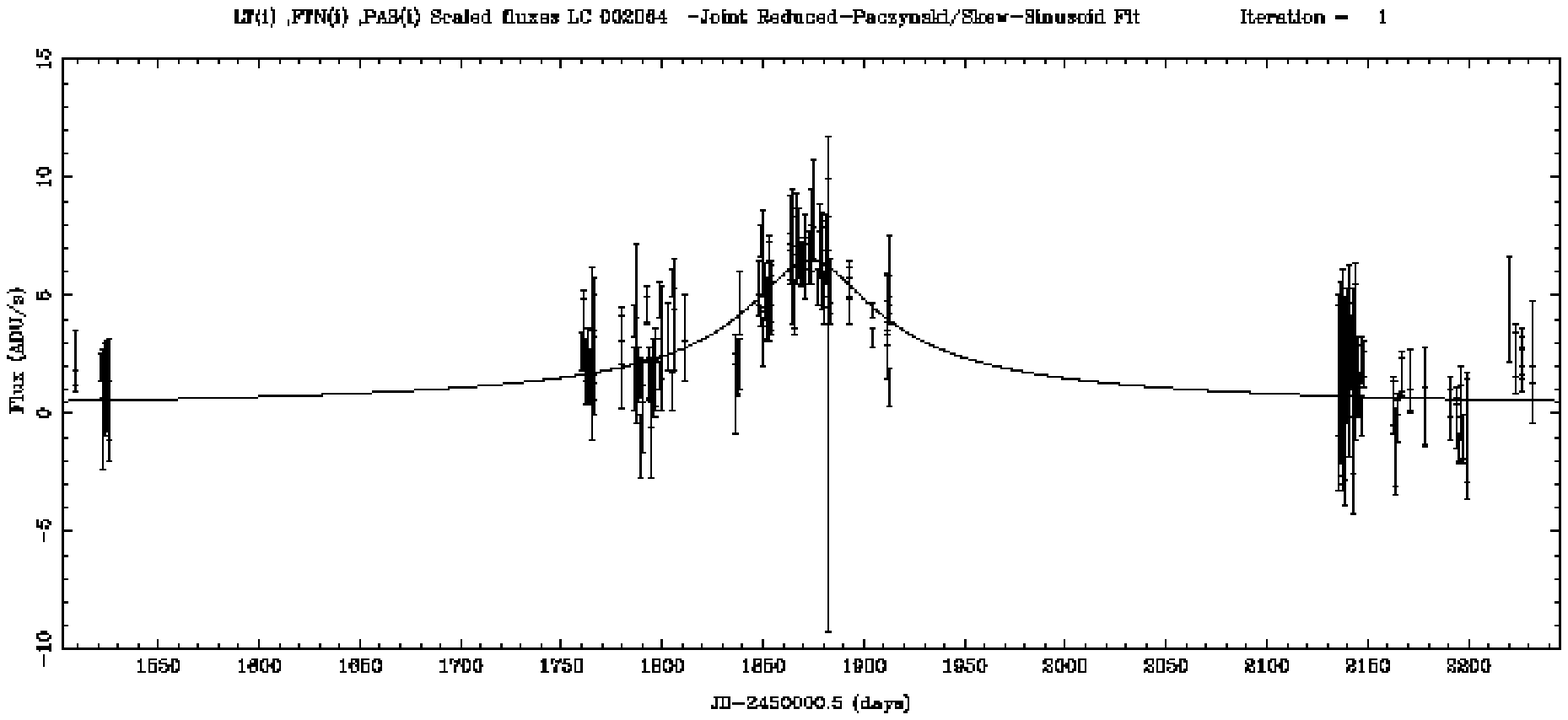} \\
\end{array}$
\caption[Lightcurve of object number $2064$ in the 2007 photometry, showing a) the original mixed fit and
b) the lensing component only, after the variable component has been subtracted.]{Lightcurve of object number $2064$ in the 2007 photometry, showing a) the original mixed fit and b) the peak region of the lensing component only, after the variable component has been subtracted.}
\end{figure}

\clearpage
\newpage

\begin{figure}[!ht]
\vspace*{10cm}
$\begin{array}{c}
\vspace*{10cm}
   \leavevmode
 \includegraphics{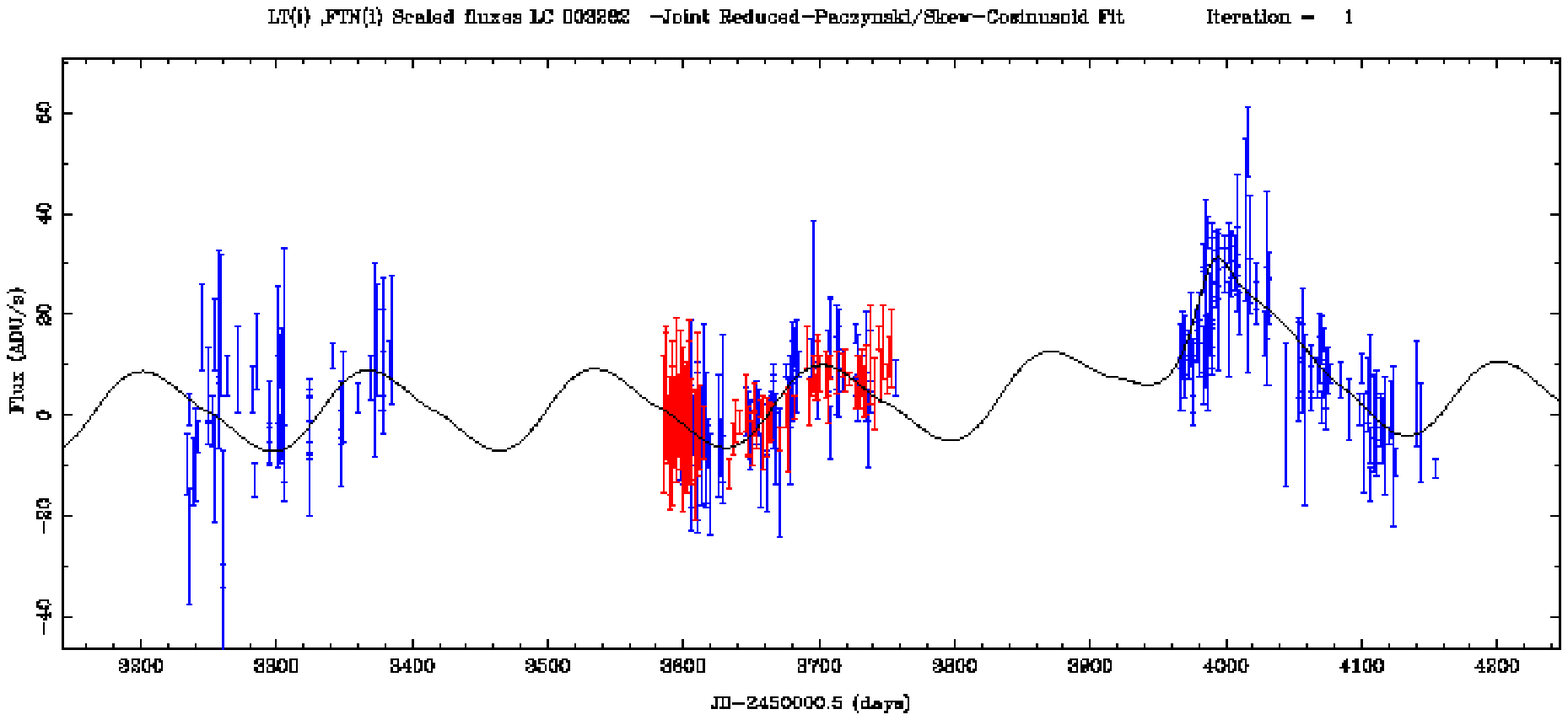} \\
\vspace*{2cm}
   \leavevmode
 \includegraphics{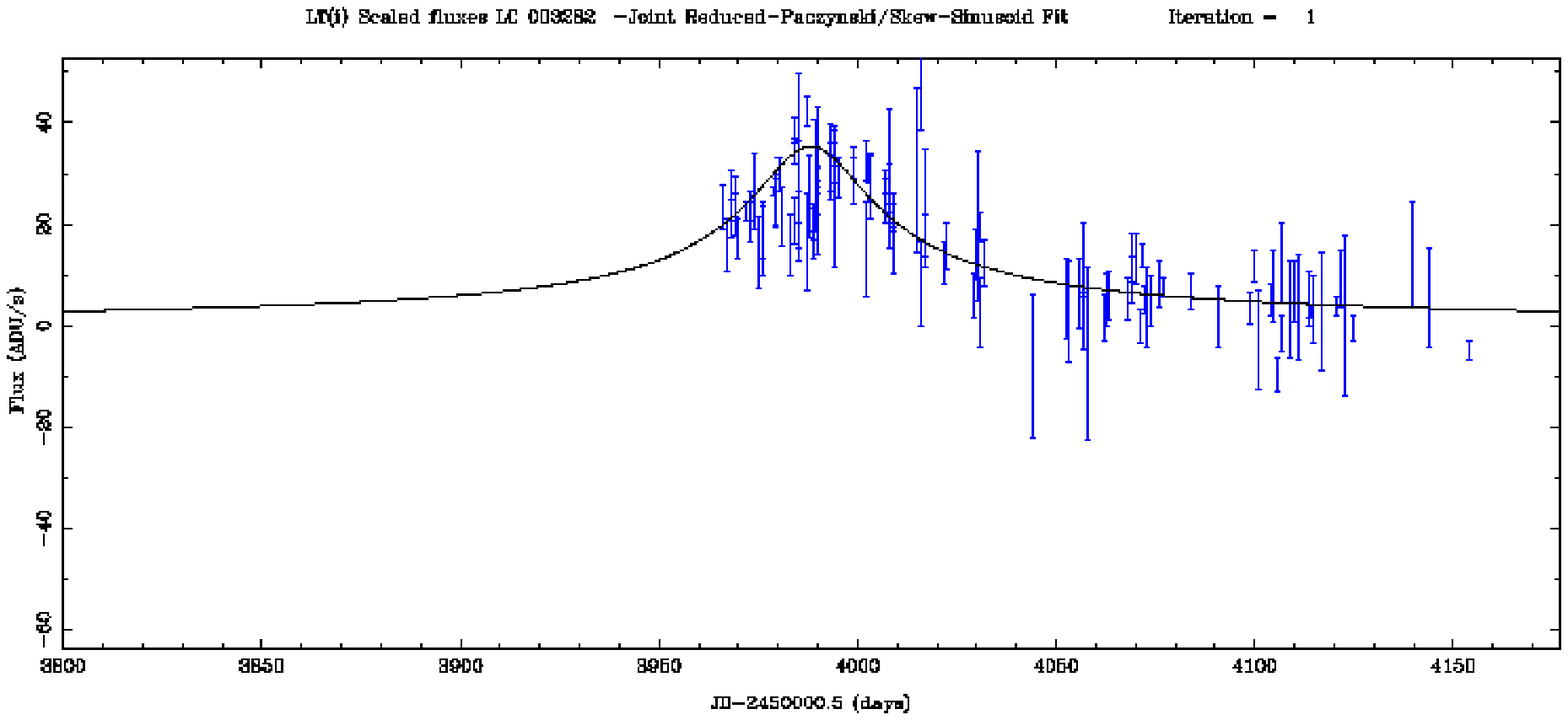} \\
\end{array}$
\caption[Lightcurve of object number $3282$ in the 2007 photometry, showing a) the original mixed fit and
b) the lensing component only, after the variable component has been subtracted.]{Lightcurve of object number $3282$ in the 2007 photometry, showing a) the original mixed fit and b) the peak region of the lensing component only, after the variable component has been subtracted.}
\label{LC_03282_2007}
\end{figure}

\clearpage

\subsubsection{Results using 2008 photometry run lightcurves}
\label{lensing_candidates_2008}

When the latest version of the Angstrom DIA photometry pipeline was complete,
new lightcurves were produced, and the selection pipeline re-run on these new data.
Only LT and PA data were available initially, and so these were used for the analysis.
The pipeline was also run on the LT data on their own, to enable the effect of adding in the PA data to be evaluated at a later date.

Once the list of lightcurves which passed the ``loose'' cuts was known, the data were analysed to calculate the quality factors, as in Section \ref{lensing_candidates_2007}.
Also as in Section \ref{lensing_candidates_2007}, the resulting list was checked for groups of lightcurves with identical quality factors (and hence fitting parameters) or very similar $t_0$ and $t_{\rm{FWHM}}$ and hence similar quality factors. Both of these sorts of groups were treated as single objects for the rest of the analysis. Any pairs of objects for which the parameters were very similar, but not similar enough to be certainly from the same object had their physical X,Y coordinates checked to firmly identify whether they were different objects or not. When the data were grouped in this way, which corresponded to an estimate of the numbers of independent astrophysical objects present in the sample, the numbers of clusters were considerably lower than the number of lightcurves, even after removal of all the identical sets. For example, at one point in the final development of the pipeline using the LT and PA run, using the ``inter'' cuts, there were initially $355$ lightcurves selected, but it seems that these represent only about $177$ actual objects. This seemed to be a startling rate of lightcurve degeneracy, being in this example roughly a $2$-$1$ ratio. The same level of repeated lightcurves was not seen in the $2007$ photometry, although they did occur.

Overall, the total numbers of objects selected using only LT or both LT and PA data were found to be of comparable order but the average effect of using the PA data in addition to the LT data was to reduce significantly the number of lightcurves selected.
 The ratio of numbers in the three categories of $\chi^2$/d.o.f. cut used above for the $2007$ data, 
(i.e. ``loose'',``inter'' and ``scaled'') 
where sufficient statistics existed, with and without PA data, was calculated to be a remarkably stable $0.36$.
That adding PA data reduced the numbers of selected lightcurves might have been predicted as the longer baselines this makes possible provide more opportunities for periodic variables to repeat, which otherwise might have been consistent with one of the two kinds of lensing fits performed in this work. Also, a greater chance exists for variable stars which do not quite fit the relatively simple model used here to exhibit variations which might not have occurred within the time range of the LT data alone, hence increasing the global $\chi^2$/d.o.f. and perhaps excluding them from selection.
Conversely, the addition of the PA data would have been expected to \emph{increase} the number of detected variable stars, if the $2007$ data were a reasonable guide. This has not yet been checked.

\textbf{Final selection of candidates}

After the analysis of the $2007$ data, Cut 12 was added to the selection procedure.
Using all $12$ cuts, the ``loose'' cuts allowed through $145$ lightcurves.
After much experimentation with various cut levels, a set of cuts was decided upon that aimed to
produce as good a set of candidates as possible without introducing any clearly
insufficiently well fitted lightcurves.  
 These cuts were: lensing ratio; $LR > 3.0$, $\chi^2$ difference ratio; $CDR
 \geq 0.4$, and the two $\chi^2$/d.o.f. upper limits were set at
 $\chi^{2}_{\rm{glob}}$/d.o.f.$ \leq 7.0$, $\chi^{2}_{\rm{local}}$/d.o.f.$ \leq 5.0$ respectively, which
 is equivalent to $\chi^{2}_{\rm{glob}}$/d.o.f.$ \leq 3.6$, $\chi^{2}_{\rm{local}}$ /d.o.f.$\leq 2.6$ for
 photon noise limited data. The values of LR and CDR were both set significantly more strictly than for the $2007$ data set to cut down the level of false positives that the pipeline could allow through. The values of $\chi^{2}_{\rm{glob}}$/d.o.f. and  $\chi^{2}_{\rm{local}}$/d.o.f. were found originally empirically by experimentation, always maintaining the value of $\chi^{2}_{\rm{local}}$/d.o.f.$\le\chi^{2}_{\rm{glob}}$/d.o.f., but once the necessary factor of $\sim2$ was known about these both also seemed to be sensible values. Using the above values, $38$ lightcurves were
 selected, which proved to correspond to $20$ physical objects,
of which $2$ were simple reduced Paczy\'nski fits. All of these were selected on the first iteration.
The fitting parameters of these $20$ lightcurves are presented in Tables \ref{2008_results_cuts_data_summary_part_I} and \ref{2008_results_cuts_data_summary_part_II}.
The parameters $t_{0}$ and $t_{\rm{FWHM}}$ which describe the shape of the lensing peaks are detailed
in the two Tables \ref{2008_results_fitting_data_summary_lensing_ratio_>3_part_I} and \ref{2008_results_fitting_data_summary_lensing_ratio_>3_part_II}.

\begin{table}
\caption[Table (Part 1) summarising the selection information of microlensing candidates 
selected for global $\chi^2$/d.o.f. $\leq 7.0$, local $\chi^2$/d.o.f. $\leq 5.0$, lensing ratio $> 3.0$, $\chi^2$ difference ratio $\geq 0.4$, using the $2008$ photometry.]
{Table (Part 1) summarising the selection information of microlensing candidates 
selected for global $\chi^2$/d.o.f. $\leq 7.0$, local $\chi^2$/d.o.f. $\leq 5.0$, lensing ratio $> 3.0$, $\chi^2$ difference ratio $\geq 0.4$ using the $2008$ photometry.                                      }
\begin{center}
\begin{tabular}{|c||c|c|c|c|c|c|c|c|c|c|}
\hline
\hline
Object No & LC & QF & iter & Fit &$\chi^{2}_{\rm{loc}}$& LSN & CDR &$\chi^{2}_{\rm{\rm{glob}}}$& BSQF & $t_{0}$ in\\
\hline
  1 &  90314 &       4.20 &          1 &   M &       2.27 &       4.47 &       1.89 &       2.15 &       0.72 &   PA \\ 
    & 6614 &       4.20 &          1 &   M &       2.27 &       4.47 &       1.89 &       2.15 &       0.72 &   PA \\ 
    & 56125 &       4.20 &          1 &   M &       2.27 &       4.47 &       1.89 &       2.15 &       0.72 &   PA \\ 
    & 16819 &       4.20 &          1 &   M &       2.27 &       4.47 &       1.89 &       2.15 &       0.72 &   PA \\ 
      &   &         &          &     &         &         &         &         &         &      \\ 
  2 &  38847 &       2.94 &          1 &   M &       3.11 &       3.46 &       1.09 &       4.19 &       1.48 &   PA \\ 
      &      &         &          &     &         &         &         &         &         &      \\ 
  3 &  87898 &       2.67 &          1 &   M &       4.90 &       6.21 &       0.87 &       4.92 &       1.13 &   LT \\ 
    & 49986 &       2.67 &          1 &   M &       4.90 &       6.21 &       0.87 &       4.92 &       1.13 &   LT \\ 
    & 29986 &       2.67 &          1 &   M &       4.90 &       6.21 &       0.87 &       4.92 &       1.13 &   LT \\ 
    & 28743 &       2.67 &          1 &   M &       4.90 &       6.21 &       0.87 &       4.92 &       1.13 &   LT \\ 
     &       &         &          &     &         &         &         &         &         &      \\ 
  4  & 82883 &       2.65 &          1 &   M &       3.24 &       3.46 &       0.46 &       3.71 &       1.81 &   PA \\ 
     &       &         &          &     &         &         &         &         &         &      \\  
  5 &   83057 &       2.30 &          1 &   M &       2.52 &       4.33 &       1.22 &       3.35 &       0.70 &   PA \\ 
     &       &         &          &     &         &         &         &         &         &      \\ 
   6 &  39599 &       1.59 &          1 &   M &       1.96 &       3.67 &       0.51 &       3.55 &       0.79 &   LT \\ 
    & 81226 &       1.59 &          1 &   M &       1.96 &       3.67 &       0.51 &       3.55 &       0.79 &   LT \\ 
    & 91968 &       1.59 &          1 &   M &       1.96 &       3.67 &       0.51 &       3.55 &       0.79 &   LT \\ 
     &       &         &          &     &         &         &         &         &         &      \\ 
  7 &   8391 &       1.49 &          1 &   M &       2.48 &       6.26 &       0.56 &       3.01 &       0.42 &   PA \\ 
     &       &         &          &     &         &         &         &         &         &      \\
  8 &  53608 &       1.46 &          1 &   M &       4.27 &       3.23 &       0.41 &       2.75 &       1.12 &   LT \\ 
     &       &         &          &     &         &         &         &         &         &      \\ 
  9 &  26795 &       1.29 &          1 &   M &       3.55 &       6.71 &       0.56 &       3.73 &       0.45 &   PA \\ 
     &       &         &          &     &         &         &         &         &         &      \\ 
  10 &   68058 &       1.27 &          1 &   M &       3.69 &       4.07 &       0.53 &       4.89 &       0.88 &   PA \\ 
    & 74112 &       1.27 &          1 &   M &       3.70 &       4.06 &       0.53 &       4.89 &       0.88 &   PA \\ 
     &       &         &          &     &         &         &         &         &         &      \\ 
\hline
\end{tabular}
\end{center}
\label{2008_results_cuts_data_summary_part_I}
\end{table}

\begin{table}
\caption[Table (Part 2) summarising the selection information of microlensing candidates 
selected for global $\chi^2$/d.o.f. $\leq 7.0$, local $\chi^2$ /d.o.f. $\leq  5.0$, lensing ratio $> 3.0$, $\chi^2$ difference ratio $\geq 0.4$, using the $2008$ photometry.]
{Table (Part 2) summarising the selection information of microlensing candidates 
selected for global $\chi^2$/d.o.f. $\leq 7.0$, local $\chi^2$/d.o.f. $\leq  5.0$, lensing ratio $> 3.0$, $\chi^2$ difference ratio $\geq 0.4$ using the $2008$ photometry.                                            
                        }
\begin{center}
\begin{tabular}{|c||c|c|c|c|c|c|c|c|c|c|}
\hline
\hline
Object No & LC & QF & iter & Fit &$\chi^{2}_{\rm{loc}}$& LSN & CDR &$\chi^{2}_{\rm{\rm{glob}}}$& BSQF & $t_{0}$ in\\
\hline 
 11 &   29137 &       1.13 &          1 &   M &       0.64 &       3.01 &       0.41 &       2.00 &       0.35 &   LT \\ 
     &       &         &          &     &         &         &         &         &         &      \\ 
 20 &   60097 &       1.10 &          1 &   M &       3.99 &       4.22 &       0.50 &       2.87 &       0.60 &   PA \\ 
     &       &         &          &     &         &         &         &         &         &      \\ 
 21 &    13139 &       1.08 &          1 &   M &       2.71 &       4.05 &       0.72 &       2.71 &       0.42 &   PA \\ 
    &  3957 &       1.08 &          1 &   M &       2.39 &       3.51 &       0.81 &       2.56 &       0.42 &   PA \\ 
    &   995 &       1.07 &          1 &   M &       2.37 &       3.48 &       0.81 &       2.56 &       0.42 &   PA \\ 
     &       &         &          &     &         &         &         &         &         &      \\ 
  22  &  2616 &       0.98 &          1 &   M &       3.55 &       3.83 &       0.43 &       0.78 &       0.39 &   PA \\ 
    & 24441 &       0.98 &          1 &   M &       3.55 &       3.83 &       0.43 &       0.78 &       0.39 &   PA \\ 
     &       &         &          &     &         &         &         &         &         &      \\ 
  23 &  42645 &       0.95 &          1 &   M &       2.91 &       5.83 &       0.62 &       3.38 &       0.31 &   PA \\ 
     &       &         &          &     &         &         &         &         &         &      \\ 
  24 &  54978 &       0.82 &          1 &   M &       4.38 &       4.24 &       0.66 &       2.27 &       0.39 &   PA \\ 
     &       &         &          &     &         &         &         &         &         &      \\ 
  25 &  77494 &       0.59 &          1 &   P &       3.34 &       3.62 &       0.81 &       3.59 &       0.31 &   LT \\ 
      &      &         &          &     &         &         &         &         &         &      \\ 
  26 & 74650 &       0.51 &          1 &   M &       0.40 &       4.62 &       1.89 &       4.62 &       0.10 &   PA \\ 
     &  19644 &       0.51 &          1 &   M &       0.40 &       4.62 &       1.89 &       4.62 &       0.10 &   PA \\ 
     &       &         &          &     &         &         &         &         &         &      \\ 
  27 & 81698 &       0.46 &          1 &   M &       4.02 &       4.67 &       0.42 &       5.00 &       0.31 &   PA \\ 
     & 28357 &       0.46 &          1 &   M &       4.00 &       4.69 &       0.42 &       5.04 &       0.31 &   PA \\ 
     &       &         &          &     &         &         &         &         &         &      \\ 
  28 &  82746 &       0.37 &          1 &   P &       1.76 &       3.31 &       0.43 &       4.51 &       0.24 &   PA \\ 
     &       &         &          &     &         &         &         &         &         &      \\ 
\hline
\end{tabular}
\end{center}
\label{2008_results_cuts_data_summary_part_II}
\end{table}

\begin{tiny}
\begin{table}
\caption[Table (Part I) summarising the information about the fitted peak of microlensing candidates 
selected for global $\chi^2$/d.o.f. $\leq 7.0$, local $\chi^2$/d.o.f. $\leq 5.0$, lensing ratio $> 3.0$,  $\chi^2$ difference ratio $\geq 0.4$ using the $2008$ photometry.]
{Table (Part I) summarising the information about the fitted peak of microlensing candidates 
selected for global $\chi^2$/d.o.f. $\leq 7.0$, local $\chi^2$/d.o.f. $\leq 5.0$, lensing ratio $> 3.0$,  $\chi^2$ difference ratio $\geq 0.4$ using the $2008$ photometry.                                                        }
\begin{center}
\begin{tabular}{|c||c|c|c|c|c|c|c|c|}
\hline
\hline
Object No. & LC & QF & iter & Fit & $t_{0}$ in & $t_{0}$ & $t_{\rm{FWHM}}$\\
\hline
 1  & 90314 &       4.20 &          1 &   M &   PA &    1918.19 &      25.15 \\ 
    &  6614 &       4.20 &          1 &   M &   PA &    1918.19 &      25.15 \\ 
    & 56125 &       4.20 &          1 &   M &   PA &    1918.19 &      25.15 \\ 
    & 16819 &       4.20 &          1 &   M &   PA &    1918.19 &      25.15 \\ 
    &        &        &          &     &      &        &        \\ 
  2 & 38847 &       2.94 &          1 &   M &   PA &    1813.39 &      64.50 \\ 
    &        &        &          &     &      &        &        \\ 
  3 & 87898 &       2.67 &          1 &   M &   LT &    4766.22 &      78.20 \\ 
    & 49986 &       2.67 &          1 &   M &   LT &    4766.22 &      78.20 \\ 
    & 29986 &       2.67 &          1 &   M &   LT &    4766.22 &      78.20 \\ 
    & 28743 &       2.67 &          1 &   M &   LT &    4766.22 &      78.20 \\ 
    &        &        &          &     &      &        &        \\
  4 & 82883 &       2.65 &          1 &   M &   PA &    1984.58 &     434.73 \\ 
    &        &        &          &     &      &        &        \\
  5 &  83057 &       2.30 &          1 &   M &   PA &    2535.37 &      45.52 \\ 
    &        &        &          &     &      &        &        \\
  6 &  39599 &       1.59 &          1 &   M &   LT &    4792.65 &      17.27 \\ 
     & 81226 &       1.59 &          1 &   M &   LT &    4792.65 &      17.27 \\ 
     & 91968 &       1.59 &          1 &   M &   LT &    4792.65 &      17.27 \\ 
     &       &        &          &     &      &        &        \\
  7 &   8391 &       1.49 &          1 &   M &   PA &    2194.58 &      10.70 \\ 
     &       &        &          &     &      &        &        \\
  8 &  53608 &       1.46 &          1 &   M &   LT &    4779.34 &      96.53 \\ 
     &       &        &          &     &      &        &        \\
  9 &  26795 &       1.29 &          1 &   M &   PA &    2636.51 &      16.51 \\ 
     &       &        &          &     &      &        &        \\
  10 &  68058 &       1.27 &          1 &   M &   PA &    2142.92 &      10.30 \\ 
      & 74112 &       1.27 &          1 &   M &   PA &    2142.93 &      10.16 \\ 
     &       &        &          &     &      &        &        \\
\hline
\end{tabular}
\end{center}
\label{2008_results_fitting_data_summary_lensing_ratio_>3_part_I}
\end{table}
\end{tiny}

\begin{tiny}
\begin{table}
\caption[Table (Part II) summarising the information about the fitted peak of microlensing candidates 
selected for global $\chi^2$/d.o.f. $\leq 7.0$, local $\chi^2$/d.o.f. $\leq 5.0$, lensing ratio $> 3.0$,  $\chi^2$ difference ratio $\geq 0.4$ using the $2008$ photometry.]
{Table (Part II) summarising the information about the fitted peak of microlensing candidates 
selected for global $\chi^2$/d.o.f. $\leq 7.0$, local $\chi^2$/d.o.f. $\leq 5.0$, lensing ratio $> 3.0$,  $\chi^2$ difference ratio $\geq 0.4$ using the $2008$ photometry.                                                        }
\begin{center}
\begin{tabular}{|c||c|c|c|c|c|c|c|c|}
\hline
\hline
Object No. & LC & QF & iter & Fit & $t_{0}$ in & $t_{0}$ & $t_{\rm{FWHM}}$\\
\hline
   11 & 29137 &       1.13 &          1 &   M &   LT &    4466.33 &      25.12 \\ 
      &      &        &          &     &      &        &        \\ 
   12 & 60097 &       1.10 &          1 &   M &   PA &    1911.02 &      14.90 \\ 
      &      &        &          &     &      &        &        \\ 
   13 & 13139 &       1.08 &          1 &   M &   PA &    2195.60 &       9.43 \\ 
      & 3957 &       1.08 &          1 &   M &   PA &    2195.59 &       9.10 \\ 
      &  995 &       1.07 &          1 &   M &   PA &    2195.59 &       9.07 \\ 
      &      &        &          &     &      &        &        \\ 
   14 &   2616 &       0.98 &          1 &   M &   PA &    1397.83 &       6.38 \\ 
      &   24441 &       0.98 &          1 &   M &   PA &    1397.83 &       6.38 \\ 
      &      &        &          &     &      &        &        \\ 
   15 &  42645 &       0.95 &          1 &   M &   PA &    1524.50 &      13.09 \\ 
      &      &        &          &     &      &        &        \\ 
   16 &  54978 &       0.82 &          1 &   M &   PA &    1396.19 &       4.30 \\ 
      &      &        &          &     &      &        &        \\ 
   17 &  77494 &       0.59 &          1 &   P &   LT &    3293.69 &      18.92 \\ 
      &      &        &          &     &      &        &        \\ 
   18 &  74650 &       0.51 &          1 &   M &   PA &    1912.95 &       6.18 \\ 
      & 19644 &       0.51 &          1 &   M &   PA &    1912.95 &       6.18 \\ 
      &      &        &          &     &      &        &        \\ 
   19 &  81698 &       0.46 &          1 &   M &   PA &    1525.44 &       8.65 \\ 
      & 28357 &       0.46 &          1 &   M &   PA &    1525.54 &       9.25 \\ 
      &      &        &          &     &      &        &        \\ 
   20 &  82746 &       0.37 &          1 &   P &   PA &    2224.60 &       3.08 \\ 
       &     &        &          &     &      &        &        \\ 
\hline
\end{tabular}
\end{center}
\label{2008_results_fitting_data_summary_lensing_ratio_>3_part_II}
\end{table}
\end{tiny}

If the ``short event'' is included in this list (since with the addition of the Maidanak data and the re-scaling of the LT error bars it would have passed all the above cuts), then the total number of events would become $21$.
\subsubsection{Spatial and timescale distributions}

\begin{table}
\begin{center}
\begin{tabular}{|c|c|}
\hline
\hline
$t_{\rm{FWHM}}$(days) & Number of lightcurves \\
\hline
$0-15$   & $10$ \\
$15-30$  & $5$ \\
$30-45$  & $0$ \\
$45-60$  & $1$ \\
$60-75$  & $1$ \\
$75-90$  & $1$ \\
$90-105$ & $1$ \\
$105+$   & $1$ \\
\hline
\end{tabular}
\end{center}
\caption[Table giving a coarsely-binned histogram of the timescales of the $20$ selected candidates from the $2008$ photometry]{Table giving a coarsely-binned histogram of the timescales of the $20$ selected candidates from the $2008$ photometry}
\label{timescale_dist}
\end{table}

By examining the information in Tables \ref{2008_results_fitting_data_summary_lensing_ratio_>3_part_I} and \ref{2008_results_fitting_data_summary_lensing_ratio_>3_part_II} it may be
seen that the majority of candidates have event timescales (as measured by their $t_{\rm{FWHM}}$) less than $30$ days. One way of binning the $20$ selected events leads to the information contained in Table \ref{timescale_dist}. While 2/3 of the 
candidates have $t_{\rm{FWHM}} < 30$ days, in actual fact $65\%$ have $t_{\rm{FWHM}} < 20$ days. However, in addition to this short duration rump is what appears to be a
longer duration tail containing the other $5$ candidates. Number statistics are clearly too low to make any more detailed comments about the timescale distribution at this stage.

\begin{figure}[!ht]
\vspace*{10cm}
$\begin{array}{c}
\vspace*{0cm}
   \leavevmode
 \includegraphics{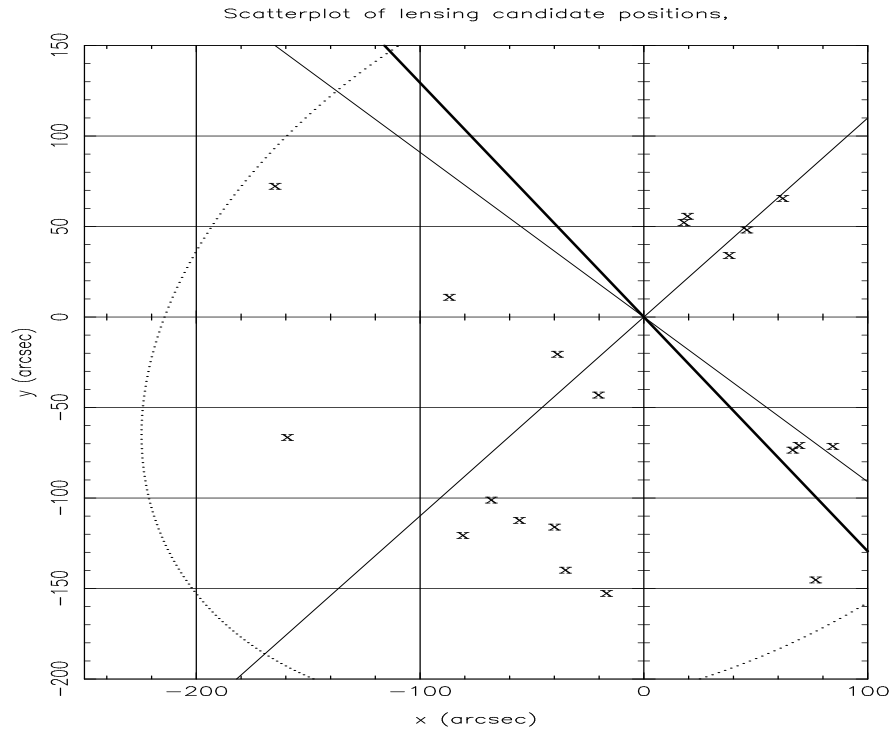} \\
\end{array}$
\caption[Scatter plot showing the spatial distribution of the $20$ selected lensing candidates]{Scatter plot showing the spatial distribution of the $20$ selected lensing candidates. The dark diagonal line shows the orientation of the M31 disk, and the lighter lines are at $10^{\circ}$ and $100^{\circ}$ from the disk direction}
\label{20_candidates_spatial_dist}
\end{figure}

In Figure \ref{20_candidates_spatial_dist} is shown the spatial
distribution of the $20$ selected candidates.
Although this plot also obviously suffers from a lack of statistics,
some qualitative observations can still be made.
Firstly, there seems to be a possible ``E-W'' asymmetry in the numbers
of candidates. For example, if the perpendicularly crossed lines which are oriented
in the direction of the previously measured bulge orientation of $10^{\circ}$ are taken as a reference, there are $14$ candidates on the Western (right) side of the N-S line, but only $6$
on the Eastern (left) side. However, only a small clockwise rotation of these lines would mean the count was $11$/$9$, which is much less significant. Secondly, there appears to be a near-far asymmetry, with the nearer part of the galaxy towards the top right having fewer candidates, although the number densities would need to be worked out using the true available areas given the shape, orientation and overlaps of the Angstrom fields before this conclusion can be made firmly.

\subsubsection{Candidate lightcurves}

Plots of the other $20$ selected events are presented below in Figures \ref{2008_selection_pac_LC_77494} to \ref{2008_selection_mixed_LC_81698}, both of the whole lightcurve with LT and/or PA data and of the Paczy\'nski peak part of the data (with the variable fit part subtracted in the case of mixed fits), plus the residuals to the chosen fits.
 Out of the groups of lightcurves with similar
fit parameters, physical coordinates and quality factors, the lightcurves chosen for the plots are either the lightcurve with the highest quality factor, or, if the lightcurves and hence quality factors were truly identical, the lightcurve with the lowest lightcurve number in the group.

\textbf{Reduced Paczy\'nski Fit Objects}
\newpage

\begin{figure}[!ht]
\vspace*{7cm}
$\begin{array}{c}
\vspace*{7.5cm}
   \leavevmode
 \includegraphics{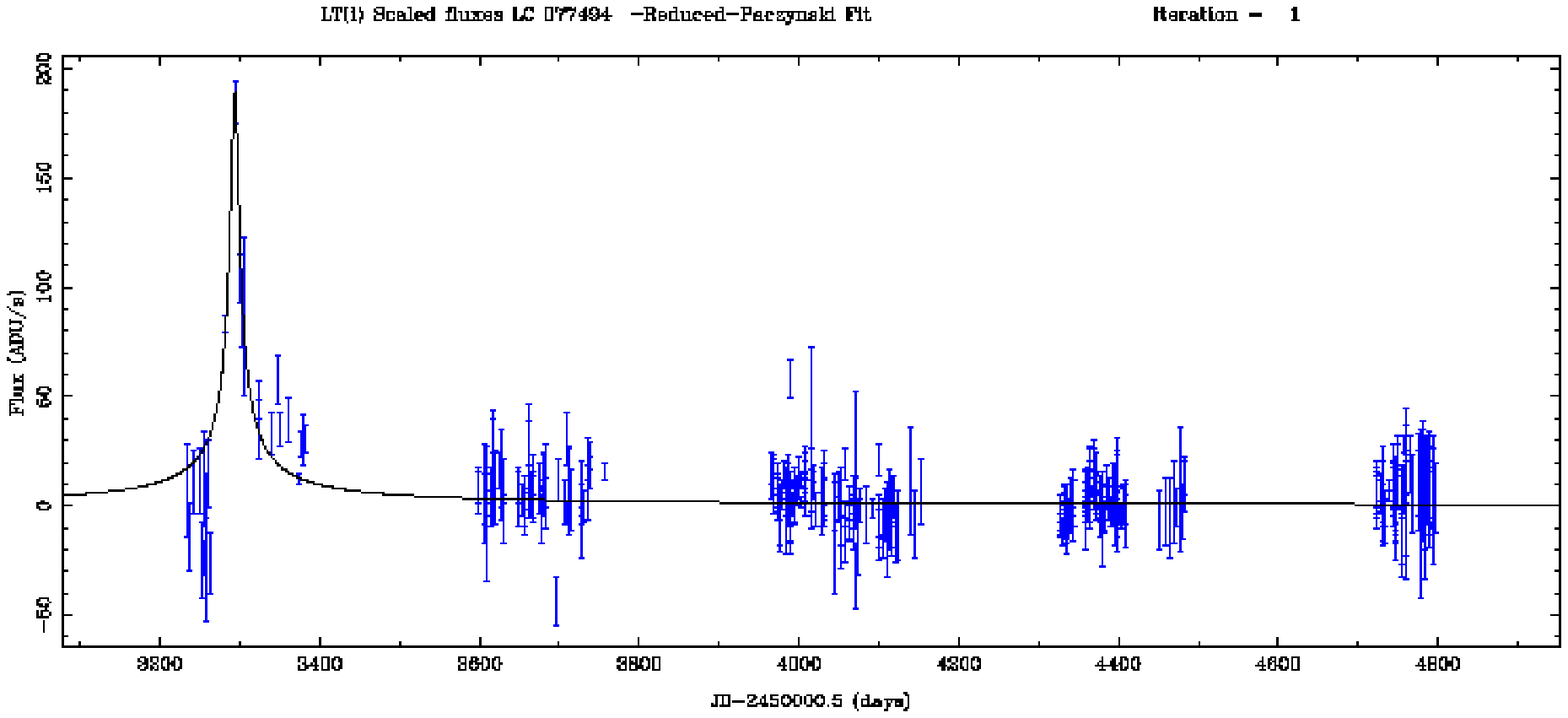} \\
\vspace*{7.5cm}
   \leavevmode
 \includegraphics{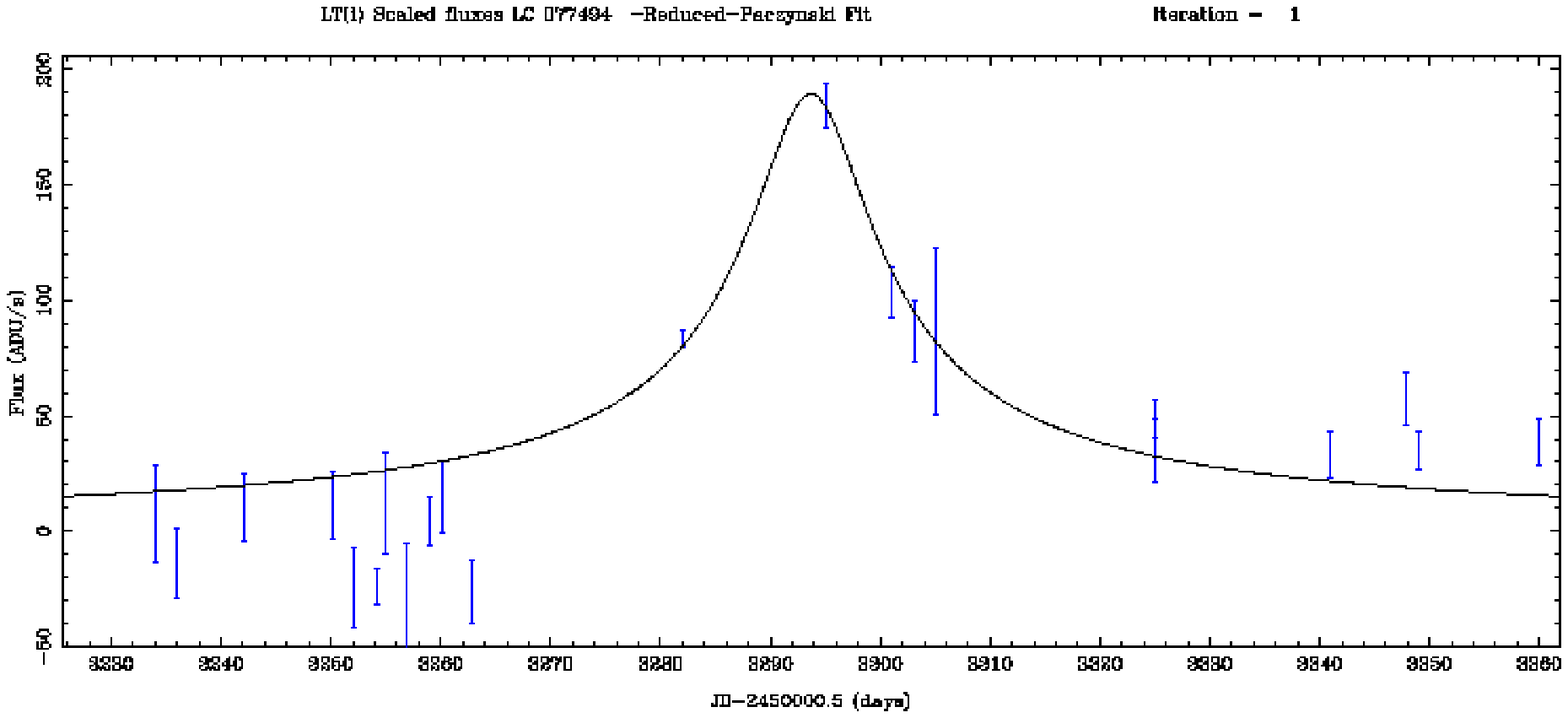} \\
\vspace*{0cm}
   \leavevmode
 \includegraphics{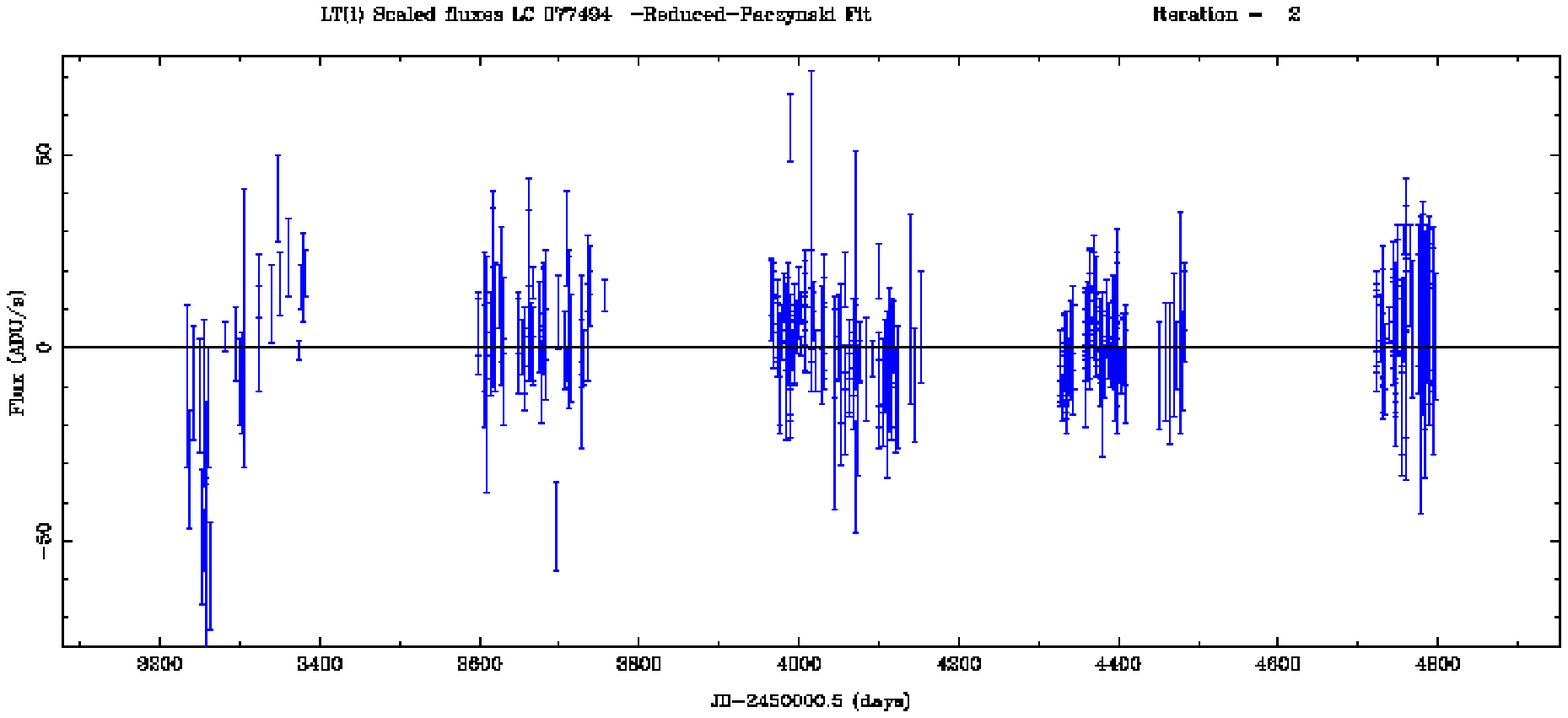} \\
\end{array}$
\caption[Lightcurve of object number $77494$ in the 2008 photometry, showing Top) the original Paczy\'nski fit,
Middle) the central peak region and Bottom) the residuals after subtraction of the fit.]{Lightcurve of object number $77494$ in the 2008 photometry, showing Top) the original Paczy\'nski fit,
Middle) the central peak region and Bottom) the residuals after subtraction of the fit.}
\label{2008_selection_pac_LC_77494}
\end{figure}

\newpage

\begin{figure}[!ht]
\vspace*{7cm}
$\begin{array}{c}
\vspace*{7.5cm}
   \leavevmode
 \includegraphics{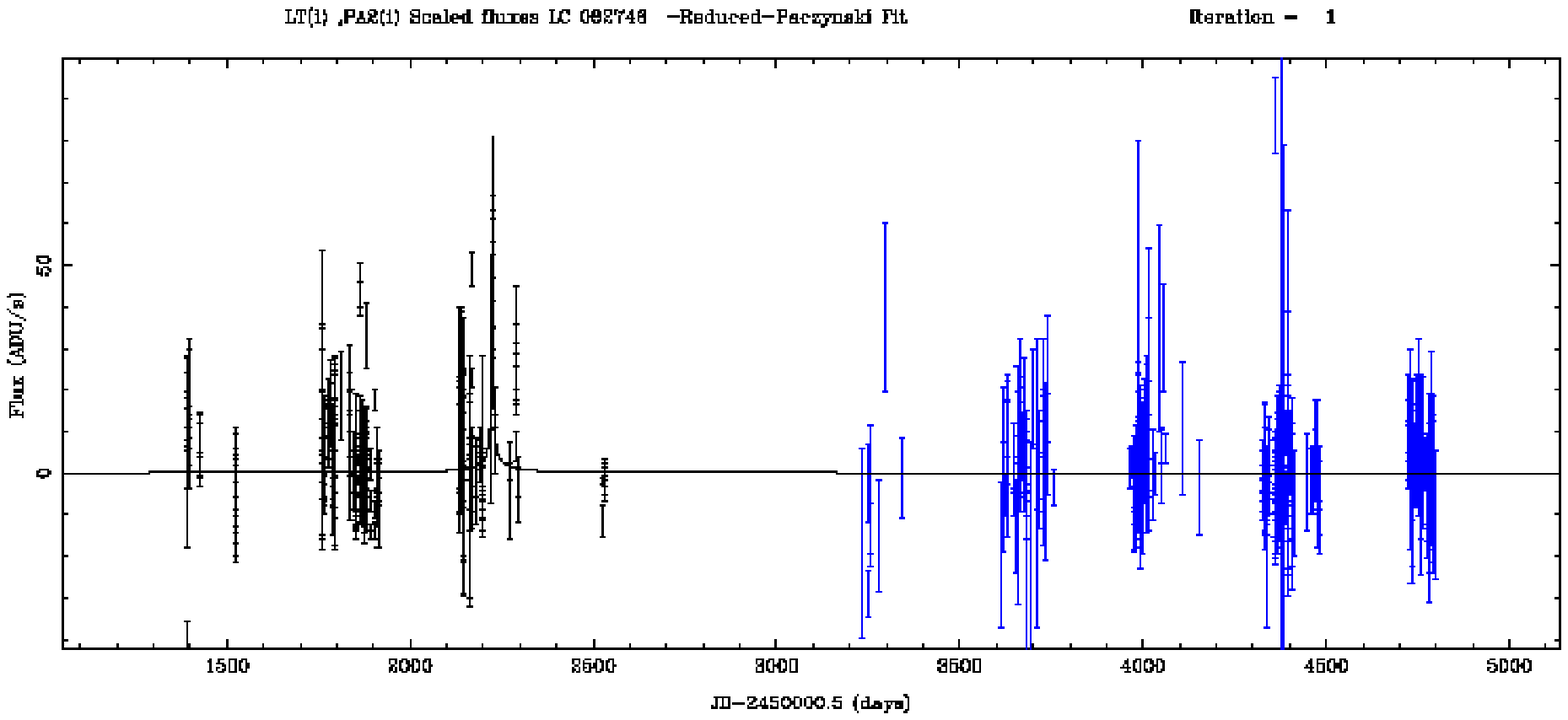} \\
\vspace*{7.5cm}
   \leavevmode
 \includegraphics{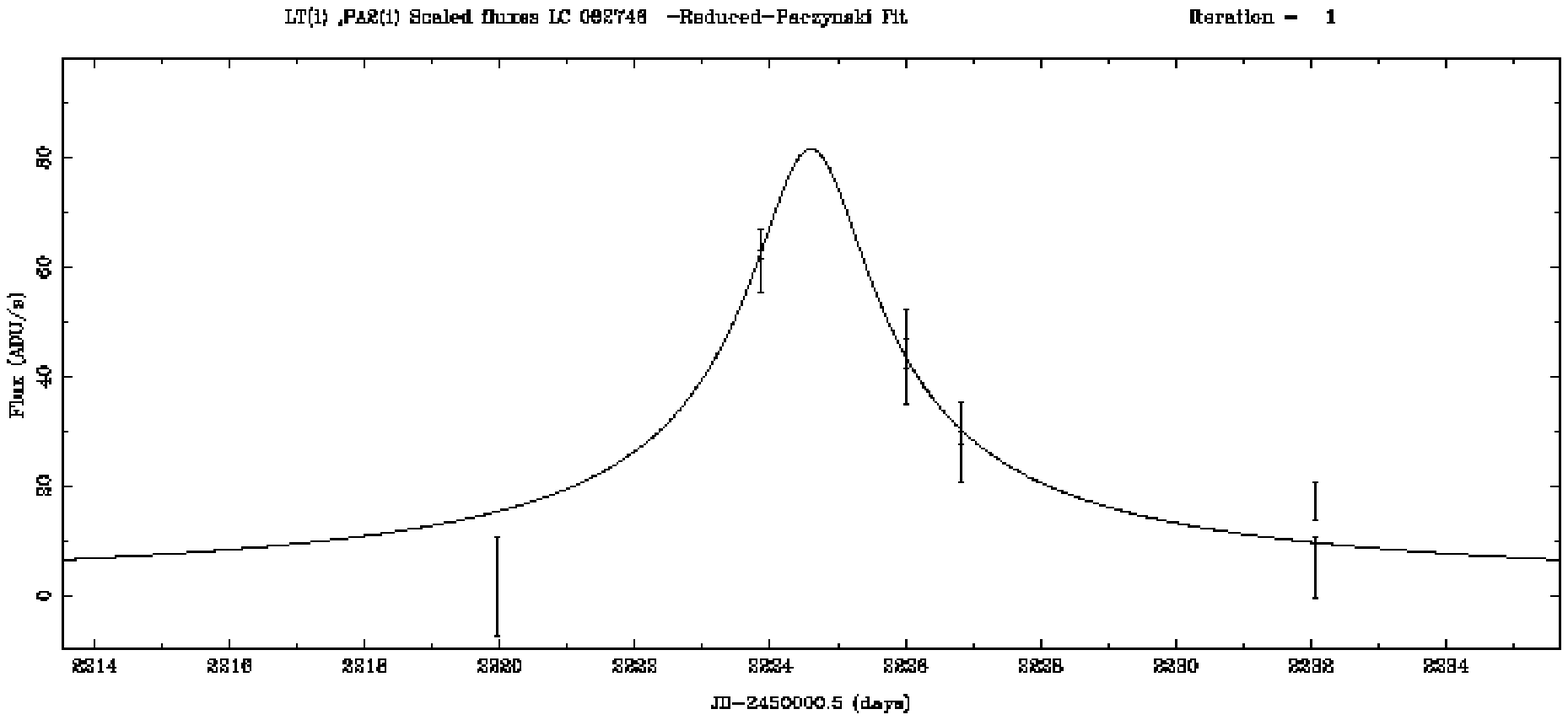} \\
\vspace*{0cm}
   \leavevmode
 \includegraphics{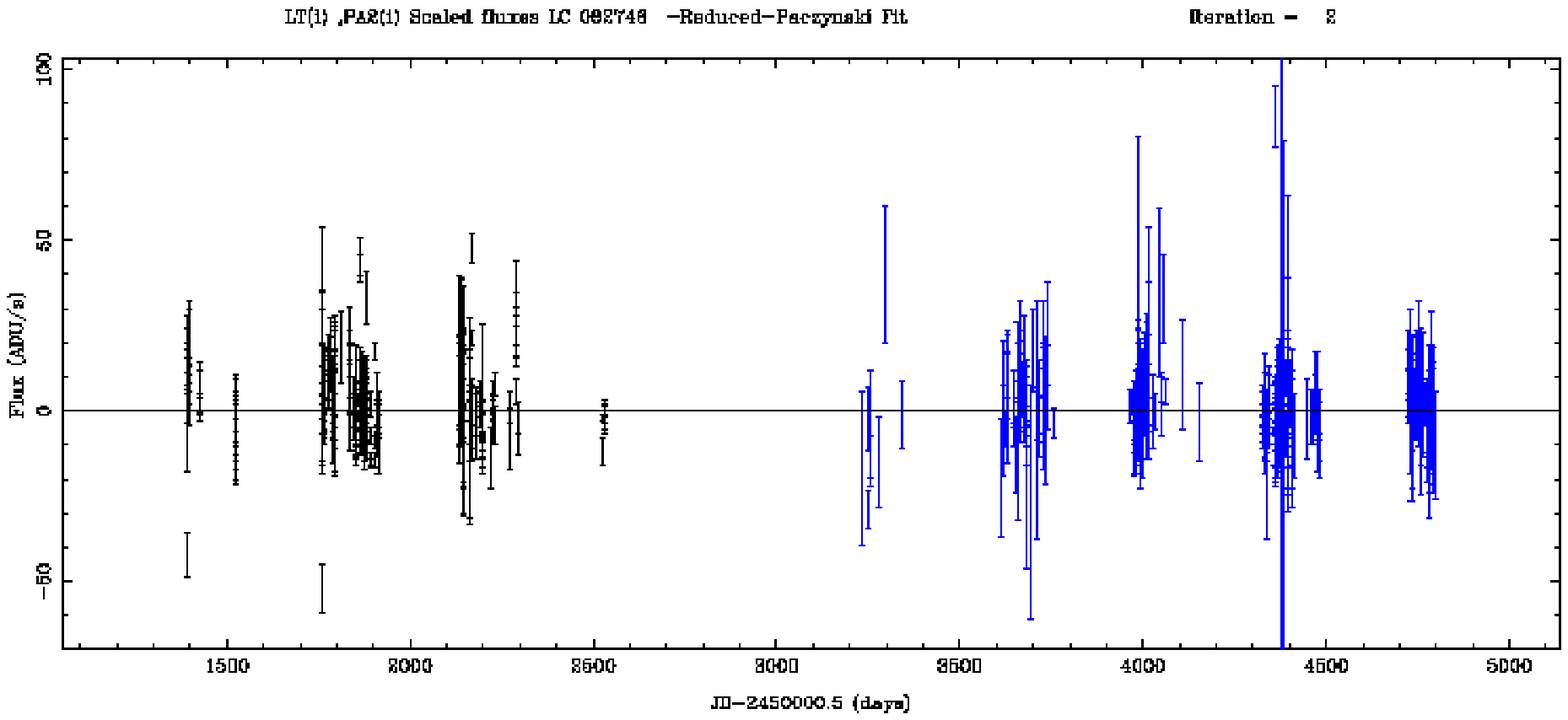} \\
\end{array}$
\caption[Lightcurve of object number $82746$ in the 2008 photometry, showing Top) the original Paczy\'nski fit,
Middle) the central peak region and Bottom) the residuals after subtraction of the fit.]{Lightcurve of object number $82746$ in the 2008 photometry, showing Top) the original Paczy\'nski fit,
Middle) the central peak region and Bottom) the residuals after subtraction of the fit.}
\label{2008_selection_pac_LC_82746}
\end{figure}

\clearpage

\textbf{Mixed Fit Objects}

\begin{figure}[!ht]
\vspace*{7cm}
$\begin{array}{c}
\vspace*{7.5cm}
   \leavevmode
 \includegraphics{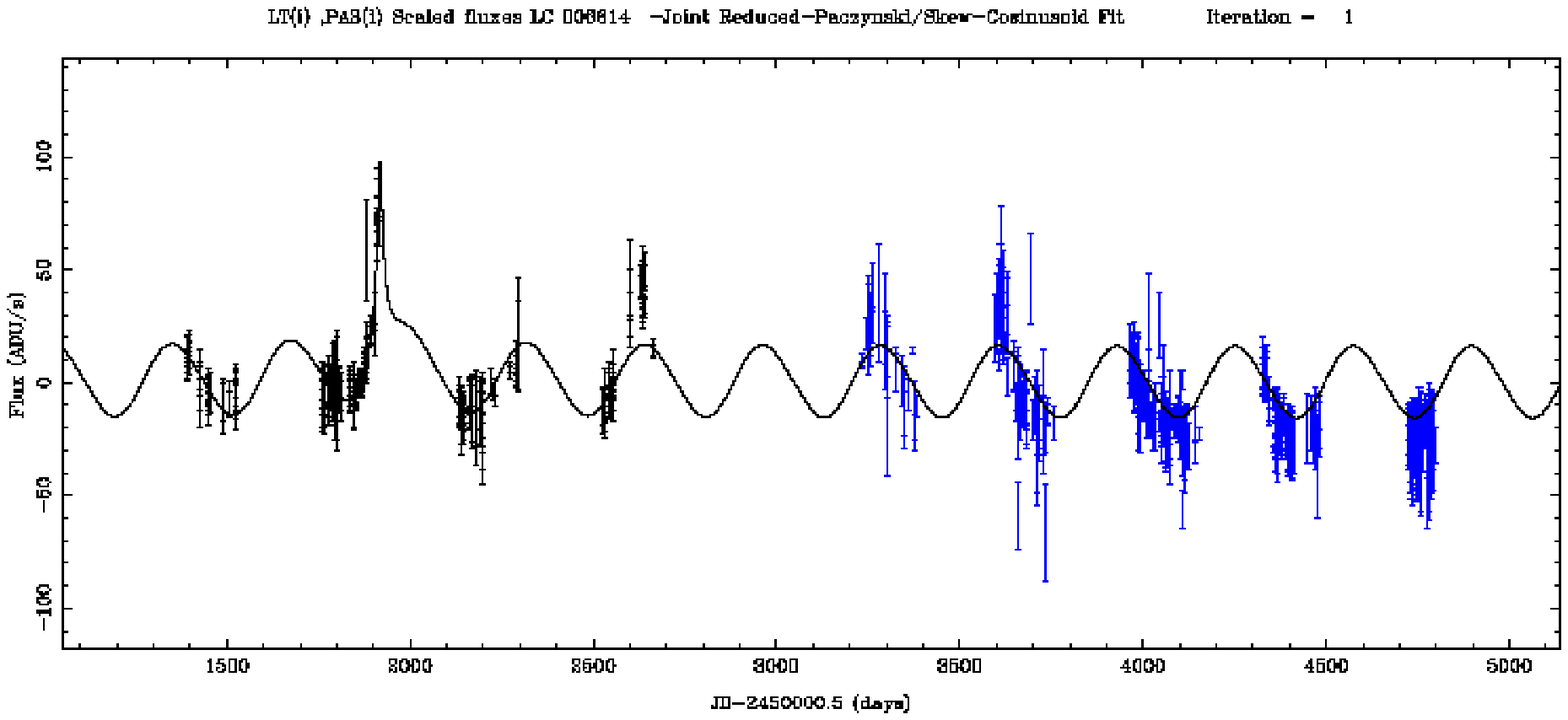} \\
\vspace*{7.5cm}
   \leavevmode
 \includegraphics{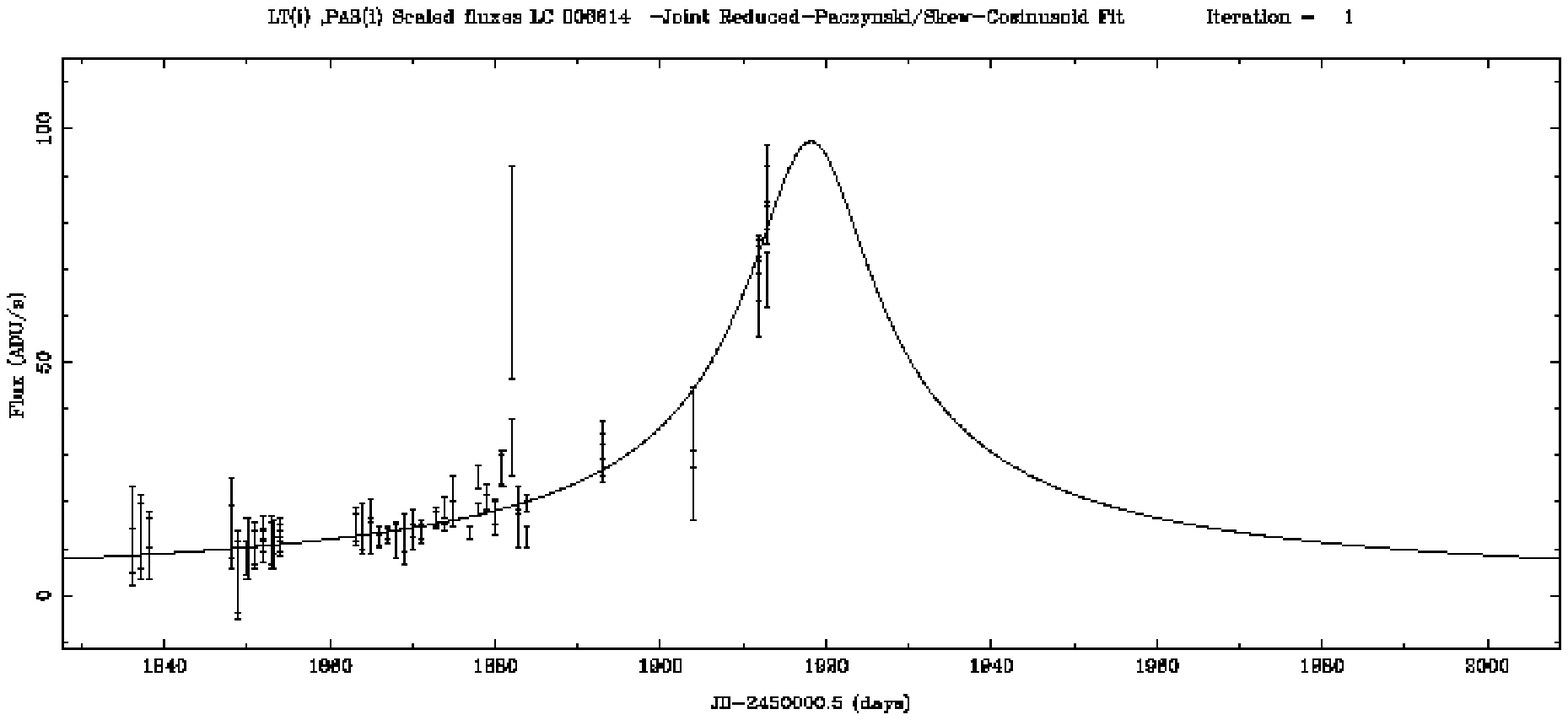} \\
  \leavevmode
 \includegraphics{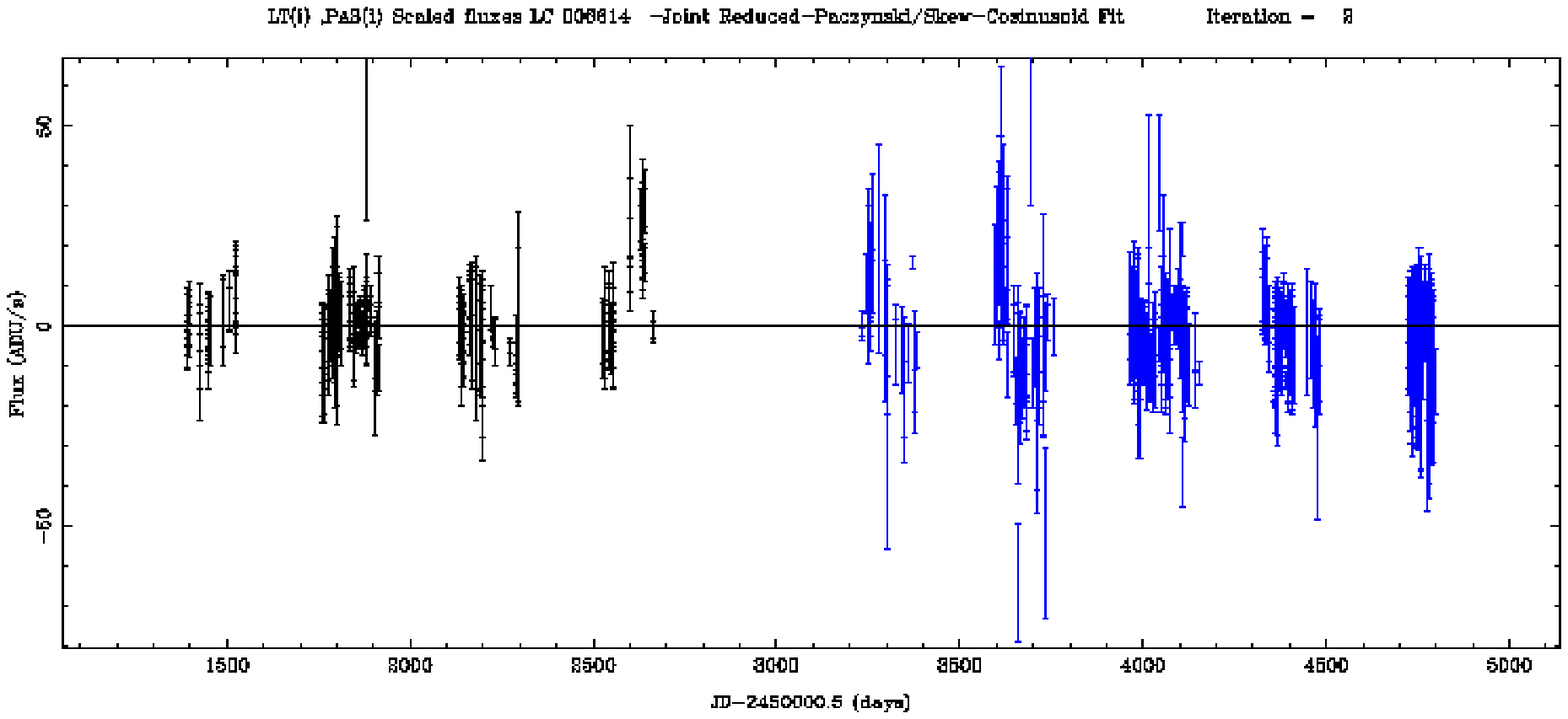} \\
\end{array}$
\caption[Lightcurve of object number $6614$ in the 2008 photometry, showing Top) the original mixed fit,
Middle) the peak region of the lensing component only, after the variable component has been subtracted, and Bottom) the residuals after subtraction of the mixed fit.]{Lightcurve of object number $6614$ in the 2008 photometry, showing Top) the original mixed fit and Middle) the peak region of the lensing component only, after the variable component has been subtracted and Bottom) the residuals after subtraction of the mixed fit.}
\label{2008_selection_mixed_LC_06614}
\end{figure}

\newpage

\begin{figure}[!ht]
\vspace*{7cm}
$\begin{array}{c}
\vspace*{7.5cm}
   \leavevmode
 \includegraphics{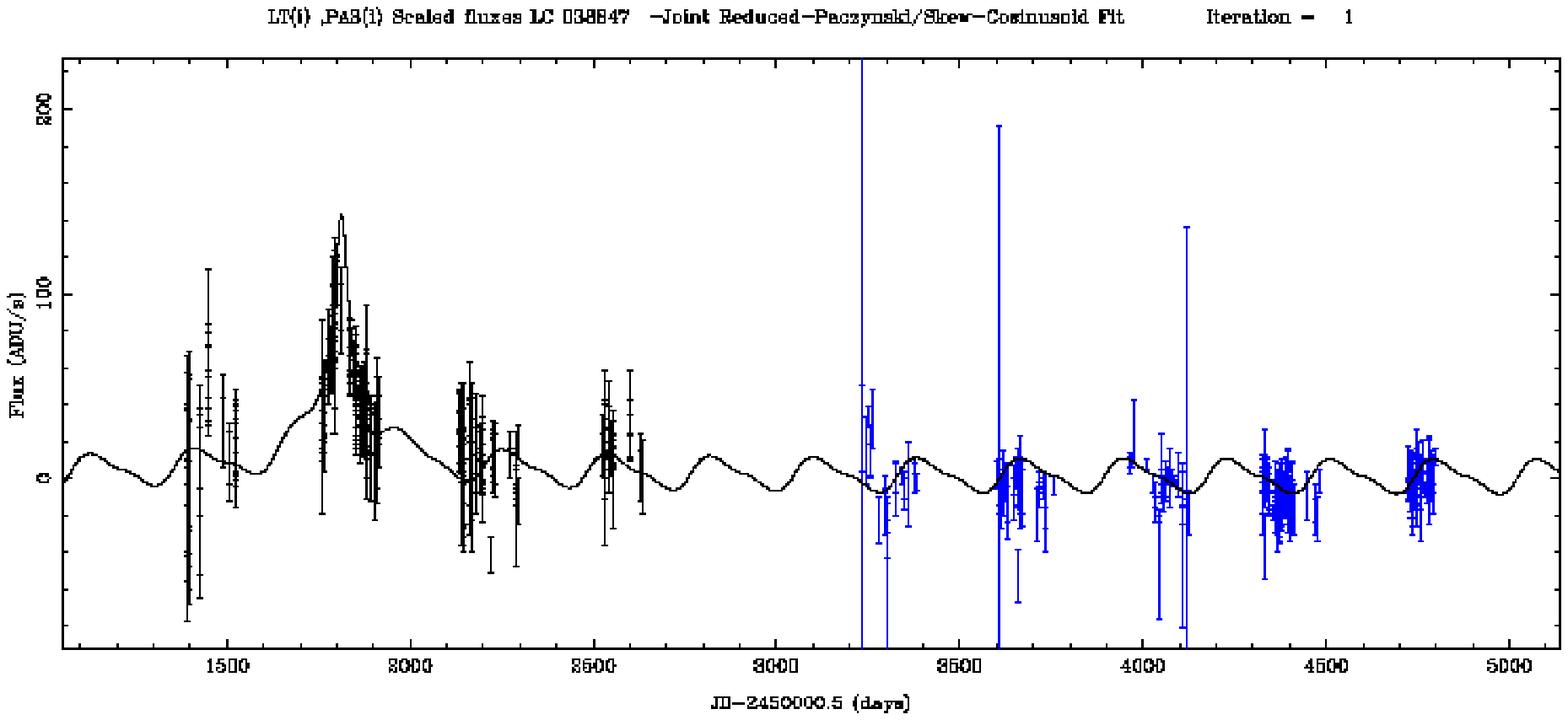} \\
\vspace*{7.5cm}
   \leavevmode
 \includegraphics{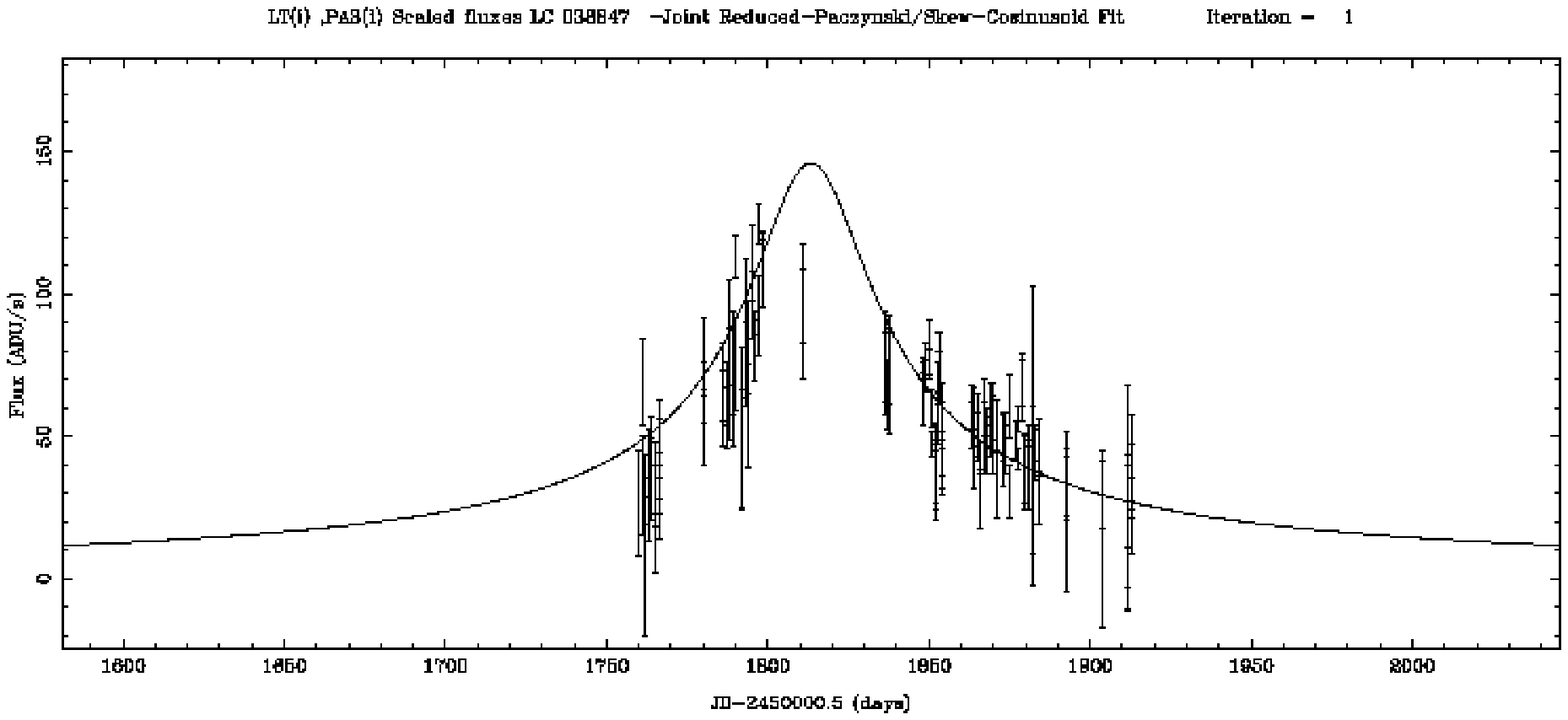} \\
  \leavevmode
 \includegraphics{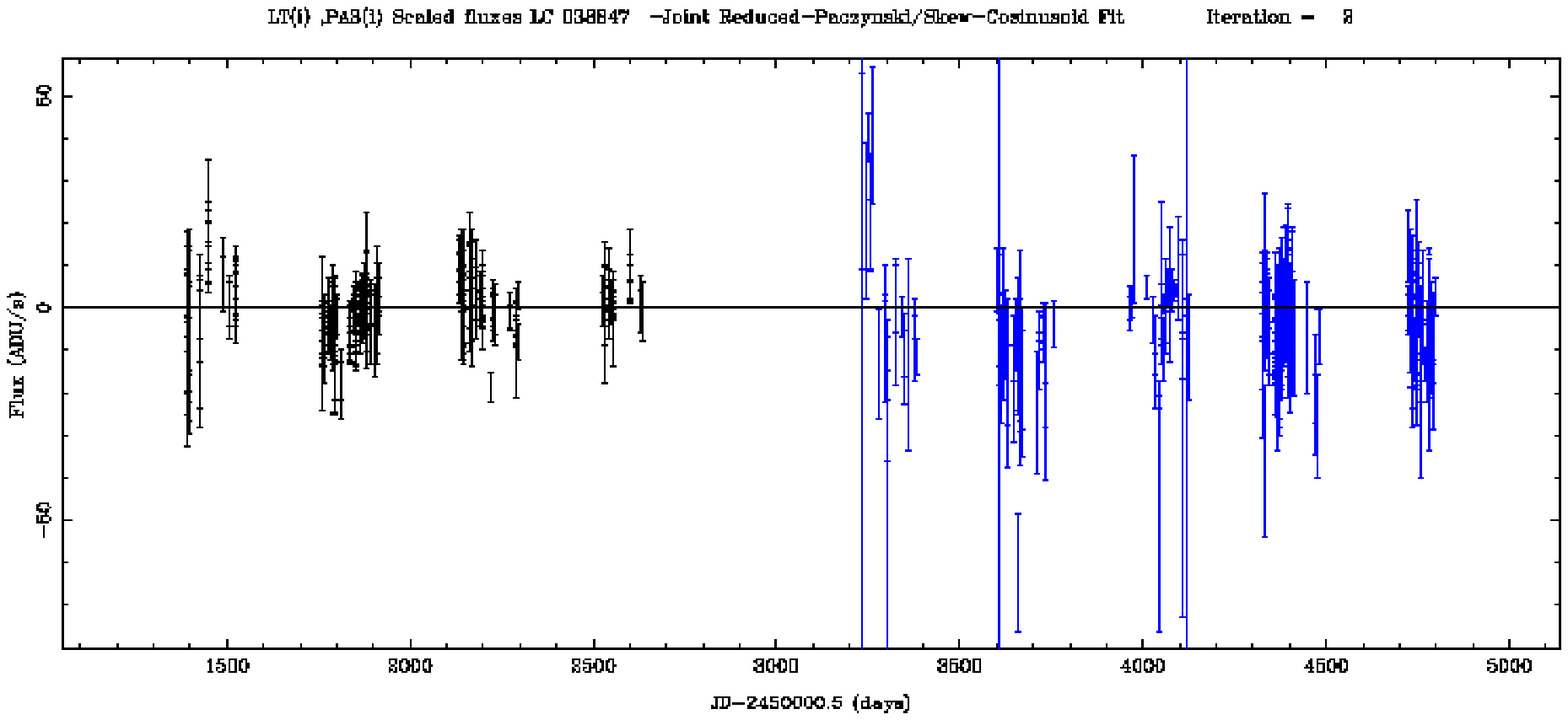} \\
\end{array}$
\caption[Lightcurve of object number $38847$ in the 2008 photometry, showing Top) the original mixed fit,
Middle) the peak region of the lensing component only, after the variable component has been subtracted, and Bottom) the residuals after subtraction of the mixed fit.]{Lightcurve of object number $38847$ in the 2008 photometry, showing Top) the original mixed fit and Middle) the peak region of the lensing component only, after the variable component has been subtracted and Bottom) the residuals after subtraction of the mixed fit.}
\label{2008_selection_mixed_LC_38847}
\end{figure}

\newpage

\begin{figure}[!ht]
\vspace*{7cm}
$\begin{array}{c}
\vspace*{7.5cm}
   \leavevmode
 \includegraphics{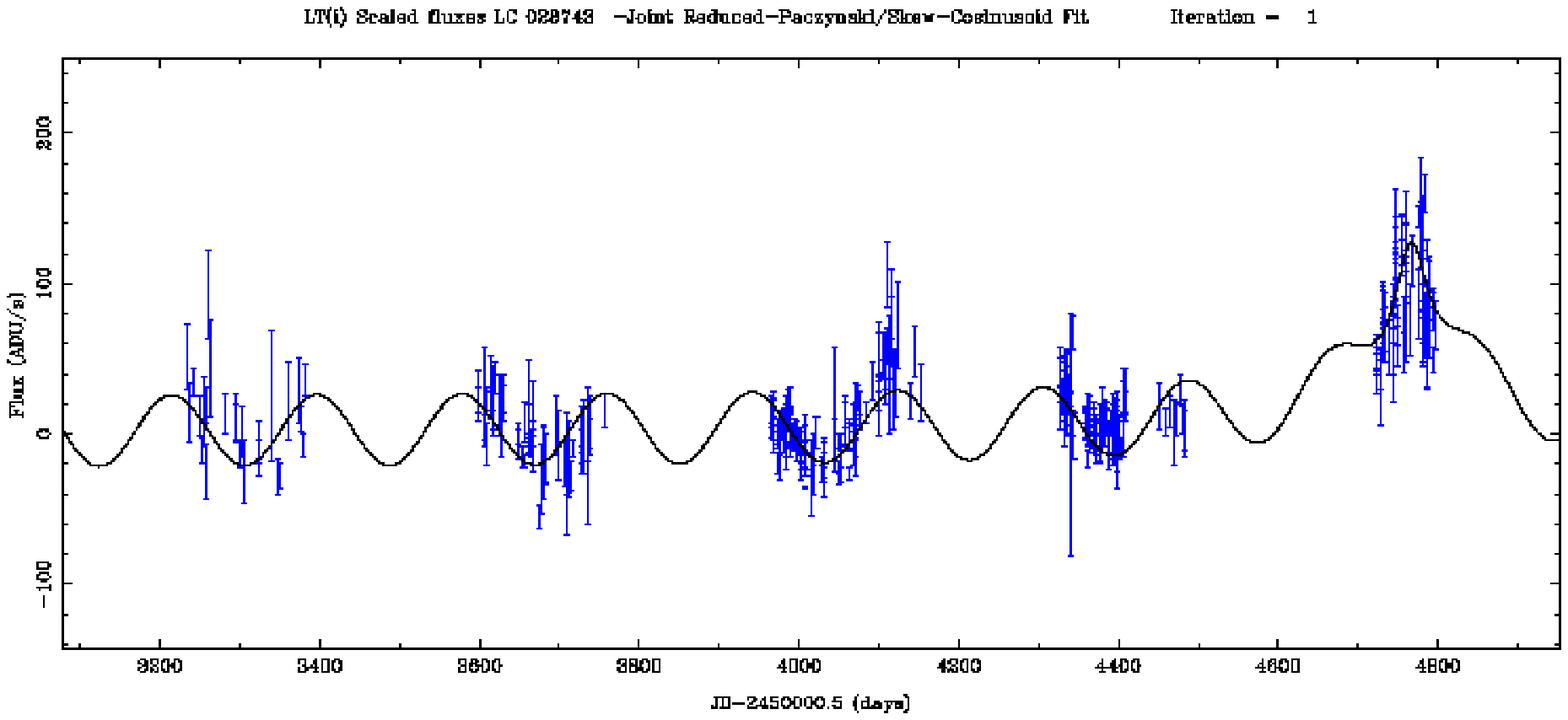} \\
\vspace*{7.5cm}
   \leavevmode
 \includegraphics{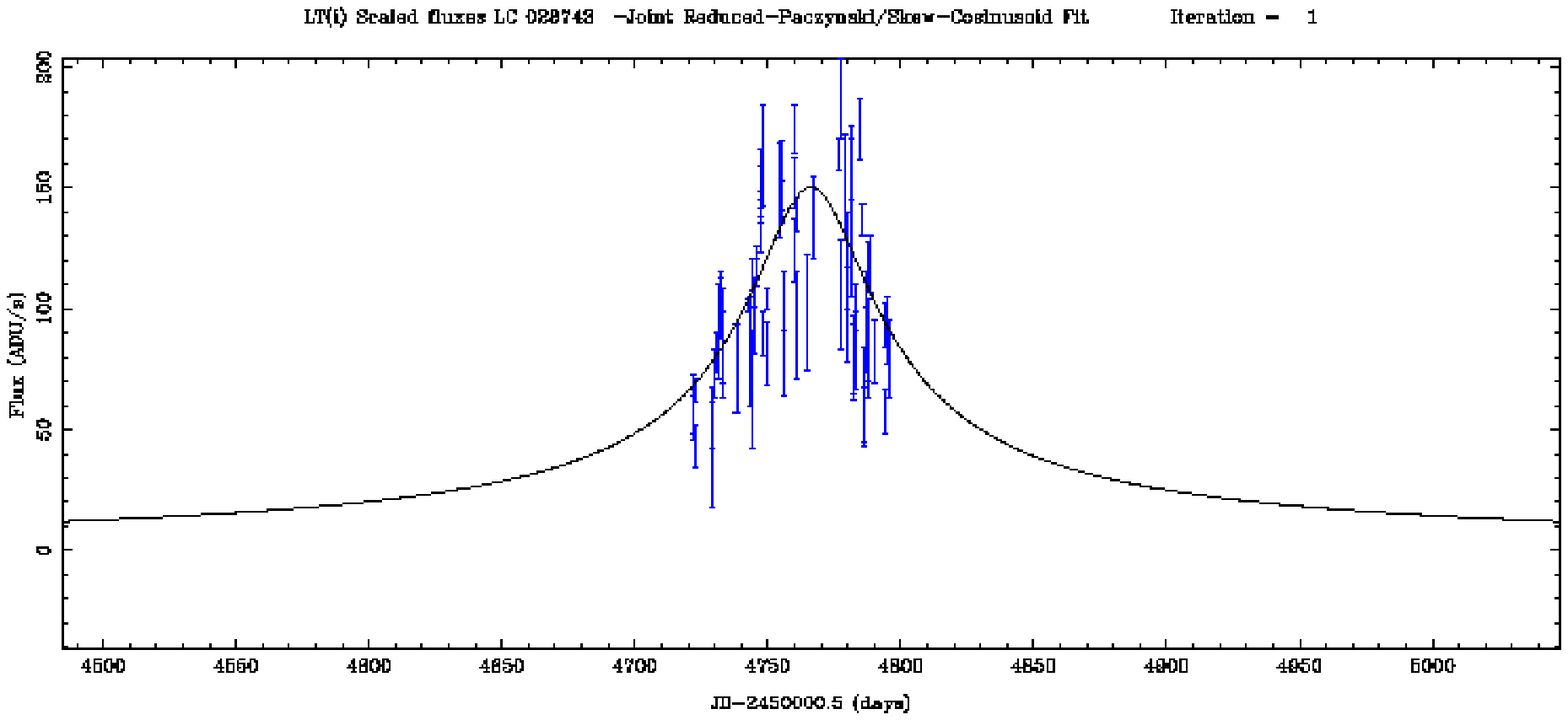} \\
  \leavevmode
 \includegraphics{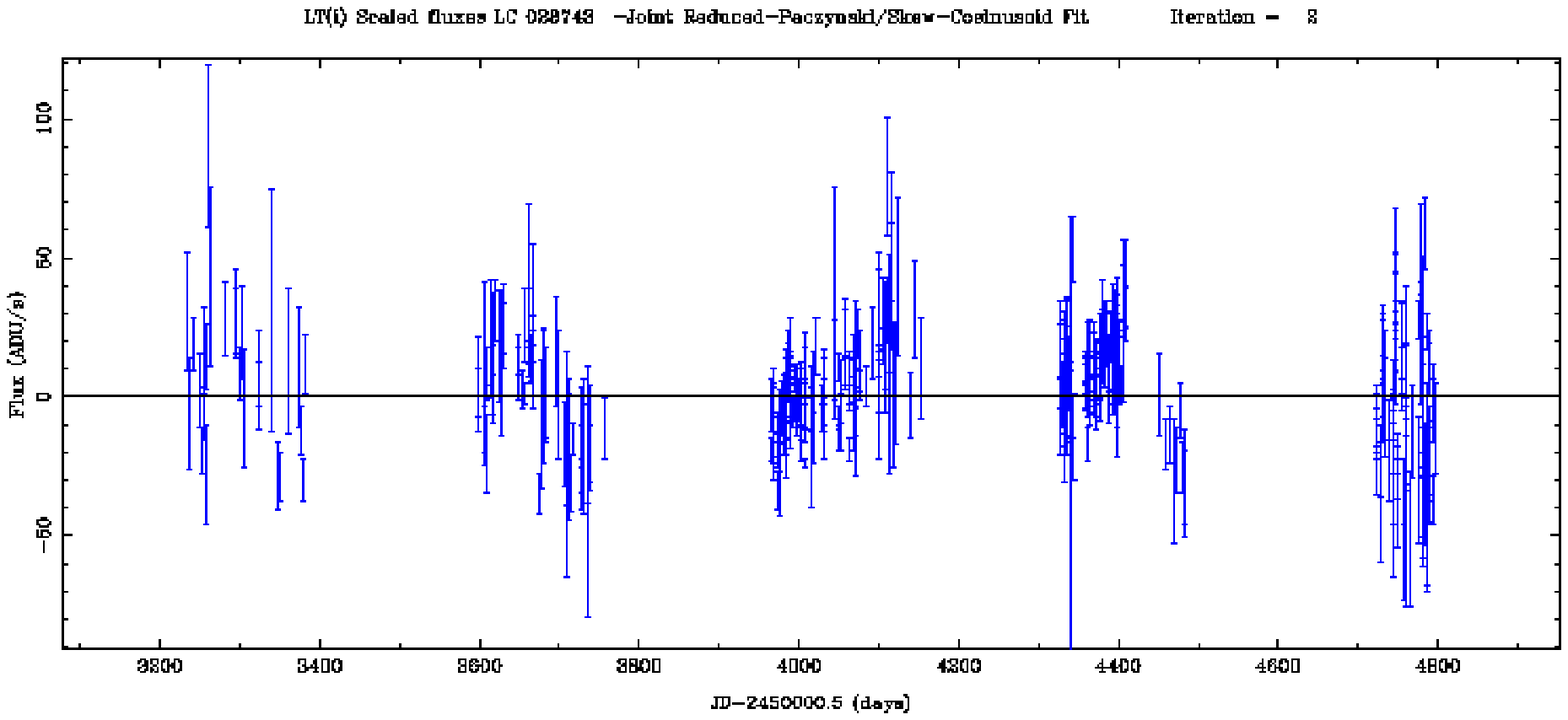} \\
\end{array}$
\caption[Lightcurve of object number $28743$ in the 2008 photometry, showing Top) the original mixed fit,
Middle) the peak region of the lensing component only, after the variable component has been subtracted, and Bottom) the residuals after subtraction of the mixed fit.]{Lightcurve of object number $28743$ in the 2008 photometry, showing Top) the original mixed fit and Middle) the peak region of the lensing component only, after the variable component has been subtracted and Bottom) the residuals after subtraction of the mixed fit.}
\label{2008_selection_mixed_LC_28743}
\end{figure}

\newpage

\begin{figure}[!ht]
\vspace*{7cm}
$\begin{array}{c}
\vspace*{7.5cm}
   \leavevmode
 \includegraphics{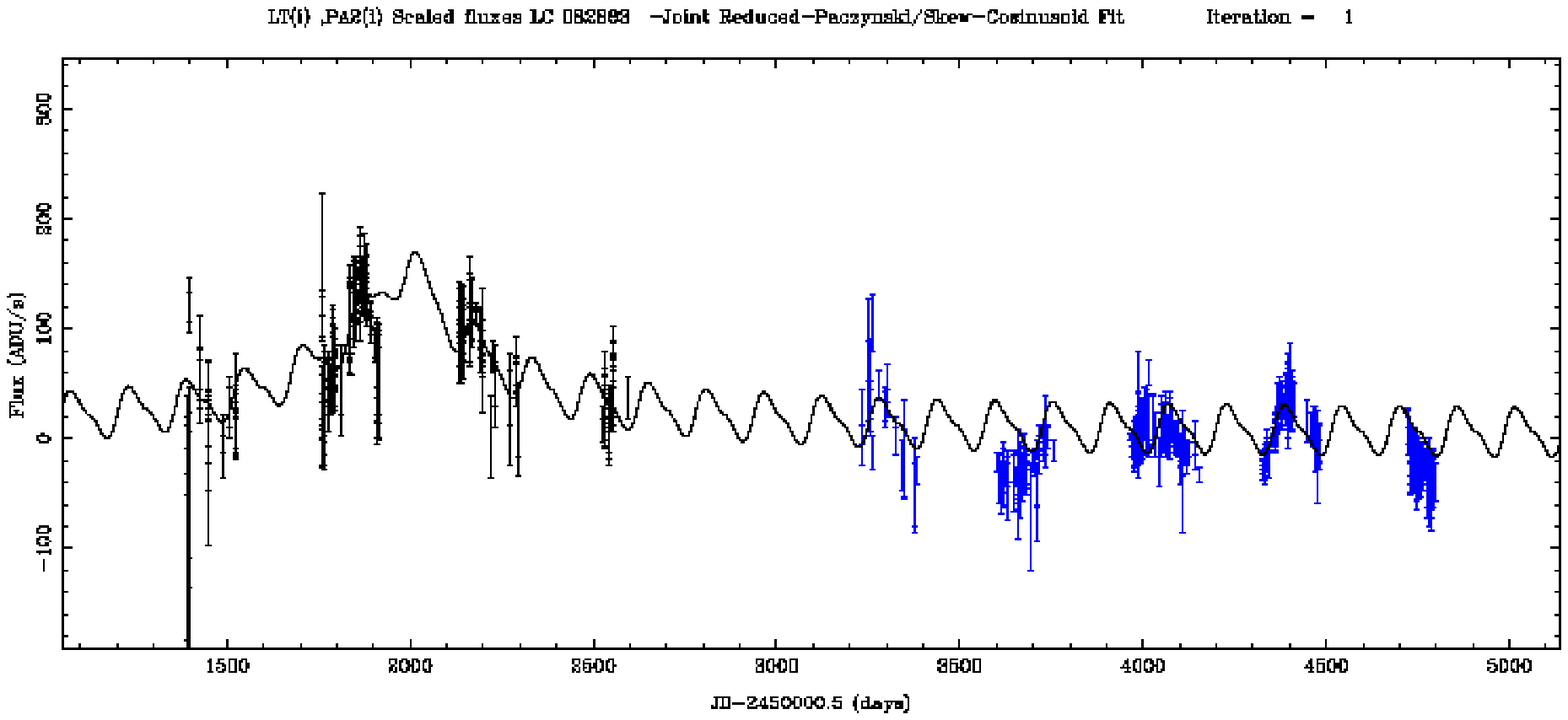} \\
\vspace*{7.5cm}
   \leavevmode
 \includegraphics{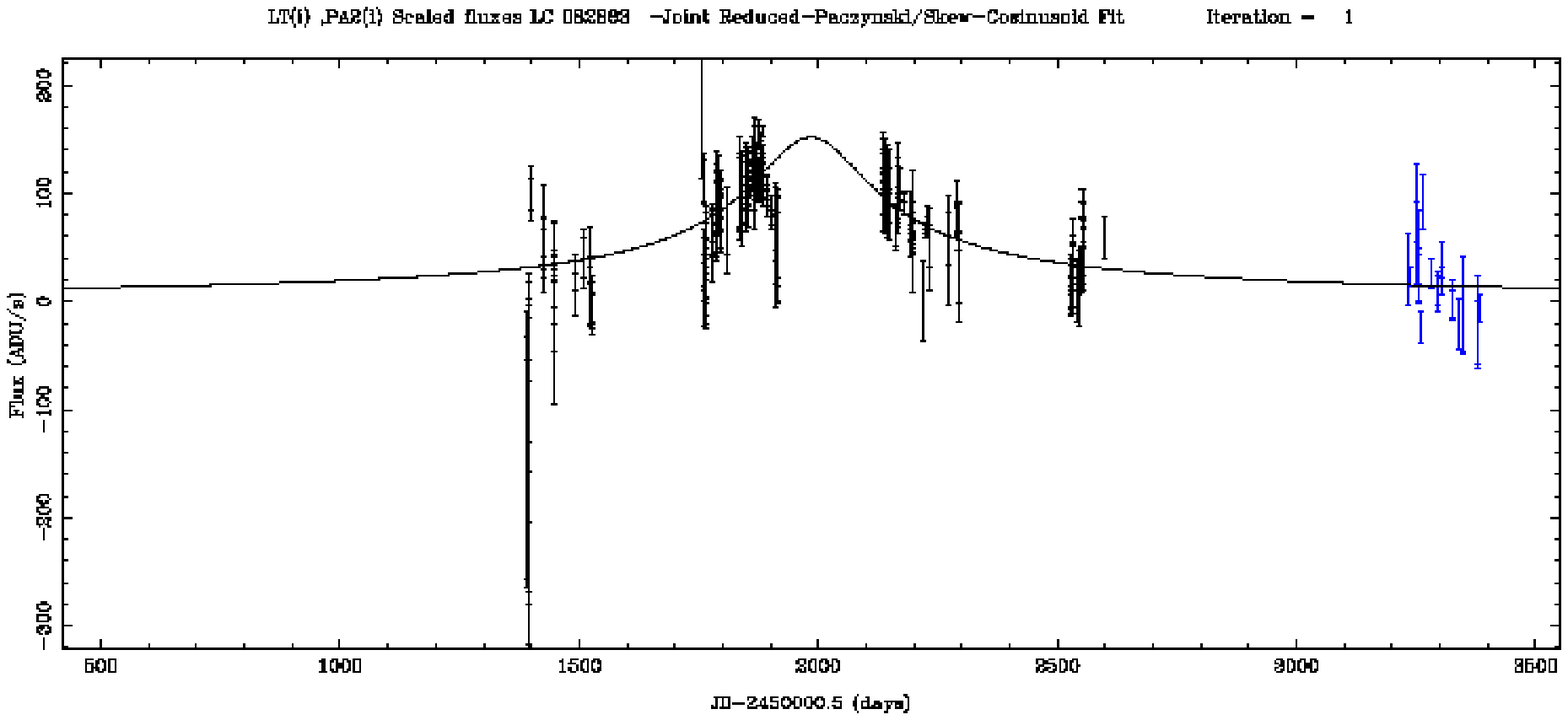} \\
  \leavevmode
 \includegraphics{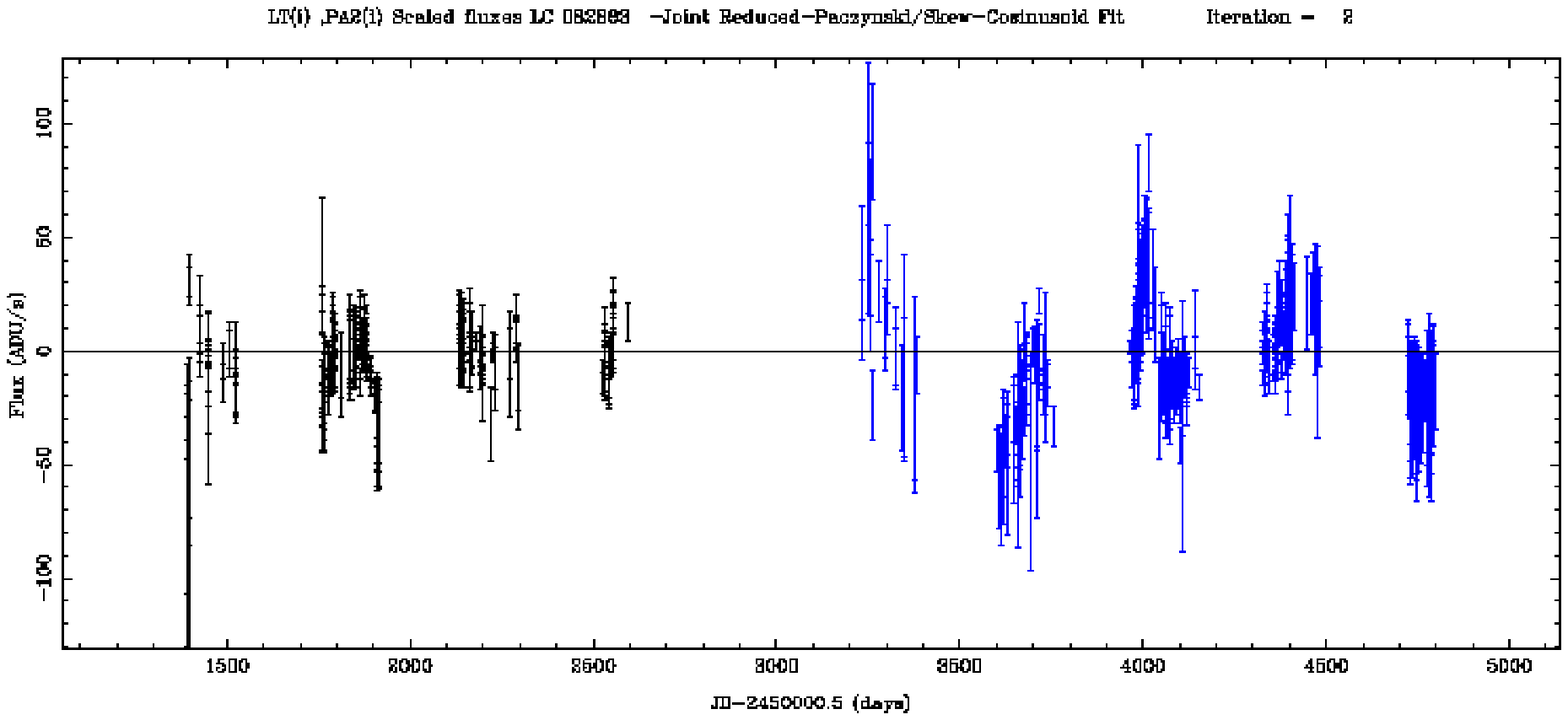} \\
\end{array}$
\caption[Lightcurve of object number $82883$ in the 2008 photometry, showing Top) the original mixed fit,
Middle) the peak region of the lensing component only, after the variable component has been subtracted, and Bottom) the residuals after subtraction of the mixed fit.]{Lightcurve of object number $82883$ in the 2008 photometry, showing Top) the original mixed fit and Middle) the peak region of the lensing component only, after the variable component has been subtracted and Bottom) the residuals after subtraction of the mixed fit.}
\label{2008_selection_mixed_LC_82883}
\end{figure}

\newpage
 
\begin{figure}[!ht]
\vspace*{7cm}
$\begin{array}{c}
\vspace*{7.5cm}
   \leavevmode
 \includegraphics{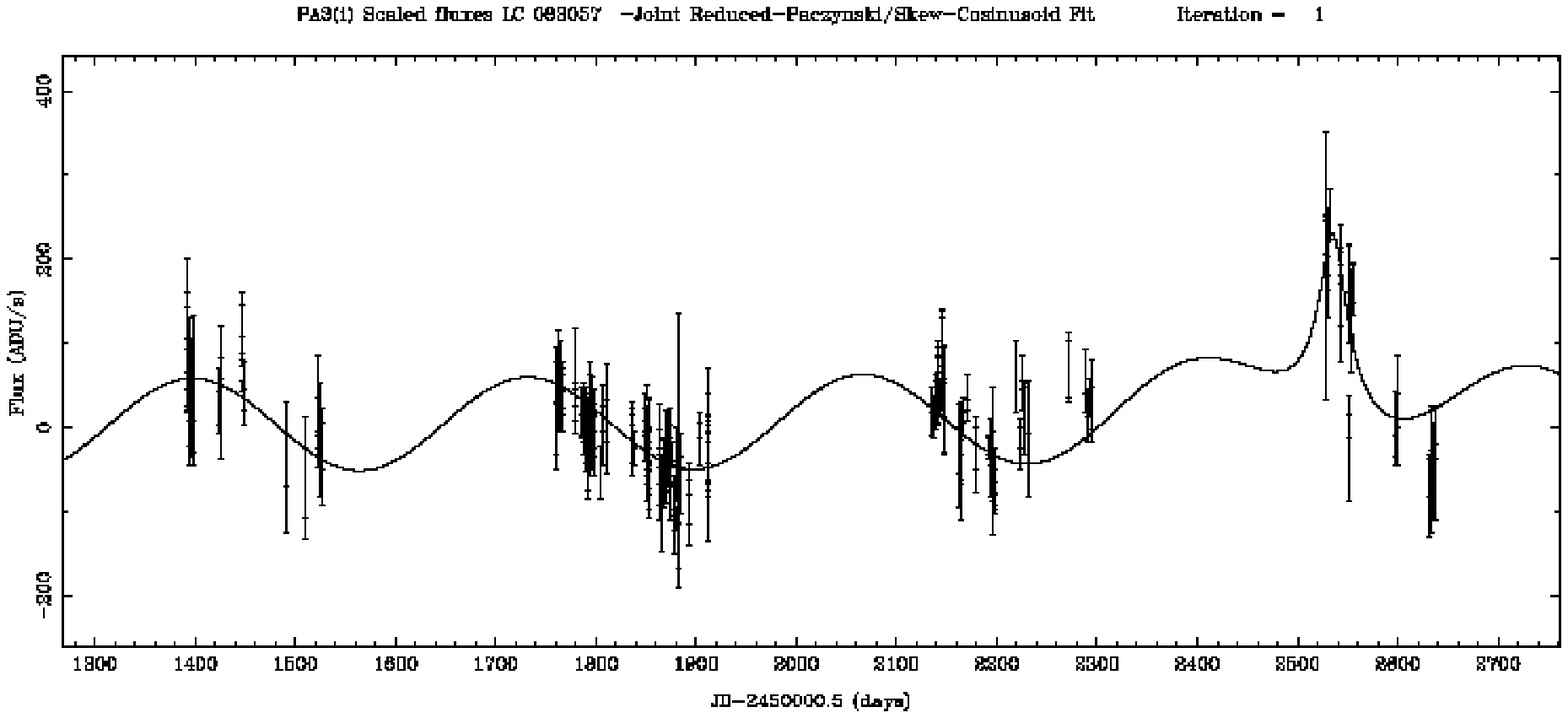} \\
\vspace*{7.5cm}
   \leavevmode
 \includegraphics{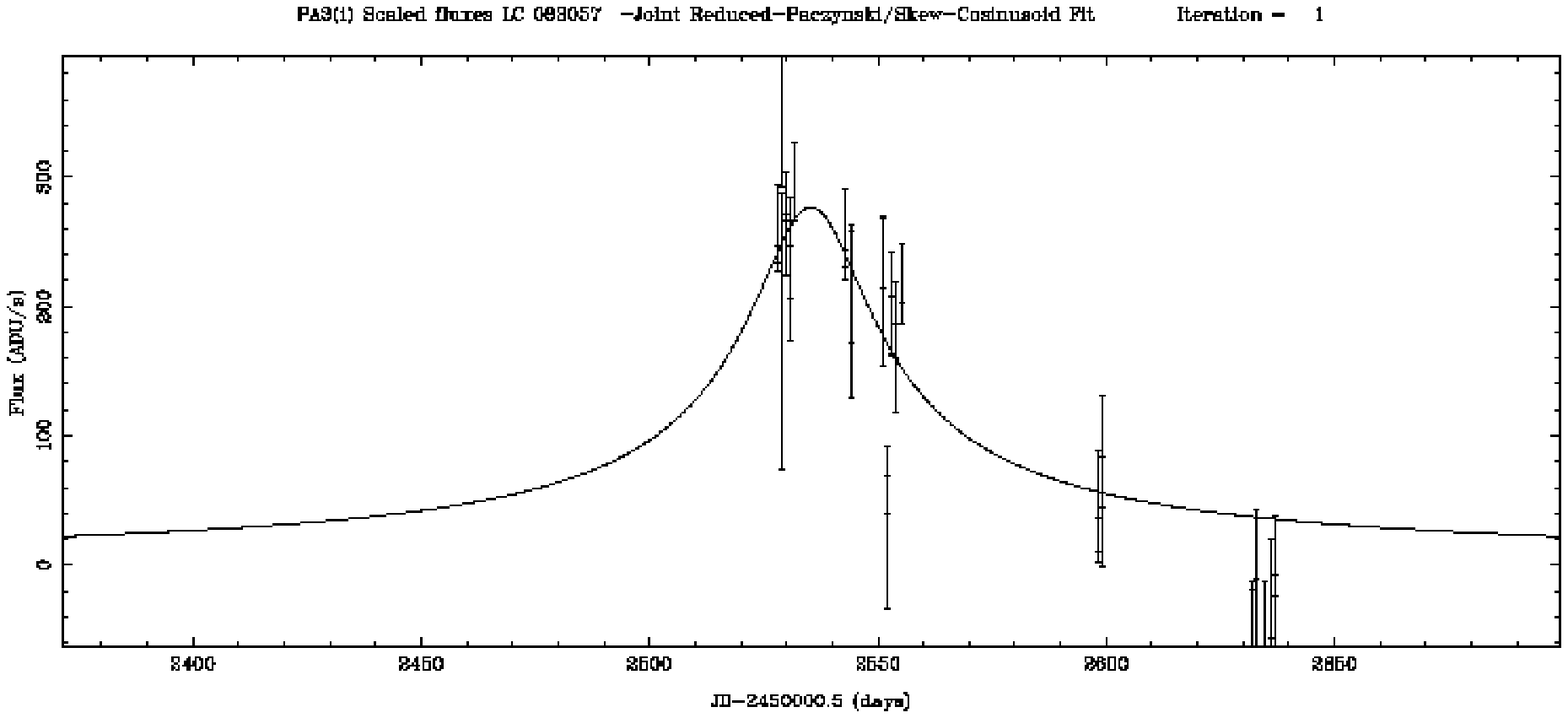} \\
  \leavevmode
 \includegraphics{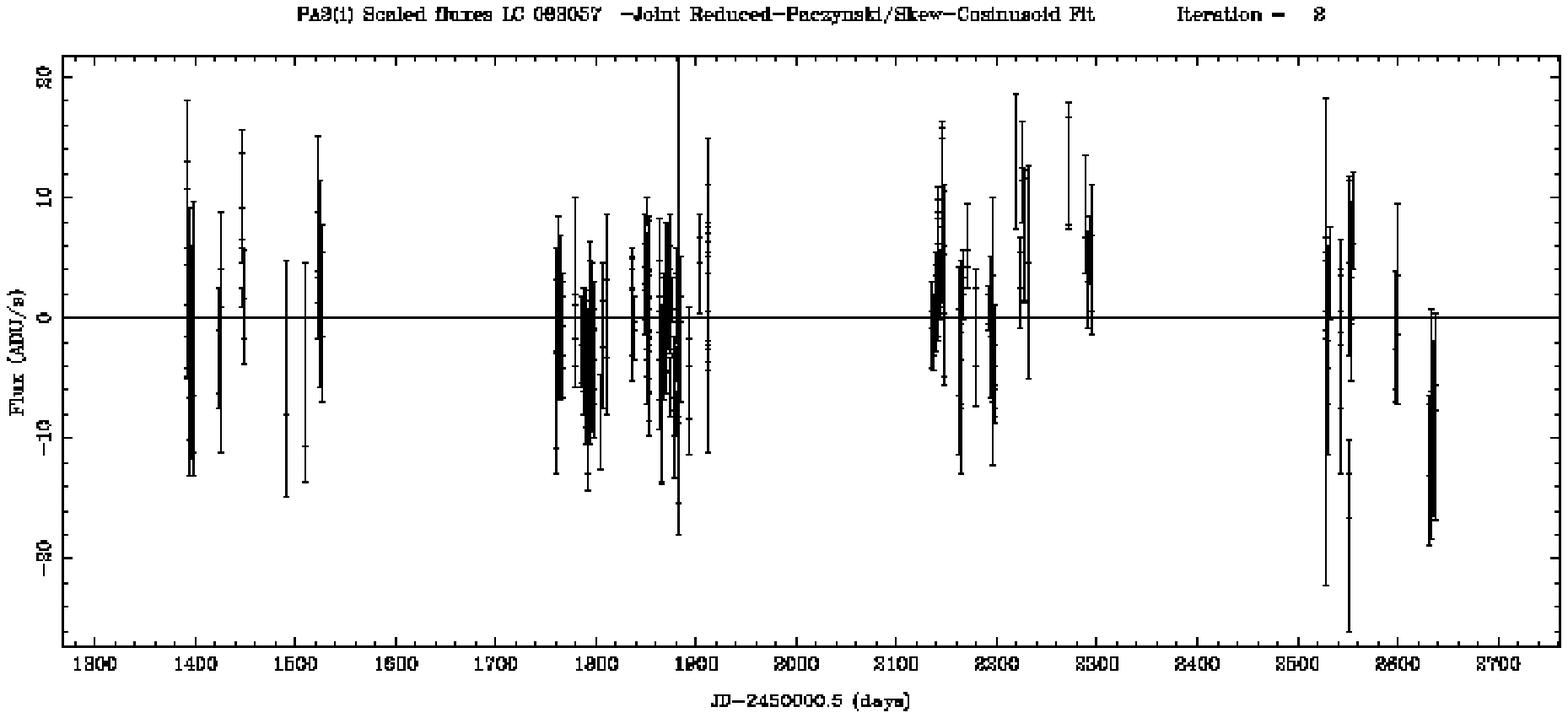} \\
\end{array}$
\caption[Lightcurve of object number $83057$ in the 2008 photometry, showing Top) the original mixed fit,
Middle) the peak region of the lensing component only, after the variable component has been subtracted, and Bottom) the residuals after subtraction of the mixed fit.]{Lightcurve of object number $83057$ in the 2008 photometry, showing Top) the original mixed fit and Middle) the peak region of the lensing component only, after the variable component has been subtracted and Bottom) the residuals after subtraction of the mixed fit.}
\label{2008_selection_mixed_LC_83057}
\end{figure}

\newpage
 
\begin{figure}[!ht]
\vspace*{7cm}
$\begin{array}{c}
\vspace*{7.5cm}
   \leavevmode
 \includegraphics{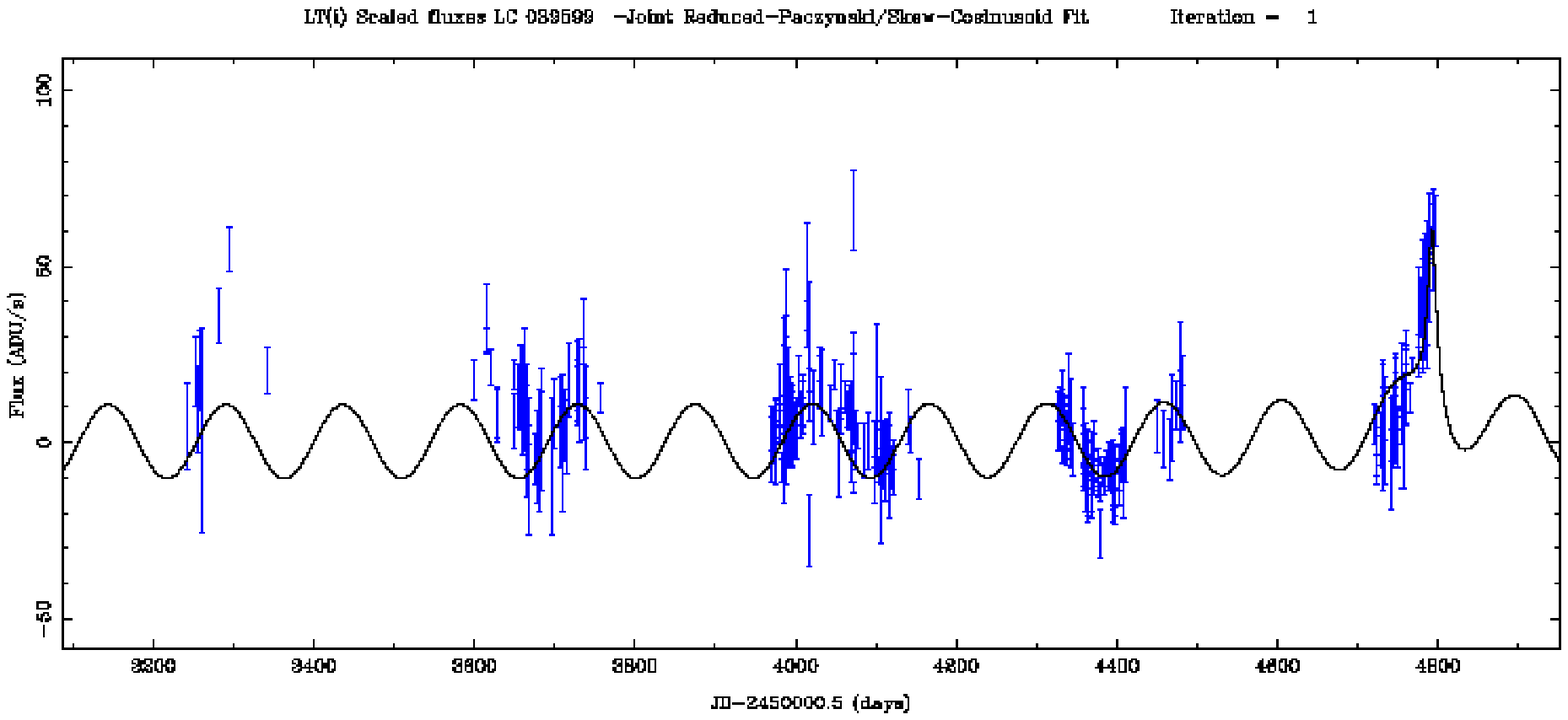} \\
\vspace*{7.5cm}
   \leavevmode
 \includegraphics{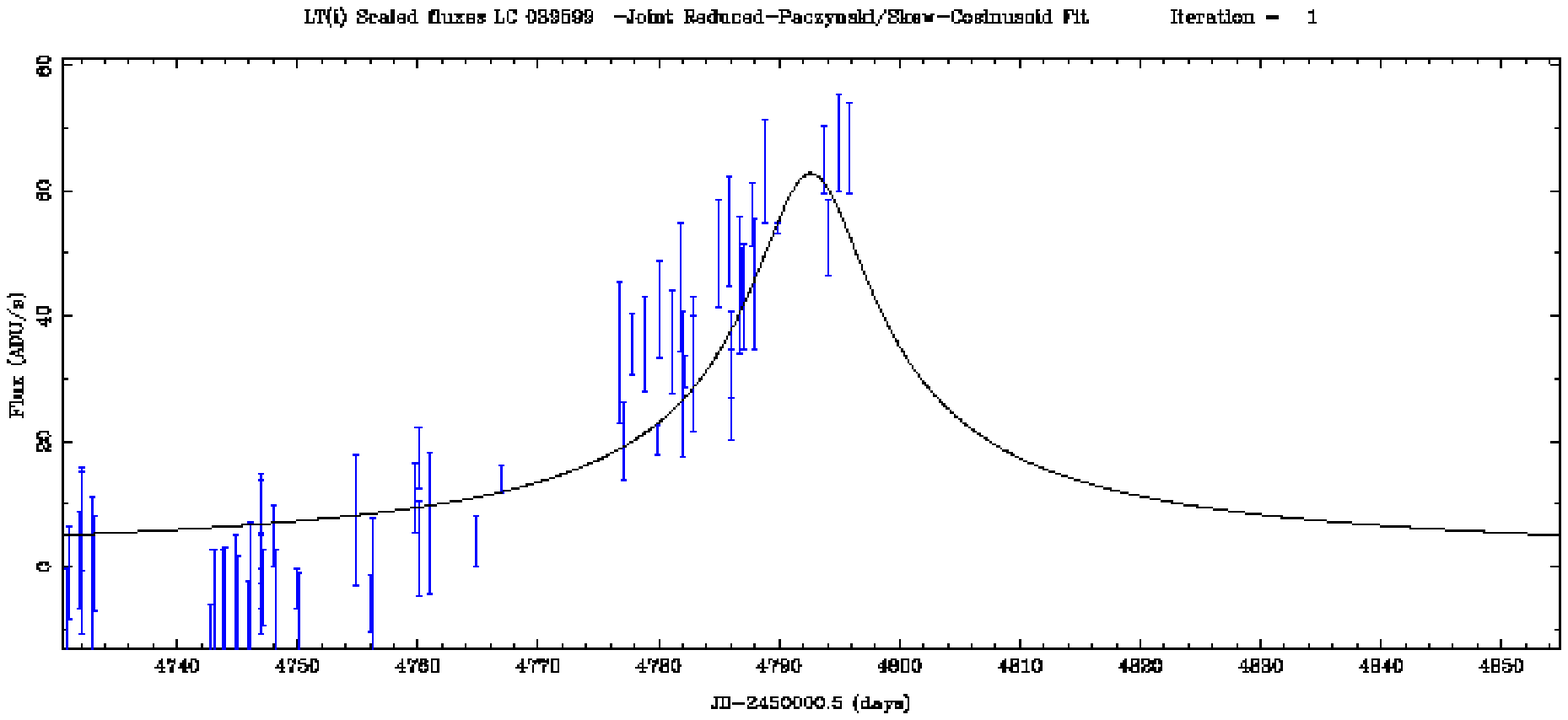} \\
  \leavevmode
 \includegraphics{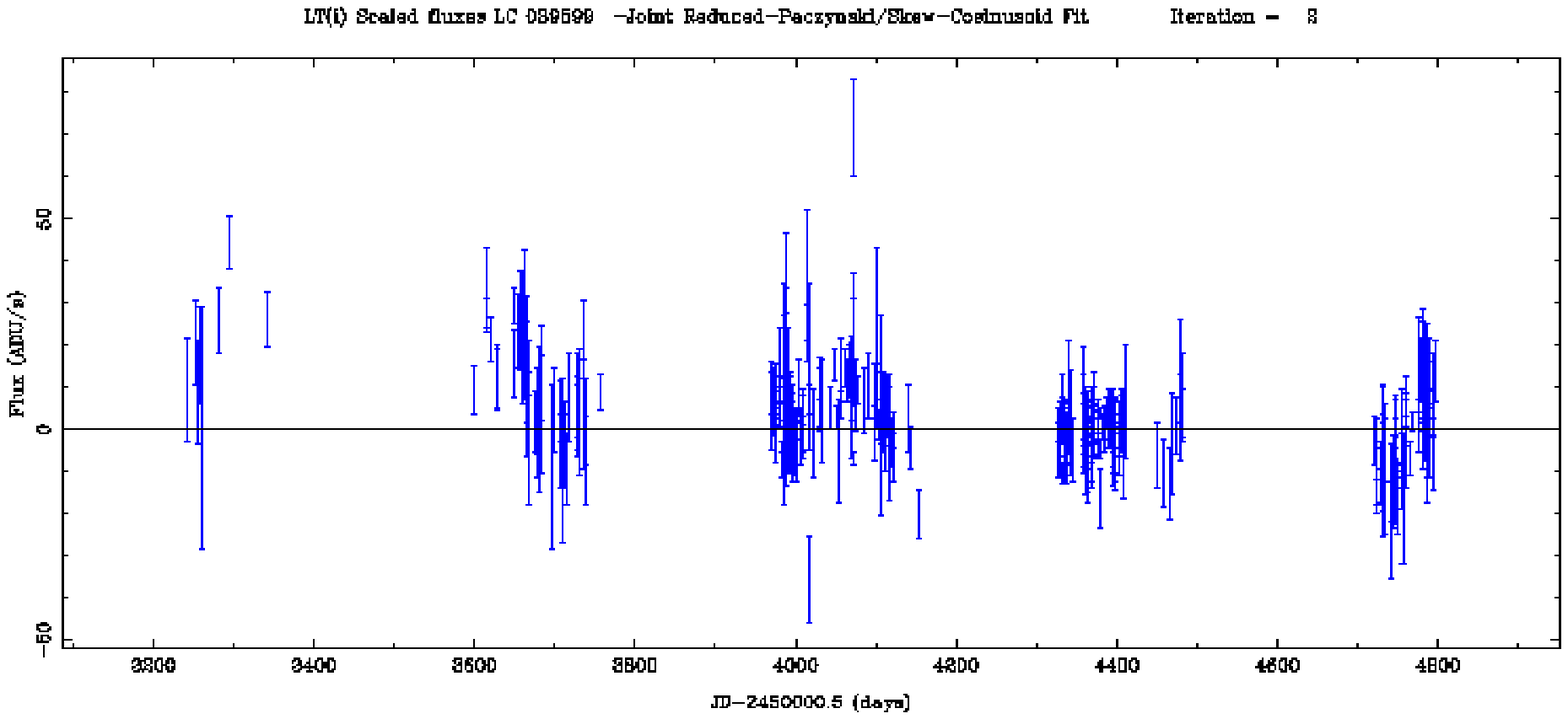} \\
\end{array}$
\caption[Lightcurve of object number $39599$ in the 2008 photometry, showing Top) the original mixed fit,
Middle) the peak region of the lensing component only, after the variable component has been subtracted, and Bottom) the residuals after subtraction of the mixed fit.]{Lightcurve of object number $39599$ in the 2008 photometry, showing Top) the original mixed fit and Middle) the peak region of the lensing component only, after the variable component has been subtracted and Bottom) the residuals after subtraction of the mixed fit.}
\label{2008_selection_mixed_LC_39599}
\end{figure}

\newpage

\begin{figure}[!ht]
\vspace*{7cm}
$\begin{array}{c}
\vspace*{7.5cm}
   \leavevmode
 \includegraphics{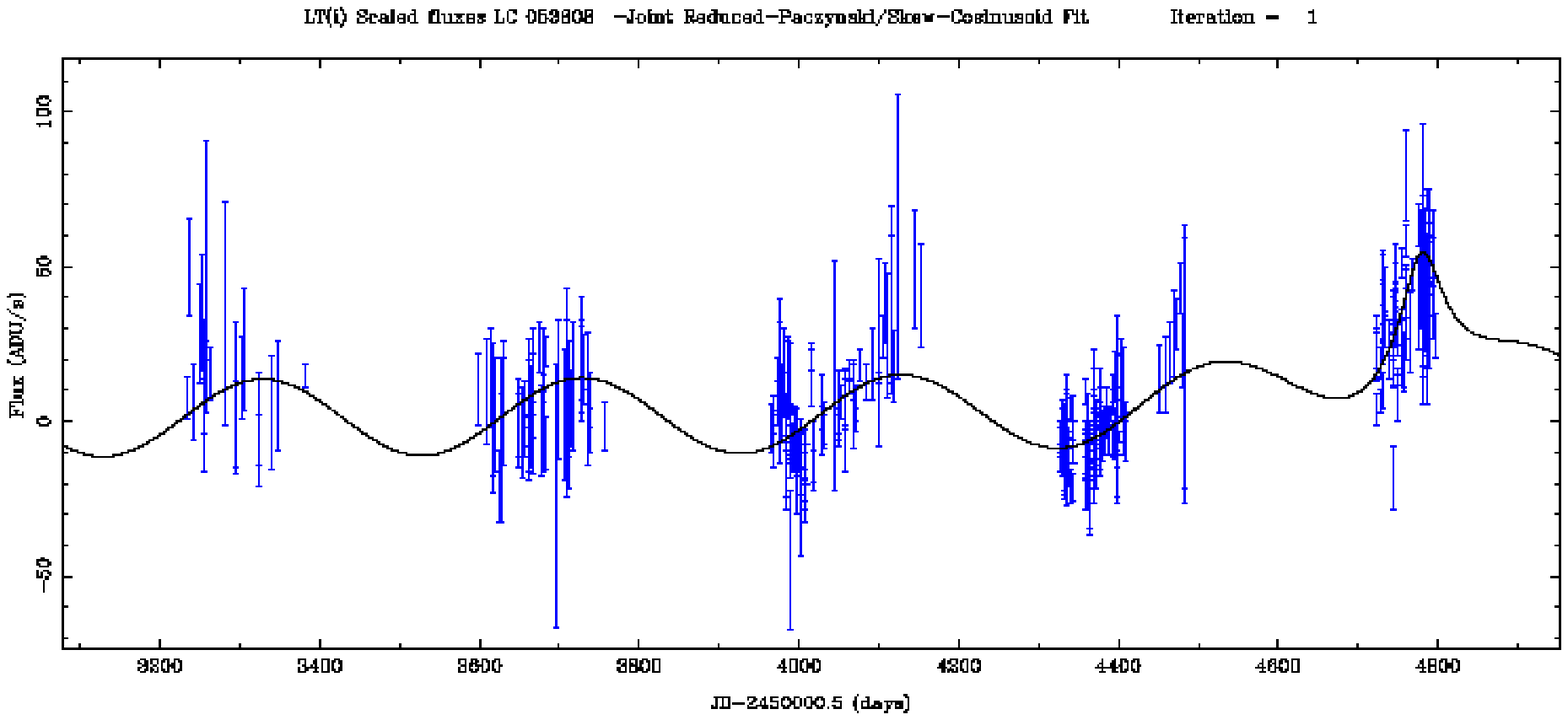} \\
\vspace*{7.5cm}
   \leavevmode
 \includegraphics{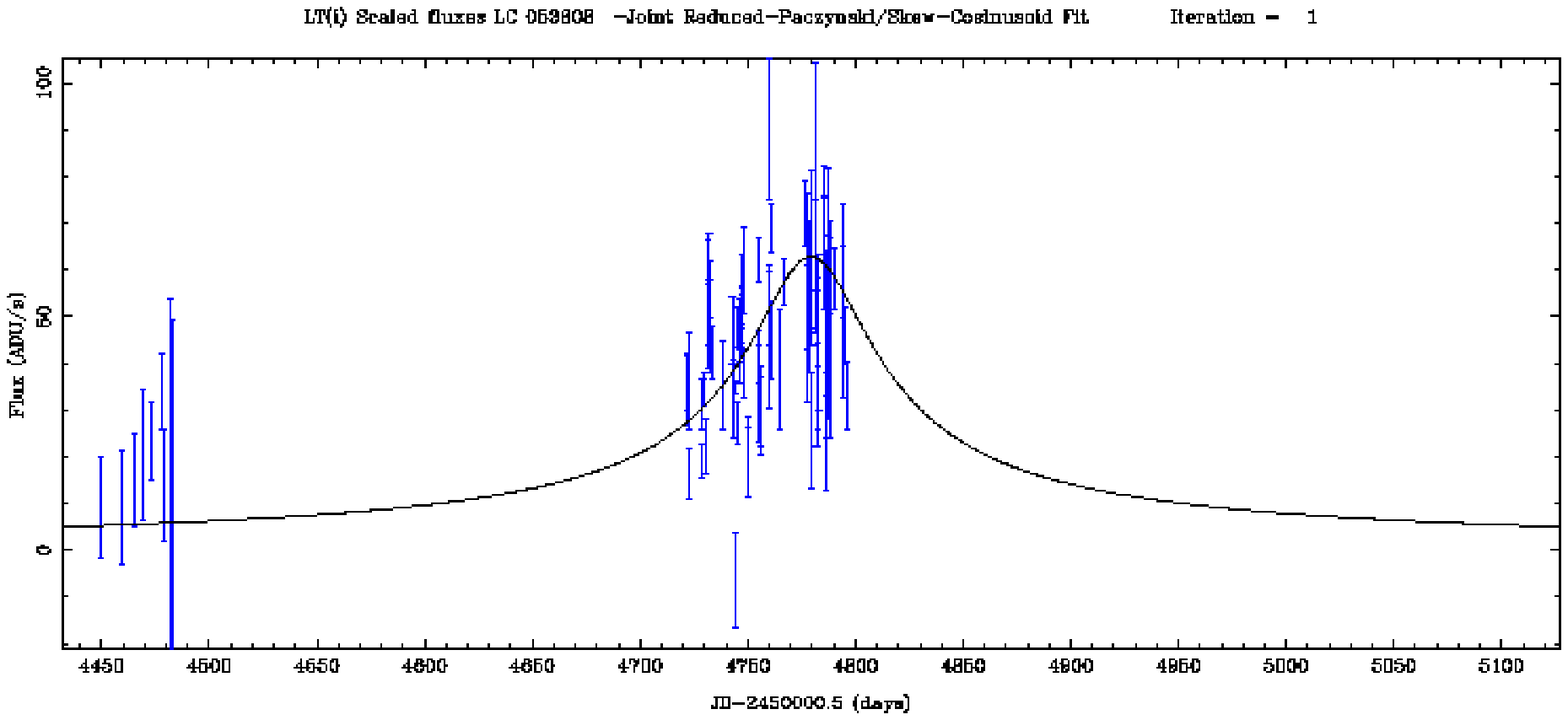} \\
  \leavevmode
 \includegraphics{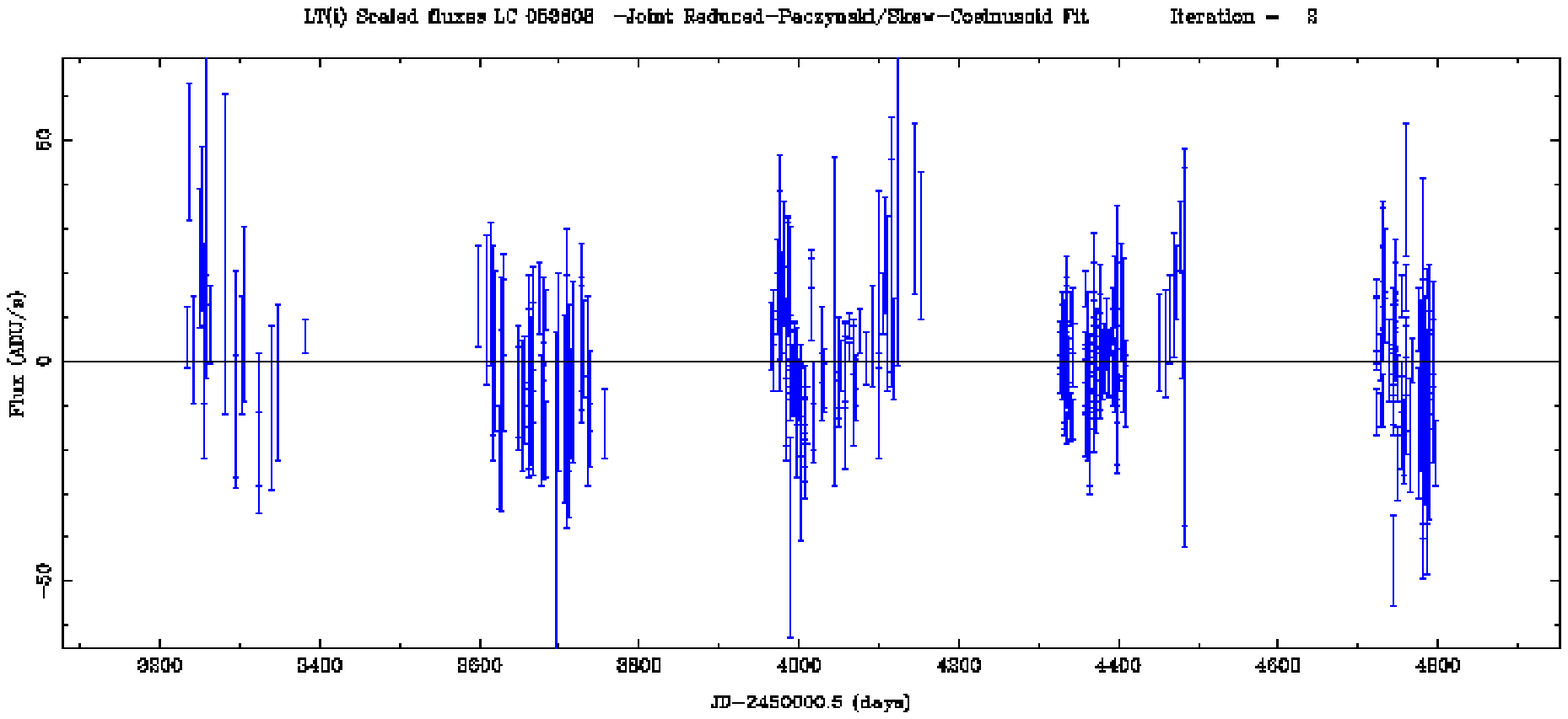} \\
\end{array}$
\caption[Lightcurve of object number $53608$ in the 2008 photometry, showing Top) the original mixed fit,
Middle) the peak region of the lensing component only, after the variable component has been subtracted, and Bottom) the residuals after subtraction of the mixed fit.]{Lightcurve of object number $53608$ in the 2008 photometry, showing Top) the original mixed fit and Middle) the peak region of the lensing component only, after the variable component has been subtracted and Bottom) the residuals after subtraction of the mixed fit.}
\label{2008_selection_mixed_LC_53608}
\end{figure}

\newpage
 
\begin{figure}[!ht]
\vspace*{7cm}
$\begin{array}{c}
\vspace*{7.5cm}
   \leavevmode
 \includegraphics{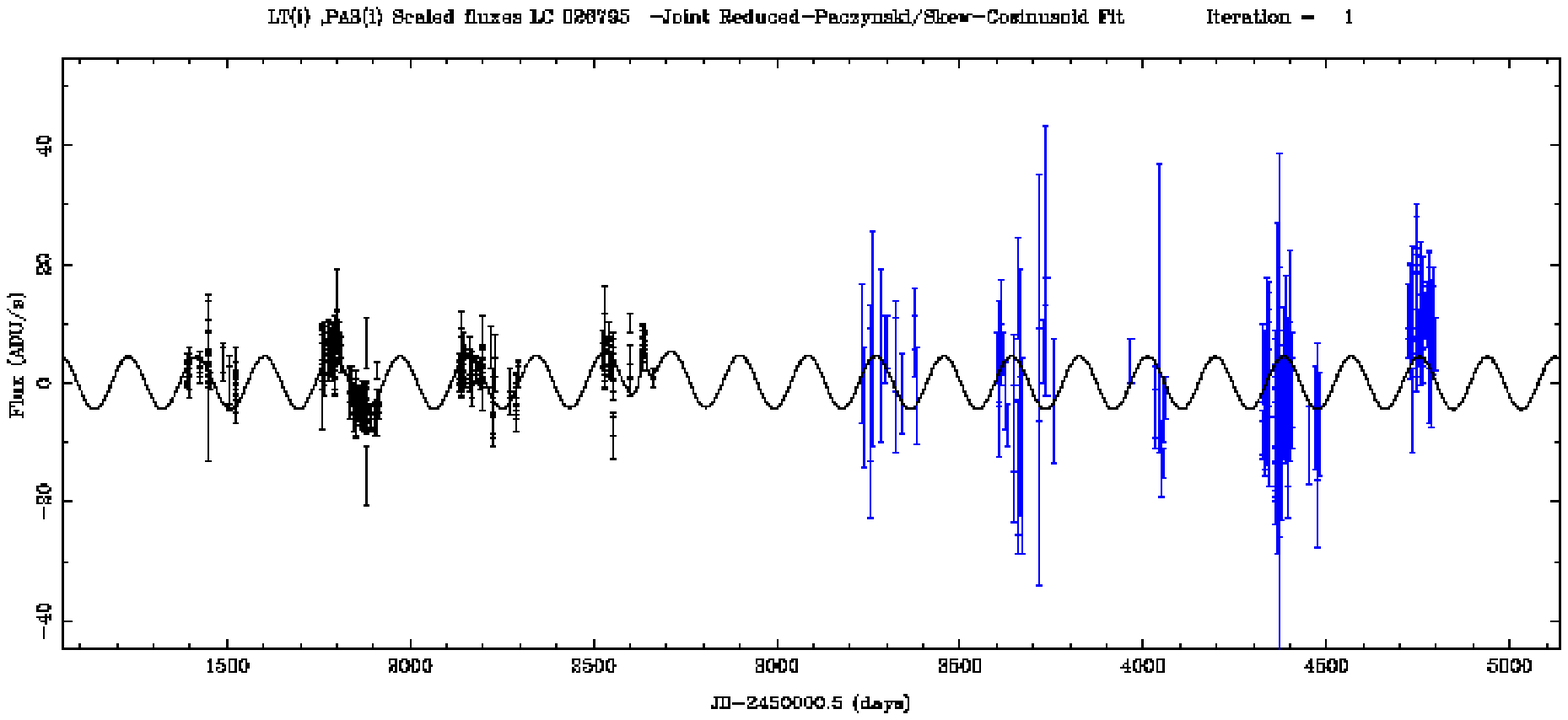} \\
\vspace*{7.5cm}
   \leavevmode
 \includegraphics{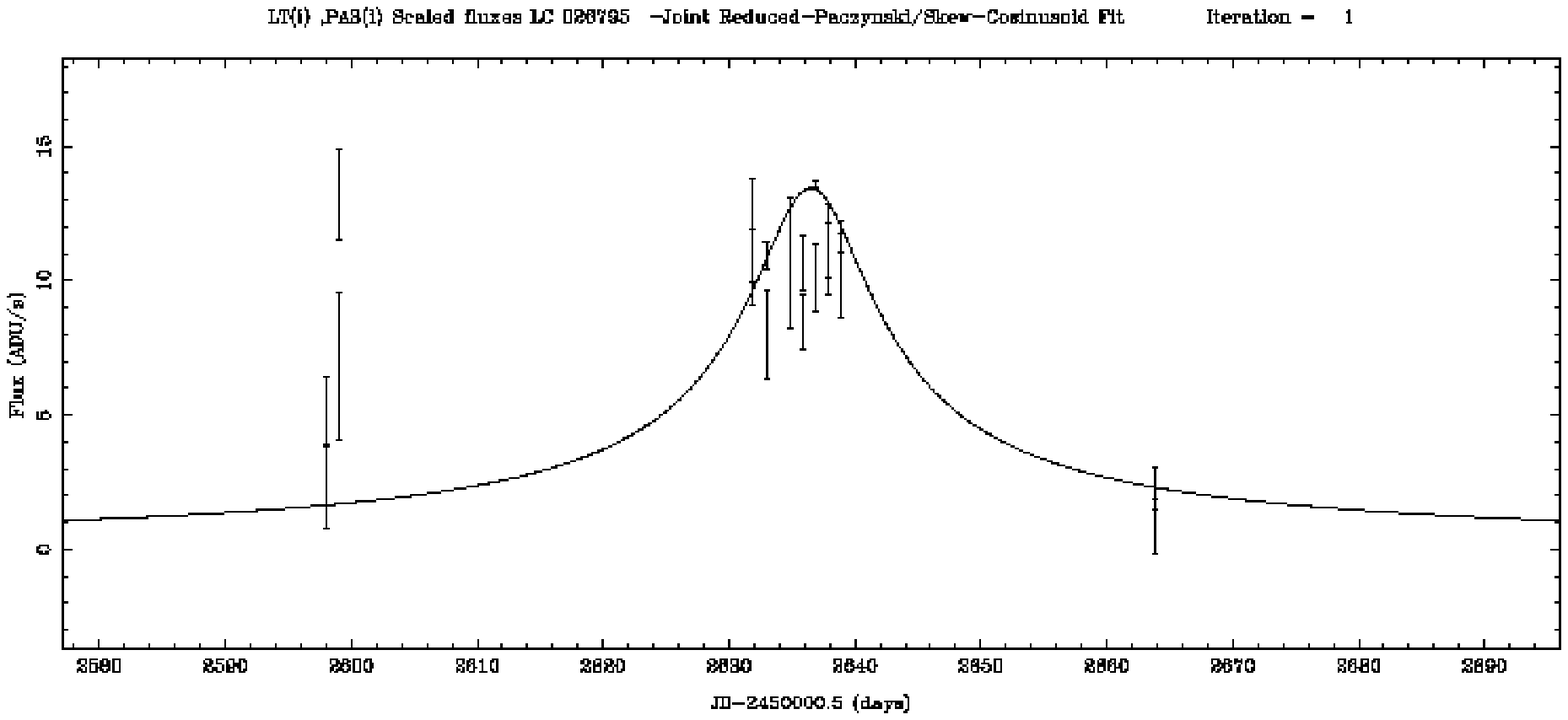} \\
  \leavevmode
 \includegraphics{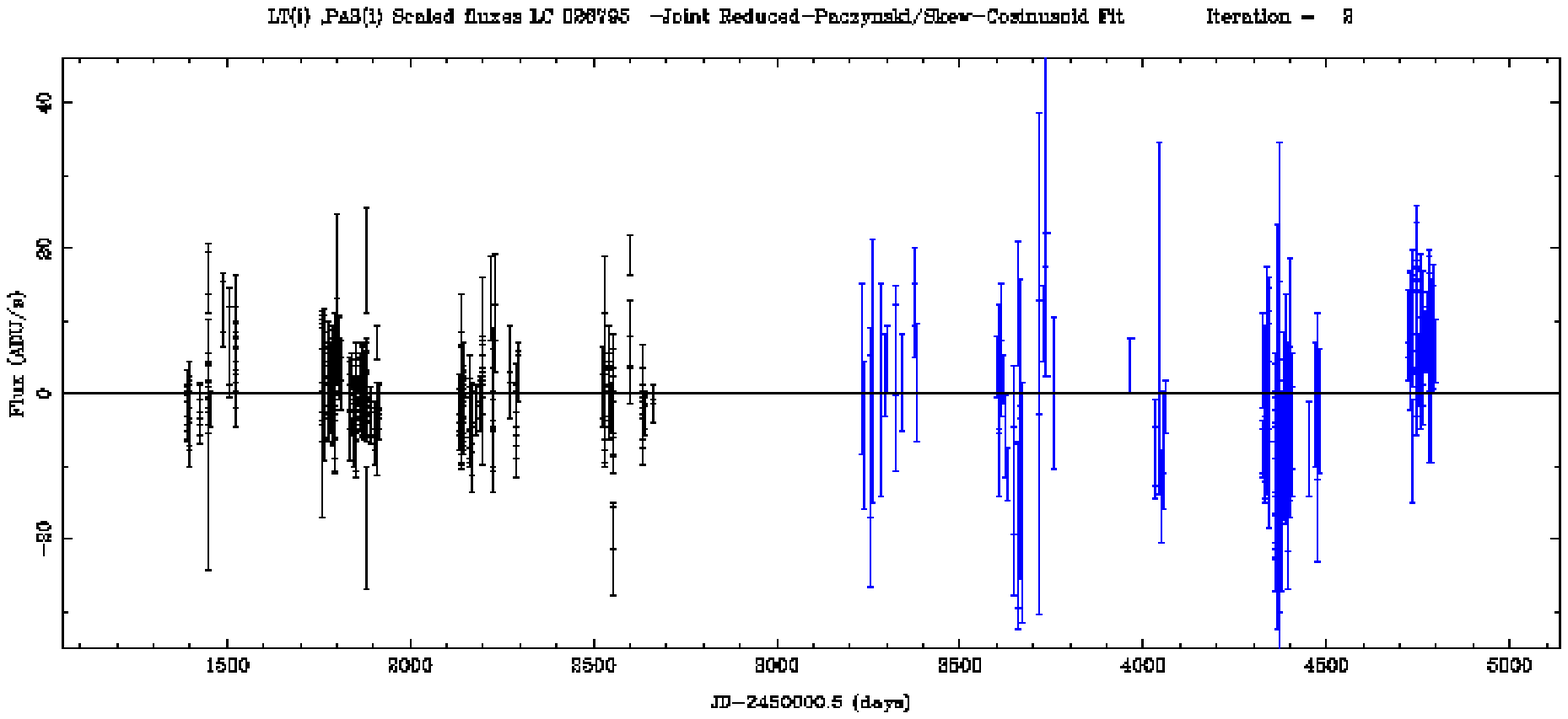} \\
\end{array}$
\caption[Lightcurve of object number $26795$ in the 2008 photometry, showing Top) the original mixed fit,
Middle) the peak region of the lensing component only, after the variable component has been subtracted, and Bottom) the residuals after subtraction of the mixed fit.]{Lightcurve of object number $26795$ in the 2008 photometry, showing Top) the original mixed fit and Middle) the peak region of the lensing component only, after the variable component has been subtracted and Bottom) the residuals after subtraction of the mixed fit.}
\label{2008_selection_mixed_LC_26795}
\end{figure}

\newpage
 
\begin{figure}[!ht]
\vspace*{7cm}
$\begin{array}{c}
\vspace*{7.5cm}
   \leavevmode
 \includegraphics{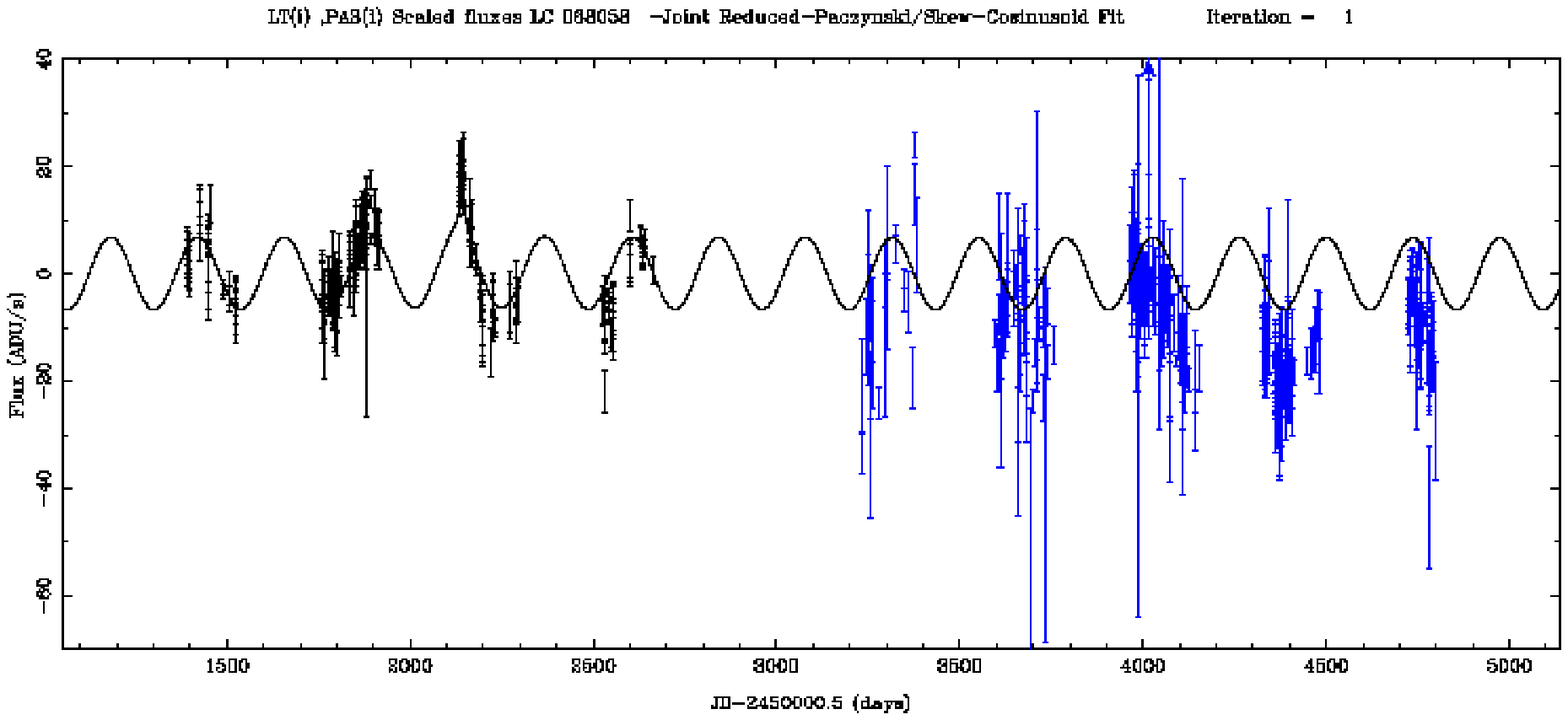} \\
\vspace*{7.5cm}
   \leavevmode
 \includegraphics{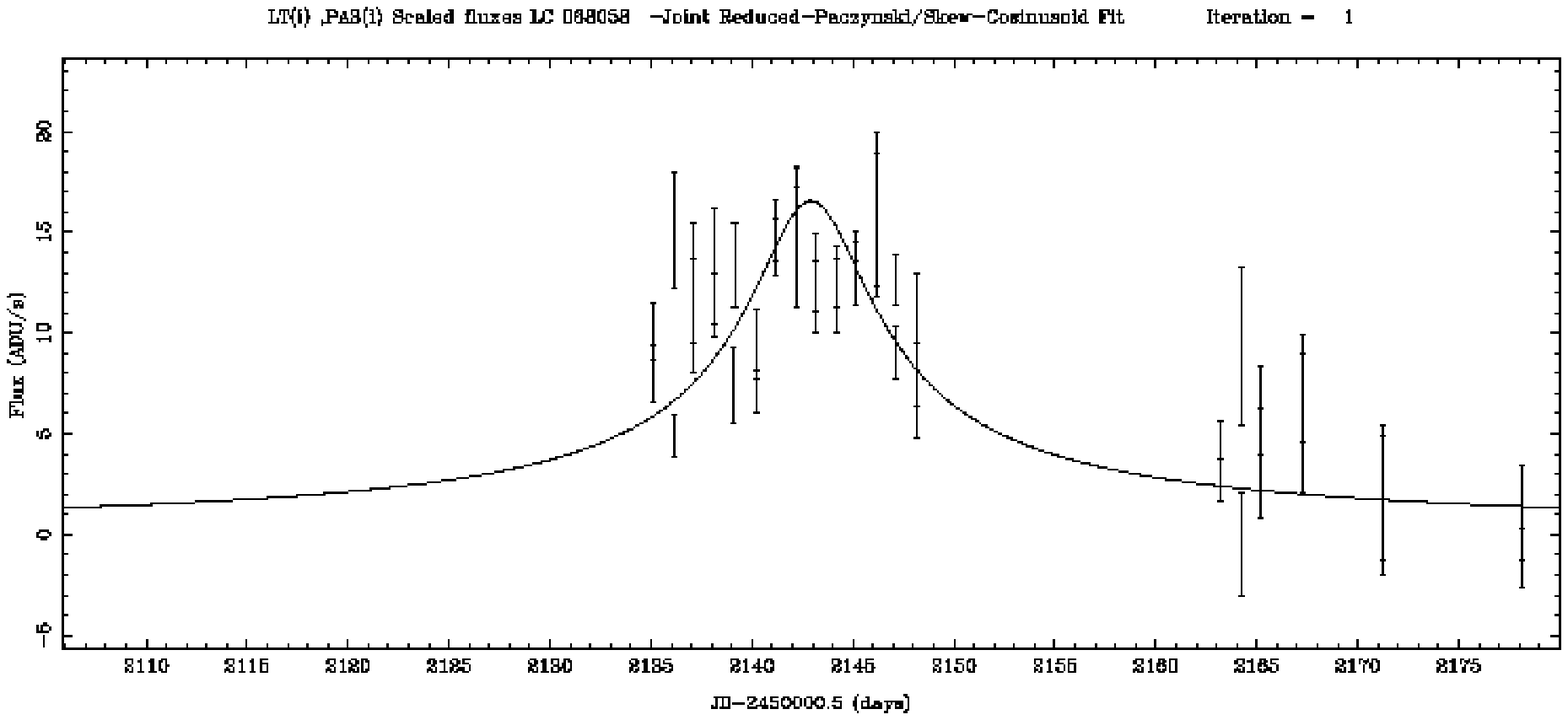} \\
  \leavevmode
 \includegraphics{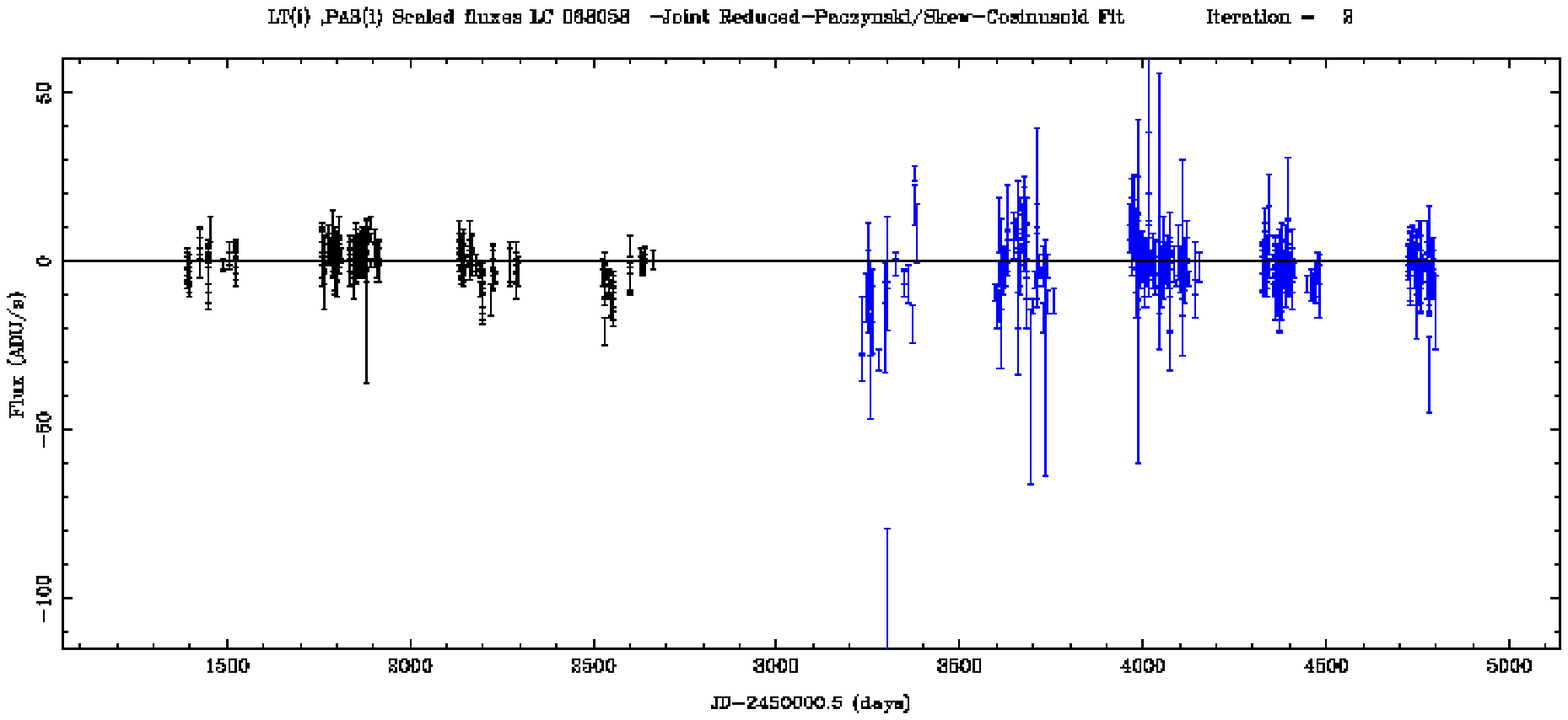} \\
\end{array}$
\caption[Lightcurve of object number $68058$ in the 2008 photometry, showing Top) the original mixed fit,
Middle) the peak region of the lensing component only, after the variable component has been subtracted, and Bottom) the residuals after subtraction of the mixed fit.]{Lightcurve of object number $68058$ in the 2008 photometry, showing Top) the original mixed fit and Middle) the peak region of the lensing component only, after the variable component has been subtracted and Bottom) the residuals after subtraction of the mixed fit.}
\label{2008_selection_mixed_LC_68058}
\end{figure}

\newpage

\begin{figure}[!ht]
\vspace*{7cm}
$\begin{array}{c}
\vspace*{7.5cm}
   \leavevmode
 \includegraphics{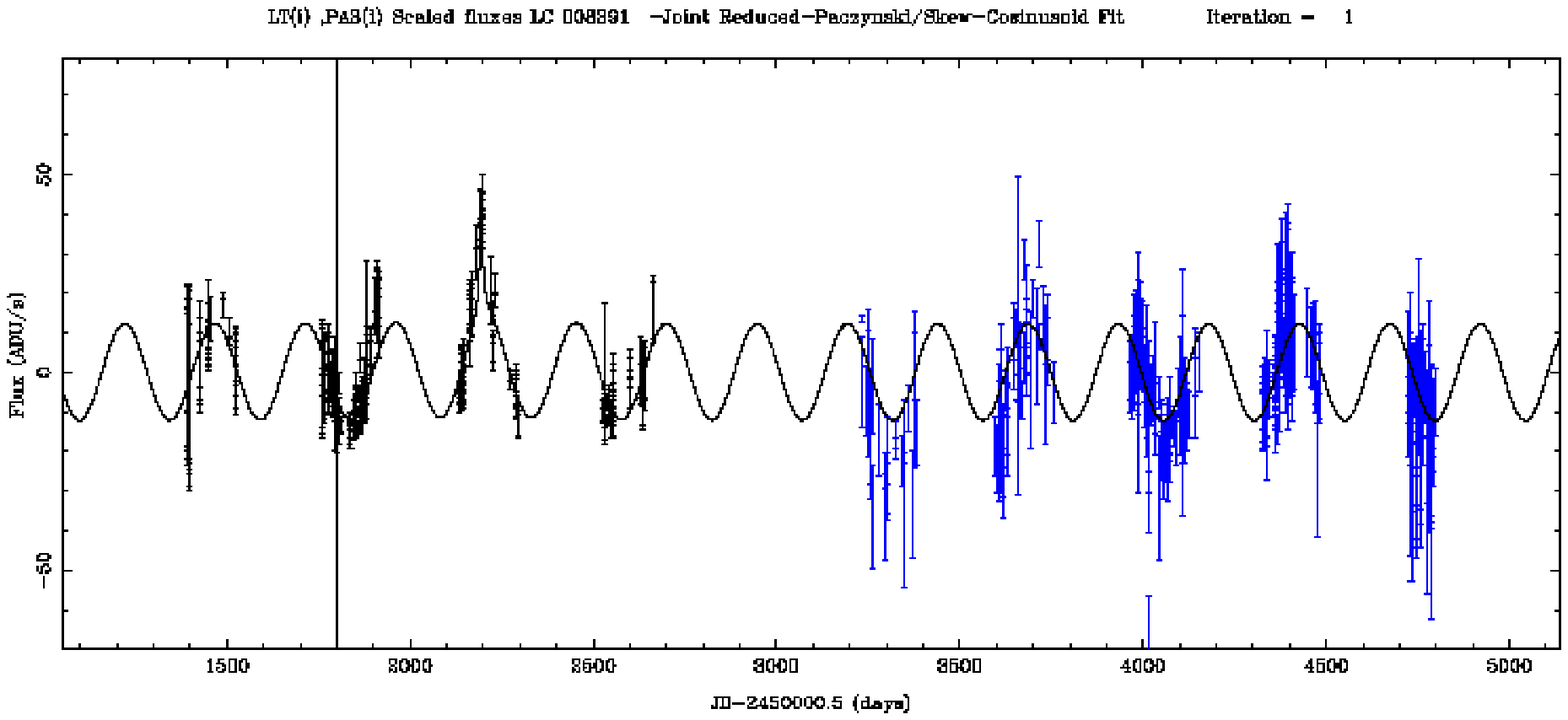} \\
\vspace*{7.5cm}
   \leavevmode
 \includegraphics{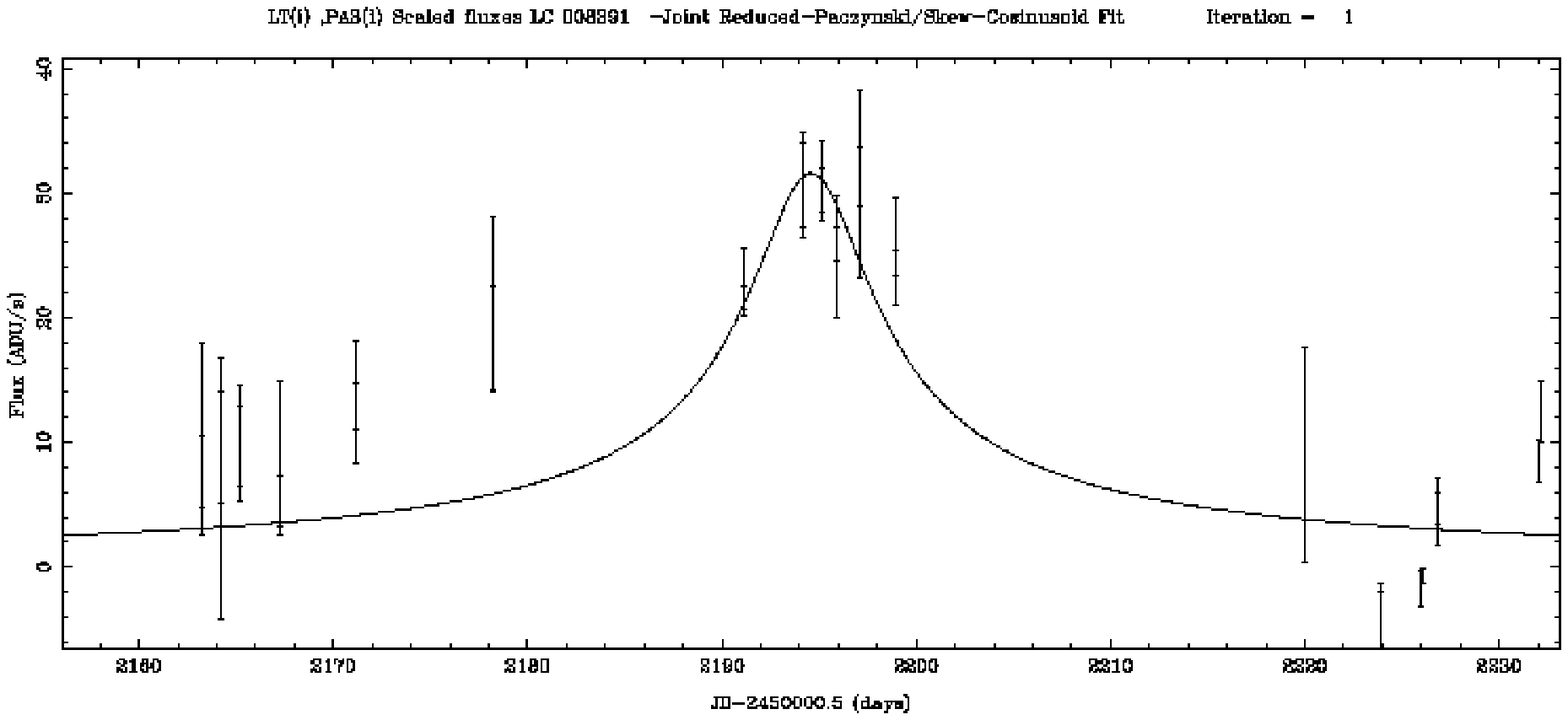} \\
  \leavevmode
 \includegraphics{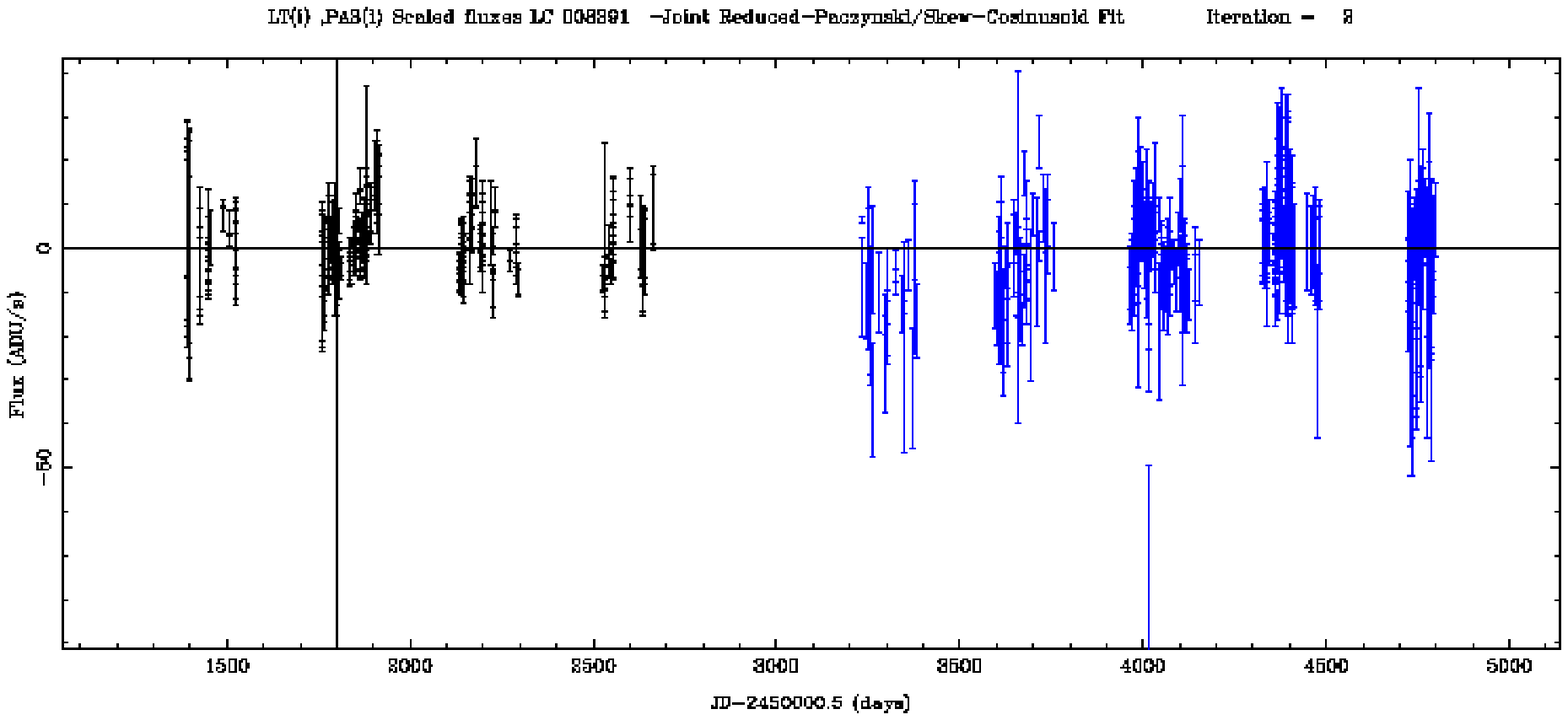} \\
\end{array}$
\caption[Lightcurve of object number $8391$ in the 2008 photometry, showing Top) the original mixed fit,
Middle) the peak region of the lensing component only, after the variable component has been subtracted, and Bottom) the residuals after subtraction of the mixed fit.]{Lightcurve of object number $8391$ in the 2008 photometry, showing Top) the original mixed fit and Middle) the peak region of the lensing component only, after the variable component has been subtracted and Bottom) the residuals after subtraction of the mixed fit.}
\label{2008_selection_mixed_LC_08391}
\end{figure}

\newpage
 
\begin{figure}[!ht]
\vspace*{7cm}
$\begin{array}{c}
\vspace*{7.5cm}
   \leavevmode
 \includegraphics{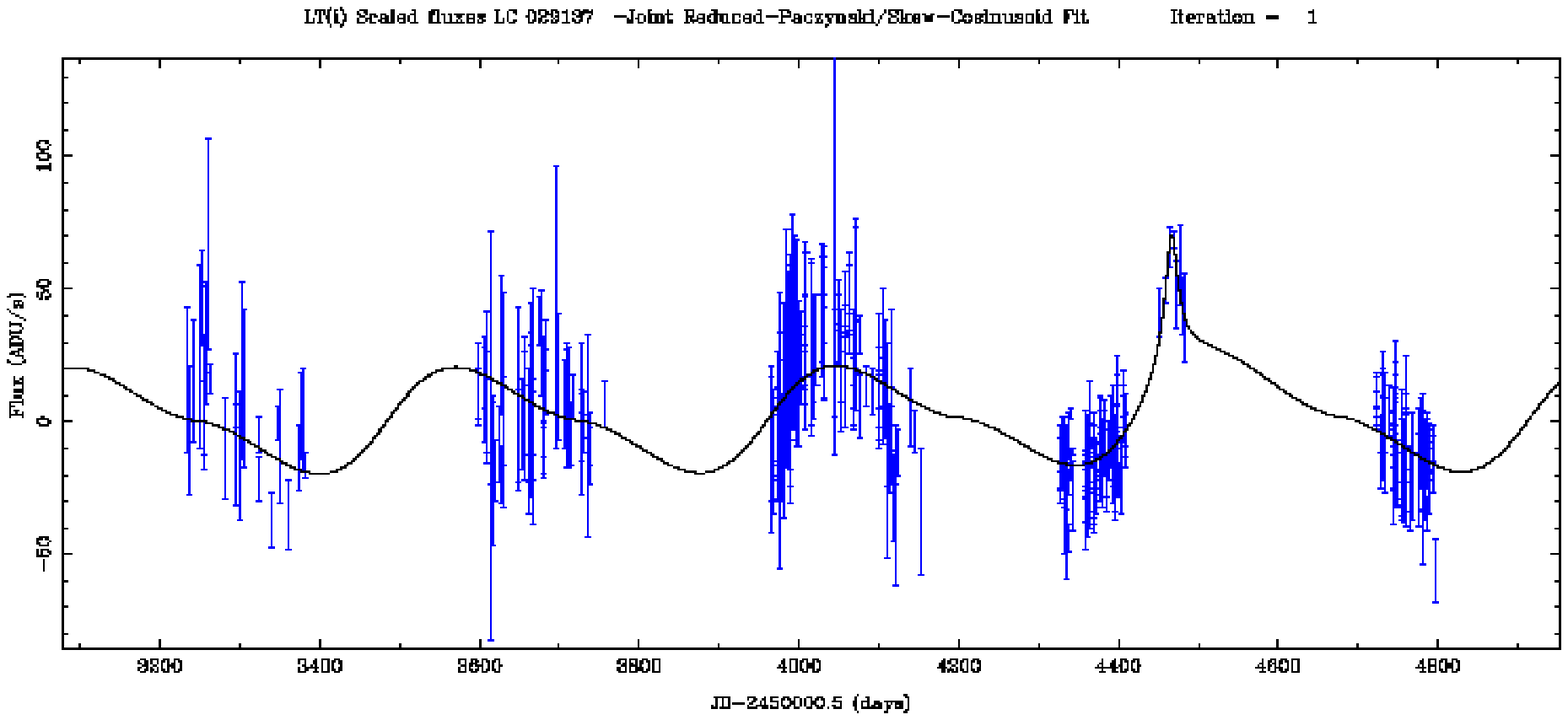} \\
\vspace*{7.5cm}
   \leavevmode
 \includegraphics{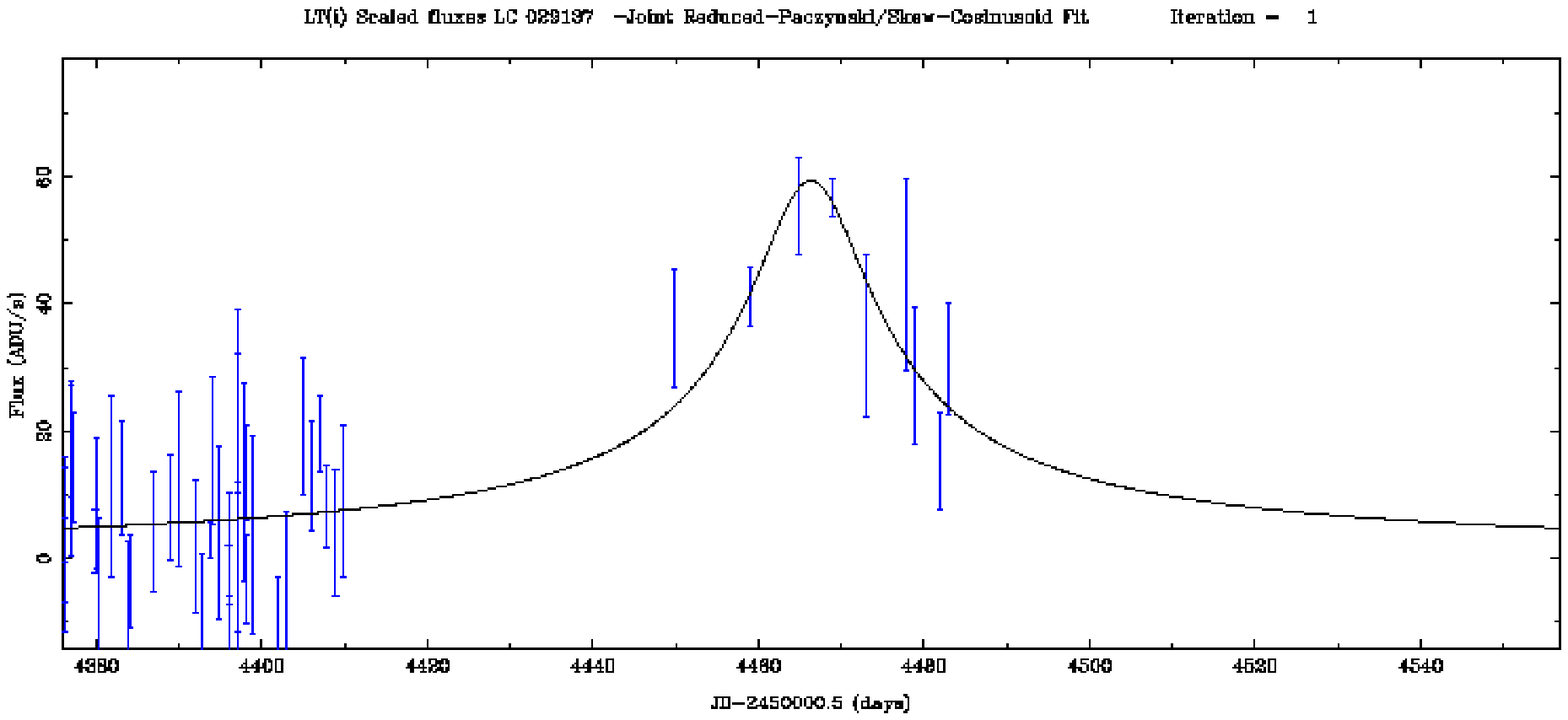} \\
  \leavevmode
 \includegraphics{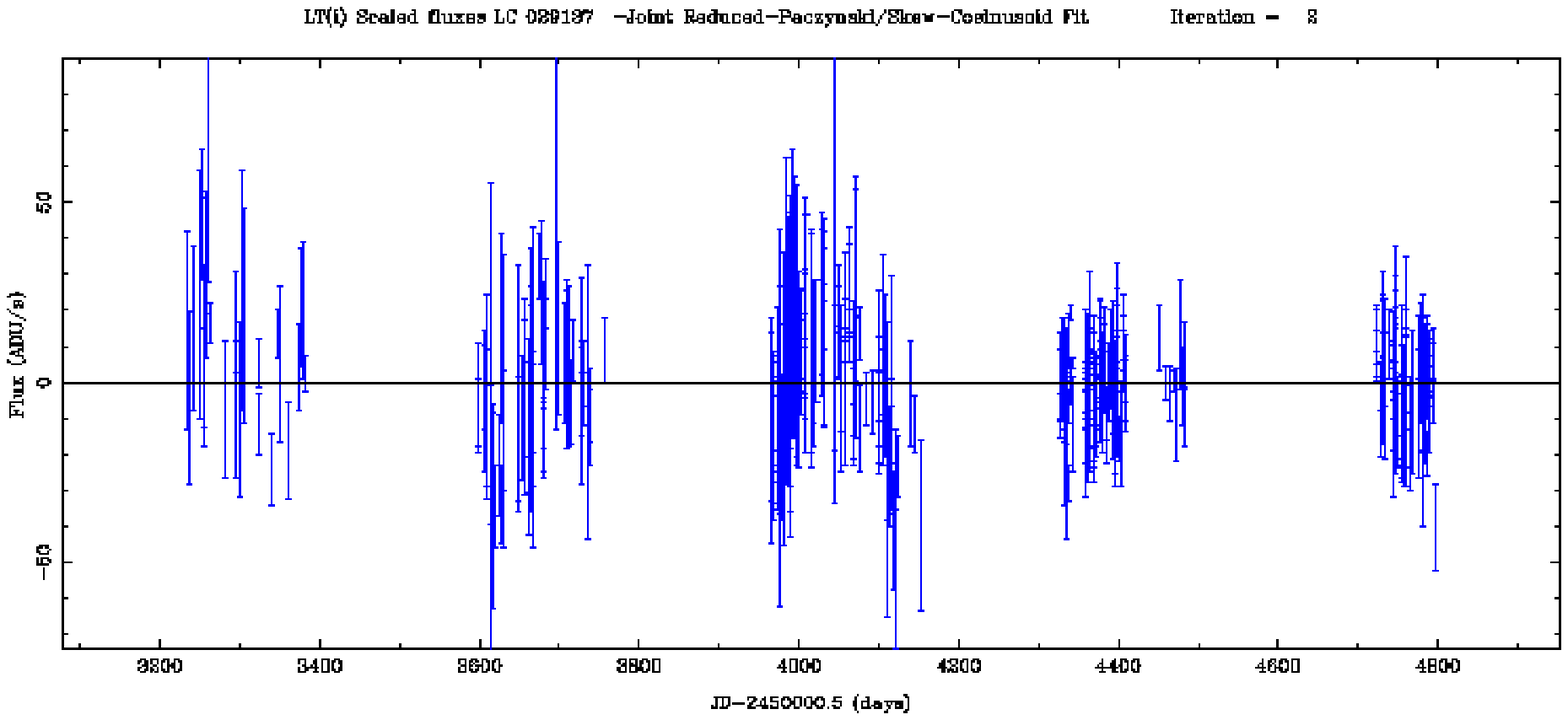} \\
\end{array}$
\caption[Lightcurve of object number $29137$ in the 2008 photometry, showing Top) the original mixed fit,
Middle) the peak region of the lensing component only, after the variable component has been subtracted, and Bottom) the residuals after subtraction of the mixed fit.]{Lightcurve of object number $29137$ in the 2008 photometry, showing Top) the original mixed fit and Middle) the peak region of the lensing component only, after the variable component has been subtracted and Bottom) the residuals after subtraction of the mixed fit.}
\label{2008_selection_mixed_LC_29137}
\end{figure}

\newpage

\begin{figure}[!ht]
\vspace*{7cm}
$\begin{array}{c}
\vspace*{7.5cm}
   \leavevmode
 \includegraphics{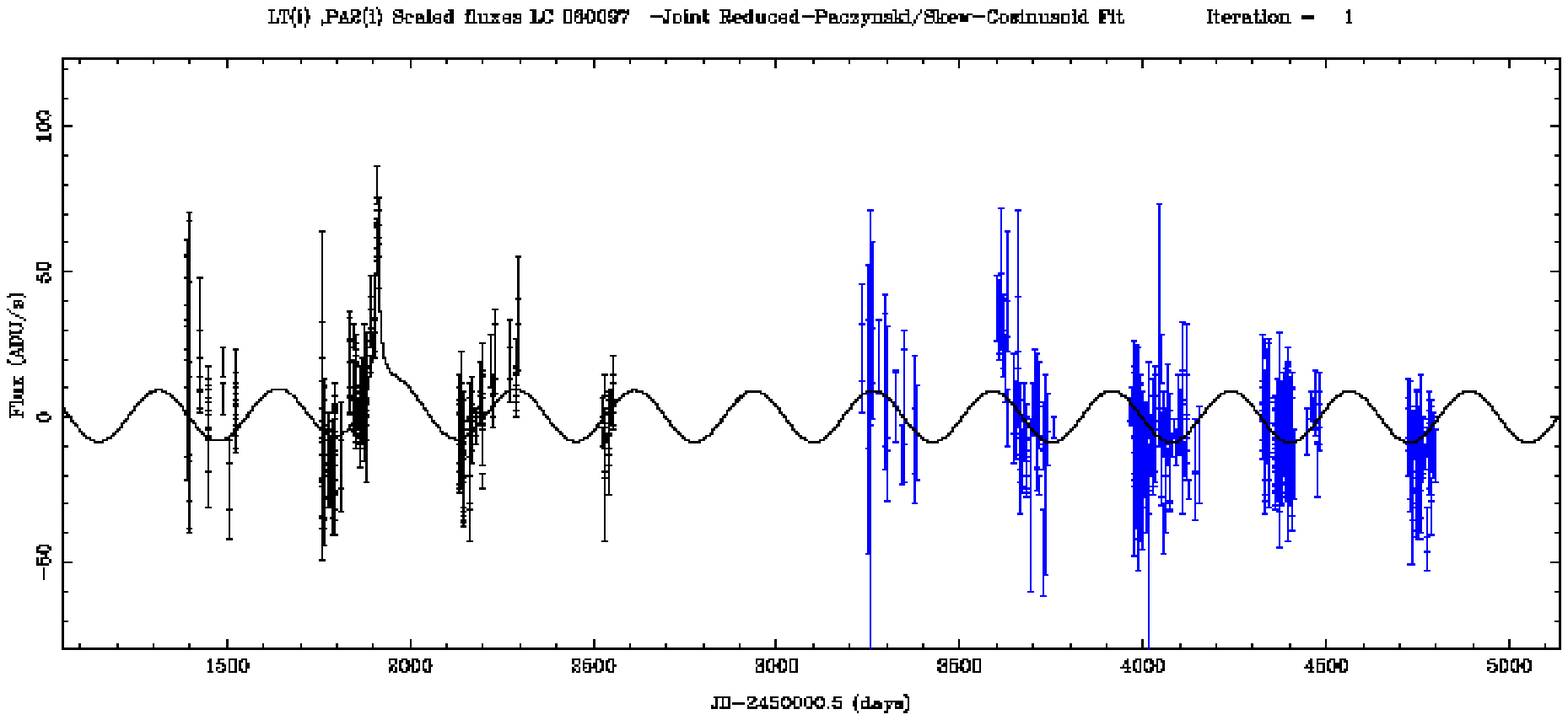} \\
\vspace*{7.5cm}
   \leavevmode
 \includegraphics{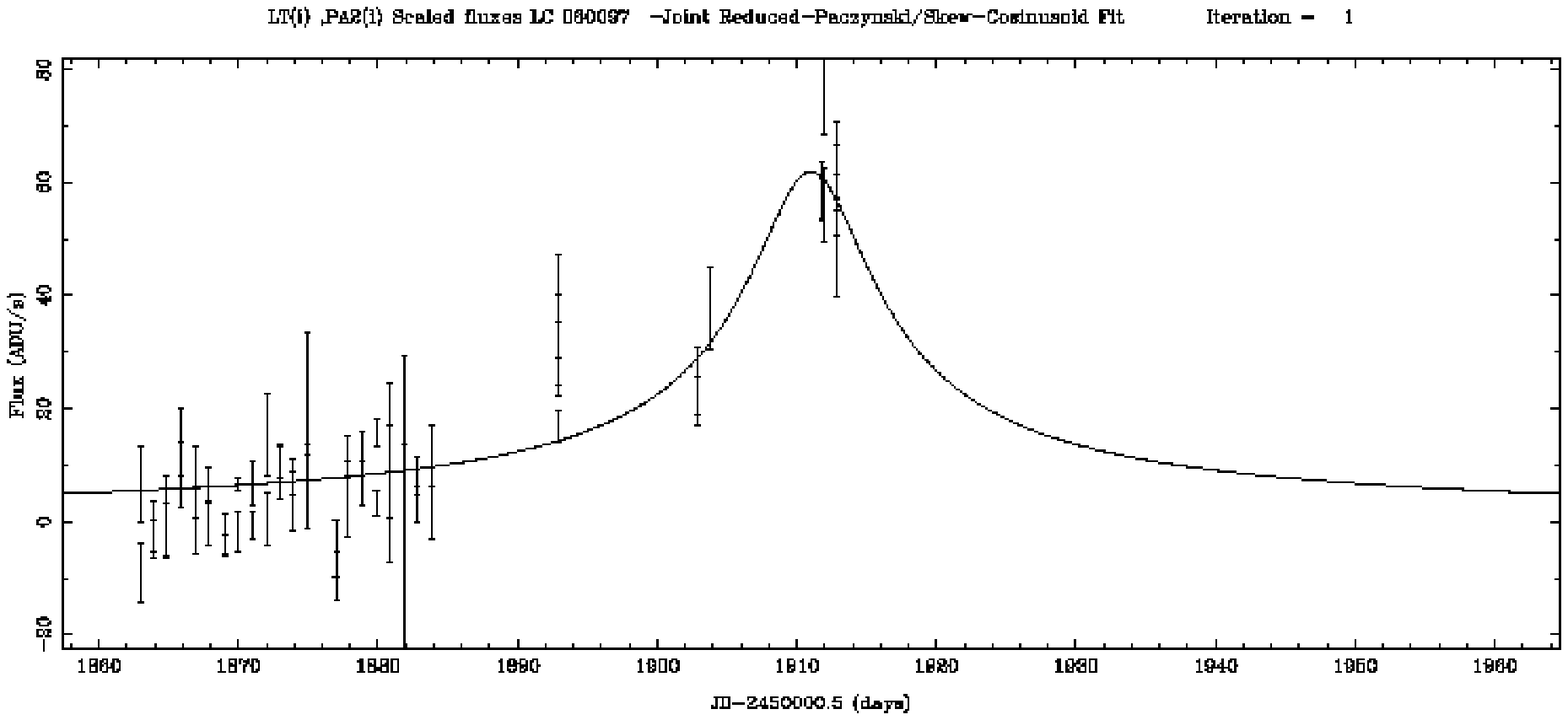} \\
  \leavevmode
 \includegraphics{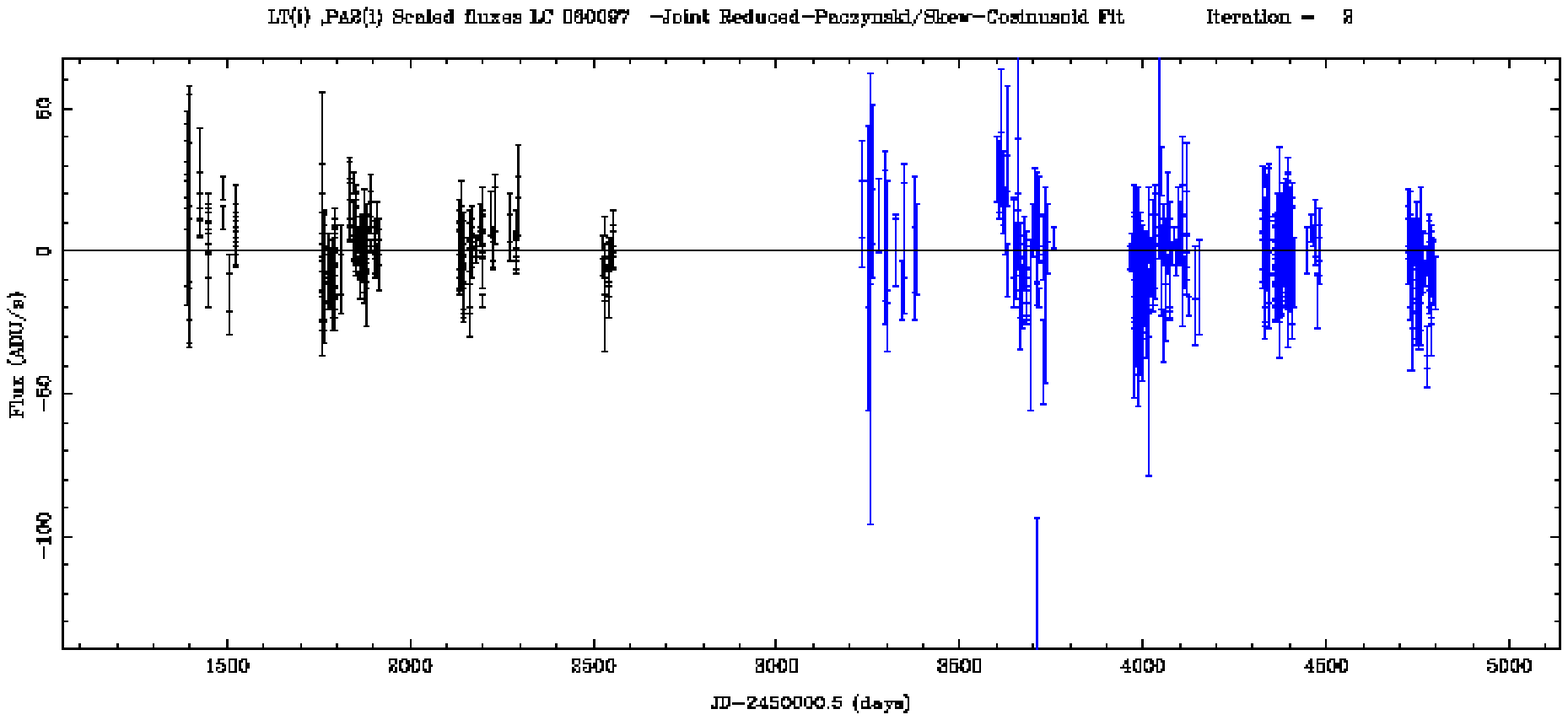} \\
\end{array}$
\caption[Lightcurve of object number $60097$ in the 2008 photometry, showing Top) the original mixed fit,
Middle) the peak region of the lensing component only, after the variable component has been subtracted, and Bottom) the residuals after subtraction of the mixed fit.]{Lightcurve of object number $60097$ in the 2008 photometry, showing Top) the original mixed fit and Middle) the peak region of the lensing component only, after the variable component has been subtracted and Bottom) the residuals after subtraction of the mixed fit.}
\label{2008_selection_mixed_LC_60097}
\end{figure}

\newpage
 
\begin{figure}[!ht]
\vspace*{7cm}
$\begin{array}{c}
\vspace*{7.5cm}
   \leavevmode
 \includegraphics{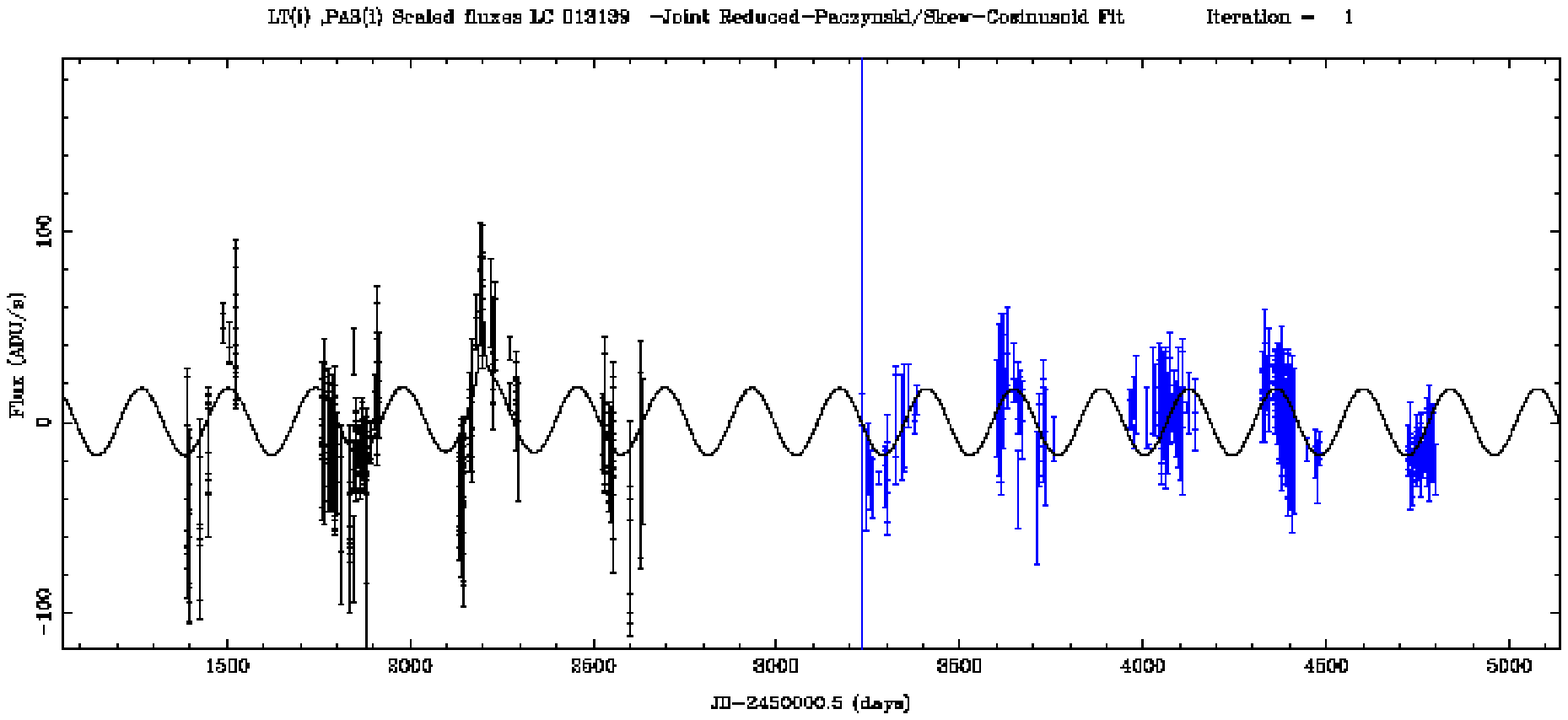} \\
\vspace*{7.5cm}
   \leavevmode
 \includegraphics{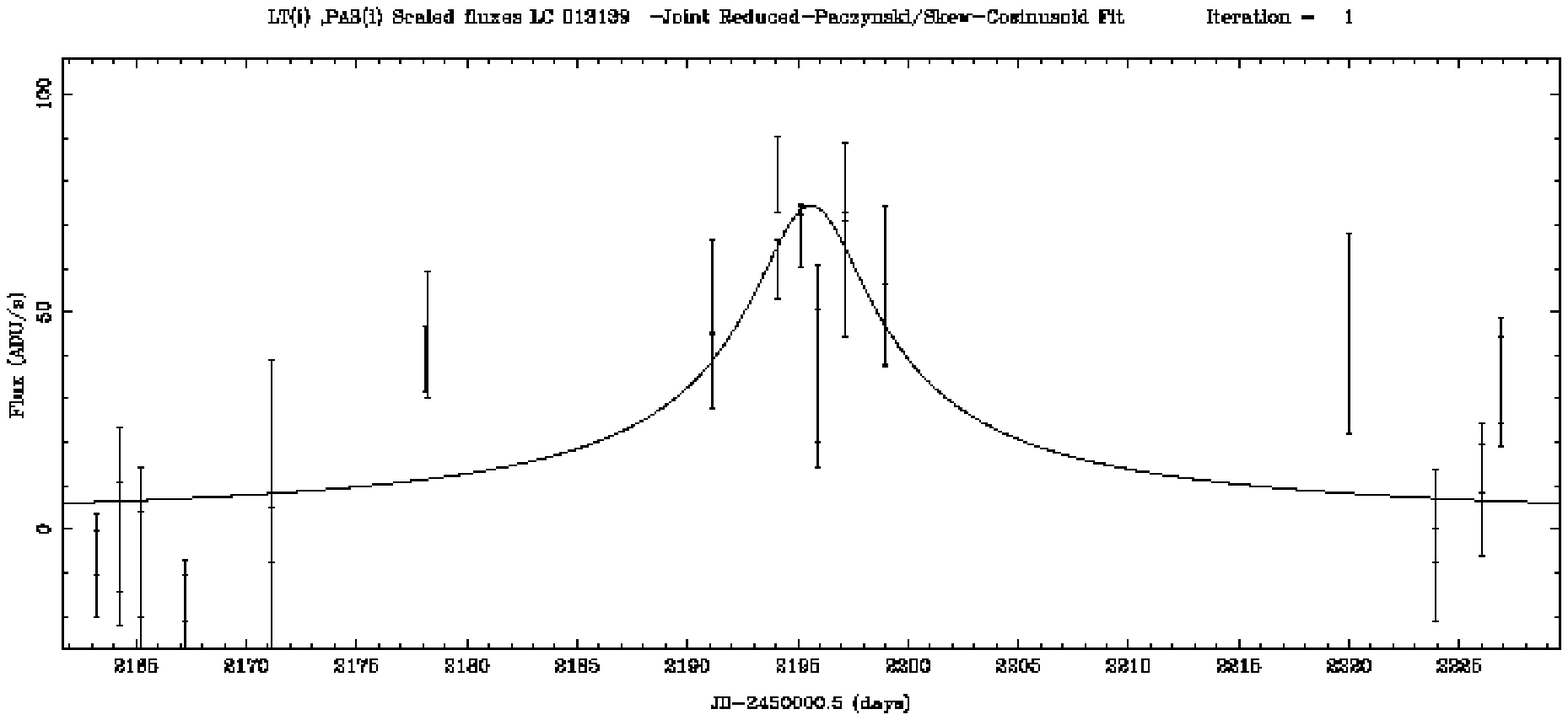} \\
  \leavevmode
 \includegraphics{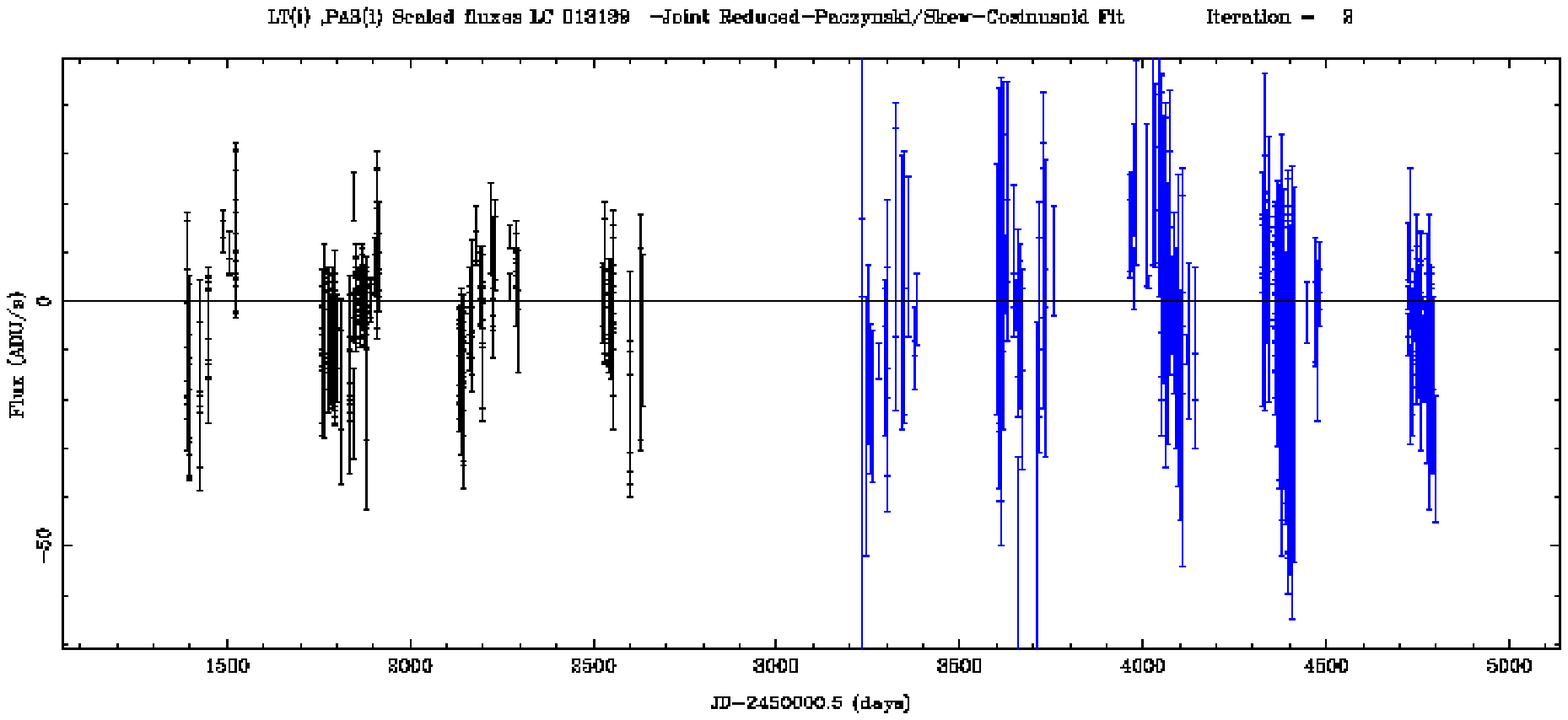} \\
\end{array}$
\caption[Lightcurve of object number $13139$ in the 2008 photometry, showing Top) the original mixed fit,
Middle) the peak region of the lensing component only, after the variable component has been subtracted, and Bottom) the residuals after subtraction of the mixed fit.]{Lightcurve of object number $13139$ in the 2008 photometry, showing Top) the original mixed fit and Middle) the peak region of the lensing component only, after the variable component has been subtracted and Bottom) the residuals after subtraction of the mixed fit.}
\label{2008_selection_mixed_LC_13139}
\end{figure}

\newpage

\begin{figure}[!ht]
\vspace*{7cm}
$\begin{array}{c}
\vspace*{7.5cm}
   \leavevmode
 \includegraphics{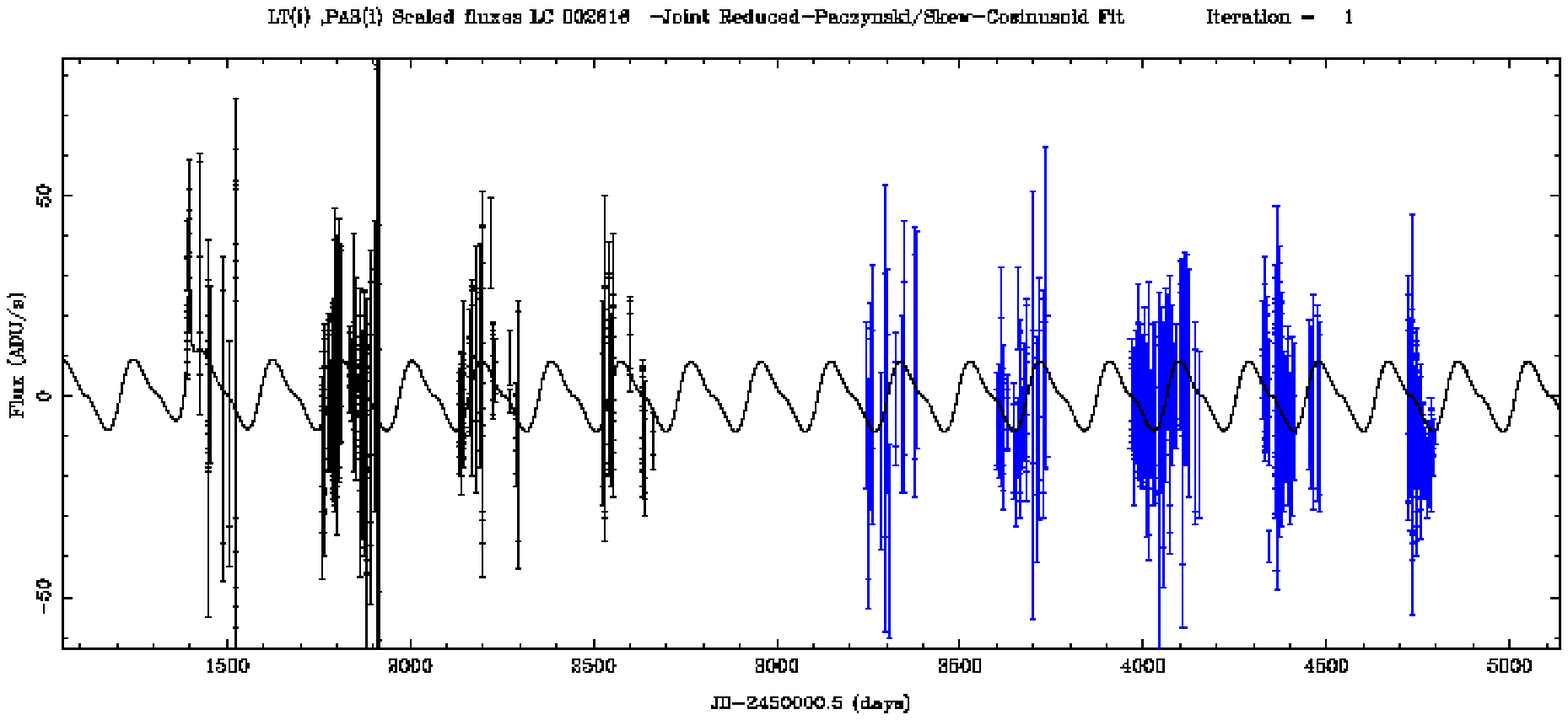} \\
\vspace*{7.5cm}
   \leavevmode
 \includegraphics{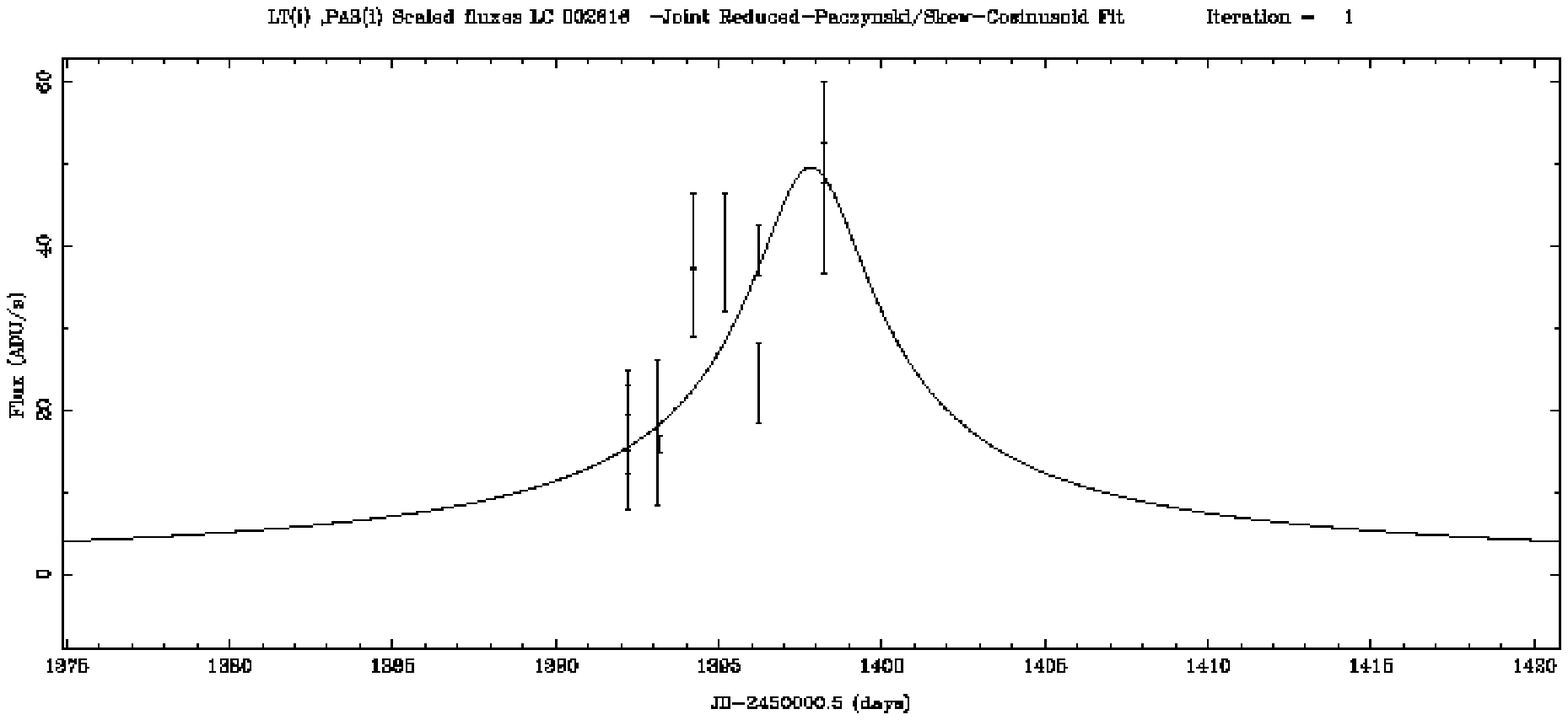} \\
  \leavevmode
 \includegraphics{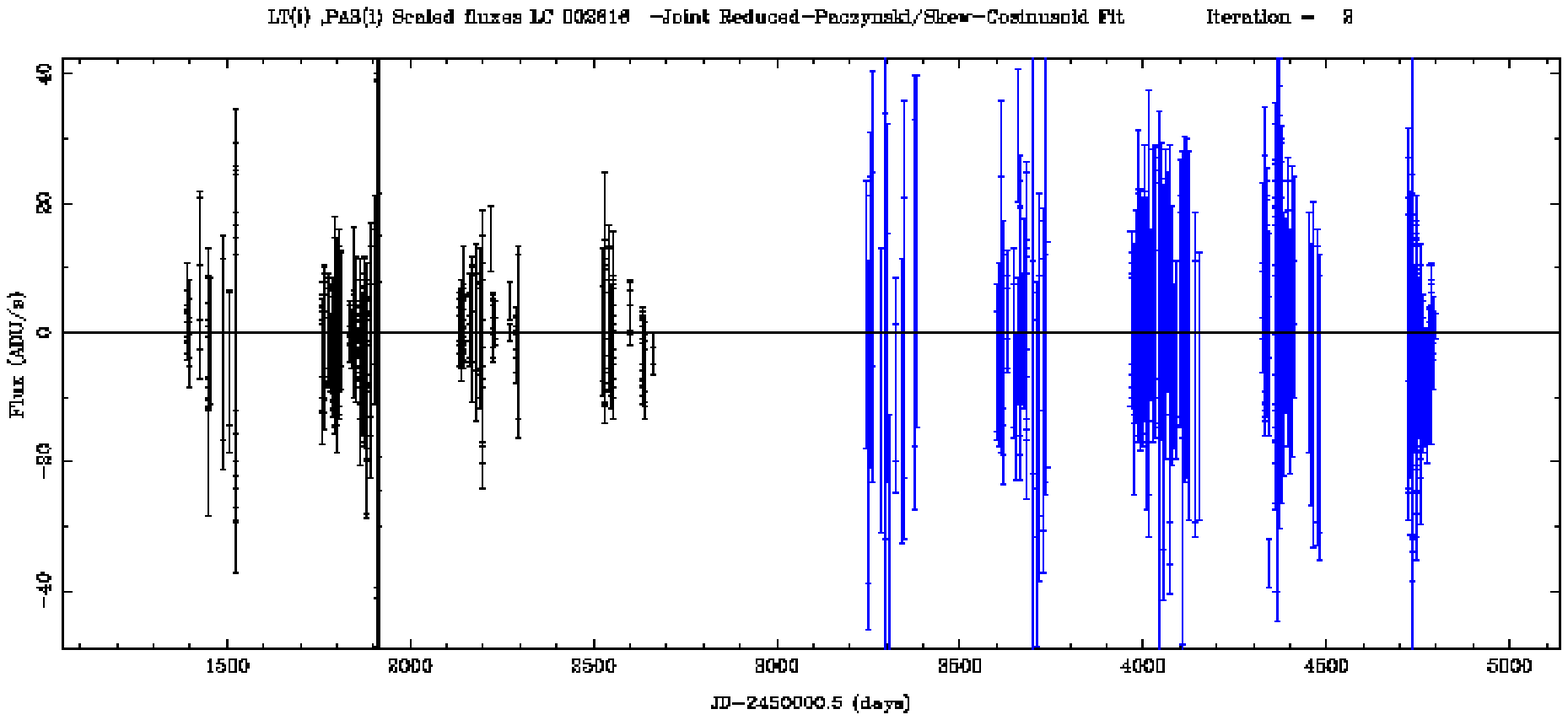} \\
\end{array}$
\caption[Lightcurve of object number $2616$ in the 2008 photometry, showing Top) the original mixed fit,
Middle) the peak region of the lensing component only, after the variable component has been subtracted, and Bottom) the residuals after subtraction of the mixed fit.]{Lightcurve of object number $2616$ in the 2008 photometry, showing Top) the original mixed fit and Middle) the peak region of the lensing component only, after the variable component has been subtracted and Bottom) the residuals after subtraction of the mixed fit.}
\label{2008_selection_mixed_LC_02616}
\end{figure}

\newpage

\begin{figure}[!ht]
\vspace*{7cm}
$\begin{array}{c}
\vspace*{7.5cm}
   \leavevmode
 \includegraphics{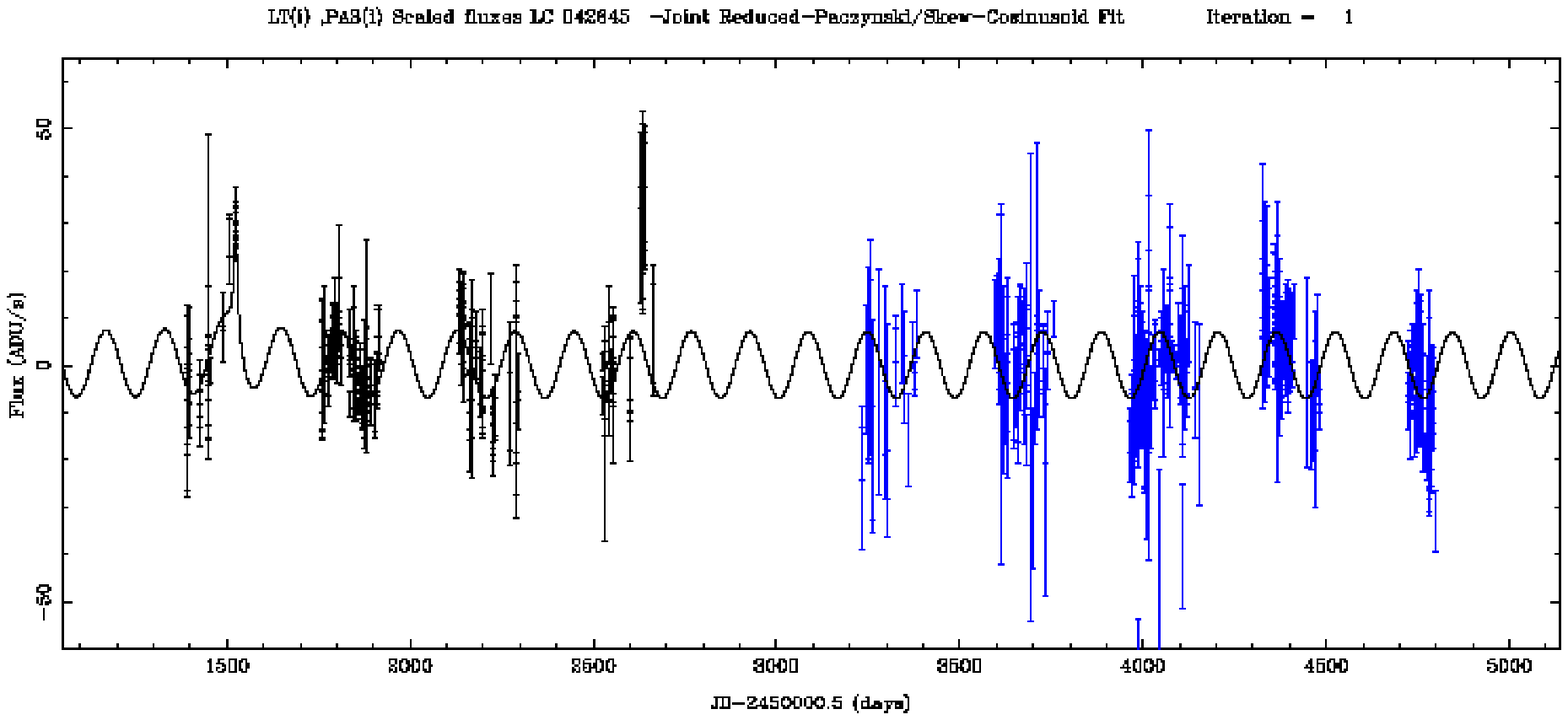} \\
\vspace*{7.5cm}
   \leavevmode
 \includegraphics{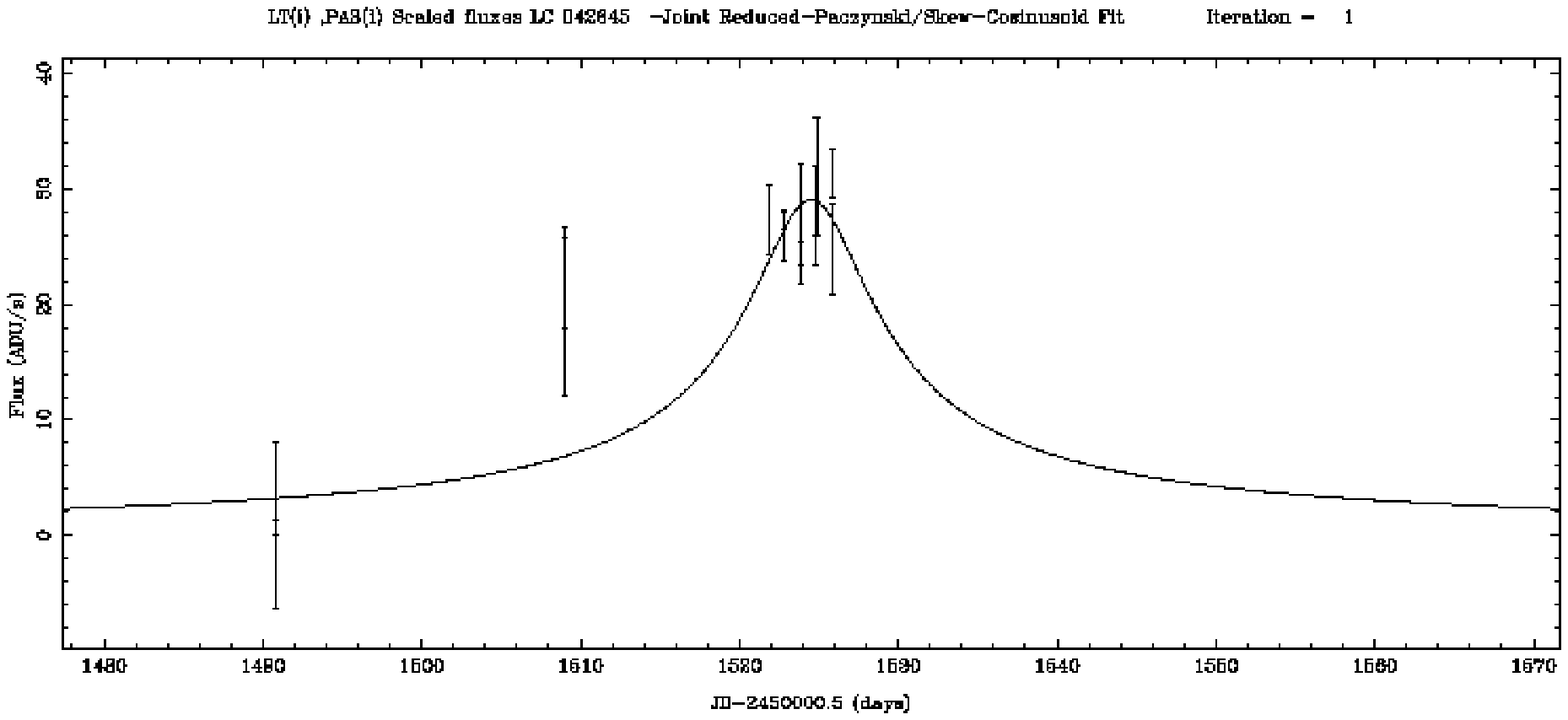} \\
  \leavevmode
 \includegraphics{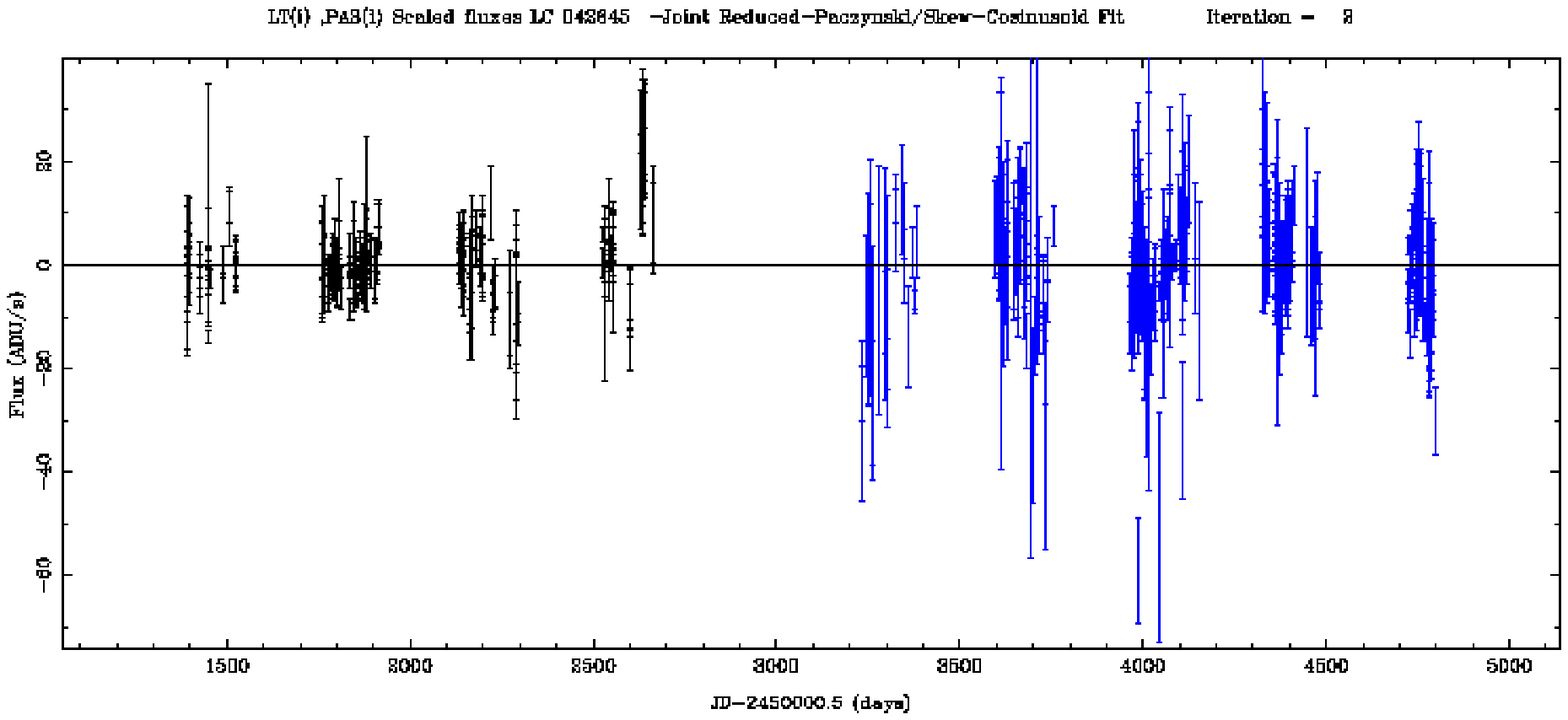} \\
\end{array}$
\caption[Lightcurve of object number $42645$ in the 2008 photometry, showing Top) the original mixed fit,
Middle) the peak region of the lensing component only, after the variable component has been subtracted, and Bottom) the residuals after subtraction of the mixed fit.]{Lightcurve of object number $42645$ in the 2008 photometry, showing Top) the original mixed fit and Middle) the peak region of the lensing component only, after the variable component has been subtracted and Bottom) the residuals after subtraction of the mixed fit.}
\label{2008_selection_mixed_LC_42645}
\end{figure}

\newpage
 
\begin{figure}[!ht]
\vspace*{7cm}
$\begin{array}{c}
\vspace*{7.5cm}
   \leavevmode
 \includegraphics{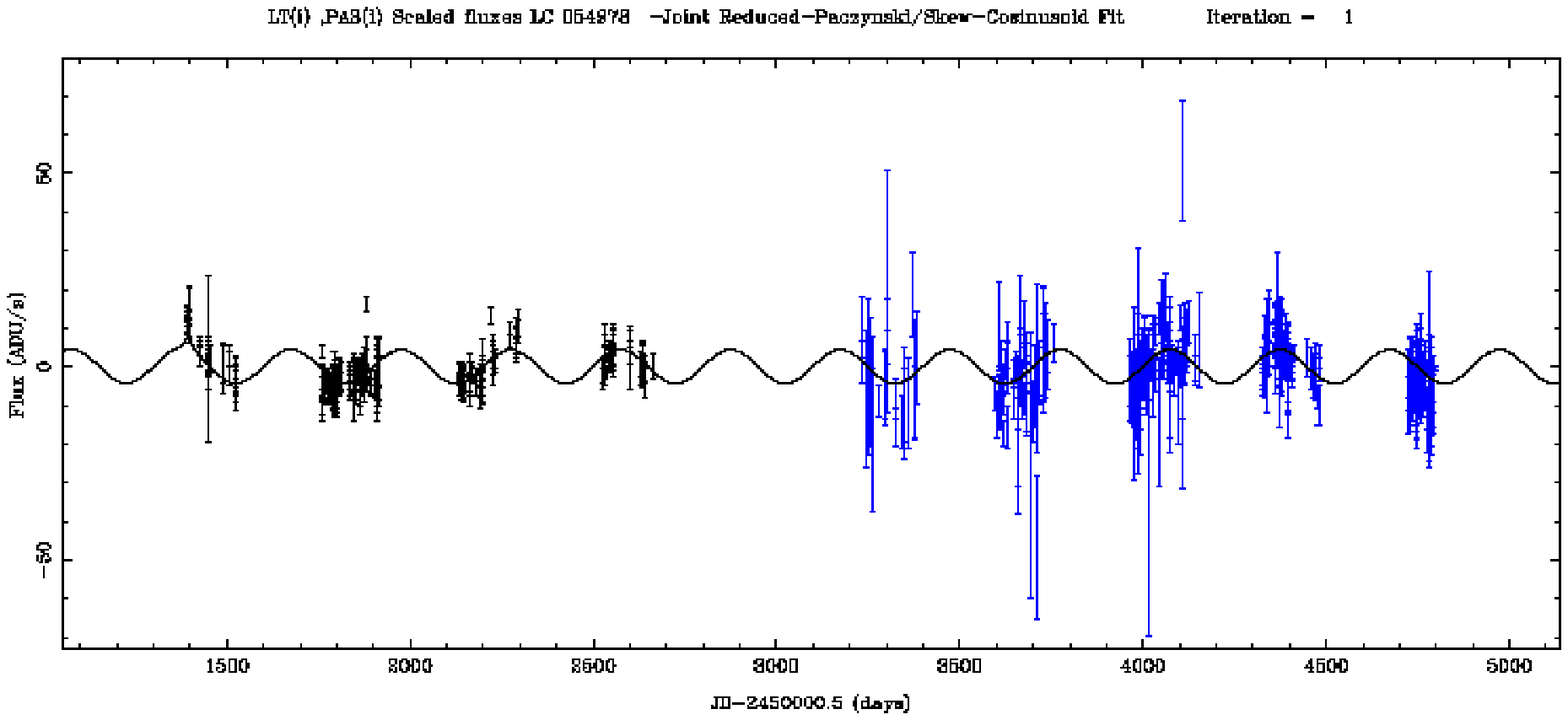} \\
\vspace*{7.5cm}
   \leavevmode
 \includegraphics{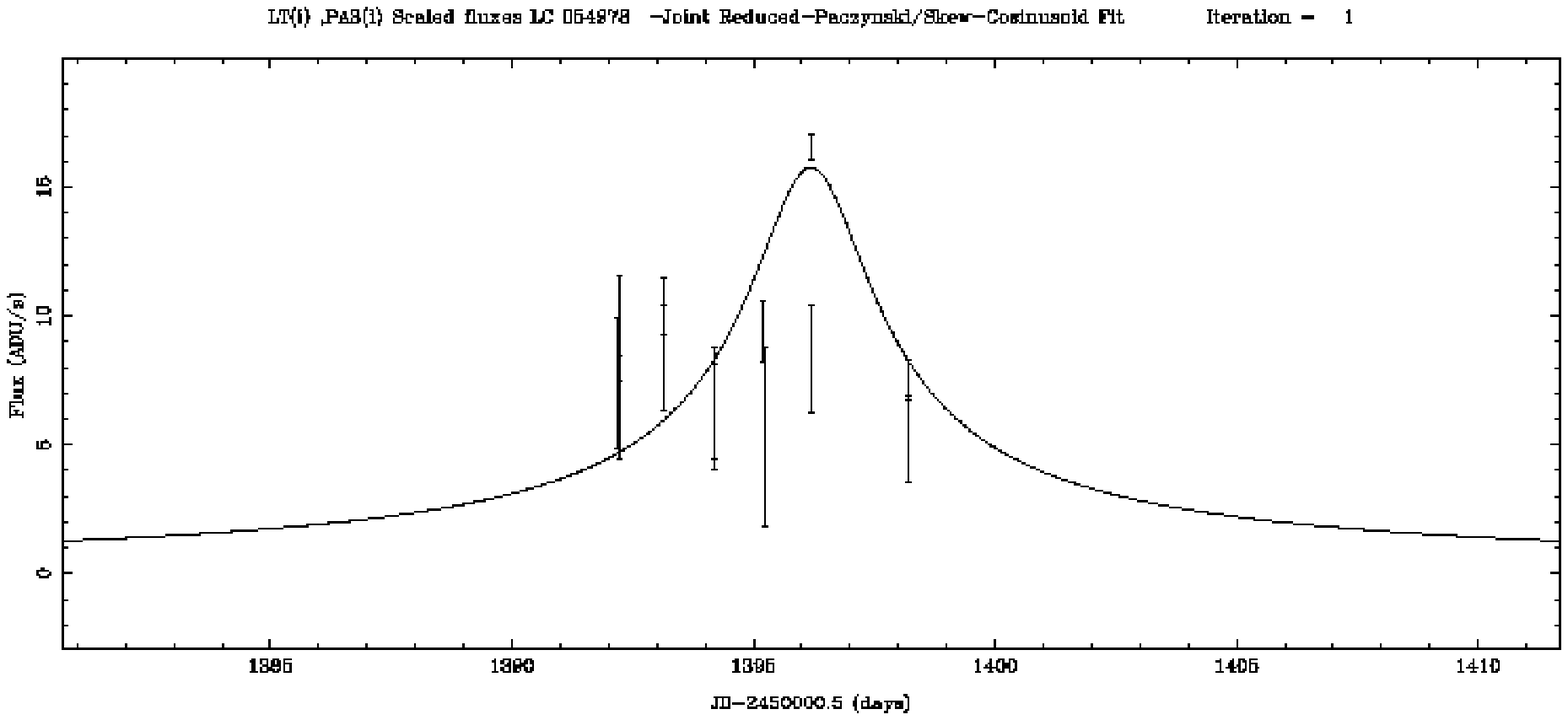} \\
  \leavevmode
 \includegraphics{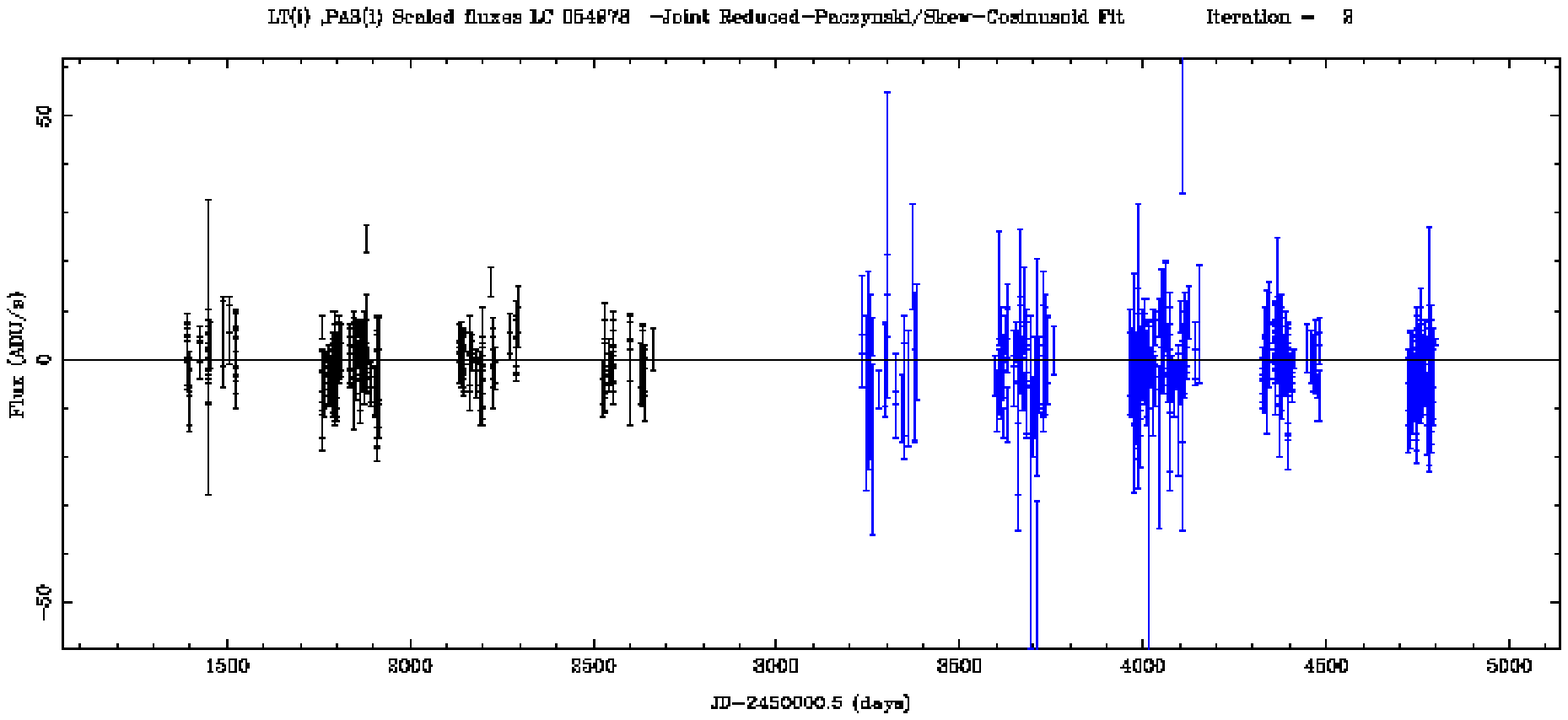} \\
\end{array}$
\caption[Lightcurve of object number $54978$ in the 2008 photometry, showing Top) the original mixed fit,
Middle) the peak region of the lensing component only, after the variable component has been subtracted, and Bottom) the residuals after subtraction of the mixed fit.]{Lightcurve of object number $54978$ in the 2008 photometry, showing Top) the original mixed fit and Middle) the peak region of the lensing component only, after the variable component has been subtracted and Bottom) the residuals after subtraction of the mixed fit.}
\label{2008_selection_mixed_LC_54978}
\end{figure}

\newpage

\begin{figure}[!ht]
\vspace*{7cm}
$\begin{array}{c}
\vspace*{7.5cm}
   \leavevmode
 \includegraphics{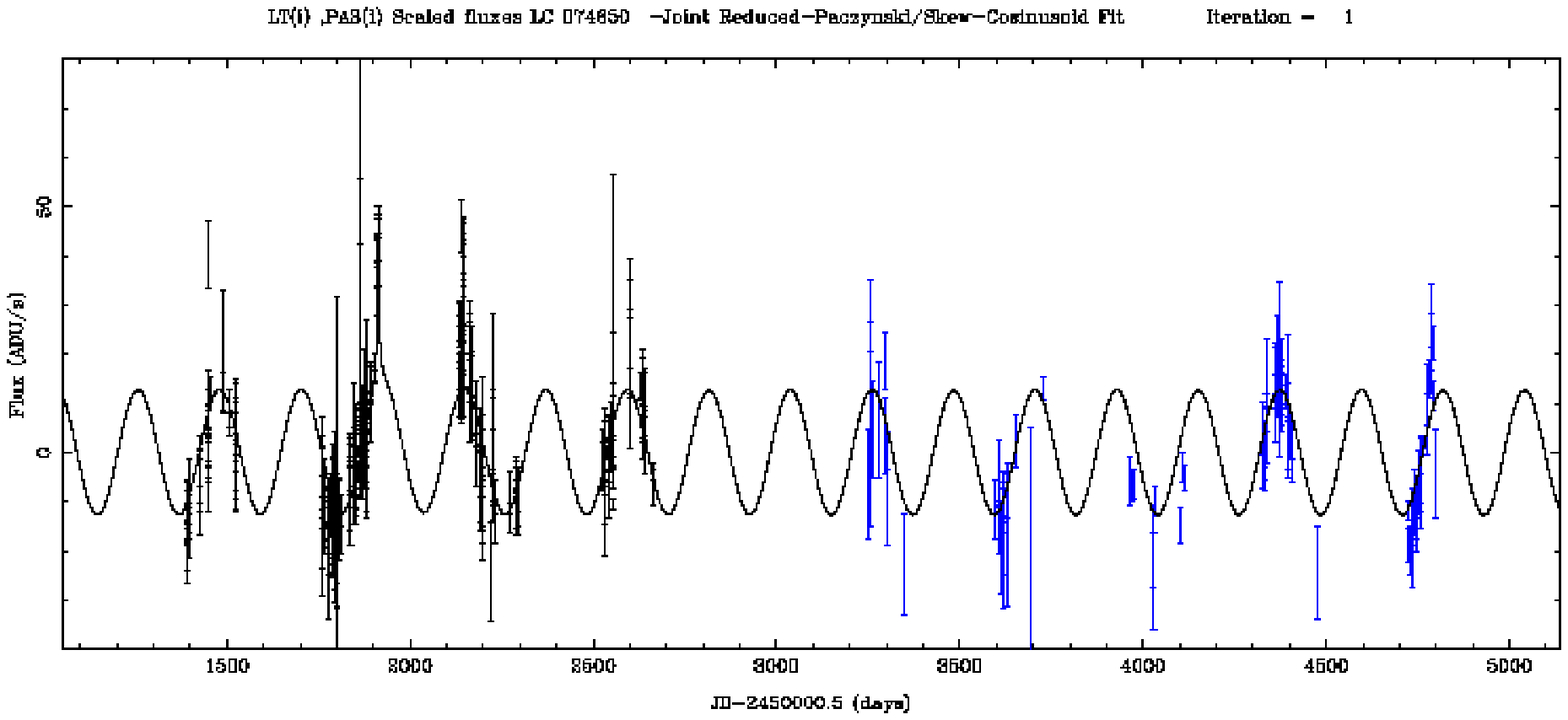} \\
\vspace*{7.5cm}
   \leavevmode
 \includegraphics{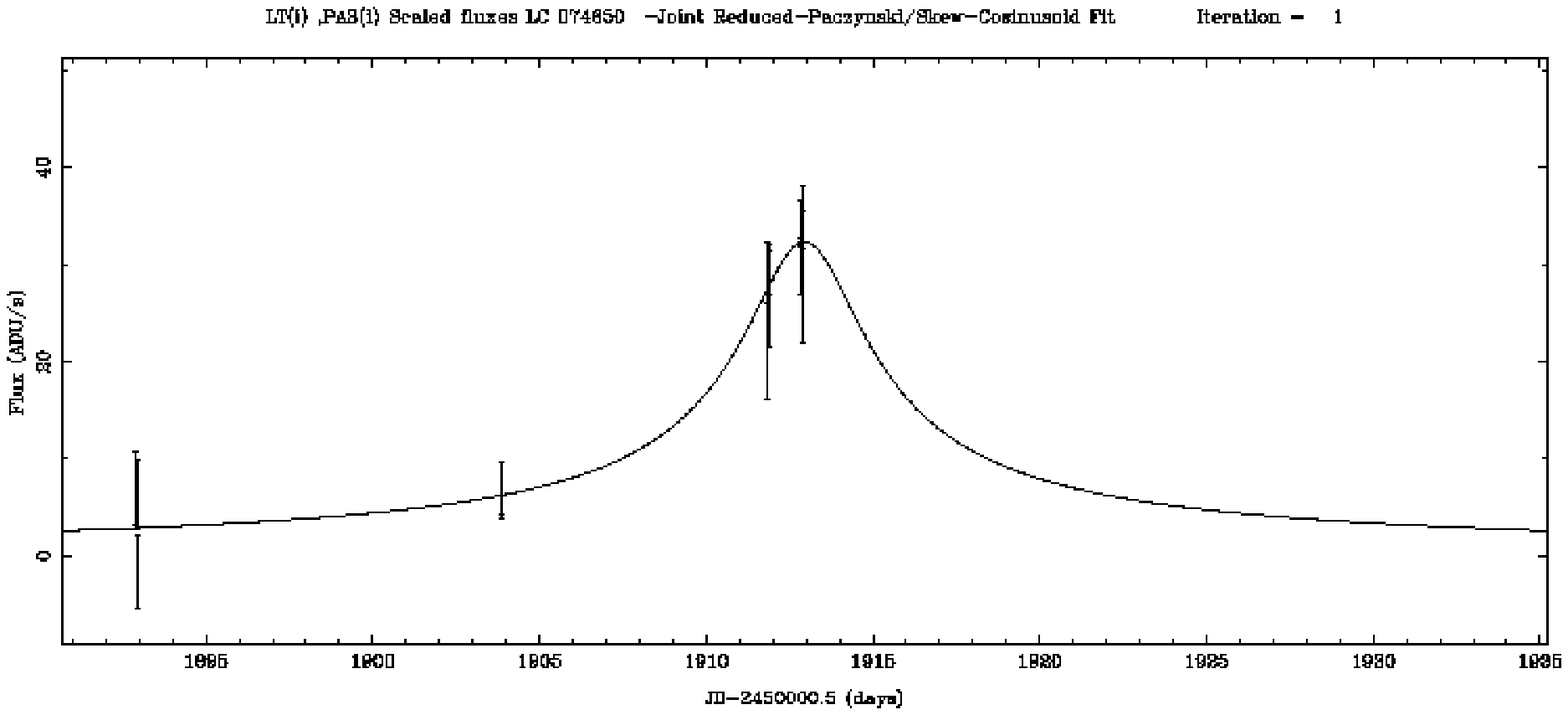} \\
  \leavevmode
 \includegraphics{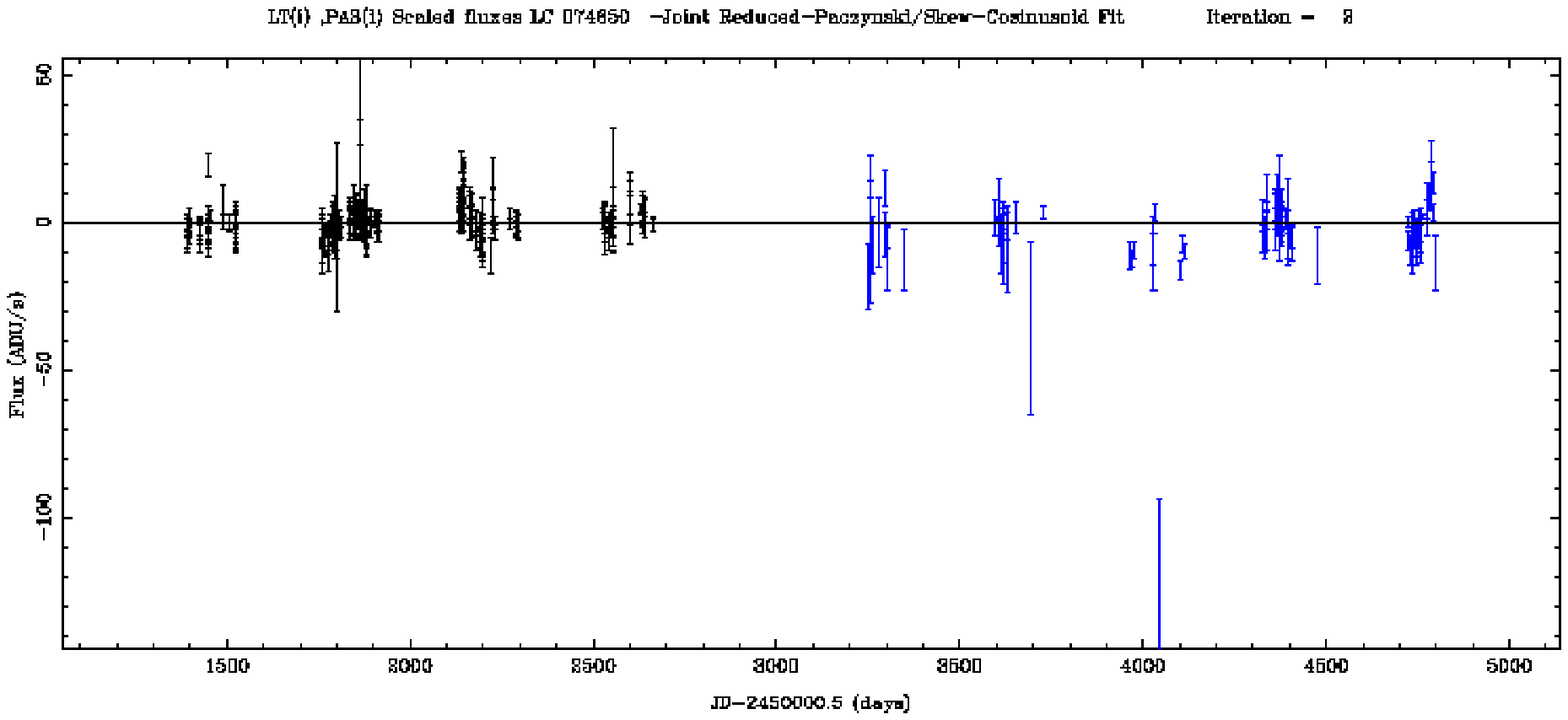} \\
\end{array}$
\caption[Lightcurve of object number $74650$ in the 2008 photometry, showing Top) the original mixed fit,
Middle) the peak region of the lensing component only, after the variable component has been subtracted, and Bottom) the residuals after subtraction of the mixed fit.]{Lightcurve of object number $74650$ in the 2008 photometry, showing Top) the original mixed fit and Middle) the peak region of the lensing component only, after the variable component has been subtracted and Bottom) the residuals after subtraction of the mixed fit.}
\label{2008_selection_mixed_LC_74650}
\end{figure}

\newpage

\begin{figure}[!ht]
\vspace*{7cm}
$\begin{array}{c}
\vspace*{7.5cm}
   \leavevmode
 \includegraphics{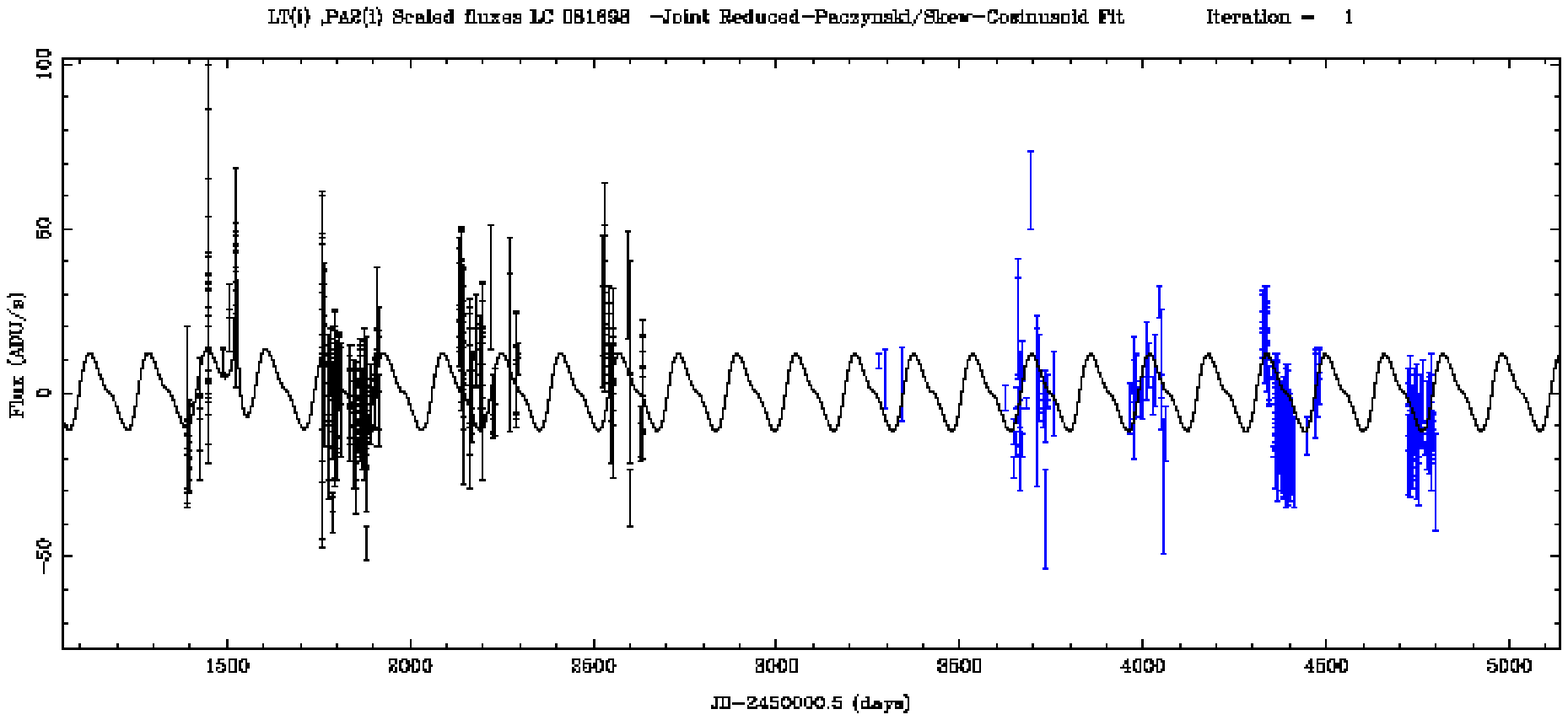} \\
\vspace*{7.5cm}
   \leavevmode
 \includegraphics{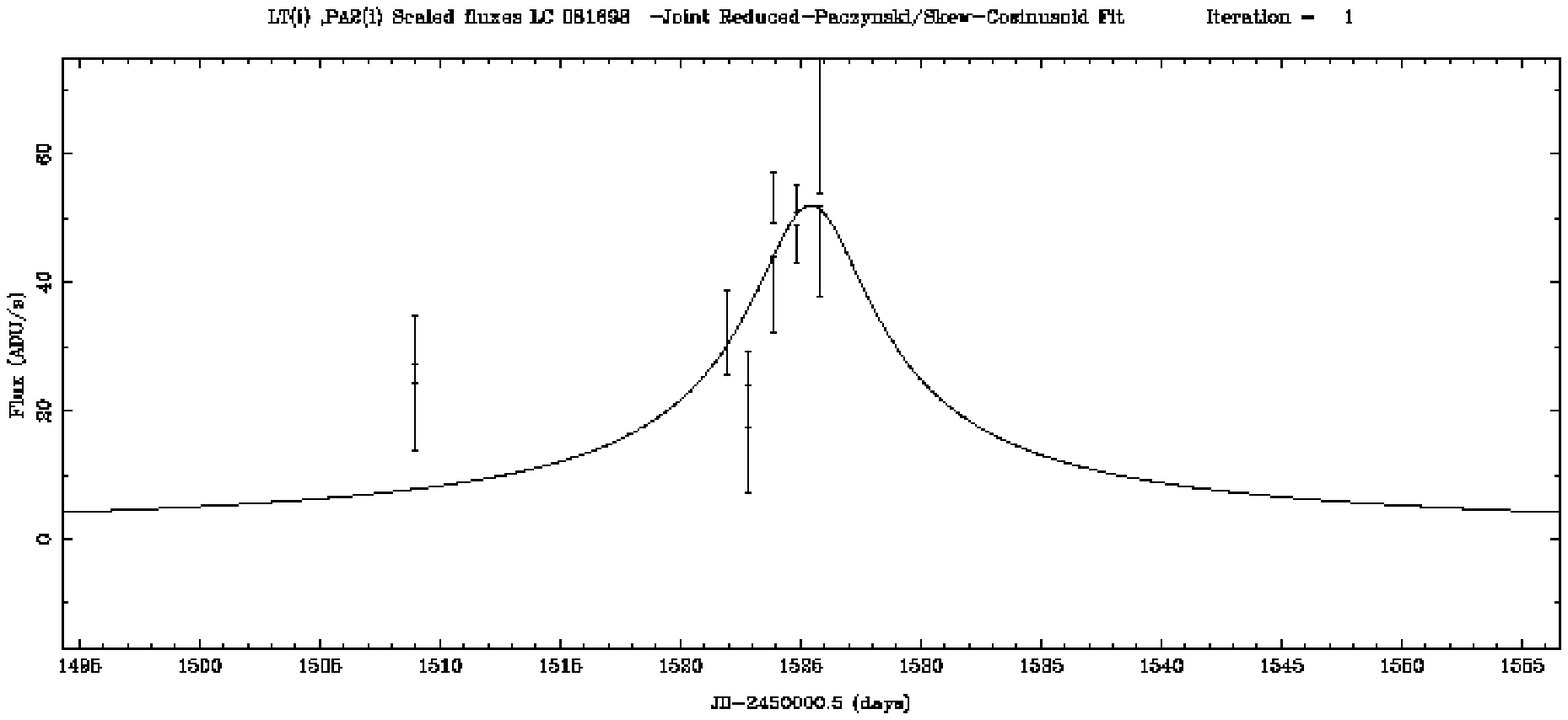} \\
  \leavevmode
 \includegraphics{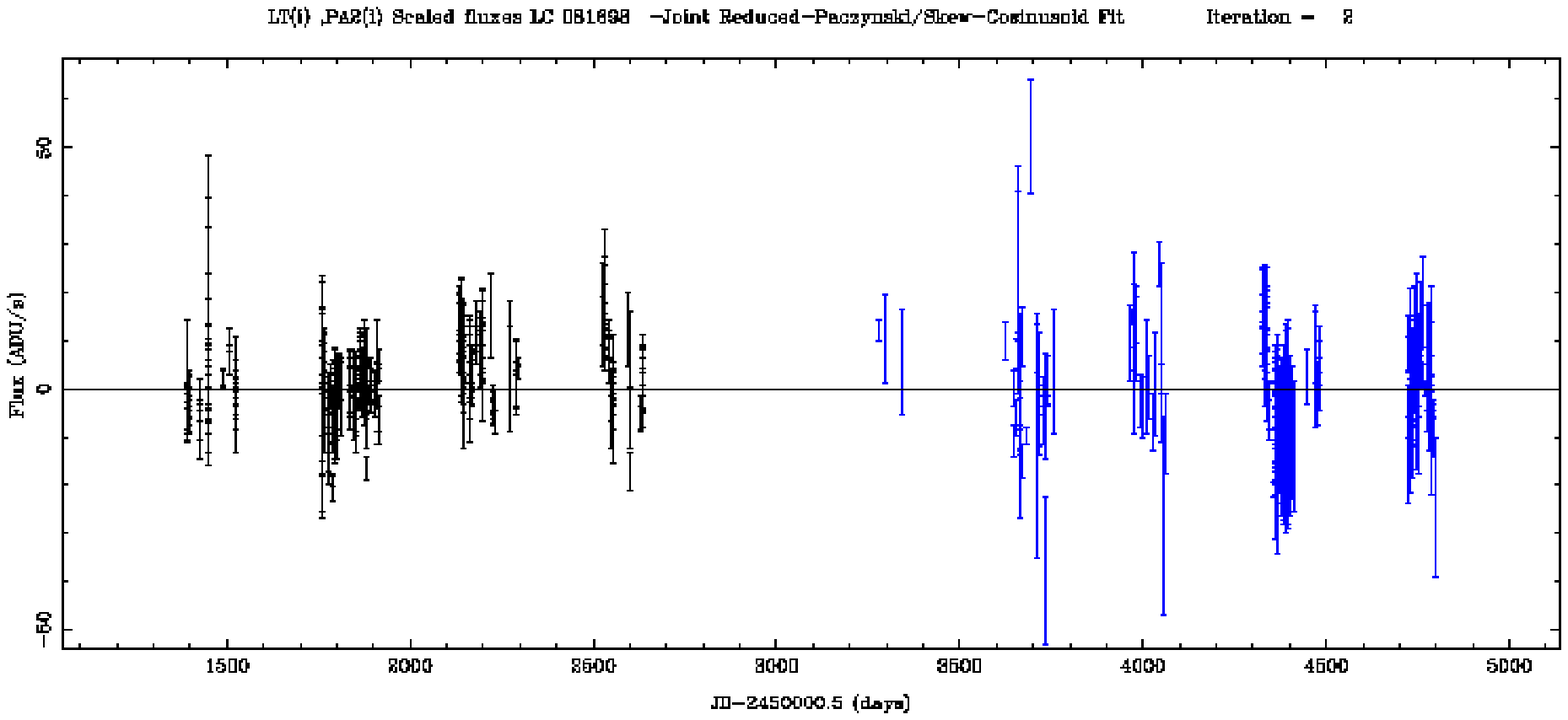} \\
\end{array}$
\caption[Lightcurve of object number $81698$ in the 2008 photometry, showing Top) the original mixed fit,
Middle) the peak region of the lensing component only, after the variable component has been subtracted, and Bottom) the residuals after subtraction of the mixed fit.]{Lightcurve of object number $81698$ in the 2008 photometry, showing Top) the original mixed fit and Middle) the peak region of the lensing component only, after the variable component has been subtracted and Bottom) the residuals after subtraction of the mixed fit.}
\label{2008_selection_mixed_LC_81698}
\end{figure}

\clearpage
\newpage
The candidate event lightcurve $39599$ is clearly still in progress when the $2008$ data end. This event has also been independently selected as an interesting transient by the Angstrom Alert System (AAS), was the subject of an alert issued by Angstrom via \cite{2008ATel.1857....1K}, and has been designated ANG-08B-M31-05. To confirm whether the continuation of the data is consistent with lensing, or is more consistent with being a nova, the data which have been collected since November $2008$ were added to the lightcurve.

\begin{figure}[!ht]
\vspace*{7cm}
$\begin{array}{c}
\vspace*{7.5cm}
   \leavevmode
 \includegraphics{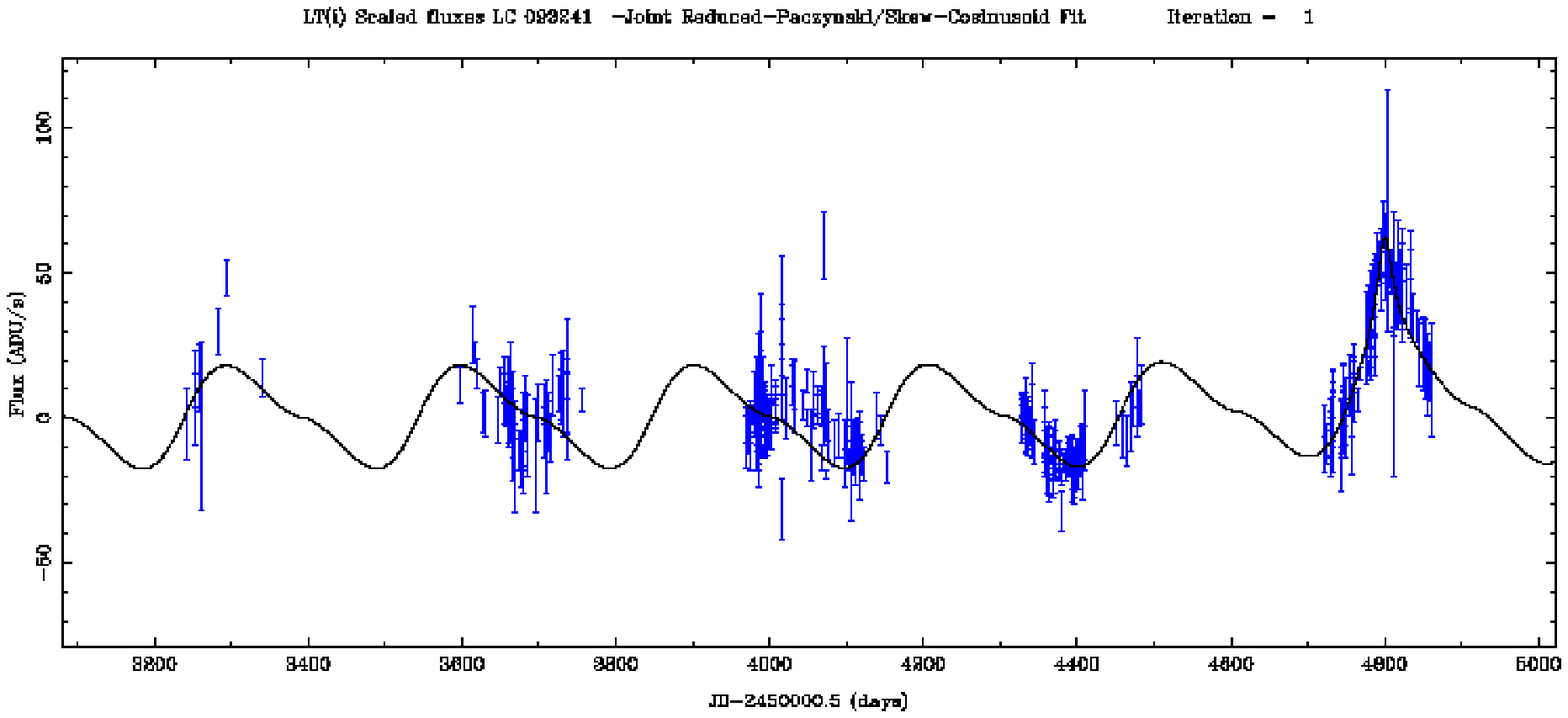} \\
\vspace*{0cm}
   \leavevmode
 \includegraphics{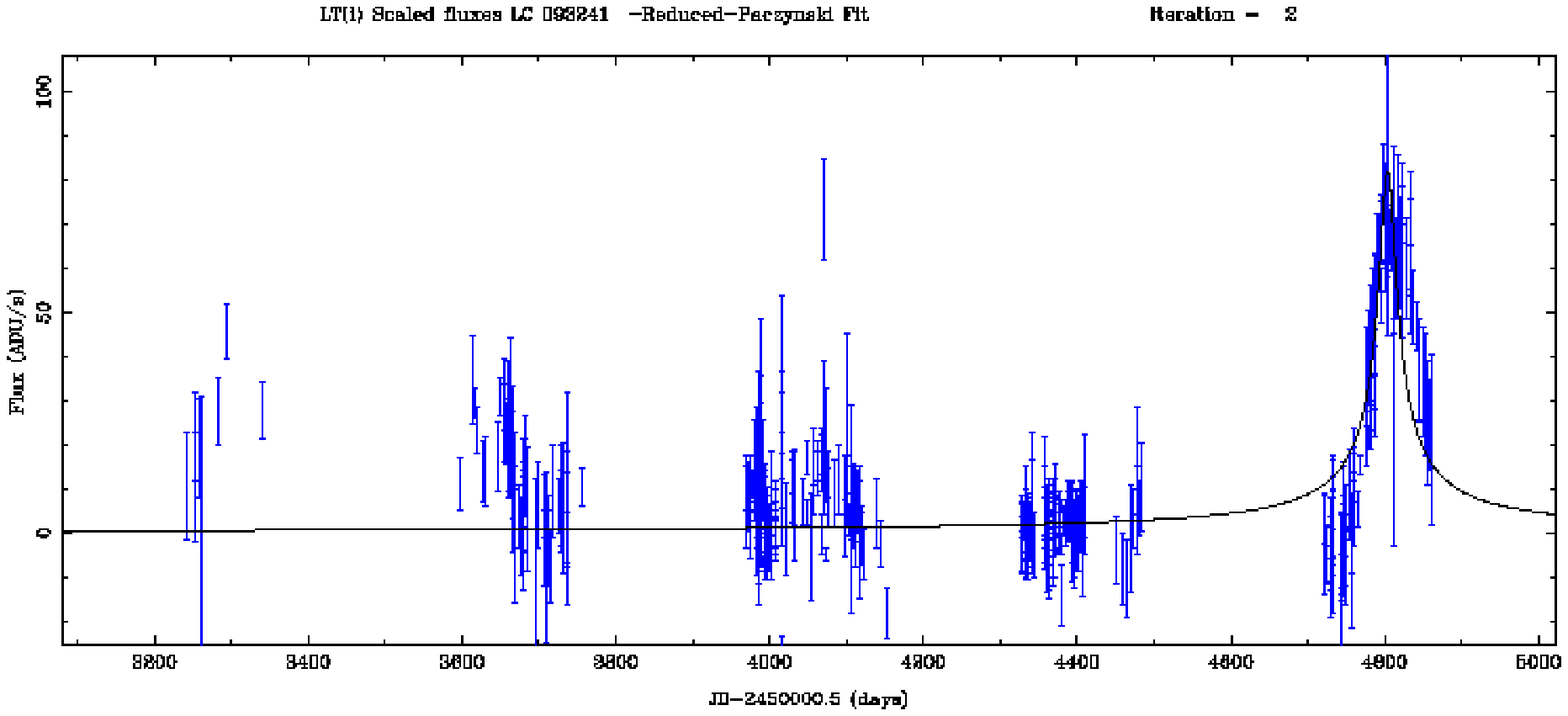} \\
\end{array}$
\caption[The extended lightcurve of object number $39599$ in the 2008 photometry, fitted in the top panel with a joint fit, and in the bottom panel with a reduced Paczy\'nski fit, after the variable part of the fit to Figure \ref{2008_selection_mixed_LC_39599} has been subtracted.]{The extended lightcurve of object number $39599$ in the 2008 photometry, fitted in the top panel with a joint fit, and in the bottom panel with a reduced Paczy\'nski fit, after the variable part of the fit to Figure \ref{2008_selection_mixed_LC_39599} has been subtracted.}
\label{2008_extended_39599_reanalysed}
\end{figure}

This plot shows the flux generally falling, as expected, but the cause of the spike is still ambiguous to the eye
when the variable background amplitude is taken into account.
Therefore, the selection pipeline was used to re-fit the extended lightcurve with a full mixed fit. A good fit to the lightcurve was still obtainable as measured by the local and global $\chi^2$/d.o.f., but in order to better fit to the peak area, the fit to the variable baseline was clearly degraded noticeably by using a less appropriate period. Also, the Lensing Ratio of the lensing event was reduced to a point where it was no longer selected according to Cut 8. Because of this, Cut 12 was not even tested, but it is suspected that this cut might also have failed due to the coherently degraded fit to the baseline.
 Since the original joint fit to the un-extended lightcurve of $39599$ had a 
better fit to the baseline, the variable part of the original joint fit was subtracted from the extended lightcurve, and then a reduced Paczy\'nski fit was attempted.
The result of this is given in the lower panel of Figure \ref{2008_extended_39599_reanalysed}, along with the mixed fit that the selection 
pipeline would normally have fitted (in the upper panel).
It can be seen in the lower panel of Figure \ref{2008_extended_39599_reanalysed} that although the baseline is substantially flatter than in the original lightcurve,
residual variations still remain, and in the peak area, the data do seem to undershoot a Paczy\'nski fit on the rise, and overshoot it on the fall, which might be consistent with a nova lightcurve. This is a fairly subtle effect, however, not much more than 1$\sigma$, and might be due to the residual variations in the baseline. It seems likely that it is these remaining ``wiggles'' in the peak that the
joint fit was attempting to reduce by sacrificing some of the baseline fit.
The fit to the Paczy\'nski model is still quite good, however.
 This lightcurve is presently judged to be a close call between a nova or 
microlensing, and at the time of submission is still an on-going event. Perhaps the 
further extension of the lightcurve with time will resolve the issue one way or the other. This event does illustrate some of the difficulties that can be caused when 
the baseline cannot be modelled simply by a one component model. Attempting to use a model that is too simple for the behaviour of the data can cause good events to be rejected. There may be a case for attempting to remove all variability in the baseline independently of the fitting of the proposed microlensing peak. This method would be computationally more difficult, however, and might introduce problems of its own, for example, in the case when the microlensing timescale was greater than the period of the variable.

\section{Testing the pipeline with fake lightcurves}
\label{montecarlo_testing}

A comprehensive Monte Carlo test of the pipeline, which would be necessary to calculate the pipeline efficiency, and hence allow a definitive estimation of the distribution of matter in our field of M31 would require relatively complex modelling of quantities such as spatial, mass, velocity and luminosity distributions, and by hence by implication the anticipated distributions of lensing timescales and amplitudes. To be a realistic model, it would also need to include variable stars, and so these would either need to be generated artificially, including their period, skewness, amplitude, and spatial distribution, or the fake lensing events seeded into the original variable object lightcurves.  
 The extent of this kind of exercise is beyond the possible scope of this work, but it is nonetheless felt that testing the pipeline in some sort of simplified way, using fake lightcurves, is a useful exercise. This testing is reported below. The results gained, due to their highly simplistic nature, cannot be taken as very quantitative, but do give a qualitative impression of the likely performance of the pipeline in semi-realistic situations. 
 
 \subsection{Selecting a baseline lightcurve}
\label{choosing_baseline}

\begin{figure}[!ht]
\vspace*{7cm}
   \leavevmode
 \includegraphics{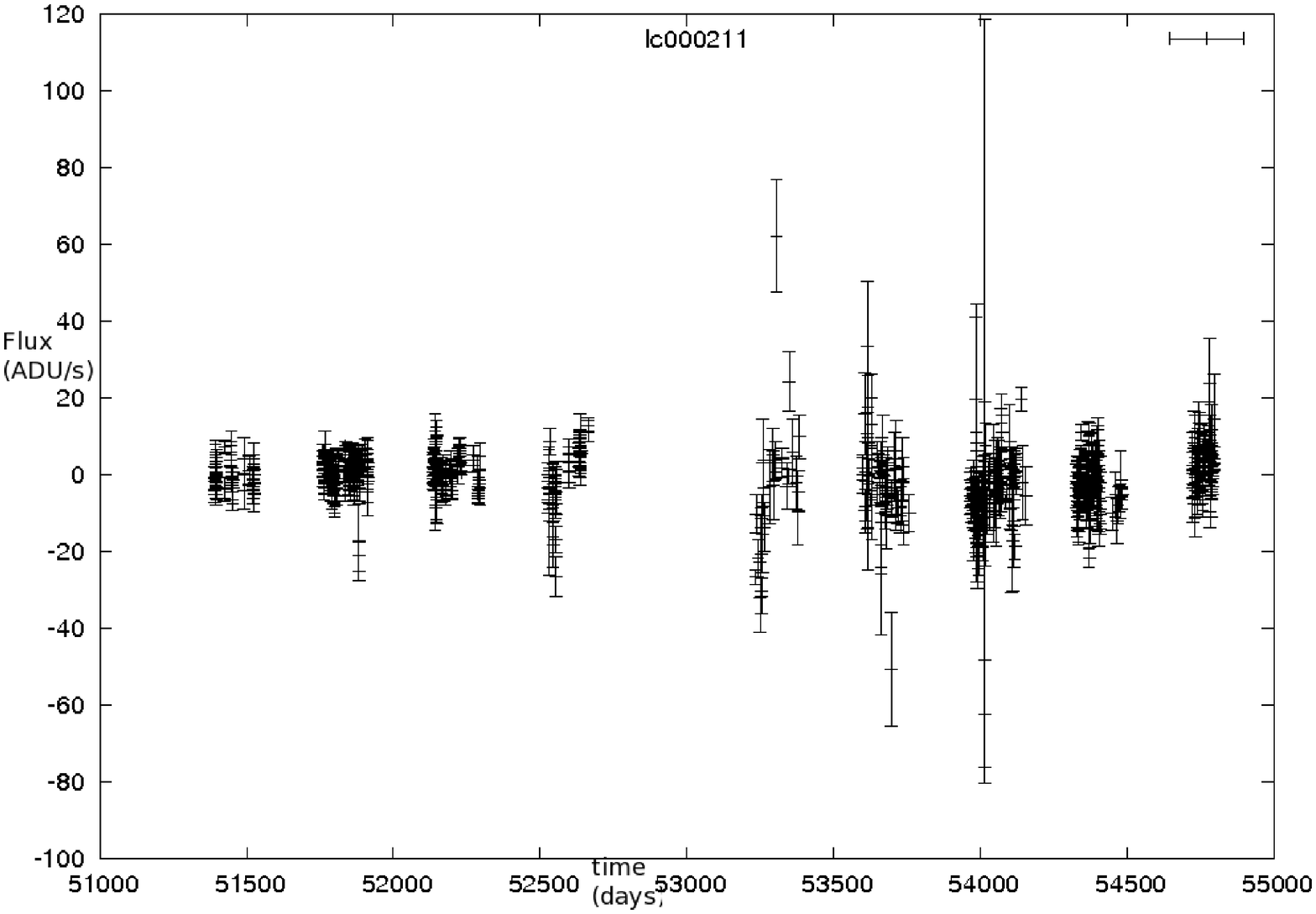} \\
\caption[Lightcurve of object number $211$ in the 2008 photometry]{Lightcurve of object number $211$ in the 2008 photometry, which is used as a baseline for the fake lightcurve test of the pipeline.}
\label{fake_baseline_LC_211}
\end{figure}

A totally flat baseline might have been chosen to add the simulated microlensing event peaks to, but it is felt that a more stringent and realistic test is to add them to a real lightcurve from the variable object database. However, to keep the test reasonably simple so that it would be expected that a
significant fraction of lightcurves would pass the test, the lightcurve selected for the baseline was chosen to be one which had previously been identified by the pipeline as a ``constant'' lightcurve. In other words, a constant fit satisfied the criterion for the $\chi^2$ of the fit. In addition, either no significant coherent bumps were found in the lightcurve, and thus no other fitting functions were fitted, or a bump or bumps were present, but the fitted $\chi^2$'s did not improve on the constant fit value.
In addition, it is desired that both LT and PA data exist, and that the number of data points in both LT and PA are close to their maximum possible values.
The first $500$ lightcurves were re-run using the current pipeline, (not fixing the fitted flux amplitude ratio of LT and PA data). Of these lightcurves, only $8$ were both classified as ``constant'' as described above and have both LT and PA data. Two of these lightcurves had global $\chi^2$ values which were almost equally the lowest of the $8$, but one has clearly more data points. This lightcurve, number $211$, is therefore chosen as the baseline for this exercise. It has a global $\chi^2$ value of $3.95$, (remembering that this should be divided by a factor of approximately $2$ to allow comparison with the expected $\chi^2$ value for a truly constant lightcurve, $\sim1$). The number of data points is $371$ in LT and $282$ in PA. The lightcurve of variable object 211 is shown in Figure \ref{fake_baseline_LC_211}. To allow easier plotting and insertion of fake events, both the flux and error values of the PA data set are divided by a scaling factor of $10$ which, as shown in Section \ref{variable_flux_ratios}, is approximately the average ratio between the PA and LT data. This process requires that the assumption be satisfied that the variations in PA flux are symmetrical on average around zero flux, which is confirmed by inspection of Figure \ref{fake_baseline_LC_211}. Inspection of this plot also shows that despite its classification by the pipeline (which may only signify that this lightcurve is not fitted very well by a simple skew-sinusoid model), lightcurve $211$ is clearly not very constant. Significant and coherent variations can be seen, especially in the final season of PA data and the first season of LT data. These are reminiscent of the kind of flux variations seen in semi-regular variable stars, but might also be the sum of the flux variations of several variable stars. Such a variable baseline was not quite the expected character of the lightcurve, but it is not thought that the variations are so extreme as to preclude its use as the baseline for inserting the fake events. 
The effect on the fake lightcurve test is to make it harder for the pipeline to find candidates.

\subsubsection{Inserting fake Paczy\'nski events}
\label{inserting_events}

After inspection of lightcurve $211$, and the observation that the variations in that lightcurve were generally of the order of $10$ ADU/s, it was decided to create fake events with Paczy\'nski peak fluxes ranging from $10$ ADU/s to $90$ ADU/s with specific peak flux values of ($10$, $30$, $50$, $70$, $90$) ADU/s. For reference the peak fitted flux of the short event is about $400$ ADU/s, so it is known that this range lies within the range of possible parameters. No additional Poisson noise is added due the extra flux. Based on the average lensing timescale in the Milky Way being of order of a month, lightcurves were created with lensing timescales $t_{\rm{FWHM}}$ = ($5$, $50$, $500$) days. The locations in time of the lightcurves were randomised in the 
range $51392.0 < t_{0} < 54796.0$ (JD-2400000.5). Ten lightcurves were generated for each set of fake event parameters in order to smooth out the random variations in event location a little. The total number of fake lightcurves generated, then, was $10\times3\times5 = 150$. 
It should be noted that this randomisation means that the expected pipeline inefficiency factor due to data only being taken for a fraction of the year has been naturally included in the overall efficiency. This ``data coverage factor'' was estimated for this particular lightcurve by working out the fraction of the above range covered by seasons of data. The fraction of the total time between the beginning and end of the data covered by data taking was found to be $0.384$. As a check, the fraction of created fake events where the $t_{0}$ value was within the ranges defined by the seasons of data was also calculated, and found to be $0.36$. These numbers are not strictly the exact efficiency factors due to the small irregular gaps in data coverage within each season, and because it is allowed and possible for the pipeline to find events with $t_{0}$ out of season as long as the specifications of Cut 6 (see Section \ref{bump_sampling}) are followed. Also, the above figure is an overestimate, as the total timespan used both starts and ends with a season of data, thus including one fewer ``off season'' than should be included. It is however consistent with the time range of the fake events themselves. It should be remembered that even with no ``noise'' on the baseline, the pipeline would not be able to find much more than the above fraction of events, depending on how many events of each timescale were included.

\subsubsection{Results of fake lightcurve fitting}
\label{fake_results}

The lightcurve numbers quoted in Tables~\ref{fake_lightcurve_original_params},~\ref{fake_lightcurve_fit_params_baseline},~\ref{fake_lightcurve_fit_params_t_0}, and \ref{fake_lightcurve_fit_params_t_FWHM} are a nominal numbering system starting after the number of the final ``real'' lightcurve, which was $93240$.

\begin{table}
\caption{Table giving the original parameters for the $13$ fake events successfully found by the candidate selection pipeline.}
\begin{center}
\begin{tabular}{|c|c|c|c|c|}
\hline
\hline
   Fake    &  $t_0$          & $t_{\rm{FWHM}}$  &  Peak Flux &   $t_0$    \\
Lightcurve & (JD-$2400000.5$)& (days)           &    (ADU/s) &  in-season?\\ 
    \hline   
    $93371$  &  $2232.92$    &       $50$       &    $30$    &   YES \\
   \hline  
    $93435$  &  $1915.26$    &       $50$       &    $70$    &   NO  \\
    $93438$  &  $1504.51$    &       $50$       &    $70$    &   YES \\
    \hline 
    $93443$  &  $3279.80$    &       $500$      &    $70$    &   YES \\
    $93444$  &  $4424.07$    &       $500$      &    $70$    &   YES \\
    $93447$  &  $1935.85$    &       $500$      &    $70$    &   NO  \\
    $93448$  &  $2116.21$    &       $500$      &    $70$    &   NO  \\
   \hline  
    $93463$  &  $1903.74$    &       $50$       &    $90$    &   YES \\
    $93468$  &  $1894.20$    &       $50$       &    $90$    &   YES \\
    \hline   
    $93471$  &  $3400.73$    &       $500$      &    $90$    &   NO  \\
    $93472$  &  $4644.21$    &       $500$      &    $90$    &   NO  \\
    $93475$  &  $4739.17$    &       $500$      &    $90$    &   YES \\    
    $93477$  &  $1874.75$    &       $500$      &    $90$    &   YES \\
\hline
\end{tabular}
\end{center}
\label{fake_lightcurve_original_params}
\end{table}

For each event found by the pipeline, it was calculated whether the \emph{original} $t_{0}$ falls inside or outside
the ranges of the data. These determinations are also recorded in the last column of Table \ref{fake_lightcurve_original_params}. 

\textbf{A benchmark figure for comparison}

 In order to put the above figures into some kind of context and shed some light on the causes of
 the overall efficiency figure in the above experiment, a simpler test is also conducted to provide a
 baseline figure with which the number above may be compared. An identical $150$ fake lensing events as before
 are generated, except in this experiment the baseline of the lightcurve is not a realistic variable lightcurve but is completely flat. To provide some statistical realism for the fitting routine, a ``jitter'' is added to the flux points corresponding to alternately raising and lowering the Paczy\'nski peak flux values by an amount equal to the size of the actual one $\sigma$ error bar (of the realistic data of lightcurve $211$).
 
  When this test is carried out, $31$ of the $150$ lightcurves are passed by the selection pipeline as having a detectable event in, of which $5$ are classed as ``Paczy\'nski'' events rather than ``Mixed'' events. Examination of the fitted lightcurves shows that the correct peaks were fitted in those events detected.
  Two of the $5$ ``Paczy\'nski'' fits have timescales of $5$ days, which are the only fake events recovered with the shortest of the three timescales, and the other three are $50$ days.
  As with the realistic baseline data set, no events were detected with peak fluxes below $30$ ADU/s.
  The above detection rate represents a percentage of $20.7\%$, which is comparable to the $\sim38\%$ data coverage fraction. There is no reason to expect a precise correspondence to this figure, as the actual number found will depend strongly on both the timescales and peak fluxes chosen, which are clearly arbitrarily chosen in this case. By comparing the results of this test with the test using lightcurve $211$ as the baseline, the effect of the variability of the realistic lightcurve used in the above test can be separated out. The deterioration in detection efficiency between these experiments caused by using a ``realistically varying'' baseline lightcurve is therefore $12\%$, corresponding to multiplying the ``flat baseline'' efficiency by a factor $0.42$. Therefore the number of detected events is reduced by $58\%$ by using this particular ``realistic'' baseline.
  
  Table \ref{comparing_baselines} gives the comparison of the statistics of detected events between the two baselines used.
  Given that there is an almost whole season more of LT data than PA, the fact that the percentage of events found within the LT data for the LC$211$ baseline is less than $50\%$, whereas using the flat baseline it is much more proportional to the relative 
  quantities of data may imply that it is harder for the pipeline to find events in the LT due to the statistical noise being greater. This seems to be consistent with the events found in the totality of lightcurves, where fewer events were also found in the LT than PA data. The percentage of events which are found with $t_0$ within a season of data is higher in both cases than the fraction of the total timespan taken up by the seasons of data themselves. This is expected, as it is much less likely
  for the specifications of Cut 6 to be satisfied if $t_0$ is out of season. The preponderance of longer timescale events found may also be explained by their relatively more even data coverage compared to their timescale, and also that there will on average be greater numbers of data points within the peak to aid the fitting routine in finding it.  
  
  \begin{table}
\caption{Table giving the numbers of events found and some statistics for comparison between the two baselines used, for the events successfully found by the candidate selection pipeline, out of the original $150$ fake events.}
\begin{center}
\begin{tabular}{|c|c|c|c|c|c|}
\hline
\hline
Baseline  & Number & Fraction &  Number found with   & Fraction IN     & Fraction     \\
  used    & found  & of total   & $t_{\rm{FWHM}}$ ($5$,$50$,$500$) days   &  data seasons         & IN LT data   \\
\hline
``Flat''  &  $31$  & $20.7\%$   &    ($2$,$11$,$18$)                &  $61.5\%$             & $58.1\%$  \\
``LC 211''&  $13$  & $8.7\%$    &    ($0$,$5$,$8$)                  &  $67.7\%$             & $38.4\%$   \\ 
  \hline 
 \hline
\end{tabular}
\end{center}
\label{comparing_baselines}
\end{table}

\textbf{Recovery of fake event microlensing parameters}

The fitted parameters of the $13$ events passed by the candidate selection pipeline were investigated to see how well the
original parameters had been recovered. The fitted parameters and the comparison to the original quantities are summarised in 
Tables \ref{fake_lightcurve_fit_params_baseline}, \ref{fake_lightcurve_fit_params_t_0}, \ref{fake_lightcurve_fit_params_t_FWHM} and \ref{fake_lightcurve_fit_params_F_peak} for the parameters $B_{\rm{LT}}$, $t_{0}$, $t_{\rm{FWHM}}$ and $\Delta{F}_{\rm{LT}}$ respectively.
On the whole, most of the original parameters are recovered quite well.
The baseline flux parameter, which is set equal to zero for the fake lightcurves is generally found to be of the order of $2$ ADU/s, compared to the magnitudes of the input flux peaks which vary from $30$ to $90$ ADU/s. The average percentage  magnitude of error in the fitted baselines, compared to the $\Delta{F}$ values, is therefore $3.0\%$. The fitted values of $t_0$ are also well recovered, being on average only $3.3\%$ of the $t_{\rm{FWHM}}$ values in error.
The peak flux values are also recovered well, with the mean error being $6.4\%$. The worst recovered of the four fitted microlensing parameters are the values of $t_{\rm{FWHM}}$. The original values of this parameter are only recovered to within $21\%$ on average. A clear tendency can also be seen among the fitted values of $t_{\rm{FWHM}}$ to underestimate the original parameters. $11$ out of $13$ events have $t_{\rm{FWHM}}$ less than the original value. One possible theory to explain this is that sometimes a better fit can be obtained to a peak in the data by combining the peak of the variable part with the Paczy\'nski part of a mixed fit, which means that some of the ``work'' is done by the variable part, leaving an inaccurate estimate of the Paczy\'nski parameters. However, given the clear variations in the lightcurve, a worst parameter error of $21\%$ and all others in single figures is not considered too bad.
Another important point to consider is that the pipeline is tuned quite heavily towards only allowing through more reliable candidates, even at the expense of rejecting events that may be quite impressive, but have an element of doubt attached to them for one reason or another. Therefore, it is good to observe that no ``false positives'' were found, even when the  Paczy\'nski peak amplitude was as low as $10$ ADU/s. It must be remembered, however, that there may be a bias against this inherent in the method used, as the baseline lightcurve was specifically selected as one that had previously been found not to have any flux deviations that were consistent with a microlensing explanation and significant enough to be found by the pipeline.

\begin{table}
\caption{Table comparing the original and fitted baseline fluxes for the $13$ fake events successfully found by the candidate selection pipeline.}
\begin{center}
\begin{tabular}{|c|c|c|c|c|c|}
\hline
\hline
 Fake        &   \footnotesize{Baseline Flux} &  Baseline   &   \footnotesize{group mean }          &   \footnotesize{group mean }                          &  \footnotesize{group mean}  \\
Lightcurve   &  \footnotesize{(original)}     &  Flux (fitted)&  \footnotesize{of $\Delta$ B.Flux}  & \footnotesize{ of $\lvert{\Delta \rm{B.Flux}}\rvert$} &  \footnotesize{of $\frac{\lvert{\Delta \rm{B.Flux}}\rvert}{\Delta{F}}(\%)$} \\
\hline
    $93371$  &   $0$   &    $-1.3\pm0.2$     &    $-1.32$    &    $1.32$  & $4.4$ \\
   \hline  
    $93435$  &   $0$   &    $-1.9\pm0.2$     &    $-0.54$  &  $1.31$ & $1.9$ \\
    $93438$  &   $0$   &    $0.8\pm0.3$       &           &  &   \\ 
    \hline 
    $93443$  &   $0$   &    $5.2\pm0.8$      &         &  &  \\
    $93444$  &   $0$   &    $-3.8\pm0.6$     &    $+0.98$  &  $2.88$ & $4.1$ \\
    $93447$  &   $0$   &    $1.2\pm0.4$      &         &  &  \\
    $93448$  &   $0$   &    $1.3\pm0.4$      &         &  &  \\
   \hline  
    $93463$  &   $0$   &    $-2.0\pm0.2$     &    $-2.06$  &  $2.06$ & $2.3$ \\
    $93468$  &   $0$   &    $-2.1\pm0.2$     &         &  &  \\
    \hline   
    $93471$  &   $0$   &    $3.2\pm0.2$      &      &  &  \\
    $93472$  &   $0$   &    $-3.6\pm0.3$     &    $0.41$  &  $2.21$ &  $2.5$\\ 
    $93475$  &   $0$   &    $1.1\pm0.2$      &        &  &  \\  
    $93477$  &   $0$   &    $0.9\pm0.5$      &        &  &  \\
   \hline
             &       &  \footnotesize{\rm{overall means}}  &  $+0.074$ &  $2.18$ &  $3.0$\\
 \hline
\end{tabular}
\end{center}
\label{fake_lightcurve_fit_params_baseline}
\end{table}

\begin{table}
\caption{Table comparing the original and fitted $t_0$ parameters for the $13$ fake events successfully found by the candidate selection pipeline.}
\begin{center}
\begin{tabular}{|c|c|c|c|c|}
\hline
\hline
 Fake          &$t_0$ (original) &   $t_0$     &   $\frac{\rm{fitted}-\rm{original}}{t_{\rm{FWHM}}}$ & \footnotesize{ group mean of}\\
 Lightcurve    & \footnotesize{(JD-$2400000.5$)}&  (fitted)   &                                            &  \footnotesize{$\lvert{\mbox{fractional ratio}}\rvert$ ($\%$)} \\
    \hline   
    $93371$  &  $2232.92$    &    $2228.3\pm0.3$     &    $-0.093$    &    $9.3$    \\
   \hline  
    $93435$  &  $1915.26$    &    $1910.8\pm0.7$     &   $-0.089$     &   $6.8$    \\   
    $93438$  &  $1504.51$    &    $1506.9\pm0.8$     &   $+0.048$     &       \\   
    \hline 
    $93443$  &  $3279.80$    &    $3325\pm4$     &   $+0.090$     &        \\
    $93444$  &  $4424.07$    &    $4428\pm4$     &   $+0.0071$     &   $2.5$     \\
    $93447$  &  $1935.85$    &    $1934\pm2$     &   $-0.0029$     &        \\ 
    $93448$  &  $2116.21$    &    $2116\pm2$     &   $-0.0012$     &       \\  
   \hline  
    $93463$  &  $1903.74$    &    $1903.0\pm0.3$     &   $-0.015$     &   $1.0$    \\   
    $93468$  &  $1894.20$    &    $1894.5\pm0.2$     &   $+0.0054$     &        \\  
    \hline   
    $93471$  &  $3400.73$    &    $3416\pm3$     &   $+0.031$     &        \\  
    $93472$  &  $4644.21$    &    $4660\pm3$     &   $+0.031$     &   $1.9$     \\     
    $93475$  &  $4739.17$    &    $4745\pm2$     &   $+0.011$     &       \\     
    $93477$  &  $1874.75$    &    $1872\pm1$     &   $-0.0049$     &       \\ 
       \hline
             &               &         &  \footnotesize{\rm{overall mean ($\%$)}}  & $3.3$ \\     
\hline
\end{tabular}
\end{center}
\label{fake_lightcurve_fit_params_t_0}
\end{table}

\begin{table}
\caption{Table comparing the original and fitted $t_{\rm{FWHM}}$ parameters for the $13$ fake events successfully found by the candidate selection pipeline.}
\begin{center}
\begin{tabular}{|c|c|c|c|c|}
\hline
\hline
 Fake       &  $t_{\rm{FWHM}}$ & $t_{\rm{FWHM}}$   &  fract. change              &  \footnotesize{group mean of} \\
 Lightcurve &  (orig.) (days) & (fitted) (days)   & $\frac{\rm{fitted}-\rm{original}}{\rm{original}}$ &  \footnotesize{$\lvert{\rm{fract. change}}\rvert$ ($\%$)}\\
 \hline
    $93371$  &     $50$       &  $26\pm2$   &  $-0.49$ &   $49$   \\
   \hline  
    $93435$  &     $50$       &  $43\pm1$   &  $-0.15$ &   $15$   \\
    $93438$  &     $50$       &  $42\pm3$   &  $-0.16$ &      \\
    \hline 
    $93443$  &     $500$      &  $240\pm10$   &  $-0.52$ &      \\
    $93444$  &     $500$      &  $567\pm3$   &  $+0.13$ &   $23$   \\
    $93447$  &     $500$      &  $430\pm10$   &  $-0.14$ &      \\
    $93448$  &     $500$      &  $440\pm10$   &  $-0.11$ &      \\
   \hline 
    $93463$  &     $50$       &  $44\pm1$   &  $-0.11$ &   $9.6$   \\
    $93468$  &     $50$       &  $46.1\pm0.9$   &  $-0.078$ &      \\
    \hline   
    $93471$  &     $500$      &  $290\pm20$   &  $-0.43$ &     \\
    $93472$  &     $500$      &  $540\pm20$   &  $+0.082$ &   $22$   \\
    $93475$  &     $500$      &  $367\pm6$   &  $-0.27$ &      \\    
    $93477$  &     $500$      &  $4453\pm8$   &  $-0.094$ &      \\
    \hline
             &               &      &  \footnotesize{\rm{overall mean ($\%$)}}  &  $21$ \\
\hline
\end{tabular}
\end{center}
\label{fake_lightcurve_fit_params_t_FWHM}
\end{table}

\begin{table}
\caption{Table comparing the original and fitted Peak Flux parameters for the $13$ fake events successfully found by the candidate selection pipeline.}
\begin{center}
\begin{tabular}{|c|c|c|c|c|}
\hline
\hline
Fake           & \footnotesize{ Peak Flux}          & Peak Flux           & fract. change                 &   \footnotesize{group mean }\\
Lightcurve     &  \footnotesize{(orig.) (ADU/s)}  &  (fitted) (ADU/s)   &  $\frac{\rm{fitted}-\rm{original}}{\rm{original}}$  &   \footnotesize{$\lvert{\rm{fract. change}}\rvert$ ($\%$)}\\
           \hline   
    $93371$  &     $30$    & $31.8\pm0.8$  & $+0.060$  & $6.0$ \\
   \hline  
    $93435$  &     $70$    & $69\pm1$  & $-0.013$  & $2.4$ \\
    $93438$  &     $70$    & $68\pm3$  & $-0.034$  &  \\
    \hline 
    $93443$  &     $70$    & $66\pm2$  & $-0.059$  &  \\
    $93444$  &     $70$    & $72.2\pm0.5$  & $+0.031$  &  $6.1$\\
    $93447$  &     $70$    & $65\pm1$  & $-0.071$  &  \\
    $93448$  &     $70$    & $64.0\pm0.9$  & $-0.085$  &  \\
   \hline  
    $93463$  &     $90$    & $95\pm1$  & $+0.058$  &  $5.4$\\
    $93468$  &     $90$    & $95\pm1$  & $+0.050$  &  \\
    \hline   
    $93471$  &     $90$    & $106\pm4$  & $+0.18$  &  \\
    $93472$  &     $90$    & $95\pm2$  & $+0.054$  &  $9.3$\\
    $93475$  &     $90$    & $96.5\pm0.6$  & $+0.072$  &  \\    
    $93477$  &     $90$    & $84.3\pm0.8$  & $-0.063$  &  \\
  \hline
             &             &     &  \footnotesize{\rm{overall mean ($\%$)}}  &  $6.4$ \\
\hline
\end{tabular}
\end{center}
\label{fake_lightcurve_fit_params_F_peak}
\end{table}

\chapter{Discussion and forward look}
\label{Chapter_6}
\section{Introduction}

In this Chapter, the main results from this Thesis are
summarised along with the main conclusions.
A brief perspective on the possible future development
of this work, and microlensing investigations in general,
is given.

\section{Summary of Thesis}

 \subsection{The Project}

    The Angstrom Project is an international collaboration which has gathered four 
seasons of data (plus a pilot season), mainly using the robotic telescopes the Liverpool Telescope and Faulkes North telescope,
 together with the BOAO $1.8$m and Maidanak $1.5$m telescopes from a survey of the central bulge region of the
 Andromeda galaxy, M31. Difference imaging has been applied to these data, using a
 modified version of the ISIS code \citep{1998ApJ...503..325A}, and variable sources identified. The
 resulting variable object lightcurves, augmented by earlier data from the POINT-AGAPE
 microlensing survey (re-analysed by us), have been
filtered using an automated candidate selection program which divides the lightcurves into microlensing candidates, variable star candidates, or no significant variation, and also classifies candidates according to signal to noise parameters. The lightcurves are also subjected to real-time analysis by the Angstrom Project Alert System (APAS) \citep{2007ApJ...661L..45D}, which is designed to detect fast-changing transients, with the aim of allowing immediate follow-up by other telescopes while the event is still in progress.

\subsection{Summary of Thesis}

   Early on in the study period, the common overlap region of image data from our pilot season was analysed, at a time when, due to a telescope fault, LT images had random
 rotations and translations. This study discovered that the area representing the overlap of all
 images was less than $50\%$ of the area of the Angstrom field. This fact informed a 
decision that it would be worthwhile altering the ``ISIS'' \citep{1998ApJ...503..325A} DIA code into what became ``AngstromISIS''.

By an analysis of comparable small regions on images from the two telescopes it was found that the flux ratio between LT and FTN in i'-band data was $1.081$ and in r band data was $0.305$. These numbers were used in the candidate selection routine to link together the fitted amplitudes of variable stars
in order to reduce the number of fitted parameters.

 The variable star candidates produced by the selection pipeline which had both PA 
 and LT data were analysed to investigate
 whether there was a constant or other stable relationship between the amplitudes of
 variability measured in the data from the two telescopes. No clear simple constant
ratio between the flux amplitude of variability in PA and LT or FTN data was found,
although there does appear to be limited range of possible values,
as shown in Figure \ref{mean_flux_ratio_with_fits}.
This investigation was only performed on the data extant in $2007$, so in the future it is intended to re-investigate the flux ratio issue as part of a general reanalysis of the variable stars in the data, using the most up to date lightcurves.
If a general rule could be found for the relationship between the flux amplitude ratio and the flux amplitude, then this would enable the use of fewer parameters in the fitting in the selection pipeline, which in turn would increase the speed and efficiency of the program.

    Throughout the duration of this work, the so-called 
``short event'' ANG-06B-M31-01 was analysed and then re-analysed several times as data processing, particularly DIA, improved and new data added to the lightcurves. This analysis was particularly aided by data from the Maidanak telescope which fortuitously covered the span of the main flux spike.
  An attempt was made at estimating the likely mass of the lens star for this event 
(assuming it was caused by microlensing). Due to our lack of knowledge of the type and hence initial flux of the
source star and of the relative source-lens velocity, and the small number of data 
points covering the flux peak, which is in turn due to the short duration of the event, an accurate estimate was not possible. However, the results indicate that the most likely type of star for the lens is a giant star.

The most time-consuming aspect of all the work performed was the writing, developing and testing of an automatic candidate selection pipeline, and the application of this pipeline to the selection of both variable star and microlensing candidates, both after the third full season $2007$ of Angstrom data and again mid-way through the fourth season, $2008$.
After the $2007$ selection pipeline run, the results were further analysed to investigate the spatial, variable period and skewness distributions of the variable star candidates.

A brief investigation was also made of the properties of the candidates selected from the $2008$ lightcurve data, but, due to their low number, few firm conclusions may yet be drawn from this. Further investigation using Monte Carlo modelling techniques will be required to illuminate the results and allow more definite conclusions to be drawn, for example about the distribution of matter in the M31 bulge.

\subsection{Conclusions}

\subsubsection{ANG-06B-M31-01}

 Initially, when the event was first detected, the fit to the main flux spike was not
sufficiently convincing to definitively classify this event as due to microlensing.  With the latest data, it now appears that the fit to a lensing lightcurve is much improved and hence the event is almost certain to be microlensing. The profile does not fit any other known phenomena and is a reasonable fit to a lensing lightcurve although the $\chi^2$/d.o.f. of the fit, either for the whole lightcurve, or more particularly in the peak region, is not as good as would ideally be desired. Some possible reasons for this are described in Section \ref{rediscovery_in_2008}.

The event was modelled using a combined reduced Paczy\'nski fit plus a variable
component.
The main parameters of the fit to this event are approximately summarised
below:

$t_{\rm{FWHM}}$:  $2.1\pm0.8$ days, taking into account the fitting using the Maidanak data, or $1.36 \pm0.1$ days using only the most recent photometry.
If confirmed, this timescale would be the shortest known microlensing event in M31.

 $t_0$ (JD-2400000.5) $= 53993.43\pm0.01$

Period of variable: $P = 245.15\pm0.14$ days

Phase of variable (JD-2400000.5) : $\phi_{0} = 54131\pm1$

The variable was almost exactly described by a perfect sinusoidal function. 

peak i-band magnitude estimate: $i = +18.07\pm0.06$

central estimate of $\beta = 0.015\pm0.015$

estimate of $t_E$ using the above value of $\beta$: $t_E = 41\pm{+44}$ days.

During the fitting process, when data from the Maidanak telescope were added, the best fit to a lensing lightcurve was obtained when the flux scaling factor between Maidanak and LT data was found to be equal to $F_{\rm{Maid}}/F_{\rm{LT}} = 1.11$. This is the first estimate (although still very approximate) that has been made of the scaling between the fluxes from these two telescopes.

In the mass estimate described above, the most likely combination of lens and source locations was assumed to be both in the bulge. Taking this estimate as a guide, the most likely central mass for the lens star is $7.5\pm{21} M_{\odot}$. This is formally consistent with zero, but the mass cannot be less than zero. So to first order, we can only conclude that 
the lens mass is likely to be less than $28.5 M_{\odot}$
and the mass of the source star is very likely to be $M_{\odot} \geq 1.44M_{\odot}$, which is a limit derived by working backwards from
the calculated peak flux magnitude and maximum possible magnification.

Other brief duration lensing candidates have been found in M31 (for example, \cite{2001ApJ...553L.137A}), but ANG-06B-M31-01 is certainly one of the shortest duration events so far seen in M31, and may be the shortest.

\subsection{Lensing candidates}

The $20$ lensing candidates selected by the candidate selection pipeline 
and presented in Section \ref{lensing_candidates_2008} clearly span a range of
``qualities'', although they all pass all twelve cuts. On the whole, although some of the lightcurves do appear noisy,
it is felt that they all have been correctly identified as containing a clear
flux anomaly which is consistent with microlensing. Whether these lightcurves actually were caused by microlensing events will probably never be known. Confirmation of whether a lensing event actually occurred, after the event, can only really currently be achieved in M31 through observations with the HST, as was done, for example, by \cite{2001ApJ...553L.137A}.
The spatial distribution of the lensing candidates is presently inconclusive. It is presented in Figure \ref{20_candidates_spatial_dist} as a matter of record. Of more interest, perhaps, at this stage is the histogram of lensing candidate timescales which does already seem to contain a certain amount of useful information. In the Milky Way, a typical lensing event is expected to last $\sim30$ days, but in this survey, half the events lie at $t_{\rm{FWHM}}< 15$ days. The Angstrom Survey has been specifically designed to have a high cadence of data collecting with a view to being better able to find short events exactly within this time range. It should also be noted that the lowest timescale found in the main sample (not including the ``short event'') was $~3$ days.
 Discovering the significance of the distribution of timescales found so far will
 have to wait until Monte Carlo modelling is performed using the same candidate
 selection pipeline. When this is complete, we will know better whether the
 timescale distribution presented here in Table \ref{timescale_dist} correctly
 reflects the actual distribution of event timescales or whether some aspects are
 affected by selection or systematic effects cause by the pipeline and/or the way 
the data have been collected or analysed.
The fact that all microlensing collaborations seem to design their own systems of cuts to perform selection, no two of which are identical, makes the process of comparison of results much harder. I feel that some kind of standardisation of methods would be very useful.
The way that the Angstrom candidate selection pipeline has been designed, to systematically
include variable baselines from the start, is a novel method. Undoubtedly this has also made filtering out ``inappropriate'' lightcurves and fits more complicated, and the process of experimentation with improving the way the cuts work should not be seen as a completed, but it does seem from the experience gained so far that the goal is an achievable one, though difficult.

\subsubsection{Testing of candidate selection pipeline using fake lightcurves}

A simple test of the candidate selection pipeline was carried out, searching for a variety of fake Paczy\'nski curves, having a range of peak amplitudes, timescales and randomised $t_{0}$'s, inserted both into a basically flat lightcurve and into one of the ``real'' variable object lightcurves (lightcurve $211$). This ``realistic'' baseline was selected as neither having a previously identified lensing candidate within it, nor a large overall variation from a flat lightcurve. The efficiency of the pipeline for recovery of the inserted events was $20.7\%$ using a flat baseline and $8.7\%$ using the lightcurve $211$ baseline. These figures should not be interpreted as the true detection efficiencies due to the unrealistic parameter distributions used in this test, but they do at least show that a significant percentage of real events are recovered, even in the case of a noisy baseline.
The original parameters of the inserted lensing peaks which were passed by the selection process were recovered within a few percent, except in the case of the lensing timescale using the lightcurve $211$ baseline, which was only recovered within about $20\%$ on average. 
Perhaps significantly, of the fake lightcurves, only two with timescales of $5$ days were recovered, and none were recovered when using the lightcurve $211$ baseline. This confirms that our completeness to these short events is still low, despite our aim of making our survey sensitive to shorter events. When the real lightcurves were searched, half the events found had timescales $<15$ days, and, if this group is subdivided into $5$ day intervals, the interval $15-10$ days has $4$ events, $10-5$ days also has $4$ events, and $5-0$ days has $2$ events. So a fall off of the numbers of events detected is apparently seen below $5$ days, but the statistics are not good enough to be sure with this time resolution. The pure numbers of events detected do not give any real information about the detection efficiency, however, as this depends on the predicted timescale distribution of actual microlensing events. In general, the number of real events are expected to rise at the low timescale end, due to the large predicted number of low mass objects, e.g. brown dwarfs, expected to exist. If this is correct then the fall off in detection efficiency at low timescale would be even steeper than shown by the timescale distribution of recovered events. 
 It would of course be possible to increase the number of low timescale events detected, for example by changing the stipulations of Cut 6 (see Section \ref{bump_sampling}), but this would be at the cost of decreasing the accuracy of the parameters recovered from some of the events, especially the peak flux (and hence the magnification). There is no easy answer to the question of where to draw the line in this situation- there will inevitably be compromises between the numbers of events detected and their security and the quality of their recovered parameters.
 The fairly low numbers of events recovered at low timescales (e.g. less than $5$ days) from the real lightcurves (although representing $10\%$ of the total found events) are a little disappointing, and further work is clearly required to make improvements in this area. As data quality generated from the ADAP improves with further development this will help as it is especially hard to find short events within noisy lightcurves.

\subsection{Variable distributions}

Several investigations of the lightcurves classified by the selection pipeline as ``variable'', which were good fits to a skew sinusoidal variable star model, were carried out, to discover what useful information could be extracted from the data. These included investigations of the reduced $\chi^2$, period and skewness distributions, along with cross-correlations between these two quantities and the with flux amplitudes of the variables,
and investigations of the spatial distribution of the variable candidates.
The results of this final investigation give much greater confidence that the variable candidates selected do belong to a physical population as their distribution has been shown to be highly non random in a self-consistent way.

In order to get better statistics, the pure variable fits were augmented by adding mixed fits in which the lensing component was too small to pass the criterion for selection as a lensing candidate. In addition, four separate pipeline runs were conducted using different, limits on the ratio between the fitted LT and PA flux amplitudes, since this was not yet well known. One of these runs allowed the ratio to ``freely'' vary (within wide limits), but in the others the ratio was fixed. Period and skewness distributions for the $16$ groups described above were investigated separately to check that they did not differ wildly between the categories and to ensure that it was therefore acceptable to combine the data into one larger group.
Two examples of these plots may be seen in Section \ref{variables}.
 When all categories were combined, the resulting data set contained $3160$ 
lightcurves, of which $80.6\%$ had reduced $\chi^2 < 5$.
The correlation between period and variable flux amplitude was investigated, and
a tight relationship was found in which the mean period at a given flux amplitude increased with flux amplitude. This relationship was fitted with a line whose equation was given in Equation \ref{logP_LT_flux_amp_relationship}.
Lower periods, below about $80$ days, did not fit this line well, and tended to
bend down steeply towards zero period without much change in flux amplitude.
It is thought that this might be due to the emergence of aliasing at shorter periods.
 When the data were transformed in order to remove this correlation, the
``true'' period distribution (at each constant flux amplitude) then emerged. This is shown in Figure \ref{logPhist_chisq_lt_5_after_trans}.
No significant correlation was found between the period and skewness distributions.

The next investigation made was of the spatial distribution of the $2546$ better fitted
 (reduced $\chi^2 < 5$) variable and mixed candidates, also subdivided by flux amplitude.
The total distribution, using all the candidates, was clearly peaked towards the centre
 of the galaxy, as would be expected given the large increase in the density of stars as
 the middle of the galaxy is approached. Some reduction of the detected spatial number
 density might have been anticipated due to the increased difficulty in detecting
 variability against the increased Poisson noise caused by the large increase in
 background light from the centre of the galaxy, and indeed subdivision of the spatial
 distribution into bins by variable flux amplitude clearly showed that as the flux
 amplitude
decreased, the central regions became relatively more sparse. This demonstrates the way
 the incompleteness for variable stars varied with flux amplitude. The cross-over point
 between the spatial distribution being centrally peaked and centrally troughed was found
 to be at around $4-5$ ADU/s.
The radial shape of the distribution appeared to be elliptical, with a ratio of semi-axis
 lengths $\sigma_{x}/\sigma_{y} \sim2$ and this was initially confirmed by plotting a
 contour plot, Figure \ref{gridded_contour_plot}, which showed that the 
ellipse was oriented approximately horizontally in the LT field, rather than either in
 the same direction as the M31 disk or the lower angle anticipated from previous
 modelling \citep{1977ApJ...213..368S}. A more detailed investigation of the spatial distribution of the objects selected as variable star candidates was also performed in Section \ref{bulge_angle}.
 The first part of the investigation used
 radial and azimuthal binning of the variable star candidates to investigate the azimuthal
distribution as a function of radius. When the derived data points were plotted and a
 constant line fitted, the best fitting angle at which the azimuthal number distributions
 peaks was $161.3^{\circ}$($\pm1.61^{\circ}$fitting error), or, alternatively, $-18.7^{\circ}$.
 
The second part of the investigation was of the purely radial distribution of variable star candidates. Dividing the area once again into concentric annuli of equal area, and adjusting for 
areas which had incomplete data due to the edge of the field, the dependence of the number density with radius was plotted. This could be fitted reasonably with a de Vaucouleurs $r^{\frac{1}{4}}$ Law within the uncertainties caused by the fairly low statistics. The number density in the central region was found to decline roughly linearly
inside $r=20^{\prime\prime}$ which may be explained both by the true incompleteness of the survey due to increasing photon noise from the galaxy and by the systematic affect of masking out the very central region in the difference imaging. By subtracting this linear fit from the
de Vaucouleurs fit above, the functional form of the completeness in the central region
could be estimated if the (large) assumption was made of this function being equal to $1$ at (r$=20^{\prime\prime}$). This was done, and was shown in Figure \ref{modelled_completeness}.

 \section{Future Work}

 \subsection{Known Problems}

The annual gaps in temporal coverage due to Andromeda being below horizon in the Northern Hemisphere are really troublesome. The duration of the off season during which Andromeda is too close to the sun to observe is annoyingly similar to the period of variability of some variable stars, and this leads to a high chance of aliasing in the fitting, as shown by Figure \ref{LTflux_v_logP}. In turn, lack of unambiguous knowledge of the baseline leads to serious doubt as to whether the fitted microlensing component is just that or rather some aspect of the variable star that was insufficiently covered by the data on previous or subsequent occurrences. There seems to be no way to improve the data coverage problem, however, without expensive solutions such as a telescope in Earth orbit diametrically opposite the Earth.

The combined (Pac.+ var) fitting function is very flexible and does not always do what would be hoped for- it has the freedom to use the microlensing part to model long low bumps which may have a moderate area under bump which are probably due to stellar variability and may leave 
some sharp spikes (low $f_{dif}^2$ or area underneath curve), which are likely to be due to microlensing, unfitted. In this work the areas of the lightcurve in which the Paczy\'nski curve may be fitted have remained unlimited, but is a possibility worth considering to limit range of fitted $t_0$ to those areas of the lightcurve where $f_{dif}$ is concentrated? This would usually require the pre-fitting of a variable star fitting function, then a re-assessment of those areas of the lightcurve which are not a good fit to this. This could be done within the same iteration.

There is a remaining issue with coherent groups of deviant points in time, where the summed deviations are not sufficient to make $\chi^2$/d.o.f. greater than a reasonable threshold. Although this issue has been partially addressed by Cut 12, a few lightcurves with secondary deviations which are low enough to escape Cut 12 but still noticeable by eye may be seen in the final sample.
Since the modelling applied in this work only allows one lensing curve to be fitted
to the lightcurve, the existence of unexplained lightcurve bumps cast doubt on the veracity/appropriateness of the model, despite passing the cuts. $\chi^2$/d.o.f. is certainly not a sufficient criterion to distinguish ``good'' lightcurves, but it is hard to know where to draw the line as regards complicating the models to fit more complicated lightcurves. Possible examples of this might be
modelling binary lensing events, or perhaps even a nova plus a lensing event plus a variable star in the same lightcurve. Given sufficient lightcurves, these relatively rare collections of deviant points will occur anyway by chance, but the level of Cut 12 is set so that much fewer than one of the cut deviations would be expected in the $93240$ lightcurves analysed.
However, there may be some variable stars which have more ``spiky'' lightcurves than are currently modelled with the cosinusoid function, and these may possibly be mistaken for an additional lensing spike in the lightcurve given sufficiently unfortunate time sampling.

 Currently, the parameters in the DIA photometry pipeline are set to favour the 
detection of as many variable events as possible, at the expense of re-detecting some events many times. This conscious choice leads to a larger number of repeat and near-repeat lightcurves. It is possible that the balance between sensitivity to detection and robustness against finding repeat lightcurves may need to be examined again in the future, and perhaps a different compromise found.

 A very important area in this work is the error estimation on data points, especially if we are relying on $\chi^2$ in the peak region to select lightcurves. 
In the 2007 photometry, based on analysis of the ``short event'', ANG-06B-M31-01 and also the analysis of the $\chi^2$/d.o.f. distribution of the variable star candidates, it seems that either the errors were underestimated or there is unexplained non-random variation in the lightcurves.
In the event ANG-06B-M31-01, upward rescaling of the error bars in the lightcurve was found to be required to make the $\chi^2$/d.o.f. $\sim1.0$, and in the case of the variable star candidates analysed in Section \ref{variables}, a similar re-scaling of order $\sqrt{2.0}$ was apparently implied by the distribution of $\chi^2$/d.o.f. values found in the fitting (see Figure \ref{reduced_chisq_dist_3160_LCs}). It has since been discovered by \cite{Kerins09etalinprep} that indeed our data are not photon noise limited, and that our $\chi^2$/d.o.f. should be adjusted downwards by a factor of $({1.4}^2 =1.96)$ if they are to be compared with photon noise limited $\chi^2$/d.o.f.'s which are $\sim1$ if a good fit is achieved.

\subsection{Suggested future extensions}

 It might be desirable at some point to add more variable star templates, for example, 
fitting functions that replicate ``saw-tooth'' Cepheids or variables with ``M'' shaped 
lightcurves, for instance RV Tauri stars or some eclipsing binaries. This would enable investigation of the relative fractions of the different 
classes of variable stars in the totality of variable stars in M31 and help us learn more about 
their physical properties, for e.g. period and skewness.
 The current sinusoidal model acts as a promising ``proof of principle'' and shows that 
fitting model lightcurves can provide real information (over and above 
that obtained from periodicity analysis) about variable stars in M31.

It would also be possible to add a routine for more clearly distinguishing classical novae which can sometimes appear to have similar lightcurves to microlensing, especially given unfortunate temporal sampling of the data. In general, nova lightcurves rise more steeply than they fall, and this asymmetry could be modelled. Currently when a Paczy\'nski curve is fitted to what may be a nova lightcurve we usually expect an flux undershoot before the peak and an overshoot after peak. This may also be tested for.

A small but important change which has not yet been made would be to introduce divisions between variable signal to noise categories which would have a 
logarithmic spacing. If this was done the spacing of the categories would get smaller when the signal to noise gets lower. This would be useful as there are many more of these lower signal to noise lightcurves than the higher signal to noise ones and with better resolution it would be possible to get better information about the signal to noise distribution.

At some point it would be useful to re-visit the work done on variable stars using the 2007 photometry, using the most up to date $2008$ photometry. If similar overall conclusions may also be drawn from a different data set, then this would strengthen
the confidence in the results. Alternatively, if the conclusions so far drawn are not supported, the reasons why not would have to be investigated.

I believe my development of, and experimentation with, the ``Quality Factor'' indicator of the quality of lensing candidate may point to a
different way of thinking about candidate selection. Often a candidate which is quite impressive by eye will fail at least one of a set of rigid cuts. Therefore, a system of cuts calibrated so that only excellent candidates pass all the cuts will throw away many
good candidates which are nevertheless worthy of further investigation. In the regime where Angstrom sits, with significant amounts of stellar blending from variable stars, this is especially true.
Often the candidates that are thrown away because they failed a particular cut will have
values for the other cut parameters which are even better than the candidates which pass all the cuts. Clearly it is not possible to go all the way down the path of using the accumulated total or product of all the cut parameters as the value upon which selection is based, as in that case one extremely good cut parameter could outweigh all the others. There must clearly remain some minimum standards. However, in this work it has been experimented with keeping these as low as possible without being unreasonable, and then using more ``fuzzy'' methods such as the Quality Factor to decide the order of priority among the remaining candidates. Another line of thought which follows on from this might be that candidate selection would be an ideal task for neural networks, which are very good at producing 
quick answers to the analysis of large quantities of data, once properly trained. This would be a way of getting away from the inevitable involvement of a human being who fairly arbitrarily decides what cut levels to set and whether a particular candidate is ``acceptable'', or ``good'', which are themselves fuzzy concepts.
It seems that in the ``black art'' of microlensing candidate selection there is no agreed set of cuts to use, and hence each collaboration uses their own magic formula
to select events. In some cases, for e.g. POINT AGAPE, the disagreement over what cuts to use is so severe that the collaboration splits into two or even three camps
which do their own independent candidate selection. It would be useful if all M31 collaborations decided a universally agreed set of cuts. This would mean that one could compare like with like, and say on a standard basis whether one collaboration had actually found more events than another, even without having to do all the subsequent Monte Carlo simulations, which are in themselves collaboration-dependent.

It may be beneficial to check for lightcurve copies before running the main pipeline.
Previously, such as in the $2007$ photometry, this was not such a big issue as repeated events
were only a small fraction of the total. In the $2008$ photometry, however,
the fraction of lightcurves selected by the ``loose'' cuts which were a copy, either exact or very similar, to another lightcurve was nearer $50\%$. Some photometric objects were found to have up to $10$ lightcurves associated with them.
This resulted in an approximate doubling of the running time for the pipeline. This in turn implies it may be now be worthwhile developing an algorithm to check for repeat or near repeat events which would be run before the main pipeline. Only the lightcurve with the most data points in the group would be kept, and then it might be possible to make a further choice based on the mean error on the flux points in the lightcurve.

\section{Future of Microlensing investigations}

\cite{1995ApJ...455...44G} performed calculations suggesting that attempting a microlensing search 
towards M87 and the Virgo cluster in general would be technically possible (currently it would be necessary to use the Wide Field and Planetary Camera on the Hubble Space Telescope). It should be feasible either to search for high magnification ``spikes'' within M87 itself or to search in the intervals between the cluster galaxies for ``Intra-cluster MACHO's" (ICMs) \citep{1995ApJ...455...44G}. No team has so far successfully proposed this experiment. For the even farther future, the Coma Cluster \citep{1996ApJ...470..201G} is an even richer cluster that would be even more interesting in which to search for ICMs, but, being about 5 times further away than the Virgo Cluster, is beyond the capabilities of any current telescope. However, with the advent of new space telescopes with larger dish sizes, such as the James Webb Space telescope, successful microlensing observations at these amazing distances may even become possible.

Another prospect for the future might be the possible observation of cosmic strings,
if they exist. These are in theory infinitely thin objects with a finite length and a finite mass, which would cause gravitational lensing if they sat, or moved between us and a stellar source. It has been proposed in \cite{2003MNRAS.343..353S} that
a cosmic string has already been seen as the most likely cause of a double galaxy.
Because strings are predicted to cause an unmagnified pair of images when sitting directly in front of a source, the lightcurve would rise linearly to an exact factor of two amplification of the original flux and then return to the original value, if the string was straight and was moving fast enough relative to the source that the changing flux could be detected. This characteristic lightcurve and the fact that they should produce a clear double image with no detectable foreground galaxy lens at all make it conceivable that cosmic strings could be detected using ``macro''
lensing or even with microlensing. This would be a very important discovery for fundamental physics.

 A more immediate prospect for the near future might be the routine detection of 
planets in another galaxy. The most obvious possibility, other than the Magellanic Clouds, is M31, as the nearest massive galaxy to the Milky Way. The possibility of this has been fairly extensively studied \citep{2006ApJ...650..432C,2007NCimB.122..471D}, and, as pointed out by \cite{2009arXiv0906.1050J}, may already have been observed by the POINT-AGAPE survey
as first described as the binary lensing candidate PA-00-S5 in \cite{2005A&A...443..911C}.
The discovery of planets and planetary systems in our own Galaxy has become relatively routine in the last few years, with over two hundred planetary systems known, each containing up to five planets. It would be very exciting if this process of discovery could also begin in the next few years in M31. There seem few technical barriers to this, except for waiting for the perfect binary lightcurve with sufficient temporal sampling to come along. Towards this end, high cadence surveys such as Angstrom have a much better chance of being able to detect and model any planetary deviations successfully, so we look forward to the future with anticipation.

\appendix
\chapter{Glossary}
\label{Glossary}

2MASS = Two Micron All Sky Survey

AGAPE = Andromeda Galaxy and Amplified Pixels Experiment

Angstrom = Andromeda Galaxy STellar Robotic Microlensing project

AAS = Angstrom Alert System

ADAP = Angstrom Data Analysis Pipeline

BSQF = Bump Sample Quality Factor

CCD = Charge Coupled Device

CDR = $\chi^2$ Difference Ratio

Colour-Magnitude Diagram = CMD

Columbia/VATT = Colombia/Vatican Advanced Technology Telescope

DIA = Difference Image Analysis

EROS = Exp\'{e}rience pour la Recherche d'Objets Sombres

FTN = Faulkes Telescope North

GLRG = Giant Luminous Red Galaxy

HST = Hubble Space Telescope

HWHM = Half-Width Half Maximum

ICM = Intra-Cluster MACHO's (see definition of ``MACHO'' below)

IMF = Initial Mass Function

JWST = James Webb Space Telescope

LC = lightcurve

LMC = Large Magellanic Cloud

LSN = Lensing Signal to Noise factor

LT = Liverpool Telescope

MACHO = MAssive Compact Halo Object 

MDM = Michigan Dartmouth M.I.T.

MS = Main Sequence
 
MEGA = Microlensing Exploration of the Galaxy and Andromeda survey

NainiTal = Astronomical survey; also a town in India

OGLE = Optical Gravitational Lensing Experiment

PA = POINT AGAPE survey of M31

POINT AGAPE = Pixel-lensing Observations with the Isaac Newton Telescope, Andromeda Galaxy Amplified Pixels Experiment

PSF = Point Spread Function

QF = Quality Factor

RATCam = Robotic Automated Telescope CAMera (on LT)

RGB = Red Giant Branch 

SDSS = Sloan Digital Sky Survey

SLOTT AGAPE = Systematic Lensing Observation at Toppo Telescope Andromeda Galaxy and Amplified Pixels Experiment

SMC = Small Magellanic Cloud

SupIRCam = SUPERNova InfraRed Camera (on LT)

WeCaPP = Wedelstein Calar alto Pixellensing Project

WFPC2 = Wide Field and Planetary Camera II (an instrument on the Hubble Space telescope)

\addtocontents{toc}{\protect\vspace{2ex}}  
\addcontentsline{toc}{chapter}{Bibliography}
\bibliographystyle{lxh}
\bibliography{bibliography.bib}
\end{document}